\documentclass[11pt]{article}
\pdfoutput=1

\usepackage[usenames,dvipsnames,svgnames,table]{xcolor} % use color
\usepackage[obeyspaces,hyphens,spaces]{url}
\usepackage{jcapmod}
\usepackage{verbatim}

\usepackage{booktabs}
\usepackage[english]{babel}
\usepackage{amsmath, amssymb, amsbsy, amstext, amsthm, simplewick}
\usepackage{hyperref}
\usepackage{graphicx}
\usepackage{amsfonts}
\usepackage{upgreek}
\usepackage{framed}
\usepackage{tensor}
\usepackage{pifont}
\usepackage{latexsym, mathrsfs}
\usepackage{array}
\usepackage{hyperref}
\usepackage{xspace}
\usepackage{longtable}
\usepackage{multirow}
\usepackage{cases}
\usepackage{empheq}

\usepackage{bm}
\usepackage{bbm}

\usepackage{afterpage}
\usepackage{setspace}
\usepackage{braket}

\usepackage[load-configurations=astronomy, range-units=brackets, range-phrase=-, per-mode=reciprocal, mode=math]{siunitx}
\usepackage{threeparttable}
\usepackage{subfig}

\usepackage{ragged2e}

\usepackage{hhline}
\usepackage{array}
\newcolumntype{P}[1]{>{\centering\arraybackslash}p{#1}}
\usepackage{booktabs}
\usepackage{tikz}
\usetikzlibrary{calc,shadings,patterns,tikzmark,fadings}
\definecolor{Blue}{rgb}{0.25, 0.41, 0.88}
\definecolor{Red}{rgb}{0.92,0.,0.}
\definecolor{darkorange}{rgb}{1.0,0.549,0.}
\definecolor{cobalt}{RGB}{44, 98, 120}
\definecolor{Mathematica1}{rgb}{0.368417, 0.506779, 0.709798}
\definecolor{Mathematica2}{rgb}{0.880722, 0.611041, 0.142051}
\definecolor{Mathematica3}{rgb}{0.560181, 0.691569, 0.194885}
\definecolor{Mathematica4}{rgb}{0.922526, 0.385626, 0.209179}
\definecolor{Mathematica5}{rgb}{0.528488, 0.470624, 0.701351}
\definecolor{Mathematica6}{rgb}{0.772079, 0.431554, 0.102387}
\definecolor{Mathematica7}{rgb}{0.363898, 0.618501, 0.782349}
\definecolor{Mathematica8}{rgb}{1, 0.75, 0}
\definecolor{Mathematica9}{rgb}{0.647624, 0.37816, 0.614037}
\definecolor{plotBlue}{RGB}{94, 130, 181}
\definecolor{plotRed}{RGB}{233, 85, 54}
\definecolor{plotGreen}{RGB}{142, 176, 50}
\definecolor{plotPurple}{RGB}{135, 120, 178}

\newcolumntype{C}[1]{>{\centering\let\newline\\\arraybackslash\hspace{0pt}}m{#1}}

\def\x{{\bf x}}

\makeatletter
\newlength{\apb@width}
\newcommand{\autoparbox}[2][c]{\settowidth{\apb@width}{#2}\parbox[#1]{\apb@width}{#2}}

\makeatother

\makeatletter
\newsavebox\myboxA
\newsavebox\myboxB
\newlength\mylenA

\newcommand*\xoverline[2][0.75]{% From Danie Els at https://tex.stackexchange.com/questions/22100/the-bar-and-overline-commands
    \sbox{\myboxA}{$\m@th#2$}%
    \setbox\myboxB\null% Phantom box
    \ht\myboxB=\ht\myboxA%
    \dp\myboxB=\dp\myboxA%
    \wd\myboxB=#1\wd\myboxA% Scale phantom
    \sbox\myboxB{$\m@th\overline{\copy\myboxB}$}%  Overlined phantom
    \setlength\mylenA{\the\wd\myboxA}%   calc width diff
    \addtolength\mylenA{-\the\wd\myboxB}%
    \ifdim\wd\myboxB<\wd\myboxA%
       \rlap{\hskip 0.5\mylenA\usebox\myboxB}{\usebox\myboxA}%
    \else
        \hskip -0.5\mylenA\rlap{\usebox\myboxA}{\hskip 0.5\mylenA\usebox\myboxB}%
    \fi}
\makeatother

\numberwithin{equation}{section}

\def\beq{\begin{equation}}
\def\eeq{\end{equation}}

\def\bea{\begin{eqnarray}}
\def\eea{\end{eqnarray}}

 \def\be{\begin{equation}}
 \def\ee{\end{equation}}
 \def\bes{\begin{eqnarray}}
 \def\ees{\end{eqnarray}}
\def\beq{\begin{equation}}
\def\eeq{\end{equation}}
\def\bea{\begin{eqnarray}}
\def\eea{\end{eqnarray}}

\DeclareRobustCommand{\SkipTocEntry}[4]{}

\usepackage{colortbl}
\definecolor{blue2}{cmyk}{1, 0.1, 0.1, 0.1}

\definecolor{pyBlue}{RGB}{31, 119, 180}
\definecolor{pyRed}{RGB}{214, 39, 40}
\definecolor{pyGreen}{RGB}{44, 160, 44}
\definecolor{pyBlue2}{RGB}{0, 111, 237}
\definecolor{pyRed2}{RGB}{224, 52, 36}

\begin{document}

\pagenumbering{roman}
\begin{titlepage}
\baselineskip=14.5pt \thispagestyle{empty}

\bigskip\

\vspace{0cm}
\begin{center}
{\fontsize{40}{40}\selectfont  \bfseries \textcolor{Sepia}{:${\cal T}$HE  ${\cal C}$OSMOLOGICAL ${\cal O}$TOC:}}\\ 
%OTOC in the Sky
\vspace{0.3cm}
{\fontsize{15.1}{12.9}\selectfont  \bfseries \textcolor{Sepia}{A new proposal for quantifying auto-correlated random non-chaotic primordial fluctuations }}
\end{center}
\vspace{0.01cm}
\begin{center}
{\fontsize{15}{15}\selectfont Sayantan Choudhury ${}^{a,b}$
		\footnote{{\it  \textcolor{blue}{ This project is the part of the non-profit virtual international research consortium
“Quantum Structures of the Space-Time \& Matter (QASTM)".}} ${}^{}$}	} 
\end{center} 

%\vspace{0.25cm}
\begin{center}
\vskip1pt
\textit{${}^{a}$School of Physical Sciences,\\
National Institute of Science Education and Research, Bhubaneswar, Odisha - 752050, India}\\
		\textit{${}^{b}$Homi Bhabha National Institute, Training School Complex, Anushakti Nagar, Mumbai - 400085, India}\\
		\vskip8pt
	\text{Email:~sayantan.choudhury@niser.ac.in, sayanphysicsisi@gmail.com }

%\vskip8pt
%\textsl{$^2$ SISSA, Via Bonomea 265, 34136 Trieste, Italy}

\end{center}

\vspace{0.09cm}
\hrule \vspace{0.09cm}
\begin{center}
\noindent {\bf Abstract}\\

\end{center} 
The underlying physical concept of computing out-of-time-ordered correlation (OTOC) is a significant new tool within the framework of quantum field theory,  which now-a-days is treated as a measure of random fluctuations.  In this paper,  by following the canonical quantization technique we demonstrate a computational method to quantify the two different types of Cosmological auto-correlated OTO functions during the epoch when the non-equilibrium features dominates in Primordial Cosmology.  In this formulation,  two distinct dynamical time scales are involved to define the quantum mechanical operators arising from cosmological perturbation scenario.  We have provided detailed explanation regarding the necessity of this new formalism to quantify any random events generated from quantum fluctuations in Primordial Cosmology.  We have performed an elaborative computation for the two types of two-point and four-point auto-correlated OTO functions in terms of the cosmological perturbation field variables and its canonically conjugate momenta to quantify random auto-correlations in the non-equilibrium regime.  For both the cases we found significantly distinguishable non-chaotic,  but random behaviour in the OTO auto-correlations,  which was not pointed before in this type of studies.  Finally,  we have also demonstrated the classical limiting behaviour of the mentioned two types of auto-correlated OTOC functions from the thermally weighted phase space averaged Poisson Brackets,  which we found exactly matches with the large time limiting behaviour of the auto-correlations in the super-horizon regime of the cosmological scalar mode fluctuation. 
 
\vskip7pt
\hrule
\vskip7pt

\text{Keywords:~~Cosmology beyond the standard model,  Quantum Dissipative Systems,}\\ \text{Stochastic Processes,  Effective Field Theories, 
Non-equilibrium Quantum Field Theory.}

\end{titlepage}

\thispagestyle{empty}
\setcounter{page}{2}
\begin{spacing}{1.03}
\tableofcontents
\end{spacing}

\clearpage
\pagenumbering{arabic}
\setcounter{page}{1}

\clearpage

%%%%%%%%%%%%%%%%
\section{Introduction}
\label{sec:0}
%%%%%%%%%%%%%%%%
The underlying physical concept of out-of-time ordered correlation (OTOC) functions \cite{Maldacena:2015waa, Hashimoto:2017oit, Chakrabarty:2018dov,Chaudhuri:2018ymp,Chaudhuri:2018ihk, Haehl:2017eob, Akutagawa:2020qbj, Bhagat:2020pcd, Romero-Bermudez:2019vej} within the framework of quantum field theory is considered to be a very strong theoretical tool to describe random phenomena in the quantum regime,  particularly the phenomena of quantum chaos. This concept was first introduced within the framework of superconductivity to explicitly compute the vertex correction factor of current in ref.~\cite{Larkin:1969}.  But for the last few years in the various contexts of gravitational paradigm this computational tool have been frequently used to describe various random-chaotic phenomena very successfully.  One can physically interpret OTO functions as a quantum analogue of the usual classical sensitivity against very small random fluctuations in the initial conditions.  

To describe this in a better way,  let us consider two quantum operators in different time scales,  $X_1(t_1)$,  $X_1(t_2)$,   its canonically conjugate momenta $\Pi_{X_1}(t_1)$ and $\Pi_{X_1}(t_2)$,  and using these operators the auto-correlated OTOs are defined as:
\bea &&\underline{\textcolor{red}{\bf Auto-Correlated~ OTOC_1}}\nonumber\\
&&\underline{\textcolor{blue}{\bf Two-point~function:}}\nonumber\\
&&~~~~~~~~~~~~~~Y_1(t_1,t_2):\equiv-\langle \left[X_1(t_1),X_1(t_2)\right]\rangle_{\beta}=-\frac{1}{Z(t_1)}~{\rm Tr}\left[\exp\left(-\beta H(t_1)\right)~\left[X_1(t_1),X_1(t_2)\right]\right]~,~~~~~~\\
&&\underline{\textcolor{blue}{\bf Four-point~function:}}\nonumber\\
&&~~~~~~~~~~~~~~C_1(t_1,t_2):\equiv-\langle \left[X_1(t_1),X_1(t_2)\right]^2\rangle_{\beta}=-\frac{1}{Z(t_1)}~{\rm Tr}\left[\exp\left(-\beta H(t_1)\right)~\left[X_1(t_1),X_1(t_2)\right]^2\right]~,~~~~~~\\
&&\underline{\textcolor{red}{\bf Auto-Correlated~ OTOC_2}}\nonumber\\
&&\underline{\textcolor{blue}{\bf Two-point~function:}}\nonumber\\&&~~~~~~~~~~~~~~Y_2(t_1,t_2):\equiv-\langle \left[\Pi_{X_1}(t_1),\Pi_{X_1}(t_2)\right]\rangle_{\beta}=-\frac{1}{Z(t_1)}~{\rm Tr}\left[\exp\left(-\beta H(t_1)\right)~\left[\Pi_{X_1}(t_1),\Pi_{X_1}(t_2)\right]\right]~,\\
&&\underline{\textcolor{blue}{\bf Four-point~function:}}\nonumber\\&&~~~~~~~~~~~~~~C_2(t_1,t_2):\equiv-\langle \left[\Pi_{X_1}(t_1),\Pi_{X_1}(t_2)\right]^2\rangle_{\beta}=-\frac{1}{Z(t_1)}~{\rm Tr}\left[\exp\left(-\beta H(t_1)\right)~\left[\Pi_{X_1}(t_1),\Pi_{X_1}(t_2)\right]^2\right],~~~~~~~\eea
where the quantum partition function at finite temperature can be expressed as:
\bea &&\underline{\textcolor{red}{\bf Quantum~Partition~function:}}\nonumber\\
&&~~~~~~~Z(t_1)={\rm Tr}\left[\exp\left(-\beta H(t_1)\right)\right]~~~~{\rm where}~~\beta=\frac{1}{T}~~~{\rm with}~~k_B=1,~~\hbar=\frac{h}{2\pi}=1~,~~~~~~\eea
where $H(t_1)$ is the Hamiltonian of the quantum system under consideration which is defined at the time scale $t_1$.  Here in the quantum regime  the effect of perturbation by the quantum operators $\Pi_{X_1}(t_1)$ and $\Pi_{X_1}(t_2)$ on the measurement of the quantum operators $X_1(t_1)$ and $X_1(t_2)$ on later time scales.  In this theoretical construction we assume that the corresponding one-point functions are trivially vanish for both of the operators:
\bea &&\underline{\textcolor{blue}{\bf One-point~function_1:}}\nonumber\\
&&~~~~~~~~~~~~~~~~~~\langle {X_1}(t_1) \rangle_{\beta}=\frac{1}{Z(t_1)}{\rm Tr}\left[\exp\left(-\beta H(t_1)\right){X_1}(t_1)\right]=0,\\
&&~~~~~~~~~~~~~~~~~~\langle {X_1}(t_2) \rangle_{\beta}=\frac{1}{Z(t_2)}{\rm Tr}\left[\exp\left(-\beta H(t_2)\right){X_1}(t_2)\right]=0.\\
 &&\underline{\textcolor{blue}{\bf One-point~function_2:}}\nonumber\\
&&~~~~~~~~~~~~~~~~~~\langle \Pi_{X_1}(t_1) \rangle_{\beta}=\frac{1}{Z(t_1)}{\rm Tr}\left[\exp\left(-\beta H(t_1)\right)\Pi_{X_1}(t_1)\right]=0,\\
&&~~~~~~~~~~~~~~~~~~\langle \Pi_{X_1}(t_2) \rangle_{\beta}=\frac{1}{Z(t_2)}{\rm Tr}\left[\exp\left(-\beta H(t_2)\right)\Pi_{X_1}(t_2)\right]=0.\eea

  The huge applicability of OTOC as a strong theoretical probe of gravity dual theories in terms of AdS/CFT \cite{Maldacena:1997re,Aharony:1999ti} has attracted a lot of attention recently in various works.  Many such examples can be given that features the importance of the OTO functions in AdS/CFT \cite{Maldacena:1997re,Aharony:1999ti}.  Particularly the study of shock waves \cite{Shenker:2013pqa,Roberts:2014isa,Shenker:2013yza,Cotler:2016fpe,Stanford:2014jda} in black hole physics is one of the remarkable examples,  which can be understood by various types of geometries within the framework of AdS/CFT.  Also it is important to note that, this particular study finally led to maximum saturation bound on Lyapunov exponent in the quantum regime.  In gravitational paradigm,  this bound is explained in terms of the well known red shift factor which is defined near the black hole event horizon having a Hawking temperature.  In this context,  the {\it Sachdev-Ye-Kitaev (SYK)} model \cite{Klebanov:2016xxf,Maldacena:2016hyu,Choudhury:2017tax,Klebanov:2018fzb,Bulycheva:2017ilt,Kim:2019upg,Gurau:2012vk,Gurau:2012vu,Gurau:2017qya,Gurau:2019qag,Benedetti:2015ara,Fu:2016vas,Witten:2016iux,Li:2017hdt,Turiaci:2017zwd,Rosenhaus:2018dtp,Gross:2017aos,Gross:2017hcz,Gross:2017vhb,Polchinski:2016xgd,Dhar:2018pii,Mandal:2017thl,Gaikwad:2018dfc,Krishnan:2016bvg,Krishnan:2017lra,Krishnan:2017ztz,Sorokhaibam:2019qho,Bhattacharya:2017vaz,Bhattacharya:2018fkq,Samui:2020jli,Das:2020kmt,Das:2017wae,Das:2017hrt,Das:2017eiw,Das:2017pif} is the most famous example which explains the quantum mechanical features of $0+1$ dimensional Majorana fermions having inherent infinitely long disorder.  From the past understanding from various works it is a very well known fact that any gravitational paradigm  having their own CFT dual are described by the strongly coupled quantum field theories.   
 
 Now,  we will point towards an unusual application of computing auto-correlations of OTO function within the framework of Primordial Cosmology,  which we have proposed in ref. ~\cite{Choudhury:2020yaa}  to describe the cosmological cross-correlated OTO function.  In Primordial Cosmology the most significant quantity that we study are the $N$-point functions of the scalar-tensor metric fluctuations.  Here we consider the {\it Friedmann-Lemaître-Robertson-Walker (FLRW)} spatially flat background metric to describe our homogeneous,  isotropic and expanding observed universe.  Using this metric cosmological perturbation theory can be studied to explain the origin of the mentioned scalar-tensor fluctuations. These quantum fluctuations are very fundamental objects from which one can compute all $N$-point functions in Primordial Cosmology \cite{Maldacena:2002vr,Maldacena:2011nz,Senatore:2009gt,Senatore:2008wk,Baumann:2019oyu,Baumann:2020dch,Baumann:2020ksv,Baumann:2009ds,Baumann:2018muz,Meerburg:2019qqi,Senatore:2016aui,Creminelli:2005hu,Creminelli:2006rz,Smith:2009jr,Senatore:2013roa,Green:2013rd,Smith:2015uia,Flauger:2016idt,Creminelli:2003iq,Creminelli:2004yq,Creminelli:2010qf,Assassi:2012zq,Behbahani:2012be,Green:2020whw,Creminelli:2006gc,Choudhury:2015yna,Bordin:2016ruc,Mirbabayi:2014zpa}.  Additionally,  these quantum fluctuations can be treated as a seed of cosmological perturbations to describe the formation of galaxy and cluster formation at present day.  Though these correlators are described in a same time scale at the late time scale of our universe and describe the equilibrium configuration.  
 \begin{figure*}[htb]
  \includegraphics[width=17cm,height=11cm]{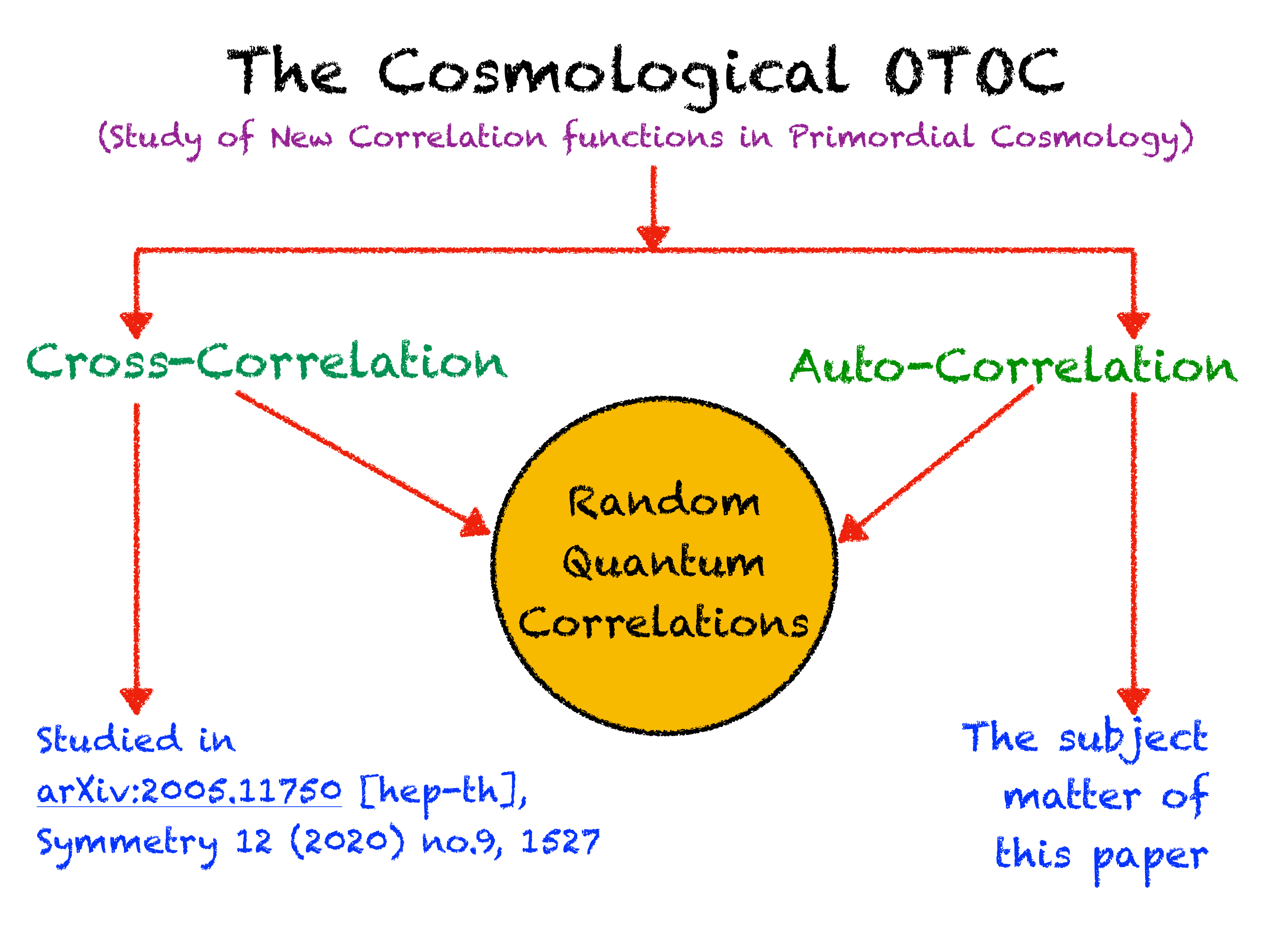}
  \caption{Schematic diagram representing the different role of Cosmological OTOC within the framework of Primordial Cosmology. } 
  \label{fig:1}
\end{figure*} 
 \begin{figure*}[htb]
  \includegraphics[width=17cm,height=11cm]{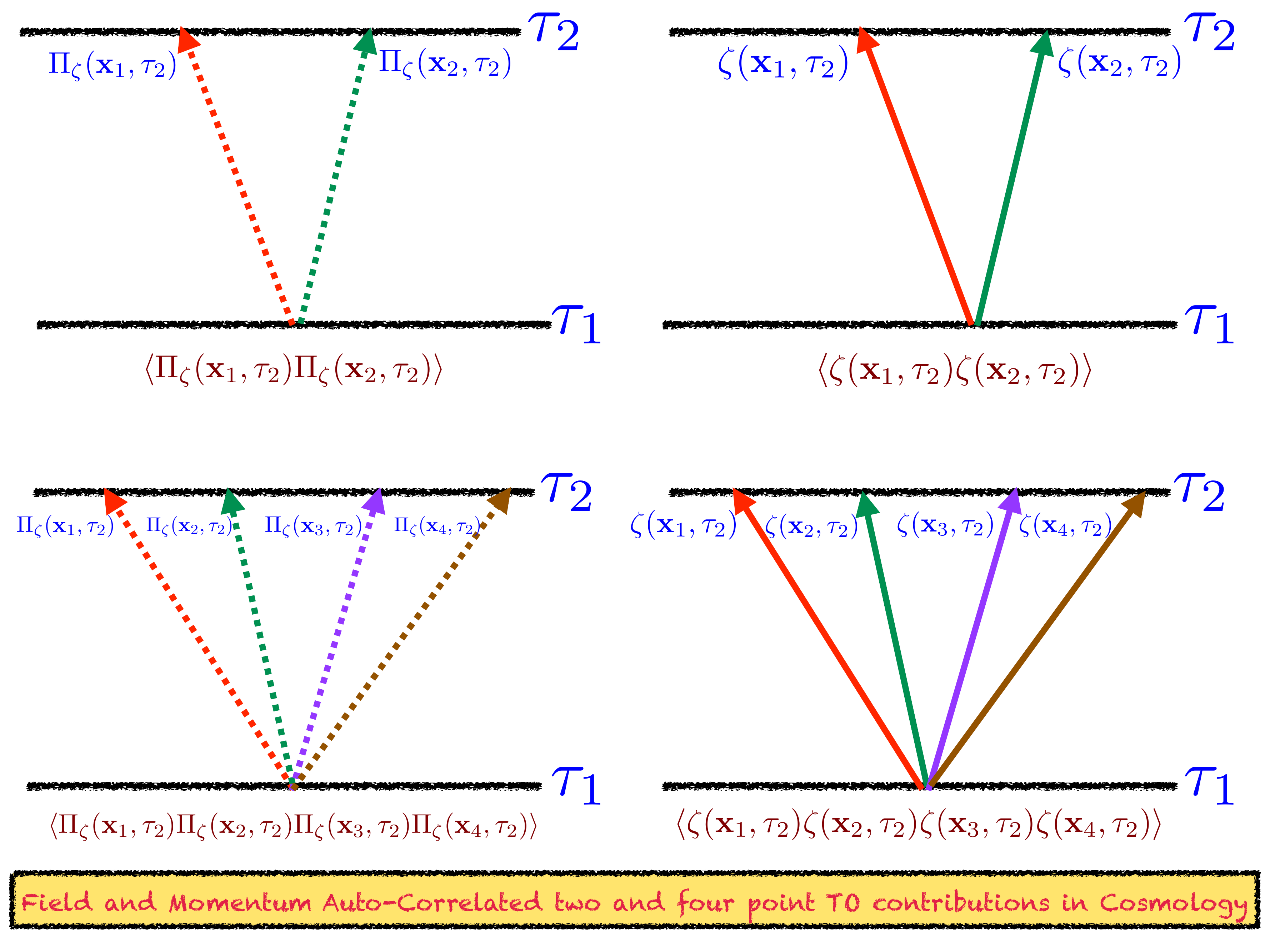}
  \caption{Diagrammatic representation of two-point \& four-point time ordered auto-correlators for the momentum and field within the framework of Primordial Cosmology. } 
  \label{fig:1a}
\end{figure*} 
\begin{figure*}[htb]
  \includegraphics[width=17cm,height=11cm]{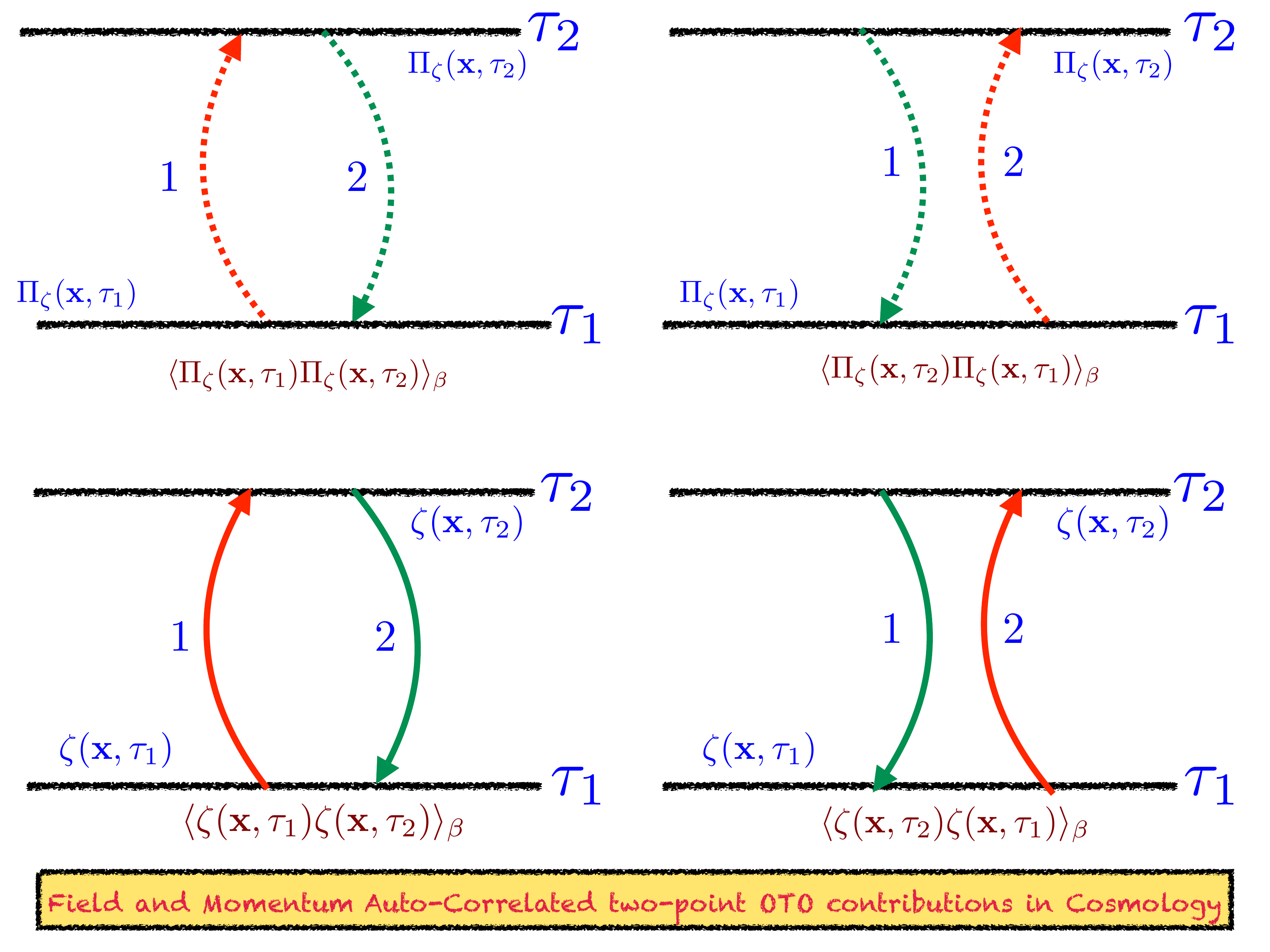}
  \caption{Diagrammatic representation of two-point OTO auto-correlators for the momentum and field within the framework of Primordial Cosmology. } 
  \label{fig:1b}
\end{figure*} 
\begin{figure*}[htb]
  \includegraphics[width=17cm,height=11cm]{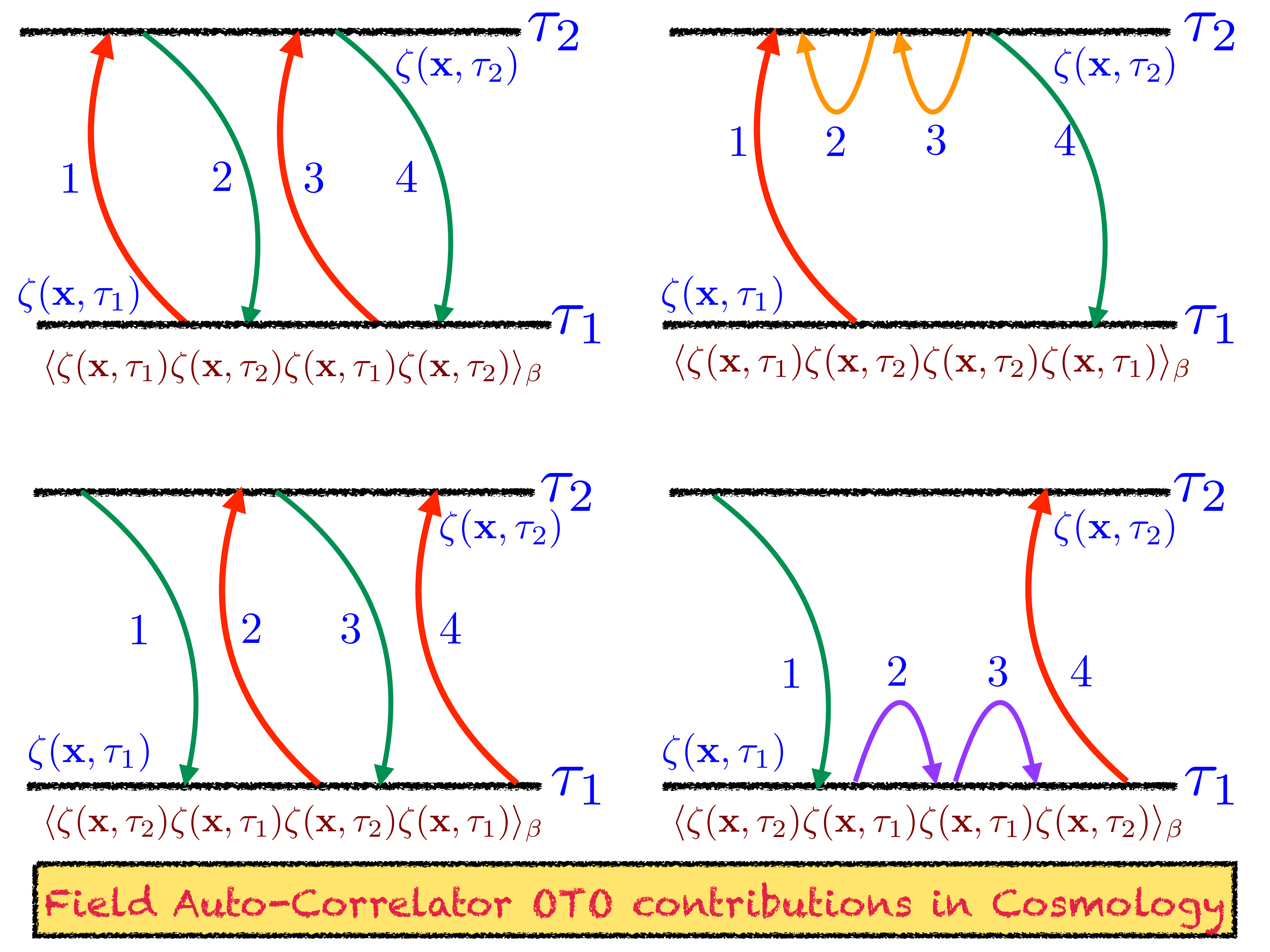}
  \caption{Diagrammatic representation of four-point OTO auto-correlators for the field within the framework of Primordial Cosmology. } 
  \label{fig:1c}
\end{figure*} 
\begin{figure*}[htb]
  \includegraphics[width=17cm,height=11cm]{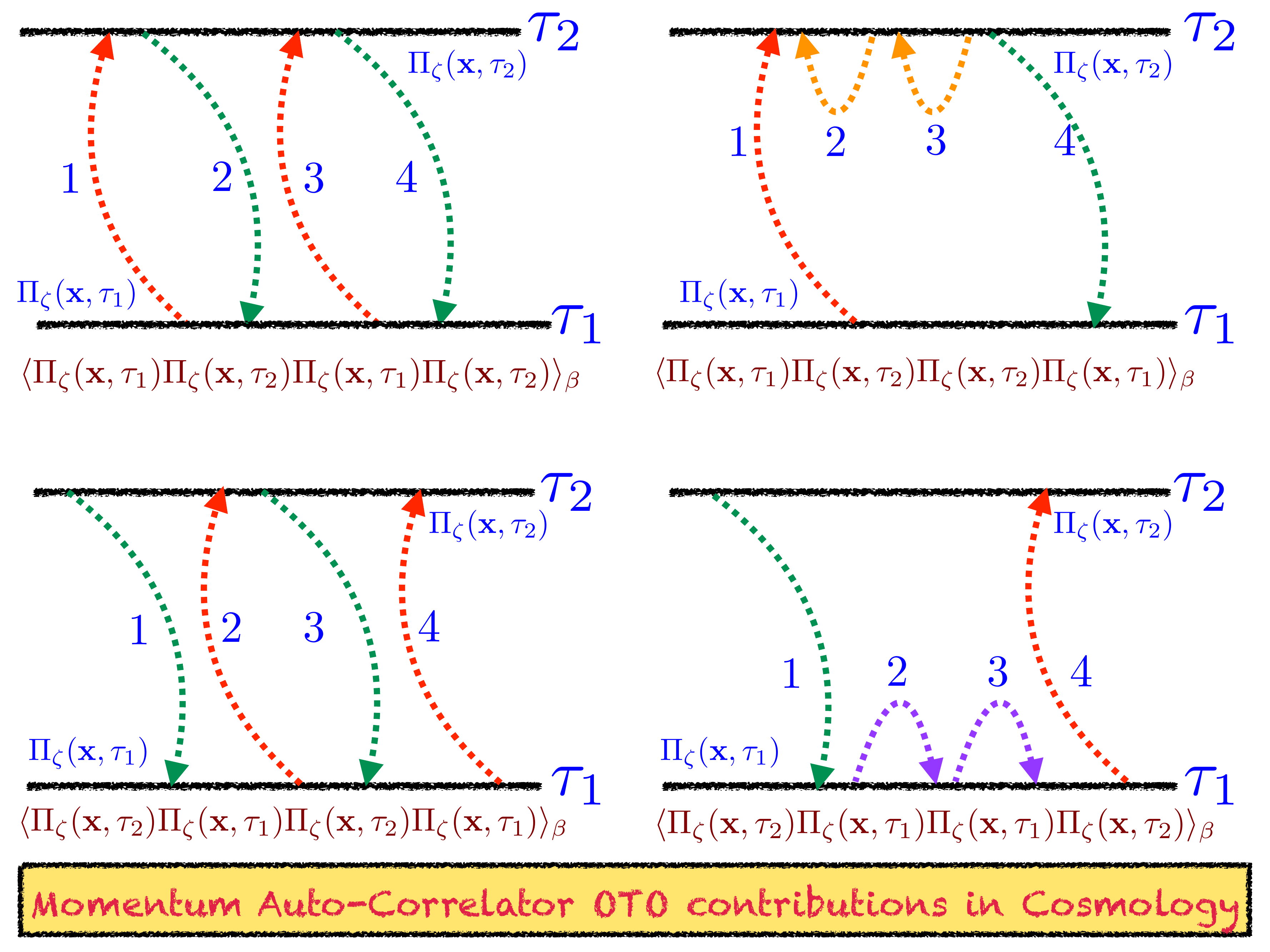}
  \caption{Diagrammatic representation of four-point OTO auto-correlators for the momentum within the framework of Primordial Cosmology. } 
  \label{fig:1d}
\end{figure*} 
Then one can immediately ask about question regarding the possible options left to explore the physics of primordial quantum fluctuations from the studies of these mentioned correlators:
 \begin{itemize}
 \item \underline{\textcolor{red}{\bf First~possibility:}}\\
 The first possibility is to include future observational aspects to verify various theoretical proposals in cosmology.  These theoretical proposals are appended below:
 \begin{enumerate}
 \item Primordial gravitational waves and tensor-to-scalar ratio \cite{Baumann:2015xxa,Baumann:2014nda,Baumann:2009mq,Choudhury:2013iaa,Choudhury:2013jya,Choudhury:2014kma,Choudhury:2014wsa,Choudhury:2014sua,Choudhury:2015pqa,Choudhury:2015hvr,Choudhury:2017glj,Choudhury:2017cos,Creminelli:2014nqa,Creminelli:2014wna,Cheung:2007st,Choudhury:2012yh,Choudhury:2003vr,Choudhury:2011jt,Choudhury:2011sq,Choudhury:2012ib,Choudhury:2013zna,Choudhury:2014sxa,Choudhury:2015yna,Mazumdar:2001mm,Panda:2007ie,Choudhury:2003vr,Ali:2010jx,Ali:2008ij,Panda:2006mw,Panda:2005sg,Chingangbam:2004ng,Panda:2010pj,Moniz:2009ax},
  \item Primordial non-Gaussianity \cite{Maldacena:2002vr,Maldacena:2011nz,Senatore:2009gt,Senatore:2008wk,Baumann:2019oyu,Baumann:2020dch,Baumann:2020ksv,Baumann:2009ds,Baumann:2018muz,Meerburg:2019qqi,Senatore:2016aui,Creminelli:2005hu,Creminelli:2006rz,Smith:2009jr,Senatore:2013roa,Green:2013rd,Smith:2015uia,Flauger:2016idt,Creminelli:2003iq,Creminelli:2004yq,Creminelli:2010qf,Assassi:2012zq,Behbahani:2012be,Green:2020whw,Creminelli:2006gc,Choudhury:2015yna,Bordin:2016ruc,Mirbabayi:2014zpa},
    \item Spectral running and scale dependence in primordial power spectrum \cite{Choudhury:2013jya,Choudhury:2013zna,Lidsey:2003cq,Zarei:2014bta,Li:2018iwg},
    \item New consistency relations \cite{Gruzinov:2004jx,Gong:2017wgx,Hui:2018cag,Choudhury:2015pqa,Choudhury:2014kma,Choudhury:2013iaa}.
    \end{enumerate}

 \item \underline{\textcolor{red}{\bf Second~possibility:}}\\
 The second possibility is probing of new physics by including significant features in Primordial Cosmology.  This can be done by incorporating the concept of out-of-equilibrium in primordial Cosmological quantum correlation functions.  Though before this work and previous work done by us in ref. ~\cite{Choudhury:2020yaa} it was not at all mentioned in the corresponding literature to how to compute and finally quantify the cosmological correlators at out-of-equilibrium.  It is also not even clear that what exact quantity one need to compute to give an estimations of these mentioned new correlators.  After this work we are hopeful that at least technically we have provided some correct estimation of these non-equilibrium quantum correlators within the framework of primordial cosmology.  Now to connect with the real cosmological scenario let us mention about some aspects which are appearing in the time line of our universe where one can implement the presented methodology:
 \begin{enumerate}
 \item  Stochastic particle production during inflation \cite{Choudhury:2018rjl,Choudhury:2018bcf,Choudhury:2020yaa,Amin:2015ftc,Garcia:2020mwi,Garcia:2019icv},
 \item  Warm~Inflationary framework \cite{Deshamukhya:2009wc,Berera:1995ie,Berera:2006xq},
 \item  Reheating \cite{Giblin:2017qjp,Kofman:1997yn,Kofman:1994rk,Choudhury:2011rz,Panda:2009ji},
\item Stochastic inflationary framework \cite{Matacz:1996gk,Pattison:2019hef,Ando:2020fjm,Vennin:2020kng,Noorbala:2018zlv},
 \item  Quantum quench \cite{Mandal:2015kxi,Kulkarni:2018ahv,Mandal:2013id,Banerjee:2019ilw,Das:2019cgl,Das:2019qaj,Caputa:2017ixa,Das:2016lla,Das:2015jka,Das:2014hqa,Das:2014jna,Basu:2013soa} in Cosmology.
 \end{enumerate} 
 \end{itemize}
 in fig.~(\ref{fig:1}),  we have shown a schematic diagram representing the different role of Cosmological OTOC within the framework of Primordial Cosmology.  Then further in fig.~(\ref{fig:1a}),  we have shown the diagrammatic representation of two-point \& four-point time ordered auto-correlators for the momentum and field.  Next in fig.~(\ref{fig:1b}),  we have shown the diagrammatic representation of two-point OTO auto-correlators for the momentum and field within the framework of Primordial Cosmology. 
Finally,  in fig.~(\ref{fig:1c}) and fig.~(\ref{fig:1d}),  we have shown the diagrammatic representation of four-point OTO auto-correlators for the field and momentum within the framework of Primordial Cosmology. 
 Now,  we mention the major highlights of our obtained results in this paper~\footnote{Note: Except the small details presented for cosmological perturbation theory for scalar modes,  the rest of the computations presented in this paper is completely new.  Though we have presented the detail in terms of the massless, partially massless and massive scalar fields which is not commonly discussed in the cosmology text book literature. Because of this reason the partial details of the computations are provided in the text portion of the paper and the rest of the details are presented in the Appendices.  We believe this will be helpful for general readers. }:
 \begin{itemize}
 \item \underline{\textcolor{red}{\bf Highlight~I:}}\\
 The computation method presented in this paper helps us to quantify the quantum auto-correlations within the framework of Primordial Cosmology with random fluctuations.  In this article we have computed the expressions for the two-point and four-point auto-correlated cosmological OTO functions in the quantum regime.  These computed expressions are completely new and the detailed discussions will be helpful to understand the underlying physical problem that we have studied in this paper.
 
 \item \underline{\textcolor{red}{\bf Highlight~II:}}\\
 We have additionally studied the classical limits of the two-point and four-point auto-correlated cosmological OTO functions in terms of the phase space thermal weighted average of classical Poisson brackets.  Most importantly this computation will provide the non-standard random behaviour,  which are perfectly consistent with the expectations from the present scenario.  Last but not the least this particular computation will be very helpful to understand the super-horizon classical limiting behaviour of the computed auto-correlated OTO functions.
 
 \item \underline{\textcolor{red}{\bf Highlight~III:}}\\
 The late time behaviour of the four-point auto-correlated OTO functions helps us to study the equilibrium feature of the quantum correlations which we have computed from the scalar cosmological perturbations.  Obviously the scalar cosmological perturbation and its related stuffs upto quantizing the Hamiltonian is very well known,  but rest of the computation in this context are completely new.  Though for better understanding purpose we have provided a small portion of the cosmological perturbation with scalar modes before starting the computation of two and four point auto-correlated OTO functions.

 \item \underline{\textcolor{red}{\bf Highlight~IV:}}\\
 We have provided the detailed computation of the normalized version of the four-point auto-correlated cosmological OTO functions which we found that are completely independent of the choice of the time dependent perturbation variable appearing in the cosmological perturbation theory.  To justify this statement we have used co-moving gauge for the simplicity.
 \end{itemize}

The plan of this paper are organized as follows:
\begin{itemize}
\item In the \underline{\bf section~(\ref{sec:1})},  we discuss the formalism of computing the auto-correlated OTO functions in the context of Primordial Cosmology.  This formalism is new which we have highlighted in this paper.

\item In the \underline{\bf section~(\ref{sec:2})},  we provide the detailed derivation of quantum two-point and four-point auto-correlated OTO amplitudes and the related OTO function within the framework of Primordial Cosmology.  This is the new calculation that we have provided in this paper.

\item  In the \underline{\bf section~(\ref{sec:3})} and \underline{\bf section~(\ref{sec:4})},  we present the numerical results from the quantum two-point and four-point auto-correlated OTO functions and also discuss about its physical interpretation.  It is important to note that the numerical predictions obtained from the mentioned computations are also new and provide a new direction in the framework of Primordial Cosmology. 

\item In the \underline{\bf section~(\ref{sec:5})},  we discuss the
classical limit of the two-point and four-point OTO amplitudes and the related implications in Cosmology.  This is another new result we have provided in this paper which gives us a better understanding of the super-horizon classical limiting behaviour of the system under consideration.

\item In the \underline{\bf Appendix~(\ref{sec:7})-Appendix~(\ref{sec:16})},  we have provided the details of the computations used in various sections of the paper. We believe this will help the general readers to understand the underlying physical problem that we have addressed in this paper.

\end{itemize}

%%%%%%%%%%%%%%%%
\section{Formulation of non-chaotic auto-correlated OTO functions in Primordial Cosmology}
\label{sec:1}

%%%%%%%%%%%%%%%% 
\subsection{Non-chaotic auto-correlated OTO functions}
Let us consider two quantum mechanical operators $X$ and $Y$ which are defined at two different time scale i.e. $X(t), X(\tau)$ and $Y(t), Y(\tau)$.  Then we will extract the information regarding the non-chaotic but random quantum mechanical correlation functions using the prescription of OTOC.  In this context,  the non-chaotic OTOC's are defined in terms of these quantum operators as:
 \bea && \textcolor{red}{\bf 2-point~OTOC_1:}~~~~  Y_1(t,\tau):=-\langle \left[X(t),X(\tau)\right]\rangle_{\beta},\\
 && \textcolor{red}{\bf 2-point~OTOC_2:}~~~~  Y_2(t,\tau):=-\langle \left[Y(t),Y(\tau)\right]\rangle_{\beta},\\
 && \textcolor{red}{\bf 4-point~OTOC_1:}~~~~  C_1(t,\tau):=-\langle \left[X(t),X(\tau)\right]^2\rangle_{\beta},\\
 && \textcolor{red}{\bf 4-point~OTOC_2:}~~~~  C_2(t,\tau):=-\langle \left[Y(t),Y(\tau)\right]^2\rangle_{\beta},\eea
 where the thermal average of any operator is defined as:
 \bea  \textcolor{red}{\bf Thermal~average:}~~~~\langle A(t)\rangle_{\beta}:=\frac{1}{Z}{\rm Tr}\left[\exp(-\beta H)~A(t)\right],\eea
 where the partition function of the system is defined as:
 \bea \textcolor{red}{\bf Partition~function:}~~~~Z={\rm Tr}\left[\exp(-\beta H)\right].\eea
 Explicitly, in terms of thermal averaging one can further write:
 \bea && \textcolor{red}{\bf 2-point~OTOC_1:}~~~~  Y_1(t,\tau):=-\frac{1}{Z}{\rm Tr}\left[\exp(-\beta H)~ \left[X(t),X(\tau)\right]\right],\\
 && \textcolor{red}{\bf 2-point~OTOC_2:}~~~~  Y_2(t,\tau):=-\frac{1}{Z}{\rm Tr}\left[\exp(-\beta H)~ \left[Y(t),Y(\tau)\right]\right],\eea
 \bea
 && \textcolor{red}{\bf 4-point~OTOC_1:}~~~~  C_1(t,\tau):=-\frac{1}{Z}{\rm Tr}\left[\exp(-\beta H)~ \left[X(t),X(\tau)\right]^2\right],\\
 && \textcolor{red}{\bf 4-point~OTOC_2:}~~~~  C_2(t,\tau):=-\frac{1}{Z}{\rm Tr}\left[\exp(-\beta H)~ \left[Y(t),Y(\tau)\right]^2\right].\eea
 One, can further write these contributions in terms of the thermal density matrix as:
  \bea && \textcolor{red}{\bf 2-point~OTOC_1:}~~~~  Y_1(t,\tau):=-{\rm Tr}\left[\rho~ \left[X(t),X(\tau)\right]\right],\\
 && \textcolor{red}{\bf 2-point~OTOC_2:}~~~~  Y_2(t,\tau):=-{\rm Tr}\left[\rho~ \left[Y(t),Y(\tau)\right]\right],\\
 && \textcolor{red}{\bf 4-point~OTOC_1:}~~~~  C_1(t,\tau):=-{\rm Tr}\left[\rho~ \left[X(t),X(\tau)\right]^2\right],\\
 && \textcolor{red}{\bf 4-point~OTOC_2:}~~~~  C_2(t,\tau):=-{\rm Tr}\left[\rho~ \left[Y(t),Y(\tau)\right]^2\right],\eea
 where in this context the thermal density matrix is defined as: 
 \bea  \textcolor{red}{\bf Thermal~density~matrix:}~~~~ \rho=\frac{1}{Z}{\rm Tr}\left[\exp(-\beta H)\right].\eea
 Now, in the large time limit the thermal average of the following four point function can be factorized as:
 \bea \langle X(\tau)X(t)X(t)X(\tau)\rangle_{\beta}&=& \underbrace{\langle X(\tau)X(\tau)\rangle_{\beta}}_{\textcolor{red}{\bf 2-point~disconnected}}~~\underbrace{\langle X(t)X(t)\rangle_{\beta}}_{\textcolor{red}{\bf 2-point~disconnected}}+\underbrace{{\cal O}(\exp(-(t+\tau)/t_d))}_{\textcolor{red}{\bf Sub-leading~contribution}},~~~~~~~~~\\
 \langle X(t)X(\tau)X(\tau)X(t)\rangle_{\beta}&=&\underbrace{\langle X(t)X(t)\rangle_{\beta}}_{\textcolor{red}{\bf 2-point~disconnected}}~~\underbrace{\langle X(\tau)X(\tau)\rangle_{\beta}}_{\textcolor{red}{\bf 2-point~disconnected}}+\underbrace{{\cal O}(\exp(-(t+\tau)/t_d))}_{\textcolor{red}{\bf Sub-leading~contribution}},\\
 \langle Y(\tau)Y(t)Y(t)Y(\tau)\rangle_{\beta}&=&\underbrace{ \langle Y(\tau)Y(\tau)\rangle_{\beta}}_{\textcolor{red}{\bf 2-point~disconnected}}~~\underbrace{\langle Y(t)Y(t)\rangle_{\beta}}_{\textcolor{red}{\bf 2-point~disconnected}}+\underbrace{{\cal O}(\exp(-(t+\tau)/t_d))}_{\textcolor{red}{\bf Sub-leading~contribution}},~~~~~~~~~\\
 \langle Y(t)Y(\tau)Y(\tau)Y(t)\rangle_{\beta}&=&\underbrace{\langle Y(t)Y(t)\rangle_{\beta}}_{\textcolor{red}{\bf 2-point~disconnected}}~~\underbrace{\langle Y(\tau)Y(\tau)\rangle_{\beta}}_{\textcolor{red}{\bf 2-point~disconnected}}+\underbrace{{\cal O}(\exp(-(t+\tau)/t_d))}_{\textcolor{red}{\bf Sub-leading~contribution}},~~~~~~~~~\eea
 where the two-point disconnected thermal two-point correlators can be expressed as:
 \bea && \textcolor{red}{\bf 2-point~correlator_1:}~~~~  \langle X(t)X(t)\rangle_{\beta}=-\frac{1}{Z}{\rm Tr}\left[\exp(-\beta H)~ X(t)X(t)\right],~~~~~~~~~~~\\
 && \textcolor{red}{\bf 2-point~correlator_2:}~~~~  \langle X(\tau)X(\tau)\rangle_{\beta}=-\frac{1}{Z}{\rm Tr}\left[\exp(-\beta H)~ X(\tau)X(\tau)\right],\\
 && \textcolor{red}{\bf 2-point~correlator_3:}~~~~  \langle Y(t)Y(t)\rangle_{\beta}=-\frac{1}{Z}{\rm Tr}\left[\exp(-\beta H)~ Y(t)Y(t)\right],\\
 && \textcolor{red}{\bf 2-point~correlator_4:}~~~~  \langle Y(\tau)Y(\tau)\rangle_{\beta}=-\frac{1}{Z}{\rm Tr}\left[\exp(-\beta H)~ Y(\tau)Y(\tau)\right],\eea
 where the time scale $t_d$ is identified as the dissipation or equilibrium time scale, which is given by:
 \bea \textcolor{red}{\bf Dissipation/equilibrium~time~scale:}~~~~ t_d\sim \beta=\frac{1}{T}~~~{\rm with}~~k_B=1.\eea  
 Here, $T$ is the equilibrium temperature of the quantum mechanical system under consideration. It is important to mention here that, the above mentioned four possible four point thermal correlation function one can able to factorize into multiplication of two distinctive disconnected two point contributions if one can wait for a large time scale. The factorization along with the sub-leading decaying contribution is actually expected from our basic understanding of quantum statistical field theory. When we give a response to a quantum system it goes to out-of-equilibrium phase and the corresponding correlation function in quantum regime starts randomly fluctuating with respect to the evolutionary time scale. During this initial time scale when the initial response is provided in terms of the initial condition to the quantum system it is not possible to factorize the present four possibilities of the four point out-of-time-ordered-correlation (OTOC) functions. But if we wait for long enough time then it is expected that the quantum random fluctuations achieve the thermodynamic equilibrium. During this time scale one can actually factorize these four possible four point OTOCs in terms of the products of two disconnected two point functions and the sub-leading contribution actually decay with respect to the very late time scale with a finite small saturation value. As a result, from this late time scale, $t_d$, which is identified to be dissipation time scale, will give the measure of the inverse temperature at thermodynamic equilibrium.
 
On the other hand, $\langle \left[X(t),X(\tau)\right]^2\rangle_{\beta}$ and $\langle \left[Y(t),Y(\tau)\right]^2\rangle_{\beta}$, can be expressed in the long time limit as:
 \bea  C_1(t,\tau)=-\langle \left[X(t),X(\tau)\right]^2\rangle_{\beta}&=&-\langle X(t)X(\tau)X(t)X(\tau)\rangle_{\beta}-\langle X(\tau)X(t)X(\tau)X(t)\rangle_{\beta}\nonumber\\
 &&~~~~~~~~~~~~~~+\langle X(\tau)X(t)X(t)X(\tau)\rangle_{\beta}+\langle X(t)X(\tau)X(\tau)X(t)\rangle_{\beta}\nonumber\\
 &=&2\left\{ \langle X(t)X(t)\rangle_{\beta} \langle X(\tau)X(\tau)\rangle_{\beta}-\Re\left[\langle X(t)X(\tau)X(t)X(\tau)\rangle_{\beta}\right]\right\}\nonumber\\
&&~~~~~~~~~~~~~~~~~~~~~~~~~~~~~~~~~~~~~~ -{\cal O}(\exp(-(t+\tau)/t_d)),~~~~~~~~\\
 C_2(t,\tau)=-\langle \left[Y(t),Y(\tau)\right]^2\rangle_{\beta}&=&-\langle Y(t)Y(\tau)Y(t)Y(\tau)\rangle_{\beta}-\langle Y(\tau)Y(t)Y(\tau)Y(t)\rangle_{\beta}\nonumber\\
 &&~~~~~~~~~~~~~~+\langle Y(\tau)Y(t)Y(t)Y(\tau)\rangle_{\beta}+\langle Y(t)Y(\tau)Y(\tau)Y(t)\rangle_{\beta}\nonumber\\
 &=&2\left\{ \langle Y(t)Y(t)\rangle_{\beta} \langle Y(\tau)Y(\tau)\rangle_{\beta}-\Re\left[\langle Y(t)Y(\tau)Y(t)Y(\tau)\rangle_{\beta}\right]\right\}\nonumber\\
&&~~~~~~~~~~~~~~~~~~~~~~~~~~~~~~~~~~~~~~ -{\cal O}(\exp(-(t+\tau)/t_d)). \eea
 Now using these results considering the large time equilibrium behaviour one can further compute the expressions for the normalized OTOCs in the present context, which are given by:
  \bea  {\cal C}_1(t,\tau)=\frac{C_1(t,\tau)}{\langle X(t)X(t)\rangle_{\beta} \langle X(\tau)X(\tau)\rangle_{\beta}}&=&\underbrace{2\left\{ 1-\frac{\Re\left[\langle X(t)X(\tau)X(t)X(\tau)\rangle_{\beta}\right]}{\langle X(t)X(t)\rangle_{\beta} \langle X(\tau)X(\tau)\rangle_{\beta}}\right\}}_{\textcolor{red}{\bf Leading~contribution}}\nonumber\\
&&~~~~~~~~~~~~~~~~~~~~~~ -\underbrace{\frac{{\cal O}(\exp(-(t+\tau)/t_d))}{\langle X(t)X(t)\rangle_{\beta} \langle X(\tau)X(\tau)\rangle_{\beta}}}_{\textcolor{red}{\bf Sub-leading~decaying~contribution}},~~~~~~~~\\ 
  {\cal C}_1(t,\tau)=\frac{C_2(t,\tau)}{\langle Y(t)Y(t)\rangle_{\beta} \langle Y(\tau)Y(\tau)\rangle_{\beta}}&=&\underbrace{2\left\{ 1-\frac{\Re\left[\langle Y(t)Y(\tau)Y(t)Y(\tau)\rangle_{\beta}\right]}{\langle Y(t)Y(t)\rangle_{\beta} \langle Y(\tau)Y(\tau)\rangle_{\beta}}\right\}}_{\textcolor{red}{\bf Leading~contribution}}\nonumber\\
&&~~~~~~~~~~~~~~~~~~~~~~ -\underbrace{\frac{{\cal O}(\exp(-(t+\tau)/t_d))}{\langle Y(t)Y(t)\rangle_{\beta} \langle Y(\tau)Y(\tau)\rangle_{\beta}}}_{\textcolor{red}{\bf Sub-leading~decaying~contribution}}. \eea
In this paper, our prime objective is to compute these leading order contributions in the context of primordial cosmology, where we chose these two quantum operators as the perturbation field operator and its canonically conjugate momentum operator as appearing in the context of cosmological perturbation theory. To remind all of us again here it is important to note down again that during the computation of OTOCs we will only consider the contributions from the same quantum operators in cosmology but defined in two separated time scale.  Here all of these sub-leading contributions will give a correct understanding of the large time limiting behaviour which helps to further comment on the equilibrium behaviour of the quantum correlation functions and to estimate the temperature at thermodynamic equilibrium. In this paper we are very very hopeful to get distinctive behaviour from the mentioned two OTOCs in the context of cosmology compared to the result obtained from the cosmological OTOC in our previous paper \cite{Choudhury:2020yaa}.  Instead of getting a random chaotic behaviour we are very hopeful to describe a general non-chaotic random behaviour out of the OTOCs that we are studying particularly in this paper within the framework of primordial cosmology.  Once we derive these mentioned OTOCs in this paper then the study of all types of random fluctuations will be completed and we strongly believe this complete study will be helpful to study various unexplored features of primordial cosmology in the quantum out-of-equilibrium regime.  In the next subsection we will talk about the eigenstate representations of these new class of OTOCs which will be very relevant for the computation of OTOCs in the context of quantum mechanical system which have the eigenstate representation of the Hamiltonian explicitly.
 \subsection{Eigenstate representation of non-chaotic OTOC in quantum statistical mechanics}
 In this subsection our prime objective is to give a simpler representation, which is known as the eigenstate representation of the OTOC. This can only be possible if the quantum system under consideration have eigenstates i.e. the system Hamiltonian have eigenstates. We will explicitly show from a general calculation that how one can express the complicated definitions of OTOCs mentioned in earlier section in a very simpler language. This representation of OTOCs are very useful for the computational purpose as it allows us separately take care of the contributions coming from the micro-canonical part of the OTOCs and the thermal Boltzmann factor, where both of them are the building blocks of the total contribution appearing in the OTOCs in the eigenstate representation. Once we compute both of these building blocks separately one can get to know the full information regarding the the quantum randomness which are appearing in the expressions for the total thermal OTOCs from a quantum mechanical system. Not only that, but also in the eigenstate representations of the OTOCs the total expressions are computed after taking over sum over all the individual contributions obtained from all possible eigenstates. In this connection, it is important to note down here that the from the final answers of the OTOCs one can actually categorize all classes of available quantum mechanical systems into two families which we are discussing in detail in the following:
 \begin{enumerate}
 \item \underline{\textcolor{red}{\bf First family of quantum systems:}}\\
 First family of quantum mechanical systems deal with only the micro-canonical part of the OTOCs.  Because, after taking the sum over all possible contributions from the eigenstates one can find out that for these class of quantum systems the cumulative contribution actually gives the expression for the thermal partition function for the class of quantum systems which will further cancels with the expression for the thermal partition function which is appearing in the denominator of these OTOCs to make the consistency with the definition of thermal average of any quantum mechanical operator at finite temperature over all possible eigenstates of the Hamiltonian of the system. This makes the final expression for OTOCs completely dependent on the micro-canonical part of the OTOCs after taking sum over all eigenstate contribution. One of the well known examples of these class of models are the Quantum Harmonic Oscillator (QHO) which can be solely represented by the contributions from the micro-canonical part after summing over all possible eigenstates of the Hamiltonian of the QHO. In this connection here it is important to note that, at the perturbation level,  cosmology with a free massive scalar field theory in a FLRW space-time can be represented as a Quantum Parametric Oscillator with a time dependent frequency in the Fourier space, but instead of having a harmonic oscillator type of  representation within the framework of quantum mechanical framework of cosmology we don't have any eigenstate~\footnote{We all know that our universe evolved with respect to the time scale once.  So if we want to explain the background framework of primordial cosmology in the quantum regime then the Euclidean vacuum state or the Bunch Davies vacuum state or the more generalized De Sitter isommetric $\alpha$-vacua cannot be treated as the eigenstate of the quantum Hamiltonian of parametric harmonic oscillator in the Fourier space representation as we cannot repeat the evolution of our universe.  So that time dependent Fourier mode integrated continuous function of energy cannot be treated as the eigen energy spectrum as appearing in the present context, where eigenstate of the Hamiltonian play significant role to determine the expression for OTOCs.  }. Instead of having a eigenstate in the context of cosmology one can define wave function of the universe and instead of having a discrete eigen energy spectrum in the context of cosmology we can find a continuous time dependent momentum integrated spectrum over all Fourier modes. 
 
 \item \underline{\textcolor{red}{\bf Second family of quantum systems:}}\\ 
 Second family of the quantum systems are those in which these defined OTOCs are quantified by the Boltzmann part as well as the micro-canonical part. Also after summing over all possible quantum states and implementing the definition of the thermal average of a quantum operator one can find out the final result is dependent on the inverse of the temperature, which can take care of the behaviour of the OTOCs when it reaches the thermodynamic equilibrium after waiting for large enough time in the late time scale and particularly this feature is completely absent in the context of the first family of the quantum mechanical systems. One of the simplest example of this class of models are particle in one dimensional potential.  Like just the previous class here also at the perturbation level cosmology with a self interacting scalar field or considering the interaction between different scalar fields in a FLRW background one can represent the total theory by a perturbation in the single Quantum Parametric Oscillator with a time dependent frequency in the Fourier space for the self interacting case.  But instead of having a very simple perturbation theory of harmonic oscillator within the framework of quantum mechanical framework of cosmology we have very complicated version because of the absence of eigenstate~\footnote{By applying the general perturbation technique one can actually able to compute the correction in the energy eigen spectrum if the Hamiltonian of the quantum system have the eigenstate representation. In this connection one needs to cite the example of simple harmonic oscillator again where one can analytically compute these corrections very easily and most of us have studied this from all quantum mechanics books exist in the literature.  On the other hand,  in the self interacting picture or for the case of the interaction between many scalar field case in the spatially flat FLRW cosmological background computation of the quantum correction factors are extremely complicated.}. 
 \end{enumerate}
 Now we will discuss in detail the construction of eigenstate representation of non-chaotic OTOCs in which we are in this paper.
  To elaborate the computation in the eigenstate representation we will talk about two canonically conjugate quantum mechanical operators in two two different times scales $t$ and $\tau$ i.e. $q(t), q(\tau)$ and $p(t), p(\tau)$ respectively. As we have already pointed and elaborately have discussed in the previous section that once we consider the OTOCs between same quantum mechanical operators in different time scale it is expected to have these classes of OTOCs which will quantify the quantum fluctuations in terms of general non-chaotic random correlation functions within the framework of quantum statistical mechanics. We will further generalize this idea to compute the cosmological OTOCs in the later part of this paper, though in the context of quantum field theory of cosmology we don't have any eigenstate representation. 
  
  In this context we are interested to compute the expressions for the following quantities, which are given by:
  \bea && \textcolor{red}{\bf 2-point~OTOC_{1}:}~~~Y_{1}(t,\tau):=-\langle \left[q(t),q(\tau)\right]\rangle_{\beta},\\
          && \textcolor{red}{\bf 2-point~OTOC_{2}:}~~~Y_{2}(t,\tau):=-\langle \left[p(t),p(\tau)\right]\rangle_{\beta},\\
          && \textcolor{red}{\bf 4-point~OTOC_{1}:}~~~C_{1}(t,\tau):=-\langle \left[q(t),q(\tau)\right]^2\rangle_{\beta},\\
          && \textcolor{red}{\bf 4-point~OTOC_{2}:}~~~C_{2}(t,\tau):=-\langle \left[p(t),p(\tau)\right]^2\rangle_{\beta},\eea
          and also the normalized version of these 4-point OTOCs can be represented by:
   \bea &&\textcolor{red}{\bf 4-point~norm.~OTOC_{1}:}~ {\cal C}_1(t,\tau):=\frac{C_{1}(t,\tau)}{\langle q(t)q(t)\rangle_{\beta}\langle q(\tau)q(\tau)\rangle_{\beta}}=-\frac{\langle \left[q(t),q(\tau)\right]^2\rangle_{\beta}}{\langle q(t)q(t)\rangle_{\beta}\langle q(\tau)q(\tau)\rangle_{\beta}},~~~~~~~~\\
          &&\textcolor{red}{\bf 4-point~norm.~OTOC_{2}:}~   {\cal C}_2(t,\tau):=\frac{C_{2}(t,\tau)}{\langle p(t)p(t)\rangle_{\beta}\langle p(\tau)p(\tau)\rangle_{\beta}}=-\frac{\langle \left[p(t),p(\tau)\right]^2\rangle_{\beta}}{\langle p(t)p(t)\rangle_{\beta}\langle p(\tau)p(\tau)\rangle_{\beta}},\eea        
    where, $\beta=1/T$ (in $k_B=1$) is the equilibrium temperature of the quantum mechanical system under consideration for the present study and it is expected that the system will achieve thermodynamic equilibrium if we wait for a very longer time in the evolutionary time scales $t$ and $\tau$ under consideration for this problem.  
    
    Next, we consider energy eigenstate $|n\rangle$ of the system time independent Hamiltonian $H$, which satisfy the following time independent Schr$\ddot{o}$inger equation:
    \bea H|n\rangle=E_{n}|n\rangle~~~~~~~~~\forall~~~n=0,1,2,\cdots\cdots\infty,\eea
    where $E_n$ represent the energy eigen values associated with the eigen energy state $|n\rangle$. Using this eigenstate representation one can further write the expressions for these classes of OTOCs in the following simpler language:
    \bea Y_1(t,\tau)=\frac{1}{Z(\beta)}\sum^{\infty}_{n=0}\exp(-\beta E_n)~{\cal E}^{(1)}_{n}(t,\tau),\\
   Y_2(t,\tau)=\frac{1}{Z(\beta)}\sum^{\infty}_{n=0}\exp(-\beta E_n)~{\cal E}^{(2)}_{n}(t,\tau),\\
   C_1(t,\tau)=\frac{1}{Z(\beta)}\sum^{\infty}_{n=0}\exp(-\beta E_n)~{\cal D}^{(1)}_{n}(t,\tau),\\
   C_2(t,\tau)=\frac{1}{Z(\beta)}\sum^{\infty}_{n=0}\exp(-\beta E_n)~{\cal D}^{(2)}_{n}(t,\tau),  \eea
   where the time dependent diagonal matrix element representing the micro-canonical part of the OTOCs are given by the following expressions:
   \bea {\cal E}^{(1)}_{n}(t,\tau):=-\langle n|\left[q(t),q(\tau)\right]|n\rangle,\\
   {\cal E}^{(2)}_{n}(t,\tau):=-\langle n|\left[p(t),p(\tau)\right]|n\rangle,\\
   {\cal D}^{(1)}_{n}(t,\tau):=-\langle n|\left[q(t),q(\tau)\right]^2|n\rangle,\\
   {\cal D}^{(2)}_{n}(t,\tau):=-\langle n|\left[p(t),p(\tau)\right]^2|n\rangle.\eea
   Also, in the normalised representation the above mentioned 4-point OTOCs further can be recast as:
   \bea {\cal C}_1(t,\tau)=\frac{1}{Z(\beta)~N_1(t,\tau)}\sum^{\infty}_{n=0}\exp(-\beta E_n)~{\cal D}^{(1)}_{n}(t,\tau),\\
   {\cal C}_2(t,\tau)=\frac{1}{Z(\beta)~N_2(t,\tau)}\sum^{\infty}_{n=0}\exp(-\beta E_n)~{\cal D}^{(2)}_{n}(t,\tau),\eea
   where the time dependent normalization factors $N_1(t,\tau)$ and $N_2(t,\tau)$ are defined as:
 \bea N_1(t,\tau):&=&\langle q(t)q(t)\rangle_{\beta}\langle q(\tau)q(\tau)\rangle_{\beta}\nonumber\\
 &=&\frac{1}{Z^2(\beta)}\left(\sum^{\infty}_{j=0}\exp(-\beta E_j)~\langle j|q(t)q(t)|j\rangle\right)\left(\sum^{\infty}_{m=0}\exp(-\beta E_m)~\langle m|q(\tau)q(\tau)|m\rangle\right),~~~~~~~\\
 N_2(t,\tau):&=&\langle p(t)p(t)\rangle_{\beta}\langle p(\tau)p(\tau)\rangle_{\beta}\nonumber\\
 &=&\frac{1}{Z^2(\beta)}\left(\sum^{\infty}_{j=0}\exp(-\beta E_j)~\langle j|p(t)p(t)|j\rangle\right)\left(\sum^{\infty}_{m=0}\exp(-\beta E_m)~\langle m|p(\tau)p(\tau)|m\rangle\right).~~~~~~~~~~~\eea
 In the present context, in the eigenstate representation of the time independent Hamiltonian of the system under consideration the partition function is defined by the following expression:
 \bea Z(\beta)=\sum^{\infty}_{i=0}\exp(-\beta E_i).~~~\eea
 The above mentioned results actually represent the eigenstate representation of OTOCs which can be decomposed into two important parts-(1) one part take care of the temperature dependence through the thermal Boltzmann factor and (2) another part will capture the temperature independent but time dependent contribution of the micro-canonical statistical ensemble average computed for any arbitrary $n$-th eigenstate and after getting the contribution for the this any $n$-th eigenstate we have to take sum over all possible eigenstates including thee thermal Boltzmann factor.  For this purpose we take all values of the number $n$, which actually runs from $0$ to $\infty$ and at the end of the day one can explicitly compute the useful eigenstate simplified representation of the previously mentioned non-chaotic class of OTOCs in the present context. 
 
 Now to further simplify the expressions for the two different types of micro-canonical OTOCs, represented by, ${\cal D}^{(1)}_{n}(t,\tau)$ and ${\cal D}^{(2)}_{n}(t,\tau)$, it would be very simpler if we can able to express them in terms of the matrix elements of the two canonically conjugate quantum mechanical operators defined in the two separate time scales $t$ and $\tau$ i.e. $q(t), q(\tau)$ and $p(t), p(\tau)$, respectively, where these matrix elements are explicitly computed by sandwiching between the any $n$-th arbitrary eigen states. To implement this simplest computational strategy we need to explicitly use the following quantum completeness condition which can be written in terms of the energy eigenstates as:
 \bea \textcolor{red}{\bf Completeness ~condition:}~~~~\sum^{\infty}_{n=0}|n\rangle\langle n|=\mathbb{I}.\eea 
 As a consequence, the micro-canonical part of the 4-point OTOCs can be expressed in terms of the required matrix elements as:
  \bea 
  && {\cal D}^{(1)}_{n}(t,\tau):=-\langle n|\left[q(t),q(\tau)\right]^2|n\rangle=\sum^{\infty}_{m=0}{\cal G}^{(1)}_{nm}(t,\tau)\left({\cal G}^{(1)}_{nm}(t,\tau)\right)^{\dagger},\\
  && {\cal D}^{(2)}_{n}(t,\tau):=-\langle n|\left[p(t),p(\tau)\right]^2|n\rangle=\sum^{\infty}_{m=0}{\cal G}^{(2)}_{nm}(t,\tau)\left({\cal G}^{(2)}_{nm}(t,\tau)\right)^{\dagger},\eea
   where the individual time dependent matrix elements, ${\cal G}^{(1)}_{nm}(t,\tau)$ and ${\cal G}^{(2)}_{nm}(t,\tau)$, can be expressed as:
   \bea {\cal G}^{(1)}_{nm}(t,\tau):=i\langle n|\left[q(t),q(\tau)\right]|m\rangle,\\
  {\cal G}^{(2)}_{nm}(t,\tau):=i\langle n|\left[p(t),p(\tau)\right]|m\rangle.  \eea
  Now we take the Hermitian conjugate of both of these matrix elements, which are given by the following expressions:
  \bea \left({\cal G}^{(1)}_{nm}(t,\tau)\right)^{\dagger}=-i\langle n|\left[q(t),q(\tau)\right]|m\rangle^{\dagger}=-i\langle m|\left[q(t),q(\tau)\right]^{\dagger}|n\rangle={\cal G}^{(1)}_{mn}(t,\tau),\\
   \left({\cal G}^{(2)}_{nm}(t,\tau)\right)^{\dagger}=-i\langle n|\left[p(t),p(\tau)\right]|m\rangle^{\dagger}=-i\langle m|\left[p(t),p(\tau)\right]^{\dagger}|n\rangle={\cal G}^{(2)}_{mn}(t,\tau),  \eea 
   where we have used the following facts:
   \bea \left[q(t),q(\tau)\right]^{\dagger}=-\left[q(t),q(\tau)\right],\\
    \left[p(t),p(\tau)\right]^{\dagger}=-\left[p(t),p(\tau)\right].\eea
    Consequently, the micro-canonical part of the 4-point OTOCs can be further simplified as:
    \bea {\cal D}^{(1)}_{n}(t,\tau):=-\langle n|\left[q(t),q(\tau)\right]^2|n\rangle=\sum^{\infty}_{m=0}{\cal G}^{(1)}_{nm}(t,\tau){\cal G}^{(1)}_{mn}(t,\tau),\\
   {\cal D}^{(2)}_{n}(t,\tau):=-\langle n|\left[p(t),p(\tau)\right]^2|n\rangle=\sum^{\infty}_{m=0}{\cal G}^{(2)}_{nm}(t,\tau){\cal G}^{(2)}_{mn}(t,\tau).\eea
   Now, further in the quantum mechanical operator representation the time dependence of the operators $q(t),q(\tau)$ and $p(t),p(\tau)$ can be expressed in the Heisenberg picture as:
   \bea q(t):=\exp(iHt)q(0)\exp(-iHt),\\
   q(\tau):=\exp(iH\tau)q(0)\exp(-iH\tau),\\
   p(t):=\exp(iHt)p(0)\exp(-iHt),\\
   p(\tau):=\exp(iH\tau)p(0)\exp(-iH\tau).\eea
   Using this representation the coefficients and the matrix elements as appearing in the expressions for the micro-canonical part of the two-point and four-point OTOCs can be further computed as:
   \bea {\cal E}^{(1)}_{n}(t,\tau)&=&i\langle n|\left[q(t),q(\tau)\right]|n\rangle=2\sum^{\infty}_{k=0}\sin(\Delta E_{kn}(t-\tau))~q_{nk}(0)q_{kn}(0),\\ {\cal E}^{(2)}_{n}(t,\tau)&=&i\langle n|\left[q(t),q(\tau)\right]|n\rangle=2\sum^{\infty}_{k=0}\sin(\Delta E_{kn}(t-\tau))~p_{nk}(0)p_{kn}(0),\\
   {\cal G}^{(1)}_{nm}(t,\tau)&=&i\langle n|\left[p(t),p(\tau)\right]|m\rangle=i\sum^{\infty}_{k=0}\left[\exp(i\Delta E_{nk}(t-\tau))-\exp(i\Delta E_{km}(t-\tau))\right]~q_{nk}(0)q_{km}(0),\\
 {\cal G}^{(2)}_{nm}(t,\tau)&=&i\langle n|\left[p(t),p(\tau)\right]|m\rangle=i\sum^{\infty}_{k=0}\left[\exp(i\Delta E_{nk}(t-\tau))-\exp(i\Delta E_{km}(t-\tau))\right]~p_{nk}(0)p_{km}(0).~~~~~~~~~~~\eea
   where, we define $\Delta E_{mn}$, $q_{mn}(0)$ and $p_{mn}(0)$ by the following simplified notations:
   \bea && \Delta E_{mn}=E_m - E_n,\\
            && q_{mn}(0)=\langle m|q(0)|n\rangle,\\
            && p_{mn}(0)=\langle m|p(0)|n\rangle.\eea
            Also for this simplification we have used the completeness condition explicitly. Additionally, it is important to note that here we have considered  a $N$ particle (many body) non-interacting Hamiltonian to describe a quantum mechanical system in the present context, which is represented by the following expression:
            \bea H(p_1,\cdots,p_N;q_1,\cdots,q_N)=\sum^{N}_{\alpha=1}\frac{p^2_{\alpha}}{2m_{\alpha}}+U(q_1,\cdots,q_N).\eea
            Now for the simplicity of the further computation we assume that each $N$ particle have the same mass, which is given by:
            \bea m_{\alpha}=\frac{1}{2}~~~~~~\forall ~~~~~\alpha=1,2,\cdots,N.\eea
            Consequently, for this simplest physical situation the $N$ particle Hamiltonian can be further simplified as:
            \bea H(p_1,\cdots,p_N;q_1,\cdots,q_N)=\sum^{N}_{\alpha=1}p^2_{\alpha}+U(q_1,\cdots,q_N).\eea  
            Using this many body $N$ particle simplest but general form of the Hamiltonian one can further simplify the required matrix elements as given by the following expressions:
             \bea {\cal E}^{(1)}_{n}(t,\tau)&=&-\langle n|\left[q(t),q(\tau)\right]|n\rangle=2i\sum^{\infty}_{k=0}\sin(\Delta E_{kn}(t-\tau))~q_{nk}(0)q_{kn}(0),\\ {\cal E}^{(2)}_{n}(t,\tau)&=&-\langle n|\left[q(t),q(\tau)\right]|n\rangle=\frac{i}{2}\sum^{\infty}_{k=0}\sin(\Delta E_{nk}(t-\tau))~\Delta E_{nk}\Delta E_{kn}~q_{nk}(0)q_{kn}(0),~~~~~~~\\  {\cal G}^{(1)}_{nm}(t,\tau)&=&i\sum^{\infty}_{k=0}\left[\exp(i\Delta E_{nk}(t-\tau))-\exp(i\Delta E_{km}(t-\tau))\right]~q_{nk}(0)q_{km}(0),
             \\
 {\cal G}^{(2)}_{nm}(t,\tau)
   &=&-\frac{i}{4}\sum^{\infty}_{k=0}\left[\exp(i\Delta E_{nk}(t-\tau))-\exp(i\Delta E_{km}(t-\tau))\right]~\Delta E_{nk}\Delta E_{km}~q_{nk}(0)q_{km}(0),~~~~~~~~~~~\eea    
   where we have explicitly used the following information to convert the matrix elements for canonically conjugate momentum defined at time $t=0$ to the matrix elements for the position operator defined at the same time:
   \bea p_{nm}(0)=\langle n|p(0)|m\rangle=\frac{i}{2}\langle n|\left[H(0),q(0)\right]|m\rangle=\frac{i}{2}\Delta E_{nm}q_{nm}(0). \eea   
      Consequently, the micro-canonical part of the OTOCs can be further simplified as:
        \bea {\cal E}^{(1)}_{n}(t,\tau)&=&2i\sum^{\infty}_{k=0}\sin(\Delta E_{kn}(t-\tau))~q_{nk}(0)q_{kn}(0),\\ 
        {\cal E}^{(2)}_{n}(t,\tau)&=&\frac{i}{2}\sum^{\infty}_{k=0}\sin(\Delta E_{nk}(t-\tau))~E_{nk}E_{kn}~q_{nk}(0)q_{kn}(0),~~~~\eea\bea {\cal D}^{(1)}_{n}(t,\tau)
        &=&\sum^{\infty}_{m=0}\sum^{\infty}_{k=0}\sum^{\infty}_{s=0}q_{nk}(0)q_{km}(0)q_{ms}(0)q_{sn}(0)\nonumber\\
        &&~\times\left[\exp(i\Delta E_{nk}(t-\tau))-\exp(i\Delta E_{km}(t-\tau))\right]\left[\exp(i\Delta E_{sn}(t-\tau))-\exp(i\Delta E_{ms}(t-\tau))\right]\nonumber\\
       &=&\sum^{\infty}_{m=0}\sum^{\infty}_{k=0}\sum^{\infty}_{s=0}q_{nk}(0)q_{km}(0)q_{ms}(0)q_{sn}(0)\nonumber\\
        &&~~~~~~~~~~~~~~~~\times\left[\exp(i\widetilde{\Delta E_{nksn}}(t-\tau))+\exp(i\widetilde{\Delta E_{kmms}}(t-\tau))\right.\nonumber\\&& \left.~~~~~~~~~~~~~~~~~~~~~~~~~~~~~ -\exp(i\widetilde{\Delta E_{kmsn}}(t-\tau))-\exp(i\widetilde{\Delta E_{nkms}}(t-\tau))\right],\\
   {\cal D}^{(2)}_{n}(t,\tau)
        &=&\frac{1}{16}\sum^{\infty}_{m=0}\sum^{\infty}_{k=0}\sum^{\infty}_{s=0}q_{nk}(0)q_{km}(0)q_{ms}(0)q_{sn}(0)~E_{nk}E_{km}E_{ms}E_{sn}\nonumber\\
        &&~\times\left[\exp(i\Delta E_{nk}(t-\tau))-\exp(i\Delta E_{km}(t-\tau))\right]\left[\exp(i\Delta E_{sn}(t-\tau))-\exp(i\Delta E_{ms}(t-\tau))\right]\nonumber\\
       &=&\frac{1}{16}\sum^{\infty}_{m=0}\sum^{\infty}_{k=0}\sum^{\infty}_{s=0}q_{nk}(0)q_{km}(0)q_{ms}(0)q_{sn}(0)~\Delta E_{nk}\Delta E_{km}\Delta E_{ms}\Delta E_{sn}\nonumber\\
        &&~~~~~~~~~~~~~~~~\times\left[\exp(i\widetilde{\Delta E_{nksn}}(t-\tau))+\exp(i\widetilde{\Delta E_{kmms}}(t-\tau))\right.\nonumber\\&& \left.~~~~~~~~~~~~~~~~~~~~~~~~~~~~~ -\exp(i\widetilde{\Delta E_{kmsn}}(t-\tau))-\exp(i\widetilde{\Delta E_{nkms}}(t-\tau))\right],\eea  
        where we have introduced a few new quantities, which are given by the following expressions:
        \bea && \widetilde{\Delta E_{nksn}}=\Delta E_{nk}+\Delta E_{sn}=E_n-E_k+E_s-E_n=E_s-E_k=\Delta E_{sk},\\
      && \widetilde{\Delta E_{kmms}}=\Delta E_{km}+\Delta E_{ms}=E_k-E_m+E_m-E_s=E_k-E_s=-\Delta E_{sk},\\
     && \widetilde{\Delta E_{kmsn}}=\Delta E_{km}+\Delta E_{sn}=E_k-E_m+E_s-E_n,\\
     && \widetilde{\Delta E_{nkms}}=\Delta E_{nk}+\Delta E_{ms}=E_n-E_k+E_m-E_s,  \eea
     which further give rise to the following properties:
     \bea && \widetilde{\Delta E_{kmms}}= -\widetilde{\Delta E_{nksn}},\\
     && \widetilde{\Delta E_{nkms}}=- \widetilde{\Delta E_{kmsn}}.\eea
     Consequently, using the Eigenstate representation one can write the expressions for the non-chaotic OTOCs as:
        \bea &&{Y_1(t,\tau)=\frac{2i}{\displaystyle\sum^{\infty}_{i=0}\exp(-\beta E_i)}\sum^{\infty}_{n=0}\sum^{\infty}_{k=0}\exp(-\beta E_n)~\sin(\Delta E_{kn}(t-\tau))~q_{nk}(0)q_{kn}(0)}~,\\
   &&{Y_2(t,\tau)=\frac{i}{2\displaystyle\sum^{\infty}_{i=0}\exp(-\beta E_i)}\sum^{\infty}_{n=0}\sum^{\infty}_{k=0}\exp(-\beta E_n)~\sin(\Delta E_{nk}(t-\tau))~\Delta E_{nk}\Delta E_{kn}~q_{nk}(0)q_{kn}(0)}~,~~~~~~~~\eea\bea
    &&{C_1(t,\tau)=\frac{4}{\displaystyle\sum^{\infty}_{i=0}\exp(-\beta E_i)}\sum^{\infty}_{m=0}\sum^{\infty}_{k=0}\sum^{\infty}_{s=0}\sum^{\infty}_{n=0}\exp(-\beta E_n)~q_{nk}(0)q_{km}(0)q_{ms}(0)q_{sn}(0)}\nonumber\\
        &&~~~~~~~~~~~~~~~{\times\sin\left(\left\{E_{s}-\frac{(E_n+E_m)}{2}\right\}(t-\tau)\right)\sin\left(\left\{E_{k}-\frac{(E_n+E_m)}{2}\right\}(t-\tau)\right)},~~~~~~~\\
   &&{C_2(t,\tau)=\frac{1}{\displaystyle 4\sum^{\infty}_{i=0}\exp(-\beta E_i)}\sum^{\infty}_{m=0}\sum^{\infty}_{k=0}\sum^{\infty}_{s=0}\sum^{\infty}_{n=0}\exp(-\beta E_n)~q_{nk}(0)q_{km}(0)q_{ms}(0)q_{sn}(0)}\nonumber\\
        &&~{\times \Delta E_{nk}\Delta E_{km}\Delta E_{ms}\Delta E_{sn}\sin\left(\left\{E_{s}-\frac{(E_n+E_m)}{2}\right\}(t-\tau)\right)\sin\left(\left\{E_{k}-\frac{(E_n+E_m)}{2}\right\}(t-\tau)\right)} \nonumber\\
        && \eea
 To compute the normalization factors the following matrix elements play significant role:
 \bea &&\langle j|q(t)q(t)|j\rangle=\sum^{\infty}_{l=0}q_{jl}(0)q_{lj}(0),\\
 &&\langle m|q(\tau)q(\tau)|m\rangle=\sum^{\infty}_{r=0}q_{mr}(0)q_{rm}(0),\\
 && \langle j|p(t)p(t)|j\rangle=\sum^{\infty}_{k=0}p_{jk}(0)p_{kj}(0)=-\frac{1}{4}\sum^{\infty}_{k=0}\Delta E_{jk}\Delta E_{kj}~q_{jk}(0)q_{jk}(0),\\
 &&\langle m|p(\tau)p(\tau)|m\rangle=\sum^{\infty}_{i=0}p_{mi}(0)p_{im}(0)=-\frac{1}{4}\sum^{\infty}_{i=0}\Delta E_{mi}\Delta E_{im}~q_{mi}(0)q_{im}(0).\eea
and using these results the normalization factors $N_1$ and $N_2$ are computed as:
 \bea &&{N_1=\frac{\displaystyle\sum^{\infty}_{j=0}\sum^{\infty}_{l=0}\sum^{\infty}_{m=0}\sum^{\infty}_{r=0}\exp(-\beta (E_j+E_m))~q_{jl}(0)q_{lj}(0)q_{mr}(0)q_{rm}(0)}{\displaystyle\left(\sum^{\infty}_{i=0}\exp(-\beta E_i)\right)^2}},~~~~~~~\\
 &&{N_2=\frac{\displaystyle\sum^{\infty}_{j=0}\sum^{\infty}_{k=0}\sum^{\infty}_{m=0}\sum^{\infty}_{i=0}\exp(-\beta (E_j+E_m))~\Delta E_{jk}\Delta E_{kj}\Delta E_{mi}\Delta E_{im}~q_{jk}(0)q_{jk}(0)q_{mi}(0)q_{im}(0)}{16~\displaystyle\left(\sum^{\infty}_{i=0}\exp(-\beta E_i)\right)^2}}.~~~~~~~~~~~\eea
  Consequently, using the Eigenstate representation one can write the expressions for the non-chaotic normalised four-point OTOCs as:
        \bea &&{{\cal C}_1(t,\tau)=\frac{4}{\left(\displaystyle\displaystyle\sum^{\infty}_{j=0}\sum^{\infty}_{l=0}\sum^{\infty}_{m=0}\sum^{\infty}_{r=0}\exp(-\beta (E_j+E_m))~q_{jl}(0)q_{lj}(0)q_{mr}(0)q_{rm}(0)\right)}}\nonumber\\
        &&\displaystyle~~~~~{\times\sum^{\infty}_{m=0}\sum^{\infty}_{k=0}\sum^{\infty}_{s=0}\sum^{\infty}_{n=0}\sum^{\infty}_{r=0}\exp(-\beta (E_n+E_r))~q_{nk}(0)q_{km}(0)q_{ms}(0)q_{sn}(0)}\nonumber\\
        &&~~~~~~~~~~~~~~~~{\times\sin\left(\left\{E_{s}-\frac{(E_n+E_m)}{2}\right\}(t-\tau)\right)\sin\left(\left\{E_{k}-\frac{(E_n+E_m)}{2}\right\}(t-\tau)\right)},~~~~~~~~~~~\\
  &&{{\cal  C}_2(t,\tau)=\frac{4}{\left(\displaystyle \displaystyle\sum^{\infty}_{j=0}\sum^{\infty}_{k=0}\sum^{\infty}_{m=0}\sum^{\infty}_{i=0}\exp(-\beta (E_j+E_m))~E_{jk}E_{kj}E_{mi}E_{im}~q_{jk}(0)q_{jk}(0)q_{mi}(0)q_{im}(0)\right)}}\nonumber\\
  &&~~~~~~~~~{\times\sum^{\infty}_{m=0}\sum^{\infty}_{k=0}\sum^{\infty}_{s=0}\sum^{\infty}_{n=0}\sum^{\infty}_{r=0}\exp(-\beta (E_n+E_r))~q_{nk}(0)q_{km}(0)q_{ms}(0)q_{sn}(0)~E_{nk}E_{km}E_{ms}E_{sn}}\nonumber\\
        &&~{\times \Delta E_{nk}\Delta E_{km}\Delta E_{ms}\Delta E_{sn}\sin\left(\left\{E_{s}-\frac{(E_n+E_m)}{2}\right\}(t-\tau)\right)\sin\left(\left\{E_{k}-\frac{(E_n+E_m)}{2}\right\}(t-\tau)\right)}. \nonumber\\
        && \eea 
        Finally, the normalised four-point non chaotic correlators can be further computed as:
        \bea &&{\frac{\Re\left[\langle q(t)q(\tau)q(t)q(\tau)\rangle_{\beta}\right]}{\langle q(t)q(t)\rangle_{\beta} \langle q(\tau)q(\tau)\rangle_{\beta}}}\nonumber\\
        &=&{1-\frac{2}{\displaystyle\displaystyle\sum^{\infty}_{j=0}\sum^{\infty}_{l=0}\sum^{\infty}_{m=0}\sum^{\infty}_{r=0}\exp(-\beta (E_j+E_m))~q_{jl}(0)q_{lj}(0)q_{mr}(0)q_{rm}(0)}}\nonumber\\
        &&\displaystyle~~~~~{\times\sum^{\infty}_{m=0}\sum^{\infty}_{k=0}\sum^{\infty}_{s=0}\sum^{\infty}_{n=0}\sum^{\infty}_{r=0}\exp(-\beta (E_n+E_r))~q_{nk}(0)q_{km}(0)q_{ms}(0)q_{sn}(0)}\nonumber\\
        &&~~~~~~~~~~~~~~~~{\times\sin\left(\left\{E_{s}-\frac{(E_n+E_m)}{2}\right\}(t-\tau)\right)\sin\left(\left\{E_{k}-\frac{(E_n+E_m)}{2}\right\}(t-\tau)\right)},~~~~~~~~~~~~\\
 &&{\frac{\Re\left[\langle p(t)p(\tau)p(t)p(\tau)\rangle_{\beta}\right]}{\langle p(t)p(t)\rangle_{\beta} \langle p(\tau)p(\tau)\rangle_{\beta}}}\nonumber\\
     &=&{1-\frac{2}{\left(\displaystyle \displaystyle 2\sum^{\infty}_{j=0}\sum^{\infty}_{k=0}\sum^{\infty}_{m=0}\sum^{\infty}_{i=0}\exp(-\beta (E_j+E_m))~E_{jk}E_{kj}E_{mi}E_{im}~q_{jk}(0)q_{jk}(0)q_{mi}(0)q_{im}(0)\right)}}\nonumber\\
  &&~~~~~~~~~{\times\sum^{\infty}_{m=0}\sum^{\infty}_{k=0}\sum^{\infty}_{s=0}\sum^{\infty}_{n=0}\sum^{\infty}_{r=0}\exp(-\beta (E_n+E_r))~q_{nk}(0)q_{km}(0)q_{ms}(0)q_{sn}(0)~E_{nk}E_{km}E_{ms}E_{sn}}\nonumber\\
        &&~{\times \Delta E_{nk}\Delta E_{km}\Delta E_{ms}\Delta E_{sn}\sin\left(\left\{E_{s}-\frac{(E_n+E_m)}{2}\right\}(t-\tau)\right)\sin\left(\left\{E_{k}-\frac{(E_n+E_m)}{2}\right\}(t-\tau)\right)}. \nonumber\\
        && \eea   
     There exists a lot of integrable and non integrable models in the context of quantum statistical mechanics which can be written in terms of its eigenstate basis. For these class of models using the above mentioned general results one can able to compute the expressions for the desired non-chaotic OTOCs as well the normalised four-point functions in the simplest eigenstate representation of the Hamiltonian of the system under consideration.  We will not explicitly compute any result for a specific quantum model in this section as our prime objective is compute the expressions for the non-chaotic OTOCs and the associated normalised four-point functions in the context of quantum field theory of curved space-time, particularly for spatially flat FLRW cosmological background space-time where the eigenstate formalism of representing OTOC is not applicable any more.  Few more things here we have to point for our understanding purpose of the derived correlators in the eigenstate representation.  It is important to note that,  the magnitudes or the amplitudes of the desired micro-canonical part of the two-point functions in the present context varying sinusoidally with both the time scale associated with any quantum mechanical system under consideration in its eigenstate representation. Further once we include the contribution from the thermal Boltzmann factor and take a sum over all eigenstates of the Hamiltonian of the quantum mechanical system we get an overall exponential fall in the amplitude of the two-point function. It is expected from these expression that at very low temperature the fall in the amplitude is very large. On the contrary, at very high temperature the suppression in the Boltzmann factor and consequently in the overall amplitude of the two-point function will be small.  We can see that all of these expressions for the canonical two point functions explicitly dependent on the information of the particular quantum mechanical system under consideration through the matrix elements of the canonically conjugate variable $q$ at time scale $t=0$ and $\tau=0$ and the difference in the energy levels of the eigenstates.  Both of these contributions are time and equilibrium temperature independent.  So this implies that the overall behaviour of these two point OTOCs are controlled by the canonical thermal Boltzmann factor and the micro-canonical time dependent correlation functions.  So it is expected that the overall cumulative contribution of the amplitude of the two-point function show decaying sinusoidal oscillating behaviour with respect to the time scales under considerations for this study. Now we will comment on the overall behaviour of the four-point functions. Like previously mentioned for the two-point function case,  in the present context also we can interpret the overall amplitude of the un-normalized OTOCs are made up of two important contributions. These are the canonical thermal Boltzmann factor and the micro-canonical part of these four-point of the correlators which are appearing from the square of the commutator bracket. From the derived expressions for the desired four-point un-normalized OTOCs one can express the micro-canonical part of the quantum correlation function in terms of the product of two sine functions with different frequencies along with the contributions from the matrix elements of the canonical quantum operator $q$ and its conjugate momentum operator $p$ at the time scales, $t=0$ and $\tau=0$. If we look into closely to the expressions then one can find that the overall amplitude of the un-normalized four-point OTOCs will be decaying due to the presence of the time independent exponential thermal Boltzmann factor and rest of the contribution is oscillatory and time dependent. This particular generic time dependent part pointing towards the fact that these pair of four-point OTOCs in which we are interested in this paper will not contribute to describe the phenomena of quantum mechanical chaos as these correlators give oscillatory decaying amplitude which will never be negative. However, once we include the contribution from normalization factor we get a non-chaotic random fluctuating dissipating behaviour from the normalized four-point OTOCs.
 \subsection{Constructing non-chaotic OTOC in Cosmology}
 
For $(3+1)$ dimensional spatially flat FLRW space time the infinitesimal line element in the conformally flat coordinate is described by:
  \bea {\textcolor{red}{\bf Spatially~flat~FLRW~cosmological~metric:}~~~~ds^2=a^2(\tau)\left(-d\tau^2+d{\bf x}^2\right)}.~~~~\eea
  Here, $a^2(\tau)$ is the overall conformal factor which is actually playing the role of scale factor in conformal coordinate system. 
  Here we have introduced the concept of conformal time which can be expressed for different patches of the FLRW universe as: 
\begin{eqnarray}
{\tau=
\int\frac{dt}{a(t)}= \large \left\{
     \begin{array}{lr}
 \displaystyle  -\frac{1}{Ha(\tau)} &~~\text{\textcolor{red}{\bf De~Sitter}}\\ 
  \displaystyle   \frac{3(1+w_{reh})}{(1+3w_{reh})}\left[a(\tau)\right]^{\frac{(1+3w_{reh})}{2}},~~~{\rm where}~~0\leq w_{reh}\leq \frac{1}{3} & \text{\textcolor{red}{\bf Reheating}}  \end{array}
   \right.}~~~~~~~~~
\end{eqnarray}
where the scale factor in the conformal coordinate can be expressed as:
\begin{eqnarray}
{a(\tau)= \large \left\{
     \begin{array}{lr}
 \displaystyle -\frac{1}{H\tau}&~~~~~~~~~~~ \text{\textcolor{red}{\bf De~Sitter}}\\ 
  \displaystyle   \left[\frac{(1+3w_{reh})}{3(1+w_{reh})}\tau\right]^{\frac{2}{(1+3w_{reh})}}~~~{\rm where}~~0\leq w_{reh}\leq \frac{1}{3} & \text{\textcolor{red}{\bf Reheating}}  \end{array}
   \right.}~~~~~~~~~
\end{eqnarray}

%%%%%%%%%%%%%%%%
\subsubsection{For massless scalar field}
%%%%%%%%%%%%%%%% 
%%%%%%%%%%%%%%%%

For massless case the scalar field action in spatially flat FLRW background can be written in the conformal coordinate as:
\bea {S=-\frac{1}{2}\int d^{4}x~\sqrt{-g}~\left(\partial\phi\right)^2=\frac{1}{2}\int d\tau~d^3{\bf x}~a^2(\tau)~\left[\left(\partial_{\tau}\phi({\bf x},\tau)\right)^2-\left(\partial_{i}\phi({\bf x},\tau)\right)^2\right]},~~~~~\eea 
from which one can further compute the Hamiltonian for the massless scalar field in spatially flat FLRW background using conformal coordinate as:
\bea {H(\tau)=\int d^3{\bf x}~\left[\frac{1}{2a^2(\tau)}\Pi^2({\bf x},\tau)+\frac{a^2(\tau)}{2}\left(\partial_{i}\phi({\bf x},\tau)\right)^2\right]~~~~{\rm where}~~\Pi({\bf x},\tau)=\partial_{\tau}\phi({\bf x},\tau)}.~~~~~~~~~~~~\eea
In this case the non-chaotic OTOC's for the massless case are given by:
\bea && {\textcolor{red}{\bf 2-point~OTOC_1:}~~~~ Y_1(t,\tau)=-\langle \left[\phi(t),\phi(\tau)\right]\rangle_{\beta}},\\
&& {\textcolor{red}{\bf 2-point~OTOC_2:}~~~~ Y_2(t,\tau)=-\langle \left[\tilde{\Pi}(t),\tilde{\Pi}(\tau)\right]\rangle_{\beta}},\\
&& {\textcolor{red}{\bf 4-point~OTOC_1:}~~~~{\cal C}_1(t,\tau)=-\frac{\langle \left[\phi(t),\phi(\tau)\right]^2\rangle_{\beta}}{\langle \phi(t) \phi(t)\rangle_{\beta}\langle\phi(\tau)\phi(\tau)\rangle_{\beta}}},\eea
\bea
&& {\textcolor{red}{\bf 4-point~OTOC_2:}~~~~{\cal C}_2(t,\tau)=-\frac{\langle \left[\widetilde{\Pi}(t),\widetilde{\Pi}(\tau)\right]^2\rangle_{\beta}}{\langle \widetilde{\Pi}(t) \widetilde{\Pi}(t)\rangle_{\beta}\langle \widetilde{\Pi}(\tau)\widetilde{\Pi}(\tau)\rangle_{\beta}}}.\eea
During inflation (when we fix $d=3$) we actually consider massless scalar field i.e. $m<<H$. In this context one needs to consider the following perturbation in the scalar field in the De Sitter background:
\bea {\textcolor{red}{\bf Field~perturbation:}~~~\phi({\bf x},\tau)=\underbrace{\phi(\tau)}_{\textcolor{red}{\bf Background~field ~in~FLRW}}+\underbrace{\delta\phi({\bf x},\tau)}_{\textcolor{red}{\bf Perturbation~in~FLRW}}}~~~~~~~~~\eea
to express the whole dynamics in terms of a gauge invariant description through a variable:
\bea{\footnotesize{\textcolor{red}{\bf Perturbation~variable:}~~~~\zeta({\bf x},\tau)=-\underbrace{\frac{{\cal H}(\tau)}{\displaystyle\left(\frac{d\phi(\tau)}{d\tau}\right)}}_{\textcolor{red}{\bf Background~contribution}}\underbrace{\delta\phi({\bf x},t)}_{\textcolor{red}{\bf Perturbation~in~FLRW}}}}~.~~~~~~~~\eea 
 
At the level of first order perturbation theory in a spatially flat FLRW background metricwe fix the following gauge constraints:
\bea {\delta\phi({\bf x},\tau)=0,~~
 g_{ij}({\bf x},\tau)= a^2(\tau)\left[\left(1+2\zeta({\bf x},\tau)\right)\delta_{ij}+h_{ij}({\bf x},\tau)\right],~~
\partial_{i}h_{ij}({\bf x},\tau)=0=h^{i}_{i}({\bf x},\tau)}.~~~~~~~~\eea
which fix the space-time re-parametrization in FLRW spatially flat background. In this gauge, the spatial curvature of constant hyper-surface vanishes, which implies curvature perturbation variable is conserved outside the horizon. 

Applying the ADM formalism one can further compute the second order perturbed action for scalar modes can be expressed by the following action after gauge fixing:
\bea {S=\frac{1}{2}\int d\tau~d^3{\bf x}~\underbrace{\frac{a^2(\tau)}{{\cal H}^2}\left(\frac{d{\phi}(\tau)}{d\tau}\right)^2}_{\textcolor{red}{\bf Overall~time~dependent~factor}}\underbrace{\left[\left(\partial_{\tau}\zeta({\bf x},\tau)\right)^2-\left(\partial_{i}\zeta({\bf x},\tau)\right)^2\right]}_{\textcolor{red}{\bf Perturbed~kinetic~term~in~first~order}}}.\eea
Now we introduce a new variable in cosmological perturbation theory, which is known as {\it Mukhanov Sasaki} variable:
\bea {\textcolor{red}{\bf Mukhanov~Sasaki~variable:}~~f({\bf x},\tau)\equiv z(\tau)\zeta({\bf x},\tau),~~{\rm with}~~z(\tau)=\frac{a(\tau)}{{\cal H}}\frac{d{\phi}(\tau)}{d\tau}},~~~~~~~~\eea 
where, ${\cal H}=d\ln a(\tau)/d\tau$ is the Hubble parameter in the conformal coordinate. Here this new perturbation field variable serves the purpose of field redefinition in this context. Translating in terms of the conformal time and applying integration by parts the above mentioned action can be recast in the following form:
\bea {S=\frac{1}{2} \int d\tau ~d^3x~\left[\left(\partial_{\tau}f({\bf x},\tau)\right)^2-\left(\partial_{i}f({\bf x},\tau)\right)^2+\frac{1}{z(\tau)}\frac{d^2z(\tau)}{d\tau^2}\left(f({\bf x},\tau)\right)^2 \right]}~~.\eea
Now we transform this problem in Fourier space by making use of the following Fourier transform on the rescaled field variable:
\bea {\textcolor{red}{\bf Fourier ~transformation~in~momentum~space:}~~~f({\bf x},\tau)=\int \frac{d^3{\bf k}}{(2\pi)^3}f_{\bf k}(\tau)\exp(i{\bf k}.{\bf x})}~,~~~~~~~~~\eea
using which the action expressed in coordinate space can be recast in the momentum space as:
\bea {\textcolor{red}{\bf Action~for~parametric~oscillator:}~~S=\frac{1}{2}\int d\tau ~d^3{\bf k}\left[\underbrace{\left|\partial_{\tau}f_{{\bf k}}(\tau)\right|^2}_{\textcolor{red}{\bf Kinetic~term}}-\underbrace{\omega^2_{\bf k}(\tau)\left|f_{\bf k}(\tau)\right|^2}_{\textcolor{red}{\bf Potential~term}}\right]}~~,~~~~~~~~\eea
where we define the effective frequency $\omega_{\bf k}(\tau)$ as:
\bea {\textcolor{red}{\bf Effective~frequency:}~~~~~~\omega^2_{\bf k}(\tau)\equiv k^2+m^2_{\rm eff}(\tau)~~~{\rm with}~~m^2_{\rm eff}(\tau)=-\frac{1}{z(\tau)}\frac{d^2z(\tau)}{d\tau^2}}~.~~~~~~~\eea
Here, $m_{\rm eff}(\tau)$ represents the conformal time dependent effective mass parameter which is appearing due to the cosmological perturbation of the cosmological spatially flat FLRW metric in the first order and correspondingly the second order contribution in the action for scalar curvature perturbation. Now one can further compute the expression for this time dependent effective mass parameter for different cosmological epochs as:
\begin{eqnarray}
&&{m^2_{\rm eff}(\tau)=-\frac{1}{z(\tau)}\frac{d^2z(\tau)}{d\tau^2}=-\frac{\left(\nu^2-\frac{1}{4}\right)}{\tau^2}= \large \left\{
     \begin{array}{lr}
\displaystyle  -\frac{2}{\tau^2} & \text{\textcolor{red}{\bf De~Sitter}}\\ \\
 \displaystyle      \frac{2(3w_{reh}-1)}{(1+3w_{reh})^2}\frac{1}{\tau^2},0\leq w_{reh}\leq \frac{1}{3} & \text{\textcolor{red}{\bf Reheating}}  \end{array}
   \right.}~~~~~~~~~
\end{eqnarray}
 Consequently, the Hamiltonian can be expressed in Fourier transformed space as:
\bea {~~H=\int d^3{\bf k}~\overbrace{\left[\underbrace{\frac{1}{2}|{\Pi}_{\bf k}(\tau)|^2}_{\textcolor{red}{\bf Kinetic~term}}+~\underbrace{\frac{1}{2}\omega^2_{\bf k}(\tau){|f_{\bf k}(\tau)|^2}}_{\textcolor{red}{\bf Potential~term}}\right]}^{\textcolor{red}{\bf Fourier~transformed~Hamiltonian~density~\equiv~{\cal H}_{\bf k}(\tau)}},~~~~{\rm where}~~\Pi_{\bf k}(\tau)=\partial_{\tau}f_{\bf k}(\tau)}~,~~~~~~~~~\eea
where, the perturbed field satisfy the constraint, 
$f_{-\bf k}(\tau)=f^{\dagger}_{\bf k}(\tau)=f^{*}_{\bf k}(\tau)$.

Now, in terms of scalar curvature perturbation variable the following interesting OTOCs can be computed explicitly: 
\bea && {\textcolor{red}{\bf 2-point~OTOC:}~~~~ Y^{\zeta}_1(\tau_1,\tau_2)=-\langle \left[\zeta(\tau_1),\zeta(\tau_2)\right]\rangle_{\beta}},\\
&& {\textcolor{red}{\bf 2-point~OTOC:}~~~~ Y^{\zeta}_2(\tau_1,\tau_2)=-\langle \left[{\Pi}_{\zeta}(\tau_1),{\Pi}_{\zeta}(\tau_2)\right]\rangle_{\beta}},\\
&&{\textcolor{red}{\bf 4-point~OTOC:}~~~~ {\cal C}^{\zeta}_1(\tau_1,\tau_2)=-\frac{\langle \left[\zeta(\tau_1),\zeta(\tau_2)\right]^2\rangle_{\beta}}{\langle \zeta(\tau_1) \zeta(\tau_1)\rangle_{\beta}\langle \zeta(\tau_2)\zeta(\tau_2)\rangle_{\beta}}},\\
&&{\textcolor{red}{\bf 4-point~OTOC:}~~~~ {\cal C}^{\zeta}_2(\tau_1,\tau_2)=-\frac{\langle \left[{\Pi}_{\zeta}(\tau_1),{\Pi}_{\zeta}(\tau_2)\right]^2\rangle_{\beta}}{\langle {\Pi}_{\zeta}(\tau_1) {\Pi}_{\zeta}(\tau_1)\rangle_{\beta}\langle {\Pi}_{\zeta}(\tau_2){\Pi}_{\zeta}(\tau_2)\rangle_{\beta}}}.\eea 
Similarly, in terms of rescaled perturbation variable the OTOCs can be recast in the following forms:
\bea && {\textcolor{red}{\bf 2-point~OTOC:}~~~~ Y^{f}_1(\tau_1,\tau_2)=-\langle \left[f(\tau_1),f(\tau_2)\right]\rangle_{\beta}},\\
&& {\textcolor{red}{\bf 2-point~OTOC:}~~~~ Y^{f}_2(\tau_1,\tau_2)=-\langle \left[{\Pi}_{f}(\tau_1),{\Pi}_{f}(\tau_2)\right]\rangle_{\beta}},\\
&& {\textcolor{red}{\bf 4-point~OTOC:}~~~~ {\cal C}^{f}_2(\tau_1,\tau_2)=-\frac{\langle \left[f(\tau_1),f(\tau_2)\right]^2\rangle_{\beta}}{\langle f(\tau_1) f(\tau_1)\rangle_{\beta}\langle f(\tau_2)f(\tau_2)\rangle_{\beta}}}\\
&& {\textcolor{red}{\bf 4-point~OTOC:}~~~~ {\cal C}^{f}_2(\tau_1,\tau_2)=-\frac{\langle \left[{\Pi}_{f}(\tau_1),{\Pi}_{f}(\tau_2)\right]^2\rangle_{\beta}}{\langle {\Pi}_{f}(\tau_1) {\Pi}_{f}(\tau_1)\rangle_{\beta}\langle {\Pi}_{f}(\tau_2){\Pi}_{f}(\tau_2)\rangle_{\beta}}}.\eea
For both the cases we use $\alpha$ vacua and Bunch Davies vacuum as quantum vacuum state.

%%%%%%%%%%%%%%%%
\subsubsection{For partially massless scalar field}

%%%%%%%%%%%%%%%%

For partially massless case the scalar field action in spatially flat FLRW background can be written in the conformal coordinate as:
\bea && {S=-\frac{1}{2}\int d^{4}x~\sqrt{-g}~\left[\left(\partial\phi\right)^2-(cH\phi)^2\right]}\nonumber\\
&&~~~{=\frac{1}{2}\int d\tau~d^3{\bf x}~a^2(\tau)~\left[\left(\partial_{\tau}\phi({\bf x},\tau)\right)^2-\left(\partial_{i}\phi({\bf x},\tau)\right)^2-\left(c{\cal H}\phi({\bf x},\tau)\right)^2\right]},~~~~~~~~\eea
and the corresponding Hamiltonian can be expressed in the conformal coordinate as:
\beq {H(\tau)=\int d^3{\bf x}~\left[\frac{1}{2a^2(\tau)}\Pi^2({\bf x},\tau)+\frac{a^2(\tau)}{2}\left\{\left(\partial_{i}\phi({\bf x},\tau\right)^2+c^2{\cal H}^2\phi^2({\bf x},\tau)\right\}\right]~{\rm where}~\Pi({\bf x},\tau)=\partial_{\tau}\phi({\bf x},\tau)}.~~~~~~~~\eeq
In this specific scenario, the conformal time dependent effect mass parameter for the partially massless scalar field in spatially flat FLRW cosmological background can be expressed as:
\begin{eqnarray}
&& {m^2_{\rm eff}(\tau)=-\frac{1}{z(\tau)}\frac{d^2z(\tau)}{d\tau^2}=-\frac{\left(\nu^2-\frac{1}{4}\right)}{\tau^2}= \large \left\{
     \begin{array}{lr}
   \displaystyle \left(c^2-2\right)\frac{1}{\tau^2},~~~{\rm where}~~c\geq \sqrt{2} &~ \text{\textcolor{red}{\bf De~Sitter}}\\ 
   \displaystyle    \frac{2(3w_{reh}-1)}{(1+3w_{reh})^2}\frac{1}{\tau^2},~0\leq w_{reh}\leq \frac{1}{3} & \text{\textcolor{red}{\bf Reheating}}  \end{array}
   \right.}~~~~~~~~~
\end{eqnarray}
Since the expression for the conformal time dependent effective mass parameter explicitly appearing in the expression for the effective conformal time dependent frequency parameter the rest of the computations will be automatically modified accordingly. Further using previously defined four set of OTOCs in the previous case here also one can compute the expression for the non-chaotic sets of quantum mechanical correlators.

%%%%%%%%%%%%%%%%
\subsubsection{For massive scalar field}

%%%%%%%%%%%%%% %%

For partially massless case the scalar field action in spatially flat FLRW background can be written in the conformal coordinate as:
\bea && S=-\frac{1}{2}\int d^{4}x~\sqrt{-g}~\left[\left(\partial\phi\right)^2-m^2_{\phi}\phi^2\right]=\frac{1}{2}\int d\tau~d^3{\bf x}~a^2(\tau)~\left[\left(\partial_{\tau}\phi({\bf x},\tau)\right)^2-\left(\partial_{i}\phi({\bf x},\tau)\right)^2-a^2(\tau)m^2_{\phi}\phi^2({\bf x},\tau)\right],\nonumber\\
&&\eea
and the corresponding Hamiltonian can be expressed in the conformal coordinate as:
\beq {H(\tau)=\int d^3{\bf x}~\left[\frac{1}{2a^2(\tau)}\Pi^2({\bf x},\tau)+\frac{a^2(\tau)}{2}\left\{\left(\partial_{i}\phi({\bf x},\tau\right)^2+m^2_{\phi}\phi^2({\bf x},\tau)\right\}\right]~~{\rm where}~~\Pi({\bf x},\tau)=\partial_{\tau}\phi({\bf x},\tau)}.~~~~~~~~\eeq
In this specific scenario, the conformal time dependent effect mass parameter for the partially massless scalar field in spatially flat FLRW cosmological background can be expressed as:
\begin{eqnarray}
&& {m^2_{\rm eff}(\tau)=-\frac{1}{z(\tau)}\frac{d^2z(\tau)}{d\tau^2}=-\frac{\left(\nu^2-\frac{1}{4}\right)}{\tau^2}= \large \left\{
     \begin{array}{lr}
  \displaystyle  \left(\frac{m^2_{\phi}}{H^2}-2\right)\frac{1}{\tau^2},~m_{\phi}\gg H &~\text{\textcolor{red}{\bf De~Sitter}}\\ 
 \displaystyle      \frac{2(3w_{reh}-1)}{(1+3w_{reh})^2}\frac{1}{\tau^2},~~0\leq w_{reh}\leq \frac{1}{3} & \text{\textcolor{red}{\bf Reheating}}  \end{array}
   \right.}~~~~~~~
\end{eqnarray}
This expression will contribute to the effective conformal time dependent frequency parameter and consequently the rest of the computations will be automatically modified accordingly.  Further using previously defined four set of OTOCs here also one can compute the expression for the non-chaotic sets of quantum mechanical correlators in the context of massive scalar field.  There might be another possibility to have a conformal time dependent profile for the massive scalar field in the spatially flat FLRW geometrical background in the context of cosmology. Now it is important to note that, it is not necessarily to include such time dependence explicitly in the mass profile of the heavy scalar field. But if we consider this possibility then it is possible to explore many more unexplored areas of theoretical physics and its underlying connection with primordial aspects of cosmology.  One can establish a connection with condensed matter physics, quantum aspects of statistical mechanics and quantum entanglement, violation of Bell's inequality and the generation of long range quantum mechanical correlation appearing in the context of quantum information aspects of cosmology.  This possibility is out of our scope of study at present in this paper.  For this reason we will only look into the massive field $m_{\phi}\gg H$, using which we will explore few other features, which is also show some new direction in the context of quantum field theory of curved space-time, particularly in the context of spatially flat FLRW cosmology,  which is quite consistent with the observational aspects as well.  Using the present set up our objective is to explore the behaviour of the quantum correlation functions in the out-of-equilibrium regime of the quantum field theory of primordial cosmology appearing in the early time scale of the evolution of our universe.  Not only we will restrict ourself to describe the out-of equilibrium feature in the early time scale of our universe, but also using the present computation and methodology we will probe the late large time equilibrium features of the mentioned quantum correlation functions in the context of primordial cosmology.  By studying the conformal time dependent behaviour of these two sets of OTOCs we can surely study the quantum mechanical random fluctuating behaviour of cosmological correlators starting from a very early time to the late time scale of our universe. Using this methodology one can further quantify the quantum correlation function in the context of reheating epoch as well as for the stochastic particle production during the epoch of inflationary paradigm, where in both the cases the phenomena of out-of-equilibrium physics play significant role at the very early time scale. on the other hand, at very late time scale of the evolution history of our universe the quantum random fluctuation in the cosmological correlation shows equilibrium feature and from this one can further estimate the corresponding equilibrium temperature. For this reason,  in the rest of our paper we will explicitly compute these results and will show that how the present methodology can be applicable to the mentioned epochs of our universe.   

%%%%%%%%%%%%%%%%
\section{Quantum non-chaotic auto-correlated OTO amplitudes and OTOC in Primordial Cosmology}
\label{sec:2}
%%%%%%%%%%%%%%%% 
In this section, our prime objective is to explicitly derive and study the physical outcomes of two different types of OTOCs which are constructed from field variable and its canonically conjugate momenta appearing in the context of cosmological perturbation theory written in spatially flat FLRW background.  We will show from our computation that these set of OTOCs play significant role to quantify the effect of random fluctuations in the quantum regime.  Most importantly, our main claim in this paper is that these set of new OTOCs in the context of primordial set up of cosmology will describe the non-chaotic time dependent behaviour with respect the conformal time scale, which is obviously a new information in this literature and have not explicitly studied earlier. So in this paper we will completely devote our full energy to justify this big claim through the present computation performed in this paper.  Before performing the detailed computation and going to the very technical details of the present topic let us first discuss the physical implications of these set of non-chaotic OTOCs in which we are interested in this paper, which are appended below point-wise:
\begin{itemize}
\item \underline{\textcolor{red}{\bf Motivation~1:}}\\
The first physical motivation of this work is to explicitly provide the result of quantum mechanical correlation function in the context of primordial cosmology when the system under consideration goes to out-of-equilibrium regime in the early time scale of the evolution history of our universe. We all know this particular regime in quantum statistical field theory is extremely complicated to probe through usual understandings cosmological correlators defined in the same late time slice.  On the other hand, in the spatially flat FLRW curved background explaining such phenomena even more harder. Our prime objective is to probe this extremely complicated physics through some basic and very simple concepts of quantum field theory prescription. This methodology is applicable to explain particularly the quantum mechanical correlation functions in the epoch of reheating and during the the stochastic particle production procedure during inflation which was not explored in this literature earlier. We are very hopeful that our computation and the derived results can perfectly able to capture the phenomena of quantum randomness and in this way one can treat the OTOCs to be the significant theoretical probe through which one can explain the out-of-equilibrium features very easily.

\item \underline{\textcolor{red}{\bf Motivation~2:}}\\
The second physical motivation of this work is to explicitly provide and understand the equilibrium behaviour of the quantum mechanical correlation functions in the context of primordial cosmology which we have a plan to derive in terms of the previously mentioned OTOCs. When we give an initial response to a quantum system in the context of cosmology it goes to the out-of-equilibrium regime. Now if we wait for long and consider the late time regime in the evolutionary scale of cosmology then it is expected from the basic understanding that the quantum mechanical system under consideration has to reach the state of thermodynamic equilibrium which can be understood in terms of the saturation of the desired set of OTOCs at the late time scale. Our expectation is that to understand this phenomena within the framework of cosmological perturbation theory in the early universe which will be also helpful to estimate the thermodynamic temperature of the equilibrium state from the saturation value of the desired OTOCs in the present context. We are very hopeful to provide the details of the equilibrium behaviour of the quantum OTOCs from cosmology at the late time scale which will be be helpful to give a physically consistent interpretation and estimation of the reheating temperature in terms of the equilibrium temperature of the quantum mechanical system under consideration. Apart from this, one can further give a physical interpretation and estimation of the equilibrium temperature at the end of the stochastic particle production procedure during the epoch of inflation. So using the late time behaviour of the cosmological desired OTOCs one cam give the estimation of reheating temperature as well the temperature of the universe at the end of inflation by following a proper physically consistent theoretical framework, which will be further helpful to determine the energy scales of that specific epoch appearing in the evolutionary time scale of our universe.

\item \underline{\textcolor{red}{\bf Motivation~3:}}\\
The third and the last physical motivation of this work is to explicitly provide and understand the intermediate behaviour of the quantum mechanical correlation functions in the context of primordial cosmology which we have a plan to derive in terms of the previously mentioned OTOCs. Our expectation is the present structure of OTOCs not only provide the physical understanding of the early time out-of-equilibrium and late time equilibrium phenomena in the context of primordial cosmological perturbation theory, but also provide the correct and physically justifiable explanation of the intermediate behaviour of the quantum correlation functions which will able to explain the phase transition from the out-of-equilibrium regime to the equilibrium regime. As far as the previous literatures in the present context are concerned people have not addressed this issue clearly as in the earlier computations implementation of this phase transition phenomena was extremely complicated. But from the present computation our one of the great expectations is to address the intermediate behaviour of the cosmological quantum correlation functions through the desired expressions for OTOCs which can further give physical interpretation during the phase transition from the out-of-equilibrium regime to the equilibrium regime and the quantum randomness behaviour in the vicinity of the transition region within the framework of primordial cosmological perturbation theory.
\end{itemize} 

\subsection{Computational strategy for non-chaotic auto-correlated OTO functions}

The steps of the specifically computing the desired non-chaotic OTOCs defined in this paper are appended below point-wise:
\begin{enumerate}
\item To compute these OTOCs at first we need find out the analytic classical solution of the equation of motion of the massless, partially massless and heavy scalar fields which can describe the super-horizon, sub-horizon limit and horizon crossing phenomena in the spatially flat FLRW cosmological background.
\item After obtaining the classical solution which can describe the physical phenomena in the asymptotic region- super-horizon and sub-horizon scale in a specific gauge of cosmological perturbation theory of early universe we further have to find out the canonically conjugate momentum of the perturbation field variable. Further using both of these classical solution our job is to promote them in the quantum regime by writing them in terms of the creation and annihilation operators. Once this job is done then using that quantum extended operators our job is to compute the square of the quantum mechanical commutator brackets appearing in the expressions for the two desired OTOCs on which we are interested in this paper. It is important to note that, such square of the commutator brackets are of course the fundamental components which will play significant role to quantify the desired OTOCs.
\item Once we know the full solution in the quantum regime, which can able to describe super-horizon and sub-horizon, the asymptotic limiting solutions in presence of a preferred choice of quantum mechanical vacuum state, which fix the initial condition for the quantum random fluctuations in the context of early universe cosmology then using that we have to explicitly compute the expression for the thermal average value of the square of the commutator bracket of the field and its canonically conjugate momenta, which can be obtained as a by-product of cosmological perturbation theory in a preferred choice of gauge. Such combinations physically signify the four-point quantum correlation function in the out-of-time ordered sense.
\item Further we have to derive the expression for another fundamental quantity, which is the partition function of the quantum mechanical system under consideration for the cosmological perturbation theoretical set up. This can be done by determining the quantum mechanical Hamiltonian which actually representing a quantum oscillator with conformal time dependent frequency. Once this is done then we need to compute the expression for the thermal trace of the thermodynamic Boltzmann factor. Here it is important to note that, in the context of quantum field theory the equivalent representation of the thermal trace operation can be presented in terms of path integral operation which can be performed in presence of wave function of the universe defined for a preferred choice of quantum mechanical vacuum state.
\item Before computing the mentioned OTOCs which are related to connected part of the four-point correlation functions we need derive the expression for the thermal two point OTOCs from the perturbation variable appearing in the preferred choice of gauge in cosmological perturbation theory of early universe~\footnote{Here it is important to note that, for the four-point OTOCs we have observed that at very large dissipation time scale one can factorize them into connected and disconnected parts. Obviously connected part carries the physical information, but the disconnected part is actually written in terms of the product of two two point functions defined at two different time scale which are separated in time scale. Now these two point functions helps us to normalize these two OTOCs which further is very significant to analyze the amplitude of the connected part of the desired four-point functions in the present context. Also, the thermal expectation value of the commutator bracket of two cosmologically relevant operators in the context of cosmological perturbation theory can be considered to be the relative difference between the thermal expectation value of two different two point functions which are appearing in the normalization of the disconnected part of the four-point functions as mentioned earlier.}. These two point functions further have used to normalize the four point desired OTOCs in the present paper and for this reason can be treated as the fundamental building block in the present computational purpose.
\item The quantum operator appearing as a by-product of cosmological perturbation theory which are the fundamental objects used to define the desired expressions for the OTOCs in this paper we first use the coordinate system basis. But to make the further simplification we transform both of the quantum operators defined in two different time scales in to Fourier space where we write both of them in the momentum basis. We particularly use this simple trick because in the context of cosmology momentum space description is very simpler and easily understandable compared to the expressions obtained in the context of coordinate space basis. The prime physical reason behind this approach is that the quantum correlation functions in the context of primordial cosmological perturbation theory preserves the overall mathematical structure under the application of $SO(1,4)$ conformal transformations~\footnote{Here $SO(1,4)$ represents the De Sitter isommetry group which are preserved under conformal transformations.}. But in the context of quasi De Sitter space this symmetry is slightly broken in the momentum basis after doing Fourier transformation from the coordinate basis and the amount of symmetry breaking is proportional to the slowly varying time dependent slow-roll parameters, which we have explicitly taken into account to get the correct physical interpretation of the final results of the desired OTOCs in our computations in this paper.
\item Another important ingredient of the present computation of the desired two OTOC is deep rooted in the cosmological perturbation theory setup used for early universe quantum mechanical set up that we are using for the present computation. In this context, the preferred choice of gauge used for the perturbation set up play a very significant role to finally quantify the mathematical structure of the OTOCs studied in this paper. For this purpose, unitary gauge, Newtonian gauge, comoving $\delta\phi=0$ gauge choices are very relevant and most commonly used. Out of them in the present computational purpose we use the comoving $\delta\phi=0$ gauge, which make the final mathematical structure and the detailed computation of the desired OTOCs very simpler compared to the performing the derivation in terms of the field $\phi$ and its canonically conjugate momentum $\Pi_{\phi}$ which are appearing directly from the ungauged version of the cosmological action in spatially flat FLRW background.

\end{enumerate}

\subsection{Classical mode functions to compute non-chaotic auto-correlated OTO functions in Cosmology}
After varying the gauged version of the second order action as appearing in the context of cosmological perturbation theory in the preferred $\delta\phi=0$ gauge and writing in terms of the redefined perturbation field variable in the momentum space after doing Fourier transformation we get the following simplified version of the equation of motion of the classical field: 
\bea &&\textcolor{red}{\bf Mukhanov~Sasaki~equation~(Classical ~EOM):}\frac{d^2f_{\bf k}(\tau)}{d\tau^2}+\underbrace{\left(k^2\overbrace{-\frac{\nu^2-\frac{1}{4}}{\tau^2}}^{\textcolor{red}{\bf \equiv m^2_{\rm eff}(\tau)}}\right)}_{\textcolor{red}{\bf \equiv~ \omega^2_{k}(\tau)}}f_{\bf k}(\tau)=0,~~~~~\eea  
which is commonly known as the {\it Mukhanov Sasaki equation}. It is actually a second order differential equation in conformal time coordinate and in momentum space $k$ plays just the role of a constant parameter. If we closely look into the equation of motion then we see that it is basically representing a parametric oscillator with conformal time dependent frequency where the conformal time is playing the role of parameter in this context. Since here the equation of motion of the Fourier mode of the rescaled scalar perturbation is second order homogeneous differential equation in conformal time coordinate then the final solution of this equation can be written in terms of the sum of two linearly independent solutions. For this reason, 
two arbitrary constants are also appearing in the total solution which can only be fixed by the initial choice of the quantum vacuum state. To serve this purpose it us very common practice in this literature to choose the well known Euclidean vacuum state, which is actually a false vacuum state in nature. In common practice this Euclidean false vacuum state is known as the Bunch Davies vacuum. There is an additional possibility regarding the choice of the vacuum state, which is the most general $\alpha$-vacua states and $\alpha=0$ can reproduce the solution for the Bunch Davies vacuum state. In this equation of motion we define the mass parameter variable $\nu$, which is defined for the massless, partially massless and massive heavy scalar field cases in De Sitter space and for reheating phase can be written as:
\begin{eqnarray}
\displaystyle{\displaystyle\nu_{ DS}:=\nu= \large \left\{ 
     \begin{array}{lr}
\displaystyle    \frac{3}{2} &~ \text{\textcolor{red}{\bf DS+massless}}\\ 
    \displaystyle \sqrt{\frac{9}{4}-c^2},~~~{\rm where}~~c\geq \sqrt{2} ~& \displaystyle\text{\textcolor{red}{\bf DS+partially~massless}}\\ 
  \displaystyle   i\sqrt{\frac{m^2_{\phi}}{H^2}-\frac{9}{4}},~~~{\rm where}~~m_{\phi}\gg H &\text{\textcolor{red}{\bf DS+heavy}} \end{array}
   \right.}~~~~~~
\\
\displaystyle{\displaystyle\nu_{reh}:=\nu= \large \left\{ 
     \begin{array}{lr}
\displaystyle  \sqrt{ \frac{1}{4}+\frac{2(1-3w_{reh})}{(1+3w_{reh})^2}}&~ \text{\textcolor{red}{\bf Reheating+massless}}\\ 
    \displaystyle \sqrt{ \frac{1}{4}+\frac{2(1-3w_{reh})}{(1+3w_{reh})^2}}~~~~~& \displaystyle\text{\textcolor{red}{\bf Reheating+partially~massless}}\\ 
  \displaystyle  \sqrt{ \frac{1}{4}+\frac{2(1-3w_{reh})}{(1+3w_{reh})^2}},~ &\text{\textcolor{red}{\bf Reheating+heavy}}  \end{array}
   \right.}~~~~~~
\end{eqnarray}
where the equation of state parameter during reheating epoch is lying within the window, $0\leq w_{reh}\leq \frac{1}{3}$. Here it is important to note that for the reheating case the expression looks exactly same for the massless, partially massless and heavy scalar field case. The expression for the equation of state parameters for these three different cases are different, which implies the physical significance of each cases are different. But the overall magnitude of the equation of state parameter will lie within the above mentioned window.

 Now to proceed further with this equation of motion of the perturbed field variable for the scalar fluctuations we transform it in a convenient mathematical form which can be easily understood. We use the following canonical field redefinition to serve this purpose:
 \beq {\textcolor{red}{\bf Canonical~field~redefinition:}~~Q(x):=\frac{1}{\sqrt{-\tau}}f_{\bf k}(\tau)~~~{\rm with}~~x\equiv-k\tau}~.\eeq
 In this new field redefinition the {\it Mukhanov Sasaki equation} can be recast as:
 \beq {\textcolor{red}{\bf Canonical~form~of~Bessel's~equation:}~~~\frac{d^2Q(x)}{dx^2}+\frac{1}{x}\frac{dQ(x)}{dx}+\left(1-\frac{\nu^2}{x^2}\right)Q(x)=0}~.~~~~~~\eeq
 The most general solution of this {\it canonical form of Bessel's equation} can be written as a linear combination of the two linearly independent solutions as given by:
 \bea {\textcolor{red}{\bf Solution~for~Bessel's~equation:}~~Q(x)=\left[{\cal D}_1~{\cal J}_{\nu}(x)+{\cal D}_2~{\cal Y}_{\nu}(x)\right]},~~~~~\eea
 where, ${\cal J}_{\nu}(x)$ and ${\cal Y}_{\nu}(x)$ are the Bessel function of first and second kind of order $\nu$ respectively. Here ${\cal D}_1$ and ${\cal D}_2$ are the two arbitrary integration constants which are fixed by the choice of the initial quantum vacuum state. Sometimes in the quantum field theory literature these time independent constants are identified to be the {\it Bogoliubov coefficients} which one can able to compute explicitly in the present context.
 
 Further one can write the the Bessel function of first and second kind of order $\nu$ in terms of the Hankel function of first and second kind of order $\nu$:
 \bea {\cal J}_{\nu}(x)=\frac{1}{2}\left[H^{(1)}_{\nu}(x)+H^{(2)}_{\nu}(x)\right],\\
 {\cal Y}_{\nu}(x)=\frac{1}{2i}\left[H^{(1)}_{\nu}(x)-H^{(2)}_{\nu}(x)\right].\eea
 Here $H^{(1)}_{\nu}(-k\tau)$ and $H^{(1)}_{\nu}(-k\tau)$ are the Hankel functions of first and second kind with order $\nu$.
 
Finally, in terms of the Hankel functions the most general solution can be expressed as:
\bea {\textcolor{red}{\bf Solution~for~Bessel's~equation:}~~Q(x)=\left[{\cal C}_1~H^{(1)}_{\nu}(x)+{\cal C}_2~H^{(2)}_{\nu}(x)\right]},~~~~~\eea
where we introduce two new arbitrary constants, ${\cal C}_1$ and ${\cal C}_2$, which are defined in terms of the previously mentioned two integration constants ${\cal D}_1$ and ${\cal D}_2$ as:
\bea {\cal C}_1=\frac{1}{2}\left({\cal D}_1-i{\cal D}_2\right),~~~~~~
       {\cal C}_2=\frac{1}{2}\left({\cal D}_1+i{\cal D}_2\right).\eea
Consequently, the most general conformal time dependent solution of the scalar mode fluctuation equation of motion for any constant mass profile (massless, partially massless and heavy scalar production during inflation and during reheating) is represented by the following expression:
\bea {\textcolor{red}{\bf Solution~for~scalar~mode~function:}~~f_{\bf k}(\tau)=\sqrt{-\tau}\left[{\cal C}_1~H^{(1)}_{\nu}(-k\tau)+{\cal C}_2~H^{(2)}_{\nu}(-k\tau)\right]},~~~~~~~~~\eea
where ${\cal C}_1$ and ${\cal C}_2$ are the previously mentioned two arbitrary integration constants which are fixed by the choice of the initial quantum vacuum state necessarily needed for this computation.

The corresponding most general canonically conjugate momentum can be further computed from this derived solution as:
\bea 
\Pi_{\bf k}(\tau)=\partial_{\tau}f_{\bf k}(\tau)=\frac{1}{2 \sqrt{-\tau }}\left[{\cal C}_1\left( k \tau  H_{\nu -1}^{(1)}(-k \tau )- H_{\nu }^{(1)}(-k \tau )- k \tau  H_{\nu +1}^{(1)}(-k \tau )\right)\nonumber\right.\\ \left.~~~~~~~~~~~~~~~~~~~~~+{\cal C}_2 \left(k \tau  H_{\nu -1}^{(2)}(-k \tau )-H_{\nu }^{(2)}(-k \tau )- k \tau  H_{\nu +1}^{(2)}(-k \tau )\right)\right].\eea
Further considering, $-k\tau\rightarrow 0$ and $-k\tau\rightarrow \infty$ asymptotic limits one can write the following simplified form of the most general solution for the perturbed field and momentum variable can be expressed as:
\bea &&{f_{\bf k}(\tau)=\frac{2^{\nu-\frac{3}{2}}(-k\tau)^{\frac{3}{2}-\nu}}{\sqrt{2}k^{\frac{3}{2}}\tau}\left|\frac{\Gamma(\nu)}{\Gamma\left(\frac{3}{2}\right)}\right|}\nonumber\\
&&~~~~~~~~~~~~~~~~\times{\left[{\cal C}_1~(1+ik\tau)~\exp\left(-i\left\{k\tau+\Delta^{-}_{\nu}\right\}\right)-{\cal C}_2~(1-ik\tau)~\exp\left(i\left\{k\tau+\Delta^{+}_{\nu}\right\}\right)\right]},~~~~~~~~~~~\\
 &&\nonumber {\Pi_{\bf k}(\tau)=\frac{2^{\nu-\frac{3}{2}}(-k\tau)^{\frac{3}{2}-\nu}}{\sqrt{2}k^{\frac{5}{2}}}\left|\frac{\Gamma(\nu)}{\Gamma\left(\frac{3}{2}\right)}\right|}\left[{{\cal C}_1~\left\{1-\vartheta_{\nu}\frac{(1+ik\tau)}{k^2\tau^2}\right\}~\exp\left(-i\left\{k\tau+\Delta^{-}_{\nu}\right\}\right)}\right.\\&& \left.~~~~~~~~~~~~~~~~~~~~~~~~~~~~~~~~~~~~~~~{-{\cal C}_2~\left\{1-\vartheta_{\nu}\frac{(1-ik\tau)}{k^2\tau^2}\right\}~\exp\left(i\left\{k\tau+\Delta^{+}_{\nu}\right\}\right)}\right],~~~~~~~~~~~
\eea 
where we define the two phase factors $\Delta^{\pm}_{\nu}$ and a new function $\vartheta_{\nu}$ as:
 \bea && \Delta^{\pm}_{\nu}=\frac{\pi}{2}\left[\left(\nu+\frac{1}{2}\right)\pm 1\right],~~~~~ \vartheta_{\nu}=\left(\nu-\frac{1}{2}\right).\eea 
These results are extremely important for the computation of the desired OTOCs in which are interested in this paper which we have derived in the the later subsections. To server this purpose we need to first of all promote both of these classical solutions of the field and momentum to the quantum level. Since we are following the canonical technique in the present context for the quantization purpose, we need to express both of these classical solutions in terms of creation and annihilation operators which we will discuss in the next subsection of this paper.  See the details of the computation in Appendix~(\ref{sec:7}).

 \subsection{Quantum mode function to compute non-chaotic auto-correlated OTO functions in Primordial Cosmology}

Here in this context, the rescaled perturbation field operator and the corresponding canonically conjugate momentum operator in the quantum regime can be expressed in terms of the classical solutions obtained in the previous section as:
\bea &&{\textcolor{red}{\bf Field~Operator:}~~~~~~~~~~~~~~\hat{f}_{\bf k}(\tau)=f_{\bf k}(\tau)~a_{\bf k}+f^{*}_{\bf -k}(\tau)~a^{\dagger}_{-\bf k}},\\
&&{\textcolor{red}{\bf Momentum~Operator:}~~~~\hat{\Pi}_{\bf k}(\tau)=\Pi_{\bf k}(\tau)~a_{\bf k}+\Pi^{*}_{\bf -k}(\tau)~a^{\dagger}_{-\bf k}}.\eea
Here $a_{\bf k}$ and $a^{\dagger}_{-\bf k}$ are the annihilation and creation operators of the quantum vacuum state, which satisfy the following canonical commutation relations:
\bea &&\textcolor{red}{\bf Canonical~commutators:}~~\left[a_{\bf k},a^{\dagger}_{-\bf k'}\right]=(2\pi)^3\delta^{3}({\bf k}+{\bf k}'),~~\left[a_{\bf k},a_{-\bf k'}\right]=0=\left[a^{\dagger}_{\bf k},a^{\dagger}_{-\bf k'}\right]~~~~.~~~~~~\eea
Consequently the curvature perturbation and the corresponding momentum operator in the quantum regime can be re-expressed as:
\bea &&{ \hat{\zeta}_{\bf k}(\tau)=\frac{\hat{f}_{\bf k}(\tau)}{z(\tau)}=\frac{f_{\bf k}(\tau)~a_{\bf k}+f^{*}_{\bf -k}(\tau)~a^{\dagger}_{-\bf k}}{z(\tau)}=\zeta_{\bf k}(\tau)~a_{\bf k}+\zeta^{*}_{\bf -k}(\tau)~a^{\dagger}_{-\bf k}},\\
&& {\hat{\Pi}_{\zeta,{\bf k}}(\tau)=\partial_{\tau}\left(\frac{\hat{f}_{\bf k}(\tau)}{z(\tau)}\right)=\frac{ \hat{\Pi}_{\bf k}(\tau)}{z(\tau)}-\frac{\hat{\zeta}_{\bf k}(\tau)}{z(\tau)}\frac{dz(\tau)}{d\tau}}\nonumber\\
&&~~~~~~~~~~{=\left[\left(\Pi_{\zeta,{\bf k}}(\tau)~a_{\bf k}+\Pi^{*}_{\zeta,{\bf -k}}(\tau)~a^{\dagger}_{-\bf k}\right)-\left(\zeta_{\bf k}(\tau)~a_{\bf k}+\zeta^{*}_{\bf -k}(\tau)~a^{\dagger}_{-\bf k}\right)\left(\frac{1}{z(\tau)}\frac{dz(\tau)}{d\tau}\right)\right]}.~~~~~~~~~\eea  
Here the last term of the above mentioned expression can be further evaluated as:
\beq {\frac{1}{z(\tau)}\frac{dz(\tau)}{d\tau}=\frac{1}{a(\tau)}\frac{da(\tau)}{d\tau}\underbrace{-\frac{1}{{\cal H}}\frac{d{\cal H}}{d\tau}+\frac{1}{\displaystyle\left(\frac{d\phi(\tau)}{d\tau}\right)}\frac{d^2\phi(\tau)}{d\tau^2}}_{\textcolor{red}{\bf Sub-leading~contribution}}}.\eeq

\subsection{Quantum operators for non-chaotic auto-correlated OTO functions} 
In this subsection, our prime objective is to promote the classical Hamiltonian to the quantum regime in terms of canonical quantum operators. This quantum mechanical Hamiltonian operator can be expressed as: 
\beq { \textcolor{red}{\bf Quantum~Hamiltonian~Operator:}~~~\hat{H}(\tau):=\int d^3{\bf k}~\overbrace{\left[\underbrace{\frac{1}{2}\hat{\Pi}^2_{\bf k}(\tau)}_{\textcolor{red}{\bf Kinetic~term}}+~\underbrace{\frac{1}{2}\omega^2_{\bf k}(\tau){\hat{f}^2_{\bf k}(\tau)}}_{\textcolor{red}{\bf Potential~term}}\right]}^{\textcolor{red}{\bf Fourier~transformed~Hamiltonian-density~operator}}}~,\eeq 
Additionally, it is important to mention that in this quantum mechanical Hamiltonian operator the following property of the perturbed field and momentum holds good perfectly in Fourier space:
\bea \hat{f}^{\dagger}_{\bf k}(\tau)&=&\left(f_{\bf k}(\tau)~a_{\bf k}+f^{*}_{\bf -k}(\tau)~a^{\dagger}_{-\bf k}\right)^{\dagger}=\left(f_{\bf -k}(\tau)~a_{-\bf k}+f^{*}_{\bf k}(\tau)~a^{\dagger}_{\bf k}\right)
=\hat{f}_{-\bf k}(\tau),\\
 \hat{\Pi}^{\dagger}_{\bf k}(\tau)&=&\left(\Pi_{\bf k}(\tau)~a_{\bf k}+\Pi^{*}_{\bf -k}(\tau)~a^{\dagger}_{-\bf k}\right)^{\dagger}=\left(\Pi_{\bf -k}(\tau)~a_{-\bf k}+\Pi^{*}_{\bf k}(\tau)~a^{\dagger}_{\bf k}\right)=\hat{\Pi}_{-\bf k}(\tau).~~~~~\eea 
Here we use the following useful facts to satisfy the above mentioned Hermiticity constraints:
\bea && a^{\dagger}_{\bf k}=a_{-{\bf k}}~~~\Longrightarrow ~~~ (a^{\dagger}_{-{\bf k}})^{\dagger}=a^{\dagger}_{\bf k},\\ 
&& f^{\dagger}_{\bf k}(\tau)=f^{*}_{\bf k}(\tau)=f_{-\bf k}(\tau)~~~\Longrightarrow ~~~(f^{*}_{-\bf k})^{\dagger}=f^{*}_{\bf k},\\
&& \Pi^{\dagger}_{\bf k}(\tau)=\Pi^{*}_{\bf k}(\tau)=\Pi_{-\bf k}(\tau)~~~\Longrightarrow ~~~(\Pi^{*}_{-\bf k})^{\dagger}=\Pi^{*}_{\bf k}.\eea
As a result, the fundamental parts of the kinetic and potential quantum operators as appearing in the expression for the quantum mechanical Hamiltonian can be further simplified as:
\bea && {\textcolor{red}{\bf Quantum~Kinetic~Operator:}}\nonumber\\  
&&~~~~~~~~~{\frac{1}{2}\hat{\Pi}^2_{\bf k}(\tau)=\frac{1}{2}\hat{\Pi}_{\bf k}(\tau)\hat{\Pi}_{-\bf k}(\tau)=\frac{1}{2}|\Pi_{\bf k}(\tau)~a_{\bf k}+\Pi^{*}_{\bf -k}(\tau)~a^{\dagger}_{-\bf k}|^2=\left(a^{\dagger}_{\bf k}a_{\bf k}+\frac{1}{2}\delta^{3}(0)\right)|\Pi_{\bf k}(\tau)|^2},~~~~~~~~~~\\
&& {\textcolor{red}{\bf Quantum~Potential~Operator:}}\nonumber\\ 
&&~~~~~~~~~{\frac{1}{2}\omega^2_{\bf k}(\tau)\hat{f}^2_{\bf k}(\tau)=\frac{1}{2}\omega^2_{\bf k}(\tau)\hat{f}_{\bf k}(\tau)\hat{f}_{-\bf k}(\tau)=\frac{1}{2}\omega^2_{\bf k}(\tau)|f_{\bf k}(\tau)~a_{\bf k}+f^{*}_{\bf -k}(\tau)~a^{\dagger}_{-\bf k}|^2}\nonumber\\
&&~~~~~~~~~~~~~~~~~~~~~~~{=\omega^2_{\bf k}(\tau)\left(a^{\dagger}_{\bf k}a_{\bf k}+\frac{1}{2}\delta^{3}(0)\right)|f_{\bf k}(\tau)|^2}.~~~~~~~~\eea 
As a result, we get the following simplified expression for the quantum mechanical Hamiltonian operator which is written in terms of the quantum number operator ($\widehat{\cal N}_{\bf k}$) in Fourier space:
\beq  { \textcolor{red}{\bf Quantum~Hamiltonian~Operator:}~~~\hat{H}(\tau)=\int d^3{\bf k}~\left(\widehat{\cal N}_{\bf k}+\frac{1}{2}\delta^{3}(0)\right)\left[|{\Pi}_{\bf k}(\tau)|^2+~\omega^2_{\bf k}(\tau)|{{f}_{\bf k}(\tau)}|^2\right]}~,\eeq
where the quantum number operator ($\widehat{\cal N}_{\bf k}$) is defined in terms of the creation and the annihilation operators of the quantum initial vacuum state as:
\bea {\textcolor{red}{\bf Quantum~ Number~Operator:}~~~\widehat{\cal N}_{\bf k}:=a^{\dagger}_{\bf k}a_{\bf k}}~.\eea
After introducing the normal ordering one can remove the contribution from the zero point energy which will give rise to the divergent contribution in the quantum mechanical Hamiltonian operator, which actually gives the divergent contribution. This further simplifies the the expression for the Hamiltonian, which is given by:
\beq  { \textcolor{red}{\bf Normal~Ordered~Hamiltonian~Operator:}~~~:\hat{H}(\tau):=\int d^3{\bf k}~\widehat{\cal N}_{\bf k}~\left[|{\Pi}_{\bf k}(\tau)|^2+~\omega^2_{\bf k}(\tau)|{{f}_{\bf k}(\tau)}|^2\right]}~.\eeq
This result is very useful for the computation of the two desired OTOCs in which we are interested in this paper as it will contribute to the thermal Boltzmann factor and the corresponding thermal partition function which is further computed out of taking the trace operation performed in terms of the path integral over the wave function of the universe. The details of this computation will be found in the later sections of this paper.  

\subsection{Cosmological two-point and four-point ``in-in" non-chaotic OTO amplitudes}
In this section, our prime objective is to explicitly derive the expressions for the two new sets of two point and four point desired OTOCs in this paper to study the non-chaotic time dependent ransom behaviour of cosmologically relevant quantum correlation functions.
\subsubsection{Non-chaotic auto-correlators in Primordial Cosmology}
 To compute this explicitly we need the following information in our hand:
\begin{enumerate}
\item \underline{\textcolor{red}{\bf Information~I:}}\\
First of all, one need to start with the quantum mechanical operators in which we are interested in the context of cosmology, those are the perturbed time dependent field and the associated canonically conjugate momenta written in Fourier transformed momentum space. This can easily be done using the obtained asymptotically viable total conformal time dependent solution of the perturbed field from the classical {\it Mukhanov Sasaki equation} which we are going to write in quantum regime as an operator by including the creation and annihilation operators of the quantum initial vacuum state.
In the framework of cosmological perturbation theory one can construct the conformal time dependent Hamiltonian density operator in Fourier space which can be made up of two important components, the kinetic operator and the potential operator, where both of them can be expressed in terms of the previously mentioned perturbed field operator and its canonically conjugate momentum operator in the quantum regime of field space. After constructing the time dependent Hamiltonian density operator in the Fourier space our next job is to integrate over all possible momenta spanned over a physically acceptable range to derive the total time dependent Hamiltonian operator frequently used for the present computation. Additionally it is important to note that, in the paper we will consider the canonical technique for the quantization purpose which we will follow throughout the paper.

\item \underline{\textcolor{red}{\bf Information~II:}}\\
In the next step, one needs to explicitly compute the expression for the commutator and square of the commutator bracket between the two perturbed field variable quantum operators and the two canonically conjugate momentum variable quantum operators defined at two different conformal time scales in the classical geometrical background of spatially flat FLRW space-time specifically in coordinate space representation. These are the crucial part which forms the building block of the desired OTOCs in the context of cosmological perturbation theory of early universe. In this section of the paper we explicitly derive the commutator and square of the commutator or these two operators
in terms of two and four parts respectively where each of the components mimics the role of scattering amplitudes in Fourier space. In the present context commutator and the square of the commutator bracket in the Fourier space involved two momenta and four momenta respectively and in both the cases two different conformal time scale appears which will finally fix the structure of the desired OTOCs which we want to study in the context of cosmological perturbation theory.
If we look into the structure of these commutator and square of the commutator quantum mechanical operator very closely then one can observe that these two different momenta and four different momenta are appearing in OTOCs as we are dealing with the product of two and quantum mechanical operators respectively in this computation. Though the OTOC computed from the square of the commutator bracket looks like $2\rightarrow 2$ scattering amplitude in the Fourier space, but technically these are actually representing the unequal time, out-of-time-ordered four point quantum correlation function in the context of cosmology. In the quantum field theory version of the trace operation one need to consider the same quantum initial vacuum state, which implies the initial and final state both are exactly same, and identified to be ``in" quantum state in the context cosmology. This further implies that within the framework of primordial cosmology we deal with in-in amplitude rather than using the usual ``in-out" amplitude in the $S$-matrix formalism, which is actually a {\it Schwinger Dyson} series. Instead of calling this quantity which we want to explicitly determine in this section as ``in-in" amplitude we call these quantities as ``in-in" quantum mechanical correlation functions within the framework of primordial cosmological perturbation theory.

\item \underline{\textcolor{red}{\bf Information~III:}}\\
Another important issue is to fix the proper definition of the trace operation within the framework of quantum field theory written in classical spatially flat FLRW curved cosmological background.
 In the context of quantum field theory of curved space-time, particularly in the context of cosmology we have to define the quantum wave function of our Universe which will help us to set up the equivalent representation of the thermal trace operation in terms of the quantum mechanical path integral formulation. To construct the wave function of the Universe as a initial condition one choose the standard definition of Euclidean false vacuum state, which is commonly known as the Bunch Davies vacuum state and it basically represents a thermal ground state in the context of primordial cosmological perturbation theory. In the quantum field theory of curved space-time literature sometimes this is identified to be the Hartle-Hawking or Cherenkov vacuum state.  In this computation, the most most generalized choice is  the $\alpha,\beta$ vacua, which is commonly known as the {\it Motta-Allen (MA) vacua} in the context of quantum field theory of De Sitter space time which are invariant
under all the conformal group $SO(1,4)$ isometries and commonly known as the $\alpha,\beta$-vacua which is CPT violating. Here $\alpha,\beta$ is a real parameter
which forms a real parameter family of continuous numbers and particularly $\beta$ 
is appearing in the phases. This phase factor
is actually the culprit for which in the quantum state CPT symmetry getting violated. Once we switch off the contribution from this phase factor by fixing $\beta = 0$, the one
can get back the CPT symmetry preserving quantum $\alpha$ vacua states. Sometimes the $\alpha$ vacua
is characterized as the squeezed quantum vacuum state. Bunch Davies vacuum state is a very special case of the generalized $\alpha$ vacua state which can be obtained by fixing $\alpha=0$ in the definition of the quantum vacuum state, which perfectly satisfy the Hadamard condition in the Green’s functions. Using Bogoliubov transformation one can express the $\alpha$ vacua in terms of the Bunch Davies state.

\item \underline{\textcolor{red}{\bf Information~IV:}}\\
Once we fix the definition of the quantum field theory version of the trace operation in terms of path integral formalism in the Euclidean formalism we can then further compute the expression for the desired two sets of OTOCs on which we are interested in this paper. Another important component which will fix the definition of the OTOC is the thermal partition function for the scalar mode fluctuations obtained from the primordial cosmological perturbation where the same trick we have to apply to define the trace operation as mentioned earlier.
In this context instead of performing a complete quantum calculation we use the semi classical approximations for the computations. The primordial perturbations in the spatially flat classical FLRW background can be written in terms of scalar, vector and tensor modes in Fourier space using the {\it SVT decomposition} from which the quantum fluctuations are generated. For this reason we will do a semi-classical (not purely quantum or classical) computation for the computation of OTOC in the framework of primordial cosmology.
\item \underline{\textcolor{red}{\bf Information~V:}}\\
Last but not at all the least, we have to fix the normalization factor of all the desired new OTOCs that we have introduced in this paper. In this connection it is important to remind ourself again that normalixzation factors of these desired OTOCs are actually made up of product of two disconnected thermal two point functions in the dissipation time scale $t=t_d\sim \beta\sim 1/T$, where $T$ is the equilibrium temperature of the quantum cosmological system under study in the present context. Whatever result we have obtained for the un-normalized OTOCs we use them and divide them the mentioned disconnected part of the correlator. This helps to treat the overall amplitude of four point OTOCs in a dimensionless fashion. Not only that such trick in normalization in OTOC also helps us to know about the exact time dependence in the quantum correlator on which we are interested in.  More precisely this can be done by making use of the previously mentioned equivalent operation of thermal trace operation in presence of $\alpha$ vacua or Bunch Davies quantum vacuum state in the context of primordial cosmological perturbation theory. We have explicitly demonstrated the detailed computation of these normalization factors in the Appendix of this paper. Please look into the technical details in the Appendix for more details on this issue.
\end{enumerate} 
\subsubsection{Fourier space representation of the commutator bracket: Application to two-point non-chaotic auto-correlated OTO functions}
Here our job is to compute the following commutator brackets, given by the following expressions:
\bea && \textcolor{red}{\bf Commutator_{1}:}~~~~\left[\hat{f}({\bf x},\tau_1),\hat{f}({\bf x},\tau_2)\right]=\underbrace{\hat{f}({\bf x},\tau_1)\hat{f}({\bf x},\tau_2)}_{\textcolor{red}{\bf \equiv \Gamma_{1}({\bf x},\tau_1,\tau_2)}}-\underbrace{\hat{f}({\bf x},\tau_2)\hat{f}({\bf x},\tau_1)}_{\textcolor{red}{\bf \equiv \Gamma_{2}({\bf x},\tau_1,\tau_2)}},\\
 && \textcolor{red}{\bf Commutator_{2}:}~~~~ \left[\hat{\Pi}({\bf x},\tau_1),\hat{\Pi}({\bf x},\tau_2)\right]=\underbrace{\hat{\Pi}({\bf x},\tau_1)\hat{\Pi}({\bf x},\tau_2)}_{\textcolor{red}{\bf \equiv \Theta_{1}({\bf x},\tau_1,\tau_2)}}-\underbrace{\hat{\Pi}({\bf x},\tau_2)\hat{\Pi}({\bf x},\tau_1)}_{\textcolor{red}{\bf \equiv \Theta_{2}({\bf x},\tau_1,\tau_2)}}.~~~~~~~\eea
 Now we use the following preferred convention for the Fourier transformation of the perturbation field and associated momentum operator, which are given by:
 \bea &&\hat{f}({\bf x},\tau_1)=\int \frac{d^3{\bf k}}{(2\pi)^3}~\exp(i{\bf k}.{\bf x})~\hat{f}_{{\bf k}}(\tau_1),\\
 &&\hat{\Pi}({\bf x},\tau_1)=\partial_{\tau_1}\hat{f}({\bf x},\tau_1)=\int \frac{d^3{\bf k}}{(2\pi)^3}~\exp(i{\bf k}.{\bf x})~\partial_{\tau_1}\hat{f}_{{\bf k}}(\tau_1)=\int \frac{d^3{\bf k}}{(2\pi)^3}~\exp(i{\bf k}.{\bf x})~\hat{\Pi}_{{\bf k}}(\tau_1),~~~~~~~~~~~~~\eea
 which we will frequently follow for rest of the computation to fix the definition of the desired OTOCs defined in this paper. 
  
   Next, we explicitly compute the expressions for these individual quantum mechanical operators in the context of primordial cosmological perturbation theory, $\Gamma_{i}({\bf x},\tau_1,\tau_2)$~$\forall~i=1,2$ and $\Theta_{i}({\bf x},\tau_1,\tau_2)$~$\forall~i=1,2$, which can be expressed in Fourier transformed space by the following simplified expressions:
 \bea \Gamma_{1}({\bf x},\tau_1,\tau_2)&=&\hat{f}({\bf x},\tau_1)\hat{f}({\bf x},\tau_2)=\int\frac{d^3{\bf k}_1}{(2\pi)^3}\int\frac{d^3{\bf k}_1}{(2\pi)^3}~\exp(i({\bf k}_1+{\bf k}_2).{\bf x})~\hat{f}_{{\bf k}_1}(\tau_1)\hat{f}_{{\bf k}_2}(\tau_2)\nonumber\\
 &=&\int\frac{d^3{\bf k}_1}{(2\pi)^3}\int\frac{d^3{\bf k}_1}{(2\pi)^3}~\exp(i({\bf k}_1+{\bf k}_2).{\bf x})~\hat{\Delta}^{(1)}_{1}({\bf k}_1,{\bf k}_2;\tau_1,\tau_2),\eea
 where we have introduced a momentum and conformal time dependent quantum mechanical operator in the context of primordial cosmological perturbation theory, $\hat{\Delta}_{1}({\bf k}_1,{\bf k}_2;\tau_1,\tau_2)$, which is defined as:
 \bea \hat{\Delta}^{(1)}_{1}({\bf k}_1,{\bf k}_2;\tau_1,\tau_2)&=&\hat{f}_{{\bf k}_1}(\tau_1)\hat{f}_{{\bf k}_2}(\tau_2)={\cal D}^{(1)}_1 ({\bf k}_1,{\bf k}_2;\tau_1,\tau_2)~a_{{\bf k}_1}a_{{\bf k}_2}+{\cal D}^{(1)}_2 ({\bf k}_1,{\bf k}_2;\tau_1,\tau_2)~a^{\dagger}_{-{\bf k}_1}a_{{\bf k}_2}\nonumber\\
 &&~~~~~~~~~~~~~~~~~~~~~~+{\cal D}^{(1)}_3 ({\bf k}_1,{\bf k}_2;\tau_1,\tau_2)~a_{{\bf k}_1}a^{\dagger}_{-{\bf k}_2}+{\cal D}^{(1)}_4 ({\bf k}_1,{\bf k}_2;\tau_1,\tau_2)~a^{\dagger}_{-{\bf k}_1}a^{\dagger}_{-{\bf k}_2},~~~~~~~~~\eea  
 where we have introduced  momentum and time dependent two-point OTO amplitudes, ${\cal D}^{(1)}_i ({\bf k}_1,{\bf k}_2;\tau_1,\tau_2)~~\forall~~i=1,2,3,4$,  which are explicitly defined in the Appendix~(\ref{sec:8})
  of this paper.
  \bea \Gamma_{2}({\bf x},\tau_1,\tau_2)&=&\hat{\Pi}({\bf x},\tau_2)\hat{f}({\bf x},\tau_1)=\int\frac{d^3{\bf k}_1}{(2\pi)^3}\int\frac{d^3{\bf k}_1}{(2\pi)^3}~\exp(i({\bf k}_1+{\bf k}_2).{\bf x})~\hat{\Pi}_{{\bf k}_1}(\tau_2)\hat{f}_{{\bf k}_2}(\tau_1)\nonumber\\
 &=&\int\frac{d^3{\bf k}_1}{(2\pi)^3}\int\frac{d^3{\bf k}_1}{(2\pi)^3}~\exp(i({\bf k}_1+{\bf k}_2).{\bf x})~\hat{\Delta}^{(1)}_{2}({\bf k}_1,{\bf k}_2;\tau_1,\tau_2),\eea
  where we have introduced a momentum and conformal time dependent quantum mechanical operator in the context of primordial cosmological perturbation theory,  $\hat{\Delta}^{(1)}_{2}({\bf k}_1,{\bf k}_2;\tau_1,\tau_2)$, which are explicitly defined in the Appendix of this paper.
 \bea \hat{\Delta}^{(1)}_{2}({\bf k}_1,{\bf k}_2;\tau_1,\tau_2)&=&\hat{f}_{{\bf k}_1}(\tau_2)\hat{f}_{{\bf k}_2}(\tau_1)={\cal L}^{(1)}_1 ({\bf k}_1,{\bf k}_2;\tau_1,\tau_2)~a_{{\bf k}_1}a_{{\bf k}_2}+{\cal L}^{(1)}_2 ({\bf k}_1,{\bf k}_2;\tau_1,\tau_2)~a^{\dagger}_{-{\bf k}_1}a_{{\bf k}_2}\nonumber\\
 &&~~~~~~~~~~~~~~~~~~~~~~+{\cal L}^{(1)}_3 ({\bf k}_1,{\bf k}_2;\tau_1,\tau_2)~a_{{\bf k}_1}a^{\dagger}_{-{\bf k}_2}+{\cal L}^{(1)}_4 ({\bf k}_1,{\bf k}_2;\tau_1,\tau_2)~a^{\dagger}_{-{\bf k}_1}a^{\dagger}_{-{\bf k}_2},~~~~~~~~~\eea  
  where we have introduced  momentum and time dependent two-point OTO amplitudes, ${\cal L}^{(1)}_i ({\bf k}_1,{\bf k}_2;\tau_1,\tau_2)~~\forall~~i=1,2,3,4$, which are explicitly defined in the Appendix of this paper.
 \bea \Theta_{1}({\bf x},\tau_1,\tau_2)&=&\hat{\Pi}({\bf x},\tau_1)\hat{\Pi}({\bf x},\tau_2)=\int\frac{d^3{\bf k}_1}{(2\pi)^3}\int\frac{d^3{\bf k}_1}{(2\pi)^3}~\exp(i({\bf k}_1+{\bf k}_2).{\bf x})~\hat{\Pi}_{{\bf k}_1}(\tau_1)\hat{\Pi}_{{\bf k}_2}(\tau_2)\nonumber\\
 &=&\int\frac{d^3{\bf k}_1}{(2\pi)^3}\int\frac{d^3{\bf k}_1}{(2\pi)^3}~\exp(i({\bf k}_1+{\bf k}_2).{\bf x})~\hat{\Delta}^{(2)}_{1}({\bf k}_1,{\bf k}_2;\tau_1,\tau_2),\eea
 where we have introduced a momentum and conformal time dependent quantum mechanical operator in the context of primordial cosmological perturbation theory, $\hat{\Delta}^{(2)}_{1}({\bf k}_1,{\bf k}_2;\tau_1,\tau_2)$, which are explicitly defined in the Appendix of this paper as:
 \bea \hat{\Delta}^{(2)}_{1}({\bf k}_1,{\bf k}_2;\tau_1,\tau_2)&=&\hat{\Pi}_{{\bf k}_1}(\tau_1)\hat{\Pi}_{{\bf k}_2}(\tau_2)={\cal D}^{(2)}_1 ({\bf k}_1,{\bf k}_2;\tau_1,\tau_2)~a_{{\bf k}_1}a_{{\bf k}_2}+{\cal D}^{(2)}_2 ({\bf k}_1,{\bf k}_2;\tau_1,\tau_2)~a^{\dagger}_{-{\bf k}_1}a_{{\bf k}_2}\nonumber\\
 &&~~~~~~~~~~~+{\cal D}^{(2)}_3 ({\bf k}_1,{\bf k}_2;\tau_1,\tau_2)~a_{{\bf k}_1}a^{\dagger}_{-{\bf k}_2}+{\cal D}^{(2)}_4 ({\bf k}_1,{\bf k}_2;\tau_1,\tau_2)~a^{\dagger}_{-{\bf k}_1}a^{\dagger}_{-{\bf k}_2},~~~~~~~~~~~~~~\eea  
 where we have introduced  momentum and time dependent two-point OTO amplitudes, ${\cal D}^{(2)}_i ({\bf k}_1,{\bf k}_2;\tau_1,\tau_2)~~\forall~~i=1,2,3,4$, which are explicitly defined in the Appendix of this paper.
  \bea \Theta_{2}({\bf x},\tau_1,\tau_2)&=&\hat{\Pi}({\bf x},\tau_2)\hat{\Pi}({\bf x},\tau_1)=\int\frac{d^3{\bf k}_1}{(2\pi)^3}\int\frac{d^3{\bf k}_1}{(2\pi)^3}~\exp(i({\bf k}_1+{\bf k}_2).{\bf x})~\hat{\Pi}_{{\bf k}_1}(\tau_2)\hat{\Pi}_{{\bf k}_2}(\tau_1)\nonumber\\
 &=&\int\frac{d^3{\bf k}_1}{(2\pi)^3}\int\frac{d^3{\bf k}_1}{(2\pi)^3}~\exp(i({\bf k}_1+{\bf k}_2).{\bf x})~\hat{\Delta}^{(2)}_{2}({\bf k}_1,{\bf k}_2;\tau_1,\tau_2),\eea
 where we have introduced a momentum and conformal time dependent quantum mechanical operator $\hat{\Delta}_{2}({\bf k}_1,{\bf k}_2;\tau_1,\tau_2)$, which is defined as:
 \bea \hat{\Delta}^{(2)}_{2}({\bf k}_1,{\bf k}_2;\tau_1,\tau_2)&=&\hat{\Pi}_{{\bf k}_1}(\tau_2)\hat{\Pi}_{{\bf k}_2}(\tau_1)\nonumber\\
 &=&{\cal L}^{(2)}_1 ({\bf k}_1,{\bf k}_2;\tau_1,\tau_2)~a_{{\bf k}_1}a_{{\bf k}_2}+{\cal L}^{(2)}_2 ({\bf k}_1,{\bf k}_2;\tau_1,\tau_2)~a^{\dagger}_{-{\bf k}_1}a_{{\bf k}_2}\nonumber\\
 &&~~~~~~~~~+{\cal L}^{(2)}_3 ({\bf k}_1,{\bf k}_2;\tau_1,\tau_2)~a_{{\bf k}_1}a^{\dagger}_{-{\bf k}_2}+{\cal L}^{(2)}_4 ({\bf k}_1,{\bf k}_2;\tau_1,\tau_2)~a^{\dagger}_{-{\bf k}_1}a^{\dagger}_{-{\bf k}_2},~~~~~~~~~~~~~~\eea  
where we have introduced  momentum and time dependent two-point OTO amplitudes, ${\cal L}^{(2)}_i ({\bf k}_1,{\bf k}_2;\tau_1,\tau_2)~~\forall~~i=1,2,3,4$, which are explicitly defined in the Appendix~(\ref{sec:8}) of this paper.
 
  This further implies that one can explicitly write down the previously mentioned two commutator brackets along with the thermal Boltzmann factor in terms of the following simplified expression, which is given by:
   \bea && e^{-\beta \widehat{H}(\tau_1)}\left[\hat{f}({\bf x},\tau_1),\hat{f}({\bf x},\tau_2)\right]\nonumber\\
   &&=e^{-\beta \widehat{H}(\tau_1)}\left[ {\Gamma}_1({\bf x},\tau_1,\tau_2)- {\Gamma}_2({\bf x},\tau_1,\tau_2)\right]\nonumber\\
 && = e^{-\beta \widehat{H}(\tau_1)}\left\{\int \frac{d^3k_1}{(2\pi)^3}\int \frac{d^3k_2}{(2\pi)^3}\exp\left[i\left({\bf k}_1+{\bf k}_2\right).{\bf x}\right]\left[\hat{\Delta}^{(1)}_1({\bf k}_1,{\bf k}_2;\tau_1,\tau_2)-\hat{\Delta}^{(1)}_2({\bf k}_1,{\bf k}_2;\tau_1,\tau_2)\right]\right\}\nonumber\\
 &&= \int \frac{d^3k_1}{(2\pi)^3}\int \frac{d^3k_2}{(2\pi)^3}\exp\left[i\left({\bf k}_1+{\bf k}_2\right).{\bf x}\right]\left[\hat{\nabla}^{(1)}_1({\bf k}_1,{\bf k}_2;\tau_1,\tau_2;\beta)-\hat{\nabla}^{(1)}_2({\bf k}_1,{\bf k}_2;\tau_1,\tau_2;\beta)\right],~~~~~~~~~~~ \eea\bea
 && e^{-\beta \widehat{H}(\tau_1)}\left[\hat{\Pi}({\bf x},\tau_1),\hat{\Pi}({\bf x},\tau_2)\right]\nonumber\\
   &&=e^{-\beta \widehat{H}(\tau_1)}\left[ {\Theta}_1({\bf x},\tau_1,\tau_2)- {\Theta}_2({\bf x},\tau_1,\tau_2)\right]\nonumber\\
 && = e^{-\beta \widehat{H}(\tau_1)}\left\{\int \frac{d^3k_1}{(2\pi)^3}\int \frac{d^3k_2}{(2\pi)^3}\exp\left[i\left({\bf k}_1+{\bf k}_2\right).{\bf x}\right]\left[\hat{\Delta}^{(2)}_1({\bf k}_1,{\bf k}_2;\tau_1,\tau_2)-\hat{\Delta}^{(2)}_2({\bf k}_1,{\bf k}_2;\tau_1,\tau_2)\right]\right\}\nonumber\\
 &&= \int \frac{d^3k_1}{(2\pi)^3}\int \frac{d^3k_2}{(2\pi)^3}\exp\left[i\left({\bf k}_1+{\bf k}_2\right).{\bf x}\right]\left[\hat{\nabla}^{(2)}_1({\bf k}_1,{\bf k}_2;\tau_1,\tau_2;\beta)-\hat{\nabla}^{(2)}_2({\bf k}_1,{\bf k}_2;\tau_1,\tau_2;\beta)\right],~~~~~~~~~~~ \eea
  where we define the new sets of quantum mechanical operators in the context of primordial cosmological perturbation theory, $:\hat{\nabla}^{(m)}_i({\bf k}_1,{\bf k}_2;\tau_1,\tau_2;\beta):~\forall~i=1,2,~~\forall~m=1,2$ as given by the following expression:
  \bea &&{\hat{\nabla}^{(m)}_i({\bf k}_1,{\bf k}_2;\tau_1,\tau_2;\beta)=e^{-\beta \hat{H}(\tau_1)}~\hat{\Delta}^{(m)}_i({\bf k}_1,{\bf k}_2;\tau_1,\tau_2)~~~~\forall ~~~i=1,2~~~~~\forall~~~m=1,2}~~~~~~~~~~~\eea
  Here the thermal Boltzmann factor can be expressed in terms of creation and annihilation operator by the following simplified expression, as given by:
  \bea && {e^{-\beta H(\tau_1)}=\exp\left(-\beta\int d^3{\bf k}~\left(\widehat{\cal N}_{\bf k}+\frac{1}{2}\delta^{3}(0)\right)E_{\bf k}(\tau_1)\right)},~~~\eea
  where we define the conformal time dependent energy spectrum as a function of Fourier modes relevant for cosmology, $E_{\bf k}(\tau_1)$ by the following expression:
  \bea E_{\bf k}(\tau_1):=\left[|\Pi_{\bf k}(\tau_1)|^2+\omega^2_{\bf k}(\tau_1)|f_{\bf k}(\tau)|^2\right],\eea 
  where the explicit expression for the conformal time dependent expression for the frequency factor at the time scale $\tau=\tau_1$, i.e. $\omega_{\bf k}(\tau_1)$ for the cosmologically relevant Fourier modes participating in the primordial cosmological perturbation theory is mentioned in the earlier half of this paper. For general readers it is further important to note that these mode frequencies for all momentum scales play significant role for the quantification of the randomness in terms of the quantum OTOCs studying in this paper.
\subsubsection{Fourier space representation of square of the commutator bracket: Application to four-point non-chaotic auto-correlated OTO functions}
  Now we explicitly compute the following combinations of the square of the commutator bracket out of only the cosmologically perturbed field operators and only with the help of the canonically conjugate momentum operators related to these cosmologically perturbed field, which are actually given by the following simplified expression:
  \bea && \left[\hat{f}({\bf x},\tau_1),\hat{f}({\bf x},\tau_2)\right]^2=\underbrace{\hat{f}({\bf x},\tau_1)\hat{f}({\bf x},\tau_2)\hat{f}({\bf x},\tau_1)\hat{f}({\bf x},\tau_2)}_{\textcolor{red}{\equiv~ {\cal K}^{(1)}_1({\bf x},\tau_1,\tau_2)}}-\underbrace{\hat{f}({\bf x},\tau_2)\hat{f}({\bf x},\tau_1)\hat{f}({\bf x},\tau_1)\hat{f}({\bf x},\tau_2)}_{\textcolor{red}{\equiv~ {\cal K}^{(1)}_2({\bf x},\tau_1,\tau_2)}}\nonumber\\ &&~~~~~~~~~~~-\underbrace{\hat{f}({\bf x},\tau_1)\hat{f}({\bf x},\tau_2)\hat{f}({\bf x},\tau_2)\hat{f}({\bf x},\tau_1)}_{\textcolor{red}{\equiv~ {\cal K}^{(1)}_3({\bf x},\tau_1,\tau_2)}}+\underbrace{\hat{f}({\bf x},\tau_2)\hat{f}({\bf x},\tau_1)\hat{f}({\bf x},\tau_2)\hat{f}({\bf x},\tau_1)}_{\textcolor{red}{\equiv~ {\cal K}^{(1)}_4({\bf x},\tau_1,\tau_2)}}, \eea\bea
   && \left[\hat{\Pi}({\bf x},\tau_1),\hat{\Pi}({\bf x},\tau_2)\right]^2=\underbrace{\hat{\Pi}({\bf x},\tau_1)\hat{\Pi}({\bf x},\tau_2)\hat{\Pi}({\bf x},\tau_1)\hat{\Pi}({\bf x},\tau_2)}_{\textcolor{red}{\equiv~ {\cal K}^{(2)}_1({\bf x},\tau_1,\tau_2)}}-\underbrace{\hat{\Pi}({\bf x},\tau_2)\hat{\Pi}({\bf x},\tau_1)\hat{\Pi}({\bf x},\tau_1)\hat{\Pi}({\bf x},\tau_2)}_{\textcolor{red}{\equiv~ {\cal K}^{(2)}_2({\bf x},\tau_1,\tau_2)}}\nonumber\\ &&~~~~~~~~~~~-\underbrace{\hat{\Pi}({\bf x},\tau_1)\hat{\Pi}({\bf x},\tau_2)\hat{\Pi}({\bf x},\tau_2)\hat{\Pi}({\bf x},\tau_1)}_{\textcolor{red}{\equiv~ {\cal K}^{(2)}_3({\bf x},\tau_1,\tau_2)}}+\underbrace{\hat{\Pi}({\bf x},\tau_2)\hat{\Pi}({\bf x},\tau_1)\hat{\Pi}({\bf x},\tau_2)\hat{\Pi}({\bf x},\tau_1)}_{\textcolor{red}{\equiv~ {\cal K}^{(2)}_4({\bf x},\tau_1,\tau_2)}} \eea
   Now we mention the explicit mathematical structure of these individual operators, ${\cal K}^{(M)}_i({\bf x},\tau_1,\tau_2)~\forall~i=1,2,3,4,~~\forall~M=1,2,3,4$, which are expressed in Fourier transformed space as:
  \bea &&{\cal K}^{(1)}_1({\bf x},\tau_1,\tau_2)
  =\hat{f}({\bf x},\tau_1)\hat{f}({\bf x},\tau_2)\hat{f}({\bf x},\tau_1)\hat{f}({\bf x},\tau_2)\nonumber\\
  &&=\int \frac{d^3k_1}{(2\pi)^3}\int \frac{d^3k_2}{(2\pi)^3}\int \frac{d^3k_3}{(2\pi)^3}\int \frac{d^3k_4}{(2\pi)^3}\exp\left[i\left({\bf k}_1+{\bf k}_2+{\bf k}_3+{\bf k}_4\right).{\bf x}\right]\widehat{\cal T}^{(1)}_1({\bf k}_1,{\bf k}_2,{\bf k}_3,{\bf k}_4;\tau_1,\tau_2),~~~~~~~~~~
 \\ &&{\cal K}^{(2)}_1({\bf x},\tau_1,\tau_2)=\hat{\Pi}({\bf x},\tau_1)\hat{\Pi}({\bf x},\tau_2)\hat{\Pi}({\bf x},\tau_1)\hat{\Pi}({\bf x},\tau_2)\nonumber\\
  &&=\int \frac{d^3k_1}{(2\pi)^3}\int \frac{d^3k_2}{(2\pi)^3}\int \frac{d^3k_3}{(2\pi)^3}\int \frac{d^3k_4}{(2\pi)^3}\exp\left[i\left({\bf k}_1+{\bf k}_2+{\bf k}_3+{\bf k}_4\right).{\bf x}\right]\widehat{\cal T}^{(2)}_1({\bf k}_1,{\bf k}_2,{\bf k}_3,{\bf k}_4;\tau_1,\tau_2),~~~~~~~~~~
 \\
  &&{\cal K}^{(1)}_2({\bf x},\tau_1,\tau_2)=\hat{f}({\bf x},\tau_2)\hat{f}({\bf x},\tau_1)\hat{f}({\bf x},\tau_1)\hat{f}({\bf x},\tau_2)\nonumber\\
  &&=\int \frac{d^3k_1}{(2\pi)^3}\int \frac{d^3k_2}{(2\pi)^3}\int \frac{d^3k_3}{(2\pi)^3}\int \frac{d^3k_4}{(2\pi)^3}\exp\left[i\left({\bf k}_1+{\bf k}_2+{\bf k}_3+{\bf k}_4\right).{\bf x}\right]\widehat{\cal T}^{(1)}_2({\bf k}_1,{\bf k}_2,{\bf k}_3,{\bf k}_4;\tau_1,\tau_2),~~~~~~~~
 \\
 &&{\cal K}^{(2)}_2({\bf x},\tau_1,\tau_2)=\hat{\Pi}({\bf x},\tau_2)\hat{\Pi}({\bf x},\tau_1)\hat{\Pi}({\bf x},\tau_1)\hat{\Pi}({\bf x},\tau_2)\nonumber\\
  &&=\int \frac{d^3k_1}{(2\pi)^3}\int \frac{d^3k_2}{(2\pi)^3}\int \frac{d^3k_3}{(2\pi)^3}\int \frac{d^3k_4}{(2\pi)^3}\exp\left[i\left({\bf k}_1+{\bf k}_2+{\bf k}_3+{\bf k}_4\right).{\bf x}\right]\widehat{\cal T}^{(2)}_2({\bf k}_1,{\bf k}_2,{\bf k}_3,{\bf k}_4;\tau_1,\tau_2),
 \\
  &&{\cal K}^{(1)}_3({\bf x},\tau_1,\tau_2)=\hat{f}({\bf x},\tau_1)\hat{f}({\bf x},\tau_2)\hat{f}({\bf x},\tau_2)\hat{f}({\bf x},\tau_1)\nonumber\\
  &&=\int \frac{d^3k_1}{(2\pi)^3}\int \frac{d^3k_2}{(2\pi)^3}\int \frac{d^3k_3}{(2\pi)^3}\int \frac{d^3k_4}{(2\pi)^3}\exp\left[i\left({\bf k}_1+{\bf k}_2+{\bf k}_3+{\bf k}_4\right).{\bf x}\right]\widehat{\cal T}^{(1)}_3({\bf k}_1,{\bf k}_2,{\bf k}_3,{\bf k}_4;\tau_1,\tau_2),
 \\
 &&{\cal K}^{(2)}_3({\bf x},\tau_1,\tau_2)=\hat{\Pi}({\bf x},\tau_1)\hat{\Pi}({\bf x},\tau_2)\hat{\Pi}({\bf x},\tau_2)\hat{\Pi}({\bf x},\tau_1)\nonumber\\
  &&=\int \frac{d^3k_1}{(2\pi)^3}\int \frac{d^3k_2}{(2\pi)^3}\int \frac{d^3k_3}{(2\pi)^3}\int \frac{d^3k_4}{(2\pi)^3}\exp\left[i\left({\bf k}_1+{\bf k}_2+{\bf k}_3+{\bf k}_4\right).{\bf x}\right]\widehat{\cal T}^{(2)}_3({\bf k}_1,{\bf k}_2,{\bf k}_3,{\bf k}_4;\tau_1,\tau_2),~~~~~~~~~~~
\\
 &&{\cal K}^{(1)}_4({\bf x},\tau_1,\tau_2)=\hat{f}({\bf x},\tau_2)\hat{f}({\bf x},\tau_1)\hat{f}({\bf x},\tau_2)\hat{f}({\bf x},\tau_1)\nonumber\\
  &&=\int \frac{d^3k_1}{(2\pi)^3}\int \frac{d^3k_2}{(2\pi)^3}\int \frac{d^3k_3}{(2\pi)^3}\int \frac{d^3k_4}{(2\pi)^3}\exp\left[i\left({\bf k}_1+{\bf k}_2+{\bf k}_3+{\bf k}_4\right).{\bf x}\right]\widehat{\cal T}^{(1)}_4({\bf k}_1,{\bf k}_2,{\bf k}_3,{\bf k}_4;\tau_1,\tau_2),
 \\
 &&{\cal K}^{(2)}_4({\bf x},\tau_1,\tau_2)=\hat{\Pi}({\bf x},\tau_2)\hat{\Pi}({\bf x},\tau_1)\hat{\Pi}({\bf x},\tau_2)\hat{\Pi}({\bf x},\tau_1)\nonumber\\
  &&=\int \frac{d^3k_1}{(2\pi)^3}\int \frac{d^3k_2}{(2\pi)^3}\int \frac{d^3k_3}{(2\pi)^3}\int \frac{d^3k_4}{(2\pi)^3}\exp\left[i\left({\bf k}_1+{\bf k}_2+{\bf k}_3+{\bf k}_4\right).{\bf x}\right]\widehat{\cal T}^{(2)}_4({\bf k}_1,{\bf k}_2,{\bf k}_3,{\bf k}_4;\tau_1,\tau_2),~~~~~~~~~~\eea 
  where the functions $\widehat{\cal T}^{(1)}_p({\bf k}_1,{\bf k}_2,{\bf k}_3,{\bf k}_4;\tau_1,\tau_2)\forall p=1,2,3,4$  and $\widehat{\cal T}^{(2)}_p({\bf k}_1,{\bf k}_2,{\bf k}_3,{\bf k}_4;\tau_1,\tau_2)\forall p=1,2,3,4$ are explicitly defined in Appendix~(\ref{sec:9}).
 
  This implies that one can write down the previously mentioned square of the commutator bracket along with the thermal Boltzmann factor as:
   \bea && e^{-\beta \widehat{H}(\tau_1)}\left[\hat{f}({\bf x},\tau_1),\hat{\Pi}({\bf x},\tau_2)\right]^2\nonumber\\ 
 &&= \int \frac{d^3k_1}{(2\pi)^3}\int \frac{d^3k_2}{(2\pi)^3}\int \frac{d^3k_3}{(2\pi)^3}\int \frac{d^3k_4}{(2\pi)^3}\exp\left[i\left({\bf k}_1+{\bf k}_2+{\bf k}_3+{\bf k}_4\right).{\bf x}\right]~~~~~~~~\nonumber\\
  &&~~~~~~~~~~~~~~~~~~~~~~~~~\left[\widehat{\cal V}^{(1)}_1({\bf k}_1,{\bf k}_2,{\bf k}_3,{\bf k}_4;\tau_1,\tau_2;\beta)-\widehat{\cal V}^{(1)}_2({\bf k}_1,{\bf k}_2,{\bf k}_3,{\bf k}_4;\tau_1,\tau_2;\beta)\right.\nonumber\\ && \left.~~~~~~~~~~~~~~~~~~~~~~~~~~~~~~ +\widehat{\cal V}^{(1)}_3({\bf k}_1,{\bf k}_2,{\bf k}_3,{\bf k}_4;\tau_1,\tau_2;\beta)-\widehat{\cal V}^{(1)}_4({\bf k}_1,{\bf k}_2,{\bf k}_3,{\bf k}_4;\tau_1,\tau_2;\beta)\right],~~~~~~~~~~~\\
  && e^{-\beta \widehat{H}(\tau_1)}\left[\hat{\Pi}({\bf x},\tau_1),\hat{\Pi}({\bf x},\tau_2)\right]^2\nonumber\\ 
 &&= \int \frac{d^3k_1}{(2\pi)^3}\int \frac{d^3k_2}{(2\pi)^3}\int \frac{d^3k_3}{(2\pi)^3}\int \frac{d^3k_4}{(2\pi)^3}\exp\left[i\left({\bf k}_1+{\bf k}_2+{\bf k}_3+{\bf k}_4\right).{\bf x}\right]~~~~~~~~\nonumber\\
  &&~~~~~~~~~~~~~~~~~~~~~~~~~\left[\widehat{\cal V}^{(2)}_1({\bf k}_1,{\bf k}_2,{\bf k}_3,{\bf k}_4;\tau_1,\tau_2;\beta)-\widehat{\cal V}^{(2)}_2({\bf k}_1,{\bf k}_2,{\bf k}_3,{\bf k}_4;\tau_1,\tau_2;\beta)\right.\nonumber\\ && \left.~~~~~~~~~~~~~~~~~~~~~~~~~~~~~~ +\widehat{\cal V}^{(2)}_3({\bf k}_1,{\bf k}_2,{\bf k}_3,{\bf k}_4;\tau_1,\tau_2;\beta)-\widehat{\cal V}^{(2)}_4({\bf k}_1,{\bf k}_2,{\bf k}_3,{\bf k}_4;\tau_1,\tau_2;\beta)\right],~~~~~~~~~~~ \eea
  where we define the new sets of quantum operators in the context of primordial cosmological perturbation theory set up, $\widehat{\cal V}^{(M)}_i({\bf k}_1,{\bf k}_2,{\bf k}_3,{\bf k}_4;\tau_1,\tau_2;\beta)~\forall~i=1,2,3,4,~~\forall~M=1,2$ as given by the following simplified expression:
  \bea &&{\widehat{\cal V}^{(M)}_i({\bf k}_1,{\bf k}_2,{\bf k}_3,{\bf k}_4;\tau_1,\tau_2;\beta)=e^{-\beta \hat{H}(\tau_1)}~\widehat{\cal T}_i({\bf k}_1,{\bf k}_2,{\bf k}_3,{\bf k}_4;\tau_1,\tau_2)~~~~\forall ~~~i=1,2,3,4,~~M=1,2}~~~~~~~~~~~\eea
where the thermal Boltzmann factor with energy dispersion is computed earlier. 

\subsection{Thermal partition function in Primordial Cosmology: Quantum version}
\subsubsection{Initial quantum state in Primordial Cosmology}
In general, one can consider an arbitrary initial quantum mechanical vacuum state which is characterised by the
two arbitrary constants ${\cal C}_1$ and ${\cal C}_2$, which are appearing in the solution of the {\it Mukhanov Sasaki equation}, which represents the classical solution of the background perturbation in the spatially flat FLRW cosmological background. Also, these arbitrary constants play very significant role to fix the definition of the quantum wave function of the universe which is dependent on a specified information of the initial quantum vacuum state for cosmology. Once this definition is fixed, using this quantum vacuum states one can further study the correlation function from the quantum fluctuation in the scalar modes which is contributing in terms of the co-moving scalar curvature perturbation or in terms of the redefined perturbed field variable. Now, if we say that ${\cal C}_{\bf k, 12}$ is the annihilation operator corresponding to the quantum state as mentioned earlier, then it satisfies the following criteria in the context of quantum field theory written for the primordial cosmological perturbation:
\bea {\cal C}_{\bf k, 12}|{\cal C}_1,{\cal C}_2\rangle=0~~\forall~ {\bf k}, ~~~{\rm with}~~|\Psi\rangle_{\bf QVac}:\equiv|{\cal C}_1,{\cal C}_2\rangle\eea 
In a most generalized prescription this arbitrary quantum vacuum state can be written in terms of the ground state, which in cosmology commonly known as the {\it Bunch Davies} Euclidean vacuum state by the following expression:
\bea |\Psi\rangle_{\bf QVac}:\equiv|{\cal C}_1,{\cal C}_2\rangle &=&\prod_{{\bf k}}\frac{1}{\sqrt{|{\cal C}_1|}}\exp\left(-\frac{i{\cal C}^{*}_2}{2{\cal C}^{*}_1}a^{\dagger}_{\bf k}a^{\dagger}_{\bf k}\right)|\Psi\rangle_{\bf BD}=\frac{1}{\sqrt{|{\cal C}_1|}}\exp\left(-\frac{i{\cal C}^{*}_2}{2{\cal C}^{*}_1}\sum_{{\bf k}}a^{\dagger}_{\bf k}a^{\dagger}_{\bf k}\right)|\Psi\rangle_{\bf BD}~~~~~~~~\eea
Now we will use the following replacement rule:
\bea \sum_{{\bf k}}\longrightarrow \int \frac{d^{3}{\bf k}}{(2\pi)^3}.\eea
Using this further one can express further the arbitrary quantum vacuum state in terms of the {\it Bunch Davies} Euclidean vacuum state as:
\bea |\Psi\rangle_{\bf QVac}:\equiv|{\cal C}_1,{\cal C}_2\rangle 
&=&\frac{1}{\sqrt{|{\cal C}_1|}}\exp\left(-\frac{i{\cal C}^{*}_2}{2{\cal C}^{*}_1}\int \frac{d^{3}{\bf k}}{(2\pi)^3}a^{\dagger}_{\bf k}a^{\dagger}_{\bf k}\right)|\Psi\rangle_{\bf BD}~~~~~~~~~ \eea
where we have actually identified the ground state as {\it Bunch Davies} Euclidean vacuum state, which is given by:
\bea |0\rangle_{\bf ground}:=|\Psi\rangle_{\bf BD}.\eea 
Additionally, it is important to note that the arbitrary quantum vacuum state, $|{\cal C}_1,{\cal C}_2\rangle $, satisfy the following constraint condition:
\bea \widehat{\cal P}_{{\cal C}_1,{\cal C}_2}|{\cal C}_1,{\cal C}_2\rangle &=&\int \frac{d^3{\bf p}}{(2\pi)^3}~{\bf p}~{\cal C}^{\dagger}_{\bf p, 12}{\cal C}_{\bf p, 12}|{\cal C}_1,{\cal C}_2\rangle \nonumber\\
&=&\prod_{{\bf k}}\int \frac{d^3{\bf p}}{(2\pi)^3}~{\bf p}~{\cal C}^{\dagger}_{\bf p, 12}{\cal C}_{\bf p, 12}\frac{1}{\sqrt{|{\cal C}_1|}}\exp\left(-\frac{i{\cal C}^{*}_2}{2{\cal C}^{*}_1}a^{\dagger}_{\bf k}a^{\dagger}_{\bf k}\right)|\Psi\rangle_{\bf BD}\nonumber\\
&=&\int \frac{d^3{\bf p}}{(2\pi)^3}~{\bf p}~{\cal C}^{\dagger}_{\bf p, 12}{\cal C}_{\bf p, 12}\frac{1}{\sqrt{|{\cal C}_1|}}\exp\left(-\frac{i{\cal C}^{*}_2}{2{\cal C}^{*}_1}\sum_{\bf k}a^{\dagger}_{\bf k}a^{\dagger}_{\bf k}\right)|\Psi\rangle_{\bf BD}\nonumber\\
&=&\int \frac{d^3{\bf p}}{(2\pi)^3}~{\bf p}~{\cal C}^{\dagger}_{\bf p, 12}{\cal C}_{\bf p, 12}\frac{1}{\sqrt{|{\cal C}_1|}}\exp\left(-\frac{i{\cal C}^{*}_2}{2{\cal C}^{*}_1}\int \frac{d^3{\bf k}}{(2\pi)^3}~a^{\dagger}_{\bf k}a^{\dagger}_{\bf k}\right)|\Psi\rangle_{\bf BD}=0. \eea
Here it is important to note that, the relationship between the annihilation and creation operator in the $\alpha$-vacua and the Bunch-Davies
vacuum is established by the following {\it Bogoliubov transformation}:
\bea && {\cal C}_{\bf k, 12}={\cal C}^{*}_1~a_{\bf k}-{\cal C}^{*}_2~a^{\dagger}_{\bf -k}~~\longrightarrow~~ {\cal C}^{\dagger}_{\bf k, 12}={\cal C}_1~a^{\dagger}_{\bf k}-{\cal C}_2~a_{\bf -k},\\
&& a_{\bf k}={\cal C}_1~ {\cal C}_{\bf k, 12}+{\cal C}^{*}_2~{\cal C}^{\dagger}_{\bf- k, 12}~~\longrightarrow~~
 a^{\dagger}_{\bf k}={\cal C}^{*}_1~ {\cal C}^{\dagger}_{\bf k, 12}+{\cal C}_2~{\cal C}_{\bf- k, 12}.\eea 
Here $({\cal C}_{\bf k, 12},{\cal C}^{\dagger}_{\bf k, 12})$ and $(a_{\bf k},a^{\dagger}_{\bf k})$, are the creation and annihilation operators of the arbitrary generalized vacua and the Bunch Davies vacuum respectively. 

In the context of quantum field theory of spatially flat FLRW cosmology,  one can define the initial quantum mechanical state corresponding to the class of all excited $SO(1,4)$ isommetric {\it Mota-Allen} or $(\alpha,\gamma)$-vacua states, which are characterized by two real parameter family $\alpha$ and $\gamma$. Now one can find that this type of vacua is not CPT symmetry preserving.  Following the previously mentioned general construction one can write down the {\it Mota-Allen} or $(\alpha,\gamma)$-vacua states in terms of the well known adiabatic Bunch Davies vacuum state by the following expression:
\bea {|\Psi_{\alpha,\gamma}\rangle=\frac{1}{\sqrt{|\cosh \alpha|}}~\exp\left(-\frac{i}{2}\exp(-i\gamma)\tanh \alpha~\int \frac{d^3{\bf k}}{(2\pi)^3}~a^{\dagger}_{\bf k}a^{\dagger}_{\bf k}\right)|\Psi_{\bf BD}\rangle}~,~~~~~~~\eea
where the integration constants ${\cal C}_1$ and ${\cal C}_2$ for {\it Mota Allen} or $(\alpha,\gamma)$ vacua can be parametrizes as: 
\bea {\cal C}_1=\cosh\alpha,~~~~{\cal C}_2=\exp(i\gamma)~\sinh\alpha, \eea
which satisfy the following normalization condition:
\bea |{\cal C}_1|^2-|{\cal C}_2|^2=1~~~~\Longrightarrow~~~~\cosh^2\alpha-\sinh^2\alpha=1~~~~\forall~~\alpha~.\eea
 Here one can easily observed that, if we fix $\alpha=0$ and $\gamma=0$, then one can easily get back the usual quantum adiabatic Bunch Davies vacuum state which are given by, ${\cal C}_1=1$ and ${\cal C}_2=0$. On the other hand if we are interested in CPT invariant quantum vacuum state then we choose only $\gamma=0$ and get CPT invariant $SO(1,4)$ isommetric $\alpha$ vacua states, which is given by:
\bea {|\Psi_{\alpha}\rangle:=|\Psi_{\alpha,0}\rangle=\frac{1}{\sqrt{|\cosh \alpha|}}~\exp\left(-\frac{i}{2}~\tanh \alpha~\int \frac{d^3{\bf k}}{(2\pi)^3}~a^{\dagger}_{\bf k}a^{\dagger}_{\bf k}\right)|\Psi_{\bf BD}\rangle}~,~~~~~~~\eea
Now, using the definition of the {\it Mota Allen} vacua or the $\alpha$ vacua or the Euclidean Bunch Davies states one can explicitly compute the expression for the desired OTOCs defined in this paper. In the following section we will derive these results explicitly.
\subsubsection{Quantum partition function in terms of rescaled perturbation field variable in Primordial Cosmology} 
In presence of these $SO(1,4)$ isommetric excited CPT violating {\it Mota Allen} or CPT preserving $\alpha$-vacua states the quantum mechanical thermal partition function can be expressed as:
\bea &&Z_{\alpha,\gamma}(\beta;\tau_1)
=\int d\Psi_{\alpha,\gamma}~\langle \Psi_{\alpha,\gamma}|e^{-\beta \hat{H}(\tau_1)}|\Psi_{\alpha,\gamma} \rangle=\frac{1}{|\cosh\alpha|}~\exp\left(-2\sin\gamma\tan\alpha\right)~Z_{\bf BD}(\beta;\tau_1),~~~~~~~~~~\\ &&Z_{\alpha}(\beta;\tau_1)
=\int d\Psi_{\alpha}~\langle \Psi_{\alpha}|e^{-\beta \hat{H}(\tau_1)}|\Psi_{\alpha} \rangle=\frac{1}{|\cosh\alpha|}Z_{\bf BD}(\beta;\tau_1),\eea
which further implies the following connecting realtionship between the thermal quantum partition functions obtained from {\it Mota Allen} vacua, $\alpha$ vacua and Bunch Davies vacuum state, which is given by:
\bea &&{Z_{\alpha,\gamma}(\beta;\tau_1)
=\exp\left(-2\sin\gamma\tan\alpha\right)~Z_{\alpha}(\beta;\tau_1)=|\cosh\alpha|^{-1}~\exp\left(-2\sin\gamma\tan\alpha\right)~Z_{\bf BD}(\beta;\tau_1)}.~~~~~~~~~~~\eea 
where $Z_{\bf BD}$ is the quantum partition function computed from adiabatic Bunch Davies vacuum as:
\bea :Z_{\bf BD}(\beta;\tau_1):&=&\int d\Psi_{\bf BD}~\langle \Psi_{\bf BD}|\exp\left(-\beta\int d^3{\bf k}~a^{\dagger}_{\bf k}a_{\bf k}~E_{\bf k}(\tau_1)\right)|\Psi_{\bf BD}\rangle\nonumber\\
&=&\exp\left(-\int d^3{\bf k}~\ln\left(2\sinh \frac{\beta E_{\bf k}(\tau_1)}{2}\right)\right).\eea.
where the contribution from the divergent Delta function can be removed after performing the normal order operation.
This implies further we get the following simplified result for the quantum thermal partition function obtained from the {\it Mota Allen} vacua, $\alpha$ vacua and Euclidean Bunch Davies vacuum state after normal ordering:
\bea &&{:Z_{\alpha,\gamma}(\beta;\tau_1):
=\frac{1}{|\cosh\alpha|}~\exp\left(-\left[2\sin\gamma\tan\alpha+\int d^3{\bf k}~\ln\left(2\sinh \frac{\beta E_{\bf k}(\tau_1)}{2}\right)\right]\right)}.~~~~~~~~~~~\eea
The detailed technical computation of these results can be found in the Appendix.   

\subsubsection{Quantum partition function in terms of cosmological scalar curvature perturbation field variable in Primordial Cosmology} 
In this subsection our prime objective is to find out the explicit expression for the quantum mechanical partition function in terms of the scalar co-moving curvature perturbation field variable for different choices for the initial vacuum states available in the context of primordial cosmology. To serve this purpose the time dependent dispersion relation can be expressed in terms of the curvature perturbation variable as:
 \bea E_{{\bf k}}(\tau_1)&=&|\Pi_{\bf k}(\tau_1)|^2+\omega^2_{\bf k}(\tau_1)|f_{\bf k}(\tau_1)|^2=z^2(\tau_1)\left(E_{{\bf k},\zeta}(\tau_1)+\Delta_{{\bf k},\zeta}(\tau_1)\right),~~~~~~\eea 
 where the additional contribution is characterized by a new momentum and time dependent function, $\Delta_{{\bf k},\zeta}(\tau_1) $, which is defined as:
 \bea \Delta_{{\bf k},\zeta}(\tau_1):=\left(\Pi^{\zeta}_{-\bf k}(\tau_1)\zeta_{\bf k}(\tau_1)+\Pi^{\zeta}_{\bf k}(\tau_1)\zeta_{-\bf k}(\tau_1)\right) \left(\frac{1}{z(\tau_1)}\frac{dz(\tau_1)}{d\tau_1}\right).\eea
where we define the conformal time dependent energy dispersion relation in terms of the co-moving curvature perturbation variable in the context of primordial cosmological perturbation variable as:
\bea E_{{\bf k},\zeta}(\tau_1):&=&\left|\Pi^{\zeta}_{\bf k}(\tau_1)\right|^2+\left(\omega^2_{\bf k}(\tau_1)+\left(\frac{1}{z(\tau_1)}\frac{dz(\tau_1)}{d\tau_1}\right)^2\right)|\zeta_{\bf k}(\tau_1)|^2.\eea
Now, the normal ordered thermal partition function function obtained from the {\it Mota Allen} vacua state can be expressed in terms of the time dependent dispersion relation for co-moving curvature perturbation as:
\bea &&{:Z^{\zeta}_{\alpha,\gamma}(\beta;\tau_1):
=\frac{1}{|\cosh\alpha|}\exp\left(-\left[2\sin\gamma\tan\alpha+\int d^3{\bf k}~\ln\left(2\sinh \frac{\beta z^2(\tau_1)\left(E_{{\bf k},\zeta}(\tau_1)+\Delta_{{\bf k},\zeta}(\tau_1)\right)}{2}\right)\right]\right)}.~~~~~~~~~~~\eea  
    \subsection{Trace of two-point ``in-in"  non-chaotic extension of  OTO amplitudes for Primordial Cosmology} 
Here we compute the numerator of the one of the $2$-point OTOCs for different quantum vacuum states, which are described by the following expressions:
    \bea && {\rm Tr}\left[e^{-\beta \widehat{H}(\tau_1)}\left[\hat{f}({\bf x},\tau_1),\hat{f}({\bf x},\tau_2)\right]\right]_{(\alpha,\gamma)}\nonumber\\
 &&= \frac{\exp\left(-2\sin\gamma\tan\alpha\right)}{|\cosh\alpha|}\int d\Psi_{\bf BD}~\prod^{2}_{j=1}\int \frac{d^3k_j}{(2\pi)^3}\exp\left[i{\bf k}_j.{\bf x}\right]\langle\Psi_{\bf BD}|\left[\sum^{2}_{i=1}\hat{\nabla}^{(1)}_i({\bf k}_1,{\bf k}_2;\tau_1,\tau_2;\beta)\right]|\Psi_{\bf BD}\rangle.~~~~~~~~~ \eea 
 which further implies the following fact:
 \bea  {\rm Tr}\left[e^{-\beta \widehat{H}(\tau_1)}\left[\hat{f}({\bf x},\tau_1),\hat{f}({\bf x},\tau_2)\right]\right]_{(\alpha,\gamma)}&=&\exp\left(-2\sin\gamma\tan\alpha\right) {\rm Tr}\left[e^{-\beta \widehat{H}(\tau_1)}\left[\hat{f}({\bf x},\tau_1),\hat{f}({\bf x},\tau_2)\right]\right]_{({\alpha})}\nonumber\\
&=& \frac{\exp\left(-2\sin\gamma\tan\alpha\right)}{|\cosh\alpha|} {\rm Tr}\left[e^{-\beta \widehat{H}(\tau_1)}\left[\hat{f}({\bf x},\tau_1),\hat{f}({\bf x},\tau_2)\right]\right]_{({\bf BD})}.\nonumber\\
&&~~~~~~~~~~~~~~~~~~~~~~
~~~~~~~~~~~~~~~~~~~~~~~~~~~~~~~~~~~~~~~~~~~~~~~~~~~~~~~~~~~~\eea
Here we compute the numerator of the other $2$-point OTOC for different quantum vacuum states, which are described by the following expressions:
    \bea && {\rm Tr}\left[e^{-\beta \widehat{H}(\tau_1)}\left[\hat{\Pi}({\bf x},\tau_1),\hat{\Pi}({\bf x},\tau_2)\right]\right]_{(\alpha,\gamma)}\nonumber\\
 &&= \frac{\exp\left(-2\sin\gamma\tan\alpha\right)}{|\cosh\alpha|}\int d\Psi_{\bf BD}~\prod^{2}_{j=1}\int \frac{d^3k_j}{(2\pi)^3}\exp\left[i{\bf k}_j.{\bf x}\right]\langle\Psi_{\bf BD}|\left[\sum^{2}_{i=1}\hat{\nabla}^{(2)}_i({\bf k}_1,{\bf k}_2;\tau_1,\tau_2;\beta)\right]|\Psi_{\bf BD}\rangle.~~~~~~~~~ \eea
 which further implies the following fact:
 \bea  {\rm Tr}\left[e^{-\beta \widehat{H}(\tau_1)}\left[\hat{f}({\bf x},\tau_1),\hat{f}({\bf x},\tau_2)\right]\right]_{(\alpha,\gamma)}&=&\exp\left(-2\sin\gamma\tan\alpha\right) {\rm Tr}\left[e^{-\beta \widehat{H}(\tau_1)}\left[\hat{f}({\bf x},\tau_1),\hat{f}({\bf x},\tau_2)\right]\right]_{({\alpha})}\nonumber\\
&=& \frac{\exp\left(-2\sin\gamma\tan\alpha\right)}{|\cosh\alpha|} {\rm Tr}\left[e^{-\beta \widehat{H}(\tau_1)}\left[\hat{f}({\bf x},\tau_1),\hat{f}({\bf x},\tau_2)\right]\right]_{({\bf BD})}.\nonumber\\
&&~~~~~~~~~~~~~~~~~~~~~~
~~~~~~~~~~~~~~~~~~~~~~~~~~~~~~~~~~~~~~~~~~~~~~~~~~~~~~~~~~~~\eea 
  Further, our aim is to compute the individual contributions which in the normal ordered form is given by the following expression and computed in Appendix:
  \bea &&\int d\Psi_{\bf BD}~\langle\Psi_{\bf BD}|:\hat{\nabla}^{(1)}_i({\bf k}_1,{\bf k}_2;\tau_1,\tau_2;\beta):|\Psi_{\bf BD}\rangle=\int d\Psi_{\bf BD}~ \langle\Psi_{\bf BD}|:e^{-\beta \hat{H}(\tau_1)}~\widehat{\Delta}^{(1)}_i({\bf k}_1,{\bf k}_2;\tau_1,\tau_2):|\Psi_{\bf BD}\rangle\nonumber\\
  &&~~~~~~~~~~~~~~~~~
 ~~~~~~~~~~~~~~~~~ ~~~~~~~~~~~~~~~~~~~~~~~~~~~~~~~~~~~~~~~~~~~~~~~~~~~~~~~~~~~\forall~i=1,2.~~~~~~~~~\\
  &&\int d\Psi_{\bf BD}~\langle\Psi_{\bf BD}|:\hat{\nabla}^{(2)}_i({\bf k}_1,{\bf k}_2;\tau_1,\tau_2;\beta):|\Psi_{\bf BD}\rangle=\int d\Psi_{\bf BD}~ \langle\Psi_{\bf BD}|:e^{-\beta \hat{H}(\tau_1)}~\widehat{\Delta}^{(2)}_i({\bf k}_1,{\bf k}_2;\tau_1,\tau_2):|\Psi_{\bf BD}\rangle\nonumber\\
  &&~~~~~~~~~~~~~~~~~
 ~~~~~~~~~~~~~~~~~ ~~~~~~~~~~~~~~~~~~~~~~~~~~~~~~~~~~~~~~~~~~~~~~~~~~~~~~~~~~~\forall~i=1,2.~~~~~~~~~\eea 
 Further, the trace of sum of these individual two sets of two-point ``in-in" OTO  amplitudes in normal ordered form can be expressed as:
    \bea &&\int d\Psi_{\bf BD}~\langle\Psi_{\bf BD}|\sum^{2}_{i=1}:\hat{\nabla}^{(1)}_i({\bf k}_1,{\bf k}_2;\tau_1,\tau_2;\beta):|\Psi_{\bf BD}\rangle=(2\pi)^3\delta^3({\bf k}_1+{\bf k}_2)~{\bf P}_{1}({\bf k}_1,{\bf k}_2;\tau_2,\tau_2;\beta).~~~~~~~~~\\
   &&\int d\Psi_{\bf BD}~\langle\Psi_{\bf BD}|\sum^{2}_{i=1}:\hat{\nabla}^{(2)}_i({\bf k}_1,{\bf k}_2;\tau_1,\tau_2;\beta):|\Psi_{\bf BD}\rangle=(2\pi)^3\delta^3({\bf k}_1+{\bf k}_2)~{\bf P}_{2}({\bf k}_1,{\bf k}_2;\tau_2,\tau_2;\beta).~~~~~~~~~\eea  
  Here we introduce, ${\bf P}_{1}({\bf k}_1,{\bf k}_2;\tau_2,\tau_2;\beta)$ and ${\bf P}_{2}({\bf k}_1,{\bf k}_2;\tau_2,\tau_2;\beta)$ are the temperature dependent two-point function, which is defined as:
  \bea &&{\bf P}_1({\bf k}_1,{\bf k}_2;\tau_2,\tau_2;\beta)=\exp\left(-\int d^3{\bf k}~\ln\left(2\sinh \frac{\beta E_{\bf k}(\tau_1)}{2}\right)\right)\nonumber\\
  &&~~~~~~~~~~~\left[ {\cal D}^{(1)}_2({\bf k}_1,{\bf k}_2;\tau_1,\tau_2)+{\cal D}^{(1)}_3({\bf k}_1,{\bf k}_2;\tau_1,\tau_2)-{\cal L}^{(1)}_2({\bf k}_1,{\bf k}_2;\tau_1,\tau_2)-{\cal L}^{(1)}_3({\bf k}_1,{\bf k}_2;\tau_1,\tau_2)\right].~~~~~~~~~~~~\\
  &&{\bf P}_2({\bf k}_1,{\bf k}_2;\tau_2,\tau_2;\beta)=\exp\left(-\int d^3{\bf k}~\ln\left(2\sinh \frac{\beta E_{\bf k}(\tau_1)}{2}\right)\right)\nonumber\\
  &&~~~~~~~~~~~\left[ {\cal D}^{(2)}_2({\bf k}_1,{\bf k}_2;\tau_1,\tau_2)+{\cal D}^{(2)}_3({\bf k}_1,{\bf k}_2;\tau_1,\tau_2)-{\cal L}^{(2)}_2({\bf k}_1,{\bf k}_2;\tau_1,\tau_2)-{\cal L}^{(2)}_3({\bf k}_1,{\bf k}_2;\tau_1,\tau_2)\right].~~~~~~~~~~~~\eea
 \subsection{New OTOCs from regularised two-point ``in-in"  non-chaotic auto-correlated OTO  amplitudes: rescaled field version}
  
The cosmological OTOC without normalization for different  quantum initial vacua can be expressed as:
    \bea && Y^{f}_1(\tau_1,\tau_2)=-\frac{1}{Z(\beta;\tau_1)}{\rm Tr}\left[e^{-\beta \widehat{H}(\tau_1)}\left[\hat{f}({\bf x},\tau_1),\hat{f}({\bf x},\tau_2)\right]\right]=-\int \frac{d^3{\bf k}_1}{(2\pi)^3}{\cal P}_1({\bf k}_1,-{\bf k}_1;\tau_1,\tau_2),~~~~~~~~~~~ \\
   && Y^{f}_2(\tau_1,\tau_2)=-\frac{1}{Z(\beta;\tau_1)}{\rm Tr}\left[e^{-\beta \widehat{H}(\tau_1)}\left[\hat{\Pi}({\bf x},\tau_1),\hat{\Pi}({\bf x},\tau_2)\right]\right]=-\int \frac{d^3{\bf k}_1}{(2\pi)^3}{\cal P}_2({\bf k}_1,-{\bf k}_1;\tau_1,\tau_2),~~~~~~~~~~~ \eea 
  where the two-point OTO  amplitude functions are explicitly given by the following expressions:
  \bea {\cal P}_1({\bf k}_1,-{\bf k}_1;\tau_1,\tau_2):&=&\left[ {\cal D}^{(1)}_2({\bf k}_1,-{\bf k}_1;\tau_1,\tau_2)+{\cal D}^{(1)}_3({\bf k}_1,-{\bf k}_1;\tau_1,\tau_2)\right.\nonumber\\
  && \left.~~~~~~~~~~~~~~~~~~~~~~~~~~~~~-{\cal L}^{(1)}_2({\bf k}_1,-{\bf k}_1;\tau_1,\tau_2)- {\cal L}^{(1)}_3({\bf k}_1,-{\bf k}_1;\tau_1,\tau_2)\right],~~~~~~~~~~\\
  {\cal P}_2({\bf k}_1,-{\bf k}_1;\tau_1,\tau_2):&=&\left[ {\cal D}^{(2)}_2({\bf k}_1,-{\bf k}_1;\tau_1,\tau_2)+{\cal D}^{(2)}_3({\bf k}_1,-{\bf k}_1;\tau_1,\tau_2)\right.\nonumber\\
  && \left.~~~~~~~~~~~~~~~~~~~~~~~~~~~~-{\cal L}^{(2)}_2({\bf k}_1,-{\bf k}_1;\tau_1,\tau_2)- {\cal L}^{(2)}_3({\bf k}_1,-{\bf k}_1;\tau_1,\tau_2)\right].~~~~~~~~~~\eea
  Here we define:
 \bea 
 {\cal D}^{(1)}_2 ({\bf k}_1,{\bf k}_2;\tau_1,\tau_2)&=&f^{*}_{{\bf -k}_1}(\tau_1)f_{{\bf k}_2}(\tau_2),\\
 {\cal D}^{(1)}_3 ({\bf k}_1,{\bf k}_2;\tau_1,\tau_2)&=&f_{{\bf k}_1}(\tau_1)f^{*}_{{\bf -k}_2}(\tau_2),\\
 {\cal L}^{(1)}_2 ({\bf k}_1,{\bf k}_2;\tau_1,\tau_2)&=&f^{*}_{{\bf -k}_1}(\tau_2)f_{{\bf k}_2}(\tau_1),\eea\bea
 {\cal L}^{(1)}_3 ({\bf k}_1,{\bf k}_2;\tau_1,\tau_2)&=&f_{{\bf k}_1}(\tau_2)f^{*}_{-{\bf k}_2}(\tau_1),\\  {\cal D}^{(2)}_2 ({\bf k}_1,{\bf k}_2;\tau_1,\tau_2)&=&\Pi^{*}_{{\bf -k}_1}(\tau_1)\Pi_{{\bf k}_2}(\tau_2),\\
 {\cal D}^{(2)}_3 ({\bf k}_1,{\bf k}_2;\tau_1,\tau_2)&=&\Pi_{{\bf k}_1}(\tau_1)\Pi^{*}_{{\bf -k}_2}(\tau_2),\\
 {\cal L}^{(2)}_2 ({\bf k}_1,{\bf k}_2;\tau_1,\tau_2)&=&\Pi^{*}_{{\bf -k}_1}(\tau_2)\Pi_{{\bf k}_2}(\tau_1),\\
 {\cal L}^{(2)}_3 ({\bf k}_1,{\bf k}_2;\tau_1,\tau_2)&=&\Pi_{{\bf k}_1}(\tau_2)\Pi^{*}_{-{\bf k}_2}(\tau_1).\eea
  Now we need to evaluate explicitly by doing the momentum integration over three volume. Now to compute this integral one can express the volume element as:
  \bea \frac{d^3 {\bf k}_1}{(2\pi)^3}=4\pi~k^2_1~dk_1~~~~0<k_1<L.~\eea
  Here we have taken care of the fact that the individual contribution appearing in the two-point OTOC momentum integral is isotropic. Also, we have introduced a momentum finite large cut-off to regulate the contribution of this integral. 
  
  Consequently, one can write the following simplified expressions for the two-point un-normalized OTOC as:
   \bea && {Y^{f}_1(\tau_1,\tau_2)=-\frac{1}{2\pi^2}{\cal B}_1(\tau_1,\tau_2)}~,~~~~~~~~~~~\\
   && {Y^{f}_2(\tau_1,\tau_2)=-\frac{1}{2\pi^2}{\cal B}_2(\tau_1,\tau_2)}~ \eea 
   where the conformal time scale dependent regularized integrals, ${\cal B}_1(\tau_1,\tau_2)$ and ${\cal B}_2(\tau_1,\tau_2)$ as appearing in the above expression, are defined as:
  \bea {\cal B}_1(\tau_1,\tau_2):&=&(-\tau_1)^{\frac{1}{2}-\nu}(-\tau_2)^{\frac{1}{2}-\nu}\left[Z^{(1)}_{(1)}(\tau_1,\tau_2)+Z^{(1)}_{(2)}(\tau_1,\tau_2)-Z^{(1)}_{(3)}(\tau_1,\tau_2)-Z^{(1)}_{(4)}(\tau_1,\tau_2)\right],~~~~~~~~~~~\\
   {\cal B}_2(\tau_1,\tau_2):&=&(-\tau_1)^{\frac{3}{2}-\nu}(-\tau_2)^{\frac{3}{2}-\nu}\left[Z^{(2)}_{(1)}(\tau_1,\tau_2)+Z^{(2)}_{(2)}(\tau_1,\tau_2)-Z^{(2)}_{(3)}(\tau_1,\tau_2)-Z^{(2)}_{(4)}(\tau_1,\tau_2)\right],~~~~~~~~~~\eea 
  where we have introduced the time dependent four individual amplitudes, $Z^{(j)}_{(i)}(\tau_1,\tau_2)~\forall~i,j=1,2,3,4$,  which are explicitly defined in the Appendix.  These amplitudes satisfy the following symmetry properties:
\bea && Z^{(l)}_{(2)}(\tau_1,\tau_2)=(-1)^{-(2\nu+1)}Z^{(l)}_{(1)}(\tau_1,\tau_2)~~~~~~\forall~~~~l=1,2,\\
&&Z^{(l)}_{(4)}(\tau_1,\tau_2)=(-1)^{-(2\nu+1)}Z^{(l)}_{(3)}(\tau_1,\tau_2)~~~~~~\forall~~~~l=1,2,\eea
using which the simplified form of the momentum integrated time dependent two-point two new desired OTOCs can be written as:
 \bea {Y^{f}_1(\tau_1,\tau_2)=-\frac{1}{2\pi^2}{\cal B}_1(\tau_1,\tau_2)=\frac{(-\tau_1)^{\frac{1}{2}-\nu}(-\tau_2)^{\frac{1}{2}-\nu}}{2\pi^2}\left[1+(-1)^{-(2\nu+1)}\right]\left(Z^{(1)}_{(3)}(\tau_1,\tau_2)-Z^{(1)}_{(2)}(\tau_1,\tau_2)\right)}.~~~~~~~~\\
 {Y^{f}_2(\tau_1,\tau_2)=-\frac{1}{2\pi^2}{\cal B}_2(\tau_1,\tau_2)=\frac{(-\tau_1)^{\frac{3}{2}-\nu}(-\tau_2)^{\frac{3}{2}-\nu}}{2\pi^2}\left[1+(-1)^{-(2\nu+1)}\right]\left(Z^{(2)}_{(3)}(\tau_1,\tau_2)-Z^{(2)}_{(2)}(\tau_1,\tau_2)\right)}.~~~~~~~~\eea
 These expression can be simplified in terms of the slowly varying time dependent phase factors as:
  \bea {Y^{f}_1(\tau_1,\tau_2)=\frac{(-\tau_1)^{\frac{1}{2}-\nu}(-\tau_2)^{\frac{1}{2}-\nu}}{2\pi^2}\left[1+\exp(-i(2\nu+1)\pi)\right]\left(Z^{(1)}_{(3)}(\tau_1,\tau_2)-Z^{(1)}_{(2)}(\tau_1,\tau_2)\right)}.~~~~~~~~\\
 {Y^{f}_2(\tau_1,\tau_2)=\frac{(-\tau_1)^{\frac{3}{2}-\nu}(-\tau_2)^{\frac{3}{2}-\nu}}{2\pi^2}\left[1+\exp(-i(2\nu+1)\pi)\right]\left(Z^{(2)}_{(3)}(\tau_1,\tau_2)-Z^{(2)}_{(2)}(\tau_1,\tau_2)\right)}.~~~~~~~~\eea
 Now, in the large mass limit we need to replace $\nu\rightarrow -i|\nu|$, for which we get the following expressions for the two-point functions:
  \bea {Y^{f}_1(\tau_1,\tau_2)=\frac{(-\tau_1)^{\frac{1}{2}+i|\nu|}(-\tau_2)^{\frac{1}{2}+i|\nu|}}{2\pi^2}\left[1+\underbrace{\exp(-2|\nu|\pi)}_{\textcolor{red}{\bf Boltzmann~suppression}}\right]\left({\cal Z}^{(1)}_{(3)}(\tau_1,\tau_2)-{\cal Z}^{(1)}_{(2)}(\tau_1,\tau_2)\right)}.~~~~~~~~\\
 {Y^{f}_2(\tau_1,\tau_2)=\frac{(-\tau_1)^{\frac{3}{2}+i|\nu|}(-\tau_2)^{\frac{3}{2}+i|\nu|}}{2\pi^2}\left[1+\underbrace{\exp(-2|\nu|\pi)}_{\textcolor{red}{\bf Boltzmann~suppression}}\right]\left({\cal Z}^{(2)}_{(3)}(\tau_1,\tau_2)-{\cal Z}^{(2)}_{(2)}(\tau_1,\tau_2)\right)},~~~~~~~~\eea
 where we define:
 \bea \lim_{\nu\rightarrow -i|\nu|}Z^{(l)}_{(i)}(\tau_1,\tau_2)\equiv {\cal Z}^{(l)}_{(i)}(\tau_1,\tau_2)~~~~~~\forall~~~~l=1,2,~~~~~i=2,3.\eea
 We need to explicitly evaluate the expression for these above mentioned momentum integrals which will going to fix the final expression for the two desired two-point  OTOCs in the context of primordial cosmological perturbation theory. We have presented the detailed computation of these all integrals in the Appendix for better understanding purpose of the two conformal time dependence of each of the contributions.   
 \subsection{New OTOCs from regularised four-point ``in-in"  non-chaotic auto-correlated OTO  amplitudes: curvature perturbation field version}

Here we need to compute the desired two-point new OTOCs  in terms of the scalar curvature perturbation and the canonically conjugate momentum, which are given by:
\bea 
&&{Y^{\zeta}_{1}(\tau_1,\tau_2)=-\frac{1}{Z^{\zeta}(\beta,\tau_1)}{\rm Tr}\left[e^{-\beta \hat{H}(\tau_1)}\left[\hat{\zeta}({\bf x},\tau_1),\hat{\Pi}({\bf x},\tau_2)\right]\right]=\frac{1}{z(\tau_1)z(\tau_2)}Y^{f}_{1}(\tau_1,\tau_2)}.~~~~~~~~~~~~~\\ && {Y^{\zeta}_{2}(\tau_1,\tau_2)=-\frac{1}{Z^{\zeta}(\beta,\tau_1)}{\rm Tr}\left[e^{-\beta \hat{H}(\tau_1)}\left[\hat{\zeta}({\bf x},\tau_1),\hat{\Pi}({\bf x},\tau_2)\right]\right]=\frac{1}{z(\tau_1)z(\tau_2)}Y^{f}_{2}(\tau_1,\tau_2)},~~~~~~~~~~~~~\eea
where as a choice of the initial quantum vacuum state we have considered {\rm Mota-Allen}, $\alpha$,~{\rm Bunch-Davies} states, which are the three popular choices of the initial vacuum states in the present computation.

Now substituting the explicit for of the two-point function that we have derived in the previous section we get the following expression:
\bea && {Y^{\zeta}_1(\tau_1,\tau_2)=-\frac{1}{2\pi^2}\frac{{\cal B}_1(\tau_1,\tau_2)}{z(\tau_1)z(\tau_2)}=\frac{(-\tau_1)^{\frac{1}{2}-\nu}(-\tau_2)^{\frac{1}{2}-\nu}}{2\pi^2~z(\tau_1)z(\tau_2)}\left[1+(-1)^{-(2\nu+1)}\right]\left(Z^{(1)}_{(3)}(\tau_1,\tau_2)-Z^{(1)}_{(2)}(\tau_1,\tau_2)\right)}\nonumber\\
&&\\ && {Y^{\zeta}_2(\tau_1,\tau_2)=-\frac{1}{2\pi^2}\frac{{\cal B}_2(\tau_1,\tau_2)}{z(\tau_1)z(\tau_2)}=\frac{(-\tau_1)^{\frac{3}{2}-\nu}(-\tau_2)^{\frac{3}{2}-\nu}}{2\pi^2~z(\tau_1)z(\tau_2)}\left[1+(-1)^{-(2\nu+1)}\right]\left(Z^{(2)}_{(3)}(\tau_1,\tau_2)-Z^{(2)}_{(2)}(\tau_1,\tau_2)\right)}\nonumber\\
&&\eea  
These above mentioned expression can be expressed in terms of the slowly varying conformal time dependent mass parameter which is appearing in the phase factor as:
\bea && {Y^{\zeta}_1(\tau_1,\tau_2)=\frac{(-\tau_1)^{\frac{1}{2}-\nu}(-\tau_2)^{\frac{1}{2}-\nu}}{2\pi^2~z(\tau_1)z(\tau_2)}\left[1+\exp(-i(2\nu+1)\pi)\right]\left(Z^{(1)}_{(3)}(\tau_1,\tau_2)-Z^{(1)}_{(2)}(\tau_1,\tau_2)\right)}~~~~~~~~~~~~\\ && {Y^{\zeta}_2(\tau_1,\tau_2)=\frac{(-\tau_1)^{\frac{3}{2}-\nu}(-\tau_2)^{\frac{3}{2}-\nu}}{2\pi^2~z(\tau_1)z(\tau_2)}\left[1+\exp(-i(2\nu+1)\pi)\right]\left(Z^{(2)}_{(3)}(\tau_1,\tau_2)-Z^{(2)}_{(2)}(\tau_1,\tau_2)\right)}~~~~~~~~~~~~\eea  
 Now, in the large mass limit we need to replace $\nu\rightarrow -i|\nu|$, for which we get the following expressions for the two-point functions:
  \bea {Y^{\zeta}_1(\tau_1,\tau_2)=\frac{(-\tau_1)^{\frac{1}{2}+i|\nu|}(-\tau_2)^{\frac{1}{2}+i|\nu|}}{2\pi^2~z(\tau_1)z(\tau_2)}\left[1+\underbrace{\exp(-2|\nu|\pi)}_{\textcolor{red}{\bf Boltzmann~suppression}}\right]\left({\cal Z}^{(1)}_{(3)}(\tau_1,\tau_2)-{\cal Z}^{(1)}_{(2)}(\tau_1,\tau_2)\right)}.~~~~~~~~\\
 {Y^{\zeta}_2(\tau_1,\tau_2)=\frac{(-\tau_1)^{\frac{3}{2}+i|\nu|}(-\tau_2)^{\frac{3}{2}+i|\nu|}}{2\pi^2~z(\tau_1)z(\tau_2)}\left[1+\underbrace{\exp(-2|\nu|\pi)}_{\textcolor{red}{\bf Boltzmann~suppression}}\right]\left({\cal Z}^{(2)}_{(3)}(\tau_1,\tau_2)-{\cal Z}^{(2)}_{(2)}(\tau_1,\tau_2)\right)}.~~~~~~~~\eea
Now additionally, few points we have to mention that from the finally obtained answer for the two-point OTOC's obtained from the two different set-ups: 
\begin{enumerate} 
\item The results obtained for two point OTOC's for different types of initial choice of the quantum vacuum states are same. No specific information of the initial condition in terms of the chosen quantum vacuum will be propagated in the overall factor of the final result of the two point OTOC's. But the information of the initial quantum vacuum will be there in the momentum integrated function $Z^{(i)}_{(2)}~\forall~i=1,2$ and $Z^{(i)}_{(3)}~\forall~i=1,2$ within a finite length cut-off $L$ as it captures the effect of the full asymptotic solution of the scalar modes and its associated momentum, and both of them are dependent on the factors ${\cal C}_1$ and ${\cal C}_2$ which carrying the required information. In this computation both the quantum and the classical contributions have been taken care of. But now if we just only concentrate on the quantum fluctuation part then the most interesting information is coming from the sub-Hubble or sub-horizon limiting region. By a very careful observation one can explicitly see that in the sub-Hubble region the final result of the two point OTOCs obtained from different types of the initial quantum states will give approximately the same answer and can solely described by the well known Euclidean Bunch Davies initial condition. This is quite interesting fact appearing in the context of two-point OTOC computation, because if we recap the similar computation of time ordered or anti time ordered correlators or any equal time two point correlators within the framework of cosmological perturbation theory, then we find that the final result, which we call in cosmologist's language power spectrum in the Fourier transformed space always captures the explicit information regarding the initial quantum vacuum state even in the super-Hubble region where the quantum effects dominates over the classical super-Hubble contribution.

\item In the presence of very heavy mass particle, which is within the framework of cosmology is identified by the limiting situation, $m\gg H$, where $H$ is the characteristic Hubble scale one need to take the analytical continuation of the mass parameter $\nu$ to the imaginary axis such that, $\nu\rightarrow -|\nu|$. As a consequence of this we get a exponential {\it Boltzmann} suppression by a factor of $\exp(-2|\nu|\pi)$, which can be be treated as a correction term in the final expressions for the two-point OTOCs. Now if the magnitude of the mass parameter $|\nu|$ is extremely large then one can simply drop this exponential {\it Boltzmann} factor in the final expression for the two-point OTOCs and can be written as:
\bea \textcolor{red}{\bf Large~|\nu|~(|\nu|\rightarrow \infty)~approximation:}~~~~~~~~~~~~~~~~~~~~~~~~~~~~\nonumber\\
{Y^{f}_1(\tau_1,\tau_2)\approx\frac{{\cal H}_1(\tau_1,\tau_2,|\nu|)}{2\pi^2} \nabla^{(1)}_{32}(\tau_1,\tau_2)}.~~~~~~~~\\
 {Y^{f}_2(\tau_1,\tau_2)\approx\frac{{\cal H}_2(\tau_1,\tau_2,|\nu|)}{2\pi^2} \nabla^{(2)}_{32}(\tau_1,\tau_2)},~~~~~~~~\\ {Y^{\zeta}_1(\tau_1,\tau_2)\approx\frac{{\cal H}_1(\tau_1,\tau_2,|\nu|)}{2\pi^2~z(\tau_1)z(\tau_2)} \nabla^{(1)}_{32}(\tau_1,\tau_2)}.~~~~~~~~\\
 {Y^{\zeta}_2(\tau_1,\tau_2)\approx\frac{{\cal H}_2(\tau_1,\tau_2,|\nu|)}{2\pi^2~z(\tau_1)z(\tau_2)} \nabla^{(2)}_{32}(\tau_1,\tau_2)},~~~~~~~~\eea 
 where we define a new function ${\cal H}_1(\tau_1,\tau_2,|\nu|)$ and ${\cal H}_2(\tau_1,\tau_2,|\nu|)$,  which are given by:
 \bea && {\cal H}_1(\tau_1,\tau_2,|\nu|):\equiv (-\tau_1)^{i|\nu|}(-\tau_2)^{i|\nu|}\left(1+\frac{1}{2}\ln(-\tau_1)\right)\left(1+\frac{1}{2}\ln(-\tau_2)\right),~~~~~~~~~~~\\
&& {\cal H}_2(\tau_1,\tau_2,|\nu|):\equiv (-\tau_1)^{i|\nu|}(-\tau_2)^{i|\nu|}\left(1+\frac{3}{2}\ln(-\tau_1)\right)\left(1+\frac{3}{2}\ln(-\tau_2)\right),~~~~~~~~~~~ \eea
 Here it is important to note that, to define this function we have used the following expansion:
\bea && \left(-\tau_k\right)^{\delta_{j}+i|\nu|}\approx  \left(-\tau_j\right)^{i|\nu|}\left[1+\delta_{j}\ln(-\tau_j)+\cdots\right],~~~\forall~~k=1,2~\&~j=1,2~(k\neq j)\nonumber\\
&&~~~~~~~~~~~~~~~~~~~~~~~~~~~~~~~~~~~~~~~~~~~~~~~~~`~~~~~~~~{\rm with}~~~\delta_{1}=\frac{1}{2},~~\delta_2=\frac{3}{2},~~~~~~~~\eea
where we have considered the fact that, $|\nu|\gg\delta_j ~~\forall ~~j=1,2$ and $|\nu|\rightarrow \infty$ to get the leading and the sub-leading or the next to leading order contribution, which is used for the further simplification of the obtained result for the two point OTOC in the large $|\nu|$ limit. We also have introduced the following symbolic representation for the sake of simplicity and clarity:
 \bea \nabla^{(i)}_{32}(\tau_1,\tau_2):\equiv \lim_{|\nu|\rightarrow \infty}\left({\cal Z}^{(i)}_{(3)}(\tau_1,\tau_2)-{\cal Z}^{(i)}_{(2)}(\tau_1,\tau_2)\right)~~~~~~\forall~~~i=1,2.\eea 
 On the other hand, if the magnitude of the mass parameter is small then one can expand the exponential {\it Boltzmann} factor in the {\it Taylor series} and keep upto the linear order term in $\nu$ in the expansion. This will give rise to an overall contribution, which is given by, $2\left(1-|\nu|\pi\right)$ and the corresponding two-point OTOCs can be approximately written as:
\bea \textcolor{red}{\bf Small~|\nu|~(|\nu|\rightarrow 0)~approximation:}~~~~~~~~~~~~~~~~~~~~~~\nonumber\\
{Y^{f}_1(\tau_1,\tau_2)=\frac{{\cal T}_1(\tau_1,\tau_2,|\nu|)}{2\pi^2}\Delta^{(1)}_{32}(\tau_1,\tau_2)}.~~~~~~~~\\
 {Y^{f}_2(\tau_1,\tau_2)=\frac{{\cal T}_2(\tau_1,\tau_2,|\nu|)}{2\pi^2}\Delta^{(2)}_{32}(\tau_1,\tau_2)},~~~~~~~~\\ {Y^{\zeta}_1(\tau_1,\tau_2)=\frac{{\cal T}_1(\tau_1,\tau_2,|\nu|)}{2\pi^2~z(\tau_1)z(\tau_2)}\Delta^{(1)}_{32}(\tau_1,\tau_2)}.~~~~~~~~\\
 {Y^{\zeta}_2(\tau_1,\tau_2)=\frac{{\cal T}_2(\tau_1,\tau_2,|\nu|)}{2\pi^2~z(\tau_1)z(\tau_2)}\Delta^{(2)}_{32}(\tau_1,\tau_2)},~~~~~~~~\eea
where we define a new function ${\cal T}_1(\tau_1,\tau_2,|\nu|)$ and ${\cal T}_2(\tau_1,\tau_2,|\nu|)$,  which are given by:
 \bea && {\cal T}_1(\tau_1,\tau_2,|\nu|):\equiv (-\tau_1)^{\frac{1}{2}}(-\tau_2)^{\frac{1}{2}}\left(1+i|\nu|\ln(-\tau_1)\right)\left(1+i|\nu|\ln(-\tau_2)\right)\left(1-|\nu|\pi\right),~~~~~~~~~~~\\ 
 && {\cal T}_2(\tau_1,\tau_2,|\nu|):\equiv (-\tau_1)^{\frac{3}{2}}(-\tau_2)^{\frac{3}{2}}\left(1+i|\nu|\ln(-\tau_1)\right)\left(1+i|\nu|\ln(-\tau_2)\right)\left(1-|\nu|\pi\right).~~~~~~~~~~~\eea
 Here it is important to note that, to define this function we have used the following expansion:
\bea &&\left(-\tau_k\right)^{\delta_{j}+i|\nu|}\approx  2\left(-\tau_k\right)^{\delta_{j}}\left[1+i|\nu|\ln(-\tau_j)+\cdots\right],~~~\forall~~j=1,2~\&~k=1,2~~(j\neq k)\nonumber\\
&&~~~~~~~~~~~~~~~~~~~~~~~~~~~~~~~~~~~~~~~~~~~~~~~~~~~~~~~~~~~~~~~~~~`{\rm with}~~~\delta_{1}=\frac{1}{2},~~\delta_2=\frac{3}{2},~~~~~~~~\eea
where we have considered the fact that, $\delta_j\gg |\nu|~~\forall ~~j=1,2$ and $|\nu|\rightarrow 0$ to get the leading and the sub-leading or the next to leading order contribution, which is used for the further simplification of the obtained result for the two point OTOC in the small $|\nu|$ limit.

 We also have introduced the following symbolic representation for the sake of simplicity and clarity:
 \bea \Delta^{(i)}_{32}(\tau_1,\tau_2):\equiv \lim_{|\nu|\rightarrow 0}\left({\cal Z}^{(i)}_{(3)}(\tau_1,\tau_2)-{\cal Z}^{(i)}_{(2)}(\tau_1,\tau_2)\right)~~~~~~\forall~~~i=1,2.\eea 
 
\item One can also consider various important cases, described by the mass parameter value $\nu=0$, $\nu=3/2$ and $\nu=1/2$. Here $\nu=3/2$ represents the massless field limiting situation and $\nu=1/2$ representing the conformally coupled case, where we have $m_{\phi}=\sqrt{2}~H$. On the other hand, $\nu=0$ represents the situation, where we have $m_{\phi}=3H/2$. Here one can treat both $\nu=0$ and $\nu=1/2$ in the partially massless field category.

\item  We have explicitly shown that the two-point OTOCs defined in this work is completely independent of the choice of the coordinate system and at the end only depend on the time scale on which the
cosmologically relevant operators in perturbation theory are  separated in time scale. To define properly we 
define these operators in terms of a specific space coordinate point but at different time scales and found that the final answer of the two-point OTOCs are only dependent on  time scale.

\item  From the present computation we found that the final expression for the two-point OTOCs are completely independent on temperature even though we start with a thermal canonical statistical ensemble in the trace formula of the two-point OTOCs. It implies that, we will get ultimately the description in terms of  statistical ensembles following the present description within the framework of cosmological perturbation theory, which is obviously a quite interesting observation to point out in the present context of computation.

\item  Now if we fix the conformal time coordinate of the two operators, $\tau_1=\tau$ and $\tau_2=\tau$, that means if both the cosmologically relevant
operators are defined at the same time coordinate (time separation vanishes)
then two-point OTOCs show divergence.
On the other hand, if we compare the obtained result with the expression for any two point equal time cosmological correlators of the perturbation field variables then the amplitude of the such correlators which we commonly identified to be the Power spectrum within the framework of cosmology will found to be finite. So this observation implies that just by converting OTOC to ETOC one cannot get correct and finite answer. This is because of the fact that, OTOC deals with completely out-of-equilibrium phenomena and ETOC deals with completely equal phenomena. It is expected that after taking very large late time limit it is possible to achieve equilibrium in any one of the time coordinates involved in the cosmological operators, but to get a complete equilibrium description in ETOC one needs to follow a completely different approach starting from the definition.

\item After doing simple computation one can explicitly show that in the present context the one-point and three point function is explicitly zero and can be written as:
\bea &&{\langle \hat{f}({\bf x},\tau) \rangle_{\beta}=\frac{1}{Z(\beta;\tau)}{\rm Tr}\left[e^{-\beta \widehat{H}(\tau)}~\hat{f}({\bf x},\tau)\right]=0}~,\\
 &&{ \langle \hat{\Pi}({\bf x},\tau) \rangle_{\beta}=\frac{1}{Z(\beta;\tau)}{\rm Tr}\left[e^{-\beta \widehat{H}(\tau)}~\hat{\Pi}({\bf x},\tau)\right]=0}~.\\  &&{\langle \hat{f}({\bf x},\tau_1) \hat{f}({\bf x},\tau_2)\hat{f}({\bf x},\tau_3)\rangle_{\beta}=\frac{1}{Z(\beta;\tau)}{\rm Tr}\left[e^{-\beta \widehat{H}(\tau)}~\hat{f}({\bf x},\tau)\hat{f}({\bf x},\tau_2)\hat{f}({\bf x},\tau_3)\right]=0}~,\\
  &&{ \langle \hat{\Pi}({\bf x},\tau_1)\hat{\Pi}({\bf x},\tau_2) \hat{\Pi}({\bf x},\tau_3)\rangle_{\beta}=\frac{1}{Z(\beta;\tau)}{\rm Tr}\left[e^{-\beta \widehat{H}(\tau)}~\hat{\Pi}({\bf x},\tau_1)\hat{\Pi}({\bf x},\tau_2) \hat{\Pi}({\bf x},\tau_3)\right]=0}~.~~~~~~~~~~~~~~
   \eea 
  Here we have used the well known {\it Kubo Martin Schwinger} condition, which basically deals with the time translational symmetry at finite temperature and using this fact one can explicitly show that the above two combinations of the three point function for the cosmological perturbation theory are explicitly vanishes in the present context. Not only these combinations, but also the other possible cross correlated three point function will become similarly zero. If we do the computation for the primordial cosmological scalar perturbations at explicitly in zero temperature then one can explicitly find non-zero but small answers for all the possible three point correlators using the concept of OTOC. Using the {\it Kubo Martin Schwinger} condition one can further show that any odd $N$-point function at finite temperature the corresponding contribution vanish due to the time translational symmetry. For these mentioned reason OTOCs are not defined in terms of any odd $N$ point function within the framework of out-of-equilibrium quantum field theory. All even $N$ point OTOC are normalised with appropriate factors which are actually appear from the expansion in the large time dissipation time scale $t\sim \beta=T^{-1}$, which is basically proportional to the equilibrium temperature of the quantum system under consideration at very late time limit. On the other hand, since one cannot decompose the any odd $N$ point thermal correlators in symmetric combinations and each correlators are trivially zero there is no normalization is possible to express these correlators. So the only physical informations are appearing from all even $N$ point correlators. For $N=2$ we get the two point OTOCs which are don't need to normalize because in this computation these are treated as the smallest building block for the computation. Any other even $N>2$ point correlators can be normalized in terms of the combinations of the $N=2$ point disconnected part of the OTOCs which are basically appearing in the large time dissipation time scale as previously we have mentioned. In our previous paper \cite{Choudhury:2020yaa}, we have provided the detailed computation of a specific type of two-point and four-point OTOC out of which one can able to extract various unknown physical information within the framework of cosmological perturbation theory described in the out of equilibrium regime of quantum field theory. In this paper we actually go beyond this understanding of the OTOC and can able to provide two new sets of two point and four point OTOCs which we believe can able to give few other informations regarding the quantum system under study within the framework of cosmological perturbation theory in the out of equilibrium regime of the quantum field theory. In the next subsections, we are going to provide the detailed computations of the ``in-in" OTO amplitudes and the related four-point OTOCs from the primordial scalar perturbations.
   
\end{enumerate}

  \subsection{Trace of four-point ``in-in" non-chaotic auto-correlated OTO amplitude for Primordial Cosmology}
Now, we will explicitly compute the numerator of the first type of four-point OTOC constructed out of cosmological perturbation field variable for different quantum vacuum states, which is given by:
  \bea && {\rm Tr}\left[e^{-\beta \widehat{H}(\tau_1)}\left[\hat{f}({\bf x},\tau_1),\hat{f}({\bf x},\tau_2)\right]^2\right]_{(\alpha,\gamma)}\nonumber\\
 &&= \frac{\exp(-2\sin\gamma\tan\alpha)}{|\cosh\alpha|}\int d\Psi_{\bf BD}~\prod^{4}_{j=1}\int \frac{d^3k_j}{(2\pi)^3}\exp\left[i{\bf k}_j.{\bf x}\right]~~~~~~~~~~~\nonumber\\
 &&~~~~~~~~~~~~~~~~~~~~~~~~~~~~~~~~~~~~~~~~~~~~~~~~~~\langle\Psi_{\bf BD}|\left[\sum^{4}_{i=1}\widehat{\cal V}^{(1)}_i({\bf k}_1,{\bf k}_2,{\bf k}_3,{\bf k}_4;\tau_1,\tau_2;\beta)\right]|\Psi_{\bf BD}\rangle.~~~~~~~~~~~  \eea  
On the other hand, we will explicitly compute the numerator of the second type of four-point OTOC constructed out of cosmological perturbation field momenta variable for different quantum vacuum states, which is given by:
  \bea && {\rm Tr}\left[e^{-\beta \widehat{H}(\tau_1)}\left[\hat{\Pi}({\bf x},\tau_1),\hat{\Pi}({\bf x},\tau_2)\right]^2\right]_{(\alpha,\gamma)}\nonumber\\
 &&= \frac{\exp(-2\sin\gamma\tan\alpha)}{|\cosh\alpha|}\int d\Psi_{\bf BD}~\prod^{4}_{j=1}\int \frac{d^3k_j}{(2\pi)^3}\exp\left[i{\bf k}_j.{\bf x}\right]~~~~~~~~~~~\nonumber\\
 &&~~~~~~~~~~~~~~~~~~~~~~~~~~~~~~~~~~~~~~~~~~~~~~~~~~\langle\Psi_{\bf BD}|\left[\sum^{4}_{i=1}\widehat{\cal V}^{(2)}_i({\bf k}_1,{\bf k}_2,{\bf k}_3,{\bf k}_4;\tau_1,\tau_2;\beta)\right]|\Psi_{\bf BD}\rangle.~~~~~~~~~~~  \eea  
  Further, our aim is to compute the individual contributions which in the normal ordered form are given by the following expression and explicitly computed in the Appendix of this paper:
  \bea &&\int d\Psi_{\bf BD}~\langle\Psi_{\bf BD}|:\widehat{\cal V}^{(1)}_i({\bf k}_1,{\bf k}_2,{\bf k}_3,{\bf k}_4;\tau_1,\tau_2;\beta):|\Psi_{\bf BD}\rangle\nonumber\\
  &&~~~~~~~~~~~~~~~~~= \int d\Psi_{\bf BD}~\langle\Psi_{\bf BD}|:e^{-\beta \hat{H}(\tau_1)}~\widehat{\cal T}^{(1)}_i({\bf k}_1,{\bf k}_2,{\bf k}_3,{\bf k}_4;\tau_1,\tau_2):|\Psi_{\bf BD}\rangle,~~~~~~~~~~~~\\
  &&\int d\Psi_{\bf BD}~\langle\Psi_{\bf BD}|:\widehat{\cal V}^{(2)}_i({\bf k}_1,{\bf k}_2,{\bf k}_3,{\bf k}_4;\tau_1,\tau_2;\beta):|\Psi_{\bf BD}\rangle\nonumber\\
  &&~~~~~~~~~~~~~~~~~= \int d\Psi_{\bf BD}~\langle\Psi_{\bf BD}|:e^{-\beta \hat{H}(\tau_1)}~\widehat{\cal T}^{(2)}_i({\bf k}_1,{\bf k}_2,{\bf k}_3,{\bf k}_4;\tau_1,\tau_2):|\Psi_{\bf BD}\rangle ~~~\forall~i=1,2,3,4.~~~~~~~~~~~~\eea 
 Here it is important to note that the dispersion relation in general has to be different for different choices of the initial quantum vacuum state.  
 
  \subsection{New OTOCs from regularised four-point ``in-in"  non-chaotic auto-correlated OTO  amplitudes: rescaled field version}
  \subsubsection{Without normalization} 
The desired cosmological OTOCs without normalization can be expressed as: 
    \bea && C^{f}_1(\tau_1,\tau_2)=-\frac{1}{Z(\beta;\tau_1)}{\rm Tr}\left[e^{-\beta \widehat{H}(\tau_1)}\left[\hat{f}({\bf x},\tau_1),\hat{f}({\bf x},\tau_2)\right]^2\right]\nonumber\\
  &&=-\int \frac{d^3k_1}{(2\pi)^3}\int \frac{d^3k_2}{(2\pi)^3}\left\{ {\cal E}^{(1)}_4({\bf k}_1,{\bf k}_2,-{\bf k}_2,-{\bf k}_1;\tau_1,\tau_2)+{\cal E}^{(1)}_4({\bf k}_1,{\bf k}_2,-{\bf k}_1,-{\bf k}_2;\tau_1,\tau_2)\right.\nonumber\\ &&\left.~~~~~~~~~~~~~~~~~~~~~~~~~~~~~+{\cal E}^{(1)}_6({\bf k}_1,{\bf k}_2,-{\bf k}_2,-{\bf k}_1;\tau_1,\tau_2)+{\cal E}^{(1)}_7({\bf k}_1,{\bf k}_2,-{\bf k}_1,-{\bf k}_2;\tau_1,\tau_2)\right.\nonumber\\ &&\left.~~~~~~~~~~~~~~~~~~~~~~~~~~~~~+{\cal E}^{(1)}_{10}({\bf k}_1,{\bf k}_2,-{\bf k}_1,-{\bf k}_2;\tau_1,\tau_2)+{\cal E}^{(1)}_{11}({\bf k}_1,{\bf k}_2,-{\bf k}_2,-{\bf k}_1;\tau_1,\tau_2)\right.\nonumber\\ &&\left.~~~~~~~~~~~~~~~~~~~~~~~~~~~~~+{\cal E}^{(1)}_{13}({\bf k}_1,{\bf k}_2,-{\bf k}_1,-{\bf k}_2;\tau_1,\tau_2)+{\cal E}^{(1)}_{13}({\bf k}_1,{\bf k}_2,-{\bf k}_2,-{\bf k}_1;\tau_1,\tau_2)\right.\nonumber\\
  &&\left.~~ +{\cal E}^{(1)}_7({\bf k}_1,-{\bf k}_1,{\bf k}_2,-{\bf k}_2;\tau_1,\tau_2)+{\cal E}^{(1)}_{10}({\bf k}_1,-{\bf k}_1,{\bf k}_2,-{\bf k}_2;\tau_1,\tau_2)+{\cal E}^{(1)}_{11}({\bf k}_1,-{\bf k}_1,{\bf k}_2,-{\bf k}_2;\tau_1,\tau_2)\right\},~~~~~~~~~ \eea
  \bea
   && C^{f}_2(\tau_1,\tau_2)=-\frac{1}{Z(\beta;\tau_1)}{\rm Tr}\left[e^{-\beta \widehat{H}(\tau_1)}\left[\hat{\Pi}({\bf x},\tau_1),\hat{\Pi}({\bf x},\tau_2)\right]^2\right]\nonumber\\
  &&=-\int \frac{d^3k_1}{(2\pi)^3}\int \frac{d^3k_2}{(2\pi)^3}\left\{ {\cal E}^{(2)}_4({\bf k}_1,{\bf k}_2,-{\bf k}_2,-{\bf k}_1;\tau_1,\tau_2)+{\cal E}^{(2)}_4({\bf k}_1,{\bf k}_2,-{\bf k}_1,-{\bf k}_2;\tau_1,\tau_2)\right.\nonumber\\ &&\left.~~~~~~~~~~~~~~~~~~~~~~~~~~~~~+{\cal E}^{(2)}_6({\bf k}_1,{\bf k}_2,-{\bf k}_2,-{\bf k}_1;\tau_1,\tau_2)+{\cal E}^{(2)}_7({\bf k}_1,{\bf k}_2,-{\bf k}_1,-{\bf k}_2;\tau_1,\tau_2)\right.\nonumber\\ &&\left.~~~~~~~~~~~~~~~~~~~~~~~~~~~~~+{\cal E}^{(2)}_{10}({\bf k}_1,{\bf k}_2,-{\bf k}_1,-{\bf k}_2;\tau_1,\tau_2)+{\cal E}^{(2)}_{11}({\bf k}_1,{\bf k}_2,-{\bf k}_2,-{\bf k}_1;\tau_1,\tau_2)\right.\nonumber\\ &&\left.~~~~~~~~~~~~~~~~~~~~~~~~~~~~~+{\cal E}^{(2)}_{13}({\bf k}_1,{\bf k}_2,-{\bf k}_1,-{\bf k}_2;\tau_1,\tau_2)+{\cal E}^{(2)}_{13}({\bf k}_1,{\bf k}_2,-{\bf k}_2,-{\bf k}_1;\tau_1,\tau_2)\right.\nonumber\\
  &&\left.~~ +{\cal E}^{(2)}_7({\bf k}_1,-{\bf k}_1,{\bf k}_2,-{\bf k}_2;\tau_1,\tau_2)+{\cal E}^{(2)}_{10}({\bf k}_1,-{\bf k}_1,{\bf k}_2,-{\bf k}_2;\tau_1,\tau_2)+{\cal E}^{(2)}_{11}({\bf k}_1,-{\bf k}_1,{\bf k}_2,-{\bf k}_2;\tau_1,\tau_2)\right\},~~~~~~~~~ \eea 
  where we have introduced new four-point OTO  amplitude functions, ${\cal E}^{(1)}_{m},{\cal E}^{(2)}_{m}~~\forall~ m=4,6,7,10,11,13$, which are defined as:
  \bea  {\cal E}^{(l)}_{m}({\bf k}_1,{\bf k}_2,{\bf k}_3,{\bf k}_4;\tau_1,\tau_2)&=& {\cal M}^{(l)}_{m}-{\cal J}^{(l)}_{m}+{\cal N}^{(l)}_{m}-{\cal Q}^{(l)}_{m}~~\forall~~l=1,2,~~~~~~~~~~~~~\eea

  All these functions signify the amplitude of the desired two types of OTOCs which are expressed in terms of the contributions from the four-point functions. For the further simplification we have to consider symmetry properties of the above mentioned amplitudes under the exchange of the momenta appearing in the third and fourth position i.e. if we replace $-{\bf k}_2 \rightarrow -{\bf k}_1$ and $-{\bf k}_1 \rightarrow -{\bf k}_2$, which are given by the following expressions:
  \bea {\cal E}^{(l)}_{4}({\bf k}_1,{\bf k}_2,-{\bf k}_2,-{\bf k}_1;\tau_1,\tau_2)={\cal E}^{(l)}_{4}({\bf k}_1,{\bf k}_2,-{\bf k}_1,-{\bf k}_2;\tau_1,\tau_2)\forall l=1,2,\\
 {\cal E}^{(l)}_{13}({\bf k}_1,{\bf k}_2,-{\bf k}_2,-{\bf k}_1;\tau_1,\tau_2)={\cal E}^{(l)}_{13}({\bf k}_1,{\bf k}_2,-{\bf k}_1,-{\bf k}_2;\tau_1,\tau_2)\forall l=1,2. \eea
  Using these exchange symmetry properties the desired OTOCs one can further simplify the results as: 
   \bea && C^{f}_l(\tau_1,\tau_2)=-\int \frac{d^3{\bf k}_1}{(2\pi)^3}\int \frac{d^3{\bf k}_2}{(2\pi)^3}\left\{ 2\left({\cal E}^{(l)}_4({\bf k}_1,{\bf k}_2,-{\bf k}_2,-{\bf k}_1;\tau_1,\tau_2)+{\cal E}^{(l)}_{13}({\bf k}_1,{\bf k}_2,-{\bf k}_1,-{\bf k}_2;\tau_1,\tau_2)\right)\right.\nonumber\\ &&\left.~~~~~~~~~~~~~~~~~~~~~~~~~~~~~+{\cal E}^{(l)}_6({\bf k}_1,{\bf k}_2,-{\bf k}_2,-{\bf k}_1;\tau_1,\tau_2)+{\cal E}^{(l)}_7({\bf k}_1,{\bf k}_2,-{\bf k}_1,-{\bf k}_2;\tau_1,\tau_2)\right.\nonumber\\ &&\left.~~~~~~~~~~~~~~~~~~~~~~~~~~~~~+{\cal E}^{(l)}_{10}({\bf k}_1,{\bf k}_2,-{\bf k}_1,-{\bf k}_2;\tau_1,\tau_2)+{\cal E}^{(l)}_{11}({\bf k}_1,{\bf k}_2,-{\bf k}_2,-{\bf k}_1;\tau_1,\tau_2)\right.\nonumber\\ 
  &&\left.~~ +{\cal E}^{(l)}_7({\bf k}_1,-{\bf k}_1,{\bf k}_2,-{\bf k}_2;\tau_1,\tau_2)+{\cal E}^{(l)}_{10}({\bf k}_1,-{\bf k}_1,{\bf k}_2,-{\bf k}_2;\tau_1,\tau_2)+{\cal E}^{(l)}_{11}({\bf k}_1,-{\bf k}_1,{\bf k}_2,-{\bf k}_2;\tau_1,\tau_2)\right\}\nonumber\\
  &&~~~~~~~~~~~~~~~~~~~~~~~~~~~~~~~~~~~~~~~~~~~~~~~~~~~~~~~~~~~~~~~~~~~~~~~~~~~~~~~~~~~~~~~~~~~~~~~~~~~~~~\forall l=1,2.~~~ \eea 
   In this computation the volume elements of the momentum integrals as given by the following expression: 
  \begin{eqnarray}
 &&\displaystyle\prod_{i=1}^{2} \frac{d^3{\bf k}_i}{(2\pi)^6}=\frac{1}{(2\pi)^6}\displaystyle\prod_{i=1}^{2}k^2_i~dk_i~\sin\theta_i~d\theta_i~d\phi_i~,\nonumber\\
 &&~~~~~~~~~~~~~~~~~~~~~~~~~~~~~~~{\rm where}~~0<k_i<\infty,~~~0<\theta_i<\pi,~~~0<\phi_i<2\pi~~~~\forall~i=1,2.~~~~~~~~~~~~
\end{eqnarray}   
 Here we need to put cut-off $0<k_i<L~~\forall~~i=1,2$ to regulate the only magnitude of the momentum dependent radial integral and consequently we can write the following regulated expressions for the desired OTOCs, as given by:
\bea && C^{f}_l(\tau_1,\tau_2)=-\frac{1}{4\pi^4}\int^{L}_{0} k^2_1~dk_1\int^{L}_{0} k^2_2~dk_2~\nonumber\\
&&~~~~~~~~~~~~~~~~~~~~~~~~~~~~~~\left\{ 2\left({\cal E}^{(l)}_4({\bf k}_1,{\bf k}_2,-{\bf k}_2,-{\bf k}_1;\tau_1,\tau_2)+{\cal E}^{(l)}_{13}({\bf k}_1,{\bf k}_2,-{\bf k}_1,-{\bf k}_2;\tau_1,\tau_2)\right)\right.\nonumber\\ &&\left.~~~~~~~~~~~~~~~~~~~~~~~~~~~~~+{\cal E}^{(l)}_6({\bf k}_1,{\bf k}_2,-{\bf k}_2,-{\bf k}_1;\tau_1,\tau_2)+{\cal E}^{(l)}_7({\bf k}_1,{\bf k}_2,-{\bf k}_1,-{\bf k}_2;\tau_1,\tau_2)\right.\nonumber\\ &&\left.~~~~~~~~~~~~~~~~~~~~~~~~~~~~~+{\cal E}^{(l)}_{10}({\bf k}_1,{\bf k}_2,-{\bf k}_1,-{\bf k}_2;\tau_1,\tau_2)+{\cal E}^{(l)}_{11}({\bf k}_1,{\bf k}_2,-{\bf k}_2,-{\bf k}_1;\tau_1,\tau_2)\right.\nonumber\\ 
  &&\left.~~~ +{\cal E}^{(l)}_7({\bf k}_1,-{\bf k}_1,{\bf k}_2,-{\bf k}_2;\tau_1,\tau_2)+{\cal E}^{(l)}_{10}({\bf k}_1,-{\bf k}_1,{\bf k}_2,-{\bf k}_2;\tau_1,\tau_2)+{\cal E}^{(l)}_{11}({\bf k}_1,-{\bf k}_1,{\bf k}_2,-{\bf k}_2;\tau_1,\tau_2)\right\},\nonumber\\
  &&~~~~~~~~~~~~~~~~~~~~~~~~~~~~~~~~~~~~~~~~~~~~~~~~~~~~~~~~~~~~~~~~~~~~~~~~~~~~~~~~~~~~~~~~~~~~~~~~~~~~~~\forall l=1,2.~~~~~~ \eea 
 Consequently, the OTOC can be expressed in terms of the four-point time dependent amplitudes as:
  \bea {C^{f}_l(\tau_1,\tau_2)=-\frac{1}{4\pi^4}\sum^{7}_{j=1}w^{(l)}_i{\cal I}^{(l)}_{j}(\tau_1,\tau_2)\quad \forall l=1,2}.\eea
  See Appendix for details where we have computed these functions ${\cal I}^{(l)}_{j}(\tau_1,\tau_2)\quad \forall l=1,2$.  Once we determine all of them then the structure of the desired OTOCs i.e. the time dependent behaviour in the OTOC will be fixed. Here it is important to note that the weight factors for each individual contributions are given by the following expression:
  \bea w^{(l)}_1=w^{(l)}_2=2,~~~~~w^{(l)}_j=1~~\forall ~~j=3,4,\cdots, 9\quad\forall l=1,2.\eea
  Further using these above mentioned results of the integrals the un-normalised OTOCs can be expressed as:
 \bea
  {C^{f}_{1}(\tau_1,\tau_2)=\left[1+(-1)^{4\nu}+\frac{7}{2}(-1)^{2\nu}\right]\frac{(-\tau_1)^{1-2\nu}(-\tau_2)^{1-2\nu}}{2\pi^4(-1)^{4\nu-1}}\sum^{4}_{i=1}X^{(1),1}_{i}(\tau_1,\tau_2).}
\\
  {C^{f}_{2}(\tau_1,\tau_2)=\left[1+(-1)^{4\nu}+\frac{7}{2}(-1)^{2\nu}\right]\frac{(-\tau_1)^{3-2\nu}(-\tau_2)^{3-2\nu}}{2\pi^4(-1)^{4\nu-1}}\sum^{4}_{i=1}X^{(2),1}_{i}(\tau_1,\tau_2).}
\eea
Here the functions $X^{(1),l}_{i}(\tau_1,\tau_2)\forall i=1,2,3,4, ~\forall l=1,2$ are defined in the Appendix.
\subsubsection{With normalization}
Further, the normalisation overall factor of OTOC, which is given by the following expression:
\bea {{\cal N}^{f}_1(\tau_1,\tau_2)=\frac{1}{\langle \hat{f}(\tau_1)\hat{f}(\tau_1)\rangle_{\beta} \langle \hat{f}(\tau_2)\hat{f}(\tau_2)\rangle_{\beta}}=\frac{\pi^4}{{\cal F}_1(\tau_1){\cal F}_1(\tau_2)}},\\
 {{\cal N}^{f}_2(\tau_1,\tau_2)=\frac{1}{\langle \hat{\Pi}(\tau_1)\hat{\Pi}(\tau_1)\rangle_{\beta} \langle \hat{\Pi}(\tau_2)\hat{\Pi}(\tau_2)\rangle_{\beta}}=\frac{\pi^4}{{\cal F}_2(\tau_1){\cal F}_2(\tau_2)}}.\eea
where the time dependent functions ${\cal F}_1(\tau_i)$ and ${\cal F}_2(\tau_i)$ for all $i=1,2$ are defined in the Appendix explicitly.  For further details please look into the Appendix for the detailed computation of the normalisation factor of both of the OTOCs.

Finally,  the normalised OTOCs in the present context can be computed as:
\bea &&{{\cal C}^{f}_1(\tau_1,\tau_2)=\frac{C^{f}_1(\tau_1,\tau_2)}{\langle \hat{f}(\tau_1)\hat{f}(\tau_1)\rangle_{\beta} \langle \hat{f}(\tau_2)\hat{f}(\tau_2)\rangle_{\beta}}}\nonumber\\
&&~~~~~~~~~~~~{=\left[1+(-1)^{4\nu}+\frac{7}{2}(-1)^{2\nu}\right]\frac{(-\tau_1)^{1-2\nu}(-\tau_2)^{1-2\nu}}{2(-1)^{4\nu-1}{\cal F}_1(\tau_1){\cal F}_1(\tau_2)}\sum^{4}_{i=1}X^{(1),1}_{i}(\tau_1,\tau_2)},~~~~~~~\\ &&{{\cal C}^{f}_2(\tau_1,\tau_2)=\frac{C^{f}_2(\tau_1,\tau_2)}{\langle \hat{\Pi}(\tau_1)\hat{\Pi}(\tau_1)\rangle_{\beta} \langle \hat{\Pi}(\tau_2)\hat{\Pi}(\tau_2)\rangle_{\beta}}}\nonumber\\
&&~~~~~~~~~~~~{=\left[1+(-1)^{4\nu}+\frac{7}{2}(-1)^{2\nu}\right]\frac{(-\tau_1)^{3-2\nu}(-\tau_2)^{3-2\nu}}{2(-1)^{4\nu-1}{\cal F}_2(\tau_1){\cal F}_2(\tau_2)}\sum^{4}_{i=1}X^{(2),1}_{i}(\tau_1,\tau_2)},~~~~~~~\eea
which is obviously a new result in the context of primordial cosmology and we are very hopeful that this result will explore various unknown physical phenomena happened in early universe. The detailed explanation of this obtained result will be discussed in the later half of this section.

Now we will discuss about two limiting results,  which is commonly known as the large mass and small mass limit of the field perturbation.  Here we will explicitly study the analytical behaviour of the two types of the derived OTOCs in the present context of discussion.  This is only because of the fact that by seeing the derived expression for the complicated structure of the OTOCs one cannot comment on the features of them without numerically plotting the behaviour.  The derived limiting results in this subsection will going to help us to understand the underlying physical features of these two types of OTOCs without explicitly performing in numerical computations using the derived full results in this section. 

First of all we discuss about the large mass limiting situation,  which can be easily obtained by performing an explicit analytical continuation in the mass parameter $\nu$ to $-i|\nu|$.  One can argue here that for this computation $\nu$ to $i|\nu|$ is not at all allowed as this will give rise to the over/excessive particle production in the four point OTOC spectra computed in the two explicit cases that we are studying in this paper.  Now after doing the mentioned allowed analytical continuation in the mass parameter we get the following results for the OTOCs that we can obtain in the large mass limiting case:
\bea &&{\widetilde{{\cal C}^{f}_1(\tau_1,\tau_2)}=\left[1+(-1)^{-4i|\nu|}+\frac{7}{2}(-1)^{-2i|\nu|}\right]\frac{(-\tau_1)^{1+2i|\nu|}(-\tau_2)^{1+2i|\nu|}}{2(-1)^{-(4i|\nu|+1)}\widetilde{{\cal F}_1(\tau_1)}\widetilde{{\cal F}_1(\tau_2)}}\sum^{4}_{i=1}\widetilde{X^{(1),1}_{i}(\tau_1,\tau_2)}},~~~~~~~~~~\\ &&{\widetilde{{\cal C}^{f}_2(\tau_1,\tau_2)}=\left[1+(-1)^{-4i|\nu|}+\frac{7}{2}(-1)^{-2i|\nu|}\right]\frac{(-\tau_1)^{3+2i|\nu|}(-\tau_2)^{3+2i|\nu|}}{2(-1)^{-(4i|\nu|+1)}\widetilde{{\cal F}_2(\tau_1)}\widetilde{{\cal F}_2(\tau_2)}}\sum^{4}_{i=1}\widetilde{X^{(2),1}_{i}(\tau_1,\tau_2)}},~~~~~~~~~~\eea
where $\widetilde{{\cal F}_1(\tau_1)}$,  $\widetilde{{\cal F}_1(\tau_2)}$,  $\widetilde{{\cal F}_2(\tau_1)}$ and $\widetilde{{\cal F}_2(\tau_2)}$ are the normalization factors for the two OTOCs for the large mass limiting situation which is obtained by taking analytical continuation in the mass parameter $\nu$ to $-i|\nu|$.  Also,  the functions $\widetilde{X^{(1),1}_{i}(\tau_1,\tau_2)}$ and $\widetilde{X^{(2),1}_{i}(\tau_1,\tau_2)}$ can be similarly obtained by doing the same analytic continuation in the large mass limiting situation.  The explicit expressions are very cumbersome to write,  so that we have only given the expressions for these functions for any general mass parameter $\nu$.  Using  these expressions by taking the above results one can explicitly compute the behaviour of these functions in the large mass limiting situation.

Now further one can simplify the above mentioned two results of the OTOCs by using the following expansion:
\bea && \left(-\tau_k\right)^{\delta_{j}+i|\nu|}\approx  \left(-\tau_j\right)^{i|\nu|}\left[1+\delta_{j}\ln(-\tau_j)+\cdots\right],~~~\forall~~k=1,2~\&~j=1,2~(k\neq j)\nonumber\\
&&~~~~~~~~~~~~~~~~~~~~~~~~~~~~~~~~~~~~~~~~~~~~~~~~~`~~~~~~~~{\rm with}~~~\delta_{1}=\frac{1}{2},~~\delta_2=\frac{3}{2},~~~~~~~~\eea
where we have considered the fact that, $|\nu|\gg\delta_j ~~\forall ~~j=1,2$ and $|\nu|\rightarrow \infty$ to get the leading and the sub-leading or the next to leading order contribution, which is used for the further simplification of the obtained results for the four point OTOCs in the large $|\nu|$ limit.  Using this result we get the following simplified results for the two types of the desired OTOCs studied in this paper:
\bea &&{\widetilde{{\cal C}^{f}_1(\tau_1,\tau_2)}\approx\left[1+(-1)^{-4i|\nu|}+\frac{7}{2}(-1)^{-2i|\nu|}\right]\left(1+\frac{1}{2}\ln(-\tau_1)\right)^2\left(1+\frac{1}{2}\ln(-\tau_2)\right)^2}~~~~~~~~~~\nonumber\\
&&~~~~~~~~~~~~~~~~~~~~~~~~~~~~~~~~~~~~~~~~~~~~~~~~~~~~{\times\frac{(-\tau_1)^{2i|\nu|}(-\tau_2)^{2i|\nu|}}{2(-1)^{-(4i|\nu|+1)}\widetilde{{\cal F}_1(\tau_1)}\widetilde{{\cal F}_1(\tau_2)}}\sum^{4}_{i=1}\widetilde{X^{(1),1}_{i}(\tau_1,\tau_2)}},~~~~~~~~~~\\ &&{\widetilde{{\cal C}^{f}_2(\tau_1,\tau_2)}\approx\left[1+(-1)^{-4i|\nu|}+\frac{7}{2}(-1)^{-2i|\nu|}\right]\left(1+\frac{3}{2}\ln(-\tau_1)\right)^2\left(1+\frac{3}{2}\ln(-\tau_2)\right)^2}~~~~~~~~~~\nonumber\\
&&~~~~~~~~~~~~~~~~~~~~~~~~~~~~~~~~~~~~~~~~~~~~~~~~~~~~{\times\frac{(-\tau_1)^{2i|\nu|}(-\tau_2)^{2i|\nu|}}{2(-1)^{-(4i|\nu|+1)}\widetilde{{\cal F}_2(\tau_1)}\widetilde{{\cal F}_2(\tau_2)}}\sum^{4}_{i=1}\widetilde{X^{(2),1}_{i}(\tau_1,\tau_2)}}.~~~~~~~~~~\eea

Next we discuss about the small mass limiting result for the two derived desired OTOCs in this paper.  Here we have use the following expansion for the case when the mass parameter is sufficiently small:
\bea &&\left(-\tau_k\right)^{\delta_{j}+i|\nu|}\approx  2\left(-\tau_k\right)^{\delta_{j}}\left[1+i|\nu|\ln(-\tau_j)+\cdots\right],~~~\forall~~j=1,2~\&~k=1,2~~(j\neq k)\nonumber\\
&&~~~~~~~~~~~~~~~~~~~~~~~~~~~~~~~~~~~~~~~~~~~~~~~~~~~~~~~~~~~~~~~~~~`{\rm with}~~~\delta_{1}=\frac{1}{2},~~\delta_2=\frac{3}{2},~~~~~~~~\eea
where we have considered the fact that, $\delta_j\gg |\nu|~~\forall ~~j=1,2$ and $|\nu|\rightarrow 0$ to get the leading and the sub-leading or the next to leading order contribution, which is used for the further simplification of the obtained result for the two point OTOC in the small $|\nu|$ limit.  In this limiting case these two desired OTOCs can be further simplified as:
\bea &&{\widehat{{\cal C}^{f}_1(\tau_1,\tau_2)}\approx\left[1+(-1)^{-4i|\nu|}+\frac{7}{2}(-1)^{-2i|\nu|}\right]\left(1+i|\nu|\ln(-\tau_1)\right)^2\left(1+i|\nu|\ln(-\tau_2)\right)^2}~~~~~~~~~~\nonumber\\
&&~~~~~~~~~~~~~~~~~~~~~~~~~~~~~~~~~~~~~~~~~~~~~~~~~~~~{\times\frac{8(-\tau_1)(-\tau_2)}{(-1)^{-(4i|\nu|+1)}\widehat{{\cal F}_1(\tau_1)}\widehat{{\cal F}_1(\tau_2)}}\sum^{4}_{i=1}\widehat{X^{(1),1}_{i}(\tau_1,\tau_2)}},~~~~~~~~~~\eea\bea &&{\widehat{{\cal C}^{f}_2(\tau_1,\tau_2)}\approx\left[1+(-1)^{-4i|\nu|}+\frac{7}{2}(-1)^{-2i|\nu|}\right]\left(1+i|\nu|\ln(-\tau_1)\right)^2\left(1+i|\nu|\ln(-\tau_2)\right)^2}~~~~~~~~~~\nonumber\\
&&~~~~~~~~~~~~~~~~~~~~~~~~~~~~~~~~~~~~~~~~~~~~~~~~~~~~{\times\frac{8(-\tau_1)^{3}(-\tau_2)^{3}}{(-1)^{-(4i|\nu|+1)}\widehat{{\cal F}_2(\tau_1)}\widehat{{\cal F}_2(\tau_2)}}\sum^{4}_{i=1}\widehat{X^{(2),1}_{i}(\tau_1,\tau_2)}}.~~~~~~~~~~\eea
Here $\widehat{{\cal F}_1(\tau_1)}$,  $\widehat{{\cal F}_1(\tau_2)}$,  $\widehat{{\cal F}_2(\tau_1)}$ and $\widehat{{\cal F}_2(\tau_2)}$ are the normalization factors for the two OTOCs for the small mass limiting situation which is obtained by taking analytical continuation in the mass parameter $\nu$ to $-i|\nu|$.  Also,  the functions $\widehat{X^{(1),1}_{i}(\tau_1,\tau_2)}$ and $\widehat{X^{(2),1}_{i}(\tau_1,\tau_2)}$ can be similarly obtained by doing the same expansion in the small mass limiting situation.  The explicit expressions are very cumbersome to write,  so that we have only given the expressions for these functions for any general mass parameter $\nu$.  Using  these expressions by taking the above expansions one can explicitly compute the behaviour of these functions in the small mass limiting situation.
  \subsection{New OTOCs from regularised four-point ``in-in"  non-chaotic auto-correlated OTO  amplitudes: curvature perturbation field version}
  \subsubsection{Without normalization}
Here we need to perform the computation for the un-normalised OTOC  in terms of the scalar curvature perturbation and the canonically conjugate momentum associated with it, which we have found that is given by the following simplified expression:
\bea && {C^{\zeta}_1(\tau_1,\tau_2)=-\frac{1}{Z^{\zeta}_{\alpha,\gamma}(\beta,\tau_1)}{\rm Tr}\left[e^{-\beta \hat{H}(\tau_1)}\left[\hat{\zeta}({\bf x},\tau_1),\hat{\zeta}({\bf x},\tau_2)\right]^2\right]_{(\alpha,\gamma)}=\frac{1}{z^2(\tau_1)z^2(\tau_2)}C^{f}_1(\tau_1,\tau_2)},~~~~~~~~~~~~~\\ && {C^{\zeta}_2(\tau_1,\tau_2)=-\frac{1}{Z^{\zeta}_{\alpha\gamma}(\beta,\tau_1)}{\rm Tr}\left[e^{-\beta \hat{H}(\tau_1)}\left[\hat{\Pi}({\bf x},\tau_1),\hat{\Pi}({\bf x},\tau_2)\right]^2\right]_{(\alpha,\gamma)}=\frac{1}{z^2(\tau_1)z^2(\tau_2)}C^{f}_2(\tau_1,\tau_2)}.~~~~~~~~~~~~~\eea
\subsubsection{With normalization}
The normalised OTOC in terms of the scalar curvature perturbation and the canonically conjugate momentum associated with it, which is basically the computation of the following normalised OTOC, in the present context:
\bea {{\cal C}^{\zeta}_1(\tau_1,\tau_2)=\frac{C^{\zeta}_1(\tau_1,\tau_2)}{\langle \zeta(\tau_1)\zeta(\tau_1)\rangle_{\beta}\langle \zeta(\tau_1)\zeta(\tau_1)\rangle_{\beta}}={\cal N}^{\zeta}_1(\tau_1,\tau_2)~C^{\zeta}_1(\tau_1,\tau_2)}~, \\ {{\cal C}^{\zeta}_2(\tau_1,\tau_2)=\frac{C^{\zeta}_2(\tau_1,\tau_2)}{\langle \Pi_{\zeta}(\tau_1)\Pi_{\zeta}(\tau_1)\rangle_{\beta}\langle \Pi_{\zeta}(\tau_1)\Pi_{\zeta}(\tau_1)\rangle_{\beta}}={\cal N}^{\zeta}_2(\tau_1,\tau_2)~C^{\zeta}_2(\tau_1,\tau_2)}~, \eea
where the normalisation factor to normalise OTOC is given by:
 \bea {{\cal N}^{\zeta}_1(\tau_1,\tau_2)=\frac{1}{\langle \zeta(\tau_1)\zeta(\tau_1)\rangle_{\beta}\langle \zeta(\tau_1)\zeta(\tau_1)\rangle_{\beta}}=\frac{\pi^4 z^2(\tau_1)z^2(\tau_2)}{{\cal F}_1(\tau_1){\cal F}_1(\tau_2)}=z^2(\tau_1)z^2(\tau_2){\cal N}^{f}_1(\tau_1,\tau_2)},~~~~~~~~\\
  {{\cal N}^{\zeta}_2(\tau_1,\tau_2)=\frac{1}{\langle \Pi_{\zeta}(\tau_1)\Pi_{\zeta}(\tau_1)\rangle_{\beta}\langle \Pi_{\zeta}(\tau_1)\Pi_{\zeta}(\tau_1)\rangle_{\beta}}=\frac{\pi^4 z^2(\tau_1)z^2(\tau_2)}{{\cal F}_2(\tau_1){\cal F}_2(\tau_2)}=z^2(\tau_1)z^2(\tau_2){\cal N}^{f}_2(\tau_1,\tau_2)}.~~~~~~~~\eea
 Consequently,  the normalised OTOC computed from the curvature perturbation variable is given by the following expression:
 \bea {{\cal C}^{\zeta}_1(\tau_1,\tau_2)={\cal C}^{f}_1(\tau_1,\tau_2)=\left[1+(-1)^{4\nu}+\frac{7}{2}(-1)^{2\nu}\right]\frac{(-\tau_1)^{1-2\nu}(-\tau_2)^{1-2\nu}}{2(-1)^{4\nu-1}{\cal F}_1(\tau_1){\cal F}_1(\tau_2)}\sum^{4}_{i=1}X^{(1),1}_{i}(\tau_1,\tau_2)}.~~~~~~~~ \\
 {{\cal C}^{\zeta}_2(\tau_1,\tau_2)={\cal C}^{f}_2(\tau_1,\tau_2)=\left[1+(-1)^{4\nu}+\frac{7}{2}(-1)^{2\nu}\right]\frac{(-\tau_1)^{3-2\nu}(-\tau_2)^{3-2\nu}}{2(-1)^{4\nu-1}{\cal F}_2(\tau_1){\cal F}_2(\tau_2)}\sum^{4}_{i=1}X^{(2),1}_{i}(\tau_1,\tau_2)}.~~~~~~~~ \eea
 Here all the unknown time dependent functions $X^{(1),1}_{i}(\tau_1,\tau_2)~\forall~i=1,2,3,4$ and $X^{(2),1}_{i}(\tau_1,\tau_2)~\forall~i=1,2,3,4$ originated from the two types of OTOC correlations are explicitly computed in the Appendix.  Here one can explicitly show that by taking the large mass and small mass limiting approximation as performed in the previous subsection we get the same results in the present context for the normalized OTOCs presented in terms of the curvature perturbation field variable and its canonically conjugate momentum variable. 
  \section{Numerical analysis~I: Interpretation of two-point  non-chaotic auto-correlated OTO functions} 
\label{sec:3}
%\begin{figure*}[htb]
 % \includegraphics[width=17cm,height=8.9cm]{YBD1.pdf}
  %\caption{Behaviour of the two-point auto-correlated field OTO function with respect to the time scale $T$ for Bunch Davies vacuum for different mass parameters.}
 % \label{fig:2}
%\end{figure*}
%\begin{figure*}[htb]
%  \includegraphics[width=17cm,height=8.9cm]{YAL2.pdf}
 % \caption{Behaviour of the two-point auto-correlated field OTO function with respect to the time scale $T$ for $\alpha$ vacua for different mass parameters.}
%  \label{fig:3}
%\end{figure*}
\begin{figure*}[htb]
  \includegraphics[width=17cm,height=7cm]{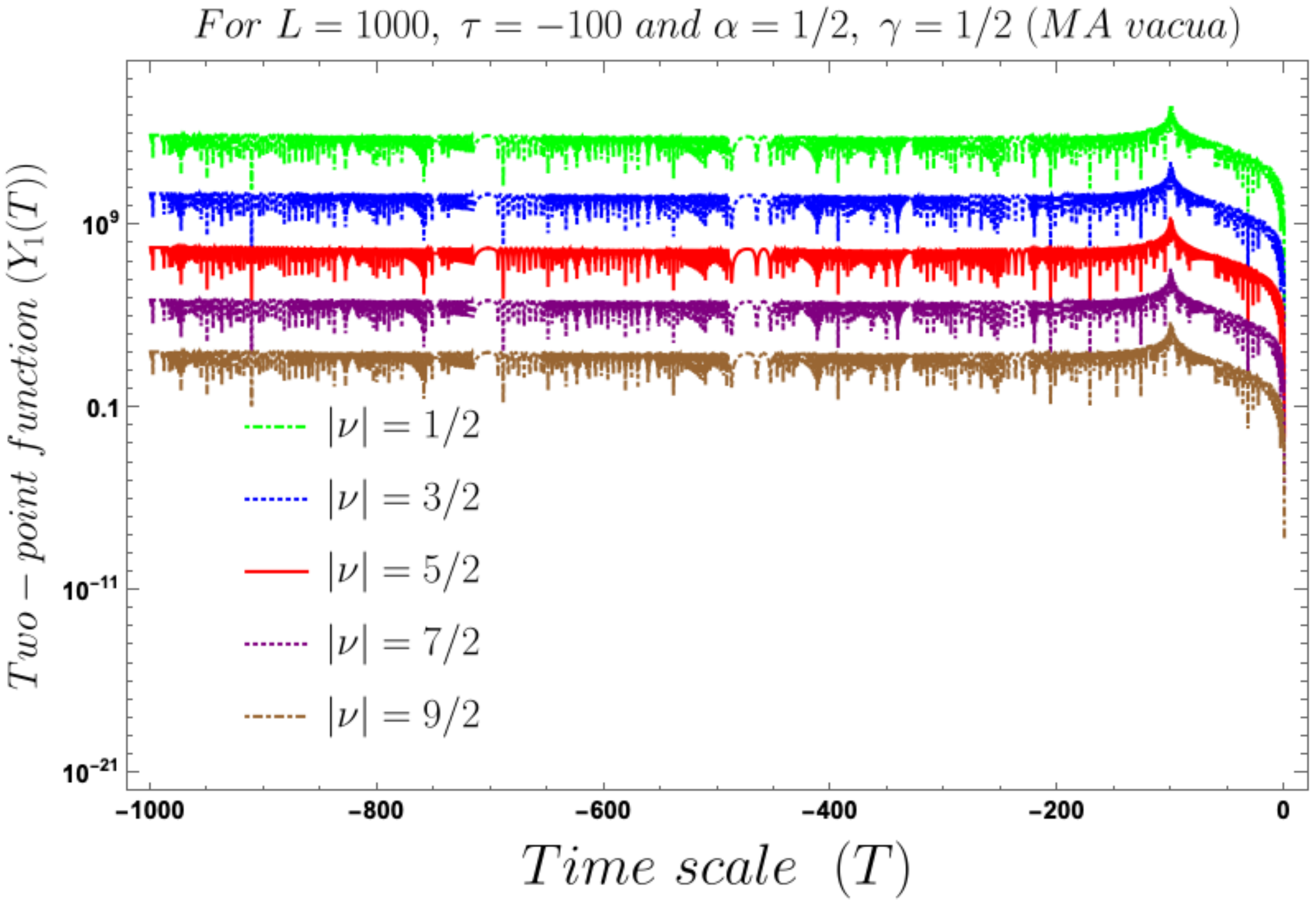}
  \caption{Behaviour of the two-point auto-correlated field OTO function with respect to the time scale $T$ for Mota Allen vacua for different mass parameters.}
  \label{fig:4}
\end{figure*}
%\begin{figure*}[htb]
%  \includegraphics[width=17cm,height=8.9cm]{YV1.pdf}
%  \caption{Behaviour of the two-point auto-correlated field OTO function with respect to the vacuum parameter $\alpha$ for $\alpha$ vacua for different mass parameters.}
 % \label{fig:5}
%\end{figure*}
\begin{figure*}[htb]
  \includegraphics[width=17cm,height=7cm]{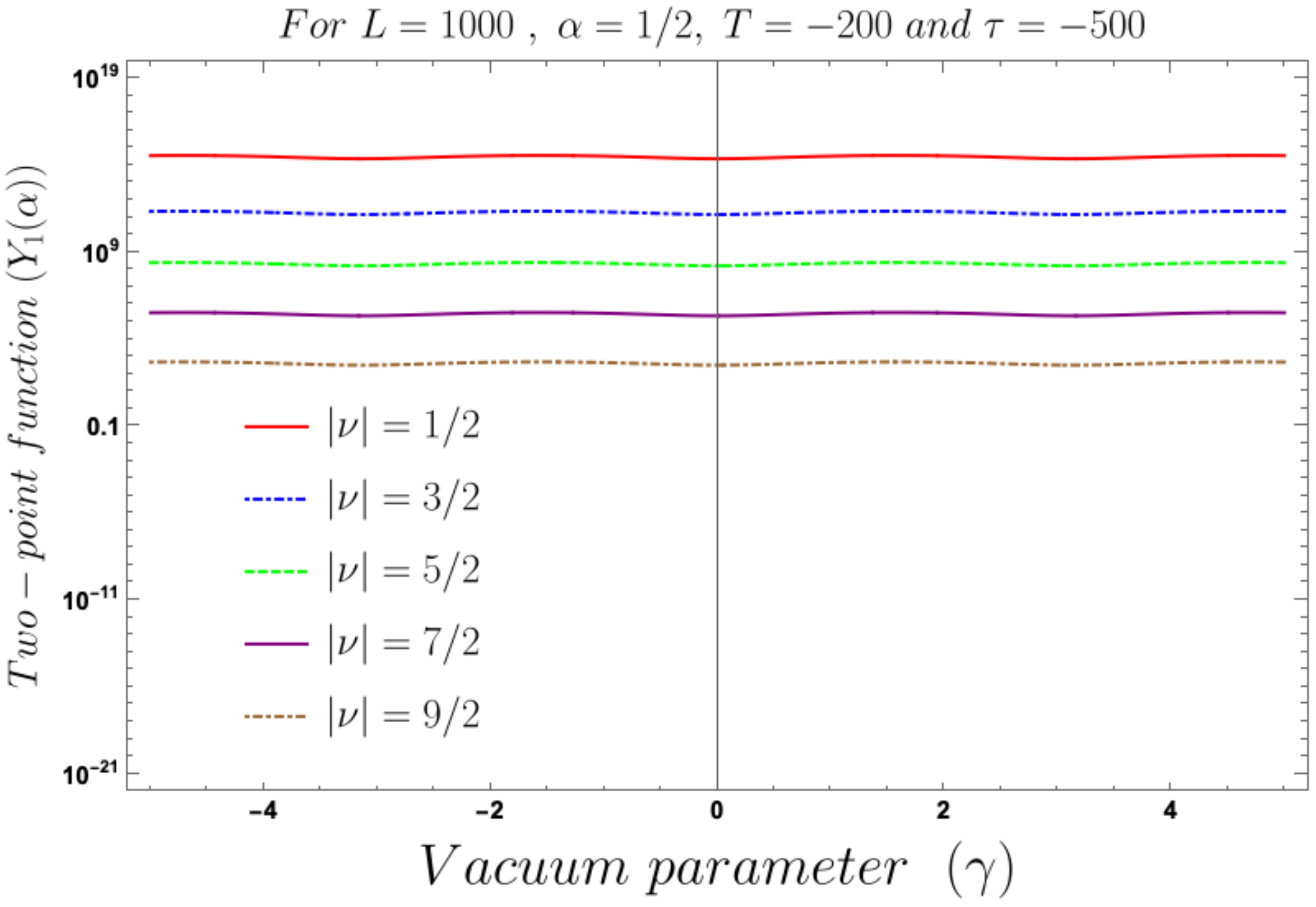}
  \caption{Behaviour of the two-point auto-correlated field OTO function with respect to the vacuum parameter $\gamma$ for Mota Allen vacua for different mass parameters.} 
  \label{fig:6}
\end{figure*}
\begin{figure*}[htb]
  \includegraphics[width=17cm,height=7cm]{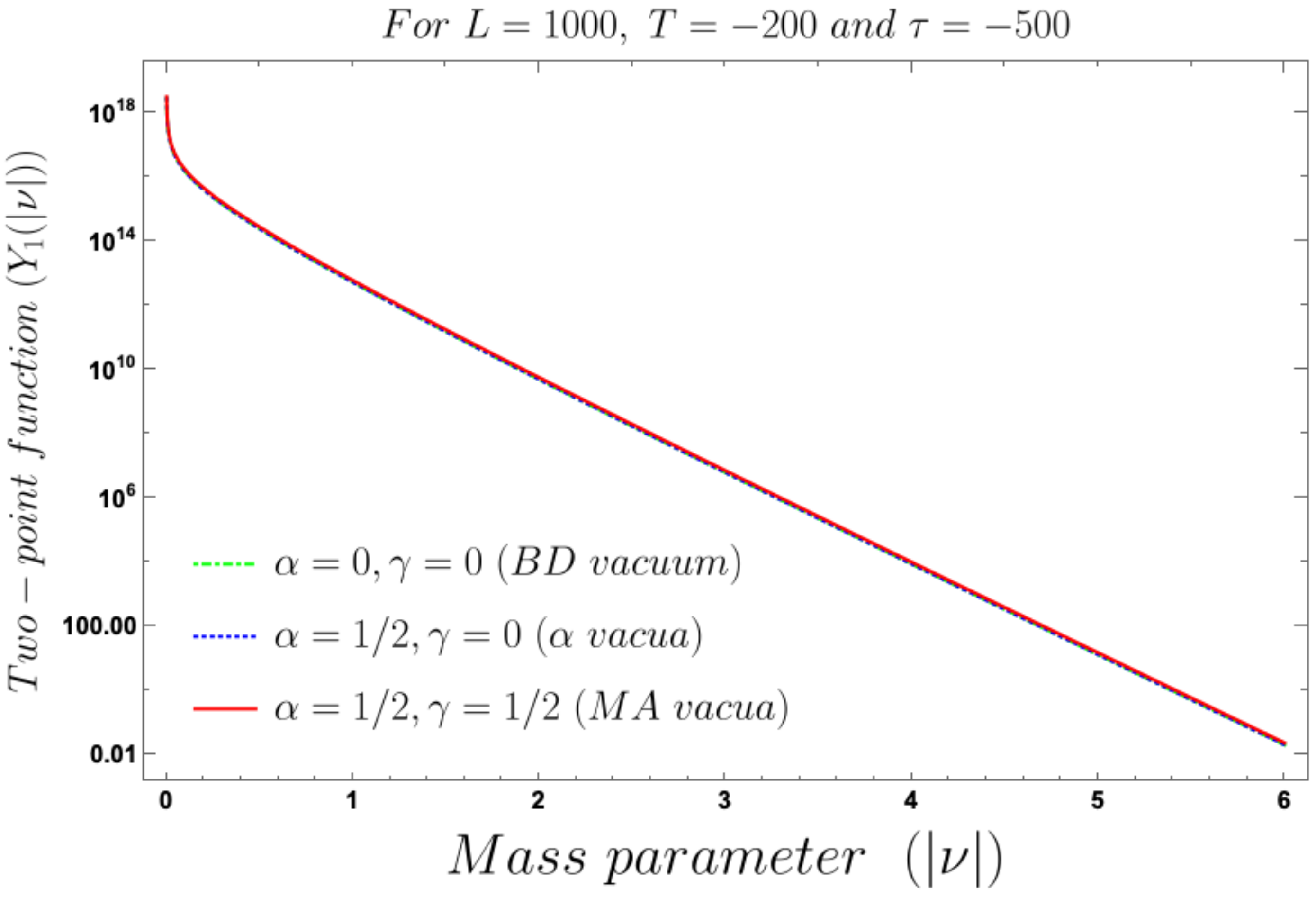}
  \caption{Behaviour of the two-point auto-correlated field OTO function with respect to the mass parameter $|\nu|$ for Mota Allen,  $\alpha$ vacua and Bunch Davies vacuum for different mass parameters.}
  \label{fig:7}
\end{figure*} 
In this section,  our prime objective is to numerically study to give interpretation of the obtained results for the two-point auto-correlated field and momentum OTO functions within the context of primordial cosmological perturbation theory studied analytically in the previous section of this paper.

\begin{figure*}[htb]
  \includegraphics[width=17cm,height=7cm]{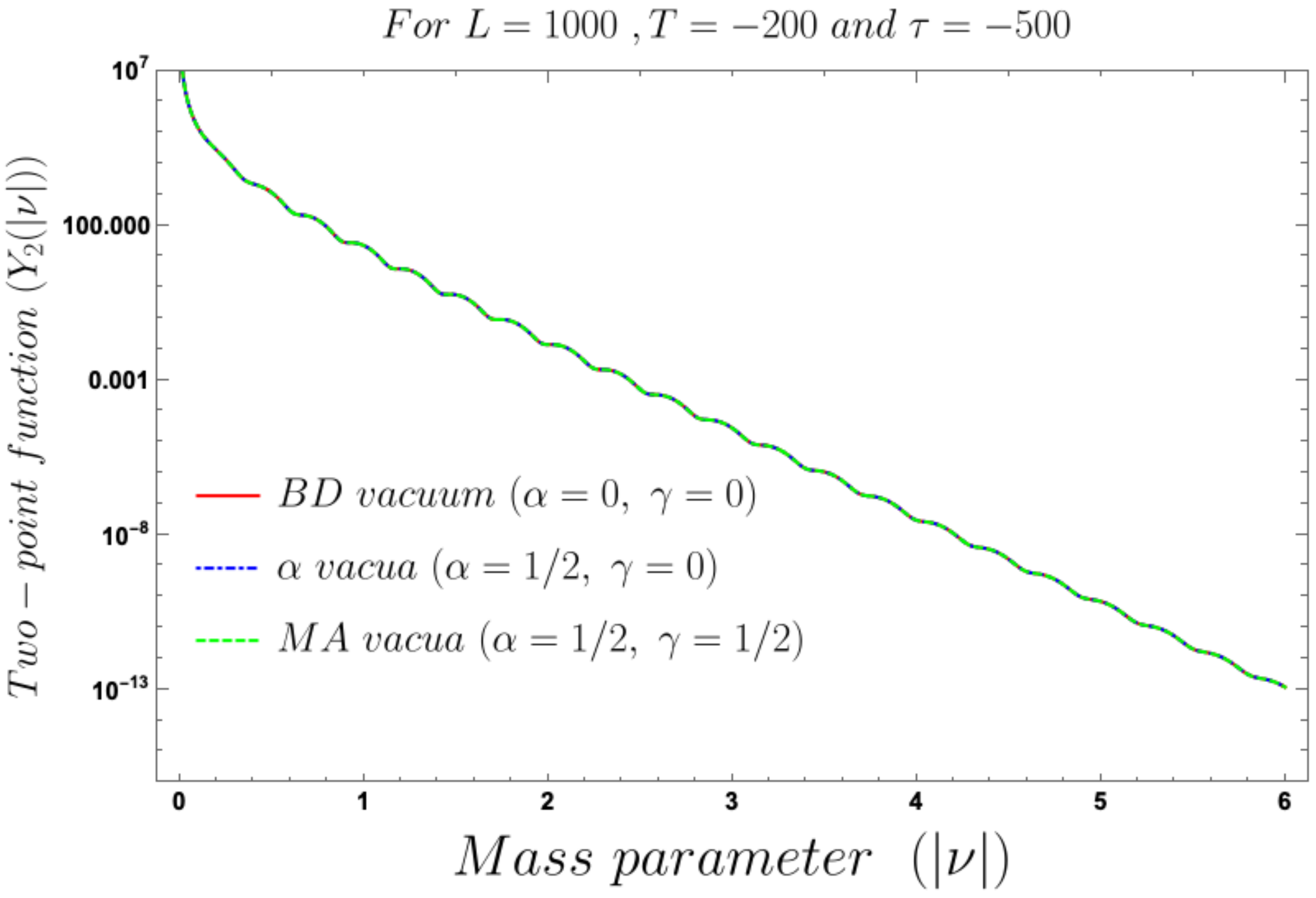}
  \caption{Behaviour of the two-point auto-correlated momentum OTO function with respect to the mass parameter $|\nu|$ for Mota Allen,  $\alpha$ vacua and Bunch Davies vacuum. }  
  \label{fig:61}
\end{figure*}

\begin{figure*}[htb]
  \includegraphics[width=17cm,height=7cm]{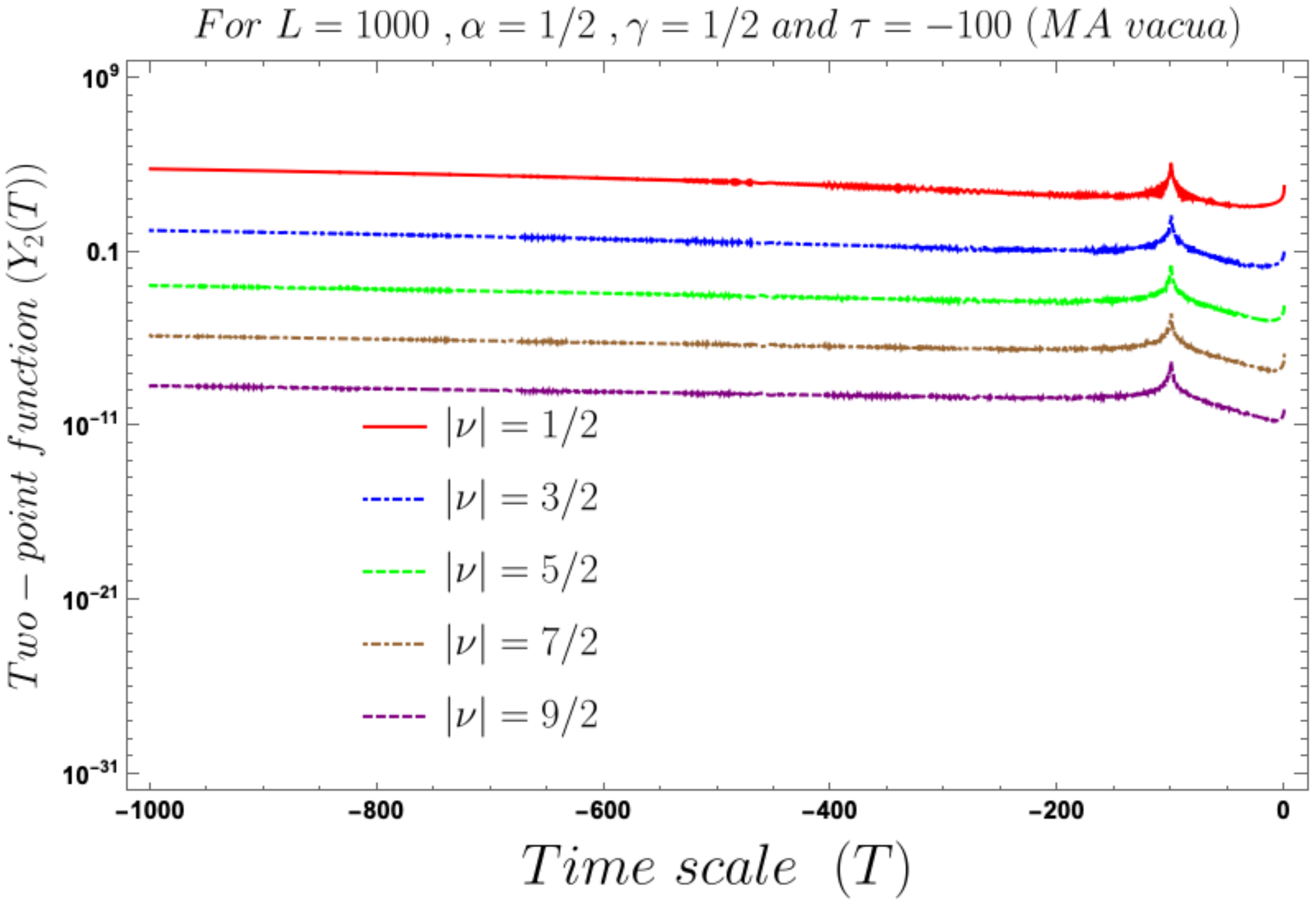}
  \caption{Behaviour of the two-point auto-correlated momentum OTO function with respect to the time scale $T$ for Mota Allen vacua. }  
  \label{fig:65}
\end{figure*}

The detailed interpretation of the two-point auto-correlated field and momentum OTO functions are appended below point-wise:
\begin{itemize}
\item  In fig.~(\ref{fig:4}) and fig.~(\ref{fig:65}),  conformal time dependent behaviour of the two-point auto-correlated field and momentum OTO functions with respect to the two time scales are explicitly shown.  From these plots it is clearly observed that with respect both the time scales the  two-point auto-correlated field and momentum OTO functions show random fluctuating behaviour.  

\item For both the cases it is observed that,  we get the significant distinguishable features in the time dependent two-point auto-correlated field and momentum OTO functions for partially massless or heavy scalar fields.  This behaviour of the two-point auto-correlated field and momentum OTO functions is explicitly depicted in fig.~(\ref{fig:7}) and fig.~(\ref{fig:61}).  From these plots one can clearly see that for the results obtained using Bunch Davies vacuum,  $\alpha$ vacua and Mota Allen vacua state initial conditions the two-point auto-correlated field and momentum OTO functions significantly decay very fast in a non-standard fashion with respect to the magnitude of the mass parameter $|\nu|$ for partially massless and heavy scalar fields in primordial cosmology.  On the other hand,  in  fig.~(\ref{fig:6}) if we change the quantum initial conditions by changing the vacuum state by introducing  the non-standard Mota Allen vacua then a clear distinctive features can be observed by changing the vacuum parameter $\alpha$ and $\gamma$ respectively.  We have explicitly have shown the behaviour of the two-point auto-correlated field and momentum OTO functions for $\alpha=1/2,\gamma=1/2$,  where it is clearly observed that the correlation decays very rapidly with the increase in the magnitude of the mass parameter $|\nu|$.  
 
\item Also we have observed from the plots that the random fluctuations with respect to the conformal time scale show decaying time dependent behaviour at very late time scale.  First of all all of them showing a saturating behaviour and then it increases at a particular value in the time scale and then these plots are showing decaying features.  This large peak value of the two-point auto correlated spectra is obtained at the scale when the two time scales of the theory becomes comparable.  Physically this is a very crucial fact as it show at this particular time scale we are actually getting zero information. from the two-point auto-correlations.  Only the preferred information can be extracted from very early time scale as well as at very late time scale after the peak.

\end{itemize}
  \section{Numerical analysis~II: Interpretation of four-point  non-chaotic auto-correlated OTO functions}
  \label{sec:4}
\begin{figure*}[htb]
  \includegraphics[width=17cm,height=7cm]{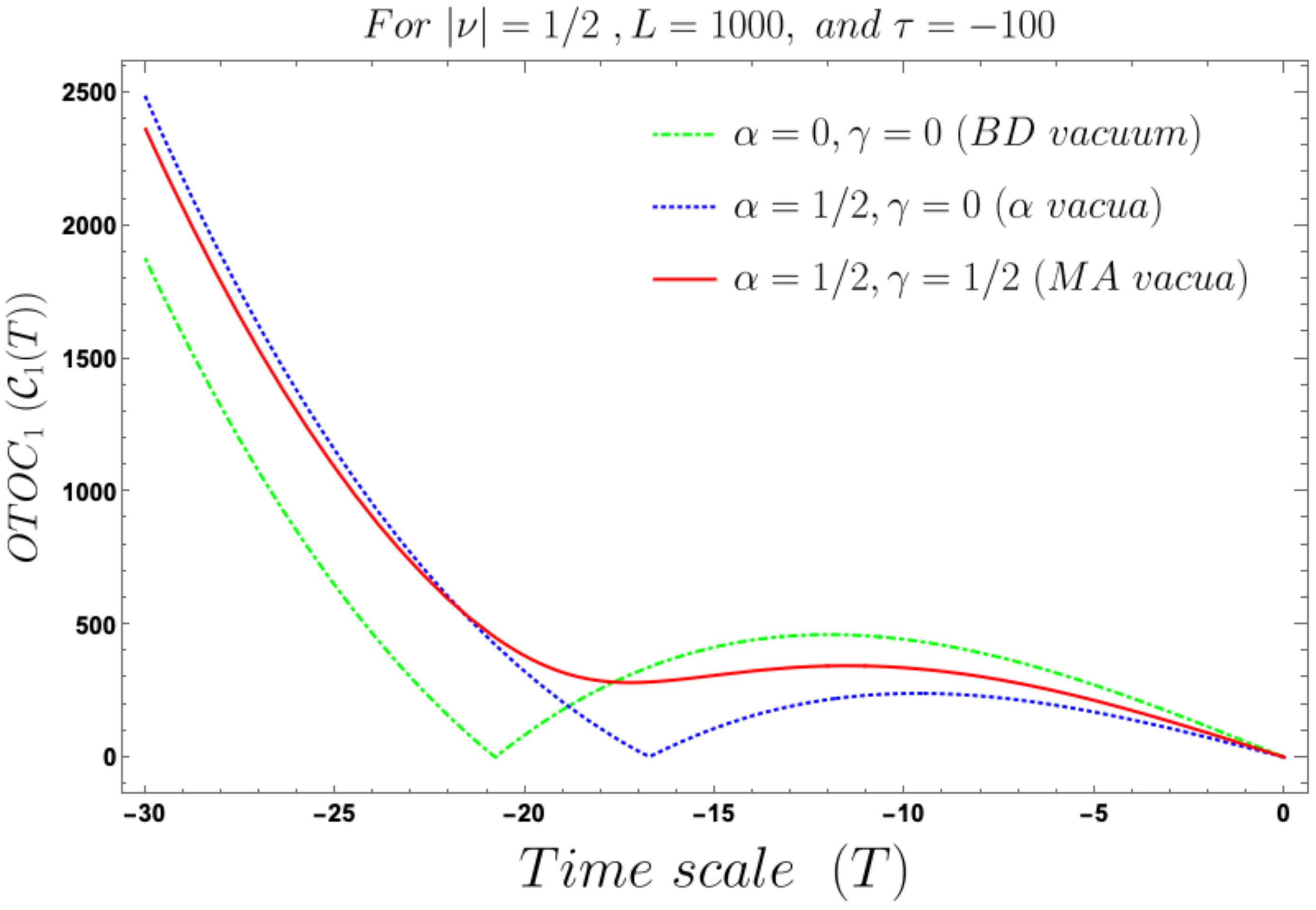}
  \caption{Behaviour of the four-point auto-correlated field OTO function with respect to the time scale $T$ for ~Mota Allen and $\alpha$ vacua and for Bunch Davies vacuum for the mass parameter $|\nu|=1/2$.}
  \label{fig:9}
\end{figure*}
\begin{figure*}[htb]
  \includegraphics[width=17cm,height=7cm]{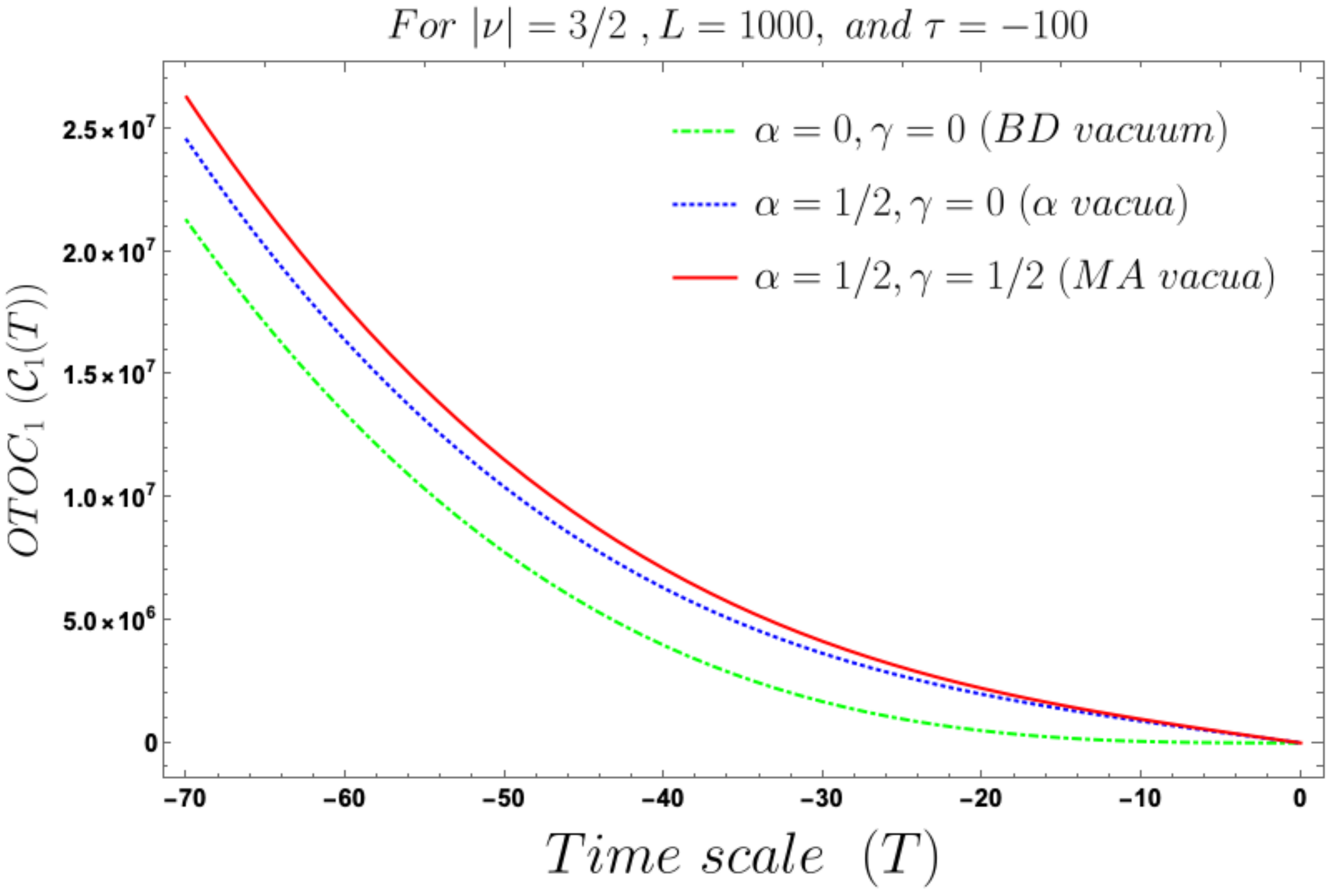}
  \caption{Behaviour of the four-point auto-correlated field OTO function with respect to the time scale $T$ for ~Mota Allen and $\alpha$ vacua and for Bunch Davies vacuum for the mass parameter $|\nu|=3/2$.}
  \label{fig:10}
\end{figure*}
%\begin{figure*}[htb]
 % \includegraphics[width=17cm,height=8.7cm]{CT3.pdf}
 % \caption{Behaviour of the four-point auto-correlated field OTO function with respect to the time scale $T$ for ~Motta Allen and $\alpha$ vacua and for Bunch Davies vacuum for the mass parameter $|\nu|=5/2$.}
%  \label{fig:11}
%\end{figure*}
\begin{figure*}[htb]
  \includegraphics[width=17cm,height=7cm]{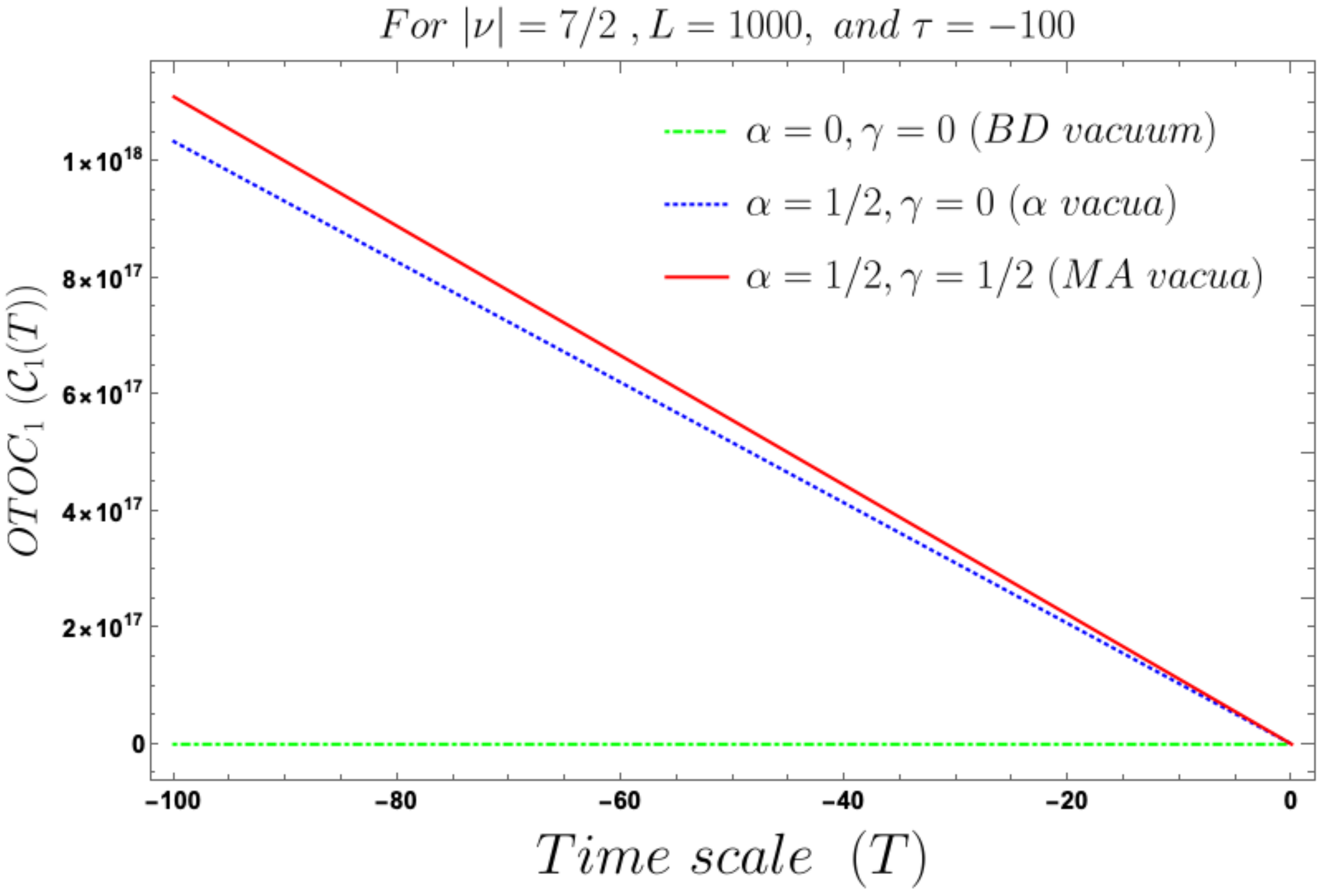}
  \caption{Behaviour of the four-point auto-correlated field OTO function with respect to the time scale $T$ for ~Mota Allen and $\alpha$ vacua and for Bunch Davies vacuum for the mass parameter $|\nu|=7/2$.}
  \label{fig:12}
\end{figure*}
%\begin{figure*}[htb]
 % \includegraphics[width=17cm,height=8.7cm]{CT5.pdf}
%  \caption{Behaviour of the four-point auto-correlated field OTO function with respect to the time scale $T$ for ~Motta Allen and $\alpha$ vacua and for Bunch Davies vacuum for the mass parameter $|\nu|=9/2$.}
 % \label{fig:13}
%\end{figure*}
 \begin{figure*}[htb]
  \includegraphics[width=17cm,height=7cm]{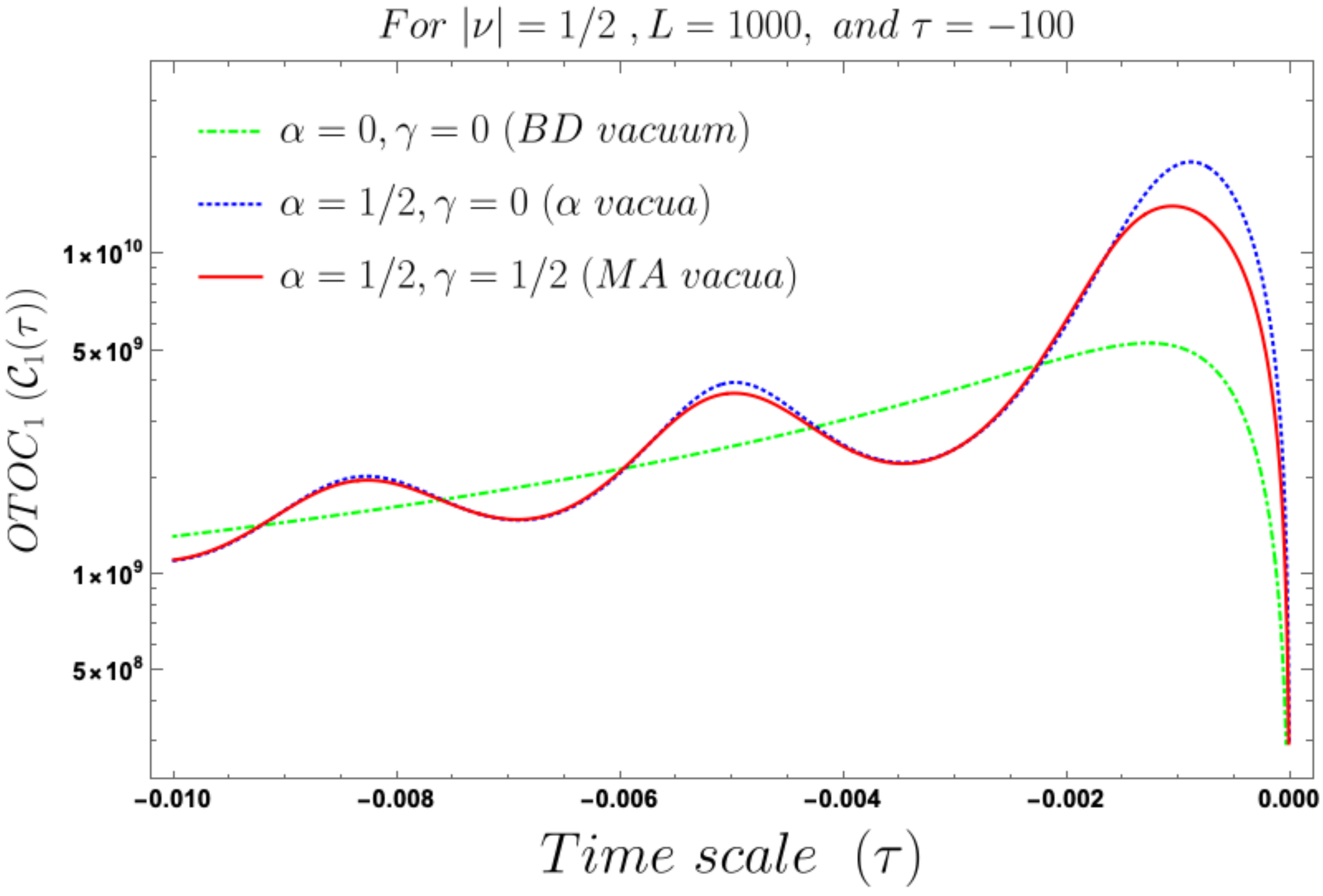}
  \caption{Behaviour of the four-point auto-correlated field OTO function with respect to the time scale $\tau$ for ~Mota Allen and $\alpha$ vacua and for Bunch Davies vacuum for the mass parameter $|\nu|=1/2$.}
  \label{fig:14}
\end{figure*}
%\begin{figure*}[htb]
%  \includegraphics[width=17cm,height=8.7cm]{CTa2.pdf}
 % \caption{Behaviour of the four-point auto-correlated field OTO function with respect to the time scale $\tau$ for ~Motta Allen and $\alpha$ vacua and for Bunch Davies vacuum for the mass parameter $|\nu|=3/2$.}
 % \label{fig:15}
%\end{figure*}
\begin{figure*}[htb]
  \includegraphics[width=17cm,height=7cm]{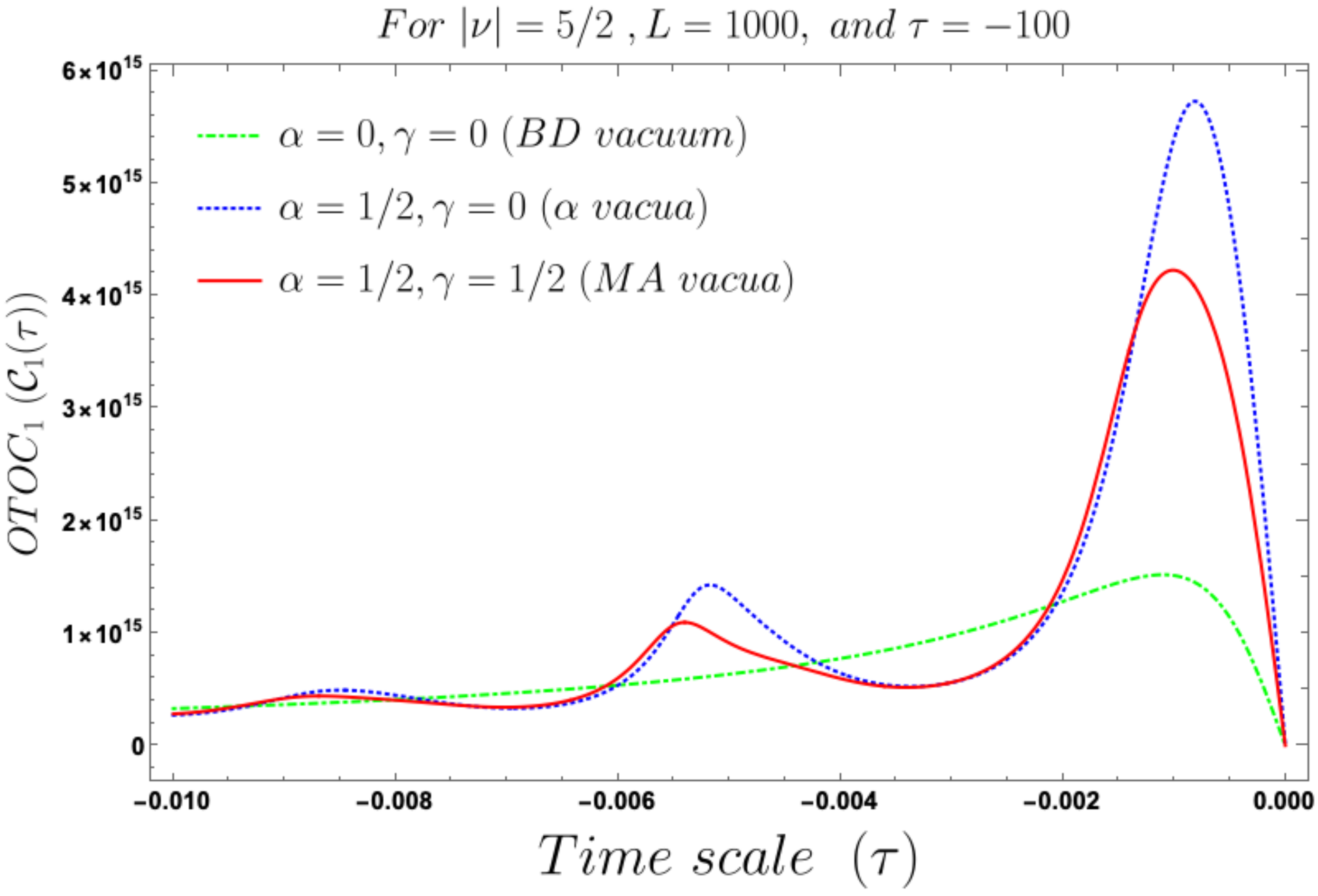}
  \caption{Behaviour of the four-point auto-correlated field OTO function with respect to the time scale $\tau$ for ~Mota Allen and $\alpha$ vacua and for Bunch Davies vacuum for the mass parameter $|\nu|=5/2$.}
  \label{fig:16}
\end{figure*}
\begin{figure*}[htb]
  \includegraphics[width=17cm,height=7cm]{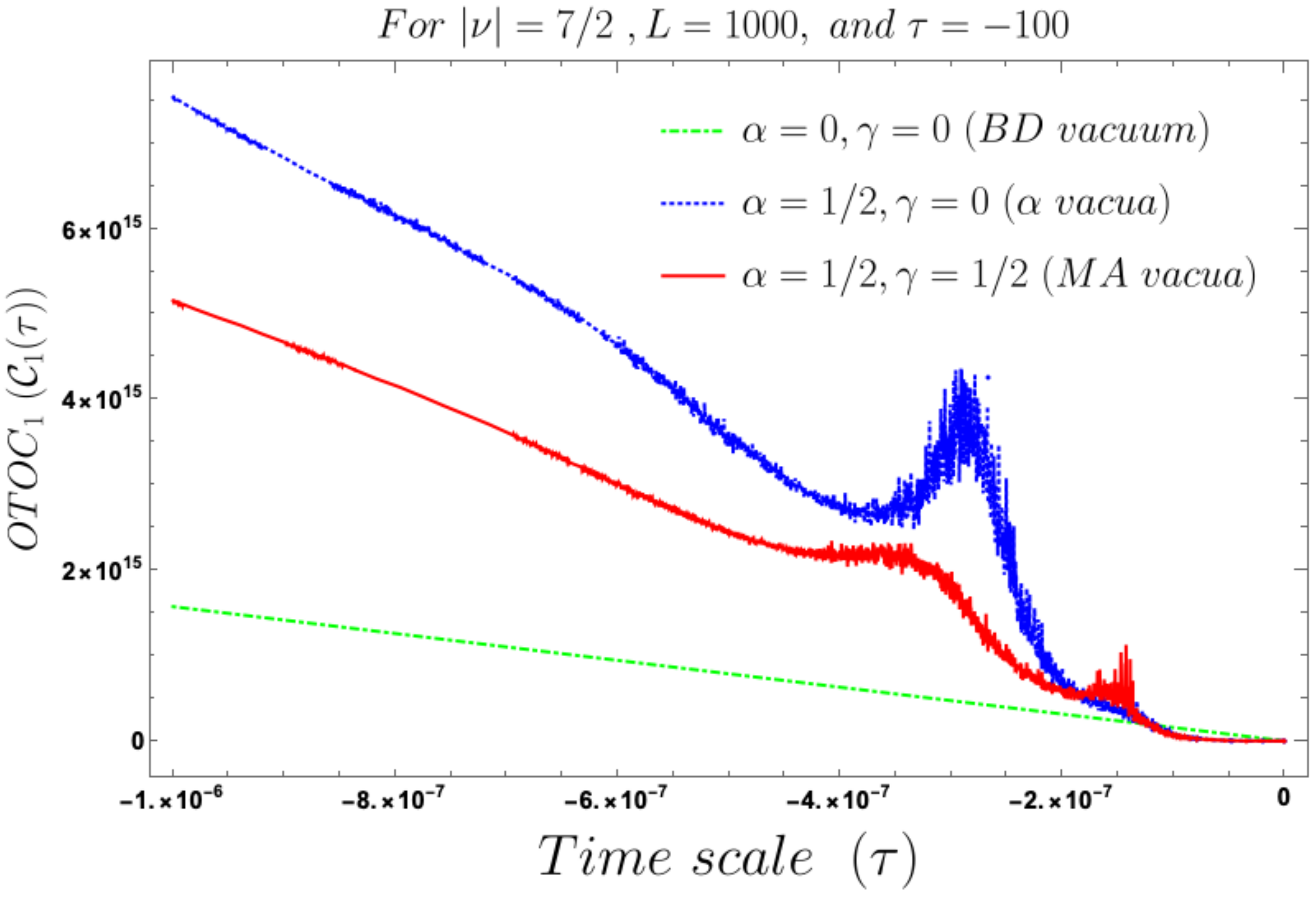}
  \caption{Behaviour of the four-point auto-correlated field OTO function with respect to the time scale $\tau$ for ~Mota Allen and $\alpha$ vacua and for Bunch Davies vacuum for the mass parameter $|\nu|=7/2$.}
  \label{fig:17}
\end{figure*}
\begin{figure*}[htb]
  \includegraphics[width=17cm,height=7cm]{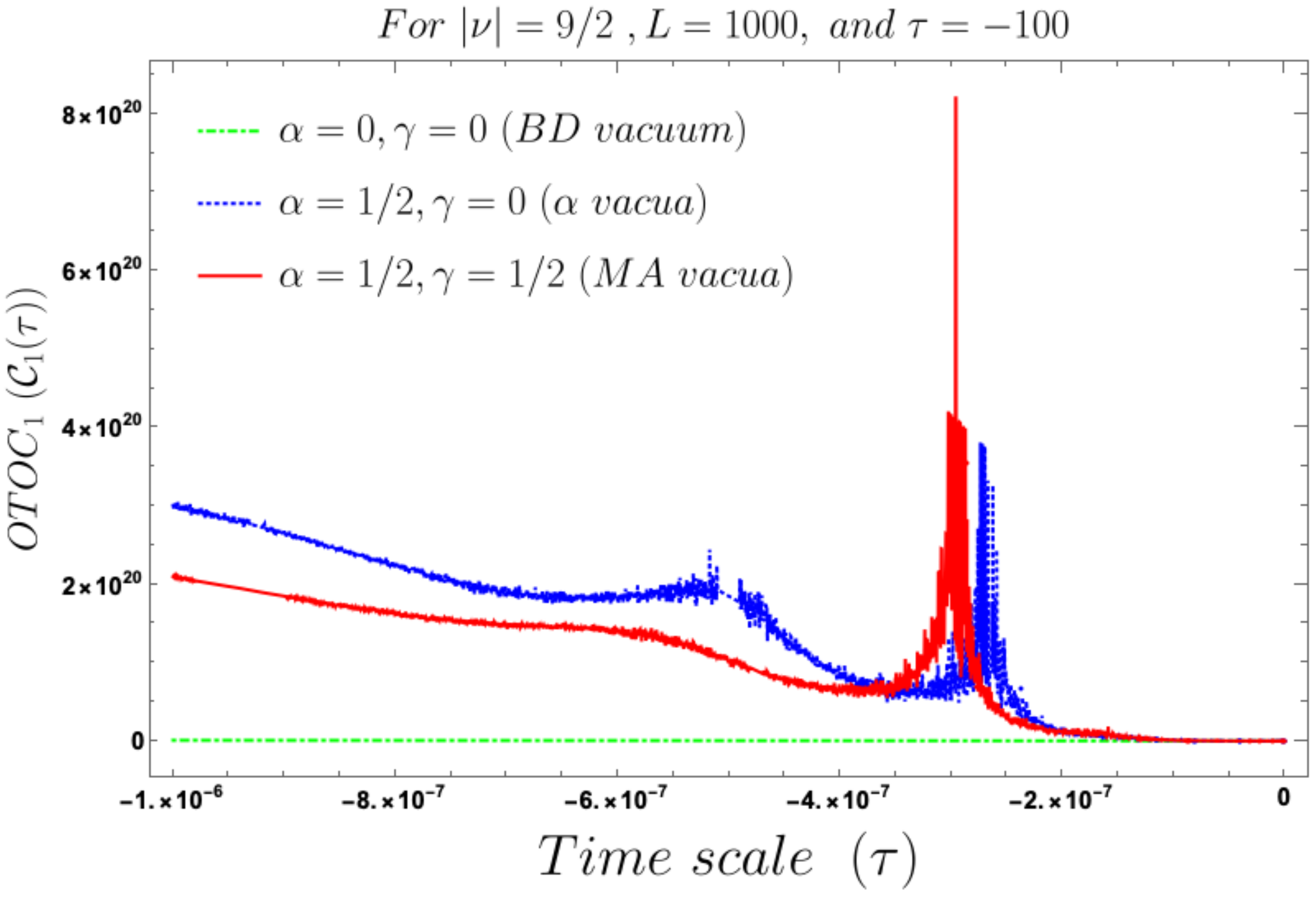}
  \caption{Behaviour of the four-point auto-correlated field OTO function with respect to the time scale $\tau$ for ~Mota Allen and $\alpha$ vacua and for Bunch Davies vacuum for the mass parameter $|\nu|=9/2$.}
  \label{fig:18}
\end{figure*}
 \begin{figure*}[htb]
  \includegraphics[width=17cm,height=7cm]{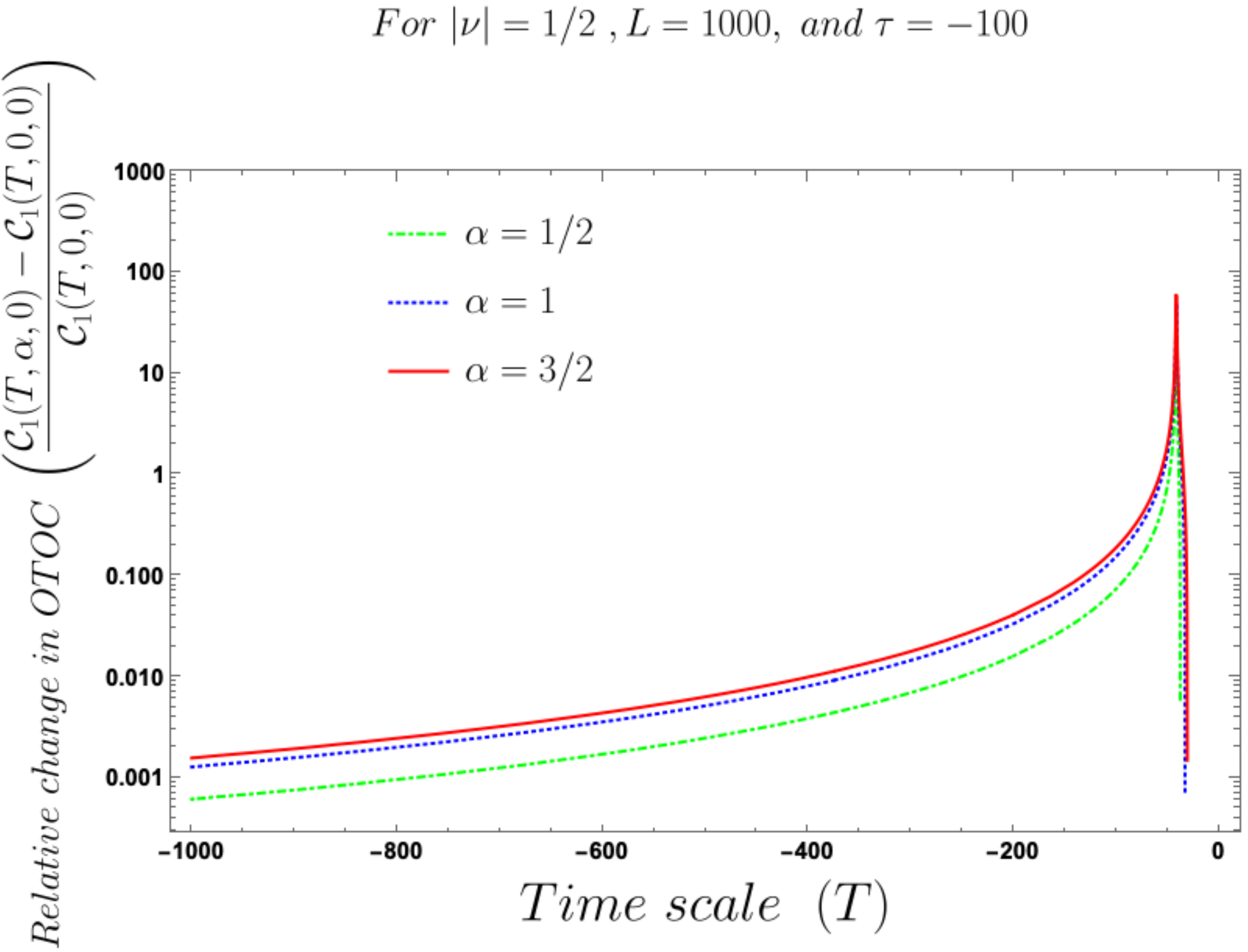}
  \caption{Behaviour of the relative change in four-point auto-correlated field OTO function with respect to the time scale $T$ for $\alpha$ vacua for the mass parameter $|\nu|=1/2$.}
  \label{fig:19}
\end{figure*}
\begin{figure*}[htb]
  \includegraphics[width=17cm,height=7cm]{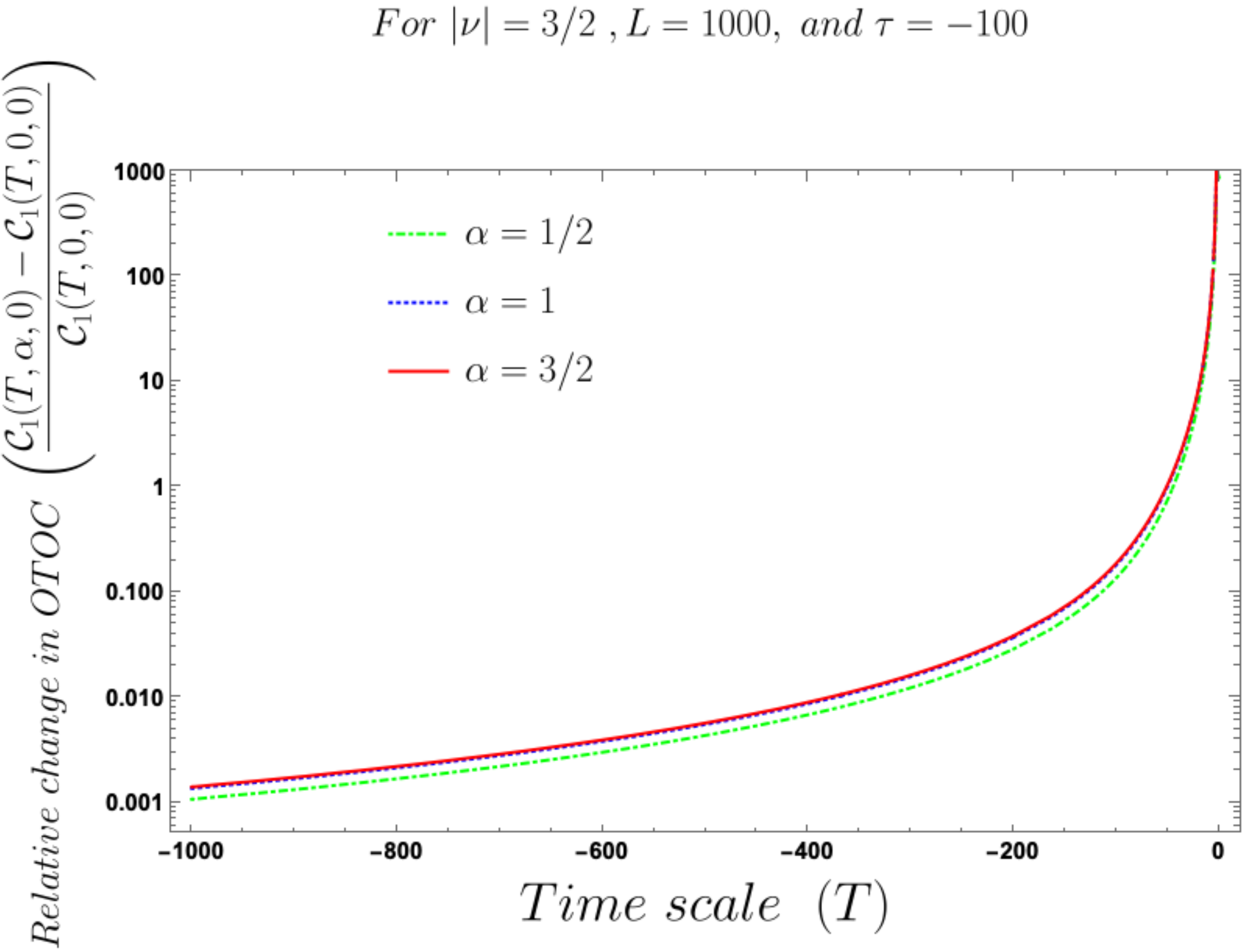}
  \caption{Behaviour of the relative change in four-point auto-correlated field OTO function with respect to the time scale $T$ for $\alpha$ vacua for the mass parameter $|\nu|=3/2$.}
  \label{fig:20}
\end{figure*}
\begin{figure*}[htb]
  \includegraphics[width=17cm,height=7cm]{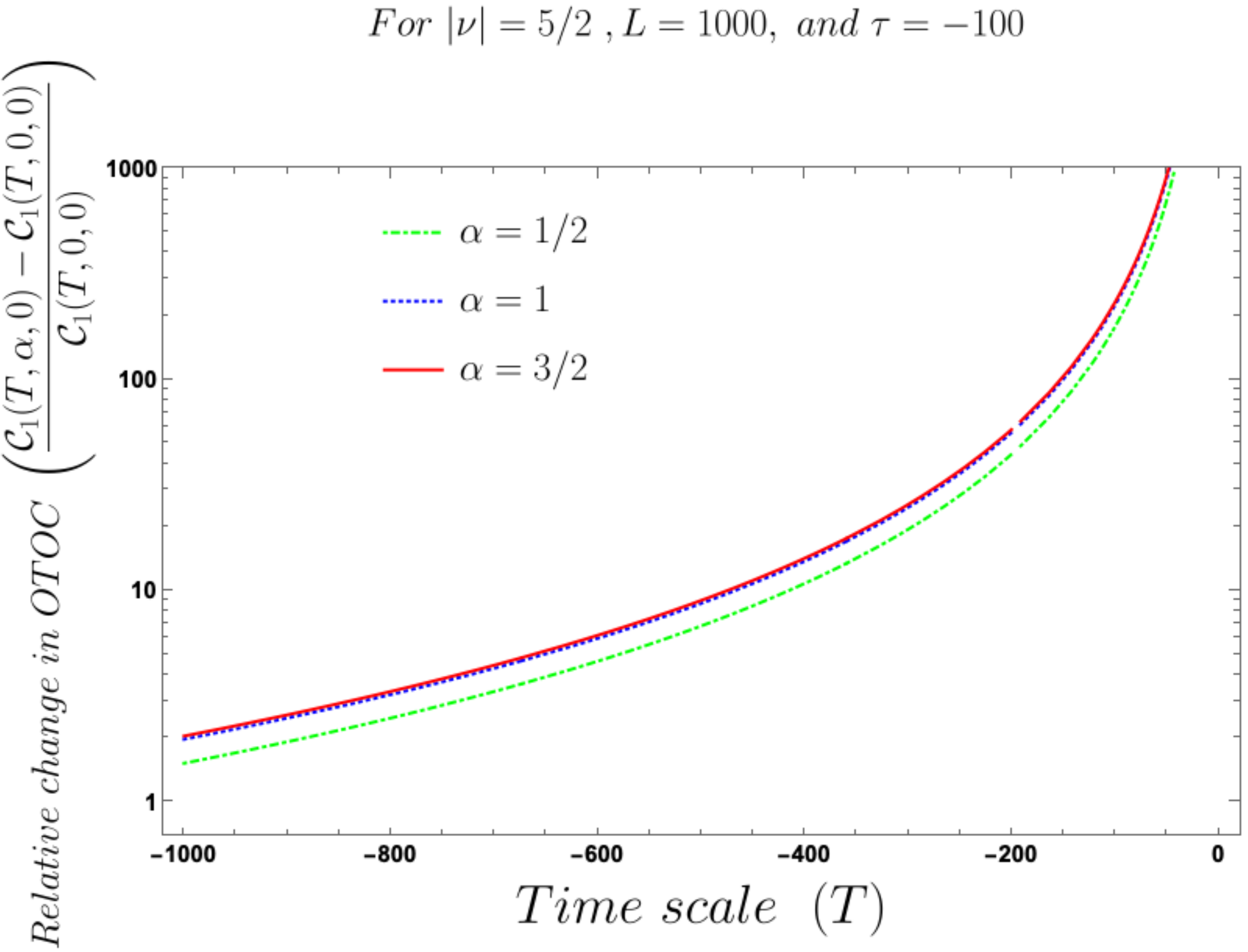}
  \caption{Behaviour of the relative change in four-point auto-correlated field OTO function with respect to the time scale $T$ for $\alpha$ vacua for the mass parameter $|\nu|=5/2$.}
  \label{fig:21}
\end{figure*}
%\begin{figure*}[htb]
%  \includegraphics[width=17cm,height=8.9cm]{R4.pdf}
 % \caption{Behaviour of the relative change in four-point auto-correlated field OTO function with respect to the time scale $T$ for $\alpha$ vacua for the mass parameter $|\nu|=7/2$.}
 % \label{fig:22}
%\end{figure*}
%\begin{figure*}[htb]
%  \includegraphics[width=17cm,height=8.9cm]{R5.pdf}
%  \caption{Behaviour of the relative change in four-point auto-correlated field OTO function with respect to the time scale $T$ for $\alpha$ vacua for the mass parameter $|\nu|=9/2$.}
%  \label{fig:23}
%\end{figure*}
\begin{figure*}[htb]
  \includegraphics[width=17cm,height=7cm]{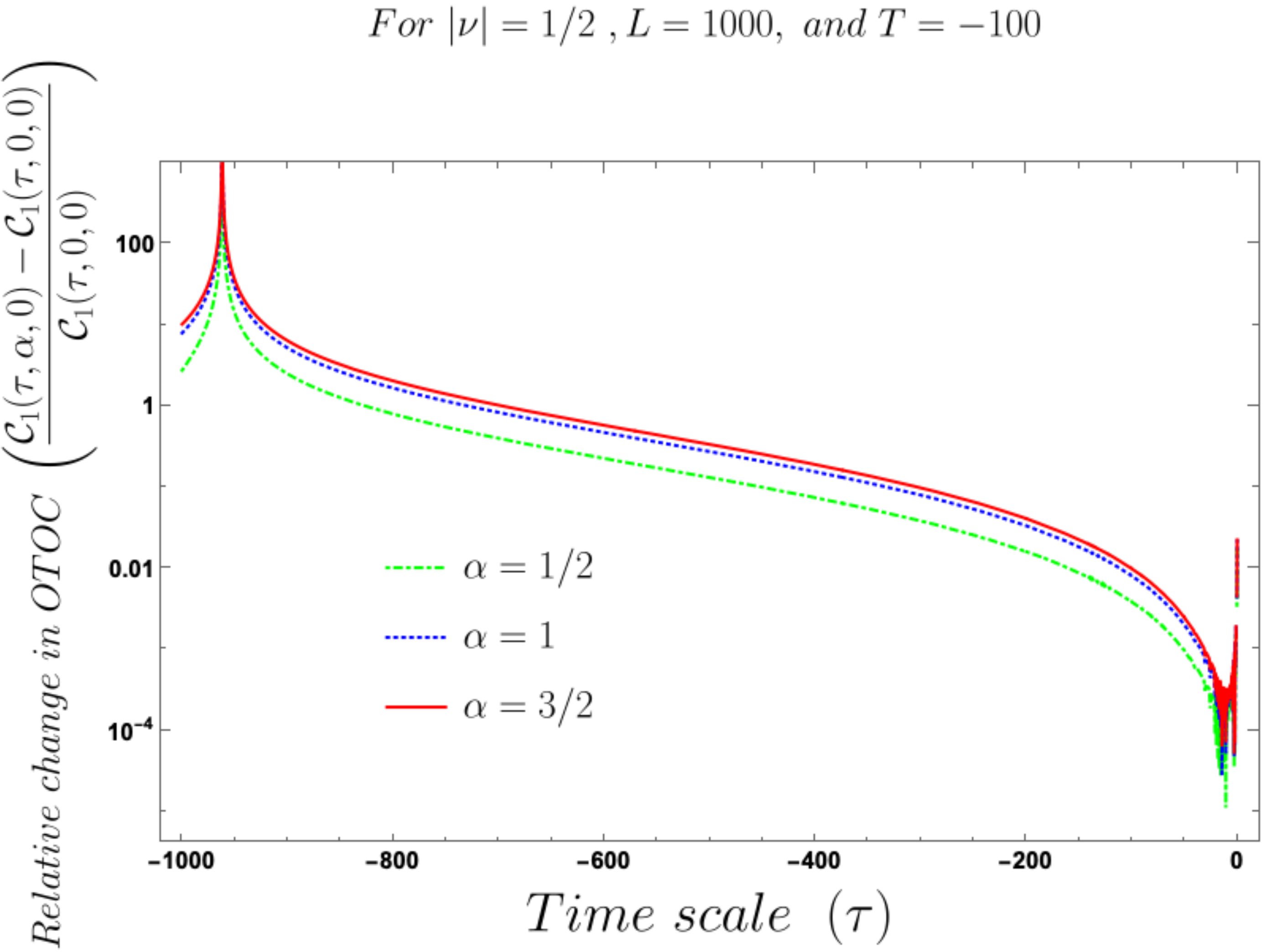}
  \caption{Behaviour of the relative change in four-point auto-correlated field OTO function with respect to the time scale $\tau$ for $\alpha$ vacua for the mass parameter $|\nu|=1/2$.}
  \label{fig:24}
\end{figure*}
\begin{figure*}[htb]
  \includegraphics[width=17cm,height=7cm]{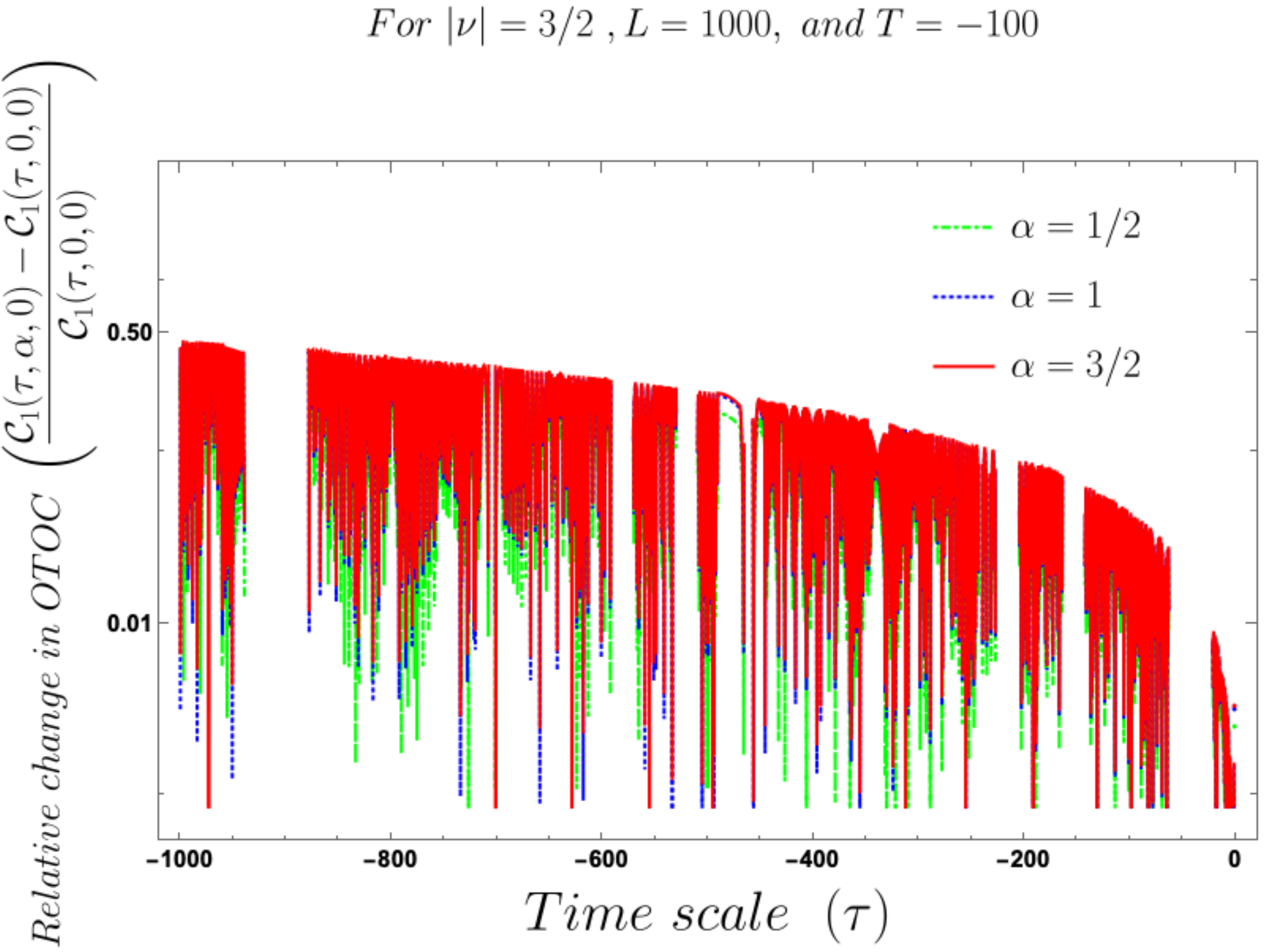}
  \caption{Behaviour of the relative change in four-point auto-correlated field OTO function with respect to the time scale $\tau$ for $\alpha$ vacua for the mass parameter $|\nu|=3/2$.}
  \label{fig:25}
\end{figure*}
%\begin{figure*}[htb]
 % \includegraphics[width=17cm,height=8.7cm]{Ra3.pdf}
 % \caption{Behaviour of the relative change in four-point auto-correlated field OTO function with respect to the time scale $\tau$ for $\alpha$ vacua for the mass parameter $|\nu|=5/2$.}
  %\label{fig:26}
%\end{figure*}
\begin{figure*}[htb]
  \includegraphics[width=17cm,height=7cm]{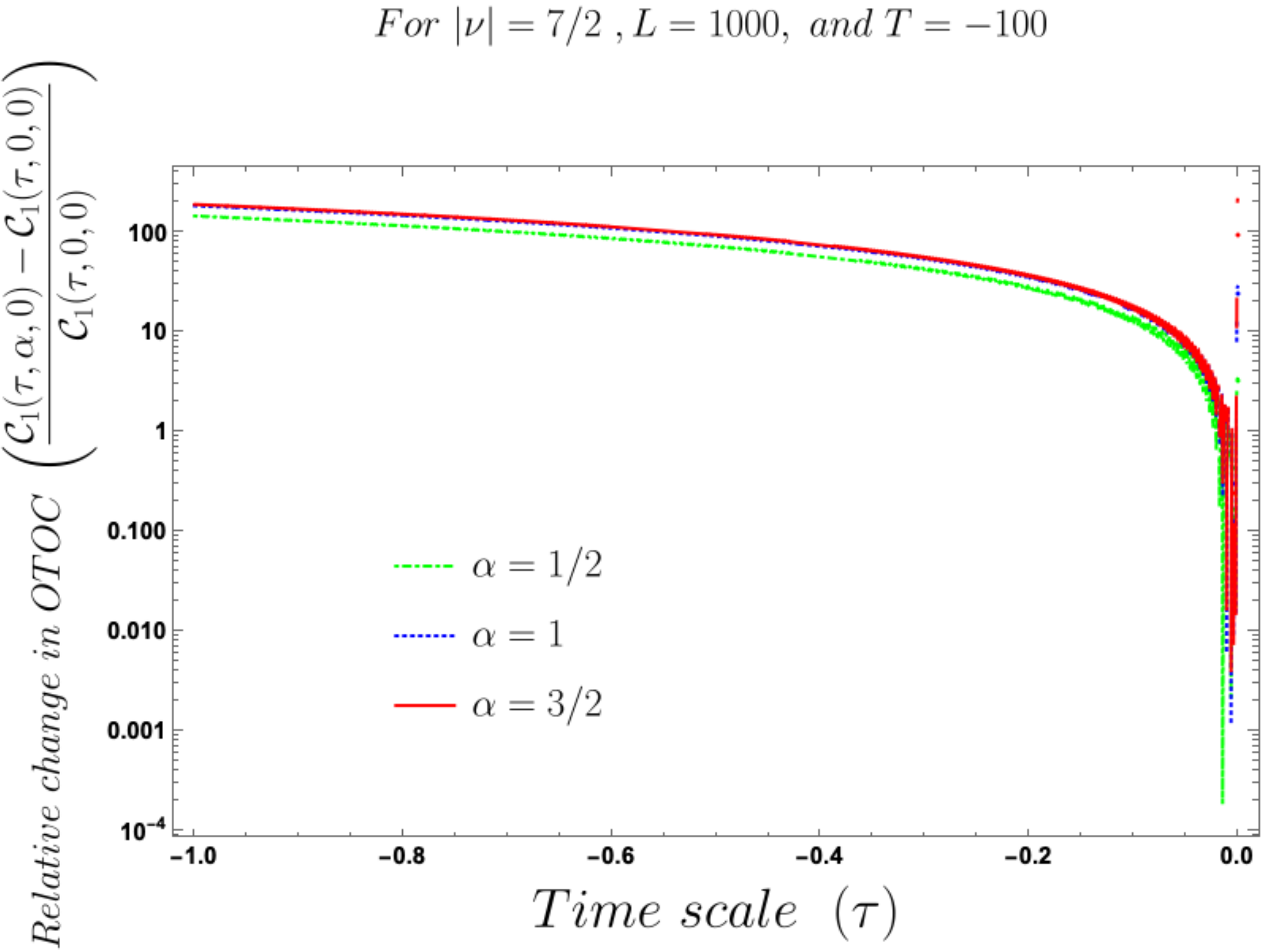}
  \caption{Behaviour of the relative change in four-point auto-correlated field OTO function with respect to the time scale $\tau$ for $\alpha$ vacua for the mass parameter $|\nu|=7/2$.}
  \label{fig:27}
\end{figure*}
%\begin{figure*}[htb]
 % \includegraphics[width=17cm,height=8.7cm]{Ra5.pdf}
 % \caption{Behaviour of the relative change in four-point auto-correlated field OTO function with respect to the time scale $\tau$ for $\alpha$ vacua for the mass parameter $|\nu|=9/2$.} 
%  \label{fig:28}
%\end{figure*}
% \begin{figure*}[htb]
 % \includegraphics[width=17cm,height=8.7cm]{CN1.pdf}
  %\caption{Behaviour of the four-point auto-correlated field OTO function with respect to the mass parameter $|\nu|$ for Motta Allen and $\alpha$ vacua and Bunch Davies vacuum.}
 % \label{fig:29}
%\end{figure*}
\begin{figure*}[htb]
  \includegraphics[width=17cm,height=7cm]{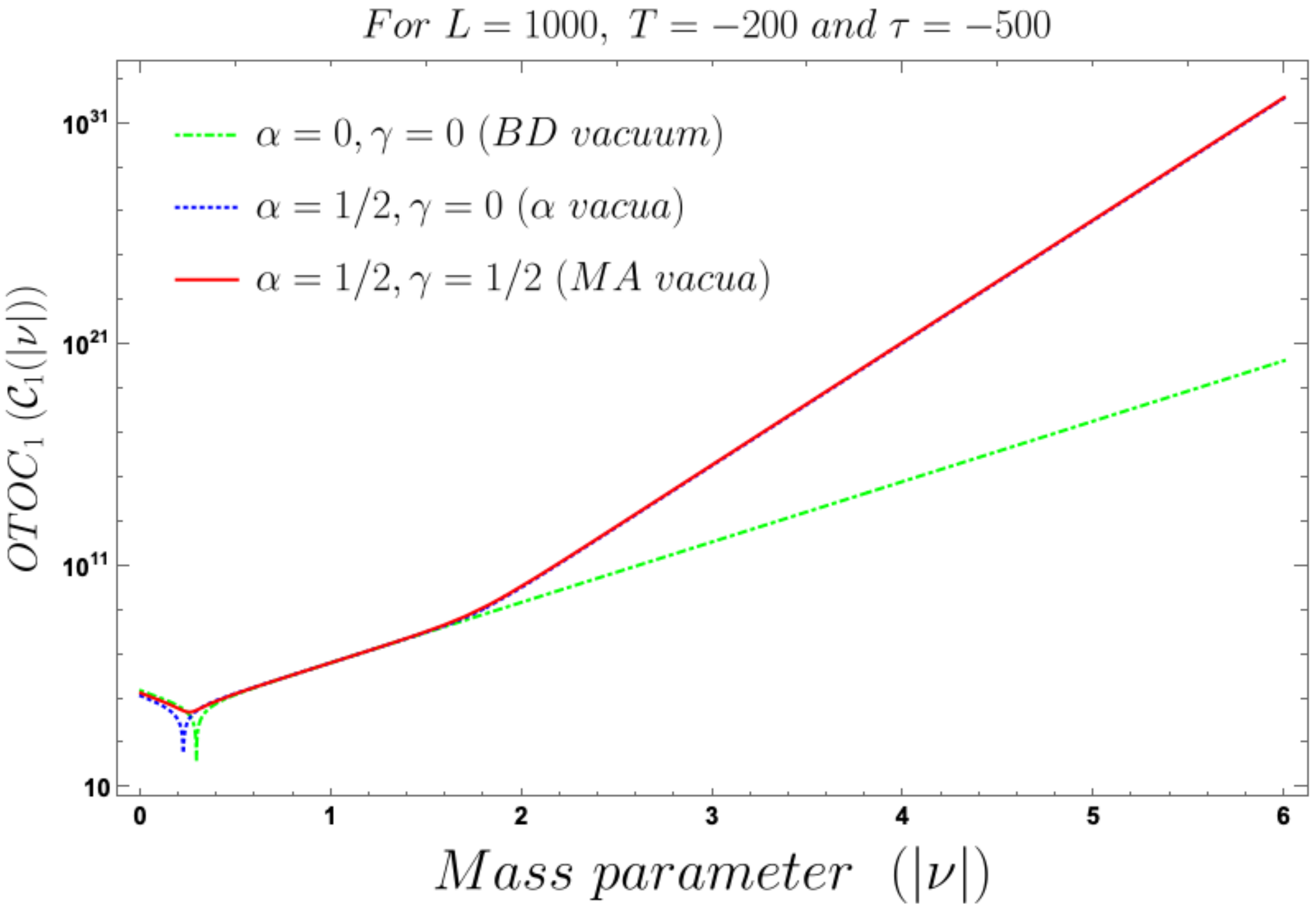}
  \caption{Behaviour of the four-point auto-correlated field OTO function with respect to the mass parameter $|\nu|$ for Mota Allen and $\alpha$ vacua and Bunch Davies vacuum.}
  \label{fig:30}
\end{figure*}
%\begin{figure*}[htb]
  %\includegraphics[width=17cm,height=8.7cm]{CN3.pdf}
  %\caption{Behaviour of the four-point auto-correlated field OTO function with respect to the mass parameter $|\nu|$ for Motta Allen and $\alpha$ vacua and Bunch Davies vacuum.}
  %\label{fig:31}
%\end{figure*}
%\begin{figure*}[htb]
 % \includegraphics[width=17cm,height=8.7cm]{CA1.pdf}
 % \caption{Behaviour of the four-point auto-correlated field OTO function with respect to the vacuum parameter $\alpha$ for $\alpha$ vacua with mass parameter $|\nu|=1/2$.}
%  \label{fig:32}
%\end{figure*}
\begin{figure*}[htb]
  \includegraphics[width=17cm,height=7cm]{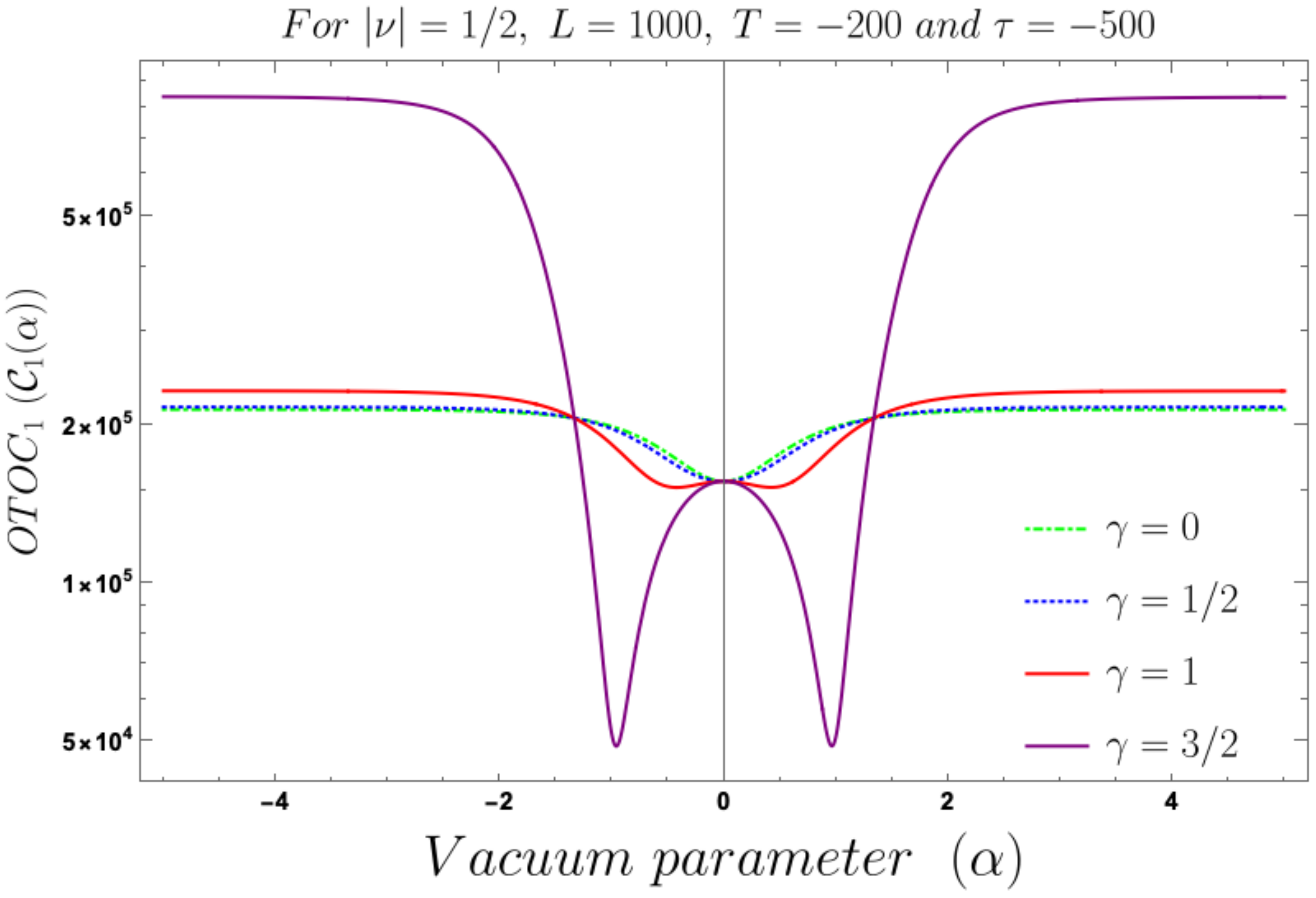}
  \caption{Behaviour of the four-point auto-correlated field OTO function with respect to the vacuum parameter $\alpha$ for $\alpha$ vacua with mass parameter $|\nu|=1/2$.}
  \label{fig:33}
\end{figure*}
%\begin{figure*}[htb]
 % \includegraphics[width=17cm,height=8.7cm]{CA3.pdf}
  %\caption{Behaviour of the four-point auto-correlated field OTO function with respect to the vacuum parameter $\alpha$ for $\alpha$ vacua with mass parameter $|\nu|=1/2$.}
 % \label{fig:34}
%\end{figure*}
 %\begin{figure*}[htb]
%  \includegraphics[width=17cm,height=8.7cm]{CG1.pdf}
%  \caption{Behaviour of the four-point auto-correlated field OTO function with respect to the vacuum parameter $\gamma$ for Motta Allen vacua with mass parameter $|\nu|=1/2$.} 
 % \label{fig:35}
%\end{figure*}
\begin{figure*}[htb]
  \includegraphics[width=17cm,height=7cm]{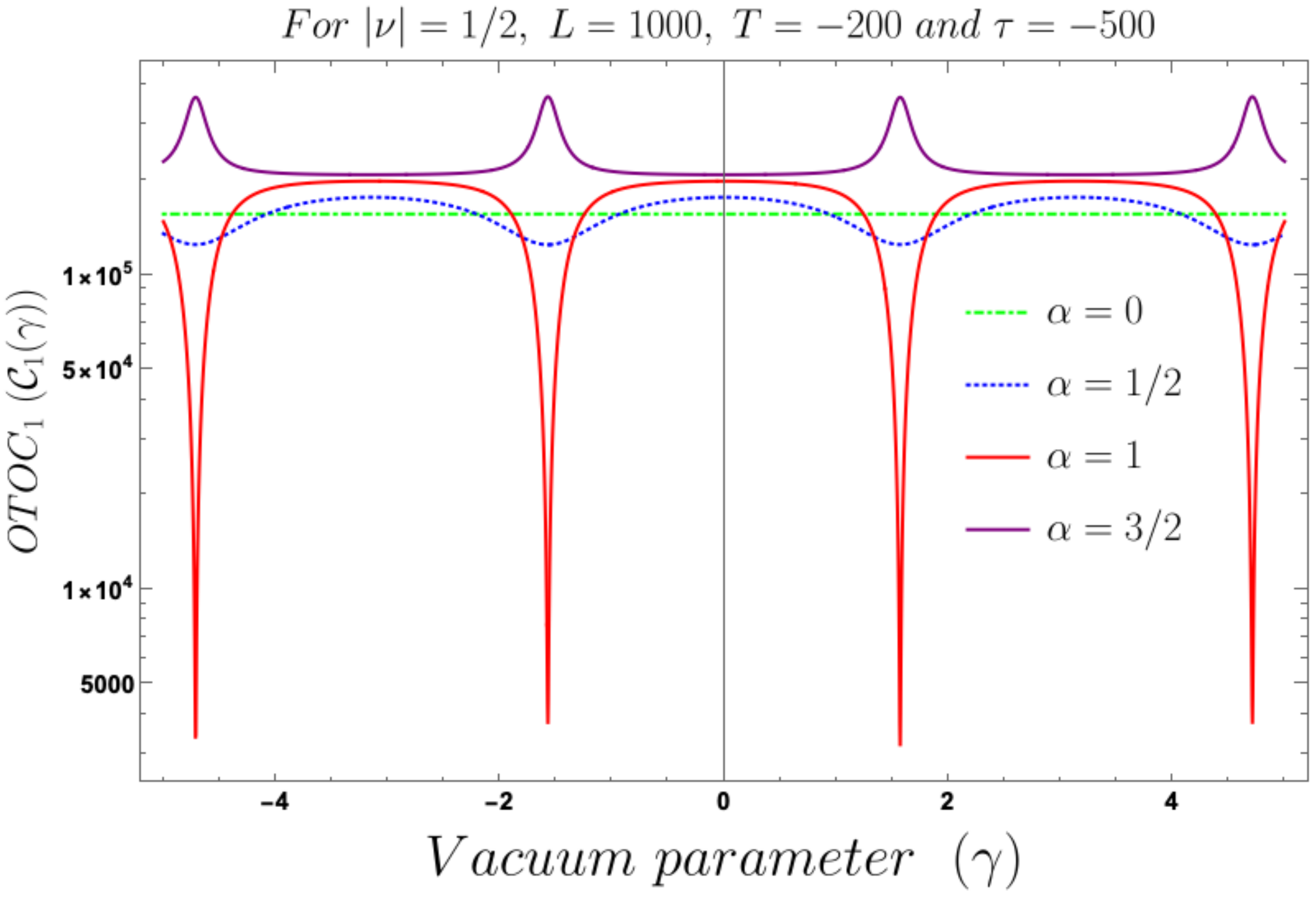}
  \caption{Behaviour of the four-point auto-correlated field OTO function with respect to the vacuum parameter $\gamma$ for Mota Allen vacua with mass parameter $|\nu|=1/2$.}
  \label{fig:36}
\end{figure*}
%\begin{figure*}[htb]
  %\includegraphics[width=17cm,height=8.7cm]{CG3.pdf}
 % \caption{Behaviour of the four-point auto-correlated field OTO function with respect to the vacuum parameter $\gamma$ for Motta Allen vacua with mass parameter $|\nu|=1/2$.}
 % \label{fig:37}
%\end{figure*} 
 \begin{figure*}[htb]
  \includegraphics[width=17cm,height=7cm]{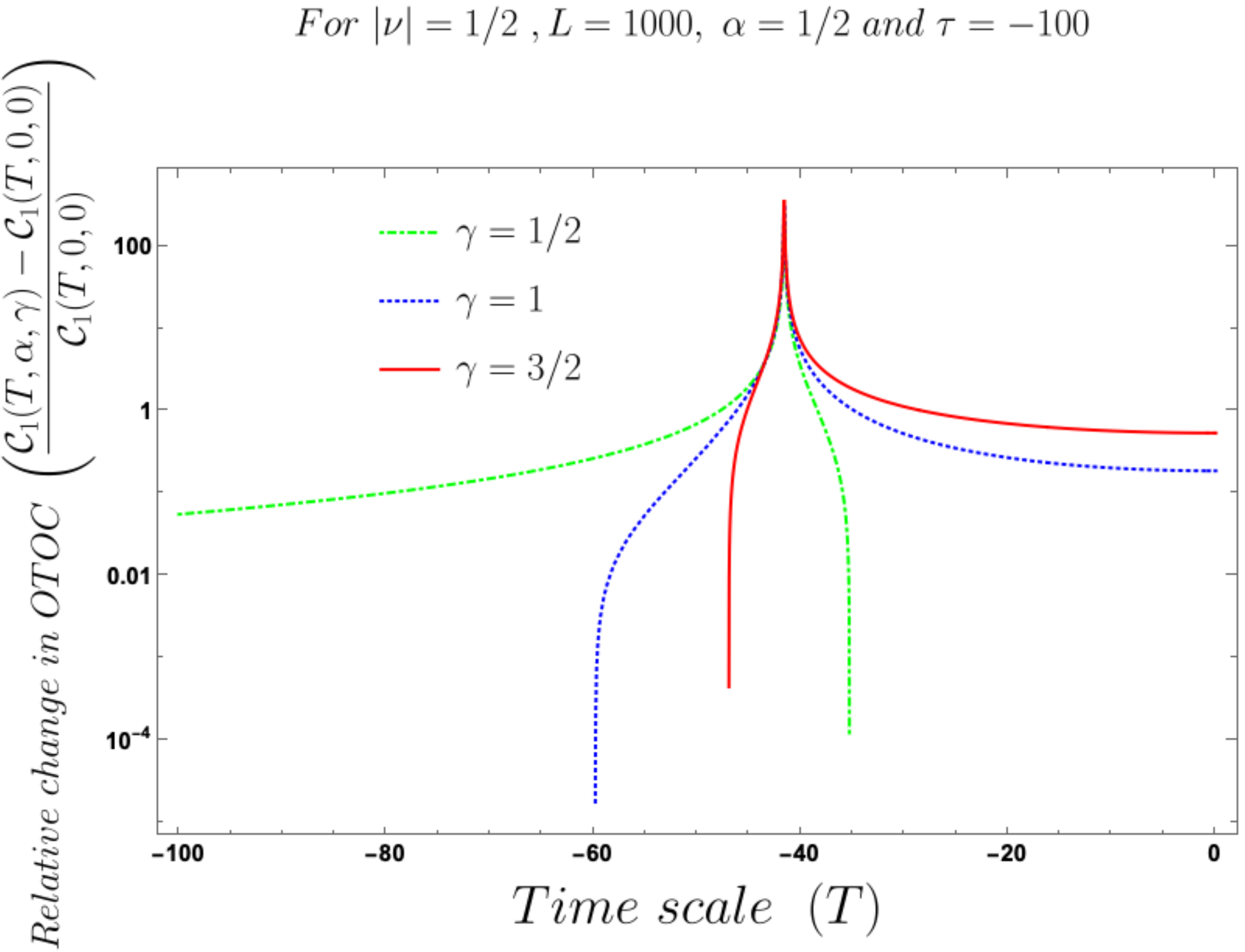}
  \caption{Behaviour of the relative change in four-point auto-correlated field OTO function with respect to the time scale $T$ for Mota Allen vacua with mass parameter $|\nu|=1/2$.} 
  \label{fig:38}
\end{figure*}
\begin{figure*}[htb]
  \includegraphics[width=17cm,height=7cm]{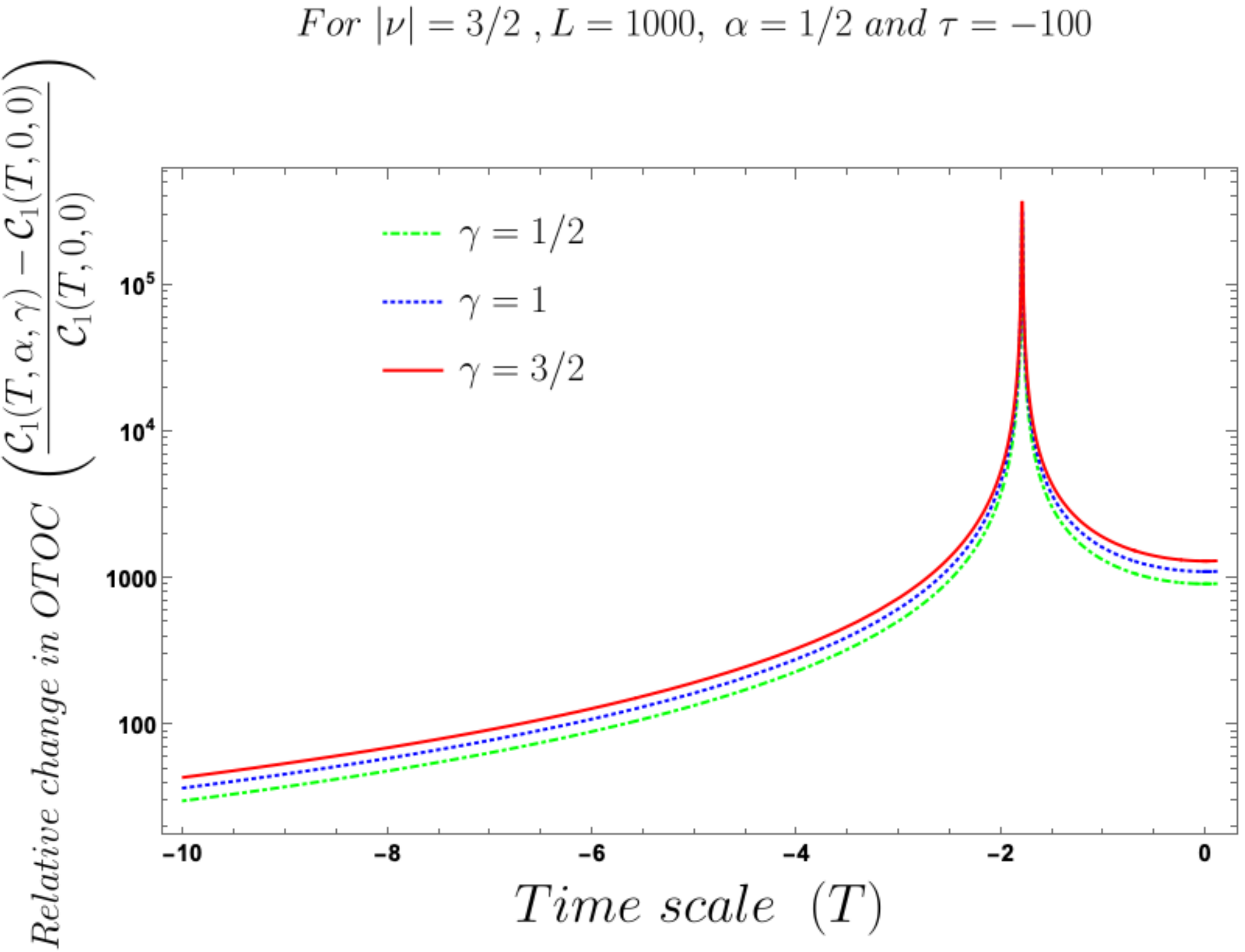}
  \caption{Behaviour of the relative change in four-point auto-correlated field OTO function with respect to the time scale $T$ for Mota Allen vacua with mass parameter $|\nu|=3/2$.}
  \label{fig:39}
\end{figure*}
 %\begin{figure*}[htb]
 % \includegraphics[width=17cm,height=8.7cm]{RGT3.pdf}
 % \caption{Behaviour of the relative change in four-point auto-correlated field OTO function with respect to the time scale $T$ for Motta Allen vacua with mass parameter $|\nu|=5/2$.}
 % \label{fig:40}
%\end{figure*}
\begin{figure*}[htb]
  \includegraphics[width=17cm,height=7cm]{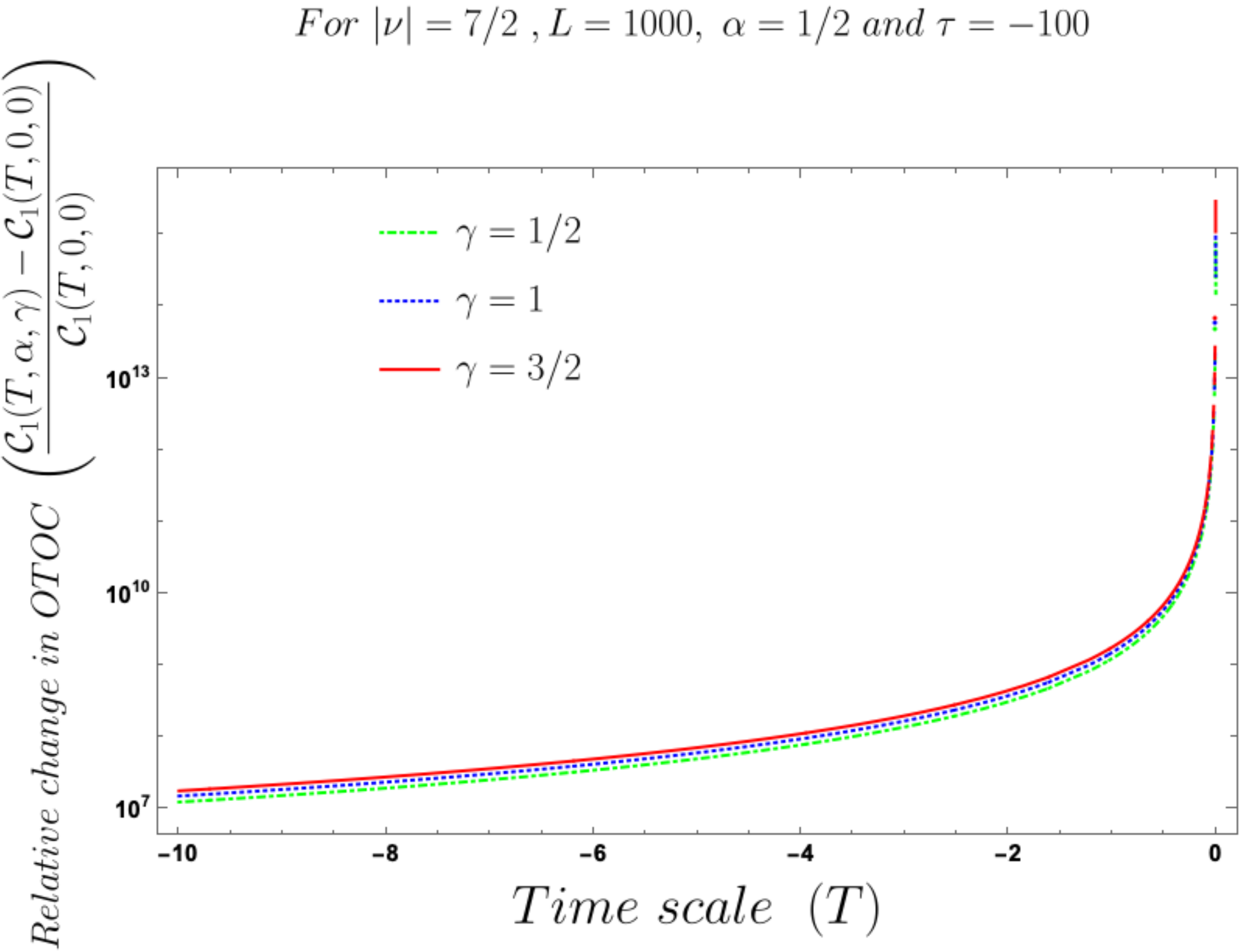}
  \caption{Behaviour of the relative change in four-point auto-correlated field OTO function with respect to the time scale $T$ for Mota Allen vacua with mass parameter $|\nu|=7/2$.}
  \label{fig:41}
\end{figure*}
%\begin{figure*}[htb]
%  \includegraphics[width=17cm,height=8.7cm]{RGT5.pdf}
 % \caption{Behaviour of the relative change in four-point auto-correlated field OTO function with respect to the time scale $T$ for Motta Allen vacua with mass parameter $|\nu|=9/2$.}
 % \label{fig:42}
%\end{figure*} 
\begin{figure*}[htb]
  \includegraphics[width=17cm,height=7cm]{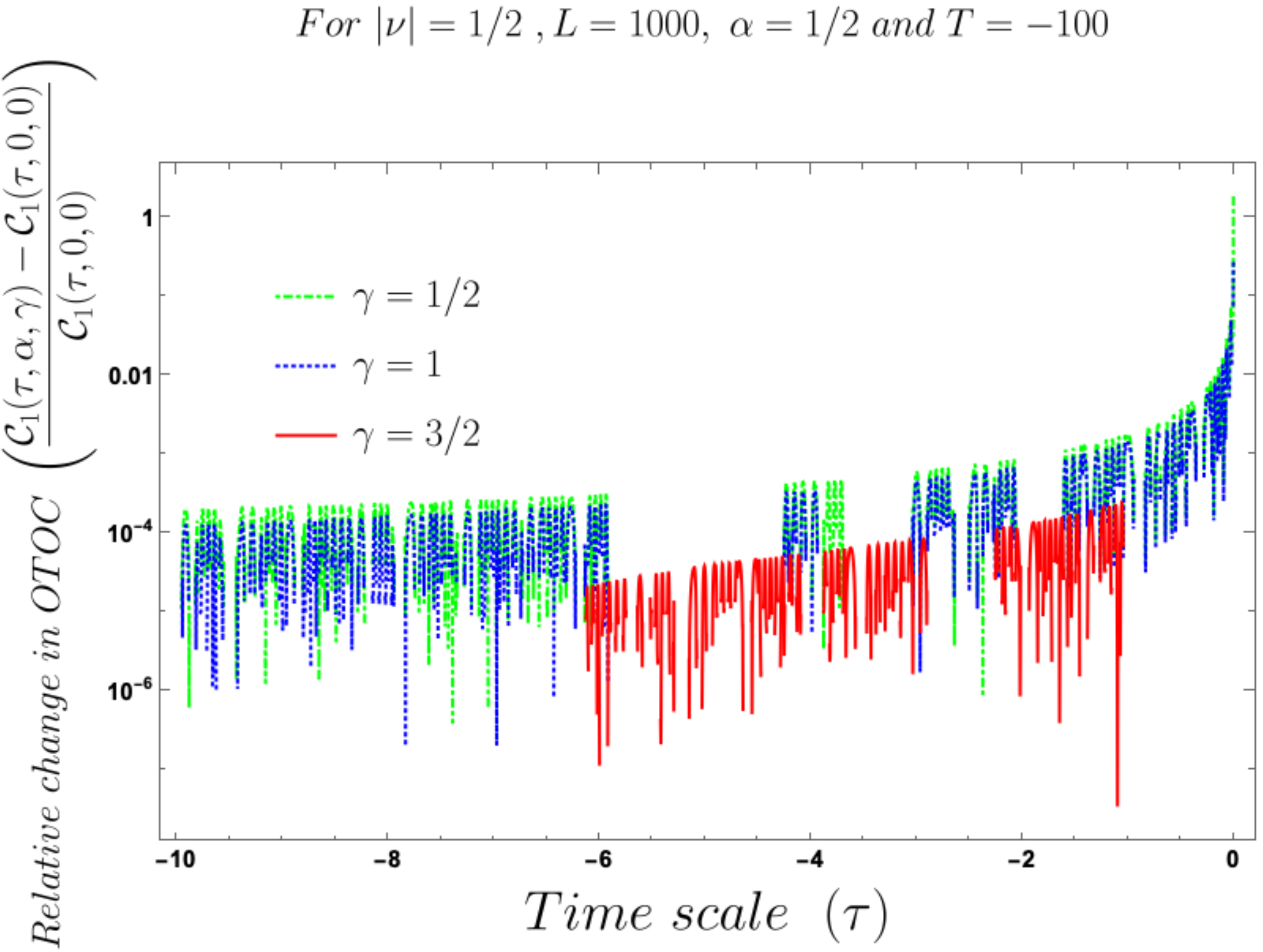}
  \caption{Behaviour of the relative change in four-point auto-correlated filed OTO function with respect to the time scale $\tau$ for Mota Allen vacua with mass parameter $|\nu|=1/2$.} 
  \label{fig:43}
\end{figure*}
\begin{figure*}[htb]
  \includegraphics[width=17cm,height=7cm]{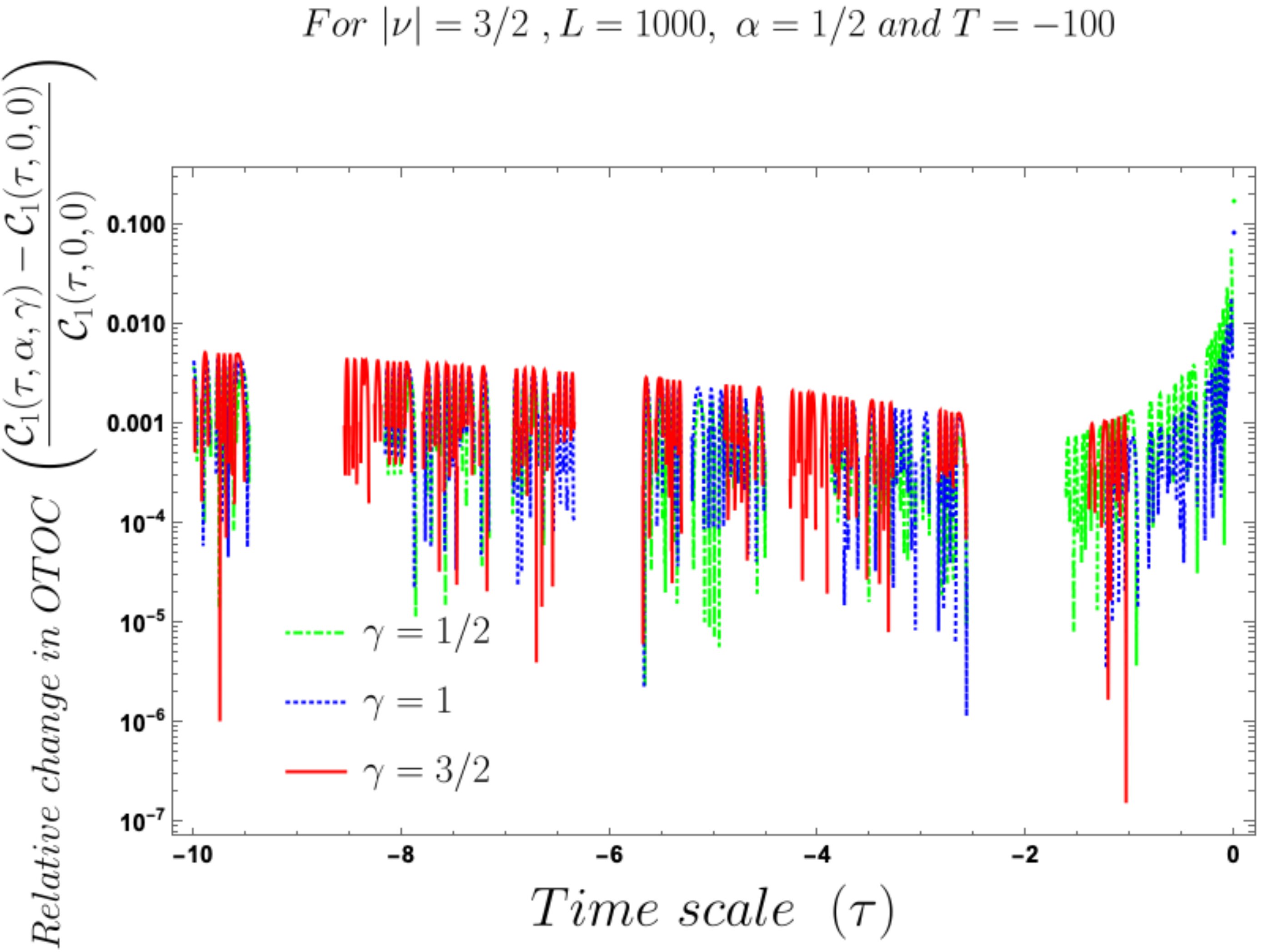}
  \caption{Behaviour of the relative change in four-point auto-correlated filed OTO function with respect to the time scale $\tau$ for Mota Allen vacua with mass parameter $|\nu|=3/2$.}
  \label{fig:44}
\end{figure*}
 %\begin{figure*}[htb]
%  \includegraphics[width=17cm,height=8.7cm]{RGTa3.pdf}
 % \caption{Behaviour of the relative change in four-point auto-correlated filed OTO function with respect to the time scale $\tau$ for Motta Allen vacua with mass parameter $|\nu|=5/2$.} 
  %\label{fig:45}
%\end{figure*}
%\begin{figure*}[htb]
 % \includegraphics[width=17cm,height=8.7cm]{RGTa4.pdf}
 % \caption{Behaviour of the relative change in four-point auto-correlated field OTO function with respect to the time scale $\tau$ for Motta Allen vacua with mass parameter $|\nu|=7/2$.}
 % \label{fig:46}
%\end{figure*}
%\begin{figure*}[htb]
%  \includegraphics[width=17cm,height=8.7cm]{RGTa5.pdf}
%  \caption{Behaviour of the relative change in four-point auto-correlated field OTO function with respect to the time scale $\tau$ for Motta Allen vacua with mass parameter $|\nu|=9/2$.}
 % \label{fig:47}
%\end{figure*}  
\begin{figure*}[htb]
  \includegraphics[width=17cm,height=7cm]{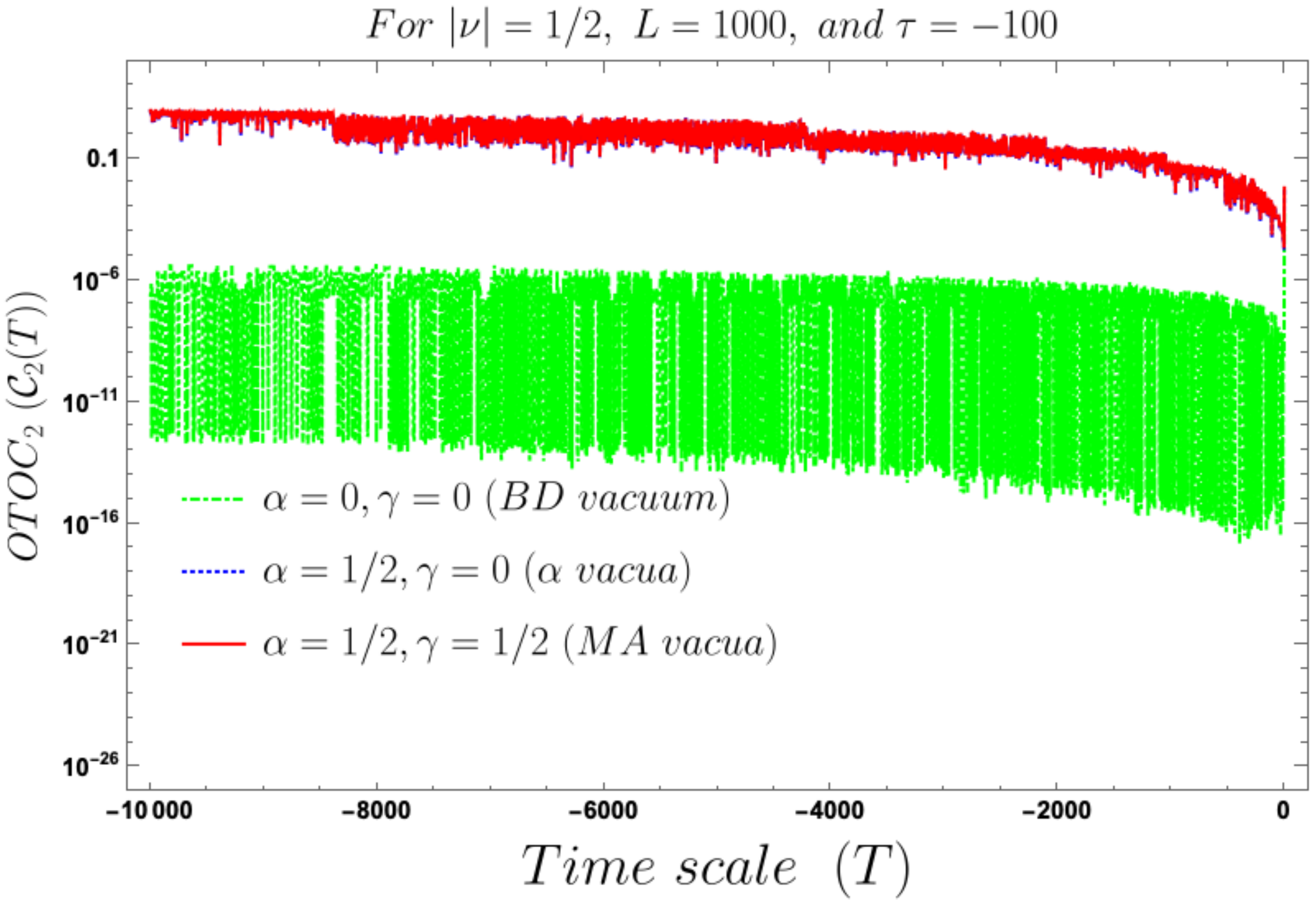}
  \caption{Behaviour of the four-point auto-correlated momentum OTO function with respect to the time scale $T$ for Mota Allen,  $\alpha$ vacua and Bunch Davies vacuum with mass parameter $|\nu|=1/2$.}
  \label{fig:48}
\end{figure*} 
\begin{figure*}[htb]
  \includegraphics[width=17cm,height=7cm]{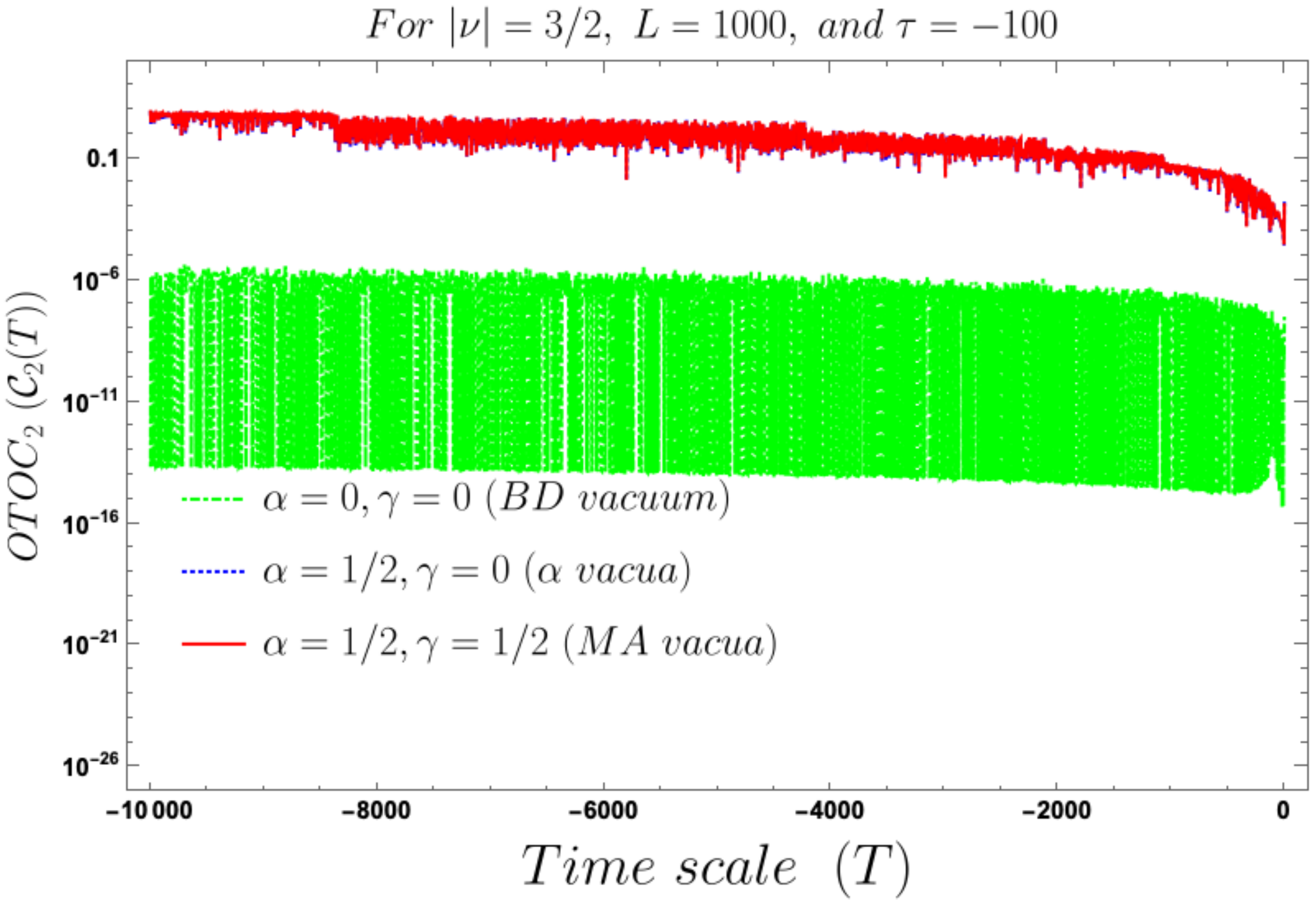}
  \caption{Behaviour of the four-point auto-correlated momentum OTO function with respect to the time scale $T$ for Mota Allen,  $\alpha$ vacua and Bunch Davies vacuum with mass parameter $|\nu|=3/2$.}
  \label{fig:49}
\end{figure*}
\begin{figure*}[htb]
  \includegraphics[width=17cm,height=7cm]{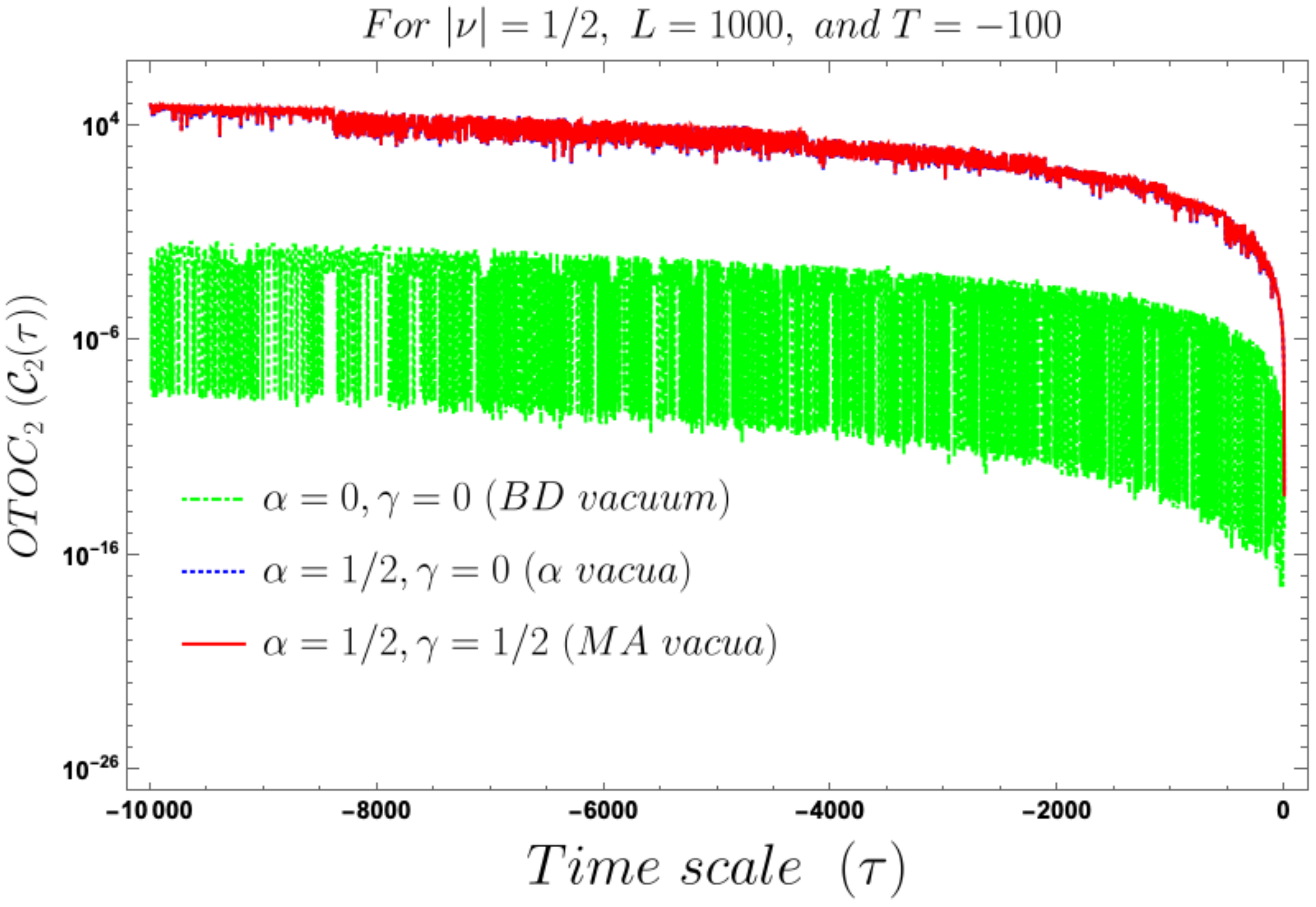}
  \caption{Behaviour of the four-point auto-correlated momentum OTO function with respect to the time scale $\tau$ for Mota Allen,  $\alpha$ vacua and Bunch Davies vacuum with mass parameter $|\nu|=1/2$.} 
  \label{fig:53}
\end{figure*}

\begin{figure*}[htb]
  \includegraphics[width=17cm,height=7cm]{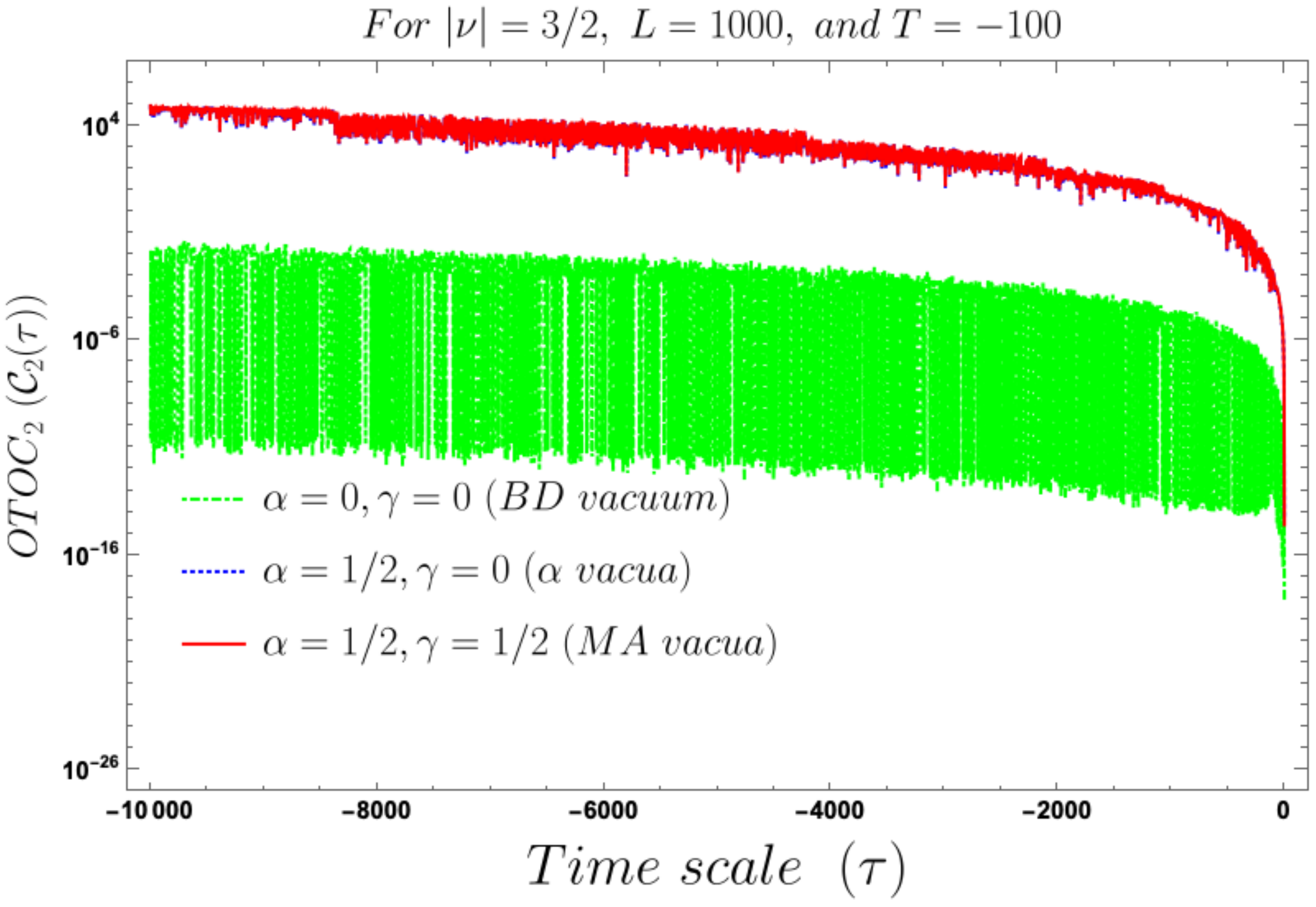}
  \caption{Behaviour of the four-point auto-correlated momentum OTO function with respect to the time scale $\tau$ for Mota Allen,  $\alpha$ vacua and Bunch Davies vacuum with mass parameter $|\nu|=3/2$.} 
  \label{fig:54}
\end{figure*}

\begin{figure*}[htb]
  \includegraphics[width=17cm,height=7cm]{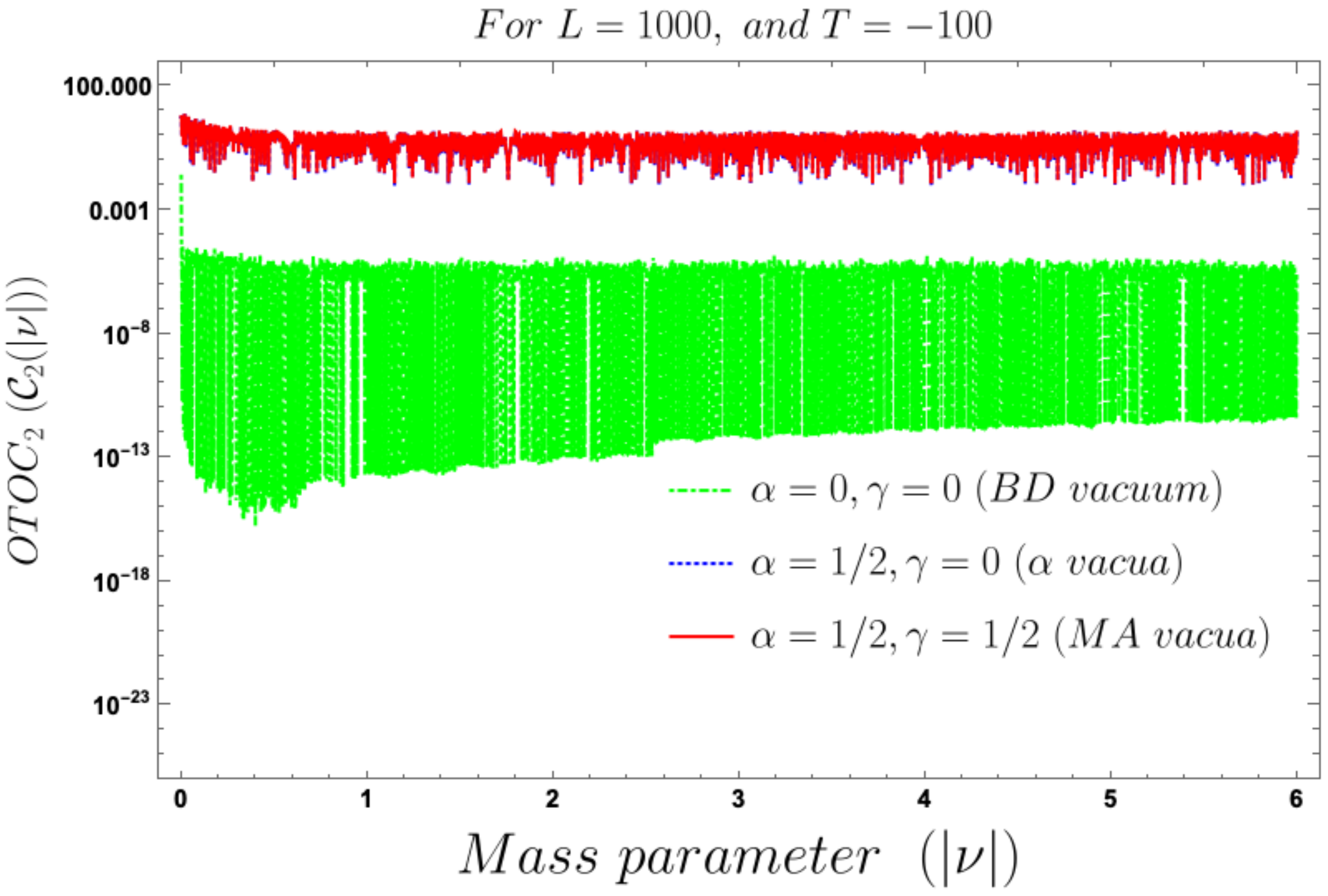}
  \caption{Behaviour of the four-point auto-correlated momentum OTO function with respect to the mass parameter $|\nu|$ for Mota Allen,  $\alpha$ vacua and Bunch Davies vacuum. }  
  \label{fig:58}
\end{figure*}

\begin{figure*}[htb]
  \includegraphics[width=17cm,height=7cm]{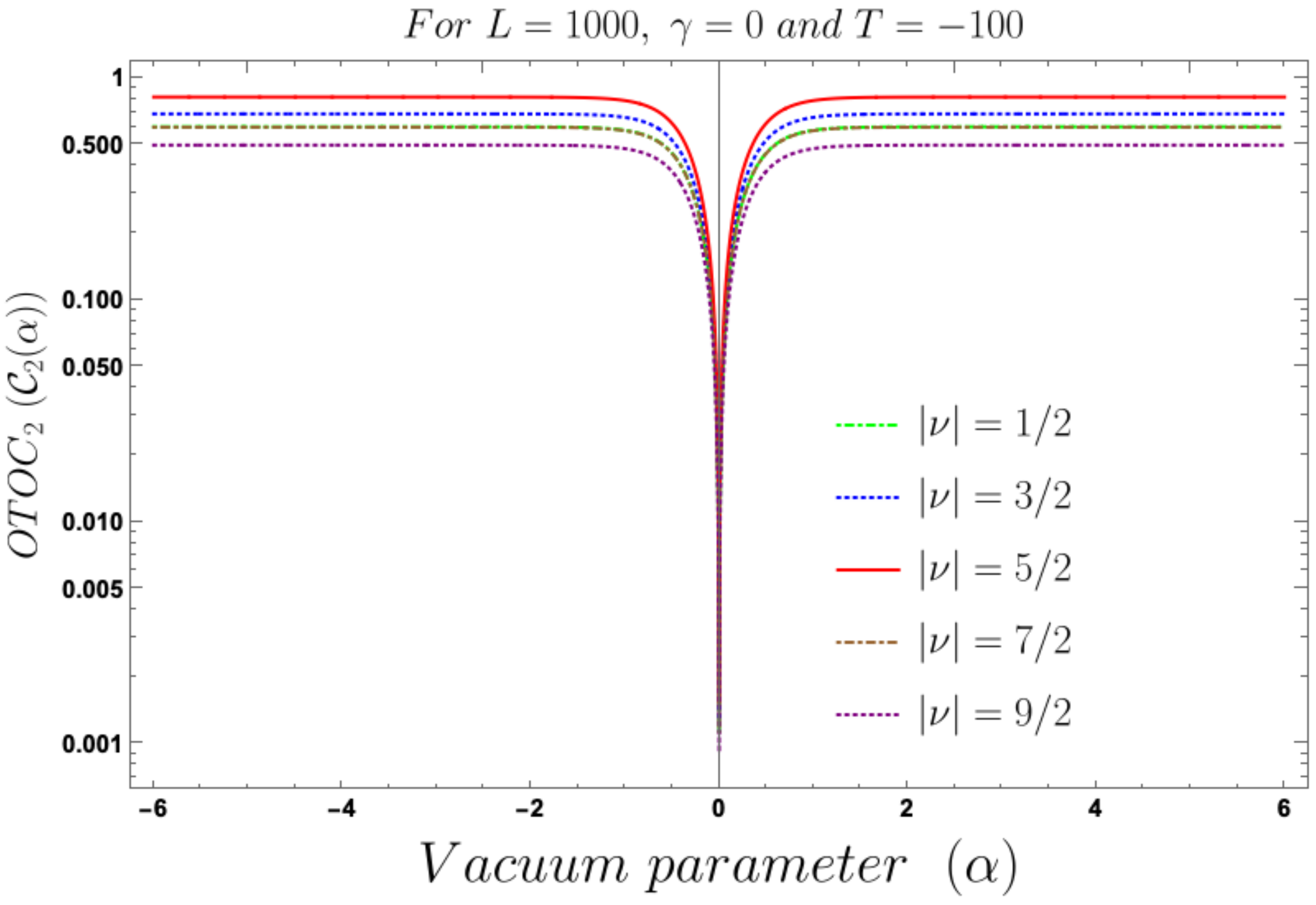}
  \caption{Behaviour of the four-point auto-correlated momentum OTO function with respect to the vacuum parameter $\alpha$ for $\alpha$ vacua. } 
  \label{fig:59}
\end{figure*}

\begin{figure*}[htb]
  \includegraphics[width=17cm,height=7cm]{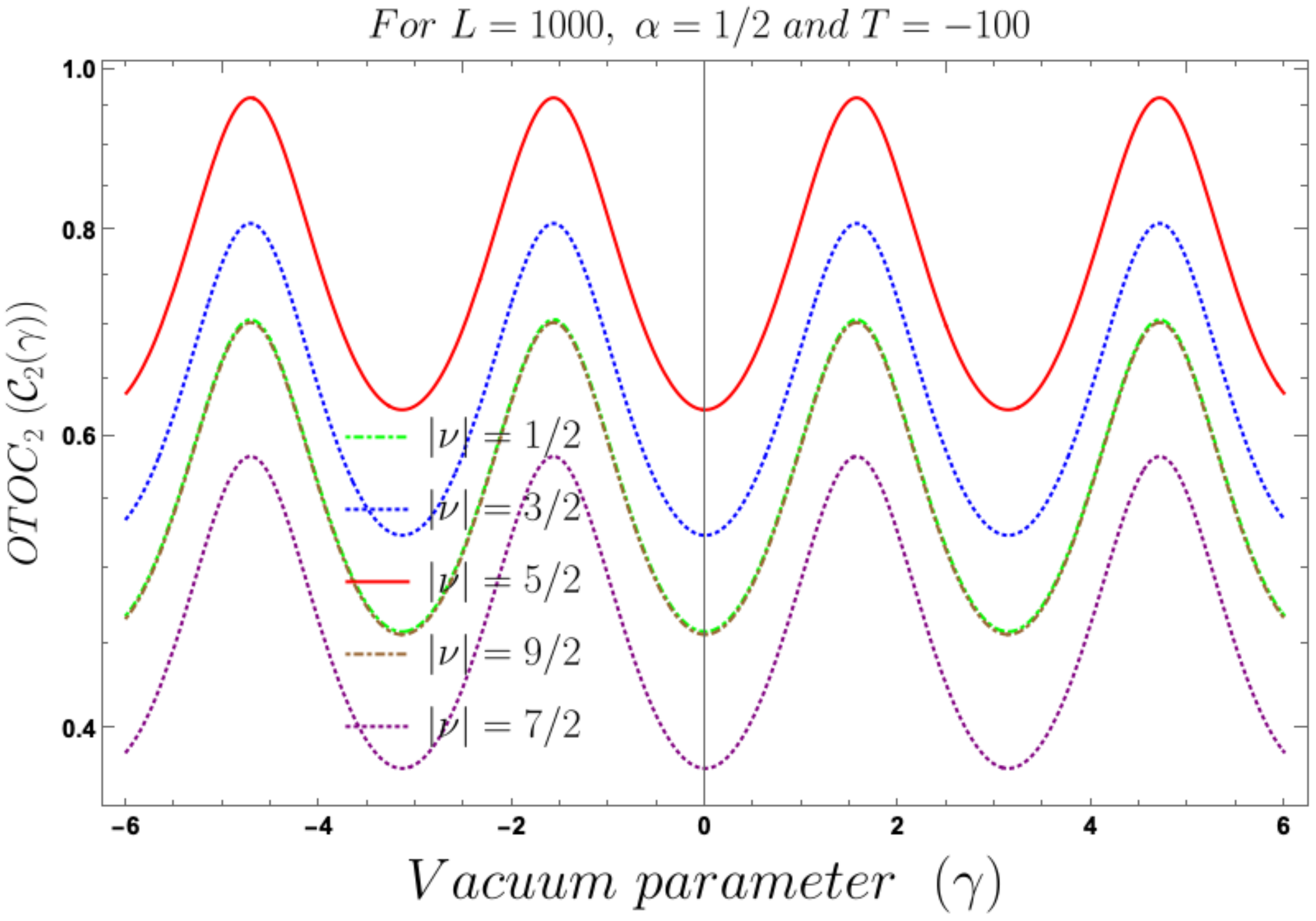}
  \caption{Behaviour of the four-point auto-correlated momentum OTO function with respect to the vacuum parameter $\gamma$ for Mota Allen vacua. }  
  \label{fig:60}
\end{figure*}

 
%\begin{figure*}[htb]
%  \includegraphics[width=17cm,height=8.7cm]{C2T3.pdf}
%  \caption{Behaviour of the four-point auto-correlated momentum OTO function with respect to the time scale $T$ for Motta Allen,  $\alpha$ vacua and Bunch Davies vacuum with mass parameter $|\nu|=5/2$.}
 % \label{fig:50}
%\end{figure*} 
%\begin{figure*}[htb]
%  \includegraphics[width=17cm,height=8.7cm]{C2T4.pdf}
 % \caption{Behaviour of the four-point auto-correlated momentum OTO function with respect to the time scale $T$ for Motta Allen,  $\alpha$ vacua and Bunch Davies vacuum with mass parameter $|\nu|=7/2$.} 
 % \label{fig:51}
%\end{figure*}  
Now comparing the results of the four-point auto-correlated field and momentum OTO functions we get the following outcomes which we have mentioned point-wise:
 \begin{enumerate}
 \item Before normalising the four-point auto-correlated field and momentum OTO functions,  the results computed from rescaled field \& momentum and the curvature perturbation field variable and its conjugate momenta the 
 sides are not at all same and connected via a time dependent {\it Mukhanov Sasaki} variable dependent conformal factor $(z(\tau_1)z(\tau_2))^{-2}$ in the final results.
 \item After normalising the four-point auto-correlated field and momentum OTO functions,  the results computed from rescaled field \& momentum and the curvature perturbation field variable and its conjugate momenta the 
 sides are not at all same.  In this case particularly for the computational purpose is the best one as we really don't need care about the origin of the quantum operators in the cosmological perturbation theory. 
 \end{enumerate}
 Next,  we give the interpretation of the obtained results for the normalised four-point auto-correlated field and momentum OTO functions computed in the present set up point-wise:
 \begin{itemize}
 \item  In our computed four-point auto-correlated field and momentum OTO functions two time scales are involved.  During the study of the behaviour of the four-point auto-correlated field and momentum OTO functions we have actually have fixed one time scale and have studied the time dependent dynamical behaviour with respect to the other time scale.  We have found that the behaviour with respect to both $\tau_1=T$ and $\tau_2=\tau$,  using both Bunch Davies,  $\alpha$ and the generalised Mota Allen vacua as initial choice of quantum vacuum state.  See fig.~(\ref{fig:9}), fig.~(\ref{fig:10}), and fig.~(\ref{fig:12}),  fig.~(\ref{fig:14}), fig.~(\ref{fig:16}), fig.~(\ref{fig:17}),  fig.~(\ref{fig:18}),   fig.~(\ref{fig:48}),  fig.~(\ref{fig:49}),  fig.~(\ref{fig:53}),  fig.~(\ref{fig:54}) and  fig.~(\ref{fig:58}) to know about the detailed conformal time dependent feature of the normalised four-point auto-correlated field and momentum OTO functions.  In the conformal time scale all of these plots varies from, $-\infty$ (Big Bang) to $0$ (present day universe),  and show significant distinctive features in the time scale.

 \item If we fix the time scale, $\tau_2=\tau$ and study the behaviour of four-point auto-correlated field and momentum OTO functions with respect to the scale $\tau_1=T$  then we found that as we approach from very early universe to the late time scale the normalised correlators show random non-standard non-chaotic behaviour.

\item Additionally we have found that we can get the desired features from the obtained results can be visible only when the mass parameter,  $\nu$ can be analytically continued to $-i\nu$.  So \underline{\textcolor{red}{massless De Sitter}} possibility ($\nu=3/2$) is not allowed.  Consequently,  we have the following options:
\begin{enumerate}
\item \underline{\textcolor{red}{Partially massless De Sitter}} case is one of the possibilities, where we can estimate the following parameter from the present understanding:
\bea {-i\nu=\sqrt{\frac{9}{4}-c^2}~~~{\rm where}~~~c\geq \sqrt{2}},\eea
which implies the following bound on the mass parameter:
\bea \nu\geq \frac{i}{2}~.\eea
This is consistent with the plots obtained for the following values of the mass parameter:
\bea \nu=\frac{i}{2},~\frac{3i}{2},~\frac{5i}{2},~\frac{7i}{2},~\frac{9i}{2}~~.\eea

\item \underline{\textcolor{red}{\bf Massive De Sitter}} case is one of the possibilities,  where we can estimate the following parameter:
\bea {\nu=i\sqrt{\frac{m^2}{H^2}-\frac{9}{4}}}~.\eea
Since we have studied the case for $\nu\geq i/2$,  we get the following bound on the mass of the heavy field:
\bea {m\geq \sqrt{\frac{5}{2}}~H}~.\eea

\item \underline{\textcolor{red}{\bf Reheating}} case is the last possibility, where we can estimate:
\bea {\frac{i}{2}\leq \nu\leq \frac{3i}{2}~~~~{\rm for}~~~~0\leq w_{reh}\leq \frac{1}{3}~}~.\eea 
\end{enumerate}

\item After fixing both the time scales at far past, the behaviour of four-point auto-correlated field and momentum OTO functions with respect to the mass parameter is depicted in the fig.~(\ref{fig:30}) and fig.~(\ref{fig:58}) for Bunch Davies,  $\alpha$ and Mota Allen vacua as quantum mechanical vacuum state. 

\item After fixing both the time scales at far past and fix $\nu=i/2$, the behaviour of four-point auto-correlated field and momentum OTO functions with respect to the vacuum parameter $\alpha$ and $\gamma$ depicted in the fig.~(\ref{fig:33}),  fig.~(\ref{fig:36}),  fig.~(\ref{fig:59}) and fig.~(\ref{fig:60}).  

\item All the momentum dependent integrals are divergent at $\infty$, for which we have introduced an UV regulator at $L=1000$ (large value).  This makes the final results of four-point auto-correlated field and momentum OTO functions numerically convergent.

\item Further we define a relative change in normalized four-point auto-correlated field and momentum OTO functions where the relativeness is defined with respect to the definition of quantum mechanical vacuum state, given by:
\bea &&{{\cal R}^{(p)}_{\alpha,\gamma}(\tau_i)=\left(\frac{\underbrace{{\cal C}_p(\tau_i,\alpha,\gamma)}_{\textcolor{red}{\bf Motta~Allen~vacua}}-\underbrace{{\cal C}_p(\tau_i,\alpha=0,\gamma=0)}_{\textcolor{red}{\bf Bunch~Davies~vacuum}}}{\underbrace{{\cal C}_p(\tau_i,\alpha=0,\gamma=0)}_{\textcolor{red}{\bf Bunch~Davies~vacuum}}}\right)~~~\forall~~ i=1 (T), 2 (\tau)}~\nonumber\\
&& ~~~~~~~~~~~\&~\forall ~p=1({\rm Field~auto~correlator}),2({\rm Momentum~auto~correlator}).~
\eea
In fig.~(\ref{fig:19}),  fig.~(\ref{fig:20}),  fig.~(\ref{fig:21}),  fig.~(\ref{fig:24}),  fig.~(\ref{fig:25}),  fig.~(\ref{fig:27}),  fig.~(\ref{fig:38}),  fig.~(\ref{fig:39}),  fig.~(\ref{fig:41}),  fig.~(\ref{fig:43}),  fig.~(\ref{fig:44}),  we have explicitly shown the behaviour of the relative change in normalized four-point auto-correlated field and momentum OTO functions with respect to the two time scales.  Here the relative change we have studies for $\alpha=1/2,\gamma=1'2$,  $\alpha=1/2,\gamma=1$,  and $\alpha=1/2,\gamma=3/2$ with respect to $\alpha=0$ for all previously allowed values of the mass parameters for partially massless and heavy scalar fields.  From this relative change one can also study the large or small deviation from the Bunch Davies result compared to the other results obtained from all non standard quantum vacuum states.  

\item Also, we have to mention here that the computed four-point auto-correlated field and momentum OTO functions are independent of any specific choice of the coordinate system.  If we go through the computation then we see that even we have started defining the four-point auto-correlated field and momentum OTO functions  in real space,  after taking Fourier transformation and applying the momentum conservation appearing in terms of Dirac Delta functions, the exponential factor which capture all the momenta and real space coordinate will be unity.  Finally,  we get a momentum integrated cut-off regulated results for the four-point auto-correlated field and momentum OTO functions which only depend on the two dynamical conformal time scales on which we have defined the rescaled perturbation variable and its canonically conjugate momenta. This is quite good outcome.  In the generalised prescriptions,  when one introduces the quantum operators in a specific time and coordinate system it always appears that in the final result it captures the effect of inhomogeneity in the space-time coordinate.  In the context of spatially flat FLRW cosmology we have found that the final result of four-point auto-correlated field and momentum OTO functions  captures the effect of inhomogeneity.  This is because in cosmology we are actually interested in the time evolution of the quantum operators which are separated in time scale,  which is one of the crucial constraint in the definition of any general four-point auto-correlated field and momentum OTO functions. But such four-point auto-correlated field and momentum OTO functions  don't capture the inhomogeneity. 

\item The final result of four-point auto-correlated field and momentum OTO functions are not explicitly dependent of the factor $\beta=1/T$ which is appearing in the expression for thermal partition function.  This further implies that the final results are also independent of the thermal partition function computed for Primordial Cosmology. 

 \end{itemize} 

 %%%%%%%%%%%%%%%%%%%%%%%%%
 \section{Classical limit of non-chaotic auto-correlated OTO amplitudes and OTOC in Primordial Cosmology}
 \label{sec:5}

\subsection{Computational strategy for non-chaotic auto-correlated OTO functions in the Classical limit}
In this section,  our prime objective is to study the classical limit of the four-point auto-correlation functions that we have explicitly derived from the quantum descriptions.  In the following subsections we will derive the classical limiting results which will be consistent with the super-horizon late time limiting behaviour of the cosmological auto-correlated functions.

In this subsection we will illustrate our computational strategy to study the classical limit of the cosmological four-point OTOC derived in this paper:
\begin{enumerate}
\item First of all,  we have to take the quantum to classical map.  For this purpose we consider the classical mode function and its canonically conjugate momentum which we have derived in the earlier section of this paper.

\item Next,  in the classical limit we use the {\it Poisson Brackets} of the classical mode function and its canonically conjugate momentum variables in cosmological perturbation theory.

\item In the classical limit the definition of quantum trace will be replaced by phase space volume measure,  $\displaystyle \frac{\displaystyle df_{\bf k}(\tau)d\Pi_{\bf k}(\tau)}{\displaystyle 2\pi}$ which mimics the role of path integral measure,  $\displaystyle\frac{{\cal D}f_{\bf k}(\tau){\cal D}\Pi_{\bf k}(\tau)}{2\pi}$  in the classical limit.  

\item Also during this computation we have to include an additional thermal Boltzmann factor which is serving the purpose of thermal weight factor during taking the phase space average over the classical micro-canonical statistical ensemble.

\item  Next,  we compute the expression for the classical thermal partition function for cosmological perturbations which is consistent with the quantum result computed from completely different formalism.  To compute the classical partition function one need to compute the expression by following the principles of classical statistical mechanics. very carefully.  Though we will show that the final expressions for the classical limit of auto-correlated OTOs are independent of the partition function for the cosmological scenario in which we are interested in.

\item Last but not the least,  we compute the expression for the normalisation factor for the classical limiting version of the auto-correlated OTO functions. by following the above mentioned general formalism.
\end{enumerate}
\subsection{Classical limit of cosmological two-point ``in-in" non-chaotic OTO  amplitudes}
To compute the classical limit we start with the {\it Poisson bracket} of these cosmologically relevant canonically conjugate operators, which are given by the following expressions:
\bea \left\{f({\bf x},\tau_1),f({\bf x},\tau_2)\right\}_{\bf PB}&=&\int\frac{d^3{\bf k}_1}{(2\pi)^3}\int\frac{d^3{\bf k}_2}{(2\pi)^3}~\exp(i({\bf k}_1+{\bf k}_2).{\bf x})~\nonumber\\
&&\left[\left\{f_{{\bf k}_1}(\tau_1),f_{{\bf k}_2}(\tau_2)\right\}_{\bf PB}+\left\{f_{{\bf k}_2}(\tau_1),f_{{\bf k}_1}(\tau_2)\right\}_{\bf PB}\right],~~~~~~~\eea\bea\left\{\Pi({\bf x},\tau_1),\Pi({\bf x},\tau_2)\right\}_{\bf PB}&=&\int\frac{d^3{\bf k}_1}{(2\pi)^3}\int\frac{d^3{\bf k}_2}{(2\pi)^3}~\exp(i({\bf k}_1+{\bf k}_2).{\bf x})~\nonumber\\
&&\left[\left\{\Pi_{{\bf k}_1}(\tau_1),\Pi_{{\bf k}_2}(\tau_2)\right\}_{\bf PB}+\left\{\Pi_{{\bf k}_2}(\tau_1),\Pi_{{\bf k}_1}(\tau_2)\right\}_{\bf PB}\right].~~~~~~~\eea
In the Fourier transformed space,  the individual {\it Poisson brackets} are given by the following expressions:
\bea \left\{f_{{\bf k}_1}(\tau_1),f_{{\bf k}_2}(\tau_2)\right\}_{\bf PB}&=&(2\pi)^3\delta^{3}({\bf k}_1+{\bf k}_2)~{\bf R}_1(\tau_1,\tau_2),\\
 \left\{f_{{\bf k}_2}(\tau_1),f_{{\bf k}_1}(\tau_2)\right\}_{\bf PB}&=&(2\pi)^3\delta^{3}({\bf k}_2+{\bf k}_1)~{\bf R}_1(\tau_1,\tau_2), \\ 
\left\{\Pi_{{\bf k}_1}(\tau_1),\Pi_{{\bf k}_2}(\tau_2)\right\}_{\bf PB}&=&(2\pi)^3\delta^{3}({\bf k}_1+{\bf k}_2)~{\bf R}_2(\tau_1,\tau_2),\\
 \left\{\Pi_{{\bf k}_2}(\tau_1),\Pi_{{\bf k}_1}(\tau_2)\right\}_{\bf PB}&=&(2\pi)^3\delta^{3}({\bf k}_2+{\bf k}_1)~{\bf R}_2(\tau_1,\tau_2),  \eea
 which further implies that following symmetric result:
 \bea \left\{f_{{\bf k}_1}(\tau_1),f_{{\bf k}_2}(\tau_2)\right\}_{\bf PB}&=&\left\{f_{{\bf k}_2}(\tau_1),f_{{\bf k}_1}(\tau_2)\right\}_{\bf PB}=(2\pi)^3\delta^{3}({\bf k}_1+{\bf k}_2)~{\bf R}_1(\tau_1,\tau_2),\\
  \left\{\Pi_{{\bf k}_1}(\tau_1),\Pi_{{\bf k}_2}(\tau_2)\right\}_{\bf PB}&=&\left\{\Pi_{{\bf k}_2}(\tau_1),\Pi_{{\bf k}_1}(\tau_2)\right\}_{\bf PB}=(2\pi)^3\delta^{3}({\bf k}_1+{\bf k}_2)~{\bf R}_2(\tau_1,\tau_2),\eea
which is appearing due to the fact that the three dimensional Dirac Delta function is symmetric under the exchange of two different momenta.
% \begin{figure*}[htb]
 % \includegraphics[width=17cm,height=8.9cm]{C2T5.pdf}
 % \caption{Behaviour of the four-point auto-correlated momentum OTO function with respect to the time scale $T$ for Motta Allen,  $\alpha$ vacua and Bunch Davies vacuum with mass parameter $|\nu|=9/2$.} 
  %\label{fig:52}
%\end{figure*} 
Now for the further purpose we define the two-point random classical correlation function ${\bf R}_1(\tau_1,\tau_2)$ and ${\bf R}_2(\tau_1,\tau_2)$ by the following expressions:
\bea {\bf R}_1(\tau_1,\tau_2):&=&{\bf W}_1(\tau_1-\tau_2)\\
{\bf R}_2(\tau_1,\tau_2):&=&{\bf W}_2(\tau_1-\tau_2),\eea 
where ${\bf W}_1(\tau_1-\tau_2)$ and ${\bf W}_2(\tau_1-\tau_2)$ are the two window functions which are defined as: 
\bea {\bf W}_1(\tau_1-\tau_2)&=&\sqrt{\langle \eta_{\bf Noise}(\tau_1)\eta_{\bf Noise}(\tau_2)\rangle},\\ 
{\bf W}_2(\tau_1-\tau_2)&=&\sqrt{\langle \Pi_{\eta_{\bf Noise}}(\tau_1)\Pi_{\eta_{\bf Noise}}(\tau_2)\rangle}.\eea
where the two-point noise and its associate canonically conjugate momentum correlation functions at the classical level are given by the following expressions:
\bea &&\langle \eta_{\bf Noise}(\tau_1)\eta_{\bf Noise}(\tau_2)\rangle:=\sqrt{{\bf G}^{(1)}_{\bf kernel}(\tau_1-\tau_2)}=\sqrt{{\bf G}^{(1)}_{\bf kernel}(\tau_2-\tau_1)}\equiv \sqrt{{\bf G}^{(1)}_{\bf kernel}(|\tau_1-\tau_2|)},~~~~~~~~~~\\ &&\langle \Pi_{\eta_{\bf Noise}}(\tau_1) \Pi_{\eta_{\bf Noise}}(\tau_2)\rangle:=\sqrt{{\bf G}^{(2)}_{\bf kernel}(\tau_1-\tau_2)}=\sqrt{{\bf G}^{(2)}_{\bf kernel}(\tau_2-\tau_1)}\equiv \sqrt{{\bf G}^{(2)}_{\bf kernel}(|\tau_1-\tau_2|)}.\eea
where ${\bf G}^{(1)}_{\bf kernel}(\tau_1-\tau_2)$ and ${\bf G}^{(2)}_{\bf kernel}(\tau_1-\tau_2)$ are known as the noise kernel which are both time translational symmetric Green's functions.  As a consequence,  we get the following symmetry properties in the classical correlation functions:
\bea {\bf R}_1(\tau_1,\tau_2)&=&{\bf R}_1(\tau_2,\tau_1)\equiv \sqrt{{\bf G}^{(1)}_{\bf kernel}(|\tau_1-\tau_2|)},\\
{\bf R}_2(\tau_1,\tau_2)&=&{\bf R}_2(\tau_2,\tau_1)\equiv \sqrt{{\bf G}^{(2)}_{\bf kernel}(|\tau_1-\tau_2|)}.\eea
%\begin{figure*}[htb]
%  \includegraphics[width=17cm,height=8.9cm]{CTaC3.pdf}
%  \caption{Behaviour of the four-point auto-correlated momentum OTO function with respect to the time scale $\tau$ for Motta Allen,  $\alpha$ vacua and Bunch Davies vacuum with mass parameter $|\nu|=5/2$.} 
 % \label{fig:55}
%\end{figure*}  
%\begin{figure*}[htb]
 % \includegraphics[width=17cm,height=8.9cm]{CTaC4.pdf}
 % \caption{Behaviour of the four-point auto-correlated momentum OTO function with respect to the time scale $\tau$ for Motta Allen,  $\alpha$ vacua and Bunch Davies vacuum with mass parameter $|\nu|=7/2$.} 
%  \label{fig:56}
%\end{figure*}  
%\begin{figure*}[htb]
  %\includegraphics[width=17cm,height=8.9cm]{CTaC5.pdf}
 % \caption{Behaviour of the four-point auto-correlated momentum OTO function with respect to the time scale $\tau$ for Motta Allen,  $\alpha$ vacua and Bunch Davies vacuum with mass parameter $|\nu|=9/2$.} 
 % \label{fig:57}
%\end{figure*}  

%\begin{figure*}[htb]
%  \includegraphics[width=17cm,height=8.9cm]{YV33.pdf}
 % \caption{Behaviour of the two-point auto-correlated momentum OTO function with respect to the mass parameter $|\nu|$ for Motta Allen vacua. }  
 % \label{fig:62}
%\end{figure*}  
%\begin{figure*}[htb]
%  \includegraphics[width=17cm,height=8.9cm]{YTV2.pdf}
%  \caption{Behaviour of the two-point auto-correlated momentum OTO function with respect to the time scale $T$ for Bunch Davies vacuum. }  
 % \label{fig:63}
%\end{figure*}  
%\begin{figure*}[htb]
 % \includegraphics[width=17cm,height=8.9cm]{YTV3.pdf}
 % \caption{Behaviour of the two-point auto-correlated momentum OTO function with respect to the time scale $T$ for $\alpha$ vacua. }  
 % \label{fig:64}
%\end{figure*}  
%\begin{figure*}[htb]
%  \includegraphics[width=17cm,height=8.9cm]{Gx1.pdf}
 % \caption{Behaviour of the two-point auto-correlated momentum OTO function with respect to the vacuum parameter $\alpha$ of $\alpha$ vacua. }  
 % \label{fig:66}
%\end{figure*}  
 %\begin{figure*}[htb]
 % \includegraphics[width=17cm,height=8.9cm]{Gx2.pdf}
%  \caption{Behaviour of the two-point auto-correlated momentum OTO function with respect to the vacuum parameter $\gamma$ of Mota Allen vacua. }  
 % \label{fig:67}
%\end{figure*}  
\subsection{Classical limit of cosmological four-point ``in-in" non-chaotic OTO  amplitudes}
Further,  in the classical limit we compute the following two types of the square of the Poisson brackets instead of computing the commutator brackets which is the key ingredient in the quantum calculation,  given by: 
  \bea  &&\left\{{f}({\bf x},\tau_1),{f}({\bf x},\tau_2)\right\}^2_{\bf PB}=\left\{{f}({\bf x},\tau_1),{f}({\bf x},\tau_2)\right\}_{\bf PB}\left\{{f}({\bf x},\tau_1),{f}({\bf x},\tau_2)\right\}_{\bf PB},\\
  && \left\{{\Pi}({\bf x},\tau_1),{\Pi}({\bf x},\tau_2)\right\}^2_{\bf PB}=\left\{{\Pi}({\bf x},\tau_1),{\Pi}({\bf x},\tau_2)\right\}_{\bf PB}\left\{{\Pi}({\bf x},\tau_1),{\Pi}({\bf x},\tau_2)\right\}_{\bf PB}.\eea 
 
To compute the above mentioned expression we need to use the following Fourier transformations:
 \bea &&\hat{f}({\bf x},\tau_1)=\int \frac{d^3{\bf k}}{(2\pi)^3}~\exp(i{\bf k}.{\bf x})~\hat{f}_{{\bf k}}(\tau_1),\\
 &&\hat{\Pi}({\bf x},\tau_1)=\partial_{\tau_1}\hat{f}({\bf x},\tau_1)=\int \frac{d^3{\bf k}}{(2\pi)^3}~\exp(i{\bf k}.{\bf x})~\partial_{\tau_1}\hat{f}_{{\bf k}}(\tau_1)=\int \frac{d^3{\bf k}}{(2\pi)^3}~\exp(i{\bf k}.{\bf x})~\hat{\Pi}_{{\bf k}}(\tau_1),~~~~~~~~~~~~~\eea
 which will be extremely useful for the computation of the classical limiting results of the previously mentioned two specific type of four-point OTOCs in terms of the square of the Poisson brackets.
 
Using these inputs we get the following simplified results in the classical limit:
 \bea && \left\{{f}({\bf x},\tau_1),{f}({\bf x},\tau_2)\right\}^2_{\bf PB}=(2\pi)^6 \int \frac{d^3{\bf k}_1}{(2\pi)^3}\int \frac{d^3{\bf k}_2}{(2\pi)^3}\int \frac{d^3{\bf k}_3}{(2\pi)^3}\int \frac{d^3{\bf k}_4}{(2\pi)^3}~\exp\left(i({\bf k}_1+{\bf k}_2+{\bf k}_3+{\bf k}_4).{\bf x}\right)\nonumber\\
 &&~~~~~~~~~~~~~~~~~~~~~~~~~\left[\delta^{3}({\bf k}_1+{\bf k}_2)\delta^3({\bf k}_3+{\bf k}_4)+\delta^{3}({\bf k}_1+{\bf k}_3)\delta^3({\bf k}_2+{\bf k}_4)\right.\nonumber\\&& \left.~~~~~~~~~~~~~~~~~~~~~~~~~+\delta^{3}({\bf k}_1+{\bf k}_4)\delta^3({\bf k}_3+{\bf k}_2)+\delta^{3}({\bf k}_2+{\bf k}_3)\delta^3({\bf k}_4+{\bf k}_1)\right.\nonumber\\&& \left.~~~~~~~~~~~~~~~~~~~~~~~~~+\delta^{3}({\bf k}_2+{\bf k}_1)\delta^3({\bf k}_4+{\bf k}_3)+\delta^{3}({\bf k}_2+{\bf k}_4)\delta^3({\bf k}_1+{\bf k}_3)\right.\nonumber\\&& \left.~~~~~~~~~~~~~~~~~~~~~~~~~+\delta^{3}({\bf k}_3+{\bf k}_1)\delta^3({\bf k}_4+{\bf k}_2)+\delta^{3}({\bf k}_3+{\bf k}_2)\delta^3({\bf k}_1+{\bf k}_4)\right.\nonumber\\&& \left.~~~~~~~~~~~~~~~~~~~~~~~~~+\delta^{3}({\bf k}_3+{\bf k}_4)\delta^3({\bf k}_1+{\bf k}_2)+\delta^{3}({\bf k}_4+{\bf k}_1)\delta^3({\bf k}_2+{\bf k}_3)\right.\nonumber\\&& \left.~~~~~~~~~~~~~~~~~~~~~~~~~+\delta^{3}({\bf k}_4+{\bf k}_2)\delta^3({\bf k}_3+{\bf k}_1)+\delta^{3}({\bf k}_4+{\bf k}_3)\delta^3({\bf k}_2+{\bf k}_1)\right]{\bf G}^{(1)}_{\bf Kernel}(|\tau_1-\tau_2|).~~~~~~~~~~~~\eea
 \bea && \left\{{\Pi}({\bf x},\tau_1),{\Pi}({\bf x},\tau_2)\right\}^2_{\bf PB}=(2\pi)^6 \int \frac{d^3{\bf k}_1}{(2\pi)^3}\int \frac{d^3{\bf k}_2}{(2\pi)^3}\int \frac{d^3{\bf k}_3}{(2\pi)^3}\int \frac{d^3{\bf k}_4}{(2\pi)^3}~\exp\left(i({\bf k}_1+{\bf k}_2+{\bf k}_3+{\bf k}_4).{\bf x}\right)\nonumber\\
 &&~~~~~~~~~~~~~~~~~~~~~~~~~\left[\delta^{3}({\bf k}_1+{\bf k}_2)\delta^3({\bf k}_3+{\bf k}_4)+\delta^{3}({\bf k}_1+{\bf k}_3)\delta^3({\bf k}_2+{\bf k}_4)\right.\nonumber\\&& \left.~~~~~~~~~~~~~~~~~~~~~~~~~+\delta^{3}({\bf k}_1+{\bf k}_4)\delta^3({\bf k}_3+{\bf k}_2)+\delta^{3}({\bf k}_2+{\bf k}_3)\delta^3({\bf k}_4+{\bf k}_1)\right.\nonumber\\&& \left.~~~~~~~~~~~~~~~~~~~~~~~~~+\delta^{3}({\bf k}_2+{\bf k}_1)\delta^3({\bf k}_4+{\bf k}_3)+\delta^{3}({\bf k}_2+{\bf k}_4)\delta^3({\bf k}_1+{\bf k}_3)\right.\nonumber\\&& \left.~~~~~~~~~~~~~~~~~~~~~~~~~+\delta^{3}({\bf k}_3+{\bf k}_1)\delta^3({\bf k}_4+{\bf k}_2)+\delta^{3}({\bf k}_3+{\bf k}_2)\delta^3({\bf k}_1+{\bf k}_4)\right.\nonumber\\&& \left.~~~~~~~~~~~~~~~~~~~~~~~~~+\delta^{3}({\bf k}_3+{\bf k}_4)\delta^3({\bf k}_1+{\bf k}_2)+\delta^{3}({\bf k}_4+{\bf k}_1)\delta^3({\bf k}_2+{\bf k}_3)\right.\nonumber\\&& \left.~~~~~~~~~~~~~~~~~~~~~~~~~+\delta^{3}({\bf k}_4+{\bf k}_2)\delta^3({\bf k}_3+{\bf k}_1)+\delta^{3}({\bf k}_4+{\bf k}_3)\delta^3({\bf k}_2+{\bf k}_1)\right]{\bf G}^{(2)}_{\bf Kernel}(|\tau_1-\tau_2|).~~~~~~~~~~\eea  
 where we have used the following crucial facts:
 \bea &&\left\{f_{{\bf k}_{i}} (\tau_1),f_{{\bf k}_{j}} (\tau_2)\right\}_{\bf PB}=(2\pi)^3\delta^{3}({\bf k}_i+{\bf k}_j)\sqrt{{\bf G}^{(1)}_{\bf Kernel}(|\tau_1-\tau_2|)},\nonumber\\
 &&\left\{\Pi_{{\bf k}_{i}} (\tau_1),\Pi_{{\bf k}_{j}} (\tau_2)\right\}_{\bf PB}=(2\pi)^3\delta^{3}({\bf k}_i+{\bf k}_j)\sqrt{{\bf G}^{(2)}_{\bf Kernel}(|\tau_1-\tau_2|)}.\nonumber\\
&&~~~~~~~~~~~~~~~~~~~~~~~~~~~~~~~~~~~~~~~~~~~~~~~~~~~~~~~~{\rm where}~~~i\neq j~ \forall~ i,j,=1,2,3,4.~~~~~~~~~~~\eea 

It is important to point that, the individual contributions appearing in the previously mentioned $12$ contributions can be evaluated as:
\bea &&\left\{f_{{\bf k}_{i}} (\tau_1),f_{{\bf k}_{j}} (\tau_2)\right\}_{\bf PB}\left\{f_{{\bf k}_{l}}(\tau_1),f_{{\bf k}_m} (\tau_2)\right\}_{\bf PB}=(2\pi)^6\delta^{3}({\bf k}_i+{\bf k}_j)\delta^3({\bf k}_l+{\bf k}_m){\bf G}^{(1)}_{\bf Kernel}(|\tau_1-\tau_2|),\nonumber\\
&&\left\{\Pi_{{\bf k}_{i}} (\tau_1),\Pi_{{\bf k}_{j}} (\tau_2)\right\}_{\bf PB}\left\{\Pi_{{\bf k}_{l}}(\tau_1),\Pi_{{\bf k}_m} (\tau_2)\right\}_{\bf PB}=(2\pi)^6\delta^{3}({\bf k}_i+{\bf k}_j)\delta^3({\bf k}_l+{\bf k}_m){\bf G}^{(2)}_{\bf Kernel}(|\tau_1-\tau_2|).\nonumber\\
&&~~~~~~~~~~~~~~~~~~~~~~~~~~~~~~~~~~~~~~~~~~~~{\rm where}~~~i\neq j \neq l \neq m~ \forall~ i,j,l,m=1,2,3,4.~~~~~~~~~~~\eea
In this computation we introduce two conformal time dependent coloured noise kernels ${\bf G}^{(1)}_{\bf Kernel}(|\tau_1-\tau_2|)$ and ${\bf G}^{(2)}_{\bf Kernel}(|\tau_1-\tau_2|)$,  which are defined as:
\bea
&&{\bf G}^{(1)}_{\bf Kernel}(|\tau_1-\tau_2|)={\bf W}^2_1(\tau_1-\tau_2),\\
&& {\bf G}^{(2)}_{\bf Kernel}(|\tau_1-\tau_2|)={\bf W}^2_2(\tau_1-\tau_2).
\eea
On the other hand,  if we consider the Gaussian white noise we have the following properties:
\begin{enumerate} 
\item For white noise two-point classical auto correlation functions are time translation invariant, which are kind of expected from this analysis.

\item Random white noise also have the following properties:
\bea &&\langle \eta_{\bf WN}(\tau_1)\rangle=0,\\
&&\langle \eta_{\bf WN}(\tau_1)\eta_{\bf WN}(\tau_2)\rangle={\bf G}^{(1)}_{\bf Kernel}(|\tau_1-\tau_2|)={\bf W}^2_1(|\tau_1-\tau_2|)\rangle={\bf B}_1~\delta(|\tau_1-\tau_2|),\\
&&\langle  \eta_{\bf WN}(\tau_1)\cdots\cdots \eta_{\bf WN}(\tau_N)  \rangle=0~~~~\forall~N=3,5,7,\cdots,\\
&&\langle  \eta_{\bf WN}(\tau_1)\cdots\cdots \eta_{\bf WN}(\tau_N)  \rangle={\bf C}^{(1)}_{\bf kernel}(\tau_1,\cdots,\tau_N)~\delta(\tau_1+\cdots\cdots+\tau_{N})~~~~\forall~N=4,6,8,\cdots.~~~~~~~~\eea
and 
\bea &&\langle \Pi_{\eta_{\bf WN}}(\tau_1)\rangle=0,\\
&&\langle \Pi_{\eta_{\bf WN}}(\tau_1)\Pi_{\eta_{\bf WN}}(\tau_2)\rangle={\bf G}^{(2)}_{\bf Kernel}(|\tau_1-\tau_2|)={\bf W}^2_2(|\tau_1-\tau_2|)\rangle={\bf B}_2~\delta(|\tau_1-\tau_2|),\\
&&\langle \Pi_{\eta_{\bf WN}}(\tau_1)\cdots\cdots \Pi_{\eta_{\bf WN}}(\tau_N)  \rangle=0~~~~\forall~N=3,5,7,\cdots,\\
&&\langle  \Pi_{\eta_{\bf WN}}(\tau_1)\cdots\cdots\Pi_{\eta_{\bf WN}}(\tau_N)  \rangle={\bf C}^{(2)}_{\bf kernel}(\tau_1,\cdots,\tau_N)~\delta(\tau_1+\cdots\cdots+\tau_{N})~~~~\forall~N=4,6,8,\cdots.~~~~~~~~\eea
where $\eta_{\bf WN}(\tau)$ and $\Pi_{\eta_{\bf WN}}$ are the conformal time dependent white noise field variable and its canonically conjugate momentum variable.  Also,  ${\bf B}_1, {\bf B}_2~(\neq {\bf B}_1)$ and ${\bf C}^{(1)}_{\bf kernel}(\tau_1,\cdots,\tau_N)$,  ${\bf C}^{(2)}_{\bf kernel}(\tau_1,\cdots,\tau_N)$ are the amplitudes of the spectra of any $N=2$ and $N=4,6,8,\cdots$ even point classical auto correlation functions.  Here all even point amplitudes are non-zero and all odd point amplitudes become zero for Gaussian white noise contributions,  which are again sourced from random fluctuations.
\end{enumerate}
From the mentioned results further we get the following features:
\begin{enumerate}

\item Here time dependent noise kernels describe the randomness of quantum mechanical fluctuations at the classical limit.

\item If we compare the obtained result in the classical limit with the quantum version of the four-point auto-correlated results then it is clearly observed that for both the cases we get same twelve contributions in the classical limit. But in the quantum case we have different individual contributions for these mentioned twelve terms in the auto-correlations.  

\item Also in both classical and quantum results momentum conservation is established via momentum dependent three dimensional Dirac Delta function contributions.
\end{enumerate}
\subsection{Cosmological  partition function: Classical version}
\subsubsection{Classical  partition function in terms of rescaled field variable} 
In this section our objective is to compute the partition function in the classical regime.  In terms of the rescaled perturbation field variable and its canonically conjugate momentum we define the following classical partition function: 
\bea Z_{\bf Classical}(\beta;\tau_1):&=&\int \int \frac{{\cal D}f{\cal D}\Pi}{2\pi}~\exp\left(-\beta H\right)\nonumber\\
&=&\prod_{{\bf k}}\exp\left(-\beta\left[\frac{E_{\bf k}(\tau_1)}{2}+\frac{1}{\beta}\ln\left(1-\exp(-\beta E_{\bf k}(\tau_1)\right)\right]\right)\nonumber\\
&=&\prod_{{\bf k}}\exp\left(-\ln\left(2\sinh\frac{\beta E_{\bf k}(\tau_1)}{2}\right)\right)\nonumber\\
&=&\exp\left(-\int d^3{\bf k} ~\ln\left(2\sinh \frac{\beta E_{\bf k}(\tau_1)}{2}\right)\right).~~~~\eea
Now if we compare the above result with the quantum result obtained in the previous section,  then we get the following interpretation:
\bea {\underbrace{Z_{\bf Classical}(\beta;\tau_1)}_{\textcolor{red}{\bf Classical~result}}=\underbrace{:Z_{\bf BD}(\beta;\tau_1):=|\cosh\alpha|~:Z_{\alpha}(\beta;\tau_1):=|\cosh\alpha|~\exp(2\sin\gamma {\rm tan}\alpha)~:Z_{\alpha,\gamma}(\beta;\tau_1):}_{\textcolor{red}{\bf Quantum~result}}}~.\nonumber\\
\eea
The good part is that,  from the above expression we found that the expression for the classical partition function and normal ordered partition function for Cosmology computed using quantum techniques are exactly same. 
This is somewhat expected from the basic understanding of the present set-up which we are considering in the present context of discussion and also helps us a lot to correctly take the classical limit of the result computed using quantum field theory techniques.  In short,  the above interpretation shows that the classical result of the partition function that we have separately computed is perfectly consistent with the quantum result.
\subsubsection{Classical  partition function in terms of curvature perturbation field variable}
In this section our objective is to compute the partition function in the classical regime.  In terms of the curvature perturbation field variable and its canonically conjugate momentum we define the following classical partition function: 
\bea Z^{\zeta}_{\bf Classical}(\beta;\tau_1):&=&\int \int \frac{{\cal D}\zeta{\cal D}\Pi_{\zeta}}{2\pi}~\exp\left(-\beta H\right)\nonumber\\
&=&\prod_{{\bf k}}\exp\left(-\beta\left[\frac{z^2(\tau_1)E^{\zeta}_{\bf k}(\tau_1)}{2}+\frac{1}{\beta}\ln\left(1-\exp(-\beta z^2(\tau_1)E^{\zeta}_{\bf k}(\tau_1)\right)\right]\right)\nonumber\\
&=&\exp\left(-\int d^3{\bf k} ~\ln\left(2\sinh \frac{\beta z^2(\tau_1)E^{\zeta}_{\bf k}(\tau_1)}{2}\right)\right).~~~~\eea
Now if we compare the above result with the quantum result obtained in the previous section,  then we get the following interpretation:
\bea &&{\underbrace{Z^{\zeta}_{\bf Classical}(\beta;\tau_1)}_{\textcolor{red}{\bf Classical~result}}=\underbrace{:Z^{\zeta}_{\bf BD}(\beta;\tau_1):=|\cosh\alpha|~:Z^{\zeta}_{\alpha}(\beta;\tau_1):=|\cosh\alpha|~\exp(2\sin\gamma {\rm tan}\alpha)~:Z^{\zeta}_{\alpha,\gamma}(\beta;\tau_1):}_{\textcolor{red}{\bf Quantum~result}}}~.\nonumber\\
&&~~~~~~~~~~~~~~~~~~~~~{\neq Z_{\bf Classical}(\beta;\tau_1)}~.\eea

\subsection{Classical limit of cosmological two-point   non-chaotic OTOC: rescaled field version}

Next,  we will explicitly compute the previously mentioned two types of two-point OTOC using the above mentioned results.  In the classical limit,  the prescribed two-point functions are given by the following expressions:
\bea Y^{f}_{1,{\bf Classical}}(\tau_1,\tau_2)&=&\frac{1}{Z_{\bf Classical}(\beta;\tau_1)}\int \int \frac{{\cal D}f {\cal D}\Pi}{2\pi}\exp\left(-\beta~H\right)~\left\{{f}({\bf x},\tau_1),{f}({\bf x},\tau_2)\right\}_{\bf PB}\nonumber\\
&=&2~{\bf W}_1(\tau_1-\tau_2)~\int \frac{d^3{\bf k}_1}{(2\pi)^3},\\
Y^{f}_{2,{\bf Classical}}(\tau_1,\tau_2)&=&\frac{1}{Z_{\bf Classical}(\beta;\tau_1)}\int \int \frac{{\cal D}f {\cal D}\Pi}{2\pi}\exp\left(-\beta~H\right)~\left\{{\Pi}({\bf x},\tau_1),{\Pi}({\bf x},\tau_2)\right\}_{\bf PB}\nonumber\\
&=&2~{\bf W}_2(\tau_1-\tau_2)~\int \frac{d^3{\bf k}_1}{(2\pi)^3}.\eea
This result is divergent because of the presence of term volume in general.  To get the finite contribution in the classical limit we regulate the momentum integrals by putting the cut-off scale $L$, for which the momentum range are given by, $0<k_1<L$. By applying this regulator we get the following regulated results:
\bea \int \frac{d^3{\bf k}_1}{(2\pi)^3}=\frac{1}{2\pi^2}\int^{L}_{k_1=0}k^2_1~dk_1=\frac{L^3}{6\pi^2}~.\eea
After substituting the above regulated volume factor finally we get:
\bea {Y^{f}_{1,{\bf Classical}}(\tau_1,\tau_2)=\frac{L^3}{3\pi^2}{\bf W}_1(\tau_1-\tau_2)=\frac{L^3}{3\pi^2}\sqrt{{\bf G}^{(1)}_{\bf Kernel}(|\tau_1-\tau_2|)}}~,\\
 {Y^{f}_{2,{\bf Classical}}(\tau_1,\tau_2)=\frac{L^3}{3\pi^2}{\bf W}_2(\tau_1-\tau_2)=\frac{L^3}{3\pi^2}\sqrt{{\bf G}^{(2)}_{\bf Kernel}(|\tau_1-\tau_2|)}}~.\eea
Here the regularised volume factor, $L^3/3\pi^2$,  is the two-point time independent amplitude of the cosmological two-point OTOCs in the classical limiting approximation.  From the above mentioned result one can consider a situation when we have $\tau_1=\tau_2=\tau$ in the classical limiting situation. In that case,  we get further simplified results for two-point functions which are given by the following expressions:
\bea {Y^{f}_{1,{\bf Classical}}(\tau,\tau)=\frac{L^3}{3\pi^2}~{\bf W}_1(0)=\frac{L^3}{3\pi^2}\sqrt{{\bf G}^{(1)}_{\bf Kernel}(0)}}~,\eea\bea
{Y^{f}_{2,{\bf Classical}}(\tau,\tau)=\frac{L^3}{3\pi^2}~{\bf W}_2(0)=\frac{L^3}{3\pi^2}\sqrt{{\bf G}^{(2)}_{\bf Kernel}(0)}}~.~\eea
This result only exists when the window functions,  ${\bf W}_1(0)$ and ${\bf W}_2(0)$, and the corresponding Green's functions ${\bf G}^{(1)}_{\bf Kernel}(0)$ and ${\bf G}^{(2)}_{\bf Kernel}(0)$ are finite in the classical limiting approximation to describe the coloured non-Gaussian noise and white Gaussian noise respectively.

Now to demonstrate the explicit role of a non-Gaussian coloured noise and Gaussian white noise in the present context of discussion one can further consider the following mathematical structures of the corresponding Green's functions/window functions:
\begin{eqnarray}
&& {{\bf W}_{1}(\tau_1-\tau_2)= \large \left\{
     \begin{array}{lr}
   \displaystyle \sqrt{\frac{{\bf A_1}}{\gamma_1}}~\exp\left(-\frac{\gamma_1}{2}|\tau_1-\tau_2|\right), & \text{\textcolor{red}{\bf Coloured~Noise}}\\  
   \displaystyle   \lim_{{\bf C}_1\rightarrow 0}\sqrt{\frac{{\bf B}_1}{|{\bf C}_1|\sqrt{\pi}}}~\exp\left(-\frac{|\tau_1-\tau_2|^2}{{{\bf C}^2_1}}\right) & \text{\textcolor{red}{\bf White~Noise}}  \end{array}
   \right.}~~~~~
\\
&& {{\bf W}_{2}(\tau_1-\tau_2)= \large \left\{
     \begin{array}{lr}
   \displaystyle \sqrt{\frac{{\bf A_2}}{\gamma_2}}~\exp\left(-\frac{\gamma_2}{2}|\tau_1-\tau_2|\right), & \text{\textcolor{red}{\bf Coloured~Noise}}\\  
   \displaystyle   \lim_{{\bf C}_2\rightarrow 0}\sqrt{\frac{{\bf B}_2}{|{\bf C}_2|\sqrt{\pi}}}~\exp\left(-\frac{|\tau_1-\tau_2|^2}{{{\bf C}^2_2}}\right) & \text{\textcolor{red}{\bf White~Noise}}  \end{array}
   \right.}~~~~~
\end{eqnarray}
where ${\bf A}_1, {\bf A}_2$ and ${\bf B}_1,{\bf B}_2$ and ${\bf C}_1,{\bf C}_2$,  represent the conformal time independent amplitudes of the coloured and white random classical noise respectively in the present context.  Also,  $\gamma_1$ and $\gamma_2$,  represent the interaction strength of the dissipation in the context of coloured noise in classical regime.

Now from the general mathematical structure of the white noise it is evident that,  ${\bf G}^{(1)}_{\bf Kernel}(0)\rightarrow \infty$ and ${\bf G}^{(2)}_{\bf Kernel}(0)\rightarrow \infty$ is giving  diverging contribution for $\tau_1=\tau_2=\tau$ case.  So appearance of the possibility of the equal time limit is completely discarded as it gives overall diverging contribution in the classical limit of the two-point functions in the present context.  On the other hand,  in the equal time limit we have:
 \bea &&{{\bf G}^{(1)}_{\bf Kernel}(0)=\frac{{\bf A}_1}{\gamma_1}={\bf W}^2_1(0)}~,\\ 
&& {{\bf G}^{(2)}_{\bf Kernel}(0)=\frac{{\bf A}_2}{\gamma_2}={\bf W}^2_2(0)}~\eea for the coloured noise case.  This implies for coloured noise equal time limit exists and one can write down the following simplified expression for the classical limit of the two-point function as:
\bea {Y^{f}_{1,{\bf Classical}}(\tau,\tau)=\frac{L^3}{3 \pi^2}\sqrt{\frac{{\bf A}_1}{\gamma_1}}}~,\\
 {Y^{f}_{2,{\bf Classical}}(\tau,\tau)=\frac{L^3}{3 \pi^2}\sqrt{\frac{{\bf A}_2}{\gamma_2}}}~.
~\eea
Similarly for the equal time limit  case with white noise profile we have:
 \bea &&{{\bf W}_{1}(0)=\lim_{{\bf C}_1\rightarrow 0}\frac{{\bf B}_1}{|{\bf C}_1|\sqrt{\pi}}}~,\\ 
&& {{\bf W}_{2}(0)=\lim_{{\bf C}_2\rightarrow 0}\frac{{\bf B}_2}{|{\bf C}_2|\sqrt{\pi}}}~\eea for the coloured noise case.  This implies for coloured noise equal time limit exists and one can write down the following simplified expression for the classical limit of the two-point function as:
\bea {Y^{f}_{1,{\bf Classical}}(\tau,\tau)=\frac{L^3}{3 \pi^2}\lim_{{\bf C}_1\rightarrow 0}\sqrt{\frac{{\bf B}_1}{|{\bf C}_1|\sqrt{\pi}}}}~,\\
 {Y^{f}_{2,{\bf Classical}}(\tau,\tau)=\frac{L^3}{3 \pi^2}\lim_{{\bf C}_2\rightarrow 0}\sqrt{\frac{{\bf B}_2}{|{\bf C}_2|\sqrt{\pi}}}}~.
~\eea
But for unequal time case both the results exist and we get:
\begin{eqnarray}
&& {Y^{f}_{1,{\bf Classical}}(\tau_1,\tau_2)= \footnotesize\left\{
     \begin{array}{lr}
   \displaystyle \frac{L^3}{3\pi^2}\sqrt{\frac{{\bf A}_1}{\gamma_1}}~\exp\left(-\frac{\gamma_1|\tau_1-\tau_2|}{2}\right)~, &~ \text{\textcolor{red}{\bf Coloured~Noise}}\\  
   \displaystyle \frac{L^3}{3\pi^2} \lim_{{\bf C}_1\rightarrow 0}\sqrt{\frac{{\bf B}_1}{|{\bf C}_1|\sqrt{\pi}}}~\exp\left(-\frac{|\tau_1-\tau_2|^2}{{{\bf C}^2_1}}\right) & \text{\textcolor{red}{\bf White~Noise}}  \end{array}
   \right.}~~~~~~~~
\\
&& {Y^{f}_{2,{\bf Classical}}(\tau_1,\tau_2)= \footnotesize\left\{
     \begin{array}{lr}
   \displaystyle \frac{L^3}{3\pi^2}\sqrt{\frac{{\bf A}_2}{\gamma_2}}~\exp\left(-\frac{\gamma_2|\tau_1-\tau_2|}{2}\right)~, &~ \text{\textcolor{red}{\bf Coloured~Noise}}\\  
   \displaystyle \frac{L^3}{3\pi^2} \lim_{{\bf C}_2\rightarrow 0}\sqrt{\frac{{\bf B}_2}{|{\bf C}_2|\sqrt{\pi}}}~\exp\left(-\frac{|\tau_1-\tau_2|^2}{{{\bf C}^2_2}}\right) & \text{\textcolor{red}{\bf White~Noise}}  \end{array}
   \right.}~~~~~~~~
\end{eqnarray} 
Since in general,  both of the two-point correlations are generated from different source,  i.e.  rescaled field variable and its canonically conjugate momenta it is expected to have.,  \\
${Y^{f}_{2,{\bf Classical}}(\tau_1,\tau_2)\neq Y^{f}_{2,{\bf Classical}}(\tau_1,\tau_2)}$.  This is simply because of the fact that,  
 ${{\bf A}_1\neq {\bf A}_2},$\\~
 ${{\bf B}_1\neq {\bf B}_2},~
 {{\bf C}_1\neq {\bf C}_2},~
  {\gamma_1\neq \gamma_2}$.
\subsection{Classical limit of cosmological two-point   non-chaotic OTOC: curvature perturbation field version}
Here we need to perform the similar type of the computation for the classical version of the two-point OTOCs that we have derived in the previous subsection,  but here we have to derive the results in terms of the scalar curvature perturbation and the canonically conjugate momentum associated with it,  instead of using the rescaled field variable and its conjugate momenta.
Here we have found the following simplified expressions:
\bea && {Y^{\zeta}_{1,{\bf Classical}}(\tau_1,\tau_2)=\frac{1}{z(\tau_1)z(\tau_2)}Y^{f}_{1,{\bf Classical}}(\tau_1,\tau_2)}~.~~~~~~~~~~~~~
\nonumber\\
&&~~~~~~~~~~~~~~~~~~~~~{= \frac{L^3}{3\pi^2 z(\tau_1)z(\tau_2)}\times\footnotesize\left\{
     \begin{array}{lr}
   \displaystyle \sqrt{\frac{{\bf A}_1}{\gamma_1}}~\exp\left(-\frac{\gamma_1|\tau_1-\tau_2|}{2}\right)~, &~ \text{\textcolor{red}{\bf Coloured~Noise}}\\  
   \displaystyle  \lim_{{\bf C}_2\rightarrow 0}\sqrt{\frac{{\bf B}_1}{|{\bf C}_1|\sqrt{\pi}}}~\exp\left(-\frac{|\tau_1-\tau_2|^2}{{{\bf C}^2_1}}\right) & \text{\textcolor{red}{\bf White~Noise}}  \end{array}
   \right.}~~~~~~~~
\eea\bea && {Y^{\zeta}_{2,{\bf Classical}}(\tau_1,\tau_2)=\frac{1}{z(\tau_1)z(\tau_2)}Y^{f}_{2,{\bf Classical}}(\tau_1,\tau_2)}~.~~~~~~~~~~~~~
\nonumber\\
&&~~~~~~~~~~~~~~~~~~~~~{= \frac{L^3}{3\pi^2 z(\tau_1)z(\tau_2)}\times\footnotesize\left\{
     \begin{array}{lr}
   \displaystyle \sqrt{\frac{{\bf A}_1}{\gamma_1}}~\exp\left(-\frac{\gamma_1|\tau_1-\tau_2|}{2}\right)~, &~ \text{\textcolor{red}{\bf Coloured~Noise}}\\  
   \displaystyle  \lim_{{\bf C}_2\rightarrow 0}\sqrt{\frac{{\bf B}_1}{|{\bf C}_1|\sqrt{\pi}}}~\exp\left(-\frac{|\tau_1-\tau_2|^2}{{{\bf C}^2_1}}\right) & \text{\textcolor{red}{\bf White~Noise}}  \end{array}
   \right.}~~~~~~~~
\eea 
Since in general,  both of the two-point correlations are generated from different source,  i.e.  rescaled field variable and its canonically conjugate momenta it is expected to have:
 \bea {Y^{\zeta}_{2,{\bf Classical}}(\tau_1,\tau_2)\neq Y^{\zeta}_{2,{\bf Classical}}(\tau_1,\tau_2)~~~\Longrightarrow ~~~Y^{f}_{2,{\bf Classical}}(\tau_1,\tau_2)\neq Y^{f}_{2,{\bf Classical}}(\tau_1,\tau_2)}.~~~~~~~~~\eea 
\subsection{Classical limit of cosmological four-point   non-chaotic OTOC: rescaled field version}

\subsubsection{Without normalization}
In this subsection,  our prime objective is to explicitly compute the classical limiting result of two un-normalized cosmological four-point OTOC in terms of the Poisson Brackets, which is given by the following simplified expressions:
\bea C^{f}_{1,{\bf Classical}}(\tau_1,\tau_2):&=&\frac{1}{Z_{\bf Classical}(\beta;\tau_1)}\int \int \frac{{\cal D}f {\cal D}\Pi}{2\pi}\exp\left(-\beta~H\right)~\left\{{f}({\bf x},\tau_1),{f}({\bf x},\tau_2)\right\}^2_{\bf PB}\nonumber\\
&=&\frac{1}{Z_{\bf Classical}(\beta;\tau_1)}\underbrace{\int \int \frac{{\cal D}f {\cal D}\Pi}{2\pi}\exp\left(-\beta~H\right)}_{\textcolor{red}{\bf \equiv ~Z_{\bf Classical}(\beta;\tau_1)}}~\nonumber\\
&&~~~~~~~~\times(2\pi)^6 \int \frac{d^3{\bf k}_1}{(2\pi)^3}\int \frac{d^3{\bf k}_2}{(2\pi)^3}\int \frac{d^3{\bf k}_3}{(2\pi)^3}\int \frac{d^3{\bf k}_4}{(2\pi)^3}~\exp\left(i({\bf k}_1+{\bf k}_2+{\bf k}_3+{\bf k}_4).{\bf x}\right)\nonumber\\
 &&~~~~~~~~~~~~~~~~~~~~~~~~~\left[\delta^{3}({\bf k}_1+{\bf k}_2)\delta^3({\bf k}_3+{\bf k}_4)+\delta^{3}({\bf k}_1+{\bf k}_3)\delta^3({\bf k}_2+{\bf k}_4)\right.\nonumber\\&& \left.~~~~~~~~~~~~~~~~~~~~~~~~~+\delta^{3}({\bf k}_1+{\bf k}_4)\delta^3({\bf k}_3+{\bf k}_2)+\delta^{3}({\bf k}_2+{\bf k}_3)\delta^3({\bf k}_4+{\bf k}_1)\right.\nonumber\\&& \left.~~~~~~~~~~~~~~~~~~~~~~~~~+\delta^{3}({\bf k}_2+{\bf k}_1)\delta^3({\bf k}_4+{\bf k}_3)+\delta^{3}({\bf k}_2+{\bf k}_4)\delta^3({\bf k}_1+{\bf k}_3)\right.\nonumber\\&& \left.~~~~~~~~~~~~~~~~~~~~~~~~~+\delta^{3}({\bf k}_3+{\bf k}_1)\delta^3({\bf k}_4+{\bf k}_2)+\delta^{3}({\bf k}_3+{\bf k}_2)\delta^3({\bf k}_1+{\bf k}_4)\right.\nonumber\\&& \left.~~~~~~~~~~~~~~~~~~~~~~~~~+\delta^{3}({\bf k}_3+{\bf k}_4)\delta^3({\bf k}_1+{\bf k}_2)+\delta^{3}({\bf k}_4+{\bf k}_1)\delta^3({\bf k}_2+{\bf k}_3)\right.\nonumber\\&& \left.~~~~~~~~~~~~~~~~~~~~~~~~~+\delta^{3}({\bf k}_4+{\bf k}_2)\delta^3({\bf k}_3+{\bf k}_1)+\delta^{3}({\bf k}_4+{\bf k}_3)\delta^3({\bf k}_2+{\bf k}_1)\right]\nonumber\\
 &&~~~~~~~~~~~~~~~~~~~~~~~~~~~~~~~~~~~~~~~~~~~~~~~~~{\bf G}^{(1)}_{\bf Kernel}(|\tau_1-\tau_2|),\eea 
 and 
 \bea C^{f}_{2,{\bf Classical}}(\tau_1,\tau_2):&=&\frac{1}{Z_{\bf Classical}(\beta;\tau_1)}\int \int \frac{{\cal D}f {\cal D}\Pi}{2\pi}\exp\left(-\beta~H\right)~\left\{{\Pi}({\bf x},\tau_1),{\Pi}({\bf x},\tau_2)\right\}^2_{\bf PB}\nonumber\\
&=&\frac{1}{Z_{\bf Classical}(\beta;\tau_1)}\underbrace{\int \int \frac{{\cal D}f {\cal D}\Pi}{2\pi}\exp\left(-\beta~H\right)}_{\textcolor{red}{\bf \equiv ~Z_{\bf Classical}(\beta;\tau_1)}}~\nonumber\\
&&~~~~~~~~\times(2\pi)^6 \int \frac{d^3{\bf k}_1}{(2\pi)^3}\int \frac{d^3{\bf k}_2}{(2\pi)^3}\int \frac{d^3{\bf k}_3}{(2\pi)^3}\int \frac{d^3{\bf k}_4}{(2\pi)^3}~\exp\left(i({\bf k}_1+{\bf k}_2+{\bf k}_3+{\bf k}_4).{\bf x}\right)\nonumber\eea
\bea
 &&~~~~~~~~~~~~~~~~~~~~~~~~~\left[\delta^{3}({\bf k}_1+{\bf k}_2)\delta^3({\bf k}_3+{\bf k}_4)+\delta^{3}({\bf k}_1+{\bf k}_3)\delta^3({\bf k}_2+{\bf k}_4)\right.\nonumber\\&& \left.~~~~~~~~~~~~~~~~~~~~~~~~~+\delta^{3}({\bf k}_1+{\bf k}_4)\delta^3({\bf k}_3+{\bf k}_2)+\delta^{3}({\bf k}_2+{\bf k}_3)\delta^3({\bf k}_4+{\bf k}_1)\right.\nonumber\\&& \left.~~~~~~~~~~~~~~~~~~~~~~~~~+\delta^{3}({\bf k}_2+{\bf k}_1)\delta^3({\bf k}_4+{\bf k}_3)+\delta^{3}({\bf k}_2+{\bf k}_4)\delta^3({\bf k}_1+{\bf k}_3)\right.\nonumber\\&& \left.~~~~~~~~~~~~~~~~~~~~~~~~~+\delta^{3}({\bf k}_3+{\bf k}_1)\delta^3({\bf k}_4+{\bf k}_2)+\delta^{3}({\bf k}_3+{\bf k}_2)\delta^3({\bf k}_1+{\bf k}_4)\right.\nonumber\\&& \left.~~~~~~~~~~~~~~~~~~~~~~~~~+\delta^{3}({\bf k}_3+{\bf k}_4)\delta^3({\bf k}_1+{\bf k}_2)+\delta^{3}({\bf k}_4+{\bf k}_1)\delta^3({\bf k}_2+{\bf k}_3)\right.\nonumber\\&& \left.~~~~~~~~~~~~~~~~~~~~~~~~~+\delta^{3}({\bf k}_4+{\bf k}_2)\delta^3({\bf k}_3+{\bf k}_1)+\delta^{3}({\bf k}_4+{\bf k}_3)\delta^3({\bf k}_2+{\bf k}_1)\right]\nonumber\\
 &&~~~~~~~~~~~~~~~~~~~~~~~~~~~~~~~~~~~~~~~~~~~~~~~~~{\bf G}^{(2)}_{\bf Kernel}(|\tau_1-\tau_2|),\eea 
where in the classical limiting version the thermal partition function is given by the following expression: 
 \bea {Z_{\bf Classical}(\beta;\tau_1)=\exp\left(-\int d^3{\bf k} ~\ln\left(2\sinh \frac{\beta E_{\bf k}(\tau_1)}{2}\right)\right)}~,\eea  
 where the individual details and derivation of this equation is given in the previous sub-section.  
 
Now,  after doing a simple computation in the present context we get the following simplified results for the un-normalized version of the regulated classical limit of OTOC,  which are given by the following expressions:
\bea {C^{f}_{1,{\bf Classical}}(\tau_1,\tau_2)=12~{\bf G}^{(1)}_{\bf Kernel}(|\tau_1-\tau_2|)\int \frac{d^3{\bf k}_1}{(2\pi)^3}\int \frac{d^3{\bf k}_2}{(2\pi)^3}}~,\\
{C^{f}_{2,{\bf Classical}}(\tau_1,\tau_2)=12~{\bf G}^{(2)}_{\bf Kernel}(|\tau_1-\tau_2|)\int \frac{d^3{\bf k}_1}{(2\pi)^3}\int \frac{d^3{\bf k}_2}{(2\pi)^3}}~.\eea 
These results are actually divergent which are appearing from the all space volume integral on the momenta.   Though the rest of the part of the classical result which is completely momentum independent is not divergent at all.  Now to get the finite sensible contribution out of the above mentioned volume integrals we regulate the momentum integrals by putting the cut-off scale $L$,  for which the momentum range are given by,  $0<k_1<L$ and $0<k_2<L$ for both the cases.  Further applying this trick we get the following volume regulated results,  which are given by:
\bea \int \frac{d^3{\bf k}_1}{(2\pi)^3}\int \frac{d^3{\bf k}_2}{(2\pi)^3}=\frac{1}{4\pi^4}\int^{L}_{k_1=0}k^2_1~dk_1\int^{L}_{k_2=0}k^2_2~dk_2=\frac{L^6}{36\pi^4}~.\eea
Further substituting the above mentioned regulated factors finally we get the following simplified results for both the classical limiting version of the correlators:
\bea {C^{f}_{1,{\bf Classical}}(\tau_1,\tau_2)=\frac{L^6}{3\pi^4}~{\bf G}^{(1)}_{\bf Kernel}(|\tau_1-\tau_2|)}~,\\  {C^{f}_{2,{\bf Classical}}(\tau_1,\tau_2)=\frac{L^6}{3\pi^4}~{\bf G}^{(2)}_{\bf Kernel}(|\tau_1-\tau_2)|}~,\eea
From the above mentioned result one can next consider a pathological situation when we have $\tau_1=\tau_2=\tau$ in the classical limit.  In that case,  we get further simplified answers,  which are given by:
\bea {C^{f}_{1,{\bf Classical}}(\tau,\tau)=\frac{L^6}{3\pi^4}~{\bf G}^{(1)}_{\bf Kernel}(0)}~,\\ {C^{f}_{2,{\bf Classical}}(\tau,\tau)=\frac{L^6}{3\pi^4}~{\bf G}^{(2)}_{\bf Kernel}(0)}~,\eea 
It is important to note that here in the above context these results only exist when ${\bf G}^{(1)}_{\bf Kernel}(0)$  and ${\bf G}^{(2)}_{\bf Kernel}(0)$ both are finite in the classical limit separately for the coloured non-Gaussian noise as well for the white Gaussian noise respectively.

Now to precisely demonstrate the explicit role of a non-Gaussian coloured noise and Gaussian white noise one can further consider the following conformal time dependent two point classical correlation functions:
\begin{eqnarray}
&& {{\bf G}^{(1)}_{\bf Kernel}(|\tau_1-\tau_2|)= \footnotesize \left\{
     \begin{array}{lr}
   \displaystyle \frac{{\bf A}_1}{\gamma_2}~\exp(-\gamma_1|\tau_1-\tau_2|)~, &~ \text{\textcolor{red}{\bf Coloured~Noise}}\\  
   \displaystyle   {\bf B}_1~\delta(\tau_1-\tau_2) & \text{\textcolor{red}{\bf White~Noise}}  \end{array}
   \right.}~~~~~~~~
\\
&& {{\bf G}^{(2)}_{\bf Kernel}(|\tau_1-\tau_2|)= \footnotesize\left\{
     \begin{array}{lr}
   \displaystyle \frac{{\bf A}_2}{\gamma_2}~\exp(-\gamma_2|\tau_1-\tau_2|)~, &~ \text{\textcolor{red}{\bf Coloured~Noise}}\\  
   \displaystyle   {\bf B}_2~\delta(\tau_1-\tau_2) & \text{\textcolor{red}{\bf White~Noise}}  \end{array}
   \right.}~~~~~~~~
\end{eqnarray}
where ${\bf A}_1, {\bf A}_2$ and ${\bf B}_1, {\bf B}_2$,  represent the overall conformal time independent contributions that appearing in the context of the coloured non-Gaussian and white Gaussian random classical noise respectively.  Also it is important to mention that,  $\gamma_1$ and $\gamma_2$ are the strength of the dissipation in the context of coloured non-Gaussian noise for the two consecutive cases respectively.

Now from the general structure of the white Gaussian noise one casn write ${\bf G}^{(1)}_{\bf Kernel}(0)={\bf B}_1~\delta(0)\rightarrow \infty$ and ${\bf G}^{(2)}_{\bf Kernel}(0)={\bf B}_2~\delta(0)\rightarrow \infty$ giving  diverging contribution for $\tau_1=\tau_2$ case. So appearance of the possibility of the equal time limit is completely discarded as it gives overall diverging contribution in the classical limit of the four-point cosmological OTOC. On the other hand, in the equal time limit we have, ${\bf G}^{(1)}_{\bf Kernel}(0)={\bf A}_1/\gamma_1$ and ${\bf G}^{(2)}_{\bf Kernel}(0)={\bf A}_2/\gamma_2$ for the non-Gaussian coloured noise case. This implies for coloured noise equal time limit exists and one can write down the following simplified expression for the classical limit of the four-point cosmological OTOC as:
\bea {C^{f}_{1,{\bf Classical}}(\tau,\tau)=\frac{{\bf A}_1~L^6}{3\gamma_1 \pi^4}}~,\\
 {C^{f}_{2,{\bf Classical}}(\tau,\tau)=\frac{{\bf A}_2~L^6}{3\gamma_2 \pi^4}}~.\eea
But for unequal time case both the results exist and we get:
\begin{eqnarray}
&& {C^{f}_{1,{\bf Classical}}(\tau_1,\tau_2)=\footnotesize \left\{
     \begin{array}{lr}
   \displaystyle \frac{{\bf A}_1 L^6}{3\gamma_1\pi^4}~\exp\left(-\gamma_1|\tau_1-\tau_2|\right)~, &~ \text{\textcolor{red}{\bf Coloured~Noise}}\\  
   \displaystyle  \frac{ {\bf B}_1 L^6}{3\pi^4}~\delta(\tau_1-\tau_2) & \text{\textcolor{red}{\bf White~Noise}}  \end{array}
   \right.}~~~~~~~~
\\
&& {C^{f}_{2,{\bf Classical}}(\tau_1,\tau_2)= \footnotesize \left\{
     \begin{array}{lr}
   \displaystyle \frac{{\bf A}_2 L^6}{3\gamma_2\pi^4}~\exp\left(-\gamma_2|\tau_1-\tau_2|\right)~, &~ \text{\textcolor{red}{\bf Coloured~Noise}}\\  
   \displaystyle  \frac{ {\bf B}_2 L^6}{3\pi^4}~\delta(\tau_1-\tau_2) & \text{\textcolor{red}{\bf White~Noise}}  \end{array}
   \right.}~~~~~~~~
\end{eqnarray}
\subsubsection{With normalization}  
The normalisation factors of classical limit of the two types of the desired OTOCs for the rescaled field variable and its canonically conjugate momenta can be computed as:
\bea &&{{\cal N}^{f}_{1,{\bf Classical}}(\tau_1,\tau_2)=\frac{36\pi^4}{L^6{\bf W}^2_1(0)}=\frac{36\pi^4}{L^6{\bf G}^{(1)}_{\bf Kernel}(0)}}~,~~~~~~\\
&&{{\cal N}^{f}_{2,{\bf Classical}}(\tau_1,\tau_2)=\frac{36\pi^4}{L^6{\bf W}^2_2(0)}=\frac{36\pi^4}{L^6{\bf G}^{(2)}_{\bf Kernel}(0)}}~.~~~~~~\eea
Now, considering the examples of non-Gaussian coloured noise and Gaussian white noise we get the following answer for the normalization factor:
\begin{eqnarray}
&& {{\cal N}^{f}_{1,{\bf Classical}}(\tau_1,\tau_2)= \footnotesize \left\{
     \begin{array}{lr}
   \displaystyle\frac{36\gamma_1 \pi^4}{L^6{\bf A}_1}~, &~ \text{\textcolor{red}{\bf Coloured~Noise}}\\  
   \displaystyle   0 & \text{\textcolor{red}{\bf White~Noise}}  \end{array}
   \right.}~~~~~~~~~~
\\
&& {{\cal N}^{f}_{2,{\bf Classical}}(\tau_1,\tau_2)= \footnotesize \left\{
     \begin{array}{lr}
   \displaystyle\frac{36\gamma_2 \pi^4}{L^6{\bf A}_2}~, &~ \text{\textcolor{red}{\bf Coloured~Noise}}\\  
   \displaystyle   0 & \text{\textcolor{red}{\bf White~Noise}}  \end{array}
   \right.}~~~~~~~~~~
\end{eqnarray}
Then the classical limiting version the normalized four-point OTOCs can be expressed in terms of the contribution of phase space averaged Poisson Bracket squared as:
\bea {{\cal C}^{f}_{1,{\bf Classical}}(\tau_1,\tau_2)={\cal N}^{f}_{1,{\bf Classical}}(\tau_1,\tau_2)C^{f}_{1,{\bf Classical}}(\tau_1,\tau_2)=12\left(\frac{{\bf G}^{(1)}_{\bf Kernel}(|\tau_1-\tau_2|)}{{\bf G}^{(1)}_{\bf Kernel}(0)}\right)}~,~~~~~~\\
{{\cal C}^{f}_{2,{\bf Classical}}(\tau_1,\tau_2)={\cal N}^{f}_{2,{\bf Classical}}(\tau_1,\tau_2)C^{f}_{2,{\bf Classical}}(\tau_1,\tau_2)=12\left(\frac{{\bf G}^{(2)}_{\bf Kernel}(|\tau_1-\tau_2|)}{{\bf G}^{(2)}_{\bf Kernel}(0)}\right)}~.~~~~~~\eea
This is a very useful result as it translates everything in terms of the time translation invariant noise field and its canonically conjugate momenta Green's functions,  ${\bf G}^{(1)}_{\bf Kernel}(|\tau_1-\tau_2|)$ and ${\bf G}^{(2)}_{\bf Kernel}(|\tau_1-\tau_2|)$ respectively.

Now,  further considering the examples of non-Gaussian coloured noise and Gaussian white noise we get the following simplified results for the classical limit of the normalized four-point OTOCs,  which are given by:
\begin{eqnarray}
&& {{\cal C}^{f}_{1,{\bf Classical}}(\tau_1,\tau_2)=\footnotesize \left\{
     \begin{array}{lr}
   \displaystyle 12~\exp\left(-\gamma_1|\tau_1-\tau_2|\right)~, &~ \text{\textcolor{red}{\bf Coloured~Noise}}\\  
   \displaystyle  0 & \text{\textcolor{red}{\bf White~Noise}}  \end{array}
   \right.}.~~~~~~~~
\\
&& {{\cal C}^{f}_{2,{\bf Classical}}(\tau_1,\tau_2)=\footnotesize \left\{
     \begin{array}{lr}
   \displaystyle 12~\exp\left(-\gamma_2|\tau_1-\tau_2|\right)~, &~ \text{\textcolor{red}{\bf Coloured~Noise}}\\  
   \displaystyle  0 & \text{\textcolor{red}{\bf White~Noise}}  \end{array}
   \right.}.~~~~~~~~
\end{eqnarray}
\subsection{Classical limit of cosmological four-point  non-chaotic OTOC: curvature perturbation field version}
 \subsubsection{Without normalization} 
Here we need to perform the computation for the classical version of the un-normalised two types of the OTOCs in terms of the scalar curvature perturbation field variable and the canonically conjugate momentum associated with it,  which we have found that for the two separate cases given by the following simplified expression:
\bea && {C^{\zeta}_{1,{\bf Classical}}(\tau_1,\tau_2)=\frac{1}{Z^{\zeta}_{\bf Classical}(\beta,\tau_1)}\int\int\frac{{\cal D}\zeta {\cal D}\Pi_{\zeta}}{2\pi}e^{-\beta H(\tau_1)}\left\{\zeta({\bf x},\tau_1),\zeta({\bf x},\tau_2)\right\}^2_{\bf PB}}\nonumber\\
&&~~~~~~~~~~~~~~~~~~~~~{=\frac{1}{z^2(\tau_1)z^2(\tau_2)}C^{f}_{1,{\bf Classical}}(\tau_1,\tau_2)}~,~~~~~~~~~~~~~\\ && {C^{\zeta}_{2,{\bf Classical}}(\tau_1,\tau_2)=\frac{1}{Z^{\zeta}_{\bf Classical}(\beta,\tau_1)}\int\int\frac{{\cal D}\zeta {\cal D}\Pi_{\zeta}}{2\pi}e^{-\beta H(\tau_1)}\left\{\Pi_{\zeta}({\bf x},\tau_1),\Pi_{\zeta}({\bf x},\tau_2)\right\}^2_{\bf PB}}\nonumber\\
&&~~~~~~~~~~~~~~~~~~~~~{=\frac{1}{z^2(\tau_1)z^2(\tau_2)}C^{f}_{2,{\bf Classical}}(\tau_1,\tau_2)}~.~~~~~~~~~~~~~\eea 

\subsubsection{With normalization}
The classical version of the normalised four-point two types of the desired OTOCs in terms of the scalar curvature perturbation field variable and its canonically conjugate momentum,  which are basically the computation of the following normalised OTOCs,  can be written as:
\bea {{\cal C}^{\zeta}_{1,{\bf Classical}}(\tau_1,\tau_2)=\frac{C^{\zeta}_{1,{\bf Classical}}(\tau_1,\tau_2)}{\langle \zeta(\tau_1)\zeta(\tau_1)\rangle_{\beta}\langle  \zeta(\tau_1) \zeta(\tau_1)\rangle_{\beta}}={\cal N}^{\zeta}_{1,{\bf Classical}}(\tau_1,\tau_2)~C^{\zeta}_{1,{\bf Classical}}(\tau_1,\tau_2)}~,~~~~~~~~ \\ {{\cal C}^{\zeta}_{2,{\bf Classical}}(\tau_1,\tau_2)=\frac{C^{\zeta}_{2,{\bf Classical}}(\tau_1,\tau_2)}{\langle \Pi_{\zeta}(\tau_1)\Pi_{\zeta}(\tau_1)\rangle_{\beta}\langle \Pi_{\zeta}(\tau_1)\Pi_{\zeta}(\tau_1)\rangle_{\beta}}={\cal N}^{\zeta}_{2,{\bf Classical}}(\tau_1,\tau_2)~C^{\zeta}_{2,{\bf Classical}}(\tau_1,\tau_2)}~,~~~~~~~~ \eea 
where the normalisation factors to normalise the classical OTOCs are given by:
 \bea &&{{\cal N}^{\zeta}_{1,{\bf Classical}}(\tau_1,\tau_2)=\frac{1}{\langle \zeta(\tau_1)\zeta(\tau_1)\rangle_{\beta}\langle \zeta(\tau_1)\zeta(\tau_1)\rangle_{\beta}}=z^2(\tau_1)z^2(\tau_2){\cal N}^{f}_{1,{\bf Classical}}(\tau_1,\tau_2)},~~~~~~~~\\
&&{{\cal N}^{\zeta}_{2,{\bf Classical}}(\tau_1,\tau_2)=\frac{1}{\langle \Pi_{\zeta}(\tau_1)\Pi_{\zeta}(\tau_1)\rangle_{\beta}\langle \Pi_{\zeta}(\tau_1)\Pi_{\zeta}(\tau_1)\rangle_{\beta}}=z^2(\tau_1)z^2(\tau_2){\cal N}^{f}_{2,{\bf Classical}}(\tau_1,\tau_2)}.~~~~~~~~ \eea
 Consequently,  the classical limit of the normalised four-point desired OTOCs computed from the curvature perturbation field variable and its canonically conjugate momentum field variable are given by the following expressions:
 \bea &&{{\cal C}^{\zeta}_{1,{\bf Classical}}(\tau_1,\tau_2)={\cal C}^{f}_{1,{\bf Classical}}(\tau_1,\tau_2)=12\left(\frac{{\bf G}^{(1)}_{\bf Kernel}(|\tau_1-\tau_2|)}{{\bf G}^{(1)}_{\bf Kernel}(0)}\right)},~~~~~~~~\\
 &&{{\cal C}^{\zeta}_{2,{\bf Classical}}(\tau_1,\tau_2)={\cal C}^{f}_{2,{\bf Classical}}(\tau_1,\tau_2)=12\left(\frac{{\bf G}^{(2)}_{\bf Kernel}(|\tau_1-\tau_2|)}{{\bf G}^{(1)}_{\bf Kernel}(0)}\right)}.~~~~~~~~ \eea  
Now,  considering the examples of non-Gaussian coloured noise and Gaussian white noise we get the following answer for the classical limit of normalized four-point two types of desired OTOCs are given by:
\begin{eqnarray}
&& {{\cal C}^{\zeta}_{1,{\bf Classical}}(\tau_1,\tau_2)={\cal C}^{f}_{1,{\bf Classical}}(\tau_1,\tau_2)= \footnotesize \left\{
     \begin{array}{lr}
   \displaystyle 12~\exp\left(-\gamma_1|\tau_1-\tau_2|\right)~, &~ \text{\textcolor{red}{\bf Coloured~Noise}}\\  
   \displaystyle  0 & \text{\textcolor{red}{\bf White~Noise}}  \end{array}
   \right.}.~~~~~~~~
\\
&& {{\cal C}^{\zeta}_{2,{\bf Classical}}(\tau_1,\tau_2)={\cal C}^{f}_{2,{\bf Classical}}(\tau_1,\tau_2)=\footnotesize \left\{
     \begin{array}{lr}
   \displaystyle 12~\exp\left(-\gamma_2|\tau_1-\tau_2|\right)~, &~ \text{\textcolor{red}{\bf Coloured~Noise}}\\  
   \displaystyle  0 & \text{\textcolor{red}{\bf White~Noise}}  \end{array}
   \right.}.~~~~~~~~
\end{eqnarray}
%%%%%%%%%%%%%%%%%%%%
\section{Summary and Outlook}
  \label{sec:6}
%%%%%%%%%%%%%%%%%%%%
To summarize,  in this work,  we have addressed the following issues to study the random non-chaotic features of Cosmological OTOCs:
\begin{itemize}
\item First of all in this paper we have provided a computation using which it now possible to derive the expressions for the two specific types of OTOCs made up of cosmological scalar perturbation field variable and its associated  canonically conjugate momentum variable auto-correlations to study the feature of randomness without having chaotic behaviour in Primordial Cosmology set up.  It is expected that our presented computation and finding for the two new types of cosmological OTOCs in this paper will surely be helpful to understand the quantum field theoretic features of random non-chaotic cosmological events in detail.  Apart from using the presented methodology in the present context of discussion,  we believe that these results will also be used in other contexts of cosmological phenomena to study similar type of random non-chaotic events appearing in the time line of the universe.

\item The computation is presented by making use of the well known Euclidean vacuum i. e.  Bunch Davies vacuum,  CPT invariant $\alpha$ vacua and CPT violating Motta Allen vacua as a choice of initial quantum mechanical vacuum state.   In each cases we get very distinctive features in the auto-correlated two types of OTO functions studied in this paper.  We have studied this issue analytically as well as numerically in detail in this paper,  which completely physically justify our prescription proposed in this paper.

\item In general the construction of OTOC's demands to have two quantum mechanical operators defined at two different time scales,  which are same operators for auto-correlations and different operators for cross-correlations.   In this paper,  we exactly follow the same strategy to define auto-correlated two different types of OTO functions within the framework of primordial cosmological perturbation theory for scalar mode fluctuations in quantum regime.  To construct these crucial auto-correlated OTO functions we use the scalar mode field variable and its associated canonically conjugate momenta which according to the mathematical construction are defined in two different conformal time scales.  To study the behaviour of the two types of the auto-correlated OTO functions we freeze one of the conformal time scales between the two and study the dynamical feature with respect to the other conformal time.  By doing this analysis we have found that the dynamical features with respect to both the time scales for the two different types of auto-correlated OTO functions are significantly different and all of them describes distinctive randomness at out-of-equilibrium without having any specific chaoticity.  For both the cases we have explicitly studied the long time behaviour of these two types of auto-correlated OTO functions.  We found that the obtained behaviour from the numerical plots are perfectly consistent with the expectation from the present set up within the framework of Primordial Cosmology.  The most interesting part of this finding is that using the present methodology it is now possible to probe various cosmological aspects at out-of-equilibrium using quantum field theory calculations in terms of quantum mechanical correlation functions.  Also we are very hopeful regarding the fact that these results can be verifiable in near future cosmological observational probes.

\item By doing the analysis we have found that at very early epoch of our universe during random cosmological events quantum fluctuations generated from the scalar perturbations goes to the out-of-equilibrium phase. Then the auto-correlators decay in a non-standard fashion with respect to the conformal time scale up to very late time scale of the universe.  After that the system reaches equilibrium and one can perform all the possible computations in quantum regime.  This computation can be applicable to describe the particle production phenomena during reheating epoch.  

\item Also, we have found that the derived cosmological OTOCs at finite temperature is dependent on two time scale and independent of any preferred choice of the coordinate system. The derived expressions for the cosmological OTOCs are homogeneous in nature with respect to the space coordinate,  or its Fourier transformed momentum coordinate.  We also have found that the final results obtained for the two different types of cosmological OTOCs are independent of the partition function which we have computed for Primordial Cosmological perturbations for scalar fluctuation.  It is important to note that, the obtained features in these auto-correlations are exactly mimics the feature obtained for OTOCs computed from inverted harmonic oscillator having a conformal time independent frequency.  Here 
one can exactly map the stochastic particle production problem in cosmology in terms of finding from inverted harmonic oscillator with time dependent frequency because both the set-up describes the same underlying physics.

\item Also we have found that the presented analysis is valid for partially massless ($m\sim H$) or massive ($m\gg H$) spin-$0$ scalar particle production in the primordial universe. 

\item Further we have studied the classical limiting behaviours of the two-point and four-point two types of auto-correlated OTO functions in terms of the phase space averaged of Poisson brackets and the square of the Poisson brackets to check the consistency with the late time behaviour in the super-horizon region of cosmological perturbations,  which supports classical behaviour.

\item Finally,  in this paper we have shown that the normalised auto-correlated OTO functions is completely independent of the choice of cosmological perturbation field variable and the associated canonically conjugate momentum in a specific gauge.

\end{itemize}

The future prospects of this work is as follows:

\begin{itemize}

\item We are very hopeful that our obtained results in this paper for the two different types of auto-correlators can be probed by future observations with significant statistical accuracy and can be treated as benchmark using which one can study the out-of-equilibrium features in primordial cosmological perturbation theory.  To know about the implications of our derived results one can further extend the present methodology for various cosmological events as well where random quantum fluctuations play significant role in the time line of our universe.  Till date this is not very well studied fact and for this reason it will be interesting to study these mentioned aspects in detail.

\item The presented methodology can also extended to derive the auto-corrected OTO functions in the context of bouncing cosmology.  It is expected to get different random feature in the context of bouncing paradigm.  But it is not even clear till date how exactly and which respect it will be different due to having lack of understanding the present formalism in broader perspective. For this reason it will be very interesting to study out-of-equilibrium features in quantum regime from bouncing cosmology framework and compare the results obtained from the presented analysis in this paper.

\item  The explicit role of quantum entanglement in the present framework is not also studied yet.  Till date the explicit role of quantum entanglement phenomena \cite{Choudhury:2017bou,Choudhury:2017qyl,Choudhury:2020ivj,Akhtar:2019qdn,Banerjee:2020ljo,Bhattacherjee:2019eml,Narayan:2015oka,Narayan:2015vda,Narayan:2016xwq,Narayan:2017xca,Narayan:2019pjl,Narayan:2020nsc,Manu:2020tty,Fernandes:2019ige,Maldacena:2012xp,Albrecht:2018prr,Arias:2019pzy,Huang:2017yjt,VanRaamsdonk:2016exw,Kanno:2017wpw,Soda:2017yzu,Kanno:2016qcc,Kanno:2016gas,Kanno:2015ewa,Kanno:2015lja,Kanno:2014bma,Kanno:2014ifa,Kanno:2014lma} have been studied within the framework of cosmological perturbation theory finding the quantum correlation function in the equilibrium regime.  But it is not studied in the out-of-equilibrium regime in presence of quantum entanglement.  So it will be interesting to investigate such possibilities in detail.  Most importantly,  the role of cosmological Bell's inequality violation \cite{Choudhury:2016cso,Choudhury:2016pfr,Martin:2017zxs,Maldacena:2015bha,Kanno:2017teu,Kanno:2017dci} can also be tested to know about the long range effect in the cosmological correlation functions.

\item It is also very important to verify the connecting relationship between the quantum circuit complexity \cite{Jefferson:2017sdb,Guo:2018kzl,Chapman:2016hwi,Caceres:2019pgf,Bhattacharyya:2018bbv,Bhattacharyya:2019kvj,Bhattacharyya:2020kgu,Bhattacharyya:2020rpy,Choudhury:2021qod,Choudhury:2020lja,Choudhury:2020hil,Bhargava:2020fhl,Khan:2018rzm} and OTO correlators \cite{Choudhury:2020yaa,Haque:2020pmp} within the framework of Primordial Cosmology setup. Till date the work have been done on both the sides separately in the cosmological framework. But no effort have been made to connect this two theoretical ideas.

\end{itemize}

	\subsection*{Acknowledgements}
	The research fellowship of SC is supported by the J.  C.  Bose National Fellowship of Sudhakar Panda.  Also SC take this opportunity to thank sincerely to Sudhakar Panda for his constant support and providing huge inspiration.  SC also would line to thank School of Physical Sciences, National Institute for Science Education and Research (NISER),  Bhubaneswar for providing the work friendly environment. SC also thank all the members of our newly formed virtual international non-profit consortium Quantum Structures of the Space-Time \& Matter (QASTM) for elaborative discussions.  Last but not the least,  we would like to acknowledge our debt to the people belonging to the various part of the world for their generous and steady support for research in natural sciences.
	\clearpage
	\appendix
\section{Asymptotic features of the scalar mode functions in cosmological perturbation theory}
\label{sec:7}
The Mukhanov Sasaki equation for the scalar modes can be expressed as:
\bea \frac{d^2f_{\bf k}(\tau)}{d\tau^2}+\left(k^2-\frac{\displaystyle \nu^2-\frac{1}{4}}{\tau^2}\right)f_{\bf k}(\tau)=0.\eea
The most general solution of the above mentioned equation of motion is given by the following expression:
\bea {f_{\bf k}(\tau)=\sqrt{-\tau}\left[{\cal C}_1~H^{(1)}_{\nu}(-k\tau)+{\cal C}_2~H^{(2)}_{\nu}(-k\tau)\right]},\eea
where ${\cal C}_1$ and ${\cal C}_2$ are two arbitrary integration constants which are fixed by the choice of the initial quantum vacuum state.  Here $H^{(1)}_{\nu}(-k\tau)$ and $H^{(1)}_{\nu}(-k\tau)$ are the Hankel functions of first and second kind with order $\nu$.  Now we take the asymptotic limits $k\tau\rightarrow 0$ and $k\tau\rightarrow- \infty$ for which we get the following simplified results for the Hankel functions:
\bea &&{\lim_{k\tau\rightarrow -\infty} H^{(1)}_{\nu}(-k\tau)=\sqrt{\frac{2}{\pi}}\frac{1}{\sqrt{-k\tau}}\exp(-ik\tau)\exp\left(-\frac{i\pi}{2}\left(\nu+\frac{1}{2}\right)\right)},\\
&&{ \lim_{k\tau\rightarrow -\infty} H^{(2)}_{\nu}(-k\tau)=-\sqrt{\frac{2}{\pi}}\frac{1}{\sqrt{-k\tau}}\exp(ik\tau)\exp\left(\frac{i\pi}{2}\left(\nu+\frac{1}{2}\right)\right)},\\
 &&{ \lim_{k\tau\rightarrow 0} H^{(1)}_{\nu}(-k\tau)=\frac{i}{\pi}\Gamma(\nu)\left(-\frac{k\tau}{2}\right)^{-\nu}},\\
   &&{\lim_{k\tau\rightarrow 0} H^{(2)}_{\nu}(-k\tau)=-\frac{i}{\pi}\Gamma(\nu)\left(-\frac{k\tau}{2}\right)^{-\nu}}.\eea
 Here $k\tau\rightarrow 0$ and $k\tau\rightarrow -\infty$ asymptotic limiting results are used to describe the super-horizon ($k\tau\ll-1$) and sub-horizon ($k\tau\gg-1$) limiting results.

Now,  in the super-horizon limit ($k\tau\ll-1$) and sub-horizon limit ($k\tau\gg-1$) the asymptotic form of the rescaled field variable and the corresponding canonically conjugate momentum computed for the arbitrary quantum initial vacuum can be expressed as:
 \bea &&\textcolor{red}{\bf \underline{Super-horizon~limiting~results~(k\tau\ll-1):\Rightarrow Classical~behaviour}}\nonumber\\
 &&{\lim_{k\tau\rightarrow 0}f_{\bf k}(\tau)=\sqrt{\frac{2}{k}}\frac{i}{\pi}\Gamma(\nu)\left(-\frac{k\tau}{2}\right)^{\frac{1}{2}-\nu}\left({\cal C}_1-{\cal C}_2\right)},\\
&&{\lim_{k\tau\rightarrow 0}\Pi_{\bf k}(\tau)=\sqrt{\frac{2}{k}}\frac{i}{2\pi k}\left(\nu-\frac{1}{2}\right)\Gamma(\nu)\left(-\frac{k\tau}{2}\right)^{-\left(\nu+\frac{1}{2}\right)}\left({\cal C}_1-{\cal C}_2\right)}, ~~~~~~~~\eea
\bea
 &&\textcolor{red}{\bf \underline{Sub-horizon~limiting~results~(k\tau\gg-1):\Rightarrow Quantum~behaviour}}\nonumber\\
 &&\lim_{k\tau\rightarrow -\infty}f_{\bf k}(\tau)=\sqrt{\frac{2}{\pi k}}\left[{\cal C}_1~\exp\left(-i\left\{k\tau+\frac{\pi}{2}\left(\nu+\frac{1}{2}\right)\right\}\right)\right.\nonumber\\
&&\left.~~~~~~~~~~~~~~~~~~~~~~~~~~~~~~~-{\cal C}_2~\exp\left(i\left\{k\tau+\frac{\pi}{2}\left(\nu+\frac{1}{2}\right)\right\}\right)\right],\\
&& \lim_{k\tau\rightarrow -\infty}\Pi_{\bf k}(\tau)=\frac{1}{i}\sqrt{\frac{2k}{\pi }}\left[{\cal C}_1~\exp\left(-i\left\{k\tau+\frac{\pi}{2}\left(\nu+\frac{1}{2}\right)\right\}\right)\right.\nonumber\\
&&\left.~~~~~~~~~~~~~~~~~~~~~~~~~~~~~~~~~~~~~+{\cal C}_2~\exp\left(i\left\{k\tau+\frac{\pi}{2}\left(\nu+\frac{1}{2}\right)\right\}\right)\right]. ~~~~~~~~\eea 
Combining the behaviour in both the super-horizon and sub-horizon limiting region we get following combined asymptotic most general solution for the rescaled field variable and momenta computed for the arbitrary quantum initial vacuum can be expressed as:
\bea &&f_{\bf k}(\tau)=2^{\nu-\frac{3}{2}}\frac{1}{i\tau}\frac{1}{\sqrt{2}k^{\frac{3}{2}}}(-k\tau)^{\frac{3}{2}-\nu}\left|\frac{\Gamma(\nu)}{\Gamma\left(\frac{3}{2}\right)}\right|\nonumber\\
&&~~~~~~~~~~~~~~~~~~~~~\times\left[{\cal C}_1~(1+ik\tau)~\exp\left(-i\left\{k\tau+\frac{\pi}{2}\left(\nu+\frac{1}{2}\right)\right\}\right)\right.\nonumber\\
&&\left.~~~~~~~~~~~~~~~~~~~~~~~~~~~~~~~~~~~~~-{\cal C}_2~(1-ik\tau)~\exp\left(i\left\{k\tau+\frac{\pi}{2}\left(\nu+\frac{1}{2}\right)\right\}\right)\right],\\
 && {\Pi_{\bf k}(\tau)=2^{\nu-\frac{3}{2}}\frac{1}{\sqrt{2}ik^{\frac{5}{2}}}(-k\tau)^{\frac{3}{2}-\nu}\left|\frac{\Gamma(\nu)}{\Gamma\left(\frac{3}{2}\right)}\right|}\nonumber\\
&&~~~~~~~~~~~~~~~~~\nonumber\times\left[{{\cal C}_1~\left\{\left(\frac{1}{2}-\nu\right)\frac{(1+ik\tau)}{k^2\tau^2}+1\right\}~\exp\left(-i\left\{k\tau+\frac{\pi}{2}\left(\nu+\frac{1}{2}\right)\right\}\right)}\right.\\&& \left.~~~~~~~~~~~~~~~~~~~~~~~~~{-{\cal C}_2~\left\{\left(\frac{1}{2}-\nu\right)\frac{(1-ik\tau)}{k^2\tau^2}+1\right\}~\exp\left(i\left\{k\tau+\frac{\pi}{2}\left(\nu+\frac{1}{2}\right)\right\}\right)}\right].~~~~~~~~~~~~ \eea
These general asymptotic expressions are extremely important to compute the expressions for the OTOC's in the later subsections. To server this purpose we need to promote both of these classical solutions to the quantum level. 
\newpage
\section{Quantum two-point OTO  amplitudes for Cosmology} 
\label{sec:8}
\subsection{Definition of  OTO amplitude $\hat{\Delta}_1({\bf k}_1,{\bf k}_2;\tau_1,\tau_2)$} 
In this subsection we define a very important momentum and conformal time dependent two-point OTO amplitude,  which is given by the following expression:
\bea \hat{\Delta}_{1}({\bf k}_1,{\bf k}_2;\tau_1,\tau_2)&=&\hat{f}_{{\bf k}_1}(\tau_1)\hat{f}_{{\bf k}_2}(\tau_2)={\cal D}_1 ({\bf k}_1,{\bf k}_2;\tau_1,\tau_2)~a_{{\bf k}_1}a_{{\bf k}_2}+{\cal D}_2 ({\bf k}_1,{\bf k}_2;\tau_1,\tau_2)~a^{\dagger}_{-{\bf k}_1}a_{{\bf k}_2}\nonumber\\
 &&~~~~~~~~+{\cal D}_3 ({\bf k}_1,{\bf k}_2;\tau_1,\tau_2)~a_{{\bf k}_1}a^{\dagger}_{-{\bf k}_2}+{\cal D}_4 ({\bf k}_1,{\bf k}_2;\tau_1,\tau_2)~a^{\dagger}_{-{\bf k}_1}a^{\dagger}_{-{\bf k}_2},~~~~~~\eea  
 where we have introduced  momentum and time dependent four individual two-point OTO amplitudes, ${\cal D}_i ({\bf k}_1,{\bf k}_2;\tau_1,\tau_2)~~\forall~~i=1,2,3,4$, which are explicitly defined as:
 \bea {\cal D}_1 ({\bf k}_1,{\bf k}_2;\tau_1,\tau_2)&=&f_{{\bf k}_1}(\tau_1)f_{{\bf k}_2}(\tau_2),\\
 {\cal D}_2 ({\bf k}_1,{\bf k}_2;\tau_1,\tau_2)&=&f^{*}_{{\bf -k}_1}(\tau_1)f_{{\bf k}_2}(\tau_2),\\
 {\cal D}_3 ({\bf k}_1,{\bf k}_2;\tau_1,\tau_2)&=&f_{{\bf k}_1}(\tau_1)f^{*}_{{\bf -k}_2}(\tau_2),\\
 {\cal D}_4 ({\bf k}_1,{\bf k}_2;\tau_1,\tau_2)&=&f^{*}_{{\bf -k}_1}(\tau_1)f^{*}_{{\bf -k}_2}(\tau_2).\eea
 These contributions are really helpful to compute the two-point  OTO amplitudes and the corresponding momentum integrated OTOC,  which we have discussed earlier in this paper. 
 \subsection{Definition of  OTO amplitude $\hat{\Delta}_2({\bf k}_1,{\bf k}_2;\tau_1,\tau_2)$}
In this subsection we define a very important momentum and conformal time dependent two-point OTO amplitude,  which is given by the following expression:
\bea \hat{\Delta}_{2}({\bf k}_1,{\bf k}_2;\tau_1,\tau_2)&=&\hat{\Pi}_{{\bf k}_1}(\tau_1)\hat{\Pi}_{{\bf k}_2}(\tau_2)={\cal L}_1 ({\bf k}_1,{\bf k}_2;\tau_1,\tau_2)~a_{{\bf k}_1}a_{{\bf k}_2}+{\cal L}_2 ({\bf k}_1,{\bf k}_2;\tau_1,\tau_2)~a^{\dagger}_{-{\bf k}_1}a_{{\bf k}_2}\nonumber\\
 &&~~~~~~~~+{\cal L}_3 ({\bf k}_1,{\bf k}_2;\tau_1,\tau_2)~a_{{\bf k}_1}a^{\dagger}_{-{\bf k}_2}+{\cal L}_4 ({\bf k}_1,{\bf k}_2;\tau_1,\tau_2)~a^{\dagger}_{-{\bf k}_1}a^{\dagger}_{-{\bf k}_2},~~~~~~\eea  
 where we have introduced  momentum and time dependent four individual two-point OTO amplitudes, ${\cal L}_i ({\bf k}_1,{\bf k}_2;\tau_1,\tau_2)~~\forall~~i=1,2,3,4$, which are explicitly defined as:
 \bea {\cal L}_1 ({\bf k}_1,{\bf k}_2;\tau_1,\tau_2)&=&\Pi_{{\bf k}_1}(\tau_2)\Pi_{{\bf k}_2}(\tau_1),\\
 {\cal L}_2 ({\bf k}_1,{\bf k}_2;\tau_1,\tau_2)&=&\Pi^{*}_{{\bf -k}_1}(\tau_2)\Pi_{{\bf k}_2}(\tau_1),\\
 {\cal L}_3 ({\bf k}_1,{\bf k}_2;\tau_1,\tau_2)&=&\Pi_{{\bf k}_1}(\tau_2)\Pi^{*}_{-{\bf k}_2}(\tau_1),\\
 {\cal L}_4 ({\bf k}_1,{\bf k}_2;\tau_1,\tau_2)&=&\Pi^{*}_{{\bf -k}_1}(\tau_2)\Pi^{*}_{{\bf -k}_2}(\tau_1).\eea
 These contributions are really helpful to compute the two-point  OTO amplitudes and the corresponding momentum integrated OTOC,  which we have discussed earlier in this paper.

 \newpage 
\section{Quantum four-point OTO  amplitudes for Cosmology} 
\label{sec:9}
\subsection{Definition of  OTO amplitude $\widehat{\cal T}^{(1)}_1({\bf k}_1,{\bf k}_2,{\bf k}_3,{\bf k}_4;\tau_1,\tau_2)$ and $\widehat{\cal T}^{(2)}_1({\bf k}_1,{\bf k}_2,{\bf k}_3,{\bf k}_4;\tau_1,\tau_2)$}
The function $\widehat{\cal T}^{(1)}_1({\bf k}_1,{\bf k}_2,{\bf k}_3,{\bf k}_4;\tau_1,\tau_2)$ is defined as:
 \bea &&\widehat{\cal T}^{(1)}_1({\bf k}_1,{\bf k}_2,{\bf k}_3,{\bf k}_4;\tau_1,\tau_2)\nonumber\\
 &&=\left[{\cal M}^{(1)}_1({\bf k}_1,{\bf k}_2,{\bf k}_3,{\bf k}_4;\tau_1,\tau_2)~a_{{\bf k}_1}a_{{\bf k}_2}a_{{\bf k}_3}a_{{\bf k}_4}\right.\nonumber\\
  && \left.+ {\cal M}^{(1)}_2({\bf k}_1,{\bf k}_2,{\bf k}_3,{\bf k}_4;\tau_1,\tau_2)~a^{\dagger}_{-{\bf k}_1}a_{{\bf k}_2}a_{{\bf k}_3}a_{{\bf k}_4}+{\cal M}^{(1)}_3({\bf k}_1,{\bf k}_2,{\bf k}_3,{\bf k}_4;\tau_1,\tau_2)~a_{{\bf k}_1}a^{\dagger}_{-{\bf k}_2}a_{{\bf k}_3}a_{{\bf k}_4}\right.\nonumber\\
  && \left.+ {\cal M}^{(1)}_4({\bf k}_1,{\bf k}_2,{\bf k}_3,{\bf k}_4;\tau_1,\tau_2)~a^{\dagger}_{-{\bf k}_1}a^{\dagger}_{-{\bf k}_2}a_{{\bf k}_3}a_{{\bf k}_4}+{\cal M}^{(1)}_5({\bf k}_1,{\bf k}_2,{\bf k}_3,{\bf k}_4;\tau_1,\tau_2)~a_{{\bf k}_1}a_{{\bf k}_2}a^{\dagger}_{-{\bf k}_3}a_{{\bf k}_4}\right.\nonumber\\
  && \left.+ {\cal M}^{(1)}_6({\bf k}_1,{\bf k}_2,{\bf k}_3,{\bf k}_4;\tau_1,\tau_2)~a^{\dagger}_{-{\bf k}_1}a_{{\bf k}_2}a^{\dagger}_{-{\bf k}_3}a_{{\bf k}_4}+{\cal M}^{(1)}_7({\bf k}_1,{\bf k}_2,{\bf k}_3,{\bf k}_4;\tau_1,\tau_2)~a_{{\bf k}_1}a^{\dagger}_{-{\bf k}_2}a^{\dagger}_{-{\bf k}_3}a_{{\bf k}_4}\right.\nonumber\\
  && \left.+ {\cal M}^{(1)}_8({\bf k}_1,{\bf k}_2,{\bf k}_3,{\bf k}_4;\tau_1,\tau_2)~a^{\dagger}_{-{\bf k}_1}a^{\dagger}_{-{\bf k}_2}a^{\dagger}_{-{\bf k}_3}a_{{\bf k}_4}+{\cal M}^{(1)}_9({\bf k}_1,{\bf k}_2,{\bf k}_3,{\bf k}_4;\tau_1,\tau_2)~a_{{\bf k}_1}a_{{\bf k}_2}a_{{\bf k}_3}a^{\dagger}_{-{\bf k}_4}\right.\nonumber\\
  && \left.+ {\cal M}^{(1)}_{10}({\bf k}_1,{\bf k}_2,{\bf k}_3,{\bf k}_4;\tau_1,\tau_2)~a^{\dagger}_{-{\bf k}_1}a_{{\bf k}_2}a_{{\bf k}_3}a^{\dagger}_{-{\bf k}_4}+{\cal M}^{(1)}_{11}({\bf k}_1,{\bf k}_2,{\bf k}_3,{\bf k}_4;\tau_1,\tau_2)~a_{{\bf k}_1}a^{\dagger}_{-{\bf k}_2}a_{{\bf k}_3}a^{\dagger}_{-{\bf k}_4}\right.\nonumber\\
  && \left.+ {\cal M}^{(1)}_{12}({\bf k}_1,{\bf k}_2,{\bf k}_3,{\bf k}_4;\tau_1,\tau_2)~a^{\dagger}_{-{\bf k}_1}a^{\dagger}_{-{\bf k}_2}a_{{\bf k}_3}a^{\dagger}_{-{\bf k}_4}\right.\nonumber\\
  && \left.+{\cal M}^{(1)}_{13}({\bf k}_1,{\bf k}_2,{\bf k}_3,{\bf k}_4;\tau_1,\tau_2)~a_{{\bf k}_1}a_{{\bf k}_2}a^{\dagger}_{-{\bf k}_3}a^{\dagger}_{-{\bf k}_4}+ {\cal M}^{(1)}_{14}({\bf k}_1,{\bf k}_2,{\bf k}_3,{\bf k}_4;\tau_1,\tau_2)~a^{\dagger}_{-{\bf k}_1}a_{{\bf k}_2}a^{\dagger}_{-{\bf k}_3}a^{\dagger}_{-{\bf k}_4}\right.\nonumber\\
  && \left.+{\cal M}^{(1)}_{15}({\bf k}_1,{\bf k}_2,{\bf k}_3,{\bf k}_4;\tau_1,\tau_2)~a_{{\bf k}_1}a^{\dagger}_{-{\bf k}_2}a^{\dagger}_{-{\bf k}_3}a^{\dagger}_{-{\bf k}_4}+ {\cal M}^{(1)}_{16}({\bf k}_1,{\bf k}_2,{\bf k}_3,{\bf k}_4;\tau_1,\tau_2)~a^{\dagger}_{-{\bf k}_1}a^{\dagger}_{-{\bf k}_2}a^{\dagger}_{-{\bf k}_3}a^{\dagger}_{-{\bf k}_4}\right],~~~~~~~~~
  \eea
  where we define new sets of functions, ${\cal M}^{(1)}_{j}({\bf k}_1,{\bf k}_2,{\bf k}_3,{\bf k}_4;\tau_1,\tau_2)~\forall~j=1,\cdots,16$, as:
  \bea &&{\cal M}^{(1)}_{1}({\bf k}_1,{\bf k}_2,{\bf k}_3,{\bf k}_4;\tau_1,\tau_2)=f_{{\bf k}_1}(\tau_1)f_{{\bf k}_2}(\tau_2)f_{{\bf k}_3}(\tau_1)f_{{\bf k}_4}(\tau_2),\\
 && {\cal M}^{(1)}_2({\bf k}_1,{\bf k}_2,{\bf k}_3,{\bf k}_4;\tau_1,\tau_2)=f^{*}_{-{\bf k}_1}(\tau_1)f_{{\bf k}_2}(\tau_2)f_{{\bf k}_3}(\tau_1)f_{{\bf k}_4}(\tau_2),\\
 && {\cal M}^{(1)}_{3}({\bf k}_1,{\bf k}_2,{\bf k}_3,{\bf k}_4;\tau_1,\tau_2)=f_{{\bf k}_1}(\tau_1)f^{*}_{-{\bf k}_2}(\tau_2)f_{{\bf k}_1}(\tau_1)f_{{\bf k}_4}(\tau_2),\\
  && {\cal M}^{(1)}_{4}({\bf k}_1,{\bf k}_2,{\bf k}_3,{\bf k}_4;\tau_1,\tau_2)=f^{*}_{-{\bf k}_1}(\tau_1)f^{*}_{-{\bf k}_2}(\tau_2)f_{{\bf k}_3}(\tau_1)f_{{\bf k}_4}(\tau_2),\\
 && {\cal M}^{(1)}_{5}({\bf k}_1,{\bf k}_2,{\bf k}_3,{\bf k}_4;\tau_1,\tau_2)=f_{{\bf k}_1}(\tau_1)f_{{\bf k}_2}(\tau_2)f^{*}_{-{\bf k}_3}(\tau_1)f_{{\bf k}_4}(\tau_2),\\
  && {\cal M}^{(1)}_{6}({\bf k}_1,{\bf k}_2,{\bf k}_3,{\bf k}_4;\tau_1,\tau_2)=f^{*}_{-{\bf k}_1}(\tau_1)f_{{\bf k}_2}(\tau_2)f^{*}_{-{\bf k}_3}(\tau_1)f_{{\bf k}_4}(\tau_2)\\
 && {\cal M}^{(1)}_{7}({\bf k}_1,{\bf k}_2,{\bf k}_3,{\bf k}_4;\tau_1,\tau_2)=f_{{\bf k}_1}(\tau_1)f^{*}_{-{\bf k}_2}(\tau_2)f^{*}_{-{\bf k}_3}(\tau_1)f_{{\bf k}_4}(\tau_2),\\
  && {\cal M}^{(1)}_{8}({\bf k}_1,{\bf k}_2,{\bf k}_3,{\bf k}_4;\tau_1,\tau_2)=f^{*}_{-{\bf k}_1}(\tau_1)f^{*}_{-{\bf k}_2}(\tau_2)f^{*}_{-{\bf k}_3}(\tau_1)f_{{\bf k}_4}(\tau_2)\\
 && {\cal M}^{(1)}_{9}({\bf k}_1,{\bf k}_2,{\bf k}_3,{\bf k}_4;\tau_1,\tau_2)=f_{{\bf k}_1}(\tau_1)f_{{\bf k}_2}(\tau_2)f_{{\bf k}_3}(\tau_1)f^{*}_{-{\bf k}_4}(\tau_2),\\
  && {\cal M}^{(1)}_{10}({\bf k}_1,{\bf k}_2,{\bf k}_3,{\bf k}_4;\tau_1,\tau_2)=f^{*}_{-{\bf k}_1}(\tau_1)f_{{\bf k}_2}(\tau_2)f_{{\bf k}_3}(\tau_1)f^{*}_{-{\bf k}_4}(\tau_2)\\
 && {\cal M}^{(1)}_{11}({\bf k}_1,{\bf k}_2,{\bf k}_3,{\bf k}_4;\tau_1,\tau_2)=f_{{\bf k}_1}(\tau_1)f^{*}_{-{\bf k}_2}(\tau_2)f_{{\bf k}_3}(\tau_1)f^{*}_{-{\bf k}_4}(\tau_2),\\
  && {\cal M}^{(1)}_{12}({\bf k}_1,{\bf k}_2,{\bf k}_3,{\bf k}_4;\tau_1,\tau_2)=f^{*}_{-{\bf k}_1}(\tau_1)f^{*}_{-{\bf k}_2}(\tau_2)f_{{\bf k}_3}(\tau_1)f^{*}_{-{\bf k}_4}(\tau_2)\\
 && {\cal M}^{(1)}_{13}({\bf k}_1,{\bf k}_2,{\bf k}_3,{\bf k}_4;\tau_1,\tau_2)=f_{{\bf k}_1}(\tau_1)f_{{\bf k}_2}(\tau_2)f^{*}_{-{\bf k}_3}(\tau_1)f^{*}_{-{\bf k}_4}(\tau_2),\eea
 \bea
  && {\cal M}^{(1)}_{14}({\bf k}_1,{\bf k}_2,{\bf k}_3,{\bf k}_4;\tau_1,\tau_2)=f^{*}_{-{\bf k}_1}(\tau_1)f_{{\bf k}_2}(\tau_2)f^{*}_{-{\bf k}_3}(\tau_1)f^{*}_{-{\bf k}_4}(\tau_2)\\
 && {\cal M}^{(1)}_{15}({\bf k}_1,{\bf k}_2,{\bf k}_3,{\bf k}_4;\tau_1,\tau_2)=f_{{\bf k}_1}(\tau_1)f^{*}_{-{\bf k}_2}(\tau_2)f^{*}_{-{\bf k}_3}(\tau_1)f^{*}_{-{\bf k}_4}(\tau_2),\\
  && {\cal M}^{(1)}_{16}({\bf k}_1,{\bf k}_2,{\bf k}_3,{\bf k}_4;\tau_1,\tau_2)=f^{*}_{-{\bf k}_1}(\tau_1)f^{*}_{-{\bf k}_2}(\tau_2)f^{*}_{-{\bf k}_3}(\tau_1)f^{*}_{-{\bf k}_4}(\tau_2)\eea
  The function $\widehat{\cal T}^{(2)}_1({\bf k}_1,{\bf k}_2,{\bf k}_3,{\bf k}_4;\tau_1,\tau_2)$ is defined as:
 \bea &&\widehat{\cal T}^{(2)}_1({\bf k}_1,{\bf k}_2,{\bf k}_3,{\bf k}_4;\tau_1,\tau_2)\nonumber\\
 &&=\left[{\cal M}^{(2)}_1({\bf k}_1,{\bf k}_2,{\bf k}_3,{\bf k}_4;\tau_1,\tau_2)~a_{{\bf k}_1}a_{{\bf k}_2}a_{{\bf k}_3}a_{{\bf k}_4}\right.\nonumber\\
  && \left.+ {\cal M}^{(2)}_2({\bf k}_1,{\bf k}_2,{\bf k}_3,{\bf k}_4;\tau_1,\tau_2)~a^{\dagger}_{-{\bf k}_1}a_{{\bf k}_2}a_{{\bf k}_3}a_{{\bf k}_4}+{\cal M}^{(2)}_3({\bf k}_1,{\bf k}_2,{\bf k}_3,{\bf k}_4;\tau_1,\tau_2)~a_{{\bf k}_1}a^{\dagger}_{-{\bf k}_2}a_{{\bf k}_3}a_{{\bf k}_4}\right.\nonumber\\
  && \left.+ {\cal M}^{(2)}_4({\bf k}_1,{\bf k}_2,{\bf k}_3,{\bf k}_4;\tau_1,\tau_2)~a^{\dagger}_{-{\bf k}_1}a^{\dagger}_{-{\bf k}_2}a_{{\bf k}_3}a_{{\bf k}_4}+{\cal M}^{(2)}_5({\bf k}_1,{\bf k}_2,{\bf k}_3,{\bf k}_4;\tau_1,\tau_2)~a_{{\bf k}_1}a_{{\bf k}_2}a^{\dagger}_{-{\bf k}_3}a_{{\bf k}_4}\right.\nonumber\\
  && \left.+ {\cal M}^{(2)}_6({\bf k}_1,{\bf k}_2,{\bf k}_3,{\bf k}_4;\tau_1,\tau_2)~a^{\dagger}_{-{\bf k}_1}a_{{\bf k}_2}a^{\dagger}_{-{\bf k}_3}a_{{\bf k}_4}+{\cal M}^{(2)}_7({\bf k}_1,{\bf k}_2,{\bf k}_3,{\bf k}_4;\tau_1,\tau_2)~a_{{\bf k}_1}a^{\dagger}_{-{\bf k}_2}a^{\dagger}_{-{\bf k}_3}a_{{\bf k}_4}\right.\nonumber\\
  && \left.+ {\cal M}^{(2)}_8({\bf k}_1,{\bf k}_2,{\bf k}_3,{\bf k}_4;\tau_1,\tau_2)~a^{\dagger}_{-{\bf k}_1}a^{\dagger}_{-{\bf k}_2}a^{\dagger}_{-{\bf k}_3}a_{{\bf k}_4}+{\cal M}^{(2)}_9({\bf k}_1,{\bf k}_2,{\bf k}_3,{\bf k}_4;\tau_1,\tau_2)~a_{{\bf k}_1}a_{{\bf k}_2}a_{{\bf k}_3}a^{\dagger}_{-{\bf k}_4}\right.\nonumber\\
  && \left.+ {\cal M}^{(2)}_{10}({\bf k}_1,{\bf k}_2,{\bf k}_3,{\bf k}_4;\tau_1,\tau_2)~a^{\dagger}_{-{\bf k}_1}a_{{\bf k}_2}a_{{\bf k}_3}a^{\dagger}_{-{\bf k}_4}+{\cal M}^{(2)}_{11}({\bf k}_1,{\bf k}_2,{\bf k}_3,{\bf k}_4;\tau_1,\tau_2)~a_{{\bf k}_1}a^{\dagger}_{-{\bf k}_2}a_{{\bf k}_3}a^{\dagger}_{-{\bf k}_4}\right.\nonumber\\
  && \left.+ {\cal M}^{(2)}_{12}({\bf k}_1,{\bf k}_2,{\bf k}_3,{\bf k}_4;\tau_1,\tau_2)~a^{\dagger}_{-{\bf k}_1}a^{\dagger}_{-{\bf k}_2}a_{{\bf k}_3}a^{\dagger}_{-{\bf k}_4}\right.\nonumber\\
  && \left.+{\cal M}^{(2)}_{13}({\bf k}_1,{\bf k}_2,{\bf k}_3,{\bf k}_4;\tau_1,\tau_2)~a_{{\bf k}_1}a_{{\bf k}_2}a^{\dagger}_{-{\bf k}_3}a^{\dagger}_{-{\bf k}_4}+ {\cal M}^{(2)}_{14}({\bf k}_1,{\bf k}_2,{\bf k}_3,{\bf k}_4;\tau_1,\tau_2)~a^{\dagger}_{-{\bf k}_1}a_{{\bf k}_2}a^{\dagger}_{-{\bf k}_3}a^{\dagger}_{-{\bf k}_4}\right.\nonumber\\
  && \left.+{\cal M}^{(2)}_{15}({\bf k}_1,{\bf k}_2,{\bf k}_3,{\bf k}_4;\tau_1,\tau_2)~a_{{\bf k}_1}a^{\dagger}_{-{\bf k}_2}a^{\dagger}_{-{\bf k}_3}a^{\dagger}_{-{\bf k}_4}+ {\cal M}^{(2)}_{16}({\bf k}_1,{\bf k}_2,{\bf k}_3,{\bf k}_4;\tau_1,\tau_2)~a^{\dagger}_{-{\bf k}_1}a^{\dagger}_{-{\bf k}_2}a^{\dagger}_{-{\bf k}_3}a^{\dagger}_{-{\bf k}_4}\right],~~~~~~~~~~~
  \eea
  where we define new sets of functions, ${\cal M}^{(2)}_{j}({\bf k}_1,{\bf k}_2,{\bf k}_3,{\bf k}_4;\tau_1,\tau_2)~\forall~j=1,\cdots,16$, as:
  \bea &&{\cal M}^{(2)}_{1}({\bf k}_1,{\bf k}_2,{\bf k}_3,{\bf k}_4;\tau_1,\tau_2)=\Pi_{{\bf k}_1}(\tau_1)\Pi_{{\bf k}_2}(\tau_2)\Pi_{{\bf k}_3}(\tau_1)\Pi_{{\bf k}_4}(\tau_2),\\
 && {\cal M}^{(2)}_2({\bf k}_1,{\bf k}_2,{\bf k}_3,{\bf k}_4;\tau_1,\tau_2)=\Pi^{*}_{-{\bf k}_1}(\tau_1)\Pi_{{\bf k}_2}(\tau_2)\Pi_{{\bf k}_3}(\tau_1)\Pi_{{\bf k}_4}(\tau_2),\\
 && {\cal M}^{(2)}_{3}({\bf k}_1,{\bf k}_2,{\bf k}_3,{\bf k}_4;\tau_1,\tau_2)=\Pi_{{\bf k}_1}(\tau_1)\Pi^{*}_{-{\bf k}_2}(\tau_2)\Pi_{{\bf k}_1}(\tau_1)\Pi_{{\bf k}_4}(\tau_2),\\
  && {\cal M}^{(2)}_{4}({\bf k}_1,{\bf k}_2,{\bf k}_3,{\bf k}_4;\tau_1,\tau_2)=\Pi^{*}_{-{\bf k}_1}(\tau_1)\Pi^{*}_{-{\bf k}_2}(\tau_2)\Pi_{{\bf k}_3}(\tau_1)\Pi_{{\bf k}_4}(\tau_2),\\
 && {\cal M}^{(2)}_{5}({\bf k}_1,{\bf k}_2,{\bf k}_3,{\bf k}_4;\tau_1,\tau_2)=\Pi_{{\bf k}_1}(\tau_1)\Pi_{{\bf k}_2}(\tau_2)\Pi^{*}_{-{\bf k}_3}(\tau_1)\Pi_{{\bf k}_4}(\tau_2),\\
  && {\cal M}^{(2)}_{6}({\bf k}_1,{\bf k}_2,{\bf k}_3,{\bf k}_4;\tau_1,\tau_2)=\Pi^{*}_{-{\bf k}_1}(\tau_1)\Pi_{{\bf k}_2}(\tau_2)\Pi^{*}_{-{\bf k}_3}(\tau_1)\Pi_{{\bf k}_4}(\tau_2)\\
 && {\cal M}_{7}({\bf k}_1,{\bf k}_2,{\bf k}_3,{\bf k}_4;\tau_1,\tau_2)=\Pi_{{\bf k}_1}(\tau_1)\Pi^{*}_{-{\bf k}_2}(\tau_2)\Pi^{*}_{-{\bf k}_3}(\tau_1)\Pi_{{\bf k}_4}(\tau_2),\\
  && {\cal M}^{(2)}_{8}({\bf k}_1,{\bf k}_2,{\bf k}_3,{\bf k}_4;\tau_1,\tau_2)=\Pi^{*}_{-{\bf k}_1}(\tau_1)\Pi^{*}_{-{\bf k}_2}(\tau_2)\Pi^{*}_{-{\bf k}_3}(\tau_1)\Pi_{{\bf k}_4}(\tau_2)\\
 && {\cal M}^{(2)}_{9}({\bf k}_1,{\bf k}_2,{\bf k}_3,{\bf k}_4;\tau_1,\tau_2)=\Pi_{{\bf k}_1}(\tau_1)\Pi_{{\bf k}_2}(\tau_2)\Pi_{{\bf k}_3}(\tau_1)\Pi^{*}_{-{\bf k}_4}(\tau_2),\\
  && {\cal M}^{(2)}_{10}({\bf k}_1,{\bf k}_2,{\bf k}_3,{\bf k}_4;\tau_1,\tau_2)=\Pi^{*}_{-{\bf k}_1}(\tau_1)\Pi_{{\bf k}_2}(\tau_2)\Pi_{{\bf k}_3}(\tau_1)\Pi^{*}_{-{\bf k}_4}(\tau_2)\\
 && {\cal M}^{(2)}_{11}({\bf k}_1,{\bf k}_2,{\bf k}_3,{\bf k}_4;\tau_1,\tau_2)=\Pi_{{\bf k}_1}(\tau_1)\Pi^{*}_{-{\bf k}_2}(\tau_2)\Pi_{{\bf k}_3}(\tau_1)\Pi^{*}_{-{\bf k}_4}(\tau_2),\\
  && {\cal M}^{(2)}_{12}({\bf k}_1,{\bf k}_2,{\bf k}_3,{\bf k}_4;\tau_1,\tau_2)=\Pi^{*}_{-{\bf k}_1}(\tau_1)\Pi^{*}_{-{\bf k}_2}(\tau_2)\Pi_{{\bf k}_3}(\tau_1)\Pi^{*}_{-{\bf k}_4}(\tau_2)\\
 && {\cal M}^{(2)}_{13}({\bf k}_1,{\bf k}_2,{\bf k}_3,{\bf k}_4;\tau_1,\tau_2)=\Pi_{{\bf k}_1}(\tau_1)\Pi_{{\bf k}_2}(\tau_2)\Pi^{*}_{-{\bf k}_3}(\tau_1)\Pi^{*}_{-{\bf k}_4}(\tau_2),\\
  && {\cal M}^{(2)}_{14}({\bf k}_1,{\bf k}_2,{\bf k}_3,{\bf k}_4;\tau_1,\tau_2)=\Pi^{*}_{-{\bf k}_1}(\tau_1)\Pi_{{\bf k}_2}(\tau_2)\Pi^{*}_{-{\bf k}_3}(\tau_1)\Pi^{*}_{-{\bf k}_4}(\tau_2)\eea\bea
 && {\cal M}^{(2)}_{15}({\bf k}_1,{\bf k}_2,{\bf k}_3,{\bf k}_4;\tau_1,\tau_2)=\Pi_{{\bf k}_1}(\tau_1)\Pi^{*}_{-{\bf k}_2}(\tau_2)\Pi^{*}_{-{\bf k}_3}(\tau_1)\Pi^{*}_{-{\bf k}_4}(\tau_2),\\
  && {\cal M}^{(2)}_{16}({\bf k}_1,{\bf k}_2,{\bf k}_3,{\bf k}_4;\tau_1,\tau_2)=\Pi^{*}_{-{\bf k}_1}(\tau_1)\Pi^{*}_{-{\bf k}_2}(\tau_2)\Pi^{*}_{-{\bf k}_3}(\tau_1)\Pi^{*}_{-{\bf k}_4}(\tau_2)\eea
\subsection{Definition of  OTO amplitude $\widehat{\cal T}^{(1)}_2({\bf k}_1,{\bf k}_2,{\bf k}_3,{\bf k}_4;\tau_1,\tau_2)$ and $\widehat{\cal T}^{(2)}_2({\bf k}_1,{\bf k}_2,{\bf k}_3,{\bf k}_4;\tau_1,\tau_2)$}
The function $\widehat{\cal T}^{(1)}_2({\bf k}_1,{\bf k}_2,{\bf k}_3,{\bf k}_4;\tau_1,\tau_2)$ is defined as:
 \bea &&\widehat{\cal T}^{(1)}_2({\bf k}_1,{\bf k}_2,{\bf k}_3,{\bf k}_4;\tau_1,\tau_2)\nonumber\\
 &&=\left[{\cal J}^{(1)}_1({\bf k}_1,{\bf k}_2,{\bf k}_3,{\bf k}_4;\tau_1,\tau_2)~a_{{\bf k}_1}a_{{\bf k}_2}a_{{\bf k}_3}a_{{\bf k}_4}+ {\cal J}^{(1)}_2({\bf k}_1,{\bf k}_2,{\bf k}_3,{\bf k}_4;\tau_1,\tau_2)~a^{\dagger}_{-{\bf k}_1}a_{{\bf k}_2}a_{{\bf k}_3}a_{{\bf k}_4}\right.\nonumber\\
  && \left.+{\cal J}^{(1)}_3({\bf k}_1,{\bf k}_2,{\bf k}_3,{\bf k}_4;\tau_1,\tau_2)~a_{{\bf k}_1}a^{\dagger}_{-{\bf k}_2}a_{{\bf k}_3}a_{{\bf k}_4}+ {\cal J}^{(1)}_4({\bf k}_1,{\bf k}_2,{\bf k}_3,{\bf k}_4;\tau_1,\tau_2)~a^{\dagger}_{-{\bf k}_1}a^{\dagger}_{-{\bf k}_2}a_{{\bf k}_3}a_{{\bf k}_4}\right.\nonumber\\
  && \left.+{\cal J}^{(1)}_5({\bf k}_1,{\bf k}_2,{\bf k}_3,{\bf k}_4;\tau_1,\tau_2)~a_{{\bf k}_1}a_{{\bf k}_2}a^{\dagger}_{-{\bf k}_3}a_{{\bf k}_4}+ {\cal J}^{(1)}_6({\bf k}_1,{\bf k}_2,{\bf k}_3,{\bf k}_4;\tau_1,\tau_2)~a^{\dagger}_{-{\bf k}_1}a_{{\bf k}_2}a^{\dagger}_{-{\bf k}_3}a_{{\bf k}_4}\right.\nonumber\\
  && \left.+{\cal J}^{(1)}_7({\bf k}_1,{\bf k}_2,{\bf k}_3,{\bf k}_4;\tau_1,\tau_2)~a_{{\bf k}_1}a^{\dagger}_{-{\bf k}_2}a^{\dagger}_{-{\bf k}_3}a_{{\bf k}_4}+ {\cal J}^{(1)}_8({\bf k}_1,{\bf k}_2,{\bf k}_3,{\bf k}_4;\tau_1,\tau_2)~a^{\dagger}_{-{\bf k}_1}a^{\dagger}_{-{\bf k}_2}a^{\dagger}_{-{\bf k}_3}a_{{\bf k}_4}\right.\nonumber\\
  && \left.+{\cal J}^{(1)}_9({\bf k}_1,{\bf k}_2,{\bf k}_3,{\bf k}_4;\tau_1,\tau_2)~a_{{\bf k}_1}a_{{\bf k}_2}a_{{\bf k}_3}a^{\dagger}_{-{\bf k}_4}+ {\cal J}^{(1)}_{10}({\bf k}_1,{\bf k}_2,{\bf k}_3,{\bf k}_4;\tau_1,\tau_2)~a^{\dagger}_{-{\bf k}_1}a_{{\bf k}_2}a_{{\bf k}_3}a^{\dagger}_{-{\bf k}_4}\right.\nonumber\\
  && \left.+{\cal J}^{(1)}_{11}({\bf k}_1,{\bf k}_2,{\bf k}_3,{\bf k}_4;\tau_1,\tau_2)~a_{{\bf k}_1}a^{\dagger}_{-{\bf k}_2}a_{{\bf k}_3}a^{\dagger}_{-{\bf k}_4}+ {\cal J}^{(1)}_{12}({\bf k}_1,{\bf k}_2,{\bf k}_3,{\bf k}_4;\tau_1,\tau_2)~a^{\dagger}_{-{\bf k}_1}a^{\dagger}_{-{\bf k}_2}a_{{\bf k}_3}a^{\dagger}_{-{\bf k}_4}\right.\nonumber\\
  && \left.+{\cal J}^{(1)}_{13}({\bf k}_1,{\bf k}_2,{\bf k}_3,{\bf k}_4;\tau_1,\tau_2)~a_{{\bf k}_1}a_{{\bf k}_2}a^{\dagger}_{-{\bf k}_3}a^{\dagger}_{-{\bf k}_4}+ {\cal J}^{(1)}_{14}({\bf k}_1,{\bf k}_2,{\bf k}_3,{\bf k}_4;\tau_1,\tau_2)~a^{\dagger}_{-{\bf k}_1}a_{{\bf k}_2}a^{\dagger}_{-{\bf k}_3}a^{\dagger}_{-{\bf k}_4}\right.\nonumber\\
  && \left.+{\cal J}^{(1)}_{15}({\bf k}_1,{\bf k}_2,{\bf k}_3,{\bf k}_4;\tau_1,\tau_2)~a_{{\bf k}_1}a^{\dagger}_{-{\bf k}_2}a^{\dagger}_{-{\bf k}_3}a^{\dagger}_{-{\bf k}_4}+ {\cal J}^{(1)}_{16}({\bf k}_1,{\bf k}_2,{\bf k}_3,{\bf k}_4;\tau_1,\tau_2)~a^{\dagger}_{-{\bf k}_1}a^{\dagger}_{-{\bf k}_2}a^{\dagger}_{-{\bf k}_3}a^{\dagger}_{-{\bf k}_4}\right],~~~~~~~~~~~
  \eea
  where we define new sets of functions, ${\cal J}^{(1)}_{j}({\bf k}_1,{\bf k}_2,{\bf k}_3,{\bf k}_4;\tau_1,\tau_2)~\forall~j=1,\cdots,16$, as:
  \bea &&{\cal J}^{(1)}_{1}({\bf k}_1,{\bf k}_2,{\bf k}_3,{\bf k}_4;\tau_1,\tau_2)=f_{{\bf k}_1}(\tau_2)f_{{\bf k}_2}(\tau_1)f_{{\bf k}_3}(\tau_1)f_{{\bf k}_4}(\tau_2),\\
 && {\cal J}^{(1)}_2({\bf k}_1,{\bf k}_2,{\bf k}_3,{\bf k}_4;\tau_1,\tau_2)=f^{*}_{-{\bf k}_1}(\tau_2)f_{{\bf k}_2}(\tau_1)f_{{\bf k}_3}(\tau_1)f_{{\bf k}_4}(\tau_2),\\
 && {\cal J}^{(1)}_{3}({\bf k}_1,{\bf k}_2,{\bf k}_3,{\bf k}_4;\tau_1,\tau_2)=f_{{\bf k}_1}(\tau_2)f^{*}_{-{\bf k}_2}(\tau_1)f_{{\bf k}_3}(\tau_1)f_{{\bf k}_4}(\tau_2),\\
  && {\cal J}^{(1)}_{4}({\bf k}_1,{\bf k}_2,{\bf k}_3,{\bf k}_4;\tau_1,\tau_2)=f^{*}_{-{\bf k}_1}(\tau_2)f^{*}_{-{\bf k}_2}(\tau_1)f_{{\bf k}_3}(\tau_1)f_{{\bf k}_4}(\tau_2),\\
 && {\cal J}^{(1)}_{5}({\bf k}_1,{\bf k}_2,{\bf k}_3,{\bf k}_4;\tau_1,\tau_2)=f_{{\bf k}_1}(\tau_2)f_{{\bf k}_2}(\tau_1)f^{*}_{-{\bf k}_3}(\tau_1)f_{{\bf k}_4}(\tau_2),\\
  && {\cal J}^{(1)}_{6}({\bf k}_1,{\bf k}_2,{\bf k}_3,{\bf k}_4;\tau_1,\tau_2)=f^{*}_{-{\bf k}_1}(\tau_2)f_{{\bf k}_2}(\tau_1)f^{*}_{-{\bf k}_3}(\tau_1)f_{{\bf k}_4}(\tau_2)\\
 && {\cal J}^{(1)}_{7}({\bf k}_1,{\bf k}_2,{\bf k}_3,{\bf k}_4;\tau_1,\tau_2)=f_{{\bf k}_1}(\tau_2)f^{*}_{-{\bf k}_2}(\tau_1)f^{*}_{-{\bf k}_3}(\tau_1)f_{{\bf k}_4}(\tau_2),\\
  && {\cal J}^{(1)}_{8}({\bf k}_1,{\bf k}_2,{\bf k}_3,{\bf k}_4;\tau_1,\tau_2)=f^{*}_{-{\bf k}_1}(\tau_2)f^{*}_{-{\bf k}_2}(\tau_1)f^{*}_{-{\bf k}_3}(\tau_1)f_{{\bf k}_4}(\tau_2)\\
 && {\cal J}^{(1)}_{9}({\bf k}_1,{\bf k}_2,{\bf k}_3,{\bf k}_4;\tau_1,\tau_2)=f_{{\bf k}_1}(\tau_2)f_{{\bf k}_2}(\tau_1)f_{{\bf k}_3}(\tau_1)f^{*}_{-{\bf k}_4}(\tau_2),\\
  && {\cal J}^{(1)}_{10}({\bf k}_1,{\bf k}_2,{\bf k}_3,{\bf k}_4;\tau_1,\tau_2)=f^{*}_{-{\bf k}_1}(\tau_2)f_{{\bf k}_2}(\tau_1)f_{{\bf k}_3}(\tau_1)f^{*}_{-{\bf k}_4}(\tau_2)\\
 && {\cal J}^{(1)}_{11}({\bf k}_1,{\bf k}_2,{\bf k}_3,{\bf k}_4;\tau_1,\tau_2)=f_{{\bf k}_1}(\tau_2)f^{*}_{-{\bf k}_2}(\tau_1)f_{{\bf k}_3}(\tau_1)f^{*}_{-{\bf k}_4}(\tau_2),\\
  && {\cal J}^{(1)}_{12}({\bf k}_1,{\bf k}_2,{\bf k}_3,{\bf k}_4;\tau_1,\tau_2)=f^{*}_{-{\bf k}_1}(\tau_2)f^{*}_{-{\bf k}_2}(\tau_1)f_{{\bf k}_3}(\tau_1)f^{*}_{-{\bf k}_4}(\tau_2)\\
 && {\cal J}^{(1)}_{13}({\bf k}_1,{\bf k}_2,{\bf k}_3,{\bf k}_4;\tau_1,\tau_2)=f_{{\bf k}_1}(\tau_2)f_{{\bf k}_2}(\tau_1)f^{*}_{-{\bf k}_3}(\tau_1)f^{*}_{-{\bf k}_4}(\tau_2),\\
  && {\cal J}^{(1)}_{14}({\bf k}_1,{\bf k}_2,{\bf k}_3,{\bf k}_4;\tau_1,\tau_2)=f^{*}_{-{\bf k}_1}(\tau_2)f_{{\bf k}_2}(\tau_1)f^{*}_{-{\bf k}_3}(\tau_1)f^{*}_{-{\bf k}_4}(\tau_2)\eea\bea
 && {\cal J}^{(1)}_{15}({\bf k}_1,{\bf k}_2,{\bf k}_3,{\bf k}_4;\tau_1,\tau_2)=f_{{\bf k}_1}(\tau_2)f^{*}_{-{\bf k}_2}(\tau_1)f^{*}_{-{\bf k}_3}(\tau_1)f^{*}_{-{\bf k}_4}(\tau_2),\\
  && {\cal J}^{(1)}_{16}({\bf k}_1,{\bf k}_2,{\bf k}_3,{\bf k}_4;\tau_1,\tau_2)=f^{*}_{-{\bf k}_1}(\tau_2)f^{*}_{-{\bf k}_2}(\tau_1)f^{*}_{-{\bf k}_3}(\tau_1)f^{*}_{-{\bf k}_4}(\tau_2)\eea
  The function $\widehat{\cal T}^{(2)}_2({\bf k}_1,{\bf k}_2,{\bf k}_3,{\bf k}_4;\tau_1,\tau_2)$ is defined as:
 \bea &&\widehat{\cal T}^{(2)}_2({\bf k}_1,{\bf k}_2,{\bf k}_3,{\bf k}_4;\tau_1,\tau_2)\nonumber\\
 &&=\left[{\cal J}^{(2)}_1({\bf k}_1,{\bf k}_2,{\bf k}_3,{\bf k}_4;\tau_1,\tau_2)~a_{{\bf k}_1}a_{{\bf k}_2}a_{{\bf k}_3}a_{{\bf k}_4}+ {\cal J}^{(2)}_2({\bf k}_1,{\bf k}_2,{\bf k}_3,{\bf k}_4;\tau_1,\tau_2)~a^{\dagger}_{-{\bf k}_1}a_{{\bf k}_2}a_{{\bf k}_3}a_{{\bf k}_4}\right.\nonumber\\
  && \left.+{\cal J}^{(2)}_3({\bf k}_1,{\bf k}_2,{\bf k}_3,{\bf k}_4;\tau_1,\tau_2)~a_{{\bf k}_1}a^{\dagger}_{-{\bf k}_2}a_{{\bf k}_3}a_{{\bf k}_4}+ {\cal J}^{(2)}_4({\bf k}_1,{\bf k}_2,{\bf k}_3,{\bf k}_4;\tau_1,\tau_2)~a^{\dagger}_{-{\bf k}_1}a^{\dagger}_{-{\bf k}_2}a_{{\bf k}_3}a_{{\bf k}_4}\right.\nonumber\\
  && \left.+{\cal J}^{(2)}_5({\bf k}_1,{\bf k}_2,{\bf k}_3,{\bf k}_4;\tau_1,\tau_2)~a_{{\bf k}_1}a_{{\bf k}_2}a^{\dagger}_{-{\bf k}_3}a_{{\bf k}_4}+ {\cal J}^{(2)}_6({\bf k}_1,{\bf k}_2,{\bf k}_3,{\bf k}_4;\tau_1,\tau_2)~a^{\dagger}_{-{\bf k}_1}a_{{\bf k}_2}a^{\dagger}_{-{\bf k}_3}a_{{\bf k}_4}\right.\nonumber\\
  && \left.+{\cal J}^{(2)}_7({\bf k}_1,{\bf k}_2,{\bf k}_3,{\bf k}_4;\tau_1,\tau_2)~a_{{\bf k}_1}a^{\dagger}_{-{\bf k}_2}a^{\dagger}_{-{\bf k}_3}a_{{\bf k}_4}+ {\cal J}^{(2)}_8({\bf k}_1,{\bf k}_2,{\bf k}_3,{\bf k}_4;\tau_1,\tau_2)~a^{\dagger}_{-{\bf k}_1}a^{\dagger}_{-{\bf k}_2}a^{\dagger}_{-{\bf k}_3}a_{{\bf k}_4}\right.\nonumber\\
  && \left.+{\cal J}^{(2)}_9({\bf k}_1,{\bf k}_2,{\bf k}_3,{\bf k}_4;\tau_1,\tau_2)~a_{{\bf k}_1}a_{{\bf k}_2}a_{{\bf k}_3}a^{\dagger}_{-{\bf k}_4}+ {\cal J}^{(2)}_{10}({\bf k}_1,{\bf k}_2,{\bf k}_3,{\bf k}_4;\tau_1,\tau_2)~a^{\dagger}_{-{\bf k}_1}a_{{\bf k}_2}a_{{\bf k}_3}a^{\dagger}_{-{\bf k}_4}\right.\nonumber\\
  && \left.+{\cal J}^{(2)}_{11}({\bf k}_1,{\bf k}_2,{\bf k}_3,{\bf k}_4;\tau_1,\tau_2)~a_{{\bf k}_1}a^{\dagger}_{-{\bf k}_2}a_{{\bf k}_3}a^{\dagger}_{-{\bf k}_4}+ {\cal J}^{(2)}_{12}({\bf k}_1,{\bf k}_2,{\bf k}_3,{\bf k}_4;\tau_1,\tau_2)~a^{\dagger}_{-{\bf k}_1}a^{\dagger}_{-{\bf k}_2}a_{{\bf k}_3}a^{\dagger}_{-{\bf k}_4}\right.\nonumber\\
  && \left.+{\cal J}^{(2)}_{13}({\bf k}_1,{\bf k}_2,{\bf k}_3,{\bf k}_4;\tau_1,\tau_2)~a_{{\bf k}_1}a_{{\bf k}_2}a^{\dagger}_{-{\bf k}_3}a^{\dagger}_{-{\bf k}_4}+ {\cal J}^{(2)}_{14}({\bf k}_1,{\bf k}_2,{\bf k}_3,{\bf k}_4;\tau_1,\tau_2)~a^{\dagger}_{-{\bf k}_1}a_{{\bf k}_2}a^{\dagger}_{-{\bf k}_3}a^{\dagger}_{-{\bf k}_4}\right.\nonumber\\
  && \left.+{\cal J}^{(2)}_{15}({\bf k}_1,{\bf k}_2,{\bf k}_3,{\bf k}_4;\tau_1,\tau_2)~a_{{\bf k}_1}a^{\dagger}_{-{\bf k}_2}a^{\dagger}_{-{\bf k}_3}a^{\dagger}_{-{\bf k}_4}+ {\cal J}^{(2)}_{16}({\bf k}_1,{\bf k}_2,{\bf k}_3,{\bf k}_4;\tau_1,\tau_2)~a^{\dagger}_{-{\bf k}_1}a^{\dagger}_{-{\bf k}_2}a^{\dagger}_{-{\bf k}_3}a^{\dagger}_{-{\bf k}_4}\right],~~~~~~~~~~~
  \eea
  where we define new sets of functions, ${\cal J}^{(2)}_{j}({\bf k}_1,{\bf k}_2,{\bf k}_3,{\bf k}_4;\tau_1,\tau_2)~\forall~j=1,\cdots,16$, as:
  \bea &&{\cal J}^{(2)}_{1}({\bf k}_1,{\bf k}_2,{\bf k}_3,{\bf k}_4;\tau_1,\tau_2)=\Pi_{{\bf k}_1}(\tau_2)\Pi_{{\bf k}_2}(\tau_1)\Pi_{{\bf k}_3}(\tau_1)\Pi_{{\bf k}_4}(\tau_2),\\
 && {\cal J}^{(2)}_2({\bf k}_1,{\bf k}_2,{\bf k}_3,{\bf k}_4;\tau_1,\tau_2)=\Pi^{*}_{-{\bf k}_1}(\tau_2)\Pi_{{\bf k}_2}(\tau_1)\Pi_{{\bf k}_3}(\tau_1)\Pi_{{\bf k}_4}(\tau_2),\\
 && {\cal J}^{(2)}_{3}({\bf k}_1,{\bf k}_2,{\bf k}_3,{\bf k}_4;\tau_1,\tau_2)=\Pi_{{\bf k}_1}(\tau_2)\Pi^{*}_{-{\bf k}_2}(\tau_1)\Pi_{{\bf k}_3}(\tau_1)\Pi_{{\bf k}_4}(\tau_2),\\
  && {\cal J}^{(2)}_{4}({\bf k}_1,{\bf k}_2,{\bf k}_3,{\bf k}_4;\tau_1,\tau_2)=\Pi^{*}_{-{\bf k}_1}(\tau_2)\Pi^{*}_{-{\bf k}_2}(\tau_1)\Pi_{{\bf k}_3}(\tau_1)\Pi_{{\bf k}_4}(\tau_2),\\
 && {\cal J}^{(2)}_{5}({\bf k}_1,{\bf k}_2,{\bf k}_3,{\bf k}_4;\tau_1,\tau_2)=\Pi_{{\bf k}_1}(\tau_2)\Pi_{{\bf k}_2}(\tau_1)\Pi^{*}_{-{\bf k}_3}(\tau_1)\Pi_{{\bf k}_4}(\tau_2),\\
  && {\cal J}^{(2)}_{6}({\bf k}_1,{\bf k}_2,{\bf k}_3,{\bf k}_4;\tau_1,\tau_2)=\Pi^{*}_{-{\bf k}_1}(\tau_2)\Pi_{{\bf k}_2}(\tau_1)\Pi^{*}_{-{\bf k}_3}(\tau_1)\Pi_{{\bf k}_4}(\tau_2)\\
 && {\cal J}^{(2)}_{7}({\bf k}_1,{\bf k}_2,{\bf k}_3,{\bf k}_4;\tau_1,\tau_2)=\Pi_{{\bf k}_1}(\tau_2)\Pi^{*}_{-{\bf k}_2}(\tau_1)\Pi^{*}_{-{\bf k}_3}(\tau_1)\Pi_{{\bf k}_4}(\tau_2),\\
  && {\cal J}^{(2)}_{8}({\bf k}_1,{\bf k}_2,{\bf k}_3,{\bf k}_4;\tau_1,\tau_2)=\Pi^{*}_{-{\bf k}_1}(\tau_2)\Pi^{*}_{-{\bf k}_2}(\tau_1)\Pi^{*}_{-{\bf k}_3}(\tau_1)\Pi_{{\bf k}_4}(\tau_2)\\
 && {\cal J}^{(1)}_{9}({\bf k}_1,{\bf k}_2,{\bf k}_3,{\bf k}_4;\tau_1,\tau_2)=\Pi_{{\bf k}_1}(\tau_2)\Pi_{{\bf k}_2}(\tau_1)\Pi_{{\bf k}_3}(\tau_1)\Pi^{*}_{-{\bf k}_4}(\tau_2),\\
  && {\cal J}^{(2)}_{10}({\bf k}_1,{\bf k}_2,{\bf k}_3,{\bf k}_4;\tau_1,\tau_2)=\Pi^{*}_{-{\bf k}_1}(\tau_2)\Pi_{{\bf k}_2}(\tau_1)\Pi_{{\bf k}_3}(\tau_1)\Pi^{*}_{-{\bf k}_4}(\tau_2)\\
 && {\cal J}^{(2)}_{11}({\bf k}_1,{\bf k}_2,{\bf k}_3,{\bf k}_4;\tau_1,\tau_2)=\Pi_{{\bf k}_1}(\tau_2)\Pi^{*}_{-{\bf k}_2}(\tau_1)\Pi_{{\bf k}_3}(\tau_1)\Pi^{*}_{-{\bf k}_4}(\tau_2),\\
  && {\cal J}^{(2)}_{12}({\bf k}_1,{\bf k}_2,{\bf k}_3,{\bf k}_4;\tau_1,\tau_2)=\Pi^{*}_{-{\bf k}_1}(\tau_2)\Pi^{*}_{-{\bf k}_2}(\tau_1)\Pi_{{\bf k}_3}(\tau_1)\Pi^{*}_{-{\bf k}_4}(\tau_2)\\
 && {\cal J}^{(1)}_{13}({\bf k}_1,{\bf k}_2,{\bf k}_3,{\bf k}_4;\tau_1,\tau_2)=\Pi_{{\bf k}_1}(\tau_2)\Pi_{{\bf k}_2}(\tau_1)f^{*}_{-{\bf k}_3}(\tau_1)\Pi^{*}_{-{\bf k}_4}(\tau_2),\\
  && {\cal J}^{(2)}_{14}({\bf k}_1,{\bf k}_2,{\bf k}_3,{\bf k}_4;\tau_1,\tau_2)=\Pi^{*}_{-{\bf k}_1}(\tau_2)\Pi_{{\bf k}_2}(\tau_1)\Pi^{*}_{-{\bf k}_3}(\tau_1)\Pi^{*}_{-{\bf k}_4}(\tau_2)\\
 && {\cal J}^{(2)}_{15}({\bf k}_1,{\bf k}_2,{\bf k}_3,{\bf k}_4;\tau_1,\tau_2)=\Pi_{{\bf k}_1}(\tau_2)\Pi^{*}_{-{\bf k}_2}(\tau_1)\Pi^{*}_{-{\bf k}_3}(\tau_1)\Pi^{*}_{-{\bf k}_4}(\tau_2),\\
  && {\cal J}^{(2)}_{16}({\bf k}_1,{\bf k}_2,{\bf k}_3,{\bf k}_4;\tau_1,\tau_2)=\Pi^{*}_{-{\bf k}_1}(\tau_2)\Pi^{*}_{-{\bf k}_2}(\tau_1)\Pi^{*}_{-{\bf k}_3}(\tau_1)\Pi^{*}_{-{\bf k}_4}(\tau_2)\eea
\subsection{Definition of  OTO amplitude $\widehat{\cal T}^{(1)}_3({\bf k}_1,{\bf k}_2,{\bf k}_3,{\bf k}_4;\tau_1,\tau_2)$ and $\widehat{\cal T}^{(2)}_3({\bf k}_1,{\bf k}_2,{\bf k}_3,{\bf k}_4;\tau_1,\tau_2)$}
  The function $\widehat{\cal T}^{(1)}_3({\bf k}_1,{\bf k}_2,{\bf k}_3,{\bf k}_4;\tau_1,\tau_2)$ is defined as:
 \bea &&\widehat{\cal T}^{(1)}_3({\bf k}_1,{\bf k}_2,{\bf k}_3,{\bf k}_4;\tau_1,\tau_2)\nonumber\\
 &&=\left[{\cal N}^{(1)}_1({\bf k}_1,{\bf k}_2,{\bf k}_3,{\bf k}_4;\tau_1,\tau_2)~a_{{\bf k}_1}a_{{\bf k}_2}a_{{\bf k}_3}a_{{\bf k}_4}+ {\cal N}^{(1)}_2({\bf k}_1,{\bf k}_2,{\bf k}_3,{\bf k}_4;\tau_1,\tau_2)~a^{\dagger}_{-{\bf k}_1}a_{{\bf k}_2}a_{{\bf k}_3}a_{{\bf k}_4}\right.\nonumber\\
  && \left.+{\cal N}^{(1)}_3({\bf k}_1,{\bf k}_2,{\bf k}_3,{\bf k}_4;\tau_1,\tau_2)~a_{{\bf k}_1}a^{\dagger}_{-{\bf k}_2}a_{{\bf k}_3}a_{{\bf k}_4}+ {\cal N}^{(1)}_4({\bf k}_1,{\bf k}_2,{\bf k}_3,{\bf k}_4;\tau_1,\tau_2)~a^{\dagger}_{-{\bf k}_1}a^{\dagger}_{-{\bf k}_2}a_{{\bf k}_3}a_{{\bf k}_4}\right.\nonumber\\
  && \left.+{\cal N}^{(1)}_5({\bf k}_1,{\bf k}_2,{\bf k}_3,{\bf k}_4;\tau_1,\tau_2)~a_{{\bf k}_1}a_{{\bf k}_2}a^{\dagger}_{-{\bf k}_3}a_{{\bf k}_4}+ {\cal N}^{(1)}_6({\bf k}_1,{\bf k}_2,{\bf k}_3,{\bf k}_4;\tau_1,\tau_2)~a^{\dagger}_{-{\bf k}_1}a_{{\bf k}_2}a^{\dagger}_{-{\bf k}_3}a_{{\bf k}_4}\right.\nonumber\\
  && \left.+{\cal N}^{(1)}_7({\bf k}_1,{\bf k}_2,{\bf k}_3,{\bf k}_4;\tau_1,\tau_2)~a_{{\bf k}_1}a^{\dagger}_{-{\bf k}_2}a^{\dagger}_{-{\bf k}_3}a_{{\bf k}_4}+ {\cal N}^{(1)}_8({\bf k}_1,{\bf k}_2,{\bf k}_3,{\bf k}_4;\tau_1,\tau_2)~a^{\dagger}_{-{\bf k}_1}a^{\dagger}_{-{\bf k}_2}a^{\dagger}_{-{\bf k}_3}a_{{\bf k}_4}\right.\nonumber\\
  && \left.+{\cal N}^{(1)}_9({\bf k}_1,{\bf k}_2,{\bf k}_3,{\bf k}_4;\tau_1,\tau_2)~a_{{\bf k}_1}a_{{\bf k}_2}a_{{\bf k}_3}a^{\dagger}_{-{\bf k}_4}+ {\cal N}^{(1)}_{10}({\bf k}_1,{\bf k}_2,{\bf k}_3,{\bf k}_4;\tau_1,\tau_2)~a^{\dagger}_{-{\bf k}_1}a_{{\bf k}_2}a_{{\bf k}_3}a^{\dagger}_{-{\bf k}_4}\right.\nonumber\\
  && \left.+{\cal N}^{(1)}_{11}({\bf k}_1,{\bf k}_2,{\bf k}_3,{\bf k}_4;\tau_1,\tau_2)~a_{{\bf k}_1}a^{\dagger}_{-{\bf k}_2}a_{{\bf k}_3}a^{\dagger}_{-{\bf k}_4}+ {\cal N}^{(1)}_{12}({\bf k}_1,{\bf k}_2,{\bf k}_3,{\bf k}_4;\tau_1,\tau_2)~a^{\dagger}_{-{\bf k}_1}a^{\dagger}_{-{\bf k}_2}a_{{\bf k}_3}a^{\dagger}_{-{\bf k}_4}\right.\nonumber\\
  && \left.+{\cal N}^{(1)}_{13}({\bf k}_1,{\bf k}_2,{\bf k}_3,{\bf k}_4;\tau_1,\tau_2)~a_{{\bf k}_1}a_{{\bf k}_2}a^{\dagger}_{-{\bf k}_3}a^{\dagger}_{-{\bf k}_4}+ {\cal N}^{(1)}_{14}({\bf k}_1,{\bf k}_2,{\bf k}_3,{\bf k}_4;\tau_1,\tau_2)~a^{\dagger}_{-{\bf k}_1}a_{{\bf k}_2}a^{\dagger}_{-{\bf k}_3}a^{\dagger}_{-{\bf k}_4}\right.\nonumber\\
  && \left.+{\cal N}^{(1)}_{15}({\bf k}_1,{\bf k}_2,{\bf k}_3,{\bf k}_4;\tau_1,\tau_2)~a_{{\bf k}_1}a^{\dagger}_{-{\bf k}_2}a^{\dagger}_{-{\bf k}_3}a^{\dagger}_{-{\bf k}_4}+ {\cal N}^{(1)}_{16}({\bf k}_1,{\bf k}_2,{\bf k}_3,{\bf k}_4;\tau_1,\tau_2)~a^{\dagger}_{-{\bf k}_1}a^{\dagger}_{-{\bf k}_2}a^{\dagger}_{-{\bf k}_3}a^{\dagger}_{-{\bf k}_4}\right],~~~~~~~~~~~
  \eea
  where we define new sets of functions, ${\cal N}^{(1)}_{j}({\bf k}_1,{\bf k}_2,{\bf k}_3,{\bf k}_4;\tau_1,\tau_2)~\forall~j=1,\cdots,16$, as:
  \bea &&{\cal N}^{(1)}_{1}({\bf k}_1,{\bf k}_2,{\bf k}_3,{\bf k}_4;\tau_1,\tau_2)=f_{{\bf k}_1}(\tau_1)f_{{\bf k}_2}(\tau_2)f_{{\bf k}_3}(\tau_2)f_{{\bf k}_4}(\tau_1),\\
 && {\cal N}^{(1)}_2({\bf k}_1,{\bf k}_2,{\bf k}_3,{\bf k}_4;\tau_1,\tau_2)=f_{{\bf k}_1}(\tau_2)f^{*}_{-{\bf k}_2}(\tau_1)f_{{\bf k}_3}(\tau_2)f_{{\bf k}_4}(\tau_1),\\
 && {\cal N}^{(1)}_{3}({\bf k}_1,{\bf k}_2,{\bf k}_3,{\bf k}_4;\tau_1,\tau_2)=f_{{\bf k}_1}(\tau_1)f^{*}_{-{\bf k}_2}(\tau_2)f_{{\bf k}_1}(\tau_2)f_{{\bf k}_4}(\tau_1),\\
  && {\cal N}^{(1)}_{4}({\bf k}_1,{\bf k}_2,{\bf k}_3,{\bf k}_4;\tau_1,\tau_2)=f^{*}_{-{\bf k}_1}(\tau_1)f^{*}_{-{\bf k}_2}(\tau_2)f_{{\bf k}_3}(\tau_2)f_{{\bf k}_4}(\tau_1),\\
 && {\cal N}^{(1)}_{5}({\bf k}_1,{\bf k}_2,{\bf k}_3,{\bf k}_4;\tau_1,\tau_2)=f_{{\bf k}_1}(\tau_1)f_{{\bf k}_2}(\tau_2)f^{*}_{-{\bf k}_3}(\tau_2)f_{{\bf k}_4}(\tau_1),\\
  && {\cal N}^{(1)}_{6}({\bf k}_1,{\bf k}_2,{\bf k}_3,{\bf k}_4;\tau_1,\tau_2)=f^{*}_{-{\bf k}_1}(\tau_1)f_{{\bf k}_2}(\tau_2)f^{*}_{-{\bf k}_3}(\tau_2)f_{{\bf k}_4}(\tau_1)\\
 && {\cal N}_{7}({\bf k}_1,{\bf k}_2,{\bf k}_3,{\bf k}_4;\tau_1,\tau_2)=f_{{\bf k}_1}(\tau_1)f^{*}_{-{\bf k}_2}(\tau_2)f^{*}_{-{\bf k}_3}(\tau_2)f_{{\bf k}_4}(\tau_1),\\
  && {\cal N}^{(1)}_{8}({\bf k}_1,{\bf k}_2,{\bf k}_3,{\bf k}_4;\tau_1,\tau_2)=f^{*}_{-{\bf k}_1}(\tau_1)f^{*}_{-{\bf k}_2}(\tau_2)f^{*}_{-{\bf k}_3}(\tau_2)f_{{\bf k}_4}(\tau_1)\\
 && {\cal N}^{(1)}_{9}({\bf k}_1,{\bf k}_2,{\bf k}_3,{\bf k}_4;\tau_1,\tau_2)=f_{{\bf k}_1}(\tau_1)f_{{\bf k}_2}(\tau_2)f_{{\bf k}_3}(\tau_2)f^{*}_{-{\bf k}_4}(\tau_1),\\
  && {\cal N}^{(1)}_{10}({\bf k}_1,{\bf k}_2,{\bf k}_3,{\bf k}_4;\tau_1,\tau_2)=f^{*}_{-{\bf k}_1}(\tau_1)f_{{\bf k}_2}(\tau_2)f_{{\bf k}_3}(\tau_2)f^{*}_{-{\bf k}_4}(\tau_1)\\
 && {\cal N}^{(1)}_{11}({\bf k}_1,{\bf k}_2,{\bf k}_3,{\bf k}_4;\tau_1,\tau_2)=f_{{\bf k}_1}(\tau_1)f^{*}_{-{\bf k}_2}(\tau_2)f_{{\bf k}_3}(\tau_2)f^{*}_{-{\bf k}_4}(\tau_1),\\
  && {\cal N}^{(1)}_{12}({\bf k}_1,{\bf k}_2,{\bf k}_3,{\bf k}_4;\tau_1,\tau_2)=f^{*}_{-{\bf k}_1}(\tau_1)f^{*}_{-{\bf k}_2}(\tau_2)f_{{\bf k}_3}(\tau_2)f^{*}_{-{\bf k}_4}(\tau_1)\\
 && {\cal N}^{(1)}_{13}({\bf k}_1,{\bf k}_2,{\bf k}_3,{\bf k}_4;\tau_1,\tau_2)=f_{{\bf k}_1}(\tau_1)f_{{\bf k}_2}(\tau_2)f^{*}_{-{\bf k}_3}(\tau_2)f^{*}_{-{\bf k}_4}(\tau_1),\\
  && {\cal N}^{(1)}_{14}({\bf k}_1,{\bf k}_2,{\bf k}_3,{\bf k}_4;\tau_1,\tau_2)=f^{*}_{-{\bf k}_1}(\tau_1)f_{{\bf k}_2}(\tau_2)f^{*}_{-{\bf k}_3}(\tau_2)f^{*}_{-{\bf k}_4}(\tau_1)\\
 && {\cal N}^{(1)}_{15}({\bf k}_1,{\bf k}_2,{\bf k}_3,{\bf k}_4;\tau_1,\tau_2)=f_{{\bf k}_1}(\tau_1)f^{*}_{-{\bf k}_2}(\tau_2)f^{*}_{-{\bf k}_3}(\tau_2)f^{*}_{-{\bf k}_4}(\tau_1),\\
  && {\cal N}^{(1)}_{16}({\bf k}_1,{\bf k}_2,{\bf k}_3,{\bf k}_4;\tau_1,\tau_2)=f^{*}_{-{\bf k}_1}(\tau_1)f^{*}_{-{\bf k}_2}(\tau_2)f^{*}_{-{\bf k}_3}(\tau_2)f^{*}_{-{\bf k}_4}(\tau_1)\eea  
    The function $\widehat{\cal T}^{(2)}_3({\bf k}_1,{\bf k}_2,{\bf k}_3,{\bf k}_4;\tau_1,\tau_2)$ is defined as:
 \bea &&\widehat{\cal T}^{(2)}_3({\bf k}_1,{\bf k}_2,{\bf k}_3,{\bf k}_4;\tau_1,\tau_2)\nonumber\\
 &&=\left[{\cal N}^{(2)}_1({\bf k}_1,{\bf k}_2,{\bf k}_3,{\bf k}_4;\tau_1,\tau_2)~a_{{\bf k}_1}a_{{\bf k}_2}a_{{\bf k}_3}a_{{\bf k}_4}+ {\cal N}^{(2)}_2({\bf k}_1,{\bf k}_2,{\bf k}_3,{\bf k}_4;\tau_1,\tau_2)~a^{\dagger}_{-{\bf k}_1}a_{{\bf k}_2}a_{{\bf k}_3}a_{{\bf k}_4}\right.\nonumber\\
  && \left.+{\cal N}^{(2)}_3({\bf k}_1,{\bf k}_2,{\bf k}_3,{\bf k}_4;\tau_1,\tau_2)~a_{{\bf k}_1}a^{\dagger}_{-{\bf k}_2}a_{{\bf k}_3}a_{{\bf k}_4}+ {\cal N}^{(2)}_4({\bf k}_1,{\bf k}_2,{\bf k}_3,{\bf k}_4;\tau_1,\tau_2)~a^{\dagger}_{-{\bf k}_1}a^{\dagger}_{-{\bf k}_2}a_{{\bf k}_3}a_{{\bf k}_4}\right.\nonumber\\
  && \left.+{\cal N}^{(2)}_5({\bf k}_1,{\bf k}_2,{\bf k}_3,{\bf k}_4;\tau_1,\tau_2)~a_{{\bf k}_1}a_{{\bf k}_2}a^{\dagger}_{-{\bf k}_3}a_{{\bf k}_4}+ {\cal N}^{(2)}_6({\bf k}_1,{\bf k}_2,{\bf k}_3,{\bf k}_4;\tau_1,\tau_2)~a^{\dagger}_{-{\bf k}_1}a_{{\bf k}_2}a^{\dagger}_{-{\bf k}_3}a_{{\bf k}_4}\right.\nonumber\\
  && \left.+{\cal N}^{(2)}_7({\bf k}_1,{\bf k}_2,{\bf k}_3,{\bf k}_4;\tau_1,\tau_2)~a_{{\bf k}_1}a^{\dagger}_{-{\bf k}_2}a^{\dagger}_{-{\bf k}_3}a_{{\bf k}_4}+ {\cal N}^{(2)}_8({\bf k}_1,{\bf k}_2,{\bf k}_3,{\bf k}_4;\tau_1,\tau_2)~a^{\dagger}_{-{\bf k}_1}a^{\dagger}_{-{\bf k}_2}a^{\dagger}_{-{\bf k}_3}a_{{\bf k}_4}\right.\nonumber\\
  && \left.+{\cal N}^{(2)}_9({\bf k}_1,{\bf k}_2,{\bf k}_3,{\bf k}_4;\tau_1,\tau_2)~a_{{\bf k}_1}a_{{\bf k}_2}a_{{\bf k}_3}a^{\dagger}_{-{\bf k}_4}+ {\cal N}^{(2)}_{10}({\bf k}_1,{\bf k}_2,{\bf k}_3,{\bf k}_4;\tau_1,\tau_2)~a^{\dagger}_{-{\bf k}_1}a_{{\bf k}_2}a_{{\bf k}_3}a^{\dagger}_{-{\bf k}_4}\right.\nonumber\\
  && \left.+{\cal N}^{(2)}_{11}({\bf k}_1,{\bf k}_2,{\bf k}_3,{\bf k}_4;\tau_1,\tau_2)~a_{{\bf k}_1}a^{\dagger}_{-{\bf k}_2}a_{{\bf k}_3}a^{\dagger}_{-{\bf k}_4}+ {\cal N}^{(2)}_{12}({\bf k}_1,{\bf k}_2,{\bf k}_3,{\bf k}_4;\tau_1,\tau_2)~a^{\dagger}_{-{\bf k}_1}a^{\dagger}_{-{\bf k}_2}a_{{\bf k}_3}a^{\dagger}_{-{\bf k}_4}\right.\nonumber\\
  && \left.+{\cal N}^{(2)}_{13}({\bf k}_1,{\bf k}_2,{\bf k}_3,{\bf k}_4;\tau_1,\tau_2)~a_{{\bf k}_1}a_{{\bf k}_2}a^{\dagger}_{-{\bf k}_3}a^{\dagger}_{-{\bf k}_4}+ {\cal N}^{(2)}_{14}({\bf k}_1,{\bf k}_2,{\bf k}_3,{\bf k}_4;\tau_1,\tau_2)~a^{\dagger}_{-{\bf k}_1}a_{{\bf k}_2}a^{\dagger}_{-{\bf k}_3}a^{\dagger}_{-{\bf k}_4}\right.\nonumber\\
  && \left.+{\cal N}^{(2)}_{15}({\bf k}_1,{\bf k}_2,{\bf k}_3,{\bf k}_4;\tau_1,\tau_2)~a_{{\bf k}_1}a^{\dagger}_{-{\bf k}_2}a^{\dagger}_{-{\bf k}_3}a^{\dagger}_{-{\bf k}_4}+ {\cal N}^{(2)}_{16}({\bf k}_1,{\bf k}_2,{\bf k}_3,{\bf k}_4;\tau_1,\tau_2)~a^{\dagger}_{-{\bf k}_1}a^{\dagger}_{-{\bf k}_2}a^{\dagger}_{-{\bf k}_3}a^{\dagger}_{-{\bf k}_4}\right],~~~~~~~~~~~
  \eea
  where we define new sets of functions, ${\cal N}^{(2)}_{j}({\bf k}_1,{\bf k}_2,{\bf k}_3,{\bf k}_4;\tau_1,\tau_2)~\forall~j=1,\cdots,16$, as:
  \bea &&{\cal N}^{(1)}_{2}({\bf k}_1,{\bf k}_2,{\bf k}_3,{\bf k}_4;\tau_1,\tau_2)=\Pi_{{\bf k}_1}(\tau_1)\Pi_{{\bf k}_2}(\tau_2)\Pi_{{\bf k}_3}(\tau_2)\Pi_{{\bf k}_4}(\tau_1),\\
 && {\cal N}^{(2)}_2({\bf k}_1,{\bf k}_2,{\bf k}_3,{\bf k}_4;\tau_1,\tau_2)=\Pi_{{\bf k}_1}(\tau_2)\Pi^{*}_{-{\bf k}_2}(\tau_1)\Pi_{{\bf k}_3}(\tau_2)\Pi_{{\bf k}_4}(\tau_1),\\
 && {\cal N}^{(2)}_{3}({\bf k}_1,{\bf k}_2,{\bf k}_3,{\bf k}_4;\tau_1,\tau_2)=\Pi_{{\bf k}_1}(\tau_1)\Pi^{*}_{-{\bf k}_2}(\tau_2)\Pi_{{\bf k}_1}(\tau_2)\Pi_{{\bf k}_4}(\tau_1),\\
  && {\cal N}^{(2)}_{4}({\bf k}_1,{\bf k}_2,{\bf k}_3,{\bf k}_4;\tau_1,\tau_2)=\Pi^{*}_{-{\bf k}_1}(\tau_1)\Pi^{*}_{-{\bf k}_2}(\tau_2)\Pi_{{\bf k}_3}(\tau_2)\Pi_{{\bf k}_4}(\tau_1),\\
 && {\cal N}^{(2)}_{5}({\bf k}_1,{\bf k}_2,{\bf k}_3,{\bf k}_4;\tau_1,\tau_2)=\Pi_{{\bf k}_1}(\tau_1)\Pi_{{\bf k}_2}(\tau_2)\Pi^{*}_{-{\bf k}_3}(\tau_2)\Pi_{{\bf k}_4}(\tau_1),\\
  && {\cal N}^{(2)}_{6}({\bf k}_1,{\bf k}_2,{\bf k}_3,{\bf k}_4;\tau_1,\tau_2)=\Pi^{*}_{-{\bf k}_1}(\tau_1)\Pi_{{\bf k}_2}(\tau_2)\Pi^{*}_{-{\bf k}_3}(\tau_2)\Pi_{{\bf k}_4}(\tau_1)\\
 && {\cal N}^{(2)}_{7}({\bf k}_1,{\bf k}_2,{\bf k}_3,{\bf k}_4;\tau_1,\tau_2)=\Pi_{{\bf k}_1}(\tau_1)\Pi^{*}_{-{\bf k}_2}(\tau_2)\Pi^{*}_{-{\bf k}_3}(\tau_2)\Pi_{{\bf k}_4}(\tau_1),\\
  && {\cal N}^{(2)}_{8}({\bf k}_1,{\bf k}_2,{\bf k}_3,{\bf k}_4;\tau_1,\tau_2)=\Pi^{*}_{-{\bf k}_1}(\tau_1)\Pi^{*}_{-{\bf k}_2}(\tau_2)\Pi^{*}_{-{\bf k}_3}(\tau_2)\Pi_{{\bf k}_4}(\tau_1)\\
 && {\cal N}^{(2)}_{9}({\bf k}_1,{\bf k}_2,{\bf k}_3,{\bf k}_4;\tau_1,\tau_2)=\Pi_{{\bf k}_1}(\tau_1)\Pi_{{\bf k}_2}(\tau_2)\Pi_{{\bf k}_3}(\tau_2)\Pi^{*}_{-{\bf k}_4}(\tau_1),\\
  && {\cal N}^{(2)}_{10}({\bf k}_1,{\bf k}_2,{\bf k}_3,{\bf k}_4;\tau_1,\tau_2)=\Pi^{*}_{-{\bf k}_1}(\tau_1)\Pi_{{\bf k}_2}(\tau_2)\Pi_{{\bf k}_3}(\tau_2)\Pi^{*}_{-{\bf k}_4}(\tau_1)\\
 && {\cal N}^{(2)}_{11}({\bf k}_1,{\bf k}_2,{\bf k}_3,{\bf k}_4;\tau_1,\tau_2)=\Pi_{{\bf k}_1}(\tau_1)\Pi^{*}_{-{\bf k}_2}(\tau_2)\Pi_{{\bf k}_3}(\tau_2)\Pi^{*}_{-{\bf k}_4}(\tau_1),\\
  && {\cal N}^{(2)}_{12}({\bf k}_1,{\bf k}_2,{\bf k}_3,{\bf k}_4;\tau_1,\tau_2)=\Pi^{*}_{-{\bf k}_1}(\tau_1)\Pi^{*}_{-{\bf k}_2}(\tau_2)\Pi_{{\bf k}_3}(\tau_2)\Pi^{*}_{-{\bf k}_4}(\tau_1)\\
 && {\cal N}^{(2)}_{13}({\bf k}_1,{\bf k}_2,{\bf k}_3,{\bf k}_4;\tau_1,\tau_2)=\Pi_{{\bf k}_1}(\tau_1)\Pi_{{\bf k}_2}(\tau_2)\Pi^{*}_{-{\bf k}_3}(\tau_2)\Pi^{*}_{-{\bf k}_4}(\tau_1),\\
  && {\cal N}^{(2)}_{14}({\bf k}_1,{\bf k}_2,{\bf k}_3,{\bf k}_4;\tau_1,\tau_2)=\Pi^{*}_{-{\bf k}_1}(\tau_1)\Pi_{{\bf k}_2}(\tau_2)\Pi^{*}_{-{\bf k}_3}(\tau_2)\Pi^{*}_{-{\bf k}_4}(\tau_1)\\
 && {\cal N}^{(2)}_{15}({\bf k}_1,{\bf k}_2,{\bf k}_3,{\bf k}_4;\tau_1,\tau_2)=\Pi_{{\bf k}_1}(\tau_1)\Pi^{*}_{-{\bf k}_2}(\tau_2)\Pi^{*}_{-{\bf k}_3}(\tau_2)\Pi^{*}_{-{\bf k}_4}(\tau_1),\\
  && {\cal N}^{(2)}_{16}({\bf k}_1,{\bf k}_2,{\bf k}_3,{\bf k}_4;\tau_1,\tau_2)=\Pi^{*}_{-{\bf k}_1}(\tau_1)\Pi^{*}_{-{\bf k}_2}(\tau_2)\Pi^{*}_{-{\bf k}_3}(\tau_2)\Pi^{*}_{-{\bf k}_4}(\tau_1)\eea  
\subsection{Definition of  OTO amplitude $\widehat{\cal T}^{(1)}_4({\bf k}_1,{\bf k}_2,{\bf k}_3,{\bf k}_4;\tau_1,\tau_2)$ and $\widehat{\cal T}^{(2)}_4({\bf k}_1,{\bf k}_2,{\bf k}_3,{\bf k}_4;\tau_1,\tau_2)$}
  The function $\widehat{\cal T}^{(1)}_4({\bf k}_1,{\bf k}_2,{\bf k}_3,{\bf k}_4;\tau_1,\tau_2)$ is defined as:
 \bea &&\widehat{\cal T}^{(1)}_4({\bf k}_1,{\bf k}_2,{\bf k}_3,{\bf k}_4;\tau_1,\tau_2)\nonumber\\
 &&=\left[{\cal Q}^{(1)}_1({\bf k}_1,{\bf k}_2,{\bf k}_3,{\bf k}_4;\tau_1,\tau_2)~a_{{\bf k}_1}a_{{\bf k}_2}a_{{\bf k}_3}a_{{\bf k}_4}+ {\cal Q}^{(1)}_2({\bf k}_1,{\bf k}_2,{\bf k}_3,{\bf k}_4;\tau_1,\tau_2)~a^{\dagger}_{-{\bf k}_1}a_{{\bf k}_2}a_{{\bf k}_3}a_{{\bf k}_4}\right.\nonumber\\
  && \left.+{\cal Q}^{(1)}_3({\bf k}_1,{\bf k}_2,{\bf k}_3,{\bf k}_4;\tau_1,\tau_2)~a_{{\bf k}_1}a^{\dagger}_{-{\bf k}_2}a_{{\bf k}_3}a_{{\bf k}_4}+ {\cal Q}^{(1)}_4({\bf k}_1,{\bf k}_2,{\bf k}_3,{\bf k}_4;\tau_1,\tau_2)~a^{\dagger}_{-{\bf k}_1}a^{\dagger}_{-{\bf k}_2}a_{{\bf k}_3}a_{{\bf k}_4}\right.\nonumber\\
  && \left.+{\cal Q}^{(1)}_5({\bf k}_1,{\bf k}_2,{\bf k}_3,{\bf k}_4;\tau_1,\tau_2)~a_{{\bf k}_1}a_{{\bf k}_2}a^{\dagger}_{-{\bf k}_3}a_{{\bf k}_4}+ {\cal Q}^{(1)}_6({\bf k}_1,{\bf k}_2,{\bf k}_3,{\bf k}_4;\tau_1,\tau_2)~a^{\dagger}_{-{\bf k}_1}a_{{\bf k}_2}a^{\dagger}_{-{\bf k}_3}a_{{\bf k}_4}\right.\nonumber\\
  && \left.+{\cal Q}^{(1)}_7({\bf k}_1,{\bf k}_2,{\bf k}_3,{\bf k}_4;\tau_1,\tau_2)~a_{{\bf k}_1}a^{\dagger}_{-{\bf k}_2}a^{\dagger}_{-{\bf k}_3}a_{{\bf k}_4}+ {\cal Q}^{(1)}_8({\bf k}_1,{\bf k}_2,{\bf k}_3,{\bf k}_4;\tau_1,\tau_2)~a^{\dagger}_{-{\bf k}_1}a^{\dagger}_{-{\bf k}_2}a^{\dagger}_{-{\bf k}_3}a_{{\bf k}_4}\right.\nonumber\\
  && \left.+{\cal Q}^{(1)}_9({\bf k}_1,{\bf k}_2,{\bf k}_3,{\bf k}_4;\tau_1,\tau_2)~a_{{\bf k}_1}a_{{\bf k}_2}a_{{\bf k}_3}a^{\dagger}_{-{\bf k}_4}+ {\cal Q}^{(1)}_{10}({\bf k}_1,{\bf k}_2,{\bf k}_3,{\bf k}_4;\tau_1,\tau_2)~a^{\dagger}_{-{\bf k}_1}a_{{\bf k}_2}a_{{\bf k}_3}a^{\dagger}_{-{\bf k}_4}\right.\nonumber\\
  && \left.+{\cal Q}^{(1)}_{11}({\bf k}_1,{\bf k}_2,{\bf k}_3,{\bf k}_4;\tau_1,\tau_2)~a_{{\bf k}_1}a^{\dagger}_{-{\bf k}_2}a_{{\bf k}_3}a^{\dagger}_{-{\bf k}_4}+ {\cal Q}^{(1)}_{12}({\bf k}_1,{\bf k}_2,{\bf k}_3,{\bf k}_4;\tau_1,\tau_2)~a^{\dagger}_{-{\bf k}_1}a^{\dagger}_{-{\bf k}_2}a_{{\bf k}_3}a^{\dagger}_{-{\bf k}_4}\right.\nonumber\\
  && \left.+{\cal Q}^{(1)}_{13}({\bf k}_1,{\bf k}_2,{\bf k}_3,{\bf k}_4;\tau_1,\tau_2)~a_{{\bf k}_1}a_{{\bf k}_2}a^{\dagger}_{-{\bf k}_3}a^{\dagger}_{-{\bf k}_4}+ {\cal Q}^{(1)}_{14}({\bf k}_1,{\bf k}_2,{\bf k}_3,{\bf k}_4;\tau_1,\tau_2)~a^{\dagger}_{-{\bf k}_1}a_{{\bf k}_2}a^{\dagger}_{-{\bf k}_3}a^{\dagger}_{-{\bf k}_4}\right.\nonumber\\
  && \left.+{\cal Q}^{(1)}_{15}({\bf k}_1,{\bf k}_2,{\bf k}_3,{\bf k}_4;\tau_1,\tau_2)~a_{{\bf k}_1}a^{\dagger}_{-{\bf k}_2}a^{\dagger}_{-{\bf k}_3}a^{\dagger}_{-{\bf k}_4}+ {\cal Q}^{(1)}_{16}({\bf k}_1,{\bf k}_2,{\bf k}_3,{\bf k}_4;\tau_1,\tau_2)~a^{\dagger}_{-{\bf k}_1}a^{\dagger}_{-{\bf k}_2}a^{\dagger}_{-{\bf k}_3}a^{\dagger}_{-{\bf k}_4}\right],~~~~~~~~~~~
  \eea
  where we define new sets of functions, ${\cal Q}^{(1)}_{j}({\bf k}_1,{\bf k}_2,{\bf k}_3,{\bf k}_4;\tau_1,\tau_2)~\forall~j=1,\cdots,16$, as:
  \bea &&{\cal Q}^{(1)}_{1}({\bf k}_1,{\bf k}_2,{\bf k}_3,{\bf k}_4;\tau_1,\tau_2)=f_{{\bf k}_1}(\tau_2)f_{{\bf k}_2}(\tau_1)f_{{\bf k}_3}(\tau_2)f_{{\bf k}_4}(\tau_1),\\
 && {\cal Q}^{(1)}_2({\bf k}_1,{\bf k}_2,{\bf k}_3,{\bf k}_4;\tau_1,\tau_2)=f^{*}_{-{\bf k}_1}(\tau_2)f_{-{\bf k}_2}(\tau_1)f_{{\bf k}_3}(\tau_2)f_{{\bf k}_4}(\tau_1),\\
 && {\cal Q}^{(1)}_{3}({\bf k}_1,{\bf k}_2,{\bf k}_3,{\bf k}_4;\tau_1,\tau_2)=f_{{\bf k}_1}(\tau_2)f^{*}_{-{\bf k}_2}(\tau_1)f_{{\bf k}_1}(\tau_2)f_{{\bf k}_4}(\tau_1),\\
  && {\cal Q}^{(1)}_{4}({\bf k}_1,{\bf k}_2,{\bf k}_3,{\bf k}_4;\tau_1,\tau_2)=f^{*}_{-{\bf k}_1}(\tau_2)f^{*}_{-{\bf k}_2}(\tau_1)f_{{\bf k}_3}(\tau_2)f_{{\bf k}_4}(\tau_1),\\
 && {\cal Q}^{(1)}_{5}({\bf k}_1,{\bf k}_2,{\bf k}_3,{\bf k}_4;\tau_1,\tau_2)=f_{{\bf k}_1}(\tau_2)f_{{\bf k}_2}(\tau_1)f^{*}_{-{\bf k}_3}(\tau_2)f_{{\bf k}_4}(\tau_1),\\
  && {\cal Q}^{(1)}_{6}({\bf k}_1,{\bf k}_2,{\bf k}_3,{\bf k}_4;\tau_1,\tau_2)=f^{*}_{-{\bf k}_1}(\tau_2)f_{{\bf k}_2}(\tau_1)f^{*}_{-{\bf k}_3}(\tau_2)f_{{\bf k}_4}(\tau_1)\\
 && {\cal Q}^{(1)}_{7}({\bf k}_1,{\bf k}_2,{\bf k}_3,{\bf k}_4;\tau_1,\tau_2)=f_{{\bf k}_1}(\tau_2)f^{*}_{-{\bf k}_2}(\tau_1)f^{*}_{-{\bf k}_3}(\tau_2)f_{{\bf k}_4}(\tau_1),\\
  && {\cal Q}^{(1)}_{8}({\bf k}_1,{\bf k}_2,{\bf k}_3,{\bf k}_4;\tau_1,\tau_2)=f^{*}_{-{\bf k}_1}(\tau_2)f^{*}_{-{\bf k}_2}(\tau_1)f^{*}_{-{\bf k}_3}(\tau_2)f_{{\bf k}_4}(\tau_1)\\
 && {\cal Q}^{(1)}_{9}({\bf k}_1,{\bf k}_2,{\bf k}_3,{\bf k}_4;\tau_1,\tau_2)=f_{{\bf k}_1}(\tau_2)f_{{\bf k}_2}(\tau_1)f_{{\bf k}_3}(\tau_2)f^{*}_{-{\bf k}_4}(\tau_1),\\
  && {\cal Q}^{(1)}_{10}({\bf k}_1,{\bf k}_2,{\bf k}_3,{\bf k}_4;\tau_1,\tau_2)=f^{*}_{-{\bf k}_1}(\tau_2)f_{{\bf k}_2}(\tau_1)f_{{\bf k}_3}(\tau_2)f^{*}_{-{\bf k}_4}(\tau_1)\\
 && {\cal Q}^{(1)}_{11}({\bf k}_1,{\bf k}_2,{\bf k}_3,{\bf k}_4;\tau_1,\tau_2)=f_{{\bf k}_1}(\tau_2)f^{*}_{-{\bf k}_2}(\tau_1)f_{{\bf k}_3}(\tau_2)f^{*}_{-{\bf k}_4}(\tau_1),\\
  && {\cal Q}^{(1)}_{12}({\bf k}_1,{\bf k}_2,{\bf k}_3,{\bf k}_4;\tau_1,\tau_2)=f^{*}_{-{\bf k}_1}(\tau_2)f^{*}_{-{\bf k}_2}(\tau_1)f_{{\bf k}_3}(\tau_2)f^{*}_{-{\bf k}_4}(\tau_1)\\
 && {\cal Q}^{(1)}_{13}({\bf k}_1,{\bf k}_2,{\bf k}_3,{\bf k}_4;\tau_1,\tau_2)=f_{{\bf k}_1}(\tau_2)f_{{\bf k}_2}(\tau_1)f^{*}_{-{\bf k}_3}(\tau_2)f^{*}_{-{\bf k}_4}(\tau_1),\\
  && {\cal Q}^{(1)}_{14}({\bf k}_1,{\bf k}_2,{\bf k}_3,{\bf k}_4;\tau_1,\tau_2)=f^{*}_{-{\bf k}_1}(\tau_2)f_{{\bf k}_2}(\tau_1)f^{*}_{-{\bf k}_3}(\tau_2)f^{*}_{-{\bf k}_4}(\tau_1)\\
 && {\cal Q}^{(1)}_{15}({\bf k}_1,{\bf k}_2,{\bf k}_3,{\bf k}_4;\tau_1,\tau_2)=f_{{\bf k}_1}(\tau_2)f^{*}_{-{\bf k}_2}(\tau_1)f^{*}_{-{\bf k}_3}(\tau_2)f^{*}_{-{\bf k}_4}(\tau_1),\\
  && {\cal Q}^{(1)}_{16}({\bf k}_1,{\bf k}_2,{\bf k}_3,{\bf k}_4;\tau_1,\tau_2)=f^{*}_{-{\bf k}_1}(\tau_2)f^{*}_{-{\bf k}_2}(\tau_1)f^{*}_{-{\bf k}_3}(\tau_2)f^{*}_{-{\bf k}_4}(\tau_1)\eea 
   The function $\widehat{\cal T}^{(2)}_4({\bf k}_1,{\bf k}_2,{\bf k}_3,{\bf k}_4;\tau_1,\tau_2)$ is defined as:
 \bea &&\widehat{\cal T}^{(2)}_4({\bf k}_1,{\bf k}_2,{\bf k}_3,{\bf k}_4;\tau_1,\tau_2)\nonumber\\
 &&=\left[{\cal Q}^{(2)}_1({\bf k}_1,{\bf k}_2,{\bf k}_3,{\bf k}_4;\tau_1,\tau_2)~a_{{\bf k}_1}a_{{\bf k}_2}a_{{\bf k}_3}a_{{\bf k}_4}+ {\cal Q}^{(2)}_2({\bf k}_1,{\bf k}_2,{\bf k}_3,{\bf k}_4;\tau_1,\tau_2)~a^{\dagger}_{-{\bf k}_1}a_{{\bf k}_2}a_{{\bf k}_3}a_{{\bf k}_4}\right.\nonumber\\
  && \left.+{\cal Q}^{(2)}_3({\bf k}_1,{\bf k}_2,{\bf k}_3,{\bf k}_4;\tau_1,\tau_2)~a_{{\bf k}_1}a^{\dagger}_{-{\bf k}_2}a_{{\bf k}_3}a_{{\bf k}_4}+ {\cal Q}^{(2)}_4({\bf k}_1,{\bf k}_2,{\bf k}_3,{\bf k}_4;\tau_1,\tau_2)~a^{\dagger}_{-{\bf k}_1}a^{\dagger}_{-{\bf k}_2}a_{{\bf k}_3}a_{{\bf k}_4}\right.\nonumber\\
  && \left.+{\cal Q}^{(2)}_5({\bf k}_1,{\bf k}_2,{\bf k}_3,{\bf k}_4;\tau_1,\tau_2)~a_{{\bf k}_1}a_{{\bf k}_2}a^{\dagger}_{-{\bf k}_3}a_{{\bf k}_4}+ {\cal Q}^{(2)}_6({\bf k}_1,{\bf k}_2,{\bf k}_3,{\bf k}_4;\tau_1,\tau_2)~a^{\dagger}_{-{\bf k}_1}a_{{\bf k}_2}a^{\dagger}_{-{\bf k}_3}a_{{\bf k}_4}\right.\nonumber\\
  && \left.+{\cal Q}^{(2)}_7({\bf k}_1,{\bf k}_2,{\bf k}_3,{\bf k}_4;\tau_1,\tau_2)~a_{{\bf k}_1}a^{\dagger}_{-{\bf k}_2}a^{\dagger}_{-{\bf k}_3}a_{{\bf k}_4}+ {\cal Q}^{(2)}_8({\bf k}_1,{\bf k}_2,{\bf k}_3,{\bf k}_4;\tau_1,\tau_2)~a^{\dagger}_{-{\bf k}_1}a^{\dagger}_{-{\bf k}_2}a^{\dagger}_{-{\bf k}_3}a_{{\bf k}_4}\right.\nonumber\\
  && \left.+{\cal Q}^{(2)}_9({\bf k}_1,{\bf k}_2,{\bf k}_3,{\bf k}_4;\tau_1,\tau_2)~a_{{\bf k}_1}a_{{\bf k}_2}a_{{\bf k}_3}a^{\dagger}_{-{\bf k}_4}+ {\cal Q}^{(2)}_{10}({\bf k}_1,{\bf k}_2,{\bf k}_3,{\bf k}_4;\tau_1,\tau_2)~a^{\dagger}_{-{\bf k}_1}a_{{\bf k}_2}a_{{\bf k}_3}a^{\dagger}_{-{\bf k}_4}\right.\nonumber\\
  && \left.+{\cal Q}^{(2)}_{11}({\bf k}_1,{\bf k}_2,{\bf k}_3,{\bf k}_4;\tau_1,\tau_2)~a_{{\bf k}_1}a^{\dagger}_{-{\bf k}_2}a_{{\bf k}_3}a^{\dagger}_{-{\bf k}_4}+ {\cal Q}^{(2)}_{12}({\bf k}_1,{\bf k}_2,{\bf k}_3,{\bf k}_4;\tau_1,\tau_2)~a^{\dagger}_{-{\bf k}_1}a^{\dagger}_{-{\bf k}_2}a_{{\bf k}_3}a^{\dagger}_{-{\bf k}_4}\right.\nonumber\\
  && \left.+{\cal Q}^{(2)}_{13}({\bf k}_1,{\bf k}_2,{\bf k}_3,{\bf k}_4;\tau_1,\tau_2)~a_{{\bf k}_1}a_{{\bf k}_2}a^{\dagger}_{-{\bf k}_3}a^{\dagger}_{-{\bf k}_4}+ {\cal Q}^{(2)}_{14}({\bf k}_1,{\bf k}_2,{\bf k}_3,{\bf k}_4;\tau_1,\tau_2)~a^{\dagger}_{-{\bf k}_1}a_{{\bf k}_2}a^{\dagger}_{-{\bf k}_3}a^{\dagger}_{-{\bf k}_4}\right.\nonumber\\
  && \left.+{\cal Q}^{(2)}_{15}({\bf k}_1,{\bf k}_2,{\bf k}_3,{\bf k}_4;\tau_1,\tau_2)~a_{{\bf k}_1}a^{\dagger}_{-{\bf k}_2}a^{\dagger}_{-{\bf k}_3}a^{\dagger}_{-{\bf k}_4}+ {\cal Q}^{(2)}_{16}({\bf k}_1,{\bf k}_2,{\bf k}_3,{\bf k}_4;\tau_1,\tau_2)~a^{\dagger}_{-{\bf k}_1}a^{\dagger}_{-{\bf k}_2}a^{\dagger}_{-{\bf k}_3}a^{\dagger}_{-{\bf k}_4}\right],~~~~~~~~~~~
  \eea
  where we define new sets of functions, ${\cal Q}^{(2)}_{j}({\bf k}_1,{\bf k}_2,{\bf k}_3,{\bf k}_4;\tau_1,\tau_2)~\forall~j=1,\cdots,16$, as:
  \bea &&{\cal Q}^{(2)}_{1}({\bf k}_1,{\bf k}_2,{\bf k}_3,{\bf k}_4;\tau_1,\tau_2)=\Pi_{{\bf k}_1}(\tau_2)\Pi_{{\bf k}_2}(\tau_1)\Pi_{{\bf k}_3}(\tau_2)\Pi_{{\bf k}_4}(\tau_1),\\
 && {\cal Q}^{(2)}_2({\bf k}_1,{\bf k}_2,{\bf k}_3,{\bf k}_4;\tau_1,\tau_2)=\Pi^{*}_{-{\bf k}_1}(\tau_2)\Pi_{-{\bf k}_2}(\tau_1)\Pi_{{\bf k}_3}(\tau_2)\Pi_{{\bf k}_4}(\tau_1),\\
 && {\cal Q}^{(2)}_{3}({\bf k}_1,{\bf k}_2,{\bf k}_3,{\bf k}_4;\tau_1,\tau_2)=\Pi_{{\bf k}_1}(\tau_2)\Pi^{*}_{-{\bf k}_2}(\tau_1)\Pi_{{\bf k}_1}(\tau_2)\Pi_{{\bf k}_4}(\tau_1),\\
  && {\cal Q}^{(2)}_{4}({\bf k}_1,{\bf k}_2,{\bf k}_3,{\bf k}_4;\tau_1,\tau_2)=\Pi^{*}_{-{\bf k}_1}(\tau_2)\Pi^{*}_{-{\bf k}_2}(\tau_1)\Pi_{{\bf k}_3}(\tau_2)\Pi_{{\bf k}_4}(\tau_1),\\
 && {\cal Q}^{(2)}_{5}({\bf k}_1,{\bf k}_2,{\bf k}_3,{\bf k}_4;\tau_1,\tau_2)=\Pi_{{\bf k}_1}(\tau_2)\Pi_{{\bf k}_2}(\tau_1)\Pi^{*}_{-{\bf k}_3}(\tau_2)\Pi_{{\bf k}_4}(\tau_1),\\
  && {\cal Q}^{(2)}_{6}({\bf k}_1,{\bf k}_2,{\bf k}_3,{\bf k}_4;\tau_1,\tau_2)=\Pi^{*}_{-{\bf k}_1}(\tau_2)\Pi_{{\bf k}_2}(\tau_1)\Pi^{*}_{-{\bf k}_3}(\tau_2)\Pi_{{\bf k}_4}(\tau_1)\\
 && {\cal Q}^{(2)}_{7}({\bf k}_1,{\bf k}_2,{\bf k}_3,{\bf k}_4;\tau_1,\tau_2)=\Pi_{{\bf k}_1}(\tau_2)\Pi^{*}_{-{\bf k}_2}(\tau_1)\Pi\Pi^{*}_{-{\bf k}_3}(\tau_2)\Pi_{{\bf k}_4}(\tau_1),\\
  && {\cal Q}^{(2)}_{8}({\bf k}_1,{\bf k}_2,{\bf k}_3,{\bf k}_4;\tau_1,\tau_2)=\Pi^{*}_{-{\bf k}_1}(\tau_2)\Pi^{*}_{-{\bf k}_2}(\tau_1)\Pi^{*}_{-{\bf k}_3}(\tau_2)\Pi_{{\bf k}_4}(\tau_1)\\
 && {\cal Q}^{(2)}_{9}({\bf k}_1,{\bf k}_2,{\bf k}_3,{\bf k}_4;\tau_1,\tau_2)=\Pi_{{\bf k}_1}(\tau_2)\Pi_{{\bf k}_2}(\tau_1)\Pi_{{\bf k}_3}(\tau_2)\Pi^{*}_{-{\bf k}_4}(\tau_1),\\
  && {\cal Q}^{(2)}_{10}({\bf k}_1,{\bf k}_2,{\bf k}_3,{\bf k}_4;\tau_1,\tau_2)=\Pi^{*}_{-{\bf k}_1}(\tau_2)\Pi_{{\bf k}_2}(\tau_1)\Pi_{{\bf k}_3}(\tau_2)\Pi^{*}_{-{\bf k}_4}(\tau_1)\\
 && {\cal Q}^{(2)}_{11}({\bf k}_1,{\bf k}_2,{\bf k}_3,{\bf k}_4;\tau_1,\tau_2)=\Pi_{{\bf k}_1}(\tau_2)\Pi^{*}_{-{\bf k}_2}(\tau_1)\Pi_{{\bf k}_3}(\tau_2)\Pi^{*}_{-{\bf k}_4}(\tau_1),\\
  && {\cal Q}^{(2)}_{12}({\bf k}_1,{\bf k}_2,{\bf k}_3,{\bf k}_4;\tau_1,\tau_2)=\Pi^{*}_{-{\bf k}_1}(\tau_2)\Pi^{*}_{-{\bf k}_2}(\tau_1)\Pi_{{\bf k}_3}(\tau_2)\Pi^{*}_{-{\bf k}_4}(\tau_1)\\
 && {\cal Q}^{(2)}_{13}({\bf k}_1,{\bf k}_2,{\bf k}_3,{\bf k}_4;\tau_1,\tau_2)=\Pi_{{\bf k}_1}(\tau_2)\Pi_{{\bf k}_2}(\tau_1)\Pi^{*}_{-{\bf k}_3}(\tau_2)\Pi^{*}_{-{\bf k}_4}(\tau_1),\\
  && {\cal Q}^{(2)}_{14}({\bf k}_1,{\bf k}_2,{\bf k}_3,{\bf k}_4;\tau_1,\tau_2)=\Pi^{*}_{-{\bf k}_1}(\tau_2)\Pi_{{\bf k}_2}(\tau_1)\Pi^{*}_{-{\bf k}_3}(\tau_2)\Pi^{*}_{-{\bf k}_4}(\tau_1)\\
 && {\cal Q}^{(2)}_{15}({\bf k}_1,{\bf k}_2,{\bf k}_3,{\bf k}_4;\tau_1,\tau_2)=\Pi_{{\bf k}_1}(\tau_2)\Pi^{*}_{-{\bf k}_2}(\tau_1)\Pi^{*}_{-{\bf k}_3}(\tau_2)\Pi^{*}_{-{\bf k}_4}(\tau_1),\\
  && {\cal Q}^{(2)}_{16}({\bf k}_1,{\bf k}_2,{\bf k}_3,{\bf k}_4;\tau_1,\tau_2)=\Pi^{*}_{-{\bf k}_1}(\tau_2)\Pi^{*}_{-{\bf k}_2}(\tau_1)\Pi^{*}_{-{\bf k}_3}(\tau_2)\Pi^{*}_{-{\bf k}_4}(\tau_1)\eea   
  
	\newpage
	\section{Computation of classical limit of four-point ``in-in" OTO  amplitudes for Cosmology} 
	\label{sec:10}
	In this section, our prime objective is to explicitly compute the classical limiting version of the four-point "in-in" OTO  amplitudes appearing in the expression or OTOC. To serve this purpose in the classical limit we explicitly compute the following square of the Poisson bracket, given by: 
  \bea  \left\{{f}({\bf x},\tau_1),{f}({\bf x},\tau_2)\right\}^2_{\bf PB}&=&\left\{{f}({\bf x},\tau_1),{f}({\bf x},\tau_2)\right\}_{\bf PB}\left\{{f}({\bf x},\tau_1),{f}({\bf x},\tau_2)\right\}_{\bf PB},\\  \left\{{f}({\bf x},\tau_1),{\Pi}({\bf x},\tau_2)\right\}^2_{\bf PB}&=&\left\{{\Pi}({\bf x},\tau_1),{\Pi}({\bf x},\tau_2)\right\}_{\bf PB}\left\{{\Pi}({\bf x},\tau_1),{\Pi}({\bf x},\tau_2)\right\}_{\bf PB}.\eea
  Now we use the following convention for the Fourier transformation, which is given by:
 \bea &&{f}({\bf x},\tau_1)=\int \frac{d^3{\bf k}}{(2\pi)^3}~\exp(i{\bf k}.{\bf x})~{f}_{{\bf k}}(\tau_1),\\
 &&{\Pi}({\bf x},\tau_1)=\partial_{\tau_1}\hat{f}({\bf x},\tau_1)=\int \frac{d^3{\bf k}}{(2\pi)^3}~\exp(i{\bf k}.{\bf x})~\partial_{\tau_1}{f}_{{\bf k}}(\tau_1)=\int \frac{d^3{\bf k}}{(2\pi)^3}~\exp(i{\bf k}.{\bf x})~{\Pi}_{{\bf k}}(\tau_1),~~~~~~~~~~~~~\eea
 which will be very useful for the computation of the classical limiting result of four-point OTOC in terms of the square of the Poisson bracket. Consequently,  we get the following simplified results:
 \bea && \left\{{f}({\bf x},\tau_1),{f}({\bf x},\tau_2)\right\}^2_{\bf PB}=\int \frac{d^3{\bf k}_1}{(2\pi)^3}\int \frac{d^3{\bf k}_2}{(2\pi)^3}\int \frac{d^3{\bf k}_3}{(2\pi)^3}\int \frac{d^3{\bf k}_4}{(2\pi)^3}~\exp\left(i({\bf k}_1+{\bf k}_2+{\bf k}_3+{\bf k}_4).{\bf x}\right)\nonumber\\
 &&~~~~~~~~~~~~\left[\left\{f_{{\bf k}_{1}} (\tau_1),f_{{\bf k}_{2}} (\tau_2)\right\}_{\bf PB}\left\{f_{{\bf k}_{3}}(\tau_1),f_{{\bf k}_4} (\tau_2)\right\}_{\bf PB}+\left\{f_{{\bf k}_{1}} (\tau_1),f_{{\bf k}_{3}} (\tau_2)\right\}_{\bf PB}\left\{f_{{\bf k}_{2}}(\tau_1),f_{{\bf k}_4} (\tau_2)\right\}_{\bf PB}\right.\nonumber\\&& \left.~~~~~~~~~+\left\{f_{{\bf k}_{1}} (\tau_1),f_{{\bf k}_{4}} (\tau_2)\right\}_{\bf PB}\left\{f_{{\bf k}_{3}}(\tau_1),f_{{\bf k}_2} (\tau_2)\right\}_{\bf PB}+\left\{f_{{\bf k}_{2}} (\tau_1),f_{{\bf k}_{3}} (\tau_2)\right\}_{\bf PB}\left\{f_{{\bf k}_{4}}(\tau_1),f_{{\bf k}_1} (\tau_2)\right\}_{\bf PB}\right.\nonumber\\&& \left.~~~~~~~~~+\left\{f_{{\bf k}_{2}} (\tau_1),f_{{\bf k}_{1}} (\tau_2)\right\}_{\bf PB}\left\{f_{{\bf k}_{4}}(\tau_1),f_{{\bf k}_3} (\tau_2)\right\}_{\bf PB}+\left\{f_{{\bf k}_{2}} (\tau_1),f_{{\bf k}_{4}} (\tau_2)\right\}_{\bf PB}\left\{f_{{\bf k}_{1}}(\tau_1),f_{{\bf k}_3} (\tau_2)\right\}_{\bf PB}\right.\nonumber\\&& \left.~~~~~~~~~+\left\{f_{{\bf k}_{3}} (\tau_1),f_{{\bf k}_{1}} (\tau_2)\right\}_{\bf PB}\left\{f_{{\bf k}_{4}}(\tau_1),f_{{\bf k}_2} (\tau_2)\right\}_{\bf PB}+\left\{f_{{\bf k}_{3}} (\tau_1),f_{{\bf k}_{2}} (\tau_2)\right\}_{\bf PB}\left\{f_{{\bf k}_{1}}(\tau_1),f_{{\bf k}_4} (\tau_2)\right\}_{\bf PB}\right.\nonumber\\&& \left.~~~~~~~~~+\left\{f_{{\bf k}_{3}} (\tau_1),f_{{\bf k}_{4}} (\tau_2)\right\}_{\bf PB}\left\{f_{{\bf k}_{1}}(\tau_1),f_{{\bf k}_2} (\tau_2)\right\}_{\bf PB}+\left\{f_{{\bf k}_{4}} (\tau_1),f_{{\bf k}_{1}} (\tau_2)\right\}_{\bf PB}\left\{f_{{\bf k}_{2}}(\tau_1),f_{{\bf k}_3} (\tau_2)\right\}_{\bf PB}\right.\nonumber\\&& \left.~~~~~~~~~+\left\{f_{{\bf k}_{4}} (\tau_1),f_{{\bf k}_{2}} (\tau_2)\right\}_{\bf PB}\left\{f_{{\bf k}_{3}}(\tau_1),f_{{\bf k}_1} (\tau_2)\right\}_{\bf PB}+\left\{f_{{\bf k}_{4}} (\tau_1),f_{{\bf k}_{3}} (\tau_2)\right\}_{\bf PB}\left\{f_{{\bf k}_{2}}(\tau_1),f_{{\bf k}_1} (\tau_2)\right\}_{\bf PB}\right].~~~~~\\ && \left\{{\Pi}({\bf x},\tau_1),{\Pi}({\bf x},\tau_2)\right\}^2_{\bf PB}=\int \frac{d^3{\bf k}_1}{(2\pi)^3}\int \frac{d^3{\bf k}_2}{(2\pi)^3}\int \frac{d^3{\bf k}_3}{(2\pi)^3}\int \frac{d^3{\bf k}_4}{(2\pi)^3}~\exp\left(i({\bf k}_1+{\bf k}_2+{\bf k}_3+{\bf k}_4).{\bf x}\right)\nonumber\\
 &&~~~~~~~~~\left[\left\{\Pi_{{\bf k}_{1}} (\tau_1),\Pi_{{\bf k}_{2}} (\tau_2)\right\}_{\bf PB}\left\{\Pi_{{\bf k}_{3}}(\tau_1),\Pi_{{\bf k}_4} (\tau_2)\right\}_{\bf PB}+\left\{\Pi_{{\bf k}_{1}} (\tau_1),\Pi_{{\bf k}_{3}} (\tau_2)\right\}_{\bf PB}\left\{\Pi_{{\bf k}_{2}}(\tau_1),\Pi_{{\bf k}_4} (\tau_2)\right\}_{\bf PB}\right.\nonumber\\&& \left.~~~~~~~+\left\{\Pi_{{\bf k}_{1}} (\tau_1),\Pi_{{\bf k}_{4}} (\tau_2)\right\}_{\bf PB}\left\{\Pi_{{\bf k}_{3}}(\tau_1),\Pi_{{\bf k}_2} (\tau_2)\right\}_{\bf PB}+\left\{\Pi_{{\bf k}_{2}} (\tau_1),\Pi_{{\bf k}_{3}} (\tau_2)\right\}_{\bf PB}\left\{\Pi_{{\bf k}_{4}}(\tau_1),\Pi_{{\bf k}_1} (\tau_2)\right\}_{\bf PB}\right.\nonumber\\&& \left.~~~~~~~+\left\{\Pi_{{\bf k}_{2}} (\tau_1),\Pi_{{\bf k}_{1}} (\tau_2)\right\}_{\bf PB}\left\{\Pi_{{\bf k}_{4}}(\tau_1),\Pi_{{\bf k}_3} (\tau_2)\right\}_{\bf PB}+\left\{\Pi_{{\bf k}_{2}} (\tau_1),\Pi_{{\bf k}_{4}} (\tau_2)\right\}_{\bf PB}\left\{\Pi_{{\bf k}_{1}}(\tau_1),\Pi_{{\bf k}_3} (\tau_2)\right\}_{\bf PB}\right.\nonumber\\&& \left.~~~~~~~+\left\{\Pi_{{\bf k}_{3}} (\tau_1),\Pi_{{\bf k}_{1}} (\tau_2)\right\}_{\bf PB}\left\{\Pi_{{\bf k}_{4}}(\tau_1),\Pi_{{\bf k}_2} (\tau_2)\right\}_{\bf PB}+\left\{\Pi_{{\bf k}_{3}} (\tau_1),\Pi_{{\bf k}_{2}} (\tau_2)\right\}_{\bf PB}\left\{\Pi_{{\bf k}_{1}}(\tau_1),\Pi_{{\bf k}_4} (\tau_2)\right\}_{\bf PB}\right.\nonumber\\&& \left.~~~~~~~+\left\{\Pi_{{\bf k}_{3}} (\tau_1),\Pi_{{\bf k}_{4}} (\tau_2)\right\}_{\bf PB}\left\{\Pi_{{\bf k}_{1}}(\tau_1),\Pi_{{\bf k}_2} (\tau_2)\right\}_{\bf PB}+\left\{\Pi_{{\bf k}_{4}} (\tau_1),\Pi_{{\bf k}_{1}} (\tau_2)\right\}_{\bf PB}\left\{\Pi_{{\bf k}_{2}}(\tau_1),\Pi_{{\bf k}_3} (\tau_2)\right\}_{\bf PB}\right.\nonumber\\&& \left.~~~~~~~+\left\{\Pi_{{\bf k}_{4}} (\tau_1),\Pi_{{\bf k}_{2}} (\tau_2)\right\}_{\bf PB}\left\{\Pi_{{\bf k}_{3}}(\tau_1),\Pi_{{\bf k}_1} (\tau_2)\right\}_{\bf PB}+\left\{\Pi_{{\bf k}_{4}} (\tau_1),\Pi_{{\bf k}_{3}} (\tau_2)\right\}_{\bf PB}\left\{\Pi_{{\bf k}_{2}}(\tau_1),\Pi_{{\bf k}_1} (\tau_2)\right\}_{\bf PB}\right].~~~~~~\eea
 Now, here our job is to explicitly compute each of the {\it Poisson brackets}, which are appearing in the above mentioned twelve terms.
 The explicit computation gives the following result:
 \bea \left\{f_{{\bf k}_{i}} (\tau_1),f_{{\bf k}_{j}} (\tau_2)\right\}_{\bf PB}&=&\left(\frac{\partial f_{{\bf k}_i}(\tau_1)}{\partial f_{{\bf k}_p}(\tau_2)}\frac{\partial f_{{\bf k}_j}(\tau_1)}{\partial \Pi_{{\bf k}_p}(\tau_2)}-\frac{\partial f_{{\bf k}_i}(\tau_1)}{\partial \Pi_{{\bf k}_p}(\tau_2)}\frac{\partial f_{{\bf k}_j}(\tau_2)}{\partial f_{{\bf k}_p}(\tau_2)}\right)\nonumber\\
 &=&(2\pi)^3\delta^{3}({\bf k}_i+{\bf k}_p)\delta^{3}({\bf k}_j+{\bf k}_p){\bf U}_1(\tau_1,\tau_2)\left(\frac{\partial_{{\bf k}_p} f_{{\bf k}_j}(\tau_1)}{\partial_{{\bf k}_j} \Pi_{{\bf k}_p}(\tau_2)}-\frac{\partial_{{\bf k}_p} f_{{\bf k}_i}(\tau_1)}{\partial_{{\bf k}_i} \Pi_{{\bf k}_p}(\tau_2)}\right)\nonumber\\
 &=&(2\pi)^3\delta^{3}({\bf k}_i+{\bf k}_j){\bf U}_1(\tau_1,\tau_2)\left(\frac{\partial_{{\bf k}_j} f_{{\bf k}_j}(\tau_1)}{\partial_{{\bf k}_j} \Pi_{{\bf k}_j}(\tau_2)}-\frac{\partial_{{\bf k}_j} f_{{\bf k}_i}(\tau_1)}{\partial_{{\bf k}_i} \Pi_{{\bf k}_j}(\tau_2)}\right)\nonumber\\
 &=&(2\pi)^3\delta^{3}({\bf k}_i+{\bf k}_j){\bf U}_1(\tau_1,\tau_2){\bf V}_1(\tau_1,\tau_2)\nonumber\\
 &=&(2\pi)^3\delta^{3}({\bf k}_i+{\bf k}_j)~{\bf R}_1(\tau_1,\tau_2).~~~~~\forall i\ne j ~~{\rm with}~~ i,j=1,2,3,4.~~~~~~~~\eea 
 Here we define the overall time dependent amplitude as:
 \bea {\bf R}_1(\tau_1,\tau_2)={\bf U}_1(\tau_1,\tau_2){\bf V}_1(\tau_1,\tau_2),\eea
 where we have used the following crucial facts:
 \bea \left(\frac{\partial f_{{\bf k}_i}(\tau_1)}{\partial f_{{\bf k}_p}(\tau_2)}\right)=(2\pi)^3\delta^{3}({\bf k}_i+{\bf k}_p){\bf U}_1(\tau_1,\tau_2),\\
\left(\frac{\partial_{{\bf k}_j} f_{{\bf k}_j}(\tau_1)}{\partial_{{\bf k}_j} \Pi_{{\bf k}_j}(\tau_2)}-\frac{\partial_{{\bf k}_j} f_{{\bf k}_i}(\tau_1)}{\partial_{{\bf k}_i} \Pi_{{\bf k}_j}(\tau_2)}\right)={\bf V}_1(\tau_1,\tau_2). \eea
Similarly,  one can compute:
  \bea \left\{\Pi_{{\bf k}_{i}} (\tau_1),\Pi_{{\bf k}_{j}} (\tau_2)\right\}_{\bf PB}&=&\left(\frac{\partial \Pi_{{\bf k}_i}(\tau_1)}{\partial f_{{\bf k}_p}(\tau_2)}\frac{\partial \Pi_{{\bf k}_j}(\tau_1)}{\partial \Pi_{{\bf k}_p}(\tau_2)}-\frac{\partial \Pi_{{\bf k}_i}(\tau_1)}{\partial \Pi_{{\bf k}_p}(\tau_2)}\frac{\partial \Pi_{{\bf k}_j}(\tau_2)}{\partial f_{{\bf k}_p}(\tau_2)}\right)\nonumber\\
 &=&(2\pi)^3\delta^{3}({\bf k}_i+{\bf k}_p)\delta^{3}({\bf k}_j+{\bf k}_p){\bf U}_2(\tau_1,\tau_2)\left(\frac{\partial_{{\bf k}_p} \Pi_{{\bf k}_j}(\tau_1)}{\partial_{{\bf k}_j} f_{{\bf k}_p}(\tau_2)}-\frac{\partial_{{\bf k}_p} \Pi_{{\bf k}_i}(\tau_1)}{\partial_{{\bf k}_i} f_{{\bf k}_p}(\tau_2)}\right)\nonumber\\
 &=&(2\pi)^3\delta^{3}({\bf k}_i+{\bf k}_j){\bf U}_2(\tau_1,\tau_2)\left(\frac{\partial_{{\bf k}_j} \Pi_{{\bf k}_j}(\tau_1)}{\partial_{{\bf k}_j} f_{{\bf k}_j}(\tau_2)}-\frac{\partial_{{\bf k}_j} \Pi_{{\bf k}_i}(\tau_1)}{\partial_{{\bf k}_i} f_{{\bf k}_j}(\tau_2)}\right)\nonumber\\
 &=&(2\pi)^3\delta^{3}({\bf k}_i+{\bf k}_j){\bf U}_2(\tau_1,\tau_2){\bf V}_2(\tau_1,\tau_2)\nonumber\\
 &=&(2\pi)^3\delta^{3}({\bf k}_i+{\bf k}_j)~{\bf R}_2(\tau_1,\tau_2).~~~~~\forall i\ne j ~~{\rm with}~~ i,j=1,2,3,4.~~~~~~~~\eea 
  Here we define the overall time dependent amplitude as:
 \bea {\bf R}_2(\tau_1,\tau_2)={\bf U}_2(\tau_1,\tau_2){\bf V}_2(\tau_1,\tau_2),\eea
 where we have used the following crucial facts:
 \bea \left(\frac{\partial \Pi_{{\bf k}_i}(\tau_1)}{\partial \Pi_{{\bf k}_p}(\tau_2)}\right)=(2\pi)^3\delta^{3}({\bf k}_i+{\bf k}_p){\bf U}_2(\tau_1,\tau_2),\\
\left(\frac{\partial_{{\bf k}_j} \Pi_{{\bf k}_j}(\tau_1)}{\partial_{{\bf k}_j} f_{{\bf k}_j}(\tau_2)}-\frac{\partial_{{\bf k}_j} \Pi_{{\bf k}_i}(\tau_1)}{\partial_{{\bf k}_i} f_{{\bf k}_j}(\tau_2)}\right)={\bf V}_2(\tau_1,\tau_2). \eea
 Here from these computed {\it Poisson brackets} we can extract the following sets of crucial information,  which will further helps us to understand more about the classical limit of the four-point OTO  amplitudes in the present computation:
 \begin{enumerate}
 \item The time dependent part as such very complicated as it contain the information regarding classical version of random stochastic quantum fluctuations in the primordial universe.  In the present context it is hypothesized by a random function ${\bf R}_1(\tau_1,\tau_2)$ and ${\bf R}_2(\tau_1,\tau_2)$ which incorporate the two conformal time scales in the results.
 
 \item Also it is important to note that,  ${\bf R}_1(\tau_1,\tau_2)$ and ${\bf R}_2(\tau_1,\tau_2)$ are homogeneous and isotropic functions,  which captures the dynamical effect of the spatially flat FLRW background.  For this reason the random functions,  ${\bf R}_1(\tau_1,\tau_2)$ and ${\bf R}_2(\tau_1,\tau_2)$ are completely $i$ and $j$ momentum index independent. This is actually the outcome of the stochastic randomness in the present context. Due to having this fact these random functions,  ${\bf R}_1(\tau_1,\tau_2)$ and ${\bf R}_2(\tau_1,\tau_2)$ are non-zero in the present computation.  This is the non-trivial result as in the usual classical field theory these two correlators gives vanishing contribution without having any random fluctuations in the theory.
 
 \item Moreover,  the interesting to point here that,  one can explicitly separately write down the contribution of the inhomogeneity and time dynamics in Fourier space after computing the classical {\it Poisson brackets}.
 
 \item Finally,  the appearance of the three dimensional {\it Dirac Delta} function confirms the momentum conservation in the Fourier space in the classical two point OTO micro-canonical amplitudes.
 \end{enumerate}
 Further,  we compute the square of the {\it Poisson brackets},  which after performing the Fourier transformation can be expressed as:
  \bea \left\{f_{{\bf k}_{i}} (\tau_1),f_{{\bf k}_{j}} (\tau_2)\right\}_{\bf PB}\left\{f_{{\bf k}_{l}} (\tau_1),f_{{\bf k}_{m}} (\tau_2)\right\}_{\bf PB}&=&(2\pi)^6\delta^{3}({\bf k}_i+{\bf k}_j)\delta^{3}({\bf k}_l+{\bf k}_m){\bf U}^2_1(\tau_1,\tau_2){\bf V}^2_1(\tau_1,\tau_2)\nonumber\\
 &=&(2\pi)^6\delta^{3}({\bf k}_i+{\bf k}_j)\delta^{3}({\bf k}_l+{\bf k}_m)~{\bf R}^2_1(\tau_1,\tau_2).~~~~~\\
  \left\{\Pi_{{\bf k}_{i}} (\tau_1),\Pi_{{\bf k}_{j}} (\tau_2)\right\}_{\bf PB}\left\{\Pi_{{\bf k}_{l}} (\tau_1),\Pi_{{\bf k}_{m}} (\tau_2)\right\}_{\bf PB}&=&
  (2\pi)^6\delta^{3}({\bf k}_i+{\bf k}_j)\delta^{3}({\bf k}_l+{\bf k}_m){\bf U}^2_2(\tau_1,\tau_2){\bf V}^2_2(\tau_1,\tau_2)\nonumber\\
 &=&(2\pi)^6\delta^{3}({\bf k}_i+{\bf k}_j)\delta^{3}({\bf k}_l+{\bf k}_m)~{\bf R}^2_2(\tau_1,\tau_2).~~~~~\nonumber\\
  &&~~~~~~~~~\forall i\neq j \neq l\neq m~~{\rm with}~~ i,j,k,l=1,2,3,4.~~~~~~~~\nonumber\\
  &&\eea
 Consequently,  we get the following simplified results:
  \bea && \left\{{f}({\bf x},\tau_1),{f}({\bf x},\tau_2)\right\}^2_{\bf PB}=(2\pi)^6 \int \frac{d^3{\bf k}_1}{(2\pi)^3}\int \frac{d^3{\bf k}_2}{(2\pi)^3}\int \frac{d^3{\bf k}_3}{(2\pi)^3}\int \frac{d^3{\bf k}_4}{(2\pi)^3}~\exp\left(i({\bf k}_1+{\bf k}_2+{\bf k}_3+{\bf k}_4).{\bf x}\right)\nonumber\\
 &&~~~~~~~~~~~~~~~~~~~~~~~~~~~~~~~~~\underbrace{\sum^{4}_{i,j,l,m=1,i\neq j\neq l \neq m}\delta^{3}({\bf k}_i+{\bf k}_j)\delta^{3}({\bf k}_l+{\bf k}_m)}_{\textcolor{red}{\bf Contribution~from~12~terms}}~{\bf R}^2_1(\tau_1,\tau_2).\eea\bea && \left\{{\Pi}({\bf x},\tau_1),{\Pi}({\bf x},\tau_2)\right\}^2_{\bf PB}=(2\pi)^6 \int \frac{d^3{\bf k}_1}{(2\pi)^3}\int \frac{d^3{\bf k}_2}{(2\pi)^3}\int \frac{d^3{\bf k}_3}{(2\pi)^3}\int \frac{d^3{\bf k}_4}{(2\pi)^3}~\exp\left(i({\bf k}_1+{\bf k}_2+{\bf k}_3+{\bf k}_4).{\bf x}\right)\nonumber\\
 &&~~~~~~~~~~~~~~~~~~~~~~~~~~~~~~~~~\underbrace{\sum^{4}_{i,j,l,m=1,i\neq j\neq l \neq m}\delta^{3}({\bf k}_i+{\bf k}_j)\delta^{3}({\bf k}_l+{\bf k}_m)}_{\textcolor{red}{\bf Contribution~from~12~terms}}~{\bf R}^2_2(\tau_1,\tau_2).\eea
 Here,  the explicit computation gives following product identity of the Dirac Delta function:
 \bea &&\sum^{4}_{i,j,l,m=1,i\neq j\neq l \neq m}\delta^{3}({\bf k}_i+{\bf k}_j)\delta^{3}({\bf k}_l+{\bf k}_m)=\left[\delta^{3}({\bf k}_1+{\bf k}_2)\delta^3({\bf k}_3+{\bf k}_4)+\delta^{3}({\bf k}_1+{\bf k}_3)\delta^3({\bf k}_2+{\bf k}_4)\right.\nonumber\\&& \left.~~~~~~~~~~~~~~~~~~~~~~~~~~~~~~
 ~~~~~~~~~~~~~~~~~~~~~+\delta^{3}({\bf k}_1+{\bf k}_4)\delta^3({\bf k}_3+{\bf k}_2)+\delta^{3}({\bf k}_2+{\bf k}_3)\delta^3({\bf k}_4+{\bf k}_1)\right.\nonumber\\&& \left.~~~~~~~~~~~~~~~~~~~~~~~~~~~~~~
 ~~~~~~~~~~~~~~~~~~~~~+\delta^{3}({\bf k}_2+{\bf k}_1)\delta^3({\bf k}_4+{\bf k}_3)+\delta^{3}({\bf k}_2+{\bf k}_4)\delta^3({\bf k}_1+{\bf k}_3)\right.\nonumber\\&& \left.~~~~~~~~~~~~~~~~~~~~~~~~~~~~~~
 ~~~~~~~~~~~~~~~~~~~~~+\delta^{3}({\bf k}_3+{\bf k}_1)\delta^3({\bf k}_4+{\bf k}_2)+\delta^{3}({\bf k}_3+{\bf k}_2)\delta^3({\bf k}_1+{\bf k}_4)\right.\nonumber\\&& \left.~~~~~~~~~~~~~~~~~~~~~~~~~~~~~~
 ~~~~~~~~~~~~~~~~~~~~~+\delta^{3}({\bf k}_3+{\bf k}_4)\delta^3({\bf k}_1+{\bf k}_2)+\delta^{3}({\bf k}_4+{\bf k}_1)\delta^3({\bf k}_2+{\bf k}_3)\right.\nonumber\\&& \left.~~~~~~~~~~~~~~~~~~~~~~~~~~~~~~
 ~~~~~~~~~~~~~~~~~~~~~+\delta^{3}({\bf k}_4+{\bf k}_2)\delta^3({\bf k}_3+{\bf k}_1)+\delta^{3}({\bf k}_4+{\bf k}_3)\delta^3({\bf k}_2+{\bf k}_1)\right].~~~~~~~~~\eea
 Now, we give the following proposal to quantify the random function ${\bf R}^2(\tau_1,\tau_2)$, which is given by the following expression:
 \bea {\bf R}^2_1(\tau_1,\tau_2):=\underbrace{\langle \eta_{\bf Noise}(\tau_1)\eta_{\bf Noise}(\tau_2)\rangle}_{\textcolor{red}{\bf Contribution~from~random~noise~field~correlation}},\\ {\bf R}^2_2(\tau_1,\tau_2):=\underbrace{\langle \Pi_{\eta_{\bf Noise}}(\tau_1) \Pi_{\eta_{\bf Noise}}(\tau_2)\rangle}_{\textcolor{red}{\bf Contribution~from~random~momentum~correlation}},\eea
 where $ \eta_{\bf Noise}(\tau_i)~\forall~i=1,2$ and $ \Pi_{\eta_{\bf Noise}}(\tau_i)~\forall~i=1,2$ represent the conformal time dependent random noise field and momentum functions.  
 
 Also it is important to note that the two consecutive noise kernels is time translation invariant,  for which we have written as: 
 \bea \langle \eta_{\bf Noise}(\tau_1)\eta_{\bf Noise}(\tau_2)\rangle={\bf G}^{(1)}_{\bf Kernel}(\tau_1,\tau_2):={\bf G}^{(1)}_{\bf Kernel}(|\tau_1-\tau_2|),\\
 \langle \Pi_{\eta_{\bf Noise}}(\tau_1)\Pi_{\eta_{\bf Noise}}(\tau_2)\rangle={\bf G}^{(2)}_{\bf Kernel}(\tau_1,\tau_2):={\bf G}^{(2)}_{\bf Kernel}(|\tau_1-\tau_2|).\eea
 Additionally, the conformal time dependent noise satisfy the following constraint conditions:
 \bea &&\langle \eta_{\bf Noise}(\tau_i)\rangle=0,~~~~~~~~~~~~~~~~~~~~~{\bf Noise=}~~\textcolor{red}{\bf Gaussian,~~Non-Gaussian},\\
 &&\langle \eta_{\bf Noise}(\tau_1)\eta_{\bf Noise}(\tau_2)\eta_{\bf Noise}(\tau_3)\rangle=0,~~~~~~~~~~~~~~~~~~~{\bf Noise=}~~\textcolor{red}{\bf Gaussian},\\
 &&\langle \eta_{\bf Noise}(\tau_1)\eta_{\bf Noise}(\tau_2).......\eta_{\bf Noise}(\tau_N)\rangle={\bf f}^{(1)}_{\bf Noise}(\tau_1,\tau_2,....,\tau_N)\neq 0~~\forall~N\geq 2,\nonumber\\
 &&~~~~~~~~~~~~~~~~~~~~~~~~~~~~~~~~~~~~~~~~~~~~~~~~~~~~~~~~~{\bf Noise=}~~\textcolor{red}{\bf Non-Gaussian}.~~\eea
 and
  \bea &&\langle \Pi_{\eta_{\bf Noise}}(\tau_i)\rangle=0,~~~~~~~~~~~~~~~~~~~~~{\bf Noise=}~~\textcolor{red}{\bf Gaussian,~~Non-Gaussian},\\
 &&\langle \Pi_{\eta_{\bf Noise}}(\tau_1)\Pi_{\eta_{\bf Noise}}(\tau_2)\eta_{\bf Noise}(\tau_3)\rangle=0,~~~~~~~~~~~~~~~~~~~{\bf Noise=}~~\textcolor{red}{\bf Gaussian},\\
 &&\langle \Pi_{\eta_{\bf Noise}}(\tau_1)\Pi_{\eta_{\bf Noise}}(\tau_2).......\Pi_{\eta_{\bf Noise}}(\tau_N)\rangle={\bf f}^{(2)}_{\bf Noise}(\tau_1,\tau_2,....,\tau_N)\neq 0~~\forall~N\geq 2,\nonumber\\
 &&~~~~~~~~~~~~~~~~~~~~~~~~~~~~~~~~~~~~~~~~~~~~~~~~~~~~~~~~~{\bf Noise=}~~\textcolor{red}{\bf Non-Gaussian}.~~\eea
 After substituting this result in the previously computed expression for the amplitude we get the following simplified expression:
  \bea && \left\{{f}({\bf x},\tau_1),{f}({\bf x},\tau_2)\right\}^2_{\bf PB}\nonumber\\
 &&=(2\pi)^6\prod^{4}_{p=1} \int \frac{d^3{\bf k}_p}{(2\pi)^3}~\exp\left(i{\bf k}_p.{\bf x}\right)\sum^{4}_{i,j,l,m=1,i\neq j\neq l \neq m}\delta^{3}({\bf k}_i+{\bf k}_j)\delta^{3}({\bf k}_l+{\bf k}_m){\bf G}^{(2)}_{\bf Kernel}(|\tau_1-\tau_2|),~~~~~~~~~~~~~~\\ && \left\{{\Pi}({\bf x},\tau_1),{\Pi}({\bf x},\tau_2)\right\}^2_{\bf PB}\nonumber\\
 &&=(2\pi)^6\prod^{4}_{q=1} \int \frac{d^3{\bf k}_q}{(2\pi)^3}~\exp\left(i{\bf k}_q.{\bf x}\right)\sum^{4}_{i,j,l,m=1,i\neq j\neq l \neq m}\delta^{3}({\bf k}_i+{\bf k}_j)\delta^{3}({\bf k}_l+{\bf k}_m){\bf G}^{(2)}_{\bf Kernel}(|\tau_1-\tau_2|).~~~~~~~~~~~~~~\eea

\section{Computation of the trace of the two-point amplitude in  OTOC}
\label{sec:11}
Now,  we will explicitly compute the numerator of the OTOC for quantum Mota Allen vacua, which is given by the following expression:
  \bea && {\rm Tr}\left[e^{-\beta \widehat{H}(\tau_1)}\left[\hat{f}({\bf x},\tau_1),\hat{f}({\bf x},\tau_2)\right]\right]_{(\alpha,\gamma)}\nonumber\\
 &&= \frac{\exp(-2\sin\gamma{\rm tan}\alpha)}{|\cosh\alpha|}\int d\Psi_{\bf BD}~\int \frac{d^3{\bf k}_1}{(2\pi)^3}\int \frac{d^3{\bf k}_2}{(2\pi)^3}\exp\left[i\left({\bf k}_1+{\bf k}_2\right).{\bf x}\right]~~~~~~~~\nonumber\\
  &&~~~~~~~~~~~~~~~~~~~~~~~~~~ \langle\Psi_{\bf BD}|\left[\hat{\nabla}^{(1)}_1({\bf k}_1,{\bf k}_2;\tau_1,\tau_2;\beta) -\hat{\nabla}^{(1)}_2({\bf k}_1,{\bf k}_2;\tau_1,\tau_2;\beta)\right]|\Psi_{\bf BD}\rangle.~~~~~~~~~~~ \eea
  and 
   \bea && {\rm Tr}\left[e^{-\beta \widehat{H}(\tau_1)}\left[\hat{\Pi}({\bf x},\tau_1),\hat{\Pi}({\bf x},\tau_2)\right]\right]_{(\alpha,\gamma)}\nonumber\\
 &&= \frac{\exp(-2\sin\gamma{\rm tan}\alpha)}{|\cosh\alpha|}\int d\Psi_{\bf BD}~\int \frac{d^3{\bf k}_1}{(2\pi)^3}\int \frac{d^3{\bf k}_2}{(2\pi)^3}\exp\left[i\left({\bf k}_1+{\bf k}_2\right).{\bf x}\right]~~~~~~~~\nonumber\\
  &&~~~~~~~~~~~~~~~~~~~~~~~~~~ \langle\Psi_{\bf BD}|\left[\hat{\nabla}^{(2)}_1({\bf k}_1,{\bf k}_2;\tau_1,\tau_2;\beta) -\hat{\nabla}^{(2)}_2({\bf k}_1,{\bf k}_2;\tau_1,\tau_2;\beta)\right]|\Psi_{\bf BD}\rangle.~~~~~~~~~~~ \eea 
  Further, our aim is to compute the individual contributions which are given by:
  \bea &&\int d\Psi_{\bf BD}~\langle\Psi_{\bf BD}|\hat{\nabla}^{(1)}_1({\bf k}_1,{\bf k}_2;\tau_1,\tau_2;\beta)|\Psi_{\bf BD}\rangle=\int d\Psi_{\bf BD}~ \langle\Psi_{\bf BD}|e^{-\beta \hat{H}(\tau_1)}~\hat{\Delta}^{(1)}_1({\bf k}_1,{\bf k}_2;\tau_1,\tau_2)|\Psi_{\bf BD}\rangle,~~~~~~~~~~\\
 &&\int d\Psi_{\bf BD}~\langle\Psi_{\bf BD}|\hat{\nabla}^{(1)}_2({\bf k}_1,{\bf k}_2;\tau_1,\tau_2;\beta)|\Psi_{\bf BD}\rangle= \int d\Psi_{\bf BD}~\langle\Psi_{\bf BD}|e^{-\beta \hat{H}(\tau_1)}~\hat{\Delta}^{(1)}_2({\bf k}_1,{\bf k}_2;\tau_1,\tau_2)|\Psi_{\bf BD}\rangle.\eea
 and 
  \bea &&\int d\Psi_{\bf BD}~\langle\Psi_{\bf BD}|\hat{\nabla}^{(2)}_1({\bf k}_1,{\bf k}_2;\tau_1,\tau_2;\beta)|\Psi_{\bf BD}\rangle=\int d\Psi_{\bf BD}~ \langle\Psi_{\bf BD}|e^{-\beta \hat{H}(\tau_1)}~\hat{\Delta}^{(2)}_1({\bf k}_1,{\bf k}_2;\tau_1,\tau_2)|\Psi_{\bf BD}\rangle,~~~~~~~~~~\\
 &&\int d\Psi_{\bf BD}~\langle\Psi_{\bf BD}|\hat{\nabla}^{(2)}_2({\bf k}_1,{\bf k}_2;\tau_1,\tau_2;\beta)|\Psi_{\bf BD}\rangle= \int d\Psi_{\bf BD}~\langle\Psi_{\bf BD}|e^{-\beta \hat{H}(\tau_1)}~\hat{\Delta}^{(2)}_2({\bf k}_1,{\bf k}_2;\tau_1,\tau_2)|\Psi_{\bf BD}\rangle.\eea
  Let us evaluate one by one each of the contributions, which are given introducing normal ordering by:
 \bea &&\int d\Psi_{\bf BD}~\langle\Psi_{\bf BD}|:e^{-\beta \hat{H}(\tau_1)}~a_{{\bf k}_1}a_{{\bf k}_2}:|\Psi_{\bf BD}\rangle=0,\\
 &&\int d\Psi_{\bf BD}~\langle\Psi_{\bf BD}|:e^{-\beta \hat{H}(\tau_1)}~a_{{\bf k}_1}a^{\dagger}_{-{\bf k}_2}:|\Psi_{\bf BD}\rangle\nonumber\\
   &&~~~~~~~~~~~~~~~~~~~~~~~~~~~~~~=(2\pi)^3\exp\left(-\int d^3{\bf k}~\ln\left(2\sinh \frac{\beta E_{\bf k}(\tau_1)}{2}\right)\right)\delta^3\left({\bf k}_1+{\bf k}_2\right),~~~~~~~~~~~\eea\bea
 &&\int d\Psi_{\bf BD}~\langle\Psi_{\bf BD}|:e^{-\beta \hat{H}(\tau_1)}~a^{\dagger}_{-{\bf k}_1}a_{{\bf k}_2}:|\Psi_{\bf BD}\rangle\nonumber\\
   &&~~~~~~~~~~~~~~~~~~~~~~~~~~~~~~=(2\pi)^3\exp\left(-\int d^3{\bf k}~\ln\left(2\sinh \frac{\beta E_{\bf k}(\tau_1)}{2}\right)\right)\delta^3\left({\bf k}_1+{\bf k}_2\right),~~~~~~~~~~~\\
   &&\int d\Psi_{\bf BD}~\langle\Psi_{\bf BD}|:e^{-\beta \hat{H}(\tau_1)}~a^{\dagger}_{-{\bf k}_1}a^{\dagger}_{-{\bf k}_2}:|\Psi_{\bf BD}\rangle=0.\eea
 Consequently, the individual contributions can be computed in the normal ordered form as: 
 \bea &&\int d\Psi_{\bf BD}~\langle\Psi_{\bf BD}|\hat{\nabla}^{(1)}_1({\bf k}_1,{\bf k}_2;\tau_1,\tau_2;\beta)|\Psi_{\bf BD}\rangle\nonumber\\
 &&=(2\pi)^3\exp\left(-\int d^3{\bf k}~\ln\left(2\sinh \frac{\beta E_{\bf k}(\tau_1)}{2}\right)\right)\delta^{3}({\bf k}_1+{\bf k}_2)\left[ {\cal D}^{(1)}_2({\bf k}_1,{\bf k}_2;\tau_1,\tau_2)+ {\cal D}^{(1)}_3({\bf k}_1,{\bf k}_2;\tau_1,\tau_2)\right],~~~~~~~~~~~\\
  &&\int d\Psi_{\bf BD}~\langle\Psi_{\bf BD}|\hat{\nabla}^{(1)}_2({\bf k}_1,{\bf k}_2;\tau_1,\tau_2;\beta)|\Psi_{\bf BD}\rangle\nonumber\\
 &&=(2\pi)^3\exp\left(-\int d^3{\bf k}~\ln\left(2\sinh \frac{\beta E_{\bf k}(\tau_1)}{2}\right)\right)\delta^{3}({\bf k}_1+{\bf k}_2)\left[ {\cal L}^{(1)}_2({\bf k}_1,{\bf k}_2;\tau_1,\tau_2)+ {\cal L}^{(1)}_3({\bf k}_1,{\bf k}_2;\tau_1,\tau_2)\right],~~~~~~~~~~~\\
 &&\int d\Psi_{\bf BD}~\langle\Psi_{\bf BD}|\hat{\nabla}^{(2)}_1({\bf k}_1,{\bf k}_2;\tau_1,\tau_2;\beta)|\Psi_{\bf BD}\rangle\nonumber\\
 &&=(2\pi)^3\exp\left(-\int d^3{\bf k}~\ln\left(2\sinh \frac{\beta E_{\bf k}(\tau_1)}{2}\right)\right)\delta^{3}({\bf k}_1+{\bf k}_2)\left[ {\cal D}^{(2)}_2({\bf k}_1,{\bf k}_2;\tau_1,\tau_2)+ {\cal D}^{(2)}_3({\bf k}_1,{\bf k}_2;\tau_1,\tau_2)\right],~~~~~~~~~~~\\
  &&\int d\Psi_{\bf BD}~\langle\Psi_{\bf BD}|\hat{\nabla}^{(2)}_2({\bf k}_1,{\bf k}_2;\tau_1,\tau_2;\beta)|\Psi_{\bf BD}\rangle\nonumber\\
 &&=(2\pi)^3\exp\left(-\int d^3{\bf k}~\ln\left(2\sinh \frac{\beta E_{\bf k}(\tau_1)}{2}\right)\right)\delta^{3}({\bf k}_1+{\bf k}_2)\left[ {\cal L}^{(2)}_2({\bf k}_1,{\bf k}_2;\tau_1,\tau_2)+ {\cal L}^{(2)}_3({\bf k}_1,{\bf k}_2;\tau_1,\tau_2)\right],~~~~~~~~~~~\eea

	\newpage
\section{Computation of the trace of the four-point amplitude in  OTOC}
\label{sec:12}
Now, we will explicitly compute the numerator of the OTOC for quantum $\alpha$ vacua,which is given by the following expression:
  \bea && {\rm Tr}\left[e^{-\beta \widehat{H}(\tau_1)}\left[\hat{f}({\bf x},\tau_1),\hat{f}({\bf x},\tau_2)\right]^2\right]_{(\alpha)}\nonumber\\
 &&= \frac{\exp(-2\sin\gamma{\rm tan}\alpha)}{|\cosh\alpha|}\int d\Psi_{\bf BD}~\int \frac{d^3{\bf k}_1}{(2\pi)^3}\int \frac{d^3{\bf k}_2}{(2\pi)^3}\int \frac{d^3{\bf k}_3}{(2\pi)^3}\int \frac{d^3{\bf k}_4}{(2\pi)^3}\nonumber\\
 &&~~~~~~~~~~~~~~~~~~~~~~~~~\exp\left[i\left({\bf k}_1+{\bf k}_2+{\bf k}_3+{\bf k}_4\right).{\bf x}\right]~~~~~~~~\nonumber\\
  &&~ \langle\Psi_{\bf BD}|\left[\widehat{\cal V}^{(1)}_1({\bf k}_1,{\bf k}_2,{\bf k}_3,{\bf k}_4;\tau_1,\tau_2;\beta)-\widehat{\cal V}^{(1)}_2({\bf k}_1,{\bf k}_2,{\bf k}_3,{\bf k}_4;\tau_1,\tau_2;\beta)\right.\nonumber\\ && \left.~~~ +\widehat{\cal V}^{(1)}_3({\bf k}_1,{\bf k}_2,{\bf k}_3,{\bf k}_4;\tau_1,\tau_2;\beta) -\widehat{\cal V}^{(1)}_4({\bf k}_1,{\bf k}_2,{\bf k}_3,{\bf k}_4;\tau_1,\tau_2;\beta)\right]|\Psi_{\bf BD}\rangle.~~~~~~~~~~~ \\
   && {\rm Tr}\left[e^{-\beta \widehat{H}(\tau_1)}\left[\hat{\Pi}({\bf x},\tau_1),\hat{\Pi}({\bf x},\tau_2)\right]^2\right]_{(\alpha)}\nonumber\\
 &&= \frac{\exp(-2\sin\gamma{\rm tan}\alpha)}{|\cosh\alpha|}\int d\Psi_{\bf BD}~\int \frac{d^3{\bf k}_1}{(2\pi)^3}\int \frac{d^3{\bf k}_2}{(2\pi)^3}\int \frac{d^3{\bf k}_3}{(2\pi)^3}\int \frac{d^3{\bf k}_4}{(2\pi)^3}\nonumber\\
 &&~~~~~~~~~~~~~~~~~~~~~~~~~\exp\left[i\left({\bf k}_1+{\bf k}_2+{\bf k}_3+{\bf k}_4\right).{\bf x}\right]~~~~~~~~\nonumber\\
  &&~\langle\Psi_{\bf BD}|\left[\widehat{\cal V}^{(2)}_1({\bf k}_1,{\bf k}_2,{\bf k}_3,{\bf k}_4;\tau_1,\tau_2;\beta) -\widehat{\cal V}^{(2)}_2({\bf k}_1,{\bf k}_2,{\bf k}_3,{\bf k}_4;\tau_1,\tau_2;\beta)\right.\nonumber\\ && \left.~~+\widehat{\cal V}^{(2)}_3({\bf k}_1,{\bf k}_2,{\bf k}_3,{\bf k}_4;\tau_1,\tau_2;\beta) -\widehat{\cal V}^{(2)}_4({\bf k}_1,{\bf k}_2,{\bf k}_3,{\bf k}_4;\tau_1,\tau_2;\beta)\right]|\Psi_{\bf BD}\rangle.~~~~~~~~~~~ \eea 
  Further, our aim is to compute the individual contributions for $l=1,2$ which are given by:
  \bea &&\int d\Psi_{\bf BD}~\langle\Psi_{\bf BD}|\widehat{\cal V}^{(l)}_1({\bf k}_1,{\bf k}_2,{\bf k}_3,{\bf k}_4;\tau_1,\tau_2;\beta)|\Psi_{\bf BD}\rangle\nonumber\\
  &&~~~~~~~~~~~~~~~~~~= \int d\Psi_{\bf BD}~\langle\Psi_{\bf BD}|e^{-\beta \hat{H}(\tau_1)}~\widehat{\cal T}^{(l)}_1({\bf k}_1,{\bf k}_2,{\bf k}_3,{\bf k}_4;\tau_1,\tau_2)|\Psi_{\bf BD}\rangle,~~~~~~~~~~~~~~~~\\
 &&\int d\Psi_{\bf BD}~\langle\Psi_{\bf BD}|\widehat{\cal V}^{(l)}_2({\bf k}_1,{\bf k}_2,{\bf k}_3,{\bf k}_4;\tau_1,\tau_2;\beta)|\Psi_{\bf BD}\rangle\nonumber\\
  &&~~~~~~~~~~~~~~~~~~=\int d\Psi_{\bf BD}~\langle\Psi_{\bf BD}|e^{-\beta \hat{H}(\tau_1)}~\widehat{\cal T}^{(l)}_2({\bf k}_1,{\bf k}_2,{\bf k}_3,{\bf k}_4;\tau_1,\tau_2)|\Psi_{\bf BD}\rangle,\\
&&\int d\Psi_{\bf BD}~\langle\Psi_{\bf BD}|\widehat{\cal V}^{(l)}_3({\bf k}_1,{\bf k}_2,{\bf k}_3,{\bf k}_4;\tau_1,\tau_2;\beta)|\Psi_{\bf BD}\rangle\nonumber\\
  &&~~~~~~~~~~~~~~~~~~= \int d\Psi_{\bf BD}~\langle\Psi_{\bf BD}|e^{-\beta \hat{H}(\tau_1)}~\widehat{\cal T}^{(l)}_3({\bf k}_1,{\bf k}_2,{\bf k}_3,{\bf k}_4;\tau_1,\tau_2)|\Psi_{\bf BD}\rangle,\\
 &&\int d\Psi_{\bf BD}~\langle\Psi_{\bf BD}|\widehat{\cal V}^{(l)}_4({\bf k}_1,{\bf k}_2,{\bf k}_3,{\bf k}_4;\tau_1,\tau_2;\beta)|\Psi_{\bf BD}\rangle\nonumber\\
  &&~~~~~~~~~~~~~~~~~~= \int d\Psi_{\bf BD}~\langle\Psi_{\bf BD}|e^{-\beta \hat{H}(\tau_1)}~\widehat{\cal T}^{(l)}_4({\bf k}_1,{\bf k}_2,{\bf k}_3,{\bf k}_4;\tau_1,\tau_2)|\Psi_{\bf BD}\rangle.\eea
  Let us evaluate one by one each of the contributions,  by introducing normal ordering for $l=1,2$,  which are given by:
    \bea &&\int d\Psi_{\bf BD}~\langle\Psi_{\bf BD}|:\widehat{\cal V}^{(l)}_1({\bf k}_1,{\bf k}_2,{\bf k}_3,{\bf k}_4;\tau_1,\tau_2;\beta):|\Psi_{\bf BD}\rangle\nonumber\\
 &&=(2\pi)^6\exp\left(-\int d^3{\bf k}~\ln\left(2\sinh \frac{\beta E_{\bf k}(\tau_1)}{2}\right)\right)\nonumber\\
 &&~~~~~~~~~~~~\left[ {\cal M}^{(l)}_4({\bf k}_1,{\bf k}_2,{\bf k}_3,{\bf k}_4;\tau_1,\tau_2)~\left\{\delta^3\left({\bf k}_1+{\bf k}_4\right)\delta^3\left({\bf k}_2+{\bf k}_3\right)+\delta^3\left({\bf k}_1+{\bf k}_3\right)\delta^3\left({\bf k}_2+{\bf k}_4\right)\right\}\right.\nonumber\\
  && \left.~~~~~~~~~~~~+ {\cal M}^{(l)}_6({\bf k}_1,{\bf k}_2,{\bf k}_3,{\bf k}_4;\tau_1,\tau_2)~\left\{\delta^3\left({\bf k}_1+{\bf k}_2\right)\delta^3\left({\bf k}_3+{\bf k}_4\right)+\delta^3\left({\bf k}_1+{\bf k}_4\right)\delta^3\left({\bf k}_2+{\bf k}_3\right)\right\}\right.\nonumber\\
  && \left.~~~~~~~~~~~~+{\cal M}^{(l)}_7({\bf k}_1,{\bf k}_2,{\bf k}_3,{\bf k}_4;\tau_1,\tau_2)~\left\{\delta^3\left({\bf k}_1+{\bf k}_2\right)\delta^3\left({\bf k}_3+{\bf k}_4\right)+\delta^3\left({\bf k}_1+{\bf k}_3\right)\delta^3\left({\bf k}_2+{\bf k}_4\right)\right\}\right.\nonumber\\
  && \left.~~~~~~~~~~~~+ {\cal M}^{(l)}_{10}({\bf k}_1,{\bf k}_2,{\bf k}_3,{\bf k}_4;\tau_1,\tau_2)~\left\{\delta^3\left({\bf k}_1+{\bf k}_2\right)\delta^3\left({\bf k}_3+{\bf k}_4\right)+\delta^3\left({\bf k}_1+{\bf k}_3\right)\delta^3\left({\bf k}_2+{\bf k}_4\right)\right\}\right.\nonumber\\
  && \left.~~~~~~~~~~~~+{\cal M}^{(l)}_{11}({\bf k}_1,{\bf k}_2,{\bf k}_3,{\bf k}_4;\tau_1,\tau_2)~\left\{\delta^3\left({\bf k}_1+{\bf k}_2\right)\delta^3\left({\bf k}_3+{\bf k}_4\right)+\delta^3\left({\bf k}_1+{\bf k}_4\right)\delta^3\left({\bf k}_2+{\bf k}_3\right)\right\}\right.\nonumber\\
  && \left.~~~~~~~~~~~~+{\cal M}^{(l)}_{13}({\bf k}_1,{\bf k}_2,{\bf k}_3,{\bf k}_4;\tau_1,\tau_2)~\left\{\delta^3\left({\bf k}_1+{\bf k}_3\right)\delta^3\left({\bf k}_2+{\bf k}_4\right)+\delta^3\left({\bf k}_1+{\bf k}_4\right)\delta^3\left({\bf k}_2+{\bf k}_3\right)\right\}\right],~~~~~~~~~~~\\
  &&\int d\Psi_{\bf BD}~\langle\Psi_{\bf BD}|:\widehat{\cal V}^{(l)}_2({\bf k}_1,{\bf k}_2,{\bf k}_3,{\bf k}_4;\tau_1,\tau_2;\beta):|\Psi_{\bf BD}\rangle\nonumber\\
 &&=(2\pi)^6\exp\left(-\int d^3{\bf k}~\ln\left(2\sinh \frac{\beta E_{\bf k}(\tau_1)}{2}\right)\right)\nonumber\\
 &&~~~~~~~~~~~~\left[ {\cal J}^{(l)}_4({\bf k}_1,{\bf k}_2,{\bf k}_3,{\bf k}_4;\tau_1,\tau_2)~\left\{\delta^3\left({\bf k}_1+{\bf k}_4\right)\delta^3\left({\bf k}_2+{\bf k}_3\right)+\delta^3\left({\bf k}_1+{\bf k}_3\right)\delta^3\left({\bf k}_2+{\bf k}_4\right)\right\}\right.\nonumber\\
  && \left.~~~~~~~~~~~~+ {\cal J}^{(l)}_6({\bf k}_1,{\bf k}_2,{\bf k}_3,{\bf k}_4;\tau_1,\tau_2)~\left\{\delta^3\left({\bf k}_1+{\bf k}_2\right)\delta^3\left({\bf k}_3+{\bf k}_4\right)+\delta^3\left({\bf k}_1+{\bf k}_4\right)\delta^3\left({\bf k}_2+{\bf k}_3\right)\right\}\right.\nonumber\\
  && \left.~~~~~~~~~~~~+{\cal J}^{(l)}_7({\bf k}_1,{\bf k}_2,{\bf k}_3,{\bf k}_4;\tau_1,\tau_2)~\left\{\delta^3\left({\bf k}_1+{\bf k}_2\right)\delta^3\left({\bf k}_3+{\bf k}_4\right)+\delta^3\left({\bf k}_1+{\bf k}_3\right)\delta^3\left({\bf k}_2+{\bf k}_4\right)\right\}\right.\nonumber\\
  && \left.~~~~~~~~~~~~+ {\cal J}^{(l)}_{10}({\bf k}_1,{\bf k}_2,{\bf k}_3,{\bf k}_4;\tau_1,\tau_2)~\left\{\delta^3\left({\bf k}_1+{\bf k}_2\right)\delta^3\left({\bf k}_3+{\bf k}_4\right)+\delta^3\left({\bf k}_1+{\bf k}_3\right)\delta^3\left({\bf k}_2+{\bf k}_4\right)\right\}\right.\nonumber\\
  && \left.~~~~~~~~~~~~+{\cal J}^{(l)}_{11}({\bf k}_1,{\bf k}_2,{\bf k}_3,{\bf k}_4;\tau_1,\tau_2)~\left\{\delta^3\left({\bf k}_1+{\bf k}_2\right)\delta^3\left({\bf k}_3+{\bf k}_4\right)+\delta^3\left({\bf k}_1+{\bf k}_4\right)\delta^3\left({\bf k}_2+{\bf k}_3\right)\right\}\right.\nonumber\\
  && \left.~~~~~~~~~~~~+{\cal J}^{(l)}_{13}({\bf k}_1,{\bf k}_2,{\bf k}_3,{\bf k}_4;\tau_1,\tau_2)~\left\{\delta^3\left({\bf k}_1+{\bf k}_3\right)\delta^3\left({\bf k}_2+{\bf k}_4\right)+\delta^3\left({\bf k}_1+{\bf k}_4\right)\delta^3\left({\bf k}_2+{\bf k}_3\right)\right\}\right],~~~~~~~~~~~\\
  &&\int d\Psi_{\bf BD}~\langle\Psi_{\bf BD}|:\widehat{\cal V}^{(l)}_3({\bf k}_1,{\bf k}_2,{\bf k}_3,{\bf k}_4;\tau_1,\tau_2;\beta):|\Psi_{\bf BD}\rangle\nonumber\\
 &&=(2\pi)^6\exp\left(-\int d^3{\bf k}~\ln\left(2\sinh \frac{\beta E_{\bf k}(\tau_1)}{2}\right)\right)\nonumber\\
 &&~~~~~~~~~~~~\left[ {\cal N}^{(l)}_4({\bf k}_1,{\bf k}_2,{\bf k}_3,{\bf k}_4;\tau_1,\tau_2)~\left\{\delta^3\left({\bf k}_1+{\bf k}_4\right)\delta^3\left({\bf k}_2+{\bf k}_3\right)+\delta^3\left({\bf k}_1+{\bf k}_3\right)\delta^3\left({\bf k}_2+{\bf k}_4\right)\right\}\right.\nonumber\\
  && \left.~~~~~~~~~~~~+ {\cal N}^{(l)}_6({\bf k}_1,{\bf k}_2,{\bf k}_3,{\bf k}_4;\tau_1,\tau_2)~\left\{\delta^3\left({\bf k}_1+{\bf k}_2\right)\delta^3\left({\bf k}_3+{\bf k}_4\right)+\delta^3\left({\bf k}_1+{\bf k}_4\right)\delta^3\left({\bf k}_2+{\bf k}_3\right)\right\}\right.\nonumber\\
  && \left.~~~~~~~~~~~~+{\cal N}^{(l)}_7({\bf k}_1,{\bf k}_2,{\bf k}_3,{\bf k}_4;\tau_1,\tau_2)~\left\{\delta^3\left({\bf k}_1+{\bf k}_2\right)\delta^3\left({\bf k}_3+{\bf k}_4\right)+\delta^3\left({\bf k}_1+{\bf k}_3\right)\delta^3\left({\bf k}_2+{\bf k}_4\right)\right\}\right.\nonumber\\
  && \left.~~~~~~~~~~~~+ {\cal N}^{(l)}_{10}({\bf k}_1,{\bf k}_2,{\bf k}_3,{\bf k}_4;\tau_1,\tau_2)~\left\{\delta^3\left({\bf k}_1+{\bf k}_2\right)\delta^3\left({\bf k}_3+{\bf k}_4\right)+\delta^3\left({\bf k}_1+{\bf k}_3\right)\delta^3\left({\bf k}_2+{\bf k}_4\right)\right\}\right.\nonumber\\
  && \left.~~~~~~~~~~~~+{\cal N}^{(l)}_{11}({\bf k}_1,{\bf k}_2,{\bf k}_3,{\bf k}_4;\tau_1,\tau_2)~\left\{\delta^3\left({\bf k}_1+{\bf k}_2\right)\delta^3\left({\bf k}_3+{\bf k}_4\right)+\delta^3\left({\bf k}_1+{\bf k}_4\right)\delta^3\left({\bf k}_2+{\bf k}_3\right)\right\}\right.\nonumber\\
  && \left.~~~~~~~~~~~~+{\cal N}^{(l)}_{13}({\bf k}_1,{\bf k}_2,{\bf k}_3,{\bf k}_4;\tau_1,\tau_2)~\left\{\delta^3\left({\bf k}_1+{\bf k}_3\right)\delta^3\left({\bf k}_2+{\bf k}_4\right)+\delta^3\left({\bf k}_1+{\bf k}_4\right)\delta^3\left({\bf k}_2+{\bf k}_3\right)\right\}\right],~~~~~~~~~~~\eea
  \bea 
  &&\int d\Psi_{\bf BD}~\langle\Psi_{\bf BD}|:\widehat{\cal V}^{(l)}_4({\bf k}_1,{\bf k}_2,{\bf k}_3,{\bf k}_4;\tau_1,\tau_2;\beta):|\Psi_{\bf BD}\rangle\nonumber\\
 &&=(2\pi)^6\exp\left(-\int d^3{\bf k}~\ln\left(2\sinh \frac{\beta E_{\bf k}(\tau_1)}{2}\right)\right)\nonumber\\
 &&~~~~~~~~~~~~\left[ {\cal Q}^{(1)}_4({\bf k}_1,{\bf k}_2,{\bf k}_3,{\bf k}_4;\tau_1,\tau_2)~\left\{\delta^3\left({\bf k}_1+{\bf k}_4\right)\delta^3\left({\bf k}_2+{\bf k}_3\right)+\delta^3\left({\bf k}_1+{\bf k}_3\right)\delta^3\left({\bf k}_2+{\bf k}_4\right)\right\}\right.\nonumber\\
  && \left.~~~~~~~~~~~~+ {\cal Q}^{(l)}_6({\bf k}_1,{\bf k}_2,{\bf k}_3,{\bf k}_4;\tau_1,\tau_2)~\left\{\delta^3\left({\bf k}_1+{\bf k}_2\right)\delta^3\left({\bf k}_3+{\bf k}_4\right)+\delta^3\left({\bf k}_1+{\bf k}_4\right)\delta^3\left({\bf k}_2+{\bf k}_3\right)\right\}\right.\nonumber\\
  && \left.~~~~~~~~~~~~+{\cal Q}^{(l)}_7({\bf k}_1,{\bf k}_2,{\bf k}_3,{\bf k}_4;\tau_1,\tau_2)~\left\{\delta^3\left({\bf k}_1+{\bf k}_2\right)\delta^3\left({\bf k}_3+{\bf k}_4\right)+\delta^3\left({\bf k}_1+{\bf k}_3\right)\delta^3\left({\bf k}_2+{\bf k}_4\right)\right\}\right.\nonumber\\
  && \left.~~~~~~~~~~~~+ {\cal Q}^{(l)}_{10}({\bf k}_1,{\bf k}_2,{\bf k}_3,{\bf k}_4;\tau_1,\tau_2)~\left\{\delta^3\left({\bf k}_1+{\bf k}_2\right)\delta^3\left({\bf k}_3+{\bf k}_4\right)+\delta^3\left({\bf k}_1+{\bf k}_3\right)\delta^3\left({\bf k}_2+{\bf k}_4\right)\right\}\right.\nonumber\\
  && \left.~~~~~~~~~~~~+{\cal Q}^{(l)}_{11}({\bf k}_1,{\bf k}_2,{\bf k}_3,{\bf k}_4;\tau_1,\tau_2)~\left\{\delta^3\left({\bf k}_1+{\bf k}_2\right)\delta^3\left({\bf k}_3+{\bf k}_4\right)+\delta^3\left({\bf k}_1+{\bf k}_4\right)\delta^3\left({\bf k}_2+{\bf k}_3\right)\right\}\right.\nonumber\\
  && \left.~~~~~~~~~~~~+{\cal Q}^{(l)}_{13}({\bf k}_1,{\bf k}_2,{\bf k}_3,{\bf k}_4;\tau_1,\tau_2)~\left\{\delta^3\left({\bf k}_1+{\bf k}_3\right)\delta^3\left({\bf k}_2+{\bf k}_4\right)+\delta^3\left({\bf k}_1+{\bf k}_4\right)\delta^3\left({\bf k}_2+{\bf k}_3\right)\right\}\right],~~~~~~~~~~~
  \eea
 After detailed computation it is possible to obtain the following OTO amplitudes which will finally contribute in the expressions for the two desired auto-correlated OTOCs which are computing in this paper.  To understand the structure of these functions more clearly one can further write them in terms of the redefined field and its canonically conjugate momenta, for the two specific types of OTOCs as:
  \bea && {\cal E}^{(1)}_{4}({\bf k}_1,{\bf k}_2,-{\bf k}_2,-{\bf k}_1;\tau_1,\tau_2)=f^{*}_{-{\bf k}_1}(\tau_1)f^{*}_{-{\bf k}_2}(\tau_2)f_{-{\bf k}_2}(\tau_1)f_{-{\bf k}_1}(\tau_2)-f^{*}_{-{\bf k}_1}(\tau_2)f^{*}_{-{\bf k}_2}(\tau_1)f_{-{\bf k}_2}(\tau_1)f_{-{\bf k}_1}(\tau_2)\nonumber\\
  &&~~~~~~~~~~~~~~~~~~~~~~~+f^{*}_{-{\bf k}_1}(\tau_1)f^{*}_{-{\bf k}_2}(\tau_2)f_{-{\bf k}_2}(\tau_2)f_{-{\bf k}_1}(\tau_1)-f^{*}_{-{\bf k}_1}(\tau_2)f^{*}_{-{\bf k}_2}(\tau_1)f_{-{\bf k}_2}(\tau_2)f_{-{\bf k}_1}(\tau_1),~~~~~~~~~~~~~\\
  && {\cal E}^{(1)}_{4}({\bf k}_1,{\bf k}_2,-{\bf k}_1,-{\bf k}_2;\tau_1,\tau_2)=f^{*}_{-{\bf k}_1}(\tau_1)f^{*}_{-{\bf k}_2}(\tau_2)f_{-{\bf k}_1}(\tau_1)f_{-{\bf k}_2}(\tau_2)-f^{*}_{-{\bf k}_1}(\tau_2)f^{*}_{-{\bf k}_2}(\tau_1)f_{-{\bf k}_1}(\tau_1)f_{-{\bf k}_2}(\tau_2)\nonumber\\
  &&~~~~~~~~~~~~~~~~~~~~~~~+f^{*}_{-{\bf k}_1}(\tau_1)f^{*}_{-{\bf k}_2}(\tau_2)f_{-{\bf k}_1}(\tau_2)f_{-{\bf k}_2}(\tau_1)-f^{*}_{-{\bf k}_1}(\tau_2)f^{*}_{-{\bf k}_2}(\tau_1)f_{-{\bf k}_1}(\tau_2)f_{-{\bf k}_2}(\tau_1),~~~~~~~~~~~~~\\
&& {\cal E}^{(1)}_{6}({\bf k}_1,{\bf k}_2,-{\bf k}_2,-{\bf k}_1;\tau_1,\tau_2)=f^{*}_{-{\bf k}_1}(\tau_1)f_{{\bf k}_2}(\tau_2)f^{*}_{{\bf k}_2}(\tau_1)f_{-{\bf k}_1}(\tau_2)-f^{*}_{-{\bf k}_1}(\tau_2)f_{{\bf k}_2}(\tau_1)f^{*}_{{\bf k}_2}(\tau_1)f_{-{\bf k}_1}(\tau_2)\nonumber\\
  &&~~~~~~~~~~~~~~~~~~~~~~~+f^{*}_{-{\bf k}_1}(\tau_1)f_{{\bf k}_2}(\tau_2)f^{*}_{{\bf k}_2}(\tau_2)f_{-{\bf k}_1}(\tau_1)-f^{*}_{-{\bf k}_1}(\tau_2)f_{{\bf k}_2}(\tau_1)f^{*}_{{\bf k}_2}(\tau_2)f_{-{\bf k}_1}(\tau_1),~~~~~~~~~~~~~\\
&& {\cal E}^{(1)}_{7}({\bf k}_1,{\bf k}_2,-{\bf k}_1,-{\bf k}_2;\tau_1,\tau_2)=f_{{\bf k}_1}(\tau_1)f^{*}_{-{\bf k}_2}(\tau_2)f^{*}_{{\bf k}_1}(\tau_1)f_{-{\bf k}_2}(\tau_2)-f_{{\bf k}_1}(\tau_2)f^{*}_{-{\bf k}_2}(\tau_1)f^{*}_{{\bf k}_1}(\tau_1)f_{-{\bf k}_2}(\tau_2)\nonumber\\
  &&~~~~~~~~~~~~~~~~~~~~~~~~~~~~~~+f_{{\bf k}_1}(\tau_1)f^{*}_{-{\bf k}_2}(\tau_2)f^{*}_{{\bf k}_1}(\tau_2)f_{-{\bf k}_2}(\tau_1)-f_{{\bf k}_1}(\tau_2)f^{*}_{-{\bf k}_2}(\tau_1)f^{*}_{{\bf k}_1}(\tau_2)f_{-{\bf k}_2}(\tau_1),~~~~~~~~~~~~~\\
&& {\cal E}^{(1)}_{10}({\bf k}_1,{\bf k}_2,-{\bf k}_1,-{\bf k}_2;\tau_1,\tau_2)=f^{*}_{-{\bf k}_1}(\tau_1)f_{{\bf k}_2}(\tau_2)f_{-{\bf k}_1}(\tau_1)f^{*}_{{\bf k}_2}(\tau_2)-f^{*}_{-{\bf k}_1}(\tau_2)f_{{\bf k}_2}(\tau_1)f_{-{\bf k}_1}(\tau_1)f^{*}_{{\bf k}_2}(\tau_2)\nonumber\\
  &&~~~~~~~~~~~~~~~~~~~~~~~~~~~~~~+f^{*}_{-{\bf k}_1}(\tau_1)f_{{\bf k}_2}(\tau_2)f_{-{\bf k}_1}(\tau_2)f^{*}_{{\bf k}_2}(\tau_1)-f^{*}_{-{\bf k}_1}(\tau_2)f_{{\bf k}_2}(\tau_1)f_{{\bf k}_3}(\tau_2)f^{*}_{-{\bf k}_4}(\tau_1),~~~~~~~~~~~~~\\
&& {\cal E}^{(1)}_{11}({\bf k}_1,{\bf k}_2,-{\bf k}_2,-{\bf k}_1;\tau_1,\tau_2)=f_{{\bf k}_1}(\tau_1)f^{*}_{-{\bf k}_2}(\tau_2)f_{-{\bf k}_2}(\tau_1)f^{*}_{{\bf k}_1}(\tau_2)-f_{{\bf k}_1}(\tau_2)f^{*}_{-{\bf k}_2}(\tau_1)f_{-{\bf k}_2}(\tau_1)f^{*}_{{\bf k}_1}(\tau_2)\nonumber\\
  &&~~~~~~~~~~~~~~~~~~~~~~~~~~~~~~+f_{{\bf k}_1}(\tau_1)f^{*}_{-{\bf k}_2}(\tau_2)f{-{\bf k}_2}(\tau_2)f^{*}_{{\bf k}_1}(\tau_1)-f_{{\bf k}_1}(\tau_2)f^{*}_{-{\bf k}_2}(\tau_1)f_{-{\bf k}_1}(\tau_2)f^{*}_{{\bf k}_2}(\tau_1),~~~~~~~~~~~~~\\
&& {\cal E}^{(1)}_{13}({\bf k}_1,{\bf k}_2,-{\bf k}_1,-{\bf k}_2;\tau_1,\tau_2)=f_{{\bf k}_1}(\tau_1)f_{{\bf k}_2}(\tau_2)f^{*}_{{\bf k}_1}(\tau_1)f^{*}_{{\bf k}_2}(\tau_2)-f_{{\bf k}_1}(\tau_2)f_{{\bf k}_2}(\tau_1)f^{*}_{{\bf k}_1}(\tau_1)f^{*}_{{\bf k}_2}(\tau_2)\nonumber\\
  &&~~~~~~~~~~~~~~~~~~~~~~~~~~~~~~+f_{{\bf k}_1}(\tau_1)f_{{\bf k}_2}(\tau_2)f^{*}_{{\bf k}_1}(\tau_2)f^{*}_{{\bf k}_2}(\tau_1)-f_{{\bf k}_1}(\tau_2)f_{{\bf k}_2}(\tau_1)f^{*}_{{\bf k}_1}(\tau_2)f^{*}_{{\bf k}_2}(\tau_1),~~~~~~~~~~~~~\\
&& {\cal E}^{(1)}_{13}({\bf k}_1,{\bf k}_2,-{\bf k}_2,-{\bf k}_1;\tau_1,\tau_2)=f_{{\bf k}_1}(\tau_1)f_{{\bf k}_2}(\tau_2)f^{*}_{{\bf k}_2}(\tau_1)f^{*}_{{\bf k}_1}(\tau_2)-f_{{\bf k}_1}(\tau_2)f_{{\bf k}_2}(\tau_1)f^{*}_{{\bf k}_2}(\tau_1)f^{*}_{{\bf k}_1}(\tau_2)\nonumber\\
  &&~~~~~~~~~~~~~~~~~~~~~~~~~~~~~~+f_{{\bf k}_1}(\tau_1)f_{{\bf k}_2}(\tau_2)f^{*}_{{\bf k}_2}(\tau_2)f^{*}_{{\bf k}_1}(\tau_1)-f_{{\bf k}_1}(\tau_2)f_{{\bf k}_2}(\tau_1)f^{*}_{{\bf k}_2}(\tau_2)f^{*}_{{\bf k}_1}(\tau_1),~~~~~~~~~~~~~\\
  && {\cal E}^{(1)}_{7}({\bf k}_1,-{\bf k}_1,{\bf k}_2,-{\bf k}_2;\tau_1,\tau_2)=f_{{\bf k}_1}(\tau_1)f^{*}_{{\bf k}_1}(\tau_2)f^{*}_{-{\bf k}_2}(\tau_1)f_{-{\bf k}_2}(\tau_2)-f_{{\bf k}_1}(\tau_2)f^{*}_{{\bf k}_1}(\tau_1)f^{*}_{-{\bf k}_2}(\tau_1)f_{-{\bf k}_2}(\tau_2)\nonumber\\
  &&~~~~~~~~~~~~~~~~~~~~~~~~~~~~~~+f_{{\bf k}_1}(\tau_1)f^{*}_{{\bf k}_1}(\tau_2)f^{*}_{-{\bf k}_2}(\tau_2)f_{-{\bf k}_2}(\tau_1)-f_{{\bf k}_1}(\tau_2)f^{*}_{{\bf k}_1}(\tau_1)f^{*}_{-{\bf k}_2}(\tau_2)f_{-{\bf k}_2}(\tau_1),~~~~~~~~~~~~~\eea\bea
&& {\cal E}^{(1)}_{10}({\bf k}_1,-{\bf k}_1,{\bf k}_2,-{\bf k}_2;\tau_1,\tau_2)=f^{*}_{-{\bf k}_1}(\tau_1)f_{-{\bf k}_1}(\tau_2)f_{{\bf k}_2}(\tau_1)f^{*}_{{\bf k}_2}(\tau_2)-f^{*}_{-{\bf k}_1}(\tau_2)f_{-{\bf k}_1}(\tau_1)f_{{\bf k}_2}(\tau_1)f^{*}_{{\bf k}_2}(\tau_2)\nonumber\\
  &&~~~~~~~~~~~~~~~~~~~~~~~~~~~~~~+f^{*}_{-{\bf k}_1}(\tau_1)f_{-{\bf k}_1}(\tau_2)f_{{\bf k}_2}(\tau_2)f^{*}_{{\bf k}_2}(\tau_1)-f^{*}_{-{\bf k}_1}(\tau_2)f_{-{\bf k}_1}(\tau_1)f_{{\bf k}_2}(\tau_2)f^{*}_{{\bf k}_2}(\tau_1),~~~~~~~~~~~~~\\
&& {\cal E}^{(1)}_{11}({\bf k}_1,-{\bf k}_1,{\bf k}_2,-{\bf k}_2;\tau_1,\tau_2)=f_{{\bf k}_1}(\tau_1)f^{*}_{{\bf k}_1}(\tau_2)f_{{\bf k}_2}(\tau_1)f^{*}_{{\bf k}_2}(\tau_2)-f_{{\bf k}_1}(\tau_2)f^{*}_{{\bf k}_1}(\tau_1)f_{{\bf k}_2}(\tau_1)f^{*}_{{\bf k}_2}(\tau_2)\nonumber\\
  &&~~~~~~~~~~~~~~~~~~~~~~~~~~~~~~+f_{{\bf k}_1}(\tau_1)f^{*}_{{\bf k}_1}(\tau_2)f_{{\bf k}_2}(\tau_2)f^{*}_{{\bf k}_2}(\tau_1)-f_{{\bf k}_1}(\tau_2)f^{*}_{{\bf k}_1}(\tau_1)f_{{\bf k}_2}(\tau_2)f^{*}_{{\bf k}_2}(\tau_1),~~~~~~~~~~~~~\eea
  and 
   \bea && {\cal E}^{(2)}_{4}({\bf k}_1,{\bf k}_2,-{\bf k}_2,-{\bf k}_1;\tau_1,\tau_2)\nonumber\\
   &&=\Pi^{*}_{-{\bf k}_1}(\tau_1)\Pi^{*}_{-{\bf k}_2}(\tau_2)\Pi_{-{\bf k}_2}(\tau_1)\Pi_{-{\bf k}_1}(\tau_2)-\Pi^{*}_{-{\bf k}_1}(\tau_2)\Pi^{*}_{-{\bf k}_2}(\tau_1)\Pi_{-{\bf k}_2}(\tau_1)\Pi_{-{\bf k}_1}(\tau_2)\nonumber\\
  &&~~~~~~~~~~~~~~~~~~+\Pi^{*}_{-{\bf k}_1}(\tau_1)\Pi^{*}_{-{\bf k}_2}(\tau_2)\Pi_{-{\bf k}_2}(\tau_2)\Pi_{-{\bf k}_1}(\tau_1)-\Pi^{*}_{-{\bf k}_1}(\tau_2)\Pi^{*}_{-{\bf k}_2}(\tau_1)\Pi_{-{\bf k}_2}(\tau_2)\Pi_{-{\bf k}_1}(\tau_1),~~~~~~~~~~~~~\\
  && {\cal E}^{(2)}_{4}({\bf k}_1,{\bf k}_2,-{\bf k}_1,-{\bf k}_2;\tau_1,\tau_2)\nonumber\\
   &&=\Pi^{*}_{-{\bf k}_1}(\tau_1)\Pi^{*}_{-{\bf k}_2}(\tau_2)\Pi_{-{\bf k}_1}(\tau_1)\Pi_{-{\bf k}_2}(\tau_2)-\Pi^{*}_{-{\bf k}_1}(\tau_2)\Pi^{*}_{-{\bf k}_2}(\tau_1)\Pi_{-{\bf k}_1}(\tau_1)\Pi_{-{\bf k}_2}(\tau_2)\nonumber\\
  &&~~~~~~~~~~~~~~~~~~+\Pi^{*}_{-{\bf k}_1}(\tau_1)\Pi^{*}_{-{\bf k}_2}(\tau_2)\Pi_{-{\bf k}_1}(\tau_2)\Pi_{-{\bf k}_2}(\tau_1)-\Pi^{*}_{-{\bf k}_1}(\tau_2)\Pi^{*}_{-{\bf k}_2}(\tau_1)\Pi_{-{\bf k}_1}(\tau_2)\Pi_{-{\bf k}_2}(\tau_1),~~~~~~~~~~~~~\\
&& {\cal E}^{(2)}_{6}({\bf k}_1,{\bf k}_2,-{\bf k}_2,-{\bf k}_1;\tau_1,\tau_2)\nonumber\\
   &&=\Pi^{*}_{-{\bf k}_1}(\tau_1)\Pi_{{\bf k}_2}(\tau_2)\Pi^{*}_{{\bf k}_2}(\tau_1)\Pi_{-{\bf k}_1}(\tau_2)-\Pi^{*}_{-{\bf k}_1}(\tau_2)\Pi_{{\bf k}_2}(\tau_1)\Pi^{*}_{{\bf k}_2}(\tau_1)\Pi_{-{\bf k}_1}(\tau_2)\nonumber\\
  &&~~~~~~~~~~~~~~~~~~+\Pi^{*}_{-{\bf k}_1}(\tau_1)\Pi_{{\bf k}_2}(\tau_2)\Pi^{*}_{{\bf k}_2}(\tau_2)\Pi_{-{\bf k}_1}(\tau_1)-\Pi^{*}_{-{\bf k}_1}(\tau_2)\Pi_{{\bf k}_2}(\tau_1)\Pi^{*}_{{\bf k}_2}(\tau_2)\Pi_{-{\bf k}_1}(\tau_1),~~~~~~~~~~~~~\\
&& {\cal E}^{(2)}_{7}({\bf k}_1,{\bf k}_2,-{\bf k}_1,-{\bf k}_2;\tau_1,\tau_2)=\Pi_{{\bf k}_1}(\tau_1)\Pi^{*}_{-{\bf k}_2}(\tau_2)\Pi^{*}_{{\bf k}_1}(\tau_1)\Pi_{-{\bf k}_2}(\tau_2)-\Pi_{{\bf k}_1}(\tau_2)\Pi^{*}_{-{\bf k}_2}(\tau_1)\Pi^{*}_{{\bf k}_1}(\tau_1)\Pi_{-{\bf k}_2}(\tau_2)\nonumber\\
  &&~~~~~~~~~~~~~~~~~~+\Pi_{{\bf k}_1}(\tau_1)\Pi^{*}_{-{\bf k}_2}(\tau_2)\Pi^{*}_{{\bf k}_1}(\tau_2)\Pi_{-{\bf k}_2}(\tau_1)-\Pi_{{\bf k}_1}(\tau_2)\Pi^{*}_{-{\bf k}_2}(\tau_1)\Pi^{*}_{{\bf k}_1}(\tau_2)\Pi_{-{\bf k}_2}(\tau_1),~~~~~~~~~~~~~\\
&& {\cal E}^{(2)}_{10}({\bf k}_1,{\bf k}_2,-{\bf k}_1,-{\bf k}_2;\tau_1,\tau_2)=\Pi^{*}_{-{\bf k}_1}(\tau_1)\Pi_{{\bf k}_2}(\tau_2)\Pi_{-{\bf k}_1}(\tau_1)\Pi^{*}_{{\bf k}_2}(\tau_2)-\Pi^{*}_{-{\bf k}_1}(\tau_2)\Pi_{{\bf k}_2}(\tau_1)\Pi_{-{\bf k}_1}(\tau_1)\Pi^{*}_{{\bf k}_2}(\tau_2)\nonumber\\
  &&~~~~~~~~~~~~~~~~~~+\Pi^{*}_{-{\bf k}_1}(\tau_1)\Pi_{{\bf k}_2}(\tau_2)\Pi_{-{\bf k}_1}(\tau_2)\Pi^{*}_{{\bf k}_2}(\tau_1)-\Pi^{*}_{-{\bf k}_1}(\tau_2)\Pi_{{\bf k}_2}(\tau_1)\Pi_{{\bf k}_3}(\tau_2)\Pi^{*}_{-{\bf k}_4}(\tau_1),~~~~~~~~~~~~~\\
&& {\cal E}^{(2)}_{11}({\bf k}_1,{\bf k}_2,-{\bf k}_2,-{\bf k}_1;\tau_1,\tau_2)=\Pi_{{\bf k}_1}(\tau_1)\Pi^{*}_{-{\bf k}_2}(\tau_2)\Pi_{-{\bf k}_2}(\tau_1)\Pi^{*}_{{\bf k}_1}(\tau_2)-\Pi_{{\bf k}_1}(\tau_2)\Pi^{*}_{-{\bf k}_2}(\tau_1)\Pi_{-{\bf k}_2}(\tau_1)\Pi^{*}_{{\bf k}_1}(\tau_2)\nonumber\\
  &&~~~~~~~~~~~~~~~~~~+\Pi_{{\bf k}_1}(\tau_1)\Pi^{*}_{-{\bf k}_2}(\tau_2)\Pi_{-{\bf k}_2}(\tau_2)\Pi^{*}_{{\bf k}_1}(\tau_1)-\Pi_{{\bf k}_1}(\tau_2)\Pi^{*}_{-{\bf k}_2}(\tau_1)\Pi_{-{\bf k}_1}(\tau_2)\Pi^{*}_{{\bf k}_2}(\tau_1),~~~~~~~~~~~~~\\
&& {\cal E}^{(2)}_{13}({\bf k}_1,{\bf k}_2,-{\bf k}_1,-{\bf k}_2;\tau_1,\tau_2)=\Pi_{{\bf k}_1}(\tau_1)\Pi_{{\bf k}_2}(\tau_2)\Pi^{*}_{{\bf k}_1}(\tau_1)\Pi^{*}_{{\bf k}_2}(\tau_2)-\Pi_{{\bf k}_1}(\tau_2)\Pi_{{\bf k}_2}(\tau_1)\Pi^{*}_{{\bf k}_1}(\tau_1)\Pi^{*}_{{\bf k}_2}(\tau_2)\nonumber\\
  &&~~~~~~~~~~~~~~~~~~+\Pi_{{\bf k}_1}(\tau_1)\Pi_{{\bf k}_2}(\tau_2)\Pi^{*}_{{\bf k}_1}(\tau_2)\Pi^{*}_{{\bf k}_2}(\tau_1)-\Pi_{{\bf k}_1}(\tau_2)\Pi_{{\bf k}_2}(\tau_1)\Pi^{*}_{{\bf k}_1}(\tau_2)\Pi^{*}_{{\bf k}_2}(\tau_1),~~~~~~~~~~~~~\\
&& {\cal E}^{(2)}_{13}({\bf k}_1,{\bf k}_2,-{\bf k}_2,-{\bf k}_1;\tau_1,\tau_2)=\Pi_{{\bf k}_1}(\tau_1)\Pi_{{\bf k}_2}(\tau_2)\Pi^{*}_{{\bf k}_2}(\tau_1)\Pi^{*}_{{\bf k}_1}(\tau_2)-\Pi_{{\bf k}_1}(\tau_2)\Pi_{{\bf k}_2}(\tau_1)\Pi^{*}_{{\bf k}_2}(\tau_1)\Pi^{*}_{{\bf k}_1}(\tau_2)\nonumber\\
  &&~~~~~~~~~~~~~~~~~~+\Pi_{{\bf k}_1}(\tau_1)\Pi_{{\bf k}_2}(\tau_2)\Pi^{*}_{{\bf k}_2}(\tau_2)\Pi^{*}_{{\bf k}_1}(\tau_1)-\Pi_{{\bf k}_1}(\tau_2)\Pi_{{\bf k}_2}(\tau_1)\Pi^{*}_{{\bf k}_2}(\tau_2)\Pi^{*}_{{\bf k}_1}(\tau_1),~~~~~~~~~~~~~\\
  && {\cal E}^{(2)}_{7}({\bf k}_1,-{\bf k}_1,{\bf k}_2,-{\bf k}_2;\tau_1,\tau_2)=\Pi_{{\bf k}_1}(\tau_1)\Pi^{*}_{{\bf k}_1}(\tau_2)\Pi^{*}_{-{\bf k}_2}(\tau_1)\Pi_{-{\bf k}_2}(\tau_2)-\Pi_{{\bf k}_1}(\tau_2)\Pi^{*}_{{\bf k}_1}(\tau_1)\Pi^{*}_{-{\bf k}_2}(\tau_1)\Pi_{-{\bf k}_2}(\tau_2)\nonumber\\
  &&~~~~~~~~~~~~~~~~~~+\Pi_{{\bf k}_1}(\tau_1)\Pi^{*}_{{\bf k}_1}(\tau_2)\Pi^{*}_{-{\bf k}_2}(\tau_2)\Pi_{-{\bf k}_2}(\tau_1)-\Pi_{{\bf k}_1}(\tau_2)\Pi^{*}_{{\bf k}_1}(\tau_1)\Pi^{*}_{-{\bf k}_2}(\tau_2)\Pi_{-{\bf k}_2}(\tau_1),~~~~~~~~~~~~~\\
&& {\cal E}^{(2)}_{10}({\bf k}_1,-{\bf k}_1,{\bf k}_2,-{\bf k}_2;\tau_1,\tau_2)=\Pi^{*}_{-{\bf k}_1}(\tau_1)\Pi_{-{\bf k}_1}(\tau_2)\Pi_{{\bf k}_2}(\tau_1)\Pi^{*}_{{\bf k}_2}(\tau_2)-\Pi^{*}_{-{\bf k}_1}(\tau_2)\Pi_{-{\bf k}_1}(\tau_1)\Pi_{{\bf k}_2}(\tau_1)\Pi^{*}_{{\bf k}_2}(\tau_2)\nonumber\\
  &&~~~~~~~~~~~~~~~~~~+\Pi^{*}_{-{\bf k}_1}(\tau_1)\Pi_{-{\bf k}_1}(\tau_2)\Pi_{{\bf k}_2}(\tau_2)\Pi^{*}_{{\bf k}_2}(\tau_1)-\Pi^{*}_{-{\bf k}_1}(\tau_2)\Pi_{-{\bf k}_1}(\tau_1)\Pi_{{\bf k}_2}(\tau_2)\Pi^{*}_{{\bf k}_2}(\tau_1),~~~~~~~~~~~~~\\
&& {\cal E}^{(2)}_{11}({\bf k}_1,-{\bf k}_1,{\bf k}_2,-{\bf k}_2;\tau_1,\tau_2)=\Pi_{{\bf k}_1}(\tau_1)\Pi^{*}_{{\bf k}_1}(\tau_2)\Pi_{{\bf k}_2}(\tau_1)\Pi^{*}_{{\bf k}_2}(\tau_2)-\Pi_{{\bf k}_1}(\tau_2)\Pi^{*}_{{\bf k}_1}(\tau_1)\Pi_{{\bf k}_2}(\tau_1)\Pi^{*}_{{\bf k}_2}(\tau_2)\nonumber\\
  &&~~~~~~~~~~~~~~~~~~+\Pi_{{\bf k}_1}(\tau_1)\Pi^{*}_{{\bf k}_1}(\tau_2)\Pi_{{\bf k}_2}(\tau_2)\Pi^{*}_{{\bf k}_2}(\tau_1)-\Pi_{{\bf k}_1}(\tau_2)\Pi^{*}_{{\bf k}_1}(\tau_1)\Pi_{{\bf k}_2}(\tau_2)\Pi^{*}_{{\bf k}_2}(\tau_1),~~~~~~~~~~~~~\eea 
\section{Time dependent two-point amplitude in   OTOC}
\label{sec:13}
We define the following momentum integrated time dependent amplitudes, which are given by:
\bea {\cal B}_1(T,\tau):&=&\int^{L}_{k_1=0}~k^2_1~dk_1~{\cal P}_1({\bf k}_1,-{\bf k}_1;T,\tau)\nonumber\\
  &=&(-T)^{\frac{1}{2}-\nu}(-\tau)^{\frac{1}{2}-\nu}\left[Z^{(1)}_{(1)}(\tau_1,\tau_2)+Z^{(1)}_{(2)}(\tau_1,\tau_2)-Z^{(1)}_{(3)}(\tau_1,\tau_2)-Z^{(1)}_{(4)}(\tau_1,\tau_2)\right],~~~~~~~~\\ {\cal B}_2(T,\tau):&=&\int^{L}_{k_1=0}~k^2_1~dk_1~{\cal P}_@({\bf k}_1,-{\bf k}_1;T,\tau)\nonumber\\
  &=&(-T)^{\frac{3}{2}-\nu}(-\tau)^{\frac{3}{2}-\nu}\left[Z^{(2)}_{(1)}(\tau_1,\tau_2)+Z^{(2)}_{(2)}(\tau_1,\tau_2)-Z^{(2)}_{(3)}(\tau_1,\tau_2)-Z^{(2)}_{(4)}(\tau_1,\tau_2)\right],~~~~~~~~\eea
  where we have introduced the time dependent four individual amplitudes, $Z^{(1)}_{(i)}(T,\tau)~\forall~i=1,2,3,4$ and $Z^{(2)}_{(i)}(T,\tau)~\forall~i=1,2,3,4$,  which are given by the following expressions:
  \bea && Z^{(1)}_{(1)}(T,\tau):=\int^{L}_{k_1=0}~k^2_1~dk_1~f_{{\bf k}_1}(T)f^{*}_{{\bf k}_1}(\tau),\\
   && Z^{(1)}_{(2)}(T,\tau)=\int^{L}_{k_1=0}~k^2_1~dk_1~ f^{*}_{{\bf -k}_1}(T)f_{-{\bf k}_1}(\tau),\\
   && Z^{(1)}_{(3)}(T,\tau):=\int^{L}_{k_1=0}~k^2_1~dk_1~ f_{{\bf k}_1}(\tau)f^{*}_{{\bf k}_1}(T),\\
   && Z^{(1)}_{(4)}(T,\tau):=\int^{L}_{k_1=0}~k^2_1~dk_1~f^{*}_{{\bf -k}_1}(\tau)f_{-{\bf k}_1}(T),\\ && Z^{(2)}_{(1)}(T,\tau):=\int^{L}_{k_1=0}~k^2_1~dk_1~\Pi_{{\bf k}_1}(T)\Pi^{*}_{{\bf k}_1}(\tau),\\
   && Z^{(2)}_{(2)}(T,\tau)=\int^{L}_{k_1=0}~k^2_1~dk_1~ \Pi^{*}_{{\bf -k}_1}(T)\Pi_{-{\bf k}_1}(\tau),\\
   && Z^{(2)}_{(3)}(T,\tau):=\int^{L}_{k_1=0}~k^2_1~dk_1~ \Pi_{{\bf k}_1}(\tau)\Pi^{*}_{{\bf k}_1}(T),\\
   && Z^{(2)}_{(4)}(T,\tau):=\int^{L}_{k_1=0}~k^2_1~dk_1~ \Pi^{*}_{{\bf -k}_1}(\tau)\Pi_{-{\bf k}_1}(T),\eea
  which we are going to explicitly evaluate in this Appendix.
  
  Now before going to evaluate the individual contributions from the symmetry properties of the momentum dependent amplitudes we have derived the following results:
\bea && Z^{(1)}_{(2)}(T,\tau)=(-1)^{-(2\nu+1)}Z^{(1)}_{(1)}(T,\tau),\\
&&Z^{(1)}_{(4)}(T,\tau)=(-1)^{-(2\nu+1)}Z^{(1)}_{(3)}(T,\tau),\\ && Z^{(2)}_{(2)}(T,\tau)=(-1)^{-(2\nu+1)}Z^{(2)}_{(1)}(T,\tau),\\
&&Z^{(2)}_{(4)}(T,\tau)=(-1)^{-(2\nu+1)}Z^{(2)}_{(3)}(T,\tau),\eea
using which the simplified form of the momentum integrated time dependent amplitudes can be written as:
\bea {\cal B}_1(T,\tau):&=&(-T)^{\frac{1}{2}-\nu}(-\tau)^{\frac{1}{2}-\nu}\left[1+(-1)^{-(2\nu+1)}\right]\left(Z_{(1)}(T,\tau)-Z_{(3)}(T,\tau)\right),\\
{\cal B}_2(T,\tau):&=&(-T)^{\frac{3}{2}-\nu}(-\tau)^{\frac{3}{2}-\nu}\left[1+(-1)^{-(2\nu+1)}\right]\left(Z_{(1)}(T,\tau)-Z_{(3)}(T,\tau)\right).\eea
Consequently,  the desired two-point OTOCs can be computed in the present context as:
\bea Y^{f}_1(T,\tau)=-\frac{1}{2\pi^2} {\cal B}_1(T,\tau)=(-T)^{\frac{1}{2}-\nu}(-\tau)^{\frac{1}{2}-\nu}\left[1+(-1)^{-(2\nu+1)}\right]\left(Z^{(1)}_{(3)}(T,\tau)-Z^{(1)}_{(1)}(T,\tau)\right),~~~~~~~\\
Y^{f}_1(T,\tau)=-\frac{1}{2\pi^2} {\cal B}_2(T,\tau)=(-T)^{\frac{3}{2}-\nu}(-\tau)^{\frac{3}{2}-\nu}\left[1+(-1)^{-(2\nu+1)}\right]\left(Z^{(2)}_{(3)}(T,\tau)-Z^{(2)}_{(1)}(T,\tau)\right).~~~~~~~\eea
The expression for $\left(Z^{(1)}_{(3)}(T,\tau)-Z^{(1)}_{(1)}(T,\tau)\right)$ is given by the following expression: 
\bea &&\left(Z^{(1)}_{(3)}(T,\tau)-Z^{(1)}_{(1)}(T,\tau)\right)=\frac{i(A^2-B^2)}{(T-\tau )^5}  L^{-2 \nu } \Biggl(-T^2 (-i L (T-\tau ))^{2 \nu } \Gamma (3-2 \nu ,-i L (T-\tau ))\nonumber\\
&&~~~~~~~-T^2 (-i L (T-\tau ))^{2 \nu } \Gamma (4-2 \nu ,-i L (T-\tau ))-T^2 (i L (T-\tau ))^{2 \nu } \Gamma (3-2 \nu ,i L (T-\tau ))\nonumber\\
&&~~~~~~~-T^2 (i L (T-\tau ))^{2 \nu } \Gamma (4-2 \nu ,i L (T-\tau ))-\tau ^2 (-i L (T-\tau ))^{2 \nu } \Gamma (3-2 \nu ,-i L (T-\tau ))\nonumber\\
&&~~~~~~~~~~-\tau ^2 (-i L (T-\tau ))^{2 \nu } \Gamma (4-2 \nu ,-i L (T-\tau ))-\tau ^2 (i L (T-\tau ))^{2 \nu } \Gamma (3-2 \nu ,i L (T-\tau ))\nonumber\\
&&~~~~~~~~~~-\tau ^2 (i L (T-\tau ))^{2 \nu } \Gamma (4-2 \nu ,i L (T-\tau ))+\tau  (-T) \Gamma (5-2 \nu ) (-i L (T-\tau ))^{2 \nu }\nonumber\\
&&+\Gamma (3-2 \nu ) (T-\tau )^2 \left((-i L (T-\tau ))^{2 \nu }+(i L (T-\tau ))^{2 \nu }\right)\nonumber\\
&&~~~~~~~~~~+\Gamma (4-2 \nu ) (T-\tau )^2 \left((-i L (T-\tau ))^{2 \nu }+(i L (T-\tau ))^{2 \nu }\right)-\tau  T \Gamma (5-2 \nu ) (i L (T-\tau ))^{2 \nu }\nonumber\\
&&~~+2 \tau  T (-i L (T-\tau ))^{2 \nu } \Gamma (3-2 \nu ,-i L (T-\tau ))+2 \tau  T (-i L (T-\tau ))^{2 \nu } \Gamma (4-2 \nu ,-i L (T-\tau ))\nonumber\\
&&~~~+\tau  T (-i L (T-\tau ))^{2 \nu } \Gamma (5-2 \nu ,-i L (T-\tau ))+2 \tau  T (i L (T-\tau ))^{2 \nu } \Gamma (3-2 \nu ,i L (T-\tau ))\nonumber\\
&&~~~+2 \tau  T (i L (T-\tau ))^{2 \nu } \Gamma (4-2 \nu ,i L (T-\tau ))+\tau  T (i L (T-\tau ))^{2 \nu } \Gamma (5-2 \nu ,i L (T-\tau ))\Biggr).~~~~~~~~~~~\eea
Similarly,  the expression for $\left(Z^{(2)}_{(3)}(T,\tau)-Z^{(2)}_{(1)}(T,\tau)\right)$ can be found.  Due to its length we are not proving the details of the result.  Though during the numerical plots the explicit detail have been taken care of.  Also for massive fields one can obtain the above mentioned results for the two types of the integrals by taking the analytic continuation from $\nu$ to $-i|\nu|$. 

\section{Time dependent four-point amplitudes in  OTOC }
\label{sec:14}
We define the following momenta integrated time dependent amplitudes for $l=1,2$,  which are given by:
  \bea && {\cal I}^{(l)}_1(\tau_1,\tau_2):=\int^{L}_{k_1=0} k^2_1~dk_1\int^{L}_{k_2=0} k^2_2~dk_2~2{\cal E}^{(l)}_4({\bf k}_1,{\bf k}_2,-{\bf k}_2,-{\bf k}_1;\tau_1,\tau_2),\\
  && {\cal I}^{(l)}_2(\tau_1,\tau_2):=\int^{L}_{k_1=0} k^2_1~dk_1\int^{L}_{k_2=0} k^2_2~dk_2~2{\cal E}^{(l)}_{13}({\bf k}_1,{\bf k}_2,-{\bf k}_2,-{\bf k}_1;\tau_1,\tau_2),\\
  && {\cal I}_3(\tau_1,\tau_2):=\int^{L}_{k_1=0} k^2_1~dk_1\int^{L}_{k_2=0} k^2_2~dk_2~{\cal E}^{(l)}_6({\bf k}_1,{\bf k}_2,-{\bf k}_2,-{\bf k}_1;\tau_1,\tau_2),\\
  && {\cal I}^{(l)}_4(\tau_1,\tau_2):=\int^{L}_{k_1=0} k^2_1~dk_1\int^{L}_{k_2=0} k^2_2~dk_2~{\cal E}^{(l)}_7({\bf k}_1,{\bf k}_2,-{\bf k}_1,-{\bf k}_2;\tau_1,\tau_2),\\
 &&  {\cal I}^{(l)}_5(\tau_1,\tau_2):=\int^{L}_{k_1=0} k^2_1~dk_1\int^{L}_{k_2=0} k^2_2~dk_2~{\cal E}^{(l)}_{10}({\bf k}_1,{\bf k}_2,-{\bf k}_1,-{\bf k}_2;\tau_1,\tau_2),\\
 && {\cal I}^{(l)}_6(\tau_1,\tau_2):=\int^{L}_{k_1=0} k^2_1~dk_1\int^{L}_{k_2=0} k^2_2~dk_2~{\cal E}^{(l)}_{11}({\bf k}_1,{\bf k}_2,-{\bf k}_2,-{\bf k}_1;\tau_1,\tau_2),\\
 && {\cal I}^{(l)}_7(\tau_1,\tau_2):=\int^{L}_{k_1=0} k^2_1~dk_1\int^{L}_{k_2=0} k^2_2~dk_2~{\cal E}^{(l)}_7({\bf k}_1,-{\bf k}_1,{\bf k}_2,-{\bf k}_2;\tau_1,\tau_2),\\
 && {\cal I}^{(l)}_8(\tau_1,\tau_2):=\int^{L}_{k_1=0} k^2_1~dk_1\int^{L}_{k_2=0} k^2_2~dk_2~{\cal E}^{(l)}_{10}({\bf k}_1,-{\bf k}_1,{\bf k}_2,-{\bf k}_2;\tau_1,\tau_2),\\
 && {\cal I}^{(l)}_9(\tau_1,\tau_2):=\int^{L}_{k_1=0} k^2_1~dk_1\int^{L}_{k_2=0} k^2_2~dk_2~{\cal E}^{(l)}_{10}({\bf k}_1,-{\bf k}_1,{\bf k}_2,-{\bf k}_2;\tau_1,\tau_2).
  \eea 
 From the symmetry properties of the momentum dependent amplitudes we have derived the following results for $l=1,2$,  which are given by:
  \bea  
  {\cal I}^{(l)}_2(\tau_1,\tau_2)&&=(-1)^{4\nu}{\cal I}^{(l)}_1(\tau_1,\tau_2)~~~~{\rm with~ weight}~~w_2=2,~~~\\
   {\cal I}^{(l)}_3(\tau_1,\tau_2)&&=(-1)^{2\nu}{\cal I}^{(l)}_1(\tau_1,\tau_2)~~~~{\rm with~ weight}~~w_3=1,~~~\\
    {\cal I}^{(l)}_4(\tau_1,\tau_2)&&=(-1)^{2\nu}{\cal I}^{(l)}_1(\tau_1,\tau_2)~~~~{\rm with~ weight}~~w_4=1,~~~\\
     {\cal I}^{(l)}_5(\tau_1,\tau_2)&&=(-1)^{2\nu}{\cal I}^{(l)}_1(\tau_1,\tau_2)~~~~{\rm with~ weight}~~w_5=1,~~~\\
      {\cal I}^{(l)}_6(\tau_1,\tau_2)&&=(-1)^{2\nu}{\cal I}^{(l)}_1(\tau_1,\tau_2)~~~~{\rm with~ weight}~~w_6=1,~~~\\
       {\cal I}^{(l)}_7(\tau_1,\tau_2)&&=(-1)^{2\nu}{\cal I}^{(l)}_1(\tau_1,\tau_2)~~~~{\rm with~ weight}~~w_7=1,~~~\\
        {\cal I}^{(l)}_8(\tau_1,\tau_2)&&=(-1)^{2\nu}{\cal I}^{(l)}_1(\tau_1,\tau_2)~~~~{\rm with~ weight}~~w_8=1,~~~\\
         {\cal I}^{(l)}_9(\tau_1,\tau_2)&&=(-1)^{2\nu}{\cal I}^{(l)}_1(\tau_1,\tau_2)~~~~{\rm with~ weight}~~w_9=1.~~~\eea
         The details of the all of these regularised four-point integral computations we have given in the following subsections. These computations are useful to construct the final expression for the cosmological OTOC.   
\subsection{Computation of ${\cal I}^{(1)}_{1}(\tau_1,\tau_2)$ and ${\cal I}^{(2)}_{1}(\tau_1,\tau_2)$}
First of all we evaluate the amplitude integral for $l=1$,  which is given by:
\bea  {{\cal I}^{(1)}_1(T,\tau)=\int^{L}_{k_1=0} k^2_1dk_1\int^{L}_{k_2=0} k^2_2dk_2{\cal E}^{(1)}_4({\bf k}_1,{\bf k}_2,-{\bf k}_2,-{\bf k}_1;T,\tau)=\frac{(-T)^{1-2\nu}(-\tau)^{1-2\nu}}{(-1)^{4\nu}}\sum^{4}_{i=1}X^{(1),1}_{i}(T,\tau)},\nonumber\\
\eea 
where we define four time dependent functions, $X^{(1),1}_{i}(T,\tau)~\forall~~i=1,2,3,4$, which are given by the following expressions:
\bea &&X^{(1),1}_1=\scriptstyle\frac{1}{32} L^{-2 \nu } \left(\frac{(A^2+B^2) L^{3-2 \nu }}{3-2 \nu }+\frac{(A^2+B^2) T^2 L^{5-2 \nu }}{5-2 \nu }\right.\nonumber\\
&&\left. \scriptstyle+A B 2^{2 \nu -5} T^2 (-i T)^{2 \nu -5} (\Gamma (5-2 \nu )-\Gamma (5-2 \nu ,-2 i L T))+\frac{A B 2^{2 \nu -3} (i T)^{2 \nu -1} (\Gamma (4-2 \nu )-\Gamma (4-2 \nu ,2 i L T))}{T^2}\right.\nonumber\\
&&\left.\scriptstyle+A B 2^{2 \nu -5} T^2 (i T)^{2 \nu -5} (\Gamma (5-2 \nu )-\Gamma (5-2 \nu ,2 i L T)) -A B 2^{2 \nu -3} (-i T)^{2 \nu -3} (\Gamma (3-2 \nu )-\Gamma (3-2 \nu ,-2 i L T))\right.\nonumber\\
&&\left. \scriptstyle-A B 2^{2 \nu -3} (i T)^{2 \nu -3} (\Gamma (3-2 \nu )-\Gamma (3-2 \nu ,2 i L T)) +\frac{i A B 2^{2 \nu -3} (-i T)^{2 \nu } (\Gamma (4-2 \nu )-\Gamma (4-2 \nu ,-2 i L T))}{T^3}\right)\nonumber\\ 
&&\times\scriptstyle\left(\frac{32 (A^2+B^2) L^5 T^2}{5-2 \nu }+\frac{32 (A^2+B^2) L^3}{3-2 \nu } -\frac{i A B 4^{\nu +1} (-i L T)^{2 \nu } \Gamma (3-2 \nu ,-2 i L T)}{T^3}\right.\nonumber\\
&&\left. \scriptstyle-\frac{i A B 4^{\nu +1} (-i L T)^{2 \nu } \Gamma (4-2 \nu ,-2 i L T)}{T^3} -\frac{i A B 4^{\nu } (-i L T)^{2 \nu } \Gamma (5-2 \nu ,-2 i L T)}{T^3} \right.\nonumber\\
&&\left.\scriptstyle+\frac{i A B 4^{\nu +1} (i L T)^{2 \nu } \Gamma (3-2 \nu ,2 i L T)}{T^3}+\frac{i A B 4^{\nu +1} (i L T)^{2 \nu } \Gamma (4-2 \nu ,2 i L T)}{T^3}\right.\nonumber\\
&&\left.\scriptstyle+\frac{i A B 4^{\nu } (i L T)^{2 \nu } \Gamma (5-2 \nu ,2 i L T)}{T^3}+\frac{i A B 4^{\nu } (2 \nu -7) \Gamma (5-2 \nu ) \left((-i L T)^{2 \nu }-(i L T)^{2 \nu }\right)}{(2 \nu -3) T^3}\right)\\
 &&X^{(1),1}_2=\scriptstyle-\left(-L^2\right)^{-2 \nu } \Biggl(\frac{B^2 (\Gamma (3-2 \nu )-\Gamma (3-2 \nu ,-i L (T-\tau ))) (-i L (T-\tau ))^{2 \nu }}{(T-\tau )^3}\scriptstyle+\frac{B^2 (T+\tau)  (\Gamma (4-2 \nu )-\Gamma (4-2 \nu ,-i L (T-\tau ))) (-i L (T-\tau ))^{2 \nu }}{(T-\tau )^4}\nonumber\\
&&\scriptstyle+\frac{B^2 T \tau  (\Gamma (5-2 \nu )-\Gamma (5-2 \nu ,-i L (T-\tau ))) (-i L (T-\tau ))^{2 \nu }}{(T-\tau )^5}+\frac{A B (i L (T+\tau ))^{2 \nu } (\Gamma (3-2 \nu )-\Gamma (3-2 \nu ,i L (T+\tau )))}{(T+\tau )^3}\nonumber\\
&&\scriptstyle+\frac{A^2 (i L (T-\tau ))^{2 \nu } \tau  (\Gamma (4-2 \nu )-\Gamma (4-2 \nu ,i L (T-\tau )))}{(T-\tau )^4}+\frac{A B (T-\tau) (i L (T+\tau ))^{2 \nu } (\Gamma (4-2 \nu )-\Gamma (4-2 \nu ,i L (T+\tau )))}{(T+\tau )^4}\nonumber\\
&&\scriptstyle+\frac{A^2 T (i L (T-\tau ))^{2 \nu } \tau  (\Gamma (5-2 \nu )-\Gamma (5-2 \nu ,i L (T-\tau )))}{(T-\tau )^5}-\frac{A^2 (i L (T-\tau ))^{2 \nu } (\Gamma (3-2 \nu )-\Gamma (3-2 \nu ,i L (T-\tau )))}{(T-\tau )^3}\nonumber\\
&&\scriptstyle-\frac{A^2 T (i L (T-\tau ))^{2 \nu } (\Gamma (4-2 \nu )-\Gamma (4-2 \nu ,i L (T-\tau )))}{(T-\tau )^4}-\frac{A B (-i L (T+\tau ))^{2 \nu } (\Gamma (3-2 \nu )-\Gamma (3-2 \nu ,-i L (T+\tau )))}{(T+\tau )^3}\nonumber\\
&&\scriptstyle-\frac{A B  (-i L (T+\tau ))^{2 \nu } (\Gamma (4-2 \nu )-\Gamma (4-2 \nu ,-i L (T+\tau )))}{(T+\tau )^3}-\frac{A B T \tau  (-i L (T+\tau ))^{2 \nu } (\Gamma (5-2 \nu )-\Gamma (5-2 \nu ,-i L (T+\tau )))}{(T+\tau )^5}\nonumber\\
&&\scriptstyle-\frac{A B T \tau  (i L (T+\tau ))^{2 \nu } (\Gamma (5-2 \nu )-\Gamma (5-2 \nu ,i L (T+\tau )))}{(T+\tau )^5}\Biggr)\scriptstyle\Biggl(\frac{A^2 (\Gamma (3-2 \nu )-\Gamma (3-2 \nu ,-i L (T-\tau ))) (-i L (T-\tau ))^{2 \nu }}{(T-\tau )^3}\nonumber\\
&&\scriptstyle+\frac{A^2  (\Gamma (4-2 \nu )-\Gamma (4-2 \nu ,-i L (T-\tau ))) (-i L (T-\tau ))^{2 \nu }}{(T-\tau )^3}-\frac{A^2 T \tau  (\Gamma (5-2 \nu )-\Gamma (5-2 \nu ,-i L (T-\tau ))) (-i L (T-\tau ))^{2 \nu }}{(T-\tau )^5}\nonumber\\
&&\scriptstyle+\frac{B^2 (i L (T-\tau ))^{2 \nu } (\Gamma (3-2 \nu )-\Gamma (3-2 \nu ,i L (T-\tau )))}{(\tau -T)^3}+\frac{A B (i L (T+\tau ))^{2 \nu } (\Gamma (3-2 \nu )-\Gamma (3-2 \nu ,i L (T+\tau )))}{(T+\tau )^3}\nonumber\\
&&\scriptstyle+\frac{B^2 (i L (T-\tau ))^{2 \nu } \tau  (\Gamma (4-2 \nu )-\Gamma (4-2 \nu ,i L (T-\tau )))}{(T-\tau )^4}+\frac{A B T (i L (T+\tau ))^{2 \nu } (\Gamma (4-2 \nu )-\Gamma (4-2 \nu ,i L (T+\tau )))}{(T+\tau )^4}\nonumber\\
&&\scriptstyle+\frac{A B \tau  (i L (T+\tau ))^{2 \nu } (\Gamma (4-2 \nu )-\Gamma (4-2 \nu ,i L (T+\tau )))}{(T+\tau )^4}+\frac{B^2 T (i L (T-\tau ))^{2 \nu } \tau  (\Gamma (5-2 \nu )-\Gamma (5-2 \nu ,i L (T-\tau )))}{(T-\tau )^5}\nonumber\\
&&\scriptstyle+\frac{A B T \tau  (i L (T+\tau ))^{2 \nu } (\Gamma (5-2 \nu )-\Gamma (5-2 \nu ,i L (T+\tau )))}{(T+\tau )^5}-\frac{B^2 T (i L (T-\tau ))^{2 \nu } (\Gamma (4-2 \nu )-\Gamma (4-2 \nu ,i L (T-\tau )))}{(T-\tau )^4}\nonumber\\
&&\scriptstyle-\frac{A B (-i L (T+\tau ))^{2 \nu } (\Gamma (3-2 \nu )-\Gamma (3-2 \nu ,-i L (T+\tau )))}{(T+\tau )^3}-\frac{A B T (-i L (T+\tau ))^{2 \nu } (\Gamma (4-2 \nu )-\Gamma (4-2 \nu ,-i L (T+\tau )))}{(T+\tau )^4}\nonumber\\
&&\scriptstyle -\frac{A B \tau  (-i L (T+\tau ))^{2 \nu } (\Gamma (4-2 \nu )-\Gamma (4-2 \nu ,-i L (T+\tau )))}{(T+\tau )^4}-\frac{A B T \tau  (-i L (T+\tau ))^{2 \nu } (\Gamma (5-2 \nu )-\Gamma (5-2 \nu ,-i L (T+\tau )))}{(T+\tau )^5}\Biggr),~~~~~~~~\eea
 
 \bea &&X^{(1),1}_3=\scriptstyle-\frac{1}{32} L^{-2 \nu } \left(\frac{A^2 L^5 \tau ^2 (-L)^{-2 \nu }}{5-2 \nu }+\frac{(A^2+B^2) L^3 (-L)^{-2 \nu }}{3-2 \nu }+\frac{B^2 \tau ^2 (-L)^{5-2 \nu }}{5-2 \nu }+\frac{i B^2 \tau  (-L)^{4-2 \nu }}{\nu -2}\right.\nonumber\\
 &&\left.\scriptstyle+\frac{A B 2^{2 \nu -3} (i \tau )^{2 \nu +1} (\Gamma (4-2 \nu )-\Gamma (4-2 \nu ,-2 i L \tau ))}{\tau ^4}+\frac{A B 2^{2 \nu -5} (i \tau )^{2 \nu +1} (\Gamma (5-2 \nu )-\Gamma (5-2 \nu ,-2 i L \tau ))}{\tau ^4}\right.\nonumber\\
 &&\left.\scriptstyle+A B 2^{2 \nu -5} \tau ^2 (-i \tau )^{2 \nu -5} (\Gamma (5-2 \nu )-\Gamma (5-2 \nu ,2 i L \tau ))+A B 2^{2 \nu -3} (i \tau )^{2 \nu -3} (\Gamma (3-2 \nu )-\Gamma (3-2 \nu ,-2 i L \tau ))\right.\nonumber\\
 &&\left.\scriptstyle+A B 2^{2 \nu -3} (-i \tau )^{2 \nu -3} (\Gamma (3-2 \nu )-\Gamma (3-2 \nu ,2 i L \tau ))\right) \scriptstyle\left(\frac{32 (A^2+B^2) L^5 T^2}{5-2 \nu }+\frac{32 (A^2+B^2) L^3}{3-2 \nu }+\frac{i A B 4^{\nu } (2 \nu -7) \Gamma (5-2 \nu ) \left((-i L T)^{2 \nu }-(i L T)^{2 \nu }\right)}{(2 \nu -3) T^3}\right.\nonumber\\
 &&\left.\scriptstyle-\frac{i A B 4^{\nu +1} (-i L T)^{2 \nu } \Gamma (3-2 \nu ,-2 i L T)}{T^3}-\frac{i A B 4^{\nu +1} (-i L T)^{2 \nu } \Gamma (4-2 \nu ,-2 i L T)}{T^3}\right.\nonumber\\
 &&\left.\scriptstyle-\frac{i A B 4^{\nu } (-i L T)^{2 \nu } \Gamma (5-2 \nu ,-2 i L T)}{T^3}+\frac{i A B 4^{\nu +1} (i L T)^{2 \nu } \Gamma (3-2 \nu ,2 i L T)}{T^3}\scriptstyle+\frac{i A B 4^{\nu +1} (i L T)^{2 \nu } \Gamma (4-2 \nu ,2 i L T)}{T^3}+\frac{i A B 4^{\nu } (i L T)^{2 \nu } \Gamma (5-2 \nu ,2 i L T)}{T^3}\right).~~~~~~~~~\\
  &&X^{(1),1}_4=\scriptstyle(-L^2)^{-2 \nu } \Biggl(\frac{B^2 (\Gamma (3-2 \nu )-\Gamma (3-2 \nu ,-i L (T-\tau ))) (-i L (T-\tau ))^{2 \nu }}{(T-\tau )^3}+\frac{B^2 T (\Gamma (4-2 \nu )-\Gamma (4-2 \nu ,-i L (T-\tau ))) (-i L (T-\tau ))^{2 \nu }}{(T-\tau )^4}\nonumber\\
 &&\scriptstyle-\frac{B^2 \tau  (\Gamma (4-2 \nu )-\Gamma (4-2 \nu ,-i L (T-\tau ))) (-i L (T-\tau ))^{2 \nu }}{(T-\tau )^4}-\frac{B^2 T \tau  (\Gamma (5-2 \nu )-\Gamma (5-2 \nu ,-i L (T-\tau ))) (-i L (T-\tau ))^{2 \nu }}{(T-\tau )^5}\nonumber\\
 &&\scriptstyle+\frac{A^2 (i L (T-\tau ))^{2 \nu } (\Gamma (3-2 \nu )-\Gamma (3-2 \nu ,i L (T-\tau )))}{(\tau -T)^3}+\frac{A B (i L (T+\tau ))^{2 \nu } (\Gamma (3-2 \nu )-\Gamma (3-2 \nu ,i L (T+\tau )))}{(T+\tau )^3}\nonumber\\
 &&\scriptstyle+\frac{A^2 (i L (T-\tau ))^{2 \nu } \tau  (\Gamma (4-2 \nu )-\Gamma (4-2 \nu ,i L (T-\tau )))}{(T-\tau )^4}+\frac{A B T (i L (T+\tau ))^{2 \nu } (\Gamma (4-2 \nu )-\Gamma (4-2 \nu ,i L (T+\tau )))}{(T+\tau )^4}\nonumber\\
 &&\scriptstyle+\frac{A B \tau  (i L (T+\tau ))^{2 \nu } (\Gamma (4-2 \nu )-\Gamma (4-2 \nu ,i L (T+\tau )))}{(T+\tau )^4}+\frac{A^2 T (i L (T-\tau ))^{2 \nu } \tau  (\Gamma (5-2 \nu )-\Gamma (5-2 \nu ,i L (T-\tau )))}{(T-\tau )^5}\nonumber\\
 &&\scriptstyle+\frac{A B T \tau  (i L (T+\tau ))^{2 \nu } (\Gamma (5-2 \nu )-\Gamma (5-2 \nu ,i L (T+\tau )))}{(T+\tau )^5}-\frac{A^2 T (i L (T-\tau ))^{2 \nu } (\Gamma (4-2 \nu )-\Gamma (4-2 \nu ,i L (T-\tau )))}{(T-\tau )^4}\nonumber\\
 &&\scriptstyle-\frac{A B (-i L (T+\tau ))^{2 \nu } (\Gamma (3-2 \nu )-\Gamma (3-2 \nu ,-i L (T+\tau )))}{(T+\tau )^3}-\frac{A B T (-i L (T+\tau ))^{2 \nu } (\Gamma (4-2 \nu )-\Gamma (4-2 \nu ,-i L (T+\tau )))}{(T+\tau )^4}\nonumber\\
 &&\scriptstyle-\frac{A B \tau  (-i L (T+\tau ))^{2 \nu } (\Gamma (4-2 \nu )-\Gamma (4-2 \nu ,-i L (T+\tau )))}{(T+\tau )^4}-\frac{A B T \tau  (-i L (T+\tau ))^{2 \nu } (\Gamma (5-2 \nu )-\Gamma (5-2 \nu ,-i L (T+\tau )))}{(T+\tau )^5}\Biggr) \nonumber\\
 &&\scriptstyle\Biggl(\frac{A^2 (\Gamma (3-2 \nu )-\Gamma (3-2 \nu ,-i L (T-\tau ))) (-i L (T-\tau ))^{2 \nu }}{(T-\tau )^3}+\frac{A^2 T (\Gamma (4-2 \nu )-\Gamma (4-2 \nu ,-i L (T-\tau ))) (-i L (T-\tau ))^{2 \nu }}{(T-\tau )^4}\nonumber\\
 &&\scriptstyle-\frac{A^2 \tau  (\Gamma (4-2 \nu )-\Gamma (4-2 \nu ,-i L (T-\tau ))) (-i L (T-\tau ))^{2 \nu }}{(T-\tau )^4}-\frac{A^2 T \tau  (\Gamma (5-2 \nu )-\Gamma (5-2 \nu ,-i L (T-\tau ))) (-i L (T-\tau ))^{2 \nu }}{(T-\tau )^5}\nonumber\\
 &&\scriptstyle+\frac{A B (i L (T+\tau ))^{2 \nu } (\Gamma (3-2 \nu )-\Gamma (3-2 \nu ,i L (T+\tau )))}{(T+\tau )^3}+\frac{B^2 (i L (T-\tau ))^{2 \nu } \tau  (\Gamma (4-2 \nu )-\Gamma (4-2 \nu ,i L (T-\tau )))}{(T-\tau )^4}\nonumber\\
 &&\scriptstyle+\frac{A B T (i L (T+\tau ))^{2 \nu } (\Gamma (4-2 \nu )-\Gamma (4-2 \nu ,i L (T+\tau )))}{(T+\tau )^4}+\frac{A B \tau  (i L (T+\tau ))^{2 \nu } (\Gamma (4-2 \nu )-\Gamma (4-2 \nu ,i L (T+\tau )))}{(T+\tau )^4}\nonumber\\
 &&\scriptstyle+\frac{B^2 T (i L (T-\tau ))^{2 \nu } \tau  (\Gamma (5-2 \nu )-\Gamma (5-2 \nu ,i L (T-\tau )))}{(T-\tau )^5}+\frac{A B T \tau  (i L (T+\tau ))^{2 \nu } (\Gamma (5-2 \nu )-\Gamma (5-2 \nu ,i L (T+\tau )))}{(T+\tau )^5}\nonumber\\
 &&\scriptstyle-\frac{B^2 (i L (T-\tau ))^{2 \nu } (\Gamma (3-2 \nu )-\Gamma (3-2 \nu ,i L (T-\tau )))}{(T-\tau )^3}-\frac{B^2 T (i L (T-\tau ))^{2 \nu } (\Gamma (4-2 \nu )-\Gamma (4-2 \nu ,i L (T-\tau )))}{(T-\tau )^4}\nonumber\\
 &&\scriptstyle-\frac{A B (-i L (T+\tau ))^{2 \nu } (\Gamma (3-2 \nu )-\Gamma (3-2 \nu ,-i L (T+\tau )))}{(T+\tau )^3}-\frac{A B T (-i L (T+\tau ))^{2 \nu } (\Gamma (4-2 \nu )-\Gamma (4-2 \nu ,-i L (T+\tau )))}{(T+\tau )^4}\nonumber\\
 &&\scriptstyle-\frac{A B \tau  (-i L (T+\tau ))^{2 \nu } (\Gamma (4-2 \nu )-\Gamma (4-2 \nu ,-i L (T+\tau )))}{(T+\tau )^4}-\frac{A B T \tau  (-i L (T+\tau ))^{2 \nu } (\Gamma (5-2 \nu )-\Gamma (5-2 \nu ,-i L (T+\tau )))}{(T+\tau )^5}\Biggr)\eea
Here we have introduced two factors $A$ and $B$ which are defined as:
 \bea  A= \frac{2^{\nu-\frac{3}{2}}}{\sqrt{2}}\left|\frac{\Gamma(\nu)}{\Gamma\left(\frac{3}{2}\right)}\right|~\exp\left(-i\left\{\frac{\pi}{2}\left(\nu+\frac{1}{2}\right)\right\}\right)C_1,~ B=\frac{2^{\nu-\frac{3}{2}}}{\sqrt{2}}\left|\frac{\Gamma(\nu)}{\Gamma\left(\frac{3}{2}\right)}\right|~\exp\left(i\left\{\frac{\pi}{2}\left(\nu+\frac{1}{2}\right)\right\}\right)C_2.~~~~~~~~~~~\eea
 Here for general Mota Allen vacua we choose,  $C_1=\cosh\alpha,~~ C_2=\exp(i\gamma)\sinh\alpha$. Now setting $\gamma=0$ and $\alpha=0=\gamma$ we get the constants for $\alpha$ vacua and Bunch Davies vacuum.

Next,  we evaluate the amplitude integral for $l=2$,  which is given by:
\bea  {{\cal I}^{(2)}_1(T,\tau)=\int^{L}_{k_1=0} k^2_1dk_1\int^{L}_{k_2=0} k^2_2dk_2{\cal E}^{(2)}_4({\bf k}_1,{\bf k}_2,-{\bf k}_2,-{\bf k}_1;T,\tau)=\frac{(-T)^{3-2\nu}(-\tau)^{3-2\nu}}{(-1)^{4\nu}}\sum^{4}_{i=1}X^{(2),1}_{i}(T,\tau)},\nonumber\\
\eea  
where we define four time dependent functions, $X^{(2),1}_{i}(T,\tau)~\forall~~i=1,2,3,4$, which can be analytically computable but due to the length of the expressions we are not providing the explicit details of the expressions here.  Though the details have been used for the numerical plots. For massive fields one can obtain the above mentioned results for the two types of the integrals by taking the analytic continuation from $\nu$ to $-i|\nu|$.
 \subsection{Computation of  ${\cal I}^{(1)}_{2}(\tau_1,\tau_2)$ and ${\cal I}^{(2)}_{2}(\tau_1,\tau_2)$}
In this context,  we have to evaluate the following integral for $l=1$,  which is given by:
\bea  {{\cal I}^{(1)}_2(T,\tau)=\int^{L}_{k_1=0} k^2_1dk_1\int^{L}_{k_2=0} k^2_2dk_2{\cal E}^{(1)}_{13}({\bf k}_1,{\bf k}_2,-{\bf k}_1,-{\bf k}_2;T,\tau)=(-T)^{1-2\nu}(-\tau)^{1-2\nu}\sum^{4}_{i=1}X^{(1),2}_{i}(T,\tau)},\nonumber\\
~\eea 
where we define four time dependent functions,  $X^{(1),2}_{i}(T,\tau)~\forall~~i=1,2,3,4$, which are given by the following expressions:
\bea X^{(1),2}_1&=&X^{(1),1}_1,~
X^{(1),2}_2=X^{(1),1}_2,~
 X^{(1),2}_3=X^{(1),1}_3,~
 X^{(1),2}_4=X^{(1),1}_4.\eea
Consequently, one can write:
\bea  {{\cal I}^{(1)}_2(T,\tau)=(-T)^{1-2\nu}(-\tau)^{1-2\nu}\sum^{4}_{i=1}X^{(1),2}_{i}(T,\tau)=(-1)^{4\nu}{\cal I}^{(1)}_1(T,\tau)}.~~~\eea
 Next,  we have to evaluate the following integral for $l=2$,  which is given by:
\bea  {{\cal I}^{(2)}_2(T,\tau)=\int^{L}_{k_1=0} k^2_1dk_1\int^{L}_{k_2=0} k^2_2dk_2{\cal E}^{(2)}_{13}({\bf k}_1,{\bf k}_2,-{\bf k}_1,-{\bf k}_2;T,\tau)=(-T)^{3-2\nu}(-\tau)^{3-2\nu}\sum^{4}_{i=1}X^{(2),2}_{i}(T,\tau)},\nonumber\\
~\eea 
where we define four time dependent functions,  $X^{(2),2}_{i}(T,\tau)~\forall~~i=1,2,3,4$, which are given by the following expressions:
\bea X^{(2),2}_1&=&X^{(2),1}_1,~
X^{(2),2}_2=X^{(2),1}_2,~
 X^{(2),2}_3=X^{(2),1}_3,~
 X^{(2),2}_4=X^{(2),1}_4.\eea
Consequently, one can write:
\bea  {{\cal I}^{(2)}_2(T,\tau)=(-T)^{3-2\nu}(-\tau)^{3-2\nu}\sum^{4}_{i=1}X^{(2),2}_{i}(T,\tau)=(-1)^{4\nu}{\cal I}^{(2)}_1(T,\tau)}.~~~\eea
 \subsection{Computation of  ${\cal I}^{(1)}_{3}(\tau_1,\tau_2)$ and ${\cal I}^{(2)}_{3}(\tau_1,\tau_2)$}
In this context,  we have to evaluate the following integral for $l=1$,  which is given by:
\bea  {{\cal I}^{(1)}_3(T,\tau)=\int^{L}_{k_1=0} k^2_1dk_1\int^{L}_{k_2=0} k^2_2dk_2{\cal E}^{(1)}_{6}({\bf k}_1,{\bf k}_2,-{\bf k}_2,-{\bf k}_1;T,\tau)=\frac{(-T)^{1-2\nu}(-\tau)^{1-2\nu}}{(-1)^{2\nu}}\sum^{4}_{i=1}X^{(1),3}_{i}(T,\tau)},\nonumber\\ \eea 
where we define four time dependent functions, $X^{(1),3}_{i}(T,\tau)~\forall~~i=1,2,3,4$, which are given by the following expressions:
\bea X^{(1),3}_1&=&X^{(1),1}_1,~
X^{(1),3}_2=X^{(1),1}_2,~
 X^{(1),3}_3=X^{(1),1}_3,~
 X^{(1),3}_4=X^{(1),1}_4.\eea
Consequently,  one can write:
\bea  {{\cal I}^{(1)}_3(T,\tau)=\frac{(-T)^{1-2\nu}(-\tau)^{1-2\nu}}{(-1)^{2\nu}}\sum^{4}_{i=1}X^{(1),3}_{i}(T,\tau)=(-1)^{2\nu}{\cal I}^{(1)}_1(T,\tau)}.~~~\eea
Next,  we have to evaluate the following integral for $l=2$,  which is given by:
\bea  {{\cal I}^{(2)}_3(T,\tau)=\int^{L}_{k_1=0} k^2_1dk_1\int^{L}_{k_2=0} k^2_2dk_2{\cal E}^{(2)}_{6}({\bf k}_1,{\bf k}_2,-{\bf k}_2,-{\bf k}_1;T,\tau)=\frac{(-T)^{3-2\nu}(-\tau)^{3-2\nu}}{(-1)^{2\nu}}\sum^{4}_{i=1}X^{(2),3}_{i}(T,\tau)},\nonumber\\ \eea 
where we define four time dependent functions, $X^{(2),3}_{i}(T,\tau)~\forall~~i=1,2,3,4$, which are given by the following expressions:
\bea X^{(2),3}_1&=&X^{(2),1}_1,~
X^{(2),3}_2=X^{(2),1}_2,~
 X^{(2),3}_3=X^{(2),1}_3,~
 X^{(2),3}_4=X^{(2),1}_4.\eea
Consequently,  one can write:
\bea  {{\cal I}^{(2)}_3(T,\tau)=\frac{(-T)^{3-2\nu}(-\tau)^{3-2\nu}}{(-1)^{2\nu}}\sum^{4}_{i=1}X^{(2),3}_{i}(T,\tau)=(-1)^{2\nu}{\cal I}^{(2)}_1(T,\tau)}.~~~\eea
 \subsection{Computation of  ${\cal I}^{(1)}_{4}(\tau_1,\tau_2)$ and ${\cal I}^{(2)}_{4}(\tau_1,\tau_2)$}
In this context,  we have to evaluate the following integral for $l=1$,  which is given by:
\bea  {{\cal I}^{(1)}_4(T,\tau)=\int^{L}_{k_1=0} k^2_1dk_1\int^{L}_{k_2=0} k^2_2dk_2{\cal E}^{(1)}_{7}({\bf k}_1,{\bf k}_2,-{\bf k}_1,-{\bf k}_2;T,\tau)=\frac{(-T)^{1-2\nu}(-\tau)^{1-2\nu}}{(-1)^{2\nu}}\sum^{4}_{i=1}X^{(1),4}_{i}(T,\tau)},\nonumber\\
\eea 
where we define four time dependent functions, $X^{(1),4}_{i}(T,\tau)~\forall~~i=1,2,3,4$, which are given by the following expressions:
\bea X^{(1),4}_1&=&X^{(1),1}_1,~
X^{(1),4}_2=X^{(1),1}_2,~
 X^{(1),4}_3=X^{(1),1}_3,~
 X^{(1),4}_4=X^{(1),1}_4.\eea
Consequently, one can write:
\bea  {{\cal I}^{(1)}_4(T,\tau)=\frac{(-T)^{1-2\nu}(-\tau)^{1-2\nu}}{(-1)^{2\nu}}\sum^{4}_{i=1}X^{(i)}_{1}(T,\tau)=(-1)^{2\nu}{\cal I}^{(1)}_1(T,\tau)}.~~~\eea
Next,  we have to evaluate the following integral for $l=2$,  which is given by:
\bea  {{\cal I}^{(2)}_4(T,\tau)=\int^{L}_{k_1=0} k^2_1dk_1\int^{L}_{k_2=0} k^2_2dk_2{\cal E}^{(2)}_{7}({\bf k}_1,{\bf k}_2,-{\bf k}_1,-{\bf k}_2;T,\tau)=\frac{(-T)^{3-2\nu}(-\tau)^{3-2\nu}}{(-1)^{2\nu}}\sum^{4}_{i=1}X^{(2),4}_{i}(T,\tau)},\nonumber\\
\eea 
where we define four time dependent functions, $X^{(2),4}_{i}(T,\tau)~\forall~~i=1,2,3,4$, which are given by the following expressions:
\bea X^{(2),4}_1&=&X^{(2),1}_1,~
X^{(2),4}_2=X^{(2),1}_2,~
 X^{(2),4}_3=X^{(2),1}_3,~
 X^{(2),4}_4=X^{(2),1}_4.\eea
Consequently, one can write:
\bea  {{\cal I}^{(2)}_4(T,\tau)=\frac{(-T)^{3-2\nu}(-\tau)^{3-2\nu}}{(-1)^{2\nu}}\sum^{4}_{i=1}X^{(2),4}_{i}(T,\tau)=(-1)^{2\nu}{\cal I}^{(2)}_1(T,\tau)}.~~~\eea
 \subsection{Computation of  ${\cal I}^{(1)}_{5}(\tau_1,\tau_2)$ and ${\cal I}^{(2)}_{5}(\tau_1,\tau_2)$}
In this context,  we have to evaluate the following integral for $l=1$,  which is given by:
\bea  {{\cal I}^{(1)}_5(T,\tau)=\int^{L}_{k_1=0} k^2_1dk_1\int^{L}_{k_2=0} k^2_2dk_2{\cal E}^{(1)}_{10}({\bf k}_1,{\bf k}_2,-{\bf k}_1,-{\bf k}_2;T,\tau)=\frac{(-T)^{1-2\nu}(-\tau)^{1-2\nu}}{(-1)^{2\nu}}\sum^{4}_{i=1}X^{(1),5}_{i}(T,\tau)},\nonumber\\ \eea 
where we define four time dependent functions, $X^{(1),5}_{i}(T,\tau)~\forall~~i=1,2,3,4$, which are given by the following expressions:
\bea X^{(1),5}_1&=&X^{(1),1}_1,~
X^{(1),5}_2=X^{(1),1}_2,~
 X^{(1),5}_3=X^{(1),1}_3,~
 X^{(1),5}_4=X^{(1),1}_4.\eea
Consequently, one can write:
\bea  {{\cal I}^{(1)}_5(T,\tau)=\frac{(-T)^{1-2\nu}(-\tau)^{1-2\nu}}{(-1)^{2\nu}}\sum^{4}_{i=1}X^{(1),5}_{i}(T,\tau)=(-1)^{2\nu}{\cal I}^{(1)}_1(T,\tau)}.~~~\eea
Next,  we have to evaluate the following integral for $l=2$,  which is given by:
\bea  {{\cal I}^{(2)}_5(T,\tau)=\int^{L}_{k_1=0} k^2_1dk_1\int^{L}_{k_2=0} k^2_2dk_2{\cal E}^{(2)}_{10}({\bf k}_1,{\bf k}_2,-{\bf k}_1,-{\bf k}_2;T,\tau)=\frac{(-T)^{3-2\nu}(-\tau)^{3-2\nu}}{(-1)^{2\nu}}\sum^{4}_{i=1}X^{(2),5}_{i}(T,\tau)},\nonumber\\ \eea 
where we define four time dependent functions, $X^{(2),5}_{i}(T,\tau)~\forall~~i=1,2,3,4$, which are given by the following expressions:
\bea X^{(2),5}_1&=&X^{(2),1}_1,~
X^{(2),5}_2=X^{(2),1}_2,~
 X^{(2),5}_3=X^{(2),1}_3,~
 X^{(2),5}_4=X^{(2),1}_4.\eea
Consequently, one can write:
\bea  {{\cal I}^{(2)}_5(T,\tau)=\frac{(-T)^{3-2\nu}(-\tau)^{3-2\nu}}{(-1)^{2\nu}}\sum^{4}_{i=1}X^{(2),5}_{i}(T,\tau)=(-1)^{2\nu}{\cal I}^{(2)}_1(T,\tau)}.~~~\eea
 \subsection{Computation of  ${\cal I}^{(1)}_{6}(\tau_1,\tau_2)$ and ${\cal I}^{(2)}_{6}(\tau_1,\tau_2)$}
In this context,  we have to evaluate the following integral for $l=1$,  which is given by:
\bea  {{\cal I}^{(1)}_6(T,\tau)=\int^{L}_{k_1=0} k^2_1dk_1\int^{L}_{k_2=0} k^2_2dk_2{\cal E}^{(1)}_{11}({\bf k}_1,{\bf k}_2,-{\bf k}_2,-{\bf k}_1;T,\tau)=\frac{(-T)^{1-2\nu}(-\tau)^{1-2\nu}}{(-1)^{2\nu}}\sum^{4}_{i=1}X^{(1),6}_{i}(T,\tau)},~~\nonumber\\\eea 
where we define four time dependent functions, $X^{(1),6}_{i}(T,\tau)~\forall~~i=1,2,3,4$, which are given by the following expressions:
\bea X^{(1),6}_1&=&X^{(1),1}_1,~
X^{(1),6}_2=X^{(1),1}_2,~
 X^{(1),6}_3=X^{(1),1}_3,~
 X^{(1),6}_4=X^{(1),1}_4.\eea
Consequently, one can write:
\bea  {{\cal I}^{(1)}_6(T,\tau)=\frac{(-T)^{1-2\nu}(-\tau)^{1-2\nu}}{(-1)^{2\nu}}\sum^{4}_{i=1}X^{(1),6}_{i}(T,\tau)=(-1)^{2\nu}{\cal I}^{(1)}_1(T,\tau)}.~~~\eea
 In this context,  we have to evaluate the following integral for $l=2$,  which is given by:
\bea  {{\cal I}^{(2)}_6(T,\tau)=\int^{L}_{k_1=0} k^2_1dk_1\int^{L}_{k_2=0} k^2_2dk_2{\cal E}^{(2)}_{11}({\bf k}_1,{\bf k}_2,-{\bf k}_2,-{\bf k}_1;T,\tau)=\frac{(-T)^{3-2\nu}(-\tau)^{3-2\nu}}{(-1)^{2\nu}}\sum^{4}_{i=1}X^{(2),6}_{i}(T,\tau)},~~\nonumber\\\eea
where we define four time dependent functions, $X^{(2),6}_{i}(T,\tau)~\forall~~i=1,2,3,4$, which are given by the following expressions:
\bea X^{(2),6}_1&=&X^{(2),1}_1,~
X^{(2),6}_2=X^{(2),1}_2,~
 X^{(2),6}_3=X^{(2),1}_3,~
 X^{(2),6}_4=X^{(2),1}_4.\eea
Consequently, one can write:
\bea  {{\cal I}^{(2)}_6(T,\tau)=\frac{(-T)^{3-2\nu}(-\tau)^{3-2\nu}}{(-1)^{2\nu}}\sum^{4}_{i=1}X^{(2),6}_{i}(T,\tau)=(-1)^{2\nu}{\cal I}^{(2)}_1(T,\tau)}.~~~\eea
 \subsection{Computation of  ${\cal I}^{(1)}_{7}(\tau_1,\tau_2)$ and ${\cal I}^{(2)}_{7}(\tau_1,\tau_2)$}
In this context,  we have to evaluate the following integral for $l=1$,  which is given by:
\bea  {{\cal I}^{(1)}_7(T,\tau)=\int^{L}_{k_1=0} k^2_1dk_1\int^{L}_{k_2=0} k^2_2dk_2{\cal E}^{(1)}_{7}({\bf k}_1,-{\bf k}_1,{\bf k}_2,-{\bf k}_2;T,\tau)=\frac{(-T)^{1-2\nu}(-\tau)^{1-2\nu}}{(-1)^{2\nu}}\sum^{4}_{i=1}X^{(1),7}_{i}(T,\tau)},~~\nonumber\\\eea 
where we define four time dependent functions, $X^{(1),7}_{i}(T,\tau)~\forall~~i=1,2,3,4$, which are given by the following expressions:
\bea X^{(1),7}_1&=&X^{(1),1}_1,~
X^{(1),7}_2=X^{(1),1}_2,~
 X^{(1),7}_3=X^{(1),1}_3,~
 X^{(1),7}_4=X^{(1),1}_4.\eea
Consequently, one can write:
\bea  {{\cal I}^{(1)}_7(T,\tau)=\frac{(-T)^{1-2\nu}(-\tau)^{1-2\nu}}{(-1)^{2\nu}}\sum^{4}_{i=1}X^{(1),7}_{i}(T,\tau)=(-1)^{2\nu}{\cal I}^{(1)}_1(T,\tau)}.~~~\eea
Next,  we have to evaluate the following integral for $l=2$,  which is given by:
\bea  {{\cal I}^{(2)}_7(T,\tau)=\int^{L}_{k_1=0} k^2_1dk_1\int^{L}_{k_2=0} k^2_2dk_2{\cal E}^{(2)}_{7}({\bf k}_1,-{\bf k}_1,{\bf k}_2,-{\bf k}_2;T,\tau)=\frac{(-T)^{3-2\nu}(-\tau)^{3-2\nu}}{(-1)^{2\nu}}\sum^{4}_{i=1}X^{(2),7}_{i}(T,\tau)},~~\nonumber\\\eea 
where we define four time dependent functions, $X^{(2),7}_{i}(T,\tau)~\forall~~i=1,2,3,4$, which are given by the following expressions:
\bea X^{(2),7}_1&=&X^{(2),1}_1,~
X^{(2),7}_2=X^{(2),1}_2,~
 X^{(2),7}_3=X^{(2),1}_3,~
 X^{(2),7}_4=X^{(2),1}_4.\eea
Consequently, one can write:
\bea  {{\cal I}^{(2)}_7(T,\tau)=\frac{(-T)^{3-2\nu}(-\tau)^{3-2\nu}}{(-1)^{2\nu}}\sum^{4}_{i=1}X^{(2),7}_{i}(T,\tau)=(-1)^{2\nu}{\cal I}^{(2)}_1(T,\tau)}.~~~\eea
\subsection{Computation of  ${\cal I}^{(1)}_{8}(\tau_1,\tau_2)$ and ${\cal I}^{(2)}_{8}(\tau_1,\tau_2)$}
In this context,  we have to evaluate the following integral for $l=1$,  which is given by:
\bea  {{\cal I}^{(1)}_8(T,\tau)=\int^{L}_{k_1=0} k^2_1dk_1\int^{L}_{k_2=0} k^2_2dk_2{\cal E}^{(1)}_{10}({\bf k}_1,-{\bf k}_1,{\bf k}_2,-{\bf k}_2;T,\tau)=\frac{(-T)^{1-2\nu}(-\tau)^{1-2\nu}}{(-1)^{2\nu}}\sum^{4}_{i=1}X^{(1),8}_{i}(T,\tau)}~~\nonumber\\\eea 
where we define four time dependent functions, $X^{(1),8}_{i}(T,\tau)~\forall~~i=1,2,3,4$, which are given by the following expressions:
\bea X^{(1),8}_1&=&X^{(1),1}_1,~
X^{(1),8}_2=X^{(1),1}_2,~
 X^{(1),8}_3=X^{(1),1}_3,~
 X^{(1),8}_4=X^{(1),1}_4.\eea
Consequently, one can write:
\bea  {{\cal I}^{(1)}_8(T,\tau)=\frac{(-T)^{1-2\nu}(-\tau)^{1-2\nu}}{(-1)^{2\nu}}\sum^{4}_{i=1}X^{(1),8}_{i}(T,\tau)=(-1)^{2\nu}{\cal I}^{(1)}_1(T,\tau)}.~~~\eea
Next,  we have to evaluate the following integral for $l=2$,  which is given by:
\bea  {{\cal I}^{(2)}_8(T,\tau)=\int^{L}_{k_1=0} k^2_1dk_1\int^{L}_{k_2=0} k^2_2dk_2{\cal E}^{(2)}_{10}({\bf k}_1,-{\bf k}_1,{\bf k}_2,-{\bf k}_2;T,\tau)=\frac{(-T)^{3-2\nu}(-\tau)^{3-2\nu}}{(-1)^{2\nu}}\sum^{4}_{i=1}X^{(2),8}_{i}(T,\tau)}~~\nonumber\\\eea 
where we define four time dependent functions, $X^{(2),8}_{i}(T,\tau)~\forall~~i=1,2,3,4$, which are given by the following expressions:
\bea X^{(2),8}_1&=&X^{(1),1}_1,~
X^{(2),8}_2=X^{(1),1}_2,~
 X^{(2),8}_3=X^{(1),1}_3,~
 X^{(2),8}_4=X^{(1),1}_4.\eea
Consequently, one can write:
\bea  {{\cal I}^{(2)}_8(T,\tau)=\frac{(-T)^{3-2\nu}(-\tau)^{3-2\nu}}{(-1)^{2\nu}}\sum^{4}_{i=1}X^{(2),8}_{i}(T,\tau)=(-1)^{2\nu}{\cal I}^{(2)}_1(T,\tau)}.~~~\eea
\subsection{Computation of  ${\cal I}^{(1)}_{9}(\tau_1,\tau_2)$ and ${\cal I}^{(2)}_{9}(\tau_1,\tau_2)$}
In this context,  we have to evaluate the following integral for $l=1$,  which is given by:
\bea  {{\cal I}^{(1)}_9(T,\tau)=\int^{L}_{k_1=0} k^2_1dk_1\int^{L}_{k_2=0} k^2_2dk_2{\cal E}^{(1)}_{11}({\bf k}_1,-{\bf k}_1,{\bf k}_2,-{\bf k}_2;T,\tau)=\frac{(-T)^{1-2\nu}(-\tau)^{1-2\nu}}{(-1)^{2\nu}}\sum^{4}_{i=1}X^{(1),9}_{i}(T,\tau)}\nonumber\\ \eea 
where we define four time dependent functions, $X^{(1),9}_{i}(T,\tau)~\forall~~i=1,2,3,4$, which are given by the following expressions:
\bea X^{(1),9}_1&=&X^{(1),1}_1,~
X^{(1),9}_2=X^{(1),1}_2,~
 X^{(1),9}_3=X^{(1),1}_3,~
 X^{(1),9}_4=X^{(1),1}_4.\eea
Consequently, one can write:
\bea  {{\cal I}^{(1)}_9(T,\tau)=\frac{(-T)^{1-2\nu}(-\tau)^{1-2\nu}}{(-1)^{2\nu}}\sum^{4}_{i=1}X^{(1),9}_{i}(T,\tau)=(-1)^{2\nu}{\cal I}^{(1)}_1(T,\tau)}.~~~\eea 
In this context,  we have to evaluate the following integral for $l=1$,  which is given by:
\bea  {{\cal I}^{(1)}_9(T,\tau)=\int^{L}_{k_1=0} k^2_1dk_1\int^{L}_{k_2=0} k^2_2dk_2{\cal E}^{(1)}_{11}({\bf k}_1,-{\bf k}_1,{\bf k}_2,-{\bf k}_2;T,\tau)=\frac{(-T)^{1-2\nu}(-\tau)^{1-2\nu}}{(-1)^{2\nu}}\sum^{4}_{i=1}X^{(1),9}_{i}(T,\tau)}\nonumber\\ \eea 
where we define four time dependent functions, $X^{(1),9}_{i}(T,\tau)~\forall~~i=1,2,3,4$, which are given by the following expressions:
\bea X^{(1),9}_1&=&X^{(1),1}_1,~
X^{(1),9}_2=X^{(1),1}_2,~
 X^{(1),9}_3=X^{(1),1}_3,~
 X^{(1),9}_4=X^{(1),1}_4.\eea
Consequently, one can write:
\bea  {{\cal I}^{(1)}_9(T,\tau)=\frac{(-T)^{1-2\nu}(-\tau)^{1-2\nu}}{(-1)^{2\nu}}\sum^{4}_{i=1}X^{(1),9}_{i}(T,\tau)=(-1)^{2\nu}{\cal I}^{(1)}_1(T,\tau)}.~~~\eea 
\section{Computation of the normalization factor in four-point  OTOC}
\label{sec:15}
\subsection{Normalization factor of four-point   OTOC computed from rescaled field variable} 
Further,  our objective is to compute the normalisation factors of two OTOCs computed from the rescaled field variable $f$ and its conjugate momentum $\Pi$,  which are given by the following expression:
\bea {{\cal N}^{f}_1(\tau_1,\tau_2):=\frac{1}{\langle \hat{f}(\tau_1)\hat{f}(\tau_1)\rangle_{\beta} \langle \hat{f}(\tau_2)\hat{f}(\tau_2)\rangle_{\beta}}},\\
 {{\cal N}^{f}_2(\tau_1,\tau_2):=\frac{1}{\langle \hat{\Pi}(\tau_1)\hat{\Pi}(\tau_1)\rangle_{\beta} \langle \hat{\Pi}(\tau_2)\hat{\Pi}(\tau_2)\rangle_{\beta}}},\eea
 for this we need to explicitly evaluate the denominator of the above mentioned expressions.
 
 Now,  the product of the two thermal two point function for general Mota Allen vacua are evaluated as:
 \bea \displaystyle {\langle \hat{f}(\tau_1)\hat{f}(\tau_1)\rangle_{\beta} =\displaystyle \frac{1}{Z_{\alpha,\gamma}(\beta;\tau_1)}{\rm Tr}\left[e^{-\beta \hat{H}(\tau_1)}\hat{f}({\bf x},\tau_1)\hat{f}({\bf x},\tau_1)\right]_{(\alpha,\gamma)}},\\ {\displaystyle \langle \hat{\Pi}(\tau_2)\hat{\Pi}(\tau_2)\rangle_{\beta} =\displaystyle \frac{1}{Z_{\alpha,\gamma}(\beta;\tau_2)}{\rm Tr}\left[e^{-\beta \hat{H}(\tau_2)}\hat{\Pi}({\bf x},\tau_2)\hat{\Pi}({\bf x},\tau_2)\right]_{(\alpha,\gamma)}},\eea
 where the thermal partition function for cosmology computed for Mota Allen vacua can be expressed as:
 \bea {Z_{\alpha,\gamma}(\beta;\tau_i)=\frac{\exp(-2\sin\gamma {\rm tan}\alpha)}{|\cosh\alpha|}\exp\left(-\left(1+\frac{1}{2}\delta^{3}(0)\right)\int d^3{\bf k}~\ln\left(2\sinh\frac{\beta E_{\bf k}(\tau_i)}{2}\right)\right)~\forall~~i=1,2}.~~~~~~\eea    
 Next, we compute the expressions for the numerators with respect to the general Mota Allen vacua,  which are given by: 
 \bea &&{\rm Tr}\left[e^{-\beta \hat{H}(\tau_1)}\hat{f}({\bf x},\tau_1)\hat{f}({\bf x},\tau_1)\right]_{(\alpha,\gamma)}\nonumber\\ 
&&=\frac{\exp(-2\sin\gamma {\rm tan}\alpha)}{|\cosh\alpha|}\int d\Psi_{\bf BD}~\langle \Psi_{\bf BD}|\left\{\exp\left(\frac{i}{2}\exp(i\gamma)\tanh \alpha~\int \frac{d^3{\bf k}_1}{(2\pi)^3}~a_{{\bf k}_1}a_{{\bf k}_1}\right)\right.\nonumber\\
&&\left.~~~~~~~~~~~~~~~~~~~~~~~~~~\exp\left(-\beta\int d^3{\bf k}~\left(a^{\dagger}_{\bf k}a_{\bf k}+\frac{1}{2}\delta^{3}(0)\right)E_{\bf k}(\tau_1)\right)\right.\nonumber\\
&&\left.~~~~~~~~~~\int\frac{d^3{\bf k}_3}{(2\pi)^3}\int\frac{d^3{\bf k}_4}{(2\pi)^3}\exp\left(\left({\bf k}_3+{\bf k}_4\right).{\bf x}\right) \left[f_{{\bf k}_3}(\tau_1)f_{{\bf k}_4}(\tau_1)~a_{{\bf k}_3}a_{{\bf k}_4}+f^{*}_{{\bf -k}_3}(\tau_1)f_{{\bf k}_4}(\tau_1)~a^{\dagger}_{-{\bf k}_3}a_{{\bf k}_4}\right.\right.\nonumber\\
&&\left.\left.~~~~~~~~~~~~~~~~~~~~~~~~~~~~~~~~~~~~~~~~~~~~~~+f_{{\bf k}_3}(\tau_1)f^{*}_{-{\bf k}_4}(\tau_1)~a_{{\bf k}_3}a^{\dagger}_{-{\bf k}_4}+f^{*}_{-{\bf k}_3}(\tau_1)f^{*}_{-{\bf k}_4}(\tau_1)~a^{\dagger}_{-{\bf k}_3}a^{\dagger}_{-{\bf k}_4}\right]\right.\nonumber\\
&&\left.~~~~~~~~~~~~~~~~~~~~~~~~~~~~~~~~~~~~~~~~~~~~~~~~~~~~~\exp\left(-\frac{i}{2}\exp(-i\gamma)\tanh \alpha~\int \frac{d^3{\bf k}_2}{(2\pi)^3}~a^{\dagger}_{{\bf k}_2}a^{\dagger}_{{\bf k}_2}\right)\right\}|\Psi_{\bf BD}\rangle.~~~~~~~~~~\eea
Now we will explicitly compute the individual contributions,  which are given by: 
\bea &&\exp\left(-\frac{i}{2}\exp(-i\gamma)\tanh \alpha~\int \frac{d^3{\bf k}_2}{(2\pi)^3}~a^{\dagger}_{{\bf k}_2}a^{\dagger}_{{\bf k}_2}\right)|\Psi_{\bf BD}\rangle \nonumber\\
&&=\sum^{\infty}_{n=0}\frac{(-1)^n}{n!}\left(\frac{i}{2}\exp(-i\gamma)\tanh \alpha~\right)^n \left(\int \frac{d^3{\bf k}_2}{(2\pi)^3}~a^{\dagger}_{{\bf k}_2}a^{\dagger}_{{\bf k}_2}\right)^n|\Psi_{\bf BD}\rangle=\exp\left(-\frac{i}{2}\exp(-i\gamma)\tanh \alpha~\right)|\Psi_{\bf BD}\rangle.,~~~~~~\\
 &&\langle \Psi_{\bf BD}|\exp\left(\frac{i}{2}\exp(i\gamma)\tanh \alpha~\int \frac{d^3{\bf k}_2}{(2\pi)^3}~a_{{\bf k}_2}a_{{\bf k}_2}\right)\nonumber\\
 &&=\left[\exp\left(-\frac{i}{2}\exp(-i\gamma)\tanh \alpha~\int \frac{d^3{\bf k}_2}{(2\pi)^3}~a^{\dagger}_{{\bf k}_2}a^{\dagger}_{{\bf k}_2}\right)|\Psi_{\bf BD}\rangle\right]^{\dagger}= \langle \Psi_{\bf BD}|\exp\left(\frac{i}{2}\exp(i\gamma)\tanh \alpha\right),\eea
and also we have used the following sets of useful results:
\bea
&&\int d\Psi_{\bf BD}~\langle \Psi_{\bf BD}|\exp\left(-\beta\int d^3{\bf k}~\left(a^{\dagger}_{\bf k}a_{\bf k}+\frac{1}{2}\delta^{3}(0)\right)E_{\bf k}(\tau_1)\right)a_{{\bf k}_3}a_{{\bf k}_4}|\Psi_{\bf BD}\rangle=0,\eea
\bea
&&\int d\Psi_{\bf BD}~\langle \Psi_{\bf BD}|\exp\left(-\beta\int d^3{\bf k}~\left(a^{\dagger}_{\bf k}a_{\bf k}+\frac{1}{2}\delta^{3}(0)\right)E_{\bf k}(\tau_1)\right)a^{\dagger}_{-{\bf k}_3}a_{{\bf k}_4}|\Psi_{\bf BD}\rangle\nonumber\\
&&~~~~~~~~~=(2\pi)^3\exp\left(-\left(1+\frac{1}{2}\delta^{3}(0)\right)\int d^3{\bf k}~\ln\left(2\sinh\frac{\beta E_{\bf k}(\tau_1)}{2}\right)\right)~\delta^{3}({\bf k}_3+{\bf k}_4),~~~~~~\\
&&\int d\Psi_{\bf BD}~\langle \Psi_{\bf BD}|\exp\left(-\beta\int d^3{\bf k}~\left(a^{\dagger}_{\bf k}a_{\bf k}+\frac{1}{2}\delta^{3}(0)\right)E_{\bf k}(\tau_1)\right)~a_{{\bf k}_3}a^{\dagger}_{-{\bf k}_4}|\Psi_{\bf BD}\rangle\nonumber\\
&&~~~~~~~~~=(2\pi)^3\exp\left(-\left(1+\frac{1}{2}\delta^{3}(0)\right)\int d^3{\bf k}~\ln\left(2\sinh\frac{\beta E_{\bf k}(\tau_1)}{2}\right)\right)~\delta^{3}({\bf k}_3+{\bf k}_4),~~~~~~\\
&&\int d\Psi_{\bf BD}~\langle \Psi_{\bf BD}|\exp\left(-\beta\int d^3{\bf k}~\left(a^{\dagger}_{\bf k}a_{\bf k}+\frac{1}{2}\delta^{3}(0)\right)E_{\bf k}(\tau_1)\right)~a^{\dagger}_{-{\bf k}_3}a^{\dagger}_{-{\bf k}_4}|\Psi_{\bf BD}\rangle=0.~~~~.\eea
Consequently,  we can simplify the final result of the previously mentioned trace as given by the following expressions:
 \bea &&{\rm Tr}\left[e^{-\beta \hat{H}(\tau_1)}\hat{f}({\bf x},\tau_1)\hat{f}({\bf x},\tau_1)\right]_{(\alpha,\gamma)}\nonumber\\ 
&&=Z_{\alpha,\gamma}(\beta;\tau_1)\int\frac{d^3{\bf k}_3}{(2\pi)^3}\int\frac{d^3{\bf k}_4}{(2\pi)^3}(2\pi)^3\delta^{3}({\bf k}_3+{\bf k}_4)\left[f_{{\bf k}_3}(\tau_1)f^{*}_{-{\bf k}_4}(\tau_1)+f^{*}_{-{\bf k}_3}(\tau_1)f_{{\bf k}_4}(\tau_1)\right]\nonumber\\
&&=\frac{Z_{\alpha,\gamma}(\beta;\tau_1)}{\pi^2}~{\cal F}^{(\alpha,\gamma)}_1(\tau_1),\\
 &&{\rm Tr}\left[e^{-\beta \hat{H}(\tau_2)}\hat{f}({\bf x},\tau_2)\hat{f}({\bf x},\tau_2)\right]_{(\alpha,\gamma)}\nonumber\\ 
&&=Z_{\alpha,\gamma}(\beta;\tau_2)\int\frac{d^3{\bf k}_3}{(2\pi)^3}\int\frac{d^3{\bf k}_4}{(2\pi)^3}(2\pi)^3\delta^{3}({\bf k}_3+{\bf k}_4)\left[f_{{\bf k}_3}(\tau_2)f^{*}_{-{\bf k}_4}(\tau_2)+f^{*}_{-{\bf k}_3}(\tau_2)f_{{\bf k}_4}(\tau_2)\right]\nonumber\\
&&=\frac{Z_{\alpha,\gamma}(\beta;\tau_2)}{\pi^2}~{\cal F}^{(\alpha,\gamma)}_1(\tau_2),\eea 
where we define a regularised time dependent functions ${\cal F}^{(\alpha,\gamma)}_1(\tau_1)$ and ${\cal F}^{(\alpha,\gamma)}_1(\tau_2)$ as:
\bea &&{\cal F}^{(\alpha,\gamma)}_1(\tau_1):=\int^{L}_{0}dk_3~k^2_3~|f_{{\bf k}_3}(\tau_1)|^2,\nonumber\\
&&~=\frac{iAB}{32\tau^3_1} \left(\frac{2}{L}\right)^{2 \nu } \left[\frac{32 (A^2+B^2)}{iAB} \tau^3_1L^3\left(\frac{L^2 \tau^2_1}{5-2 \nu }+\frac{1}{3-2 \nu }\right)+\frac{(2 \nu -7) \Gamma (5-2 \nu ) \left((-i L \tau_1)^{2 \nu }-(i L \tau_1)^{2 \nu }\right)}{(2 \nu -3)}\right.\nonumber\\&& \left.~~~~~~~~~~~~~~~~~~+(i L \tau_1)^{2 \nu } \left\{ 4\Gamma (3-2 \nu ,2 i L \tau_1)+\Gamma (5-2 \nu ,2 i L \tau_1)+4 \Gamma (4-2 \nu ,2 i L \tau_1)\right\}\right.\nonumber\\&& \left.~~~~~~~~~~~~~~~-(-i L \tau_1)^{2 \nu } \left\{  4\Gamma (3-2 \nu ,2 i L \tau_1)+\Gamma (5-2 \nu ,-2 i L \tau_1)+4\Gamma (4-2 \nu ,-2 i L \tau_1)\right\}\right],~~~~~~~~~~~~\eea
Here the constants $A$ and $B$ are in general dependent on the mass parameter $\nu$ and the vacuum parameters $\alpha$ and $\gamma$ for the Mota Allen vacua.  Replacing $\tau_1$ with $\tau_2$ one can write down the expression for ${\cal F}^{(\alpha,\gamma)}_1(\tau_2)$.

Similarly,  following the same logical arguments one can show that:
\bea &&{\rm Tr}\left[e^{-\beta \hat{H}(\tau_1)}\hat{\Pi}({\bf x},\tau_1)\hat{\Pi}({\bf x},\tau_1)\right]_{(\alpha,\gamma)}\nonumber\\ 
&&=Z_{\alpha,\gamma}(\beta;\tau_1)\int\frac{d^3{\bf k}_3}{(2\pi)^3}\int\frac{d^3{\bf k}_4}{(2\pi)^3}(2\pi)^3\delta^{3}({\bf k}_3+{\bf k}_4)\left[\Pi_{{\bf k}_3}(\tau_1)\Pi^{*}_{-{\bf k}_4}(\tau_1)+\Pi^{*}_{-{\bf k}_3}(\tau_1)\Pi_{{\bf k}_4}(\tau_1)\right]\nonumber\\
&&=\frac{Z_{\alpha,\gamma}(\beta;\tau_1)}{\pi^2}~{\cal F}^{(\alpha,\gamma)}_2(\tau_2),\\ &&{\rm Tr}\left[e^{-\beta \hat{H}(\tau_2)}\hat{\Pi}({\bf x},\tau_2)\hat{\Pi}({\bf x},\tau_2)\right]_{(\alpha,\gamma)}\nonumber\\ 
&&=Z_{\alpha}(\beta;\tau_2)\int\frac{d^3{\bf k}_3}{(2\pi)^3}\int\frac{d^3{\bf k}_4}{(2\pi)^3}(2\pi)^3\delta^{3}({\bf k}_3+{\bf k}_4)\left[\Pi_{{\bf k}_3}(\tau_2)\Pi^{*}_{-{\bf k}_4}(\tau_2)+\Pi^{*}_{-{\bf k}_3}(\tau_2)\Pi_{{\bf k}_4}(\tau_2)\right]\nonumber\\
&&=\frac{Z_{\alpha}(\beta;\tau_2)}{\pi^2}~{\cal F}^{(\alpha)}_2(\tau_2),\eea 
where we define a regularised time dependent function ${\cal F}^{(\alpha,\gamma)}_2(\tau_1)$ and ${\cal F}^{(\alpha,\gamma)}_2(\tau_2)$ as:
\bea &&{\cal F}^{(\alpha,\gamma)}_2(\tau_1):=\int^{L}_{0}dk_3~k^2_3~|\Pi_{{\bf k}_3}(\tau_1)|^2,\nonumber\\
&&=\frac{1}{4 \tau ^4_1}L^{-2 \nu } \left[-i 4^{\nu +1} A B \tau ^3_1 \Gamma (-2 \nu ,-2 i L \tau_1 ) (-i L \tau_1 )^{2 \nu }+2^{2 \nu +3} A B i \nu  \tau ^3_1 \Gamma (-2 \nu ,-2 i L \tau_1 ) (-i L \tau_1 )^{2 \nu }\right.\nonumber\\ &&\left.+2^{2 \nu +5} A B i \nu ^2 \tau ^3_1 \Gamma (-2 (\nu +1),-2 i L \tau_1 ) (-i L \tau_1 )^{2 \nu }+2^{2 \nu +3} A B i \tau ^3_2 \Gamma (-2 (\nu +1),-2 i L \tau_1 ) (-i L \tau_1 )^{2 \nu }\right.\nonumber\\ &&\left.-i 2^{2 \nu +5} A B \nu  \tau ^3_2 \Gamma (-2 (\nu +1),-2 i L \tau_1 ) (-i L \tau_1 )^{2 \nu }+2^{2 \nu +3} A B i (1-2 \nu )^2 \tau ^3_1 \Gamma (-2 \nu -3,-2 i L \tau_1 ) (-i L \tau_1 )^{2 \nu }\right.\nonumber\\ &&\left.+2^{2 \nu +3} A B i \nu ^2 \tau ^3_1 \Gamma (-2 \nu -1,-2 i L \tau_1 ) (-i L \tau_1 )^{2 \nu }-3 i 2^{2 \nu +1} A B \tau ^3_1 \Gamma (-2 \nu -1,-2 i L \tau_1 ) (-i L \tau_1 )^{2 \nu }\right.\nonumber\\ &&\left.+2^{2 \nu +3} A B i \nu  \tau ^3_1 \Gamma (-2 \nu -1,-2 i L \tau_1 ) (-i L \tau_1 )^{2 \nu }+2^{2 \nu +1} A B i \tau ^3_1 \Gamma (1-2 \nu ,-2 i L \tau_1 ) (-i L \tau_1 )^{2 \nu }\right.\nonumber\\ &&\left.+\frac{4 (A^2+B^2) L \tau ^4_1}{1-2 \nu }-4 i B^2 \tau ^3_1+\frac{2 B^2 i \tau ^3_1}{\nu }\right]\nonumber\\
&&+\frac{1}{4 \tau ^4_1}L^{-2 \nu }\left[\frac{\tau ^2_1[4\nu^2(2A^2+B^2)+B^2(4\nu-3)-5A^2]}{L(2 \nu+1)}+\frac{B^2 i \tau_2[4\nu(\nu-1)+1]}{L^2 (\nu +1)}\right.\nonumber\\ &&\left.+4^{\nu +1} A B i \tau ^3_1 \left(7 (-i L \tau_1 )^{2 \nu }-13 (i L \tau_1 )^{2 \nu }+4 \nu ^3 \left((-i L \tau_1 )^{2 \nu }+7 (i L \tau_1 )^{2 \nu }\right)\right.\right.\nonumber\\ &&\left.\left.+\nu  \left(9 (i L \tau_1 )^{2 \nu }-25 (-i L \tau_1 )^{2 \nu }\right)+\nu ^2 \left(68 (i L \tau_1 )^{2 \nu }-28 (-i L \tau_1 )^{2 \nu }\right)\right) \Gamma (-2 \nu -3)\right.\nonumber\\ &&\left.-i 2^{2 \nu +5} A B \nu ^2 \tau ^3_1 (i L \tau_2 )^{2 \nu } \Gamma (-2 \nu -3,2 i L \tau_2 )-i 2^{2 \nu +3} A B \tau ^3_2 (i L \tau_2 )^{2 \nu } \Gamma (-2 \nu -3,2 i L \tau_1 )\right.\nonumber\\ &&\left.+2^{2 \nu +5} A B i \nu  \tau ^3_1 (i L \tau_1 )^{2 \nu } \Gamma (-2 \nu -3,2 i L \tau_1 )+2^{2 \nu +3} A B i \nu ^2 \tau ^3_1 (i L \tau_1 )^{2 \nu } \Gamma (-2 \nu -1,2 i L \tau_1 )\right.\nonumber\\ &&\left.+5\ 2^{2 \nu +1} A B i \tau ^3_1 (i L \tau_1 )^{2 \nu } \Gamma (-2 \nu -1,2 i L \tau_1 )-3 i 2^{2 \nu +3} A B \nu  \tau ^3_2 (i L \tau_1 )^{2 \nu } \Gamma (-2 \nu -1,2 i L \tau_1 )\right.\nonumber\\ &&\left.-i 2^{2 \nu +1} A B \tau ^3_1 (i L \tau_2 )^{2 \nu } \Gamma (1-2 \nu ,2 i L \tau_1 )+\frac{4 (A^2+B^2) [\nu(1-\nu)-1] }{L^3 (2 \nu +3)}\right],\eea
Replacing $\tau_1$ with $\tau_2$ one can write down the expression for ${\cal F}^{(\alpha,\gamma)}_2(\tau_2)$.

Then we have found the following expression:
  \bea &&{\displaystyle \langle \hat{f}(\tau_1)\hat{f}(\tau_1)\rangle_{\beta} =\displaystyle \frac{1}{Z_{\alpha,\gamma}(\beta;\tau_1)}{\rm Tr}\left[e^{-\beta \hat{H}(\tau_1)}\hat{f}({\bf x},\tau_1)\hat{f}({\bf x},\tau_1)\right]_{(\alpha)}=\frac{1}{\pi^2}{\cal F}^{(\alpha,\gamma)}_1(\tau_1)},\eea\bea
&&{\displaystyle \langle \hat{f}(\tau_2)\hat{f}(\tau_2)\rangle_{\beta} =\displaystyle \frac{1}{Z_{\alpha,\gamma}(\beta;\tau_2)}{\rm Tr}\left[e^{-\beta \hat{H}(\tau_2)}\hat{f}({\bf x},\tau_2)\hat{f}({\bf x},\tau_2)\right]_{(\alpha)}=\frac{1}{\pi^2}{\cal F}^{(\alpha,\gamma)}_1(\tau_2)},\\
 &&{\displaystyle \langle \hat{\Pi}(\tau_1)\hat{\Pi}(\tau_1)\rangle_{\beta} =\displaystyle \frac{1}{Z_{\alpha,\gamma}(\beta;\tau_1)}{\rm Tr}\left[e^{-\beta \hat{H}(\tau_1)}\hat{\Pi}({\bf x},\tau_1)\hat{\Pi}({\bf x},\tau_1)\right]_{(\alpha,\gamma)}=\frac{1}{\pi^2}{\cal F}^{(\alpha,\gamma)}_2(\tau_1)},\\
 &&{\displaystyle \langle \hat{\Pi}(\tau_2)\hat{\Pi}(\tau_2)\rangle_{\beta} =\displaystyle \frac{1}{Z_{\alpha,\gamma}(\beta;\tau_2)}{\rm Tr}\left[e^{-\beta \hat{H}(\tau_2)}\hat{\Pi}({\bf x},\tau_2)\hat{\Pi}({\bf x},\tau_2)\right]_{(\alpha,\gamma)}=\frac{1}{\pi^2}{\cal F}^{(\alpha,\gamma)}_2(\tau_2)} .\eea
Consequently,  the normalisation factors of the previously defined two types of auto correlated OTO functions for the rescaled field variable and its canonically conjugate momenta operators can be computed as:
\bea {{\cal N}^{f}_1(\tau_1,\tau_2)=\frac{1}{\langle \hat{f}(\tau_1)\hat{f}(\tau_1)\rangle_{\beta} \langle \hat{f}(\tau_2)\hat{f}(\tau_2)\rangle_{\beta}}=\frac{\pi^4}{{\cal F}^{(\alpha,\gamma)}_1(\tau_1){\cal F}^{(\alpha)}_1(\tau_2)}},\\
{{\cal N}^{f}_2(\tau_1,\tau_2)=\frac{1}{\langle \hat{\Pi}(\tau_1)\hat{\Pi}(\tau_1)\rangle_{\beta} \langle \hat{\Pi}(\tau_2)\hat{\Pi}(\tau_2)\rangle_{\beta}}=\frac{\pi^4}{{\cal F}^{(\alpha,\gamma)}_2(\tau_1){\cal F}^{(\alpha,\gamma)}_2(\tau_2)}}.\eea
For further computation we frequently drop the exponent $(\alpha,\gamma)$ as appearing in the normalization factors.  So in this simple notation one can write:
\bea {{\cal N}^{f}_1(\tau_1,\tau_2)=\frac{\pi^4}{{\cal F}_1(\tau_1){\cal F}_1(\tau_2)}},\\
{{\cal N}^{f}_2(\tau_1,\tau_2)=\frac{\pi^4}{{\cal F}_2(\tau_1){\cal F}_2(\tau_2)}}.\eea 
 \subsection{Normalization factor of four-point  OTOC computed from curvature perturbation field variable}
 Now,  we are going to perform the similar computations when we express the normalisation factors of the two types of the desired OTOCs written in terms of the scalar curvature perturbation field variable and its canonical conjugate momenta:
  \bea {{\cal N}^{\zeta}_1(\tau_1,\tau_2):=\frac{1}{\langle \hat{\zeta}(\tau_1)\hat{\zeta}(\tau_1)\rangle_{\beta} \langle \hat{\zeta}(\tau_2)\hat{\zeta}(\tau_2)\rangle_{\beta}}},\\
  {{\cal N}^{\zeta}_2(\tau_1,\tau_2):=\frac{1}{\langle  \hat{\Pi}_{\zeta}(\tau_1) \hat{\Pi}_{\zeta}(\tau_1)\rangle_{\beta} \langle \hat{\Pi}_{\zeta}(\tau_2)\hat{\Pi}_{\zeta}(\tau_2)\rangle_{\beta}}}.
  \eea
 for this we need to explicitly evaluate the denominator of the above mentioned expression.
 
 Now, the product of the two thermal two point function written in terms of curvature perturbation and its canonically conjugate momenta are evaluated for the generalized Mota Allen quantum vacua as:
 \bea {\displaystyle \langle \hat{\zeta}(\tau_1)\hat{\zeta}(\tau_1)\rangle_{\beta} =\displaystyle \frac{1}{Z_{\alpha,\gamma}(\beta;\tau_1)}{\rm Tr}\left[e^{-\beta \hat{H}(\tau_1)}\hat{\zeta}({\bf x},\tau_1)\hat{\zeta}({\bf x},\tau_1)\right]_{(\alpha,\gamma)}},\\
{\displaystyle \langle  \hat{\Pi}_{\zeta}((\tau_2) \hat{\Pi}_{\zeta}((\tau_1)\rangle_{\beta} =\displaystyle \frac{1}{Z_{\alpha,\gamma}(\beta;\tau_2)}{\rm Tr}\left[e^{-\beta \hat{H}(\tau_2)} \hat{\Pi}_{\zeta}({\bf x},\tau_2) \hat{\Pi}_{\zeta}({\bf x},\tau_2)\right]_{(\alpha,\gamma)}},\eea
  where the thermal partition function for primordial cosmology in terms of curvature perturbation field variable is computed for generalised Mota Allen vacua can be expressed as:
 \bea {Z^{\zeta}_{\alpha,\gamma}(\beta;\tau_i)=\frac{\exp(-2\sin\gamma{\rm tan}\alpha)~Z^{\zeta}_{\bf BD}(\beta;\tau_i)}{|\cosh\alpha|}~~~~~~~\forall~~i=1,2}.~~~~~\eea
 Here the thermal partition function for primordial cosmology in terms of curvature perturbation field variable computed for the Bunch Davies quantum vacuum as:
 \bea {Z^{\zeta}_{\bf BD}(\beta;\tau_i)\approx\exp\left(-\left(1+\frac{1}{2}\delta^{3}(0)\right)\int d^3{\bf k}~\ln\left(2\sinh\frac{\beta z^2(\tau_i)E_{{\bf k},\zeta}(\tau_i)}{2}\right)\right)~\forall~~i=1,2.}~~~~~~~~~~\eea
Here we define the conformal time dependent energy dispersion relation in terms of the curvature perturbation field variable and its canonically conjugate momenta as:
\bea E_{{\bf k},\zeta}(\tau_i):&=&\left|\Pi^{\zeta}_{\bf k}(\tau_i)\right|^2+\left(k^2-\frac{1}{z(\tau_i)}\frac{d^2z(\tau_i)}{d\tau^2_i}+\left(\frac{1}{z(\tau_i)}\frac{dz(\tau_i)}{d\tau_i}\right)^2\right)|\zeta_{\bf k}(\tau_i)|^2~\forall~~i=1,2~~~~~~~~~~\eea
Consequently, we can simplify the final result of the previously mentioned trace in terms of the curvature perturbation and its canonically conjugate momenta as given by the following expression:
\bea {\rm Tr}\left[e^{-\beta \hat{H}(\tau_1)}\hat{\zeta}({\bf x},\tau_1)\hat{\zeta}({\bf x},\tau_1)\right]_{(\alpha,\gamma)}&=&\frac{Z^{\zeta}_{\alpha,\gamma}(\beta;\tau_1)}{\pi^2z^2(\tau_1)}~{\cal F}^{(\alpha,\gamma)}_1(\tau_1),\\
{\rm Tr}\left[e^{-\beta \hat{H}(\tau_2)}\hat{\zeta}({\bf x},\tau_2)\hat{\zeta}({\bf x},\tau_2)\right]_{(\alpha,\gamma)}&=&\frac{Z^{\zeta}_{\alpha,\gamma}(\beta;\tau_2)}{\pi^2z^2(\tau_2)}~{\cal F}^{(\alpha,\gamma)}_1(\tau_2),\\
{\rm Tr}\left[e^{-\beta \hat{H}(\tau_1)}\hat{\Pi}_{\zeta}({\bf x},\tau_1)\hat{\Pi}_{\zeta}({\bf x},\tau_1)\right]_{(\alpha,\gamma)}&=&\frac{Z^{\zeta}_{\alpha}(\beta;\tau_2)}{\pi^2z^2(\tau_2)}~{\cal F}^{(\alpha,\gamma)}_2(\tau_1),
\\
{\rm Tr}\left[e^{-\beta \hat{H}(\tau_2)}\hat{\Pi}_{\zeta}({\bf x},\tau_2)\hat{\Pi}_{\zeta}({\bf x},\tau_2)\right]_{(\alpha)}&=&\frac{Z^{\zeta}_{\alpha}(\beta;\tau_2)}{\pi^2z^2(\tau_2)}~{\cal F}^{(\alpha,\gamma)}_2(\tau_2).\eea 
Then we have found the following expressions for the thermal two point functions:
  \bea &&{\displaystyle \langle \hat{\zeta}(\tau_1)\hat{\zeta}(\tau_1)\rangle_{\beta} =\displaystyle \frac{1}{Z^{\zeta}_{\alpha}(\beta;\tau_1)}{\rm Tr}\left[e^{-\beta \hat{H}(\tau_1)}\hat{\zeta}({\bf x},\tau_1)\hat{\zeta}({\bf x},\tau_1)\right]_{(\alpha)}=\frac{1}{\pi^2z^2(\tau_1)}{\cal F}^{(\alpha)}_1(\tau_1)},\\
  &&{\displaystyle \langle \hat{\zeta}(\tau_1)\hat{\zeta}(\tau_1)\rangle_{\beta} =\displaystyle \frac{1}{Z^{\zeta}_{\alpha}(\beta;\tau_1)}{\rm Tr}\left[e^{-\beta \hat{H}(\tau_1)}\hat{\zeta}({\bf x},\tau_1)\hat{\zeta}({\bf x},\tau_1)\right]_{(\alpha)}=\frac{1}{\pi^2z^2(\tau_1)}{\cal F}^{(\alpha)}_1(\tau_1)},\\
 &&{\displaystyle \langle \hat{\Pi}_{\zeta}(\tau_1)\hat{\Pi}_{\zeta}(\tau_1)\rangle_{\beta} =\displaystyle \frac{1}{Z^{\zeta}_{\alpha,\gamma}(\beta;\tau_2)}{\rm Tr}\left[e^{-\beta \hat{H}(\tau_1)}\hat{\Pi}_{\zeta}({\bf x},\tau_1)\hat{\Pi}_{\zeta}({\bf x},\tau_1)\right]_{(\alpha,\gamma)}=\frac{1}{\pi^2z^2(\tau_1)}{\cal F}^{(\alpha,\gamma)}_2(\tau_1)}~,~~~~~~~~~~\\
 &&{\displaystyle \langle \hat{\Pi}_{\zeta}(\tau_2)\hat{\Pi}_{\zeta}(\tau_2)\rangle_{\beta} =\displaystyle \frac{1}{Z^{\zeta}_{\alpha,\gamma}(\beta;\tau_2)}{\rm Tr}\left[e^{-\beta \hat{H}(\tau_2)}\hat{\Pi}_{\zeta}({\bf x},\tau_2)\hat{\Pi}_{\zeta}({\bf x},\tau_2)\right]_{(\alpha,\gamma)}=\frac{1}{\pi^2z^2(\tau_2)}{\cal F}^{(\alpha,\gamma)}_2(\tau_2)}~.~~~~~~~~~~\eea
 This further implies that the connecting relation between the two point thermal correlation functions computed from the rescaled field variable and curvature perturbation variable and their conjugate momenta are given by:
   \bea &&{\displaystyle \langle \hat{f}(\tau_1)\hat{f}(\tau_1)\rangle_{\beta}=z^2(\tau_1)\langle \hat{\zeta}(\tau_1)\hat{\zeta}(\tau_1)\rangle_{\beta}},\\
   &&{\displaystyle \langle \hat{f}(\tau_2)\hat{f}(\tau_2)\rangle_{\beta}=z^2(\tau_2)\langle \hat{\zeta}(\tau_2)\hat{\zeta}(\tau_2)\rangle_{\beta}},\\
&& {\displaystyle \langle \hat{\Pi}(\tau_1)\hat{\Pi}(\tau_1)\rangle_{\beta}=z^2(\tau_1)\langle \hat{\Pi}_{\zeta}(\tau_1)\hat{\Pi}_{\zeta}(\tau_1)\rangle_{\beta}},~~~~~~~\\
&& {\displaystyle \langle \hat{\Pi}(\tau_2)\hat{\Pi}(\tau_2)\rangle_{\beta}=z^2(\tau_2)\langle \hat{\Pi}_{\zeta}(\tau_2)\hat{\Pi}_{\zeta}(\tau_2)\rangle_{\beta}}.\eea
   Consequently,  the normalisation factors of the two desired auto-correlated OTOs for the curvature perturbation variable and in terms of the canonically conjugate momenta can be finally computed as:
  \bea &&{ {\cal N}^{\zeta}_1(\tau_1,\tau_2)=\frac{1}{\langle \hat{\zeta}(\tau_1)\hat{\zeta}(\tau_1)\rangle_{\beta} \langle \hat{\zeta}(\tau_2)\hat{\zeta}(\tau_2)\rangle_{\beta}}=\frac{\pi^4z^2(\tau_1)z^2(\tau_2)}{{\cal F}^{(\alpha,\gamma)}_1(\tau_1){\cal F}^{(\alpha,\gamma)}_1(\tau_2)}=z^2(\tau_1)z^2(\tau_2){\cal N}^{f}_1(\tau_1,\tau_2)},~~~~~~~~~~~\\
 && { {\cal N}^{\zeta}_2(\tau_1,\tau_2)=\frac{1}{\langle \hat{\Pi}_{\zeta}(\tau_1)\hat{\Pi}_{\zeta}(\tau_1)\rangle_{\beta} \langle \hat{\Pi}_{\zeta}(\tau_2)\hat{\Pi}_{\zeta}(\tau_2)\rangle_{\beta}}=\frac{\pi^4z^2(\tau_1)z^2(\tau_2)}{{\cal F}^{(\alpha,\gamma)}_2(\tau_1){\cal F}^{(\alpha,\gamma)}_2(\tau_2)}=z^2(\tau_1)z^2(\tau_2){\cal N}^{f}_2(\tau_1,\tau_2)}.~~~~~~~~~~~\eea  

  \section{Computation of the normalization factor in classical limit of four-point  OTOC}
 \label{sec:16}
\subsection{Normalization factor of the classical version of four-point  OTOC computed from rescaled field variable} 
Further, our aim is to compute the normalisation factor of the classical version of OTOC computed from the rescaled field variable $f$, which is given by the following expression: 
\bea {{\cal N}^{f}_{1,{\bf Classical}}(\tau_1,\tau_2):=\frac{1}{\langle f(\tau_1)f(\tau_1)\rangle_{\beta} \langle{\Pi}(\tau_2{\Pi}(\tau_2)\rangle_{\beta}}},\\
{{\cal N}^{f}_{1,{\bf Classical}}(\tau_1,\tau_2):=\frac{1}{\langle f(\tau_1)f(\tau_1)\rangle_{\beta} \langle{\Pi}(\tau_2{\Pi}(\tau_2)\rangle_{\beta}}},
\eea
 for this we need to explicitly evaluate the denominators of the above mentioned expressions.  In the classical limiting case there is no physical notion of vacuum state exist using which we can take the thermal average over a statistical ensemble.  To avoid such confusion in the classical case the commutator brackets are replaced by the usual Poisson brackets and the thermal tracing operation to find the OTOC will be replaced by the phase space measure $\displaystyle \frac{{\cal D}f{\cal D}\Pi}{2\pi}$ over which we have to perform the average at the end.
 
 Now,  the two thermal two point functions in the classical limiting case are evaluated in terms of the phase space averaged Poisson Brackets as:
 \bea \displaystyle {\langle{f}(\tau_1){f}(\tau_1)\rangle_{\beta} =\displaystyle \frac{1}{Z_{\bf Classical}(\beta;\tau_1)}\int\int\frac{{\cal D}f{\cal D}\Pi}{2\pi}e^{-\beta {H}(\tau_1)}\left\{{f}({\bf x},\tau_1),{f}({\bf x},\tau_1)\right\}_{\bf PB}},\\
 \displaystyle {\langle{f}(\tau_2){f}(\tau_2)\rangle_{\beta} =\displaystyle \frac{1}{Z_{\bf Classical}(\beta;\tau_1)}\int\int\frac{{\cal D}f{\cal D}\Pi}{2\pi}e^{-\beta {H}(\tau_2)}\left\{{f}({\bf x},\tau_2),{f}({\bf x},\tau_2)\right\}_{\bf PB}},\eea
 \bea
 {\displaystyle \langle {\Pi}(\tau_1){\Pi}(\tau_1)\rangle_{\beta} =\displaystyle \frac{1}{Z_{\bf Classical}(\beta;\tau_1)}\int\int\frac{{\cal D}f{\cal D}\Pi}{2\pi}e^{-\beta {H}(\tau_1)}\left\{{\Pi}({\bf x},\tau_1),{\Pi}({\bf x},\tau_1)\right\}_{\bf PB}},\\
  {\displaystyle \langle {\Pi}(\tau_2){\Pi}(\tau_2)\rangle_{\beta} =\displaystyle \frac{1}{Z_{\bf Classical}(\beta;\tau_2)}\int\int\frac{{\cal D}f{\cal D}\Pi}{2\pi}e^{-\beta {H}(\tau_2)}\left\{{\Pi}({\bf x},\tau_2),{\Pi}({\bf x},\tau_2)\right\}_{\bf PB}}.\eea
 where the thermal partition function for cosmology in the classical limit is computed as:
 \bea {Z_{\bf Classical}(\beta;\tau_i)=\exp\left(-\int d^3{\bf k}~\ln\left(2\sinh\frac{\beta E_{\bf k}(\tau_i)}{2}\right)\right)~\forall~~i=1,2}.~~~~~\eea
 Now we compute the Poission brackets as:
\bea \left\{{f}({\bf x},\tau_1),{f}({\bf x},\tau_1)\right\}_{\bf PB}&=&\int\frac{d^3{\bf k}_1}{(2\pi)^3}\int\frac{d^3{\bf k}_2}{(2\pi)^3}~\exp(i({\bf k}_1+{\bf k}_2).{\bf x})\left\{f_{{\bf k}_1}(\tau_1),f_{{\bf k}_2}(\tau_1)\right\}_{\bf PB}\nonumber\\
&=&
\frac{{\bf W}_1(0)}{2\pi^2}\int^{L}_{k_1=0} k^2_1dk_1=\frac{L^3}{6\pi^2}{\bf W}_1(0),\\
 \left\{{f}({\bf x},\tau_2),{f}({\bf x},\tau_2)\right\}_{\bf PB}&=&\int\frac{d^3{\bf k}_1}{(2\pi)^3}\int\frac{d^3{\bf k}_2}{(2\pi)^3}~\exp(i({\bf k}_1+{\bf k}_2).{\bf x})\left\{f_{{\bf k}_1}(\tau_2),f_{{\bf k}_2}(\tau_2)\right\}_{\bf PB}\nonumber\\
&=&
\frac{{\bf W}_1(0)}{2\pi^2}\int^{L}_{k_1=0} k^2_1dk_1=\frac{L^3}{6\pi^2}{\bf W}_1(0),\\
 \left\{{\Pi}({\bf x},\tau_1),{\Pi}({\bf x},\tau_1)\right\}_{\bf PB}&=&\int\frac{d^3{\bf k}_1}{(2\pi)^3}\int\frac{d^3{\bf k}_2}{(2\pi)^3}~\exp(i({\bf k}_1+{\bf k}_2).{\bf x})\left\{\Pi_{{\bf k}_1}(\tau_2),\Pi_{{\bf k}_2}(\tau_2)\right\}_{\bf PB}\nonumber\\
&=&
\frac{{\bf W}_2(0)}{2\pi^2}\int^{L}_{k_1=0} k^2_1dk_1=\frac{L^3}{6\pi^2}{\bf W}_2(0),\\
 \left\{{\Pi}({\bf x},\tau_2),{\Pi}({\bf x},\tau_2)\right\}_{\bf PB}&=&\int\frac{d^3{\bf k}_1}{(2\pi)^3}\int\frac{d^3{\bf k}_2}{(2\pi)^3}~\exp(i({\bf k}_1+{\bf k}_2).{\bf x})\left\{\Pi_{{\bf k}_1}(\tau_2),\Pi_{{\bf k}_2}(\tau_2)\right\}_{\bf PB}\nonumber\\
&=&
\frac{{\bf W}_2(0)}{2\pi^2}\int^{L}_{k_1=0} k^2_1dk_1=\frac{L^3}{6\pi^2}{\bf W}_2(0).\eea
Here it is important to note that, 
${\bf W}_1(0)\neq{\bf W}_2(0)$.
Then we have found the following expression:
  \bea &&\displaystyle \langle {f}(\tau_1){f}(\tau_1)\rangle_{\beta} =\displaystyle \frac{L^3}{6\pi^2}{\bf W}_1(0)=\langle {f}(\tau_2){f}(\tau_2)\rangle_{\beta} ,\\
 &&{\displaystyle \langle {\Pi}(\tau_1){\Pi}(\tau_1)\rangle_{\beta} =\displaystyle \frac{L^3}{6\pi^2}{\bf W}_2(0)}=\langle {\Pi}(\tau_2){\Pi}(\tau_2)\rangle_{\beta}.\eea
Consequently,  the normalisation factors of classical limit of the two types of the desired OTOCs for the rescaled field variable and its associated canonical conjugate momenta can be computed as:
\bea {{\cal N}^{f}_{1,{\bf Classical}}(\tau_1,\tau_2)=\frac{36\pi^4}{L^6{\bf W}^2_1(0)}=\frac{36\pi^4}{L^6{\bf G}^{(1)}_{\bf Kernel}(0)}}~,~~~~~~\\
{{\cal N}^{f}_{2,{\bf Classical}}(\tau_1,\tau_2)=\frac{36\pi^4}{L^6{\bf W}^2_2(0)}=\frac{36\pi^4}{L^6{\bf G}^{(2)}_{\bf Kernel}(0)}}~,~~~~~~\eea
which further implies, 
${\cal N}^{f}_{1,{\bf Classical}}(\tau_1,\tau_2)\neq {\cal N}^{f}_{2,{\bf Classical}}(\tau_1,\tau_2)$.
Now,  considering the examples of non-Gaussian coloured noise and Gaussian white noise we get the following simplified result for the normalization factors in the classical limit:
\bea
&& {\cal N}^{f}_{1,{\bf Classical}}(\tau_1,\tau_2)= \large \left\{
     \begin{array}{lr}
   \displaystyle\frac{36\gamma_1 \pi^4}{L^6{\bf A}_1}~, ~~~~~~~~~~~~~&~ \text{\textcolor{red}{\bf Coloured~Noise}}\\  
   \displaystyle   0 & \text{\textcolor{red}{\bf White~Noise}}  \end{array}
   \right.~~~~~~~~~~
\\
&& {\cal N}^{f}_{2,{\bf Classical}}(\tau_1,\tau_2)= \large \left\{
     \begin{array}{lr}
   \displaystyle\frac{36\gamma_2 \pi^4}{L^6{\bf A}_2}~, ~~~~~~~~~~~~~&~ \text{\textcolor{red}{\bf Coloured~Noise}}\\  
   \displaystyle   0 & \text{\textcolor{red}{\bf White~Noise}}  \end{array}
   \right.~~~~~~~~~~
\eea
where for the non-Gaussian coloured noise we have,  
$ \gamma_1\neq \gamma_2,
{\bf A}_1 \neq {\bf A}_2$. 
 \subsection{Normalization factor of the classical version of four-point  OTOC computed from curvature perturbation field variable} 
 Now,  we are going to perform the similar type of computation when we express the normalisation factors of the two types of the desired OTOCs in the classical limit written in terms of the scalar curvature perturbation field variable and its canonically conjugate momentum variable as:
  \bea &&{{\cal N}^{\zeta}_{1,{\bf Classical}}(\tau_1,\tau_2):=\frac{1}{\langle {\zeta}(\tau_1){\zeta}(\tau_1)\rangle_{\beta} \langle {\zeta}(\tau_2){\zeta}(\tau_2)\rangle_{\beta}}},\\
  &&{{\cal N}^{\zeta}_{2,{\bf Classical}}(\tau_1,\tau_2):=\frac{1}{\langle {\Pi}_{\zeta}(\tau_1){\Pi}_{\zeta}(\tau_1)\rangle_{\beta} \langle {\Pi}_{\zeta}(\tau_2){\Pi}_{\zeta}(\tau_2)\rangle_{\beta}}},\eea
 for this we need to explicitly evaluate the denominators of the above mentioned expressions.
 
 Now, the product of the two thermal two point function written in terms of curvature perturbation and its canonically conjugate momenta are evaluated as: 
 \bea {\displaystyle \langle {\zeta}(\tau_1){\zeta}(\tau_1)\rangle_{\beta} =\displaystyle \frac{1}{Z_{\bf Classical}(\beta;\tau_1)}\int\int\frac{{\cal D}f{\cal D}\Pi}{2\pi}e^{-\beta {H}(\tau_1)}\left\{{\zeta}({\bf x},\tau_1),{\zeta}({\bf x},\tau_1)\right\}_{\bf PB}},\\
 {\displaystyle \langle {\zeta}(\tau_2){\zeta}(\tau_2)\rangle_{\beta} =\displaystyle \frac{1}{Z_{\bf Classical}(\beta;\tau_2)}\int\int\frac{{\cal D}f{\cal D}\Pi}{2\pi}e^{-\beta {H}(\tau_2)}\left\{{\zeta}({\bf x},\tau_2),{\zeta}({\bf x},\tau_2)\right\}_{\bf PB}},\\
{\displaystyle \langle \hat{\Pi}(\tau_1)\hat{\Pi}(\tau_1)\rangle_{\beta} =\displaystyle \frac{1}{Z_{\bf Classical}(\beta;\tau_1)}\int\int\frac{{\cal D}f{\cal D}\Pi}{2\pi}e^{-\beta {H}(\tau_2)}\left\{{\Pi}({\bf x},\tau_1),{\Pi}({\bf x},\tau_1)\right\}_{\bf PB}},\\
{\displaystyle \langle \hat{\Pi}(\tau_2)\hat{\Pi}(\tau_2)\rangle_{\beta} =\displaystyle \frac{1}{Z_{\bf Classical}(\beta;\tau_2)}\int\int\frac{{\cal D}f{\cal D}\Pi}{2\pi}e^{-\beta {H}(\tau_2)}\left\{{\Pi}({\bf x},\tau_2),{\Pi}({\bf x},\tau_2)\right\}_{\bf PB}},\eea
  where the thermal partition function for cosmology in the classical limit in terms of curvature perturbation field variable can be computed as:
 \bea {Z^{\zeta}_{\bf Classical}(\beta;\tau_i)=\exp\left(-\int d^3{\bf k}~\ln\left(2\sinh\frac{\beta z^2(\tau_i)E_{{\bf k},\zeta}(\tau_i)}{2}\right)\right)~\forall~~i=1,2,}~~~~~~~~~~\eea
Here we define the time dependent energy dispersion relation in terms of the curvature perturbation field variable and the its canonically conjugate momentum as:
\bea E_{{\bf k},\zeta}(\tau_i):&=&\left|\Pi^{\zeta}_{\bf k}(\tau_i)\right|^2+\left(k^2-\frac{1}{z(\tau_i)}\frac{d^2z(\tau_i)}{d\tau^2_i}+\left(\frac{1}{z(\tau_i)}\frac{dz(\tau_i)}{d\tau_i}\right)^2\right)|\zeta_{\bf k}(\tau_i)|^2~\forall~~i=1,2~~~~~~~~~~\eea
 Now we compute the classical Poission brackets as:
\bea \left\{{\zeta}({\bf x},\tau_1),{\zeta}({\bf x},\tau_1)\right\}_{\bf PB}&=&\int\frac{d^3{\bf k}_1}{(2\pi)^3}\int\frac{d^3{\bf k}_2}{(2\pi)^3}~\exp(i({\bf k}_1+{\bf k}_2).{\bf x})\left\{\zeta_{{\bf k}_1}(\tau_1),\zeta_{{\bf k}_2}(\tau_1)\right\}_{\bf PB},~~~~~~~\\
\left\{{\zeta}({\bf x},\tau_2),{\zeta}({\bf x},\tau_2)\right\}_{\bf PB}&=&\int\frac{d^3{\bf k}_1}{(2\pi)^3}\int\frac{d^3{\bf k}_2}{(2\pi)^3}~\exp(i({\bf k}_1+{\bf k}_2).{\bf x})\left\{\zeta_{{\bf k}_1}(\tau_1),\zeta_{{\bf k}_2}(\tau_1)\right\}_{\bf PB},~~~~~~~~\\
 \left\{{\Pi}_{\zeta}({\bf x},\tau_1),{\Pi}_{\zeta}({\bf x},\tau_1)\right\}_{\bf PB}&=&\int\frac{d^3{\bf k}_1}{(2\pi)^3}\int\frac{d^3{\bf k}_2}{(2\pi)^3}~\exp(i({\bf k}_1+{\bf k}_2).{\bf x})\left\{\Pi_{\zeta;{\bf k}_1}(\tau_2),\Pi_{\zeta;{\bf k}_2}(\tau_2)\right\}_{\bf PB},~~~~~~~~~~\\
 \left\{{\Pi}_{\zeta}({\bf x},\tau_2),{\Pi}_{\zeta}({\bf x},\tau_2)\right\}_{\bf PB}&=&\int\frac{d^3{\bf k}_1}{(2\pi)^3}\int\frac{d^3{\bf k}_2}{(2\pi)^3}~\exp(i({\bf k}_1+{\bf k}_2).{\bf x})\left\{\Pi_{\zeta;{\bf k}_1}(\tau_2),\Pi_{\zeta;{\bf k}_2}(\tau_2)\right\}_{\bf PB}.\eea 
Then we have found the following expressions for the thermal two point auto correlation functions:
  \bea &&{\displaystyle \langle{\zeta}(\tau_1){\zeta}(\tau_1)\rangle_{\beta} =\displaystyle \frac{L^3}{6\pi^2 z^2(\tau_1)}{\bf W}_1(0)},\\
  &&{\displaystyle \langle{\zeta}(\tau_2){\zeta}(\tau_2)\rangle_{\beta} =\displaystyle \frac{L^3}{6\pi^2 z^2(\tau_2)}{\bf W}_1(0)},\\
 &&{\displaystyle \langle {\Pi}_{\zeta}(\tau_1){\Pi}_{\zeta}(\tau_1)\rangle_{\beta} =\displaystyle \frac{L^3}{6\pi^2 z^2(\tau_1)}{\bf W}_2(0)}~,\\
 &&{\displaystyle \langle {\Pi}_{\zeta}(\tau_2){\Pi}_{\zeta}(\tau_2)\rangle_{\beta} =\displaystyle \frac{L^3}{6\pi^2 z^2(\tau_2)}{\bf W}_2(0)}~.~~~~~~~~~~\eea
 This further implies the following connecting relations between the two point thermal auto correlation functions computed from the rescaled field variable and curvature perturbation field variable and their conjugate momenta in the classical limit are given by:
   \bea &&{\displaystyle \langle {f}(\tau_1){f}(\tau_1)\rangle_{\beta}=z^2(\tau_1)\langle{\zeta}(\tau_1){\zeta}(\tau_1)\rangle_{\beta}},\\
   &&{\displaystyle \langle {f}(\tau_2){f}(\tau_2)\rangle_{\beta}=z^2(\tau_2)\langle{\zeta}(\tau_2){\zeta}(\tau_2)\rangle_{\beta}},\\
&& {\displaystyle \langle {\Pi}(\tau_1){\Pi}(\tau_1)\rangle_{\beta}=z^2(\tau_1)\langle {\Pi}_{\zeta}(\tau_1){\Pi}_{\zeta}(\tau_1)\rangle_{\beta}}, \\
&& {\displaystyle \langle {\Pi}(\tau_2){\Pi}(\tau_2)\rangle_{\beta}=z^2(\tau_2)\langle {\Pi}_{\zeta}(\tau_2){\Pi}_{\zeta}(\tau_2)\rangle_{\beta}}.~~~~~~~\eea
   Consequently, the normalisation factors of auto-correlated OTO functions in the classical limit for the curvature perturbation variable and its canonically conjugate momenta can be computed as:
  \bea &&{ {\cal N}^{\zeta}_{1,{\bf Classical}}(\tau_1,\tau_2)=\frac{1}{\langle {\zeta}(\tau_1){\zeta}(\tau_1)\rangle_{\beta} \langle {\zeta}(\tau_2){\zeta}(\tau_2)\rangle_{\beta}}=z^2(\tau_1)z^2(\tau_2){\cal N}^{f}_{\bf Classical}(\tau_1,\tau_2)},~~~~~~~~~~~\\
  &&{ {\cal N}^{\zeta}_{2,{\bf Classical}}(\tau_1,\tau_2)=\frac{1}{\langle {\Pi}_{\zeta}(\tau_1){\Pi}_{\zeta}(\tau_1)\rangle_{\beta} \langle {\Pi}_{\zeta}(\tau_2){\Pi}_{\zeta}(\tau_2)\rangle_{\beta}}=z^2(\tau_1)z^2(\tau_2){\cal N}^{f}_{\bf Classical}(\tau_1,\tau_2)}.~~~~~~~~~~~\eea

\newpage
\phantomsection
\addcontentsline{toc}{section}{References}
\bibliographystyle{utphys}
\bibliography{references_coco}
%\begin{thebibliography}{99} %weiss,sbbook,javid,sin2beta,icts

%\end{thebibliography}

\end{document}